 \documentclass[twocolumn]{aastex63}

\shorttitle{Color Evolution of Classical Novae}
\shortauthors{Hachisu \& Kato}

\begin{document}

\title{The $UBV$ Color Evolution of Classical Novae. IV. 
Time-Stretched $(U-B)_0$-$(M_B-2.5\log f_{\rm s})$ and
$(V-I)_0$-$(M_I-2.5\log f_{\rm s})$ Color-Magnitude Diagrams 
of Novae in Outburst}


\correspondingauthor{Izumi Hachisu}

\author[0000-0002-0884-7404]{Izumi Hachisu}
\affil{Department of Earth Science and Astronomy, 
College of Arts and Sciences, The University of Tokyo,
3-8-1 Komaba, Meguro-ku, Tokyo 153-8902, Japan} 
\email{izumi.hachisu@outlook.jp}


\author[0000-0002-8522-8033]{Mariko Kato}
\affil{Department of Astronomy, Keio University, 
Hiyoshi, Kouhoku-ku, Yokohama 223-8521, Japan} 

%


%
%



\begin{abstract}
Light curves and color evolutions of two classical novae can be largely
overlapped if we properly squeeze or stretch the timescale of a target
nova against that of a template nova by $t'=t/f_{\rm s}$.
Then the brightness of the target nova is related to the brightness of
the template nova by $(M[t])_{\rm template} = (M[t/f_{\rm s}] - 2.5 
\log f_{\rm s})_{\rm target}$, where $M[t]$ is the absolute magnitude and
a function of time $t$, and $f_{\rm s}$ is the ratio of timescales between
the target and template novae.  In the previous papers of this series,
we show that many novae broadly overlap in the time-stretched 
$(B-V)_0$-$(M_V-2.5 \log f_{\rm s})$ color-magnitude diagram.
In the present paper, we propose two other 
$(U-B)_0$-$(M_B-2.5\log f_{\rm s})$ and $(V-I)_0$-$(M_I-2.5\log f_{\rm s})$
diagrams, and show that their tracks overlap for 16 novae
and for 52 novae, respectively.
Here, $(U-B)_0$, $(B-V)_0$, and $(V-I)_0$ are the intrinsic $U-B$,
$B-V$, and $V-I$ colors and not changed by the time-stretch, 
and $M_B$, $M_V$, and $M_I$ are the absolute $B$, $V$, and $I$ magnitudes.
Using these properties, we considerably refine the previous estimates 
of their distance and reddening.  The obtained distances are in 
reasonable agreement with those of {\it Gaia} Data Release 2 catalogue. 
\end{abstract}




\section{Introduction}
\label{introduction}
A nova is a thermonuclear runaway event on a mass-accreting 
white dwarf (WD).  Hydrogen burning explosion releases nuclear energy 
that results in an expansion of a hydrogen-rich envelope to a giant size. 
Strong winds are accelerated deep inside the photosphere owing to
the iron peak of the opacity.  They are called optically thick winds
\citep[e.g.,][]{kat94h}.  The wind mass-loss rate increases as 
the photosphere is expanding.  
The nova brightness also increases,
if we assume that free-free emission dominates the spectra of novae 
\citep[e.g.,][]{enn77, gal76, kra84, nai09},
because the brightness of free-free emission 
depends mainly on the wind mass-loss rate \citep[e.g.,][]{hac06kb, hac15k}.
The nova brightness begins to decline after the maximum wind mass-loss rate
is attained \citep[e.g.,][]{hac17k}.
After a significant part of the hydrogen-rich envelope is blown in the wind,
the winds stop.  The nova enters the supersoft X-ray source
(SSS) phase.  The hydrogen nuclear burning extinguishes 
and the nova ends \citep[e.g.,][]{kat14shn, kat20h}.

\citet{kat94h} calculated optically thick winds to follow nova
evolutions in the decay phase and obtained the photospheric 
radius $R_{\rm ph}$, photospheric temperature $T_{\rm ph}$, 
photospheric luminosity $L_{\rm ph}$, photospheric velocity $v_{\rm ph}$, 
and wind mass-loss rate $\dot M_{\rm wind}$, for hydrogen-rich envelopes 
of various WD masses and chemical compositions.  
Using these physical quantities, \citet{hac06kb} calculated many 
free-free emission model light curves with $F_\nu \propto 
\dot M_{\rm wind}^2 / (v_{\rm ph}^2 R_{\rm ph})$,
where $F_\nu$ is the flux at the frequency $\nu$.
The free-free flux is almost independent of the frequency, so the
light curve shapes are almost the same among broad band light curves
such as $M_B$, $M_V$, and $M_I$, where $M_B$, $M_V$, and $M_I$ are
the absolute magnitudes for $B$, $V$, and $I$ bands, respectively.

The theoretical light curves also show a homologous shape independent of
the WD mass and chemical composition.  They called this property of
nova model light curves ``the universal decline law.''
These properties, i.e., homologous and frequency independent shapes of
light curves, are extremely useful for analysis of novae.
For the light curve shapes, \citet{hac10k} found that the 
time-stretched absolute $V$ brightnesses of the model light curves, 
$M_V-2.5\log f_{\rm s}$,
overlap each other in the $(t/f_{\rm s})$-$(M_V-2.5\log f_{\rm s})$
light curve \citep[see, e.g., Figures 48 and 49 of][]{hac18kb}.  
They explained this property in more detail in Appendix B of
\citet{hac19kb}, that is, if the $V$ light curve of a template nova
overlaps with that of a target nova by squeezing its timescale 
as $t'=t/f_{\rm s}$, we have the relation
\begin{eqnarray}
\left( M_V[t] \right)_{\rm template} 
&=& \left( M'_V[t'] \right)_{\rm target} \cr
&=& \left( M_V[t/f_{\rm s}]-2.5\log f_{\rm s} \right)_{\rm target},
\label{time-stretching_general}
\end{eqnarray}
where $M_V[t]$ is the original absolute $V$ brightness and $M'_V[t']$ 
is the time-stretched brightness after time-stretch of $t'=t/f_{\rm s}$
\citep[see also ][]{hac20skhs}.

The similarity among different broad band light curves such as $M_U$,
$M_B$, $M_V$, and $M_I$ strongly suggests a common path (or track) of
nova outburst evolutions in the color-magnitude diagram.
In the first paper of this series \citep[][hereafter Paper I]{hac14k},
they found that many nova outbursts follow a part of the common path
in the $(B-V)_0$-$(U-B)_0$ color-color diagram.
Here, $(B-V)_0$ and $(U-B)_0$ are the intrinsic $B-V$ and $U-B$ colors,
respectively.  They called this common path ``the nova-giant sequence.''
In the second paper of this series \citep[][hereafter Paper II]{hac16kb},
they studied 40 novae in outburst in the $(B-V)_0$-$M_V$ color-magnitude 
diagram.  They proposed six different types of tracks in the 
$(B-V)_0$-$M_V$ diagram \citep[see Figure 12 of][]{hac16kb}.

In the third paper of this series \citep[][hereafter Paper III]{hac19ka},
they revised Hachisu \& Kato's (2016b) naive method and
studied 20 novae in outburst in the time-stretched color-magnitude diagram,
$(B-V)_0$-$(M_V - 2.5\log f_{\rm s})$.
They found that many novae follow one of the two template paths in the
$(B-V)_0$-$(M_V - 2.5\log f_{\rm s})$ diagram.
\citet{hac19kb} further applied this new method to 32 recent novae.
This paper, the fourth of this series, extends our time-stretched
color-magnitude diagram method to $UBVI$ bands, i.e.,
the $(U-B)_0$-$(M_B-2.5 \log f_{\rm s})$
and $(V-I)_0$-$(M_I-2.5 \log f_{\rm s})$ diagrams.
We try to find similar common tracks even in these time-stretched
color-magnitude diagrams. Moreover, we show that, newly including
the $U$, $B$, and $I$ data in our analysis, we can determine the color excess
$E(B-V)$, distance moduli $(m-M)_U$, $(m-M)_B$, $(m-M)_V$, 
$(m-M)_I$, and timescaling factor $f_{\rm s}$, more precisely.

Our paper is organized as follows.  First we propose
the time-stretched $(U-B)_0$-$(M_B-2.5 \log f_{\rm s})$ diagrams for
16 novae in Section \ref{ub_color_magnitude_diagram} and then the
time-stretched $(V-I)_0$-$(M_I-2.5 \log f_{\rm s})$ diagrams for 
52 novae in Sections \ref{vi_color_magnitude_diagram_lv_vul},
\ref{vi_color_magnitude_diagram_v1500_cyg},
\ref{exceptional_type_novae_vi}, and \ref{other_novae_vi}.  
Our conclusions are given in Section \ref{conclusions}.  
Appendix \ref{light_curves} introduces our time-stretching method
to obtain the distances and extinctions of each nova. Here, we
reanalyzed four novae newly including $I$ magnitude data. 
Other 44 novae, which are once analyzed in our previous papers,
are reanalyzed in Appendix \ref{revised_analysis}.

It should be noted that our theoretical model light curves are calculated
from free-free emission and do not include the effects of bound-bound
and free-bound emissions.  Therefore, if such effects contribute 
significantly to the spectra of novae, our model light curves deviate
from the observation.  Such an example is the $V$ band light curve
in the nebular phase, in which strong emission lines such as [\ion{O}{3}]
contribute to the $V$ band.  We already discussed such deviations in
our previous papers \citep[e.g.][]{hac06kb, hac10k, hac14k, hac15k, 
hac16k, hac18kb, hac19ka, hac19kb}.  Our model light curves reproduce
reasonably well such nova $V$ light curves except for the nebular phase
as already shown in our previous papers and also in Appendixes 
\ref{light_curves} and \ref{revised_analysis} of the present paper.
As concerns to the free-bound emission in the continuum, this can
affect mainly the $U$ magnitude due to the Balmer jump in emission
\citep[see, e.g., Figure 3 of][]{sko19}.

To formulate the time-stretched color-magnitude diagram,
we use the nature of nova light curves described by Equation 
(\ref{time-stretching_general}), that is, time-stretched nova light curves
overlap each other with the time-scaling factor of $f_{\rm s}$.
This nature can be derived from the universal decline law of novae.
The universal decline law was derived from the nova model light curves
based on optically thick wind and free-free emission.
This relation has been calibrated on many novae \citep[e.g.,][]{hac16k,
hac16kb, hac18kb, hac19ka, hac19kb}. 
Once we accept Equation (\ref{time-stretching_general}), 
we do not use the universal decline law or nova model light curves,
but directly compare template novae with target novae.
In this circumstance, it is obvious that 
each template or target nova includes all effects
that are significant in the spectra.

\begin{figure*}
\plotone{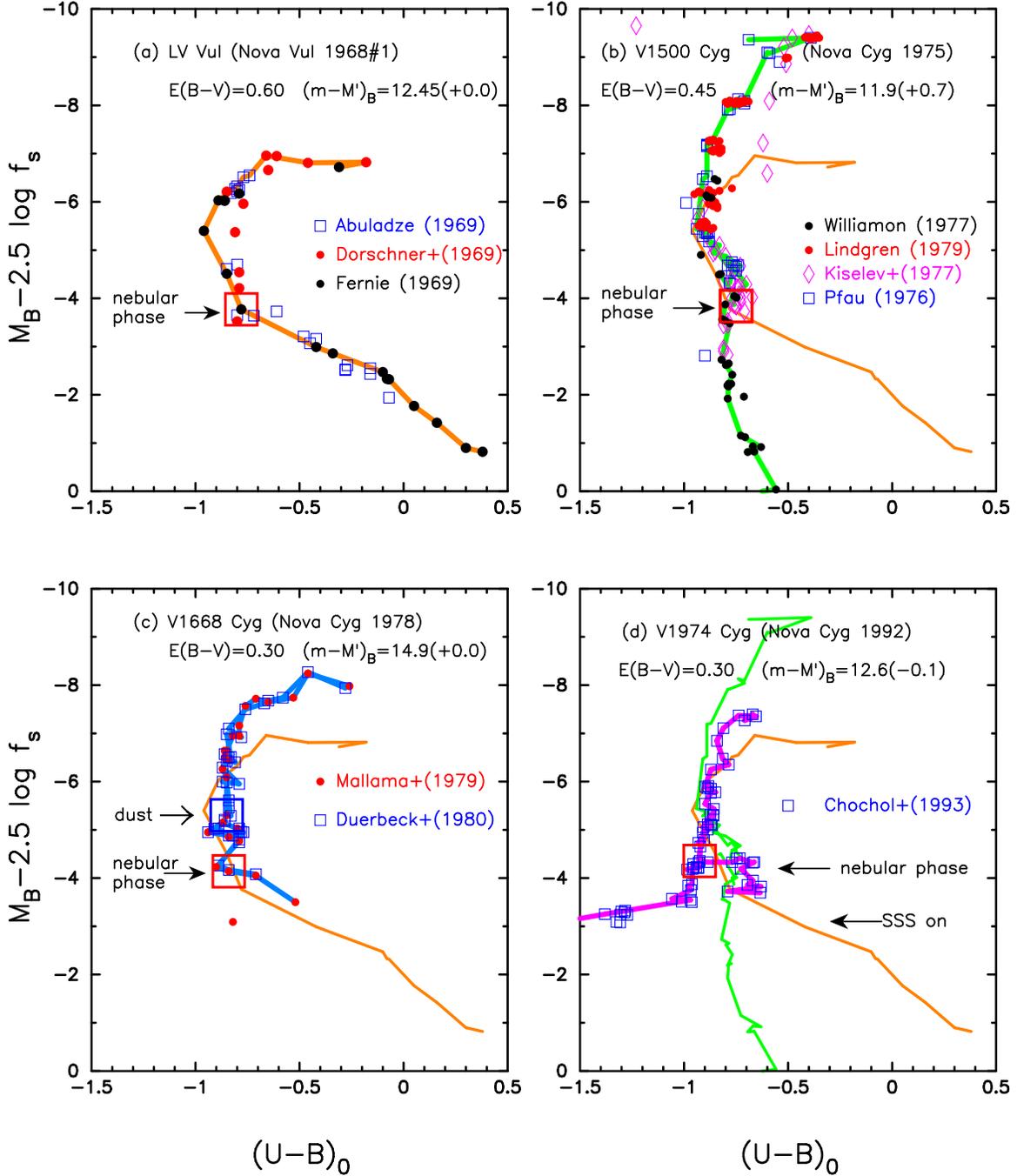}
\caption{
Time-stretched $(U-B)_0$-$(M_B-2.5\log f_{\rm s})$ color-magnitude
diagrams for (a) LV~Vul, (b) V1500~Cyg, (c) V1668~Cyg, and (d) V1974~Cyg.
We obtain template tracks for LV~Vul (thick orange line),
V1500~Cyg (thick green line), V1668~Cyg (thick cyan-blue line), and
V1974~Cyg (thick magenta line), all from the observed data.
The large unfilled red squares denote the positions at the start of
nebular phase.  The large unfilled blue square indicates the
start of dust shell formation.  In panel (d), the black arrow labeled
``SSS on'' indicates the start of the supersoft X-ray source (SSS) phase.
Each smaller symbol shows the source of data.  
See the text for more details.
\label{hr_diagram_lv_vul_v1500_cyg_v1668_cyg_v1974_cyg_outburst_ub}}
\end{figure*}


\begin{figure*}
\plotone{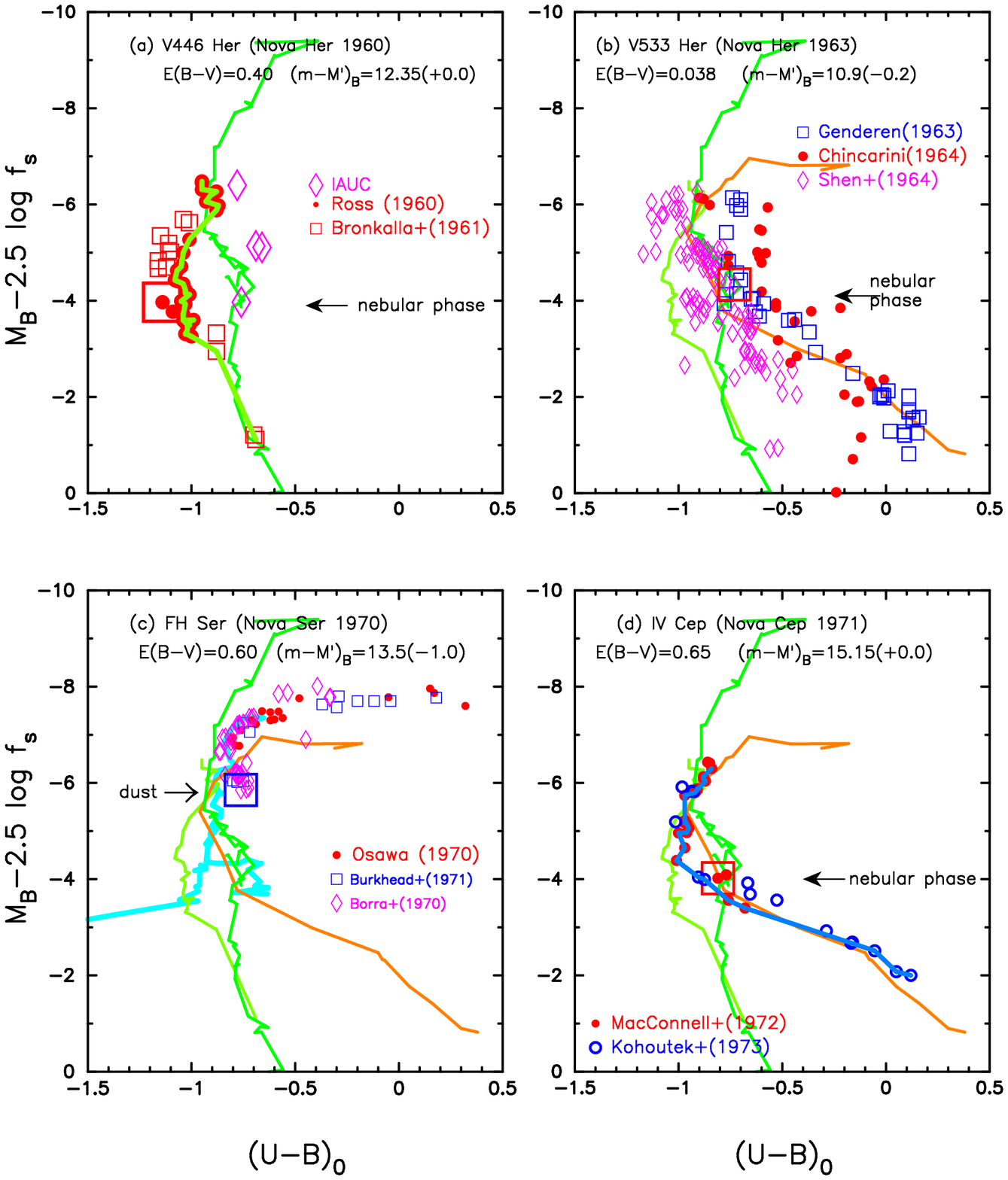}
\caption{
Same as Figure
\ref{hr_diagram_lv_vul_v1500_cyg_v1668_cyg_v1974_cyg_outburst_ub},
but for (a) V446~Her, (b) V533~Her, (c) FH~Ser, and (d) IV~Cep.
The thick light-green lines denote the template track of V446~Her. 
In panel (c), we add the template track of V1974~Cyg (thick cyan line).
In panel (d), the thick cyan-blue line denotes the template track of IV~Cep. 
\label{hr_diagram_v446_her_v533_her_fh_ser_iv_cep_outburst_ub}}
\end{figure*}

\section{Time-Stretched $(U-B)_0$-$(M_B-2.5\log \lowercase{f}_{\rm 
\lowercase{s}})$ Color-Magnitude Diagram}
\label{ub_color_magnitude_diagram}
We examine $(U-B)_0$-$(M_B-2.5\log f_{\rm s})$ diagram in the first time
of this series of papers.  In recent CCD photometric observations,
$U$ band filters are not frequently used.  We adopt 16 novae having
rich $U$ data.  Most of them are already examined in our previous papers
\citep{hac16kb, hac18kb, hac19ka, hac19kb}, but without information of the
$(U-B)_0$-$(M_B-2.5\log f_{\rm s})$ diagram.  Their timescaling factors,
distances, color excesses, and absolute $V$ magnitudes are well determined.
In what follows, we obtain $(U-B)_0$-$(M_B-2.5\log f_{\rm s})$ diagrams of
16 novae and plot their color-magnitude diagram in Figures
\ref{hr_diagram_lv_vul_v1500_cyg_v1668_cyg_v1974_cyg_outburst_ub}-\ref{hr_diagram_u_sco_t_pyx_v2659_cyg_pu_vul_outburst_ub}.
Among them, LV~Vul, V1500~Cyg, V1668~Cyg, V1974~Cyg, FH~Ser, and
PU~Vul are the six template novae proposed by \citet{hac16kb}.

We obtain the intrinsic colors via
\begin{equation}
(B-V)_0 = (B-V) - E(B-V),
\label{dereddening_eq_bv}
\end{equation}
and
\begin{equation}
(U-B)_0 = (U-B) - 0.64 E(B-V),
\label{dereddening_eq_ub}
\end{equation}
where the factor of $0.64$ is taken from \citet{rie85}.
The distance modulus in $B$ band, $(m-M)_B$, is usually
taken from \citet{hac19ka, hac19kb}, Appendix \ref{light_curves},
or Appendix \ref{revised_analysis}.  Then, its time-stretched value 
is calculated from
\begin{equation}
(m-M')_B \equiv (m - (M - 2.5 \log f_{\rm s}))_B.
\label{time-stretched_distance_modulus_b}
\end{equation}

We adopt the relations between the distance moduli in $V$, $B$,
and $U$, distance, and color excess to a nova, i.e.,
\begin{eqnarray}
(m-M)_V = 3.1 E(B-V) + 5 \log (d/10~{\rm pc}),
\label{distance_modulus_rv}
\end{eqnarray}
\begin{eqnarray}
(m-M)_B = 4.1 E(B-V) + 5 \log (d/10~{\rm pc}),
\label{distance_modulus_rb}
\end{eqnarray}
and
\begin{eqnarray}
(m-M)_U = 4.75 E(B-V) + 5 \log (d/10~{\rm pc}),
\label{distance_modulus_ru}
\end{eqnarray}
where the factor $R_V=A_V/E(B-V)=3.1$, $A_B/E(B-V)=4.1$, and
$A_U/E(B-V)=4.75$ are the ratios of total to selective extinctions
\citep{rie85}.

\subsection{LV~Vul 1968\#1}
\label{lv_vul_ub}
LV~Vul is a typical classical nova and its various properties were
discussed in \citet{hac14k, hac16kb}.  The peak $V$ brightness is 
$M_V= -7.15$.
\citet{hac19ka, hac19kb} adopted LV~Vul as a template nova, 
against which they measured the timescaling factor $f_{\rm s}$
of a target nova.  Based on the time-stretching method \citep{hac10k},
\citet{hac19ka} obtained $E(B-V)=0.60$, $(m-M)_V=11.85$, $d=1.0$~kpc,
and $\log f_{\rm s}= 0.0$ for LV~Vul.  Recent distance determination
based on {\it Gaia} Data Release2 ({\it Gaia} DR2) trigonometric 
parallax shows similar values of $d= 0.91\pm0.12$~kpc \citep{tap20}.
These values are summarized
in Table 1 of \citet{hac19kb}.  We have $(m-M)_B= 12.45$ and
$(m-M')_B= 12.45 + 0.0 = 12.45$ from \citet{hac19ka}.  
We plot the $(U-B)_0$-$(M_B-2.5\log f_{\rm s})$ diagram in Figure 
\ref{hr_diagram_lv_vul_v1500_cyg_v1668_cyg_v1974_cyg_outburst_ub}(a).
Here, we adopt the $UBV$ data from \citet{abu69}, \citet{dor69}, and
\citet{fer69}.  We have the peak $B$ brightness of 
$M'_B\equiv M_B-2.5\log f_{\rm s}= -7.0 - 0.0 = -7.0$.  In Figure 
\ref{hr_diagram_lv_vul_v1500_cyg_v1668_cyg_v1974_cyg_outburst_ub}(a),
the text ``(m-M')$_{\rm B}$=12.45(+0.0)'' 
means that $(m-M')_B = 12.45$ and $(m-M)_B = 12.45 + 0.0 =12.45$.

We define the template track of LV~Vul by the thick solid orange line.
In the rising phase of $B$ magnitude, the $(U-B)_0$ color goes
toward the red (right) until $(U-B)_0\sim -0.2$
and then turns back to the blue (left).
It horizontally moves blueward near the peak and then goes down.
After the start of the nebular phase (large unfilled red square),
it turns to the red.
This is because strong emission lines contribute much more 
to the $B$ band than to the $U$ band in the nebular phase.

\subsection{V1500~Cyg 1975}
\label{v1500_cyg_ub}
V1500~Cyg is also a well-examined nova in this series of papers.
\citet{hac19ka} proposed V1500~Cyg as another template nova.
We have reanalyzed this nova including the $(V-I)_0$ color curves in
Appendix \ref{v1500_cyg_ubvi} and revised the timescaling factor
from $\log f_{\rm s}= -0.22$ \citep{hac19ka} to the present
$\log f_{\rm s}= -0.28$ to overlap the $(V-I)_0$ color curves.
Correspondingly, the distance modulus in $V$ band, $(m-M)_V$, is
slightly changed from the old value of $(m-M)_V=12.3$ \citep{hac19ka}
to a new value of $(m-M)_V=12.15$.
We finally obtain $E(B-V)=0.45$ and $d=1.4$~kpc
(see Figure \ref{distance_reddening_v1500_cyg_xxxxxx}(b) in Appendix
\ref{revised_analysis}).   Recent distance determination based on
{\it Gaia} DR2 
trigonometric parallaxes shows a similar value of $d= 1.29\pm 0.31$~kpc
\citep{del20}.  These new parameters are listed in 
Tables \ref{extinction_various_novae} and \ref{wd_mass_novae}.
Then, we plot the $(U-B)_0$-$(M_B-2.5\log f_{\rm s})$ 
diagram in Figure 
\ref{hr_diagram_lv_vul_v1500_cyg_v1668_cyg_v1974_cyg_outburst_ub}(b).
Here, we adopt the $UBV$ data from \citet{kis77}, \citet{lin79}, 
\citet{pfa76}, and \citet{wil77}.  

V1500~Cyg is a super-bright nova and its peak magnitudes are
$M_V= -10.3$ and $M_B= -10.1$.  Here, we adopt the distance modulus
in $B$ band, $(m-M)_B= 12.6$ from Appendix \ref{v1500_cyg_ubvi}.
We have $(m-M')_B=12.6 - 0.7= 11.9$.  In Figure 
\ref{hr_diagram_lv_vul_v1500_cyg_v1668_cyg_v1974_cyg_outburst_ub}(b),
the time-stretched peak $B$ brightness is 
$M'_B\equiv M_B-2.5\log f_{\rm s}= -10.1 + 0.7 = -9.4$.

We define the template track of V1500~Cyg by the thick solid green line.
It moves redward until $(U-B)_0\sim -0.35$
before the peak and comes back toward the blue.
After the peak it goes down almost straight.
The track overlaps with that of LV~Vul between 
$M'_B= M_B-2.5\log f_{\rm s}\sim -6$ and $-4$.
After the nebular phase starts, the track departs from LV~Vul.
This is because strong [\ion{Ne}{3}] and [\ion{Ne}{5}] 
lines contribute to the $U$ band in the nebular phase.
The two tracks of V1500~Cyg and LV~Vul are closely located only 
in their middle phases, because their peak brightnesses are 
very different in the early phase and strong emission line 
contributions are different in the later nebular phase.

\subsection{V1668~Cyg 1978}
\label{v1668_cyg_ub}
V1668~Cyg is a well-observed nova and its peak $V$ brightness is $M_V=-8.6$.  
\citet[][Paper III]{hac19ka} obtained $E(B-V)=0.30$, $(m-M)_V=14.6$,
$d=5.4$~kpc, and $\log f_{\rm s}= 0.0$.  We plot the 
$(U-B)_0$-$(M_B-2.5\log f_{\rm s})$ diagram in Figure 
\ref{hr_diagram_lv_vul_v1500_cyg_v1668_cyg_v1974_cyg_outburst_ub}(c).
Here, the $UBV$ data are taken from \citet{due80} and \citet{mal79}.

The distance modulus in $B$ band, $(m-M)_B=14.9$, is taken from
\citet{hac19ka}.  Then, we have $(m-M')_B=14.9+0.0=14.9$.
The time-stretched peak $B$ brightness is 
$M'_B= M_B-2.5\log f_{\rm s}= -8.3 + 0.0 = -8.3$ as shown in Figure 
\ref{hr_diagram_lv_vul_v1500_cyg_v1668_cyg_v1974_cyg_outburst_ub}(c).
We define the template track of V1668~Cyg by the thick cyan-blue line.
It moves blueward near the peak, goes down almost straight, and then 
turns to the red after the start of nebular phase.
The shape of the track is very similar to that of LV~Vul except for
the early phase because of the brighter peak by $-1.3$ mag than that of
LV~Vul.

\subsection{V1974~Cyg 1992}
\label{v1974_cyg_ub}
V1974~Cyg is a well-observed neon nova
and its peak $V$ brightness is $M_V=-8.0$.  
\citet[][Paper III]{hac19ka} obtained $E(B-V)=0.30$, $(m-M)_V=12.2$,
$d=1.8$~kpc, and $\log f_{\rm s}= +0.03$ for V1974~Cyg.  Recent distance 
determination based on {\it Gaia} DR2 shows a similar value of 
$d= 1631^{+261}_{-131}$~pc \citep{schaefer18}.
We plot the $(U-B)_0$-$(M_B-2.5\log f_{\rm s})$ diagram in Figure 
\ref{hr_diagram_lv_vul_v1500_cyg_v1668_cyg_v1974_cyg_outburst_ub}(d).
Here, the $UBV$ data are taken from \citet{cho93}.

The distance modulus in $B$ band, $(m-M)_B=12.5$, is taken from
\citet{hac19ka}.  Then, we have $(m-M')_B=12.5+0.075=12.6$.
The time-stretched peak $B$ brightness is 
$M'_B= M_B-2.5\log f_{\rm s}= -7.3 - 0.075 = -7.4$ as shown in Figure 
\ref{hr_diagram_lv_vul_v1500_cyg_v1668_cyg_v1974_cyg_outburst_ub}(d).
We define the template track of V1974~Cyg by the thick solid magenta line.
It moves blueward near the peak, goes down almost straight, and then
splits into two branches after the start of nebular phase.
This is because there are small differences between the $B$ filters
at their blue edges and strong emission lines such as [\ion{Ne}{3}]
contribute differently to their $B$ magnitudes. 
The shape of the right track is similar to that of LV~Vul or V1668~Cyg
except for the peak $B$ brightness, while the left branch moves greatly
to the blue due to the strong contributions of [\ion{Ne}{3}] 
and [\ion{Ne}{5}] emission lines to the $U$ band than
the contribution of [\ion{Ne}{3}] emission to the $B$ band.

The difference in the inclinations of tracks after the start of SSS phase in
Figure \ref{hr_diagram_lv_vul_v1500_cyg_v1668_cyg_v1974_cyg_outburst_ub}(d)
could partly originate from the following effects: Just prior to and 
during the SSS phase, a rise of the WD temperature increases also 
electron temperature of the ionized ejecta, which makes the nebular
continuum steeper, and thus bluer indices. Following the SSS phase,
the situation can be opposite, although the effect of emission lines
is probably dominant (see, Figures
\ref{v5114_sgr_lv_vul_v1668_cyg_iv_cep_v_bv_ub_logscale},
\ref{v5114_sgr_lv_vul_v1668_cyg_iv_cep_b_ub_logscale},
\ref{v2362_cyg_lv_vul_v1500_cyg_v_bv_ub_color_logscale_no2},
\ref{v1065_cen_v1419_aql_v1668_cyg_v1974_cyg_v_bv_ub_color_logscale_no2},
\ref{v959_mon_v1668_cyg_lv_vul_v_bv_logscale_no4}, etc.).

\subsection{V446~Her 1960}
\label{v446_her_ub}
\citet[][Paper III]{hac19ka} obtained $E(B-V)=0.40$, $(m-M)_V=11.95$,
$d=1.38$~kpc, and $\log f_{\rm s}= 0.0$ for V446~Her.  Recent distance 
determination based on {\it Gaia} DR2 shows similar values of 
$d= 1361 ^{+185}_{-100}$~pc \citep{schaefer18}
and $d= 1308\pm 130$~pc \citep{sel19}.  We plot the 
$(U-B)_0$-$(M_B-2.5\log f_{\rm s})$ diagram in Figure 
\ref{hr_diagram_v446_her_v533_her_fh_ser_iv_cep_outburst_ub}(a).
Here, we adopt the $UBV$ data from IAU Circular No. 7176, 7179, 
7196, 7209, 7216, 7226, 7232, 7238, 7277, and
\citet{ross60} and \citet{bro61}.
From Equation (\ref{distance_modulus_rb}), we have $(m-M)_B=12.35$.
Then, we have $(m-M')_B=12.35+0.0=12.35$ as shown in Figure 
\ref{hr_diagram_v446_her_v533_her_fh_ser_iv_cep_outburst_ub}(a).

We define the template track of V446~Her by the thick solid light-green line.
The peak $B$ brightness was probably missed.
The track goes down almost straight and
splits into two branches near $M'_B=M_B- 2.5 \log f_{\rm s} \approx -6$. 
The left branch (light-green) follows the data of \citet{ross60}
and \citet{bro61} while the right branch (green) follows 
the data of IAU Circular.
This split is caused by small differences at blue edges 
of the $B$ filters together with strong emission lines such 
as [\ion{Ne}{3}] that contribute differently to their $B$ magnitudes. 
We define the track of V446~Her by the left branch (light-green line).
Comparing with V1500~Cyg (green line), the middle part of the track is 
$\Delta (U-B) \sim 0.3$ mag bluer whereas the early and late parts
overlap with the V1500~Cyg track.

\subsection{V533~Her 1963}
\label{v533_her_ub}
\citet[][Paper III]{hac19ka} obtained $E(B-V)=0.038$, $(m-M)_V=10.65$,
$d=1.28$~kpc, and $\log f_{\rm s}= +0.08$ for V533~Her.  Recent distance 
determination based on {\it Gaia} DR2 shows similar values of 
$d= 1202 ^{+51}_{-42}$~pc \citep{schaefer18}
and $d= 1165\pm 44$~pc \citep{sel19}.  
From Equation (\ref{distance_modulus_rb}), we have $(m-M)_B=10.7$
and $(m-M')_B=10.7+0.2=10.9$.
We plot the $(U-B)_0$-$(M_B-2.5\log f_{\rm s})$ diagram in Figure 
\ref{hr_diagram_v446_her_v533_her_fh_ser_iv_cep_outburst_ub}(b).
The $UBV$ data are taken from \citet{gen63}, \citet{chi64},
and \citet{she64}.

The peak $B$ brightness was missed.
The data points are so scattered among the three observers, but
broadly follow the LV~Vul track (orange line).


\begin{figure*}
\plottwo{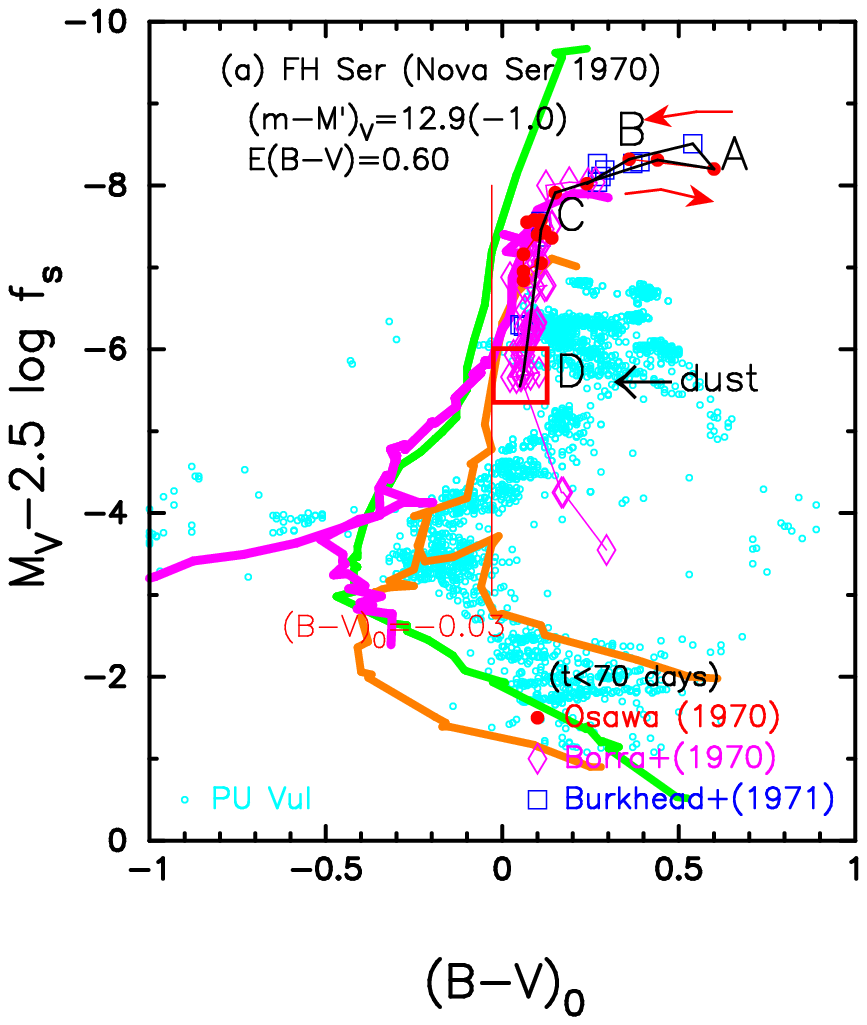}{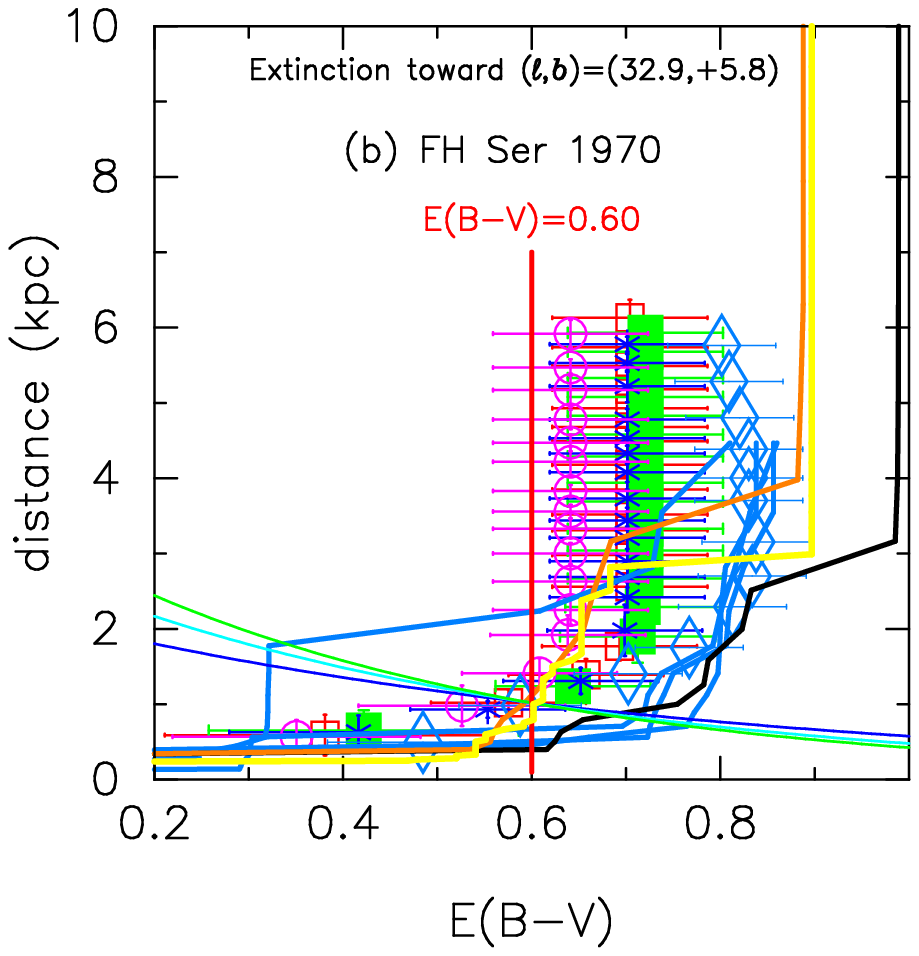}
\caption{
Time-stretched $(B-V)_0$-$(M_V-2.5 \log f_{\rm s})$ color-magnitude 
diagram of FH~Ser (a) and various distance-reddening relations 
toward FH~Ser (b).  In panel (a), the solid black line represents 
the template track of FH~Ser while the thick solid green, orange, 
and magenta lines denote the template tracks of V1500~Cyg, LV~Vul, and 
V1974~Cyg, respectively.  The points A, B, C, and D on the black line
correspond to the same points in Figure 2 of \citet{hac14k}.  
In panel (b), the three thin 
solid lines of green, cyan, and blue denote the distance-reddening
relations given by $(m-M)_U=12.9$, $(m-M)_B=12.5$, and $(m-M)_V=11.9$,
respectively.  We add four distance-reddening relations 
(thick cyan-blue lines) of \citet{chen19}, which correspond to four
nearby directions toward FH~Ser, i.e., the galactic coordinates of 
$(\ell, b)= (32\fdg85, +5\fdg75)$, $(32\fdg85, +5\fdg85)$,
$(32\fdg95, +5\fdg85)$, and $(32\fdg95, +5\fdg75)$.
We also add the distance-reddening relation (thick solid yellow line)
toward FH~Ser given by \citet{gre19}.
Other symbols and lines are the same as those in
Figure \ref{distance_reddening_v5114_sgr_v2362_cyg_v1065_cen_v959_mon}.
\label{distance_reddening_fh_ser_ubv}}
\end{figure*}


\begin{figure*}
\plotone{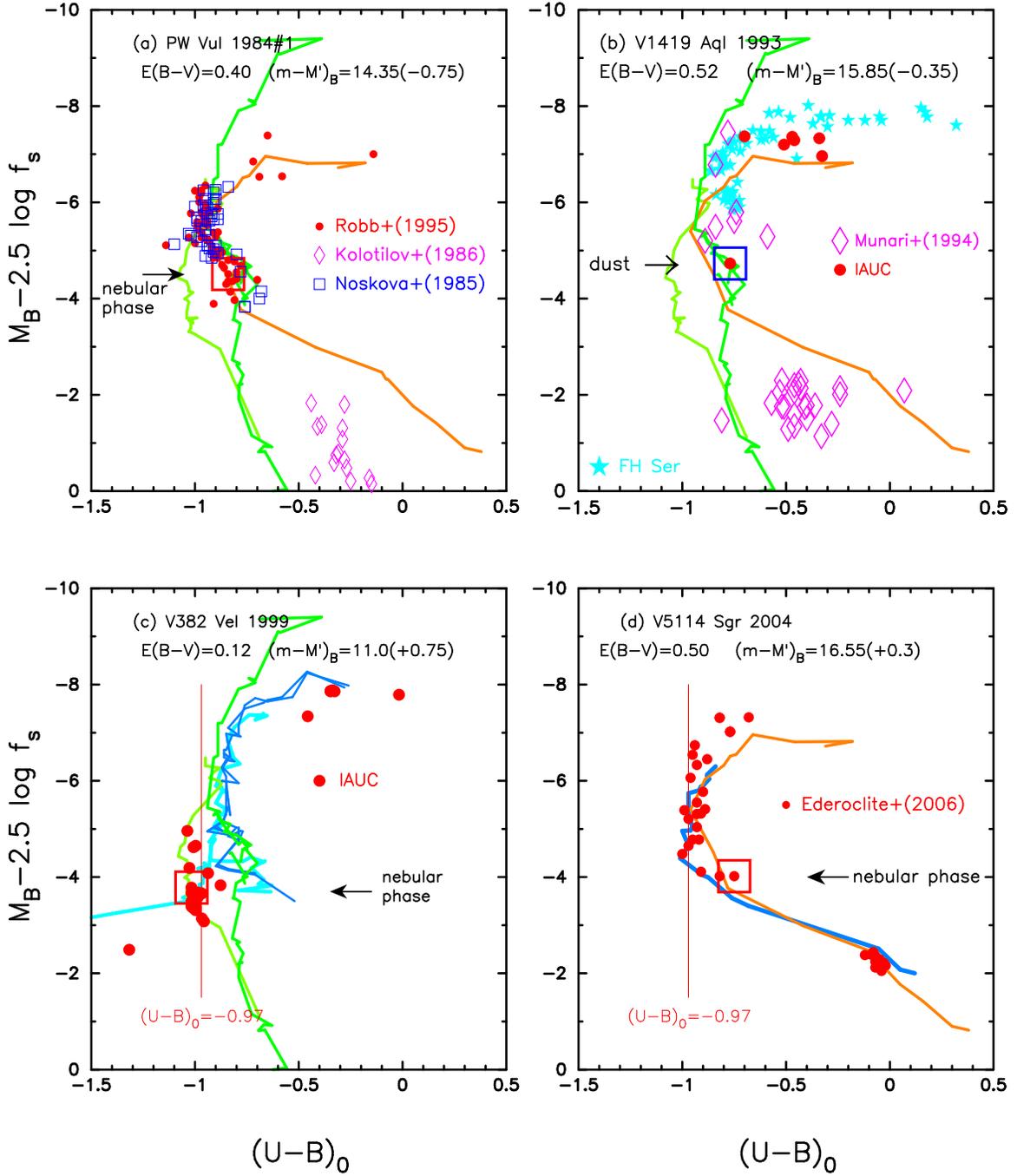}
\caption{
Same as Figure
\ref{hr_diagram_lv_vul_v1500_cyg_v1668_cyg_v1974_cyg_outburst_ub},
but for (a) PW~Vul, (b) V1419~Aql, (c) V382~Vel, and (d) V5114~Sgr.
In panel (b), we add the track of FH~Ser (filled cyan stars).  
In panels (c) and (d), the vertical solid red lines show the intrinsic
$U-B$ color of optically-thick free-free emission, that is,
$(U-B)_0= -0.97$ \citep{hac14k}.
In panel (c), we add the tracks of V1974~Cyg (thick solid cyan lines)
and V1668~Cyg (cyan-blue lines).  In panel (d), we add the track of
IV~Cep (cyan-blue line).
\label{hr_diagram_pw_vul_v1419_aql_v382_vel_v5114_sgr_outburst_ub}}
\end{figure*}


\begin{figure*}
\plotone{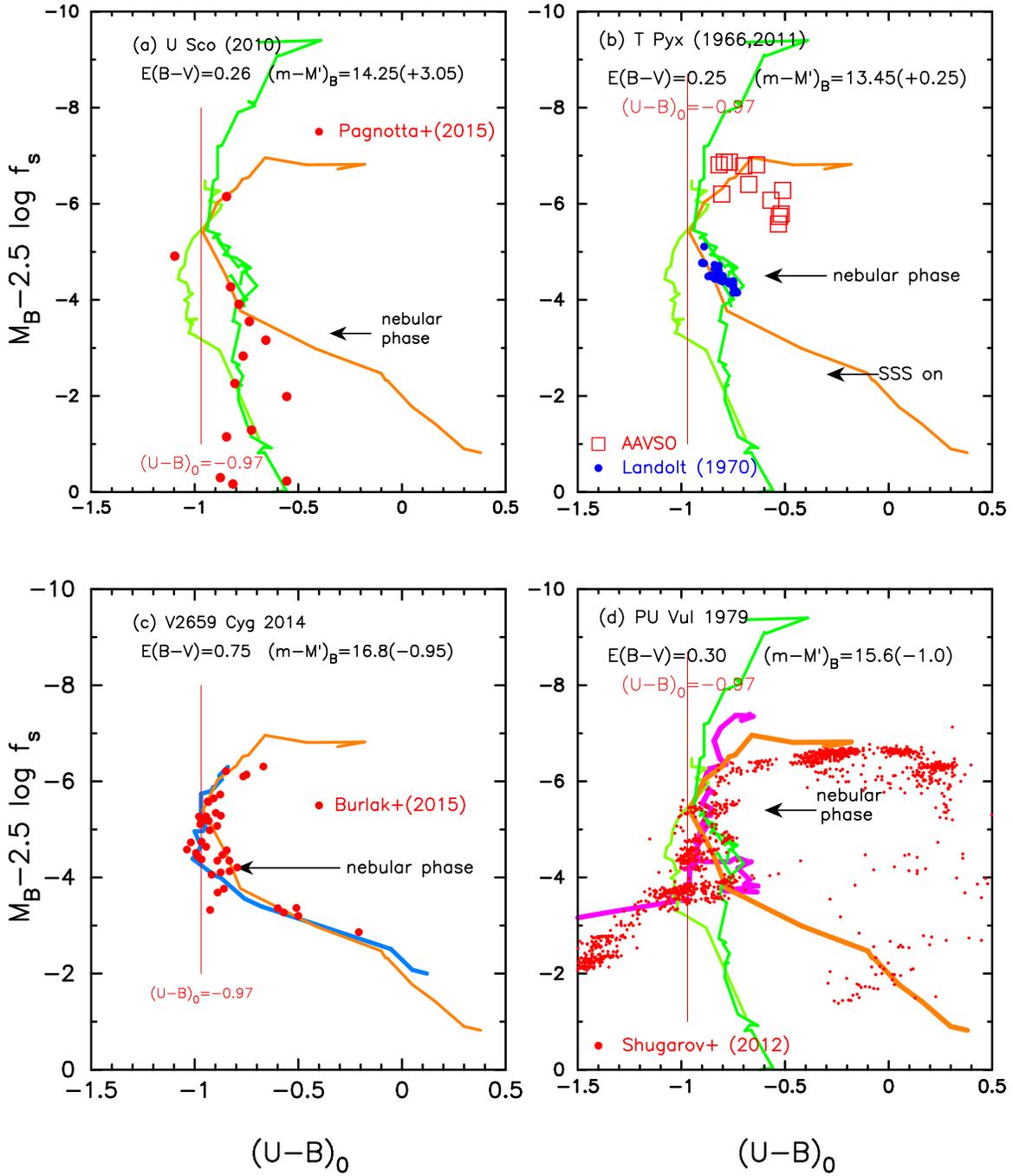}
\caption{
Same as Figure
\ref{hr_diagram_lv_vul_v1500_cyg_v1668_cyg_v1974_cyg_outburst_ub},
but for (a) U~Sco, (b) T~Pyx, (c) V2659~Cyg, and (d) PU~Vul.
In panel (c), we add the template track of IV~Cep (cyan-blue line).
In panel (d), the thick solid magenta line denotes the template track of
V1974~Cyg. 
\label{hr_diagram_u_sco_t_pyx_v2659_cyg_pu_vul_outburst_ub}}
\end{figure*}


\begin{figure*}
\plotone{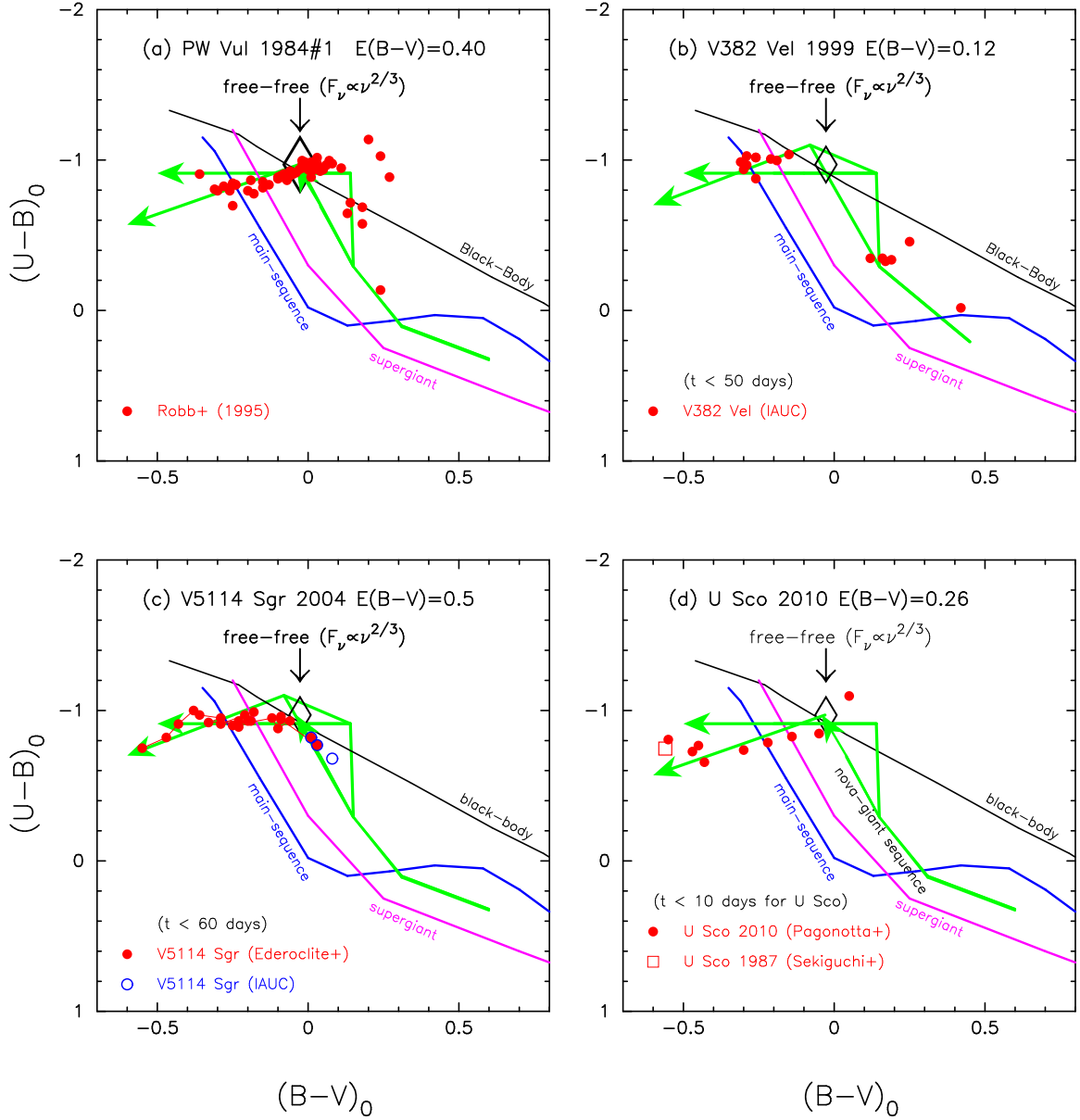}
\caption{
Color-color diagrams of (a) PW~Vul, (b) V382~Vel, (c) V5114~Sgr, and (d) U~Sco.
The thick solid green lines denote the track of nova-giant sequence
defined by \citet{hac14k}. 
We also show the three other sequences, i.e., blackbody,
main-sequence, and supergiant sequences \citep[see, e.g.,][]{hac14k}. 
The unfilled black diamond denotes the position of optically thick 
free-free emission \citep[$F_\nu\propto \nu^{2/3}$, ][]{wri75},
whose colors are $(B-V)_0= -0.03$ and $(U-B)_0= -0.97$ \citep{hac14k}.
See the text for the sources of observational data.
\label{color_color_diagram_pw_vul_v382_vel_v5114_sgr_u_sco_bv_ub}}
\end{figure*}


\begin{figure*}
\plotone{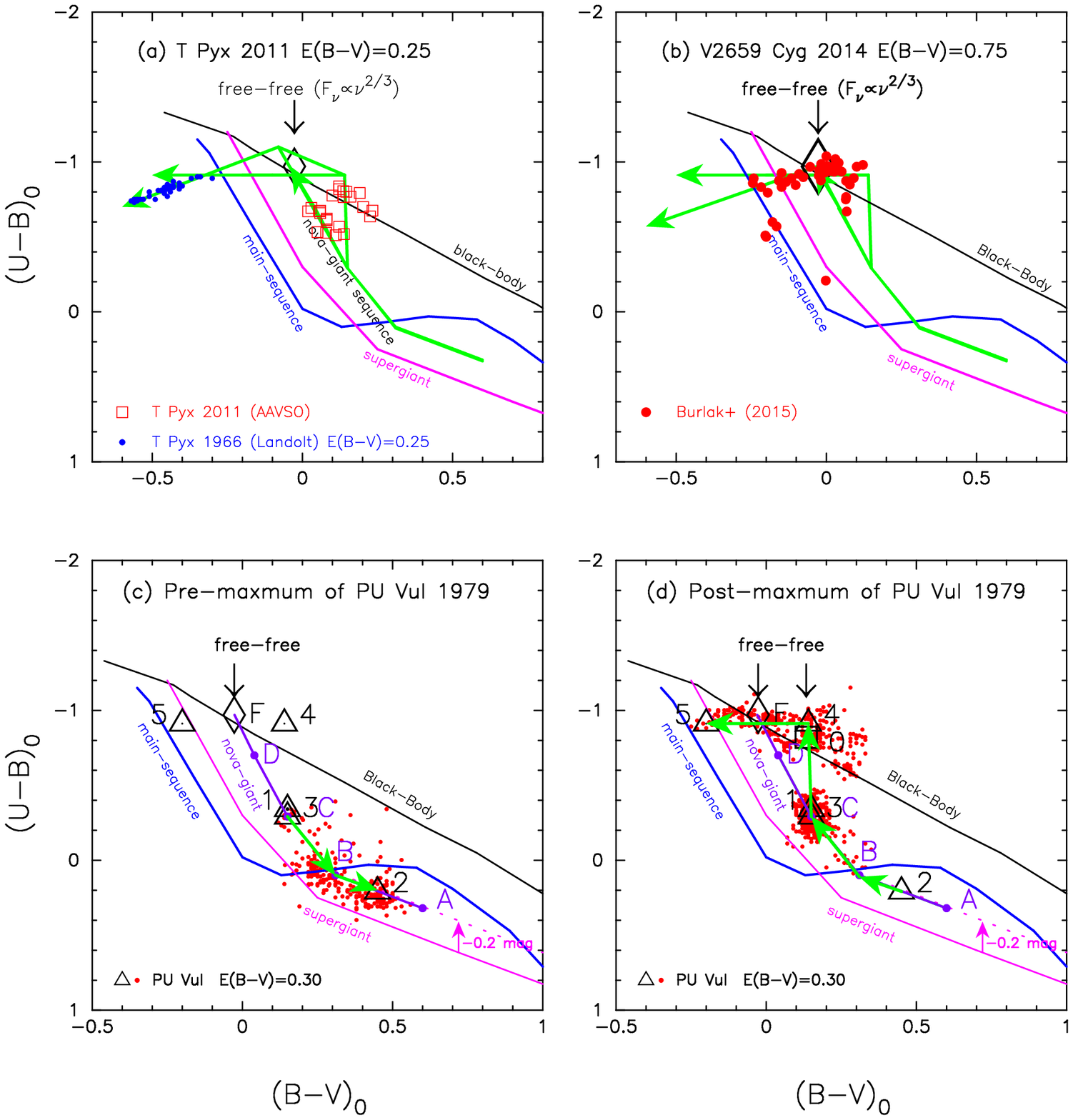}
\caption{
Same as Figure
\ref{color_color_diagram_pw_vul_v382_vel_v5114_sgr_u_sco_bv_ub}, but
for (a) T~Pyx, (b) V2659~Cyg, (c) Pre-maximum phase of PU~Vul,
and (d) Post-maximum phase of PU~Vul.  In panels (c) and (d),
the attached numbers (1-5) to the unfilled triangles denote the epochs
of light curve of PU~Vul in Figure 15 of \citet{hac14k}, and
the attached alphabets (A-D) to the filled purple circles represent
the epochs of light curve of FH~Ser in Figure 2 of \citet{hac14k}
(see also Figure \ref{distance_reddening_fh_ser_ubv}(a)). 
The solid purple line corresponds to the track of FH~Ser, which is
designated the nova-giant sequence by \citet{hac14k}.  The points 
labeled ``F'' (open black diamond) and ``O'' (open black square)
are the positions of optically-thick and
optically-thin free-free emission, respectively.  
\label{color_color_diagram_t_pyx_v2659_cyg_bv_ub}}
\end{figure*}

\subsection{FH~Ser 1970}
\label{fh_ser_ub_bv}
\citet[][Paper II]{hac16kb} obtained $E(B-V)=0.60$, $(m-M)_V=11.7$,
and $d=0.93$~kpc.  Recent distance determination based on
{\it Gaia} DR2 shows a similar value of $d= 1060 ^{+112}_{-68}$~pc 
\citep{schaefer18}.  We have determined the timescaling factor of
$\log f_{\rm s}= +0.40$ as well as $(m-M)_U=12.9$, $(m-M)_B=12.5$,
$(m-M)_V=11.9$, and $d=1.0$~kpc from Figures 
\ref{v5579_sgr_lv_vul_v1668_cyg_v1535_sco_v_bv_ub_color_logscale_no2},
\ref{v2659_cyg_v5114_sgr_v1668_cyg_lv_vul_b_ub_color_logscale_no2},
and \ref{distance_reddening_xxxxxx_v2659_cyg}(a) in Appendix
\ref{revised_analysis}.  These new parameters are listed in 
Tables \ref{extinction_various_novae} and \ref{wd_mass_novae}.
Then, we have $(m-M')_B=12.5+1.0=13.5$.  The time-stretched peak $B$
brightness is $M'_B= M_B-2.5\log f_{\rm s}= -7.0 - 1.0 = -8.0$.
We plot the $(U-B)_0$-$(M_B-2.5\log f_{\rm s})$ diagram in Figure 
\ref{hr_diagram_v446_her_v533_her_fh_ser_iv_cep_outburst_ub}(c).
The $UBV$ data are taken from \citet{bor70}, \citet{osa70},
and \citet{bur71}.  The track almost horizontally moves blueward
near the peak. Then, it almost vertically goes down up to the formation
of an optically-thick dust shell, which clouds the following evolution.
The track is located close to the track of V1974~Cyg (cyan line) in
the early phase.

We also plot the $(B-V)_0$-$(M_V-2.5\log f_{\rm s})$ diagram in Figure 
\ref{distance_reddening_fh_ser_ubv}(a).  The track of FH~Ser is located close
to the track of V1974~Cyg (magenta line) until the dust blackout started. 
The solid black line represents the template track of FH~Ser and
four points, A, B, C, and D are the same points as those in Figure 2
of \citet{hac14k}.
We obtain $(m-M)_U=12.9$, $(m-M)_B=12.5$, and $(m-M)_V=11.9$ from
our time-stretching method in Figures
\ref{distance_reddening_xxxxxx_v2659_cyg}(a),
\ref{v2659_cyg_v5114_sgr_v1668_cyg_lv_vul_b_ub_color_logscale_no2}, and
\ref{v5579_sgr_lv_vul_v1668_cyg_v1535_sco_v_bv_ub_color_logscale_no2},
respectively.  We plot their distance-reddening relations, i.e.,
Equations (\ref{distance_modulus_ru}), (\ref{distance_modulus_rb}), 
and (\ref{distance_modulus_rv}), respectively, in 
Figure \ref{distance_reddening_fh_ser_ubv}(b).
These three lines broadly cross at $d= 1.0$~kpc and $E(B-V)= 0.60$.
This crossing point is consistent with the distance-reddening relations
given by \citet[][thick solid orange line]{gre18}, 
\citet[][thick solid yellow line]{gre19}, 
\citet[][unfilled cyan-blue diamonds]{ozd18}, and
\citet[][unfilled open magenta circles, blue asterisks, filled green squares,
and unfilled red squares]{mar06}.

\subsection{IV~Cep 1971}
\label{iv_cep_ub}
\citet{hac18k} obtained $E(B-V)=0.65$, $(m-M)_V=14.5$, $d=3.1$~kpc,
and $\log f_{\rm s}= 0.0$ for IV~Cep.  
From Equation (\ref{distance_modulus_rb}), we have
$(m-M)_B=15.15$ and then have $(m-M')_B=15.15+0.0=15.15$.
We plot the $(U-B)_0$-$(M_B-2.5\log f_{\rm s})$ diagram in Figure 
\ref{hr_diagram_v446_her_v533_her_fh_ser_iv_cep_outburst_ub}(d).
The $UBV$ data are taken from \citet{mac72} and \citet{koh73}.

The peak in the $B$ band was missed.
The track goes down almost straight and then turns to the right (redward)
just after the start of nebular phase.
It almost follows the track of LV~Vul (orange line).

\subsection{PW~Vul 1984\#1}
\label{pw_vul_ub}
\citet[][Paper III]{hac19ka} obtained $E(B-V)=0.57$, $(m-M)_V=13.0$,
$d=1.8$~kpc, and $\log f_{\rm s}= +0.35$ for PW~Vul.  Recent distance 
determination based on {\it Gaia} DR2 shows a slightly larger value
of $d= 2420 ^{+1337}_{-277}$~pc \citep{schaefer18}.  We have reanalyzed
the $UBVI$ data of PW~Vul based on the time-stretching
method in Appendix \ref{pw_vul_ubvi}, including
the $(t/f_{\rm s})$-$(M_I - 2.5 \log f_{\rm s})$ light curve 
and $(t/f_{\rm s})$-$(V-I)_0$ color curve fitting.
We have obtained a new parameter set of $E(B-V)=0.40$, $(m-M)_U=13.88$,
$(m-M)_B=13.6$, $(m-M)_V=13.2$, $(m-M)_I=12.55$,
$d=2.46$~kpc, and $\log f_{\rm s}= +0.30$.  
These new parameters are listed in 
Tables \ref{extinction_various_novae} and \ref{wd_mass_novae}.
The new distance estimate is consistent with the distance of
{\it Gaia} DR2.  The main difference is the color excess.  
The new color excess of $E(B-V)= 0.40$ is consistent with
several distance-reddening relations as shown in Figure
\ref{distance_reddening_pw_vul_xxxxxx}(b) of Appendix 
\ref{pw_vul_ubvi}.  
We plot the $(B-V)_0$-$(U-B)_0$ color-color diagram of PW~Vul in Figure 
\ref{color_color_diagram_pw_vul_v382_vel_v5114_sgr_u_sco_bv_ub}(a).
The new track of PW~Vul (filled red circles) is quite consistent with
the template tracks of nova-giant sequence (green lines)
defined by \citet{hac14k}.  

In the $(B-V)_0$-$(U-B)_0$ diagram, we denote the position of
optically-thick free-free emission (unfilled black diamond).
This position is accidentally coincident with the cross of
blackbody sequence (black line) and nova-giant sequence (green line). 
It should be noted that the color indices $(U-B)_0 = -0.97$, 
$(B-V)_0 = -0.03$, and $(V-I)_0 = +0.22$ of free-free emission are derived
from the continuum flux of $F_\nu\propto\nu^{2/3}$ for the radio and
infrared domains.  This approximation corresponds to $h\nu\ll k T_{\rm e}$
and is, strictly speaking, not valid in the optical region.
Here, $h$, $\nu$, $k$, and $T_{\rm e}$ are the Planck constant,
frequency, Boltzmann constant, and electron temperature
\citep[see, e.g., Equation (7) of][]{hac14k}.
In the optical region, these three color indices can be
a result of different contributions from different sources
of radiation to the continuum during some stages of the nova evolution.
Usually, the optical continuum is given by superposition of the blackbody
(or atmospheric model) and the nebular radiation (or free-free emission).
\citet{hac14k} showed that the continuum flux $F_\nu\propto\nu^{2/3}$
in the optical region of PW~Vul is well reproduced with the summation of
the blackbody and optically-thick free-free emissions at this epoch 
(see their Figure 9).  
In this sense, the color indices $(U-B)_0 = -0.97$, $(B-V)_0 = -0.03$,
and $(V-I)_0 = +0.22$ need not to represent the pure optically-thick
free-free emission in the optical region.

We have $(m-M')_B= 13.6+0.75= 14.35$.  The time-stretched
peak $B$ brightness is $M'_B= M_B-2.5\log f_{\rm s}= -6.64 - 0.75 = -7.39$.
We plot the $(U-B)_0$-$(M_B-2.5\log f_{\rm s})$ diagram in Figure 
\ref{hr_diagram_pw_vul_v1419_aql_v382_vel_v5114_sgr_outburst_ub}(a).
Here, we adopt the $UBV$ data from \citet{nos85}, \citet{kol86},
and \citet{rob95}.  The track almost follows the LV~Vul template track 
(orange line) until the nebular phase started.
The overlapping of PW~Vul and LV~Vul 
suggests a reasonable value of the color excess, $E(B-V)= 0.40$,
because the horizontal shift of $U-B$ color determines the
color excess of $E(U-B)= 0.64 E(B-V)$.
The overlapping of PW~Vul and LV~Vul in the time-stretched
color-magnitude diagram also supports the vertical fit with
$(m-M)_B=13.6$, $d=2.46$~kpc, and $\log f_{\rm s}= +0.30$.

\subsection{V1419~Aql 1993}
\label{v1419_aql_ub}
\citet[][Paper III]{hac19ka} obtained $E(B-V)=0.52$, $(m-M)_V=15.0$,
$d=4.7$~kpc, and $\log f_{\rm s}= +0.15$ for V1419~Aql.  
The distance modulus in $B$ band, $(m-M)_B=15.5$, is calculated
from Equation (\ref{distance_modulus_rb}).  Then, we have 
$(m-M')_B=15.5+0.35=15.85$.  The time-stretched peak $B$ brightness is
$M'_B= M_B-2.5\log f_{\rm s}= -7.1 - 0.35 = -7.45$.
We plot the $(U-B)_0$-$(M_B-2.5\log f_{\rm s})$ diagram in Figure 
\ref{hr_diagram_pw_vul_v1419_aql_v382_vel_v5114_sgr_outburst_ub}(b).
Here, we adopt the $UBV$ data from IAU Circular No. 5794, 5802, 5807,
5829, and \citet{mun94a}.
The track broadly follows LV~Vul (orange line) until an optically thick
dust shell formed.  We also add the track of FH~Ser (filled cyan stars).
The V1419~Aql track almost overlaps with the track of FH~Ser until the dust
blackout of FH~Ser.

\subsection{V382~Vel 1999}
\label{v382_vel_ub}
\citet[][Paper III]{hac19ka} obtained $E(B-V)=0.25$, $(m-M)_V=11.5$,
$d=1.4$~kpc, and $\log f_{\rm s}= -0.29$ for V382~Vel.  Recent distance 
determination based on {\it Gaia} DR2 shows a slightly larger value of 
$d= 1800 ^{+243}_{-133}$~pc \citep{schaefer18}.  We have reanalyzed
the $UBVI$ data of V382~Vel based on the time-stretching method
in Appendix \ref{v382_vel_ubvi}, including
the $(t/f_{\rm s})$-$(M_I - 2.5 \log f_{\rm s})$ light curve 
and $(t/f_{\rm s})$-$(V-I)_0$ color curve fitting.
We have obtained a new parameter set of $E(B-V)=0.12$, $(m-M)_U=11.78$,
$(m-M)_B=11.71$, $(m-M)_V=11.6$, $(m-M)_I=11.41$,
$d=1.76\pm0.2$~kpc, and $\log f_{\rm s}= -0.29$.  
These new parameters are listed in 
Tables \ref{extinction_various_novae} and \ref{wd_mass_novae}.
The new distance estimate is consistent with the distance of {\it Gaia} DR2. 
The main difference is the color excess.  With the $U$, $B$, and $I$ data
being newly included in our analysis, we can determine
the color excess more precisely.   The new color excess of 
$E(B-V)= 0.12$ is consistent with the $(B-V)_0$-$(U-B)_0$ color-color
diagram of nova-giant sequence as shown in Figure
\ref{color_color_diagram_pw_vul_v382_vel_v5114_sgr_u_sco_bv_ub}(b) and
the distance-reddening relation given by \citet{chen19} as shown in Figure
\ref{color_color_distance_reddening_v382_vel_xxxxxx}(b) of Appendix 
\ref{v382_vel_ubvi}.

The distance modulus in $B$ band is $(m-M)_B=11.71$.
Then, we have $(m-M')_B=11.71-0.725=10.99$.  The peak $B$ brightness is 
$M'_B= M_B-2.5\log f_{\rm s}= -8.75 + 0.725 = -8.0$.
We plot the $(U-B)_0$-$(M_B-2.5\log f_{\rm s})$ diagram in Figure 
\ref{hr_diagram_pw_vul_v1419_aql_v382_vel_v5114_sgr_outburst_ub}(c).
Here, we adopt the $UBV$ data from IAU Circular No. 7176, 7179, 7196,
7209, 7216, 7226, 7232, 7238, and 7277.
The track almost follows V1668~Cyg (blue line) in the very early phase.
Then, it follows the template track of V446~Her (light-green line). 
The overlapping of V382~Vel and V446~Her in the middle phase
suggests that the color excess, $E(B-V)= 0.12$, is reasonable
because the horizontal shift of $U-B$ color determines the
color excess of $E(U-B)= 0.64 E(B-V)$.

The V382~Vel track also overlaps with that of the LV~Vul template track
in the $(B-V)_0$-$(M_V-2.5 \log f_{\rm s})$ diagram, that is, Figure 
\ref{hr_diagram_pw_vul_v382_vel_v5117_sgr_v2362_cyg_outburst}(b).
The V382~Vel track also overlaps with the V496~Sct/V959~Mon
template tracks (orange lines) in 
the $(V-I)_0$-$(M_I-2.5 \log f_{\rm s})$ diagram, that is, Figure 
\ref{hr_diagram_pw_vul_v382_vel_v5117_sgr_v2362_cyg_outburst_vi}(b).
These overlappings also support the reddening value of
$E(B-V)= 0.12$, the distance moduli of $(m-M)_B= 11.71$,
$(m-M)_V= 11.6$, and $(m-M)_I= 11.41$, and 
the timescaling factor of $\log f_{\rm s}= -0.29$.

\subsection{V5114~Sgr 2004}
\label{v5114_sgr_ub}
\citet[][Paper III]{hac19ka} obtained $E(B-V)=0.47$, $(m-M)_V=16.65$,
$d=10.9$~kpc, and $\log f_{\rm s}= -0.12$ for V5114~Sgr.  
We have reanalyzed the $UBVI_{\rm C}$ data of V5114~Sgr in Appendix
\ref{v5114_sgr_ubvi} and obtained
a slightly different set of parameters, that is, $E(B-V)=0.5$,
$(m-M)_V=16.35$, $d=9.1$~kpc, and $\log f_{\rm s}= -0.12$.
The track of V5114~Sgr with the new color-excess $E(B-V)=0.5$ 
is quite consistent with the nova-giant sequence
in the $(B-V)_0$-$(U-B)_0$ color-color diagram, as shown in Figure
\ref{color_color_diagram_pw_vul_v382_vel_v5114_sgr_u_sco_bv_ub}(c).
These new parameters are listed in 
Tables \ref{extinction_various_novae} and \ref{wd_mass_novae}.

We adopt the distance modulus in $B$ band, $(m-M)_B=16.85$, 
from Appendix \ref{v5114_sgr_ubvi}.  Then, we have $(m-M')_B=16.85-0.3=16.55$.
The peak $B$ brightness is $M'_B= M_B-2.5\log f_{\rm s}= -7.65 + 0.3 = -7.35$.
We plot the $(U-B)_0$-$(M_B-2.5\log f_{\rm s})$ diagram in Figure 
\ref{hr_diagram_pw_vul_v1419_aql_v382_vel_v5114_sgr_outburst_ub}(d).
Here, we adopt the $UBV$ data from \citet{ede06}.
The track almost follows the LV~Vul (orange line) or IV~Cep (cyan-blue line).
The overlapping of V5114~Sgr and LV~Vul (or IV~Cep) in the time-stretched
color-magnitude diagram may support the results of
$E(B-V)=0.50$, $(m-M)_B=16.85$, $d=9.1$~kpc,
and $\log f_{\rm s}= -0.12$ for V5114~Sgr.

\subsection{U~Sco 2010}
\label{u_sco_ub}
We have reanalyzed the $UBVI_{\rm C}$ data of U~Sco (2010) in Appendix
\ref{u_sco_ubvik} and obtained a new parameter set of 
$E(B-V)=0.26$, $(m-M)_U=17.3$, $(m-M)_B=17.1$, $(m-M)_V=16.85$,
$(m-M)_I=16.45$, $d=16.2$~kpc, and $\log f_{\rm s}= -1.22$.
The color excess of $E(B-V)= 0.26$ is consistent with
the $(B-V)_0$-$(U-B)_0$ color-color diagram of nova-giant sequence,
as shown in Figure
\ref{color_color_diagram_pw_vul_v382_vel_v5114_sgr_u_sco_bv_ub}(d).
These new parameters are listed in 
Tables \ref{extinction_various_novae} and \ref{wd_mass_novae}.
We plot the $(U-B)_0$-$(M_B-2.5\log f_{\rm s})$ diagram in Figure 
\ref{hr_diagram_u_sco_t_pyx_v2659_cyg_pu_vul_outburst_ub}(a).
The track of U~Sco broadly follows 
the LV~Vul track (orange line) in the early phase and then transfers to
the V1500~Cyg track (green line) in the middle and later phases.

\subsection{T~Pyx 2011}
\label{t_pyx_ub}
We have reanalyzed the $UBVI$ data of T~Pyx
and obtained a new parameter set of $E(B-V)=0.25$, 
$(m-M)_B=13.70$, $(m-M)_V=13.45$, $(m-M)_I=13.05$, $d=3.4$~kpc, 
and $\log f_{\rm s}= -0.10$ in Appendix \ref{t_pyx_ubvi}.  
Recent distance determination based on {\it Gaia} DR2 shows a similar value of 
$d= 3185 ^{+607}_{-283}$~pc \citep{schaefer18}.  
The color excess of $E(B-V)= 0.25$ is consistent with
the $(B-V)_0$-$(U-B)_0$ color-color diagram of nova-giant sequence,
as shown in Figure \ref{color_color_diagram_t_pyx_v2659_cyg_bv_ub}(a).
These new parameters are listed in 
Tables \ref{extinction_various_novae} and \ref{wd_mass_novae}.
This value of $E(B-V)= 0.25$ is the same as that estimated with
the color-color diagram method by \citet{hac14k}.

We plot the $(U-B)_0$-$(M_B-2.5\log f_{\rm s})$ diagram in Figure 
\ref{hr_diagram_u_sco_t_pyx_v2659_cyg_pu_vul_outburst_ub}(b).
The $(U-B)_0$-$(M_B-2.5\log f_{\rm s})$ track almost follows 
the LV~Vul track (orange lines) in the early and middle phases.
The overlappings of T~Pyx with the LV~Vul track 
suggests that the new values of $E(B-V)= 0.25$, $(m-M)_B=13.70$, 
$d=3.4$~kpc, and $\log f_{\rm s}= -0.10$ are reasonable.

\subsection{V2659~Cyg 2014}
\label{v2659_cyg_ub}
\citet{hac19kb} obtained $E(B-V)=0.80$, $(m-M)_V=15.7$, $d=4.4$~kpc,
and $\log f_{\rm s}= +0.52$ for V2659~Cyg based on the $BVI_{\rm C}$
light curves.   The $UBV$ data are now available in \citet{bur15}.
Therefore, we have reanalyzed the $UBVI_{\rm C}$ light curves and
obtained new parameters of $E(B-V)=0.75$, $(m-M)_V=15.1$, $d=3.6$~kpc,
and $\log f_{\rm s}= +0.37$ for V2659~Cyg in Appendix \ref{v2659_cyg_ubvi}.
The main difference is the timescaling factor of $\log f_{\rm s}= +0.37$ and,
as a result, the distance modulus in $V$ band of $(m-M)_V=15.1$.
These new parameters are listed in 
Tables \ref{extinction_various_novae} and \ref{wd_mass_novae}.
We first plot the $(B-V)_0$-$(U-B)_0$ color-color diagram
in Figure \ref{color_color_diagram_t_pyx_v2659_cyg_bv_ub}(b).
The data (filled red circles) are located on the thick solid green
lines, which represent the typical tracks of the nova-giant sequence
\citep{hac14k}.
This overlapping of the V2659~Cyg track and the nova-giant sequence
supports the color excess of $E(B-V)=0.75$ \citep[see Figure 8 
of][for more details]{hac14k}.

We adopt the distance modulus in $B$ band of $(m-M)_B=15.85$
from Appendix \ref{v2659_cyg_ubvi}.
Then, we have $(m-M')_B=15.85+0.95=16.8$.
The peak $B$ brightness is $M'_B= M_B-2.5\log f_{\rm s}= -5.4 - 0.95 = -6.35$.

We next plot the $(U-B)_0$-$(M_B-2.5\log f_{\rm s})$ diagram in Figure 
\ref{hr_diagram_u_sco_t_pyx_v2659_cyg_pu_vul_outburst_ub}(c).
The $UBV$ data are taken from \citet{bur15}.
The track almost follows the template track (cyan-blue line) of
IV~Cep as well as the template track of LV~Vul (orange line).
This overlapping of V2659~Cyg and IV~Cep (or LV~Vul)
supports the new values of
$E(B-V)=0.75$, $(m-M)_B=15.85$, $d=3.6$~kpc,
and $\log f_{\rm s}= +0.37$ for V2659~Cyg.

\subsection{PU~Vul 1979}
\label{pu_vul_ub}
We adopt $E(B-V)= 0.30$ and
$d= 4.7$~kpc after \citet{kat12mh}.   PU~Vul shows no signature of
wind mass-loss in the early phase.  Therefore, we do not determine
the timescaling factor by comparing the PU~Vul light curve with
those of other normal novae.
Instead, we determine a common value of $f_{\rm s}$ by fitting
the PU~Vul track with those of other template novae in the three
time-stretched color-magnitude diagrams.  
We obtain $\log f_{\rm s}= +0.40$ against that of LV~Vul,    
and plot the $(U-B)_0$-$(M_B-2.5\log f_{\rm s})$ track in Figure
\ref{hr_diagram_u_sco_t_pyx_v2659_cyg_pu_vul_outburst_ub}(d),
$(B-V)_0$-$(M_V-2.5\log f_{\rm s})$ track in Figure
\ref{hr_diagram_v1500_cyg_pu_vul_v5114_sgr_v574_pup_outburst}(b), and
$(V-I)_0$-$(M_I-2.5\log f_{\rm s})$ track in Figure
\ref{hr_diagram_v1500_cyg_pu_vul_v5114_sgr_v574_pup_outburst_vi}(b).
These parameters are listed in 
Tables \ref{extinction_various_novae} and \ref{wd_mass_novae}.
The $UBVI_{\rm C}$ data of PU~Vul are taken from \citet{hen08} and
\citet{shu12}.

The $U-B$ color evolution of PU~Vul was summarized in \citet{hac14k}. 
Before the $B$ maximum, the $B$ brightness goes up almost vertically
along $(U-B)_0\sim +0.3$ in Figure
\ref{hr_diagram_u_sco_t_pyx_v2659_cyg_pu_vul_outburst_ub}(d).
Then, it horizontally moves blueward near the $B$ peak and then goes 
down vertically along $(U-B)_0\sim -0.97$ (vertical solid red line),
which is the intrinsic color of free-free emission.  When the track
goes down to $M'_B=M_B-2.5\log f_{\rm s} \sim -4$, it turns to the blue.
This is because strong emission lines contribute much more 
to the $U$ band than to the $B$ band in the nebular phase
like in V1974~Cyg (magenta line).

During the early phase of flat peak, PU~Vul shows pure absorption spectra
for F-type supergiants \citep{kan91a}, which \citet{hac14k} attributed to 
the photospheric emission of PU~Vul.  
These spectra showed no signature of wind mass-loss.
The $U-B$ color becomes bluer with the photospheric temperature increasing
like in normal supergiants.  The track of PU~Vul in the $(B-V)_0$-$(U-B)_0$
color-color diagram is parallel to, but $\Delta (U-B) \approx -0.2$
mag bluer than, the supergiant sequence as shown in Figure 
\ref{color_color_diagram_t_pyx_v2659_cyg_bv_ub}(c) and (d).
This part of the track coincides perfectly with 
the nova-giant sequence (solid purple line) defined by \citet{hac14k}.

This bluer position in the $(B-V)_0$-$(U-B)_0$ diagram 
can be seen in the spectra of PU~Vul obtained
by \citet{bel89}, who reported, ``There is a good
agreement of the energy distribution of PU~Vul and normal supergiants
in the spectrum region from 3000\AA\  to 7000\AA.
In the region from 3200\AA\  to 3800\AA, the UV excess is clearly seen.
The amount of this excess in 1983 agrees well with the photometric estimate
$\Delta (U-B)=-0.2$ mag.''  (See the spectrum in Figure 6
of \citet{bel89} for more details.)   

The reddest $(U-B)_0\sim +0.4$ color in PU~Vul comes from the faintness of
$U$ band brightness in the Balmer jump, the depth of which is shallower
than that of normal supergiants.  This depth becomes shallower 
and shallower and finally disappears along the flat $B$ peak in Figure
\ref{hr_diagram_u_sco_t_pyx_v2659_cyg_pu_vul_outburst_ub}(d)
and, at the same time, along the nova-giant sequence
in the $(B-V)_0$-$(U-B)_0$ diagram (Figure
\ref{color_color_diagram_t_pyx_v2659_cyg_bv_ub}(d)).
The evolution of $(U-B)_0$ color from $+0.4$ to $-0.97$ originates
significantly from this change in the depth of
Balmer jump \citep[e.g.,][]{bel89}.

Finally, we obtain a parameter set of $E(B-V)=0.30$, $(m-M)_U=14.8$,
$(m-M)_B=14.6$, $(m-M)_V=14.3$, $(m-M)_I=13.82$,
$d=4.7$~kpc, and $\log f_{\rm s}= +0.40$ for PU~Vul.
The $(U-B)_0$-$(M_B-2.5\log f_{\rm s})$ track in Figure
\ref{hr_diagram_u_sco_t_pyx_v2659_cyg_pu_vul_outburst_ub}(d)
almost follows the V1974~Cyg track (magenta lines).
The $(B-V)_0$-$(M_V-2.5\log f_{\rm s})$ track broadly follows 
the LV~Vul track (orange lines) in Figure
\ref{hr_diagram_v1500_cyg_pu_vul_v5114_sgr_v574_pup_outburst}(b).
The $(V-I)_0$-$(M_I-2.5\log f_{\rm s})$ track follows 
the V5114~Sgr track (green line) in Figure
\ref{hr_diagram_v1500_cyg_pu_vul_v5114_sgr_v574_pup_outburst_vi}(b).
These overlappings suggests that the values of $E(B-V)= 0.30$, 
$d=4.7$~kpc, and $\log f_{\rm s}= +0.40$ are reasonable.

\subsection{Summary of $(U-B)_0$-$(M_B-2.5 \log f_{\rm s})$ Diagram}
\label{summary_ub_b_diagram}
We plot the 16 novae in outburst in the $(U-B)_0$-$(M_B-2.5 \log f_{\rm s})$
diagram (Figures 
\ref{hr_diagram_lv_vul_v1500_cyg_v1668_cyg_v1974_cyg_outburst_ub},
\ref{hr_diagram_v446_her_v533_her_fh_ser_iv_cep_outburst_ub},
\ref{hr_diagram_pw_vul_v1419_aql_v382_vel_v5114_sgr_outburst_ub}, and
\ref{hr_diagram_u_sco_t_pyx_v2659_cyg_pu_vul_outburst_ub}).
From these figures, we deduce the following properties:
\begin{itemize}
\item[(1)] In the rising phase of $B$ magnitude, $(U-B)_0$ color
moves redward (rightward) as shown in LV~Vul, V1500~Cyg, and PU~Vul.
Before the $B$ peak, it turns toward the blue (left) and goes 
almost horizontally up to $(U-B)_0\sim -0.7$.  This turning point from
toward the red to toward the blue in 
the $(U-B)_0$ color before the $B$ peak is different from nova to nova, i.e.,
$(U-B)_0\sim +0.4$ in PU~Vul, $(U-B)_0\sim -0.2$ in LV~Vul, and
$(U-B)_0\sim -0.4$ in V1500~Cyg.

\item[(2)] The tracks begin to go down at/around $(U-B)_0\sim -0.7$.
We observe that the tracks split into two paths; one is redder than
the other by $\Delta (U-B)_0 \sim 0.3$.  For example, 
the tracks of LV~Vul and V1500~Cyg belong to the redder group.
The tracks of V446~Her and V382~Vel belong to the bluer group.

\item[(3)] The nebular phase starts at/around $M'_B= M_B - 2.5\log f_{\rm s}
\sim -4$ except for PU~Vul.  PU~Vul entered the nebular phase at
$M'_B= M_B - 2.5\log f_{\rm s} \sim -5.5$. 
The tracks further split into three paths.
For example, the tracks of V1974~Cyg and PU~Vul turn to the blue
(leftward). The track of V1500~Cyg goes almost straight down.
The tracks of LV~Vul, IV~Cep, and V2659~Cyg turn to the red (rightward). 

\item[(4)]  The reason of split into two or three (possibly more) 
paths in the nebular phase is the effect of strong emission lines 
near the edges of $U$ and $B$ filters.  
For example, one of the V1974~Cyg tracks turns to
the blue (leftward) just after the nebular phase starts.
This is because strong emission lines such as [\ion{Ne}{5}] contribute
to the $U$ band.  Moreover, there are small differences
among the $B$ filters at their blue edges and strong emission lines
such as [\ion{Ne}{3}] contribute differently to their $B$ magnitudes. 
The shape of the rightside track of V1974~Cyg is similar to that of
LV~Vul or V1668~Cyg except for the peak $B$ brightness, 
while the leftside branch moves greatly to the blue.  
\end{itemize}

The most important property is that each track undergoes 
a similar path at $(U-B)_0\sim -0.8$ between 
$M'_B= M_B-2.5\log f_{\rm s}= -6$ and $-4$, and then
splits into several paths after the nebular phase starts.
These tracks almost overlap with each other, one of
the template tracks.  
If we adopt a different color excess $E(B-V)$ or timescaling factor
$f_{\rm s}$, we cannot obtain such overlappings.   In other words,
the overlappings of many novae in the time-stretched color-magnitude diagram
may support our results of the color excess $E(B-V)$, distance modulus
in $B$ band $(m-M)_B$, and timescaling factor $f_{\rm s}$.


\startlongtable
\begin{deluxetable*}{llllllrll}
\tabletypesize{\scriptsize}
\tablecaption{Extinctions, distance moduli, and distances for revised novae
\label{extinction_various_novae}}
\tablewidth{0pt}
\tablehead{
\colhead{Object} & \colhead{Outburst} & \colhead{$E(B-V)$} 
& \colhead{$(m-M)_V$} & 
\colhead{$d$} & \colhead{$d_{\rm {\it Gaia}}$\tablenotemark{a}} & 
\colhead{$\log f_{\rm s}$\tablenotemark{b}} &
\colhead{$(m-M')_V$} & \colhead{References\tablenotemark{c}} \\
  & (year) &  &  &  (kpc) &  (kpc) &  &  & 
} 
\startdata
OS~And & 1986 & 0.15 & 14.8 & 7.3 & & $-0.15$ & 14.4 & 2 \\
V1663~Aql & 2005 & 1.88 & 18.6 & 3.6 & & $-0.10$ & 18.35 & 6,7 \\
V679~Car & 2008 & 0.59 & 16.05 & 7.0 & & $-0.10$ & 15.8 & 5,7 \\
V834~Car & 2012 & 0.50 & 17.25 & 14 & & $-0.02$ & 17.2 & 6,7 \\
V1065~Cen & 2007 & 0.48 & 15.0 & 5.0 & & $0.0$ & 15.0 & 3,7 \\
V1213~Cen & 2009 & 0.68 & 16.8 & 8.6 & & 0.05 & 16.9 & 6,7 \\
V1368~Cen & 2012 & 0.98 & 17.6 & 8.2 & & $0.07$ & 17.8 & 6,7 \\
V962~Cep & 2014 & 1.10 & 18.6 & 10.9 & & $0.12$ & 18.9 & 6,7 \\
V1500~Cyg & 1975 & 0.45 & 12.15 & 1.4 & $1.29\pm0.31$ & $-0.28$ & 11.45 & 5,7 \\
V2362~Cyg & 2006 & 0.56 & 15.4 & 5.4 & & $0.25$ & 16.0 & 5,7 \\
V2468~Cyg & 2008 & 0.80 & 15.55 & 4.1 & & $-0.06$ & 15.4 & 5,7 \\
V2491~Cyg & 2008 & 0.40 & 16.65 & 12.1 & & $-0.40$ & 15.65 & 5,7 \\
V2659~Cyg & 2014 & 0.75 & 15.1 & 3.6 & & $0.37$ & 16.05 & 6,7 \\
PR~Lup & 2011 & 0.70 & 16.4 & 7.0 & & $0.20$ & 16.9 & 6,7 \\
V959~Mon & 2012 & 0.38 & 13.15 & 2.5 & & $0.18$ & 13.6 & 3,7 \\
V390~Nor & 2007 & 1.0 & 15.6 & 3.2 & & $0.14$ & 15.95 & 6,7 \\
V2575~Oph  & 2006\#1 & 1.43 & 17.9 & 4.9 & & $0.18$ & 18.35 & 6,7 \\
V2615~Oph  & 2007 & 1.05 & 15.9 & 3.4 & & $0.04$ & 16.0 & 5,7 \\
V2670~Oph  & 2008\#1 & 0.90 & 16.15 & 4.7 & & 0.59 & 17.65 & 6,7 \\
V2677~Oph  & 2012\#2 & 1.40 & 19.2 & 9.4 & & $-0.17$ & 18.75 & 6 \\
V2944~Oph  & 2015 & 0.62 & 15.8 & 6.0 & & $0.14$ & 16.15 & 6,7 \\
V574~Pup  & 2004 & 0.40 & 14.7 & 5.0 & & $0.0$ & 14.7 & 5,7 \\
V597~Pup  & 2007 & 0.24 & 16.5 & 14.2 & & $-0.18$ & 16.05 & 6,7 \\
T~Pyx  & 2011 & 0.25 & 13.45 & 3.4 & $3.185^{+0.607}_{-0.283}$ & $-0.10$ & 13.2 & 2,7 \\
V5114~Sgr & 2004 & 0.50 & 16.35 & 9.1 & & $-0.12$ & 16.05 & 5,7 \\
V5116~Sgr & 2005\#2 & 0.28 & 15.45 & 8.2 & & $-0.14$ & 15.1 & 6,7 \\
V5117~Sgr & 2006 & 0.35 & 16.05 & 9.8 & & $0.10$ & 16.3 & 6,7 \\
V5579~Sgr & 2008 & 0.56 & 14.55 & 3.6 & & $0.24$ & 15.15 & 6,7 \\
V5584~Sgr & 2009\#4 & 0.75 & 16.9 & 8.2 & & $0.10$ & 17.15 & 6,7 \\
V5585~Sgr & 2010 & 0.47 & 16.8 & 11.6 & & $0.0$ & 16.8 & 6,7 \\
V5589~Sgr & 2012\#1 & 0.80 & 17.7 & 11.0 & & $-0.67$ & 16.0 & 6,7 \\
V5666~Sgr & 2014 & 0.50 & 15.5 & 6.2 & & $0.20$ & 16.0 & 5,7 \\
V5667~Sgr & 2015\#1 & 0.63 & 15.1 & 4.3 & & $0.36$ & 16.0 & 6,7 \\
V5668~Sgr & 2015\#2 & 0.20 & 10.8 & 1.1 & & $0.27$ & 11.45 & 6,7 \\
U~Sco & 2010 & 0.26 & 16.85 & 16.2 & & $-1.22$ & 13.8 & 4,7 \\
V1281~Sco & 2007\#2 & 0.76 & 17.4 & 10.1 & & $-0.07$ & 17.2 & 6,7 \\
V1313~Sco & 2011\#2 & 1.1 & 18.65 & 11.2 & & $-0.22$ & 18.1 & 6,7 \\
V1324~Sco & 2012 & 1.32 & 16.95 & 3.7 & & $0.32$ & 17.75 & 6,7 \\
V1535~Sco & 2015 & 0.78 & 17.95 & 12.8 & & $-0.26$ & 17.3 & 6,7 \\
V496~Sct & 2009 & 0.45 & 13.6 & 2.8 & & 0.30 & 14.35 & 5,7 \\
FH~Ser & 1970 & 0.60 & 11.9 & 1.0 & $1.06^{+0.112}_{-0.068}$ & $0.40$ & 12.9 & 2,7\\
NR~TrA & 2008 & 0.19 & 14.25 & 5.4 & & $0.55$ & 15.6 & 6,7 \\
V382~Vel & 1999 & 0.12 & 11.6 & 1.76 & $1.8^{+0.243}_{-0.133}$ & $-0.29$ & 10.85 & 3,5,7 \\
PU~Vul & 1979 & 0.30 & 14.3 & 4.7 & & $0.40$ & 15.3 & 1,7 \\
PW~Vul & 1984\#1 & 0.40 & 13.2 & 2.5 & $2.42^{+1.337}_{-0.277}$ & $0.30$ & 13.95 & 5,7 \\
QU~Vul & 1984\#2 & 0.40 & 12.55 & 1.83 & $1.786^{+3.495}_{-0.196}$ & 0.59 & 14.0 & 2,7 \\
V459~Vul & 2007\#2 & 0.85 & 15.55 & 3.8 & & $-0.04$ & 15.45 & 6,7 \\  
\enddata
\tablenotetext{a}{
The {\it Gaia} DR2 distances except for V1500~Cyg are taken from 
\citet{schaefer18}
and that of V1500~Cyg is taken from \citet{del20}.}
\tablenotetext{b}{
$f_{\rm s}$ is the timescale against that of LV~Vul, that is, 
$f_{\rm s}=1$ for LV~Vul.}
\tablenotetext{c}{
(1) \citet{kat12mh},
(2) \citet{hac16kb},
(3) \citet{hac18k},
(4) \citet{hac18kb},
(5) \citet{hac19ka},
(6) \citet{hac19kb},
(7) present paper.
} 
\end{deluxetable*}


\startlongtable
\begin{deluxetable*}{lllllll}
\tabletypesize{\scriptsize}
\tablecaption{White dwarf masses of revised novae
\label{wd_mass_novae}}
\tablewidth{0pt}
\tablehead{
\colhead{Object} & \colhead{$\log f_{\rm s}$}
& \colhead{$M_{\rm WD}$}
& \colhead{$M_{\rm WD}$}
& \colhead{$M_{\rm WD}$}
& \colhead{$M_{\rm WD}$}
& \colhead{Chem. Comp.}
\\
\colhead{} & \colhead{}
& \colhead{$f_{\rm s}$\tablenotemark{a}}
& \colhead{UV~1455~\AA\tablenotemark{b}}
& \colhead{$t_{\rm SSS-on}$\tablenotemark{c}}
& \colhead{$t_{\rm SSS-off}$\tablenotemark{d}}
& \colhead{$(X,Y,Z)$ or}
\\
\colhead{} & \colhead{} & \colhead{($M_\sun$)} & \colhead{($M_\sun$)}
& \colhead{($M_\sun$)} & \colhead{($M_\sun$)} & \colhead{template}
}
\startdata
V1663~Aql & $-0.10$ & 1.05 & --- & --- & --- & CO3 \\
V679~Car & $-0.10$ & 1.05 & --- & --- & --- & CO3 \\
V834~Car & $-0.02$ & 1.0 & --- & --- & --- & CO3 \\
V1065~Cen & $+0.0$ & 0.98 & --- & --- & --- & CO3 \\
V1213~Cen & $+0.05$ & 1.0 & --- & --- & --- & Ne2 \\
V1368~Cen & $+0.07$ & 0.95 & --- & --- & --- & CO3 \\
V962~Cep & $+0.12$ & 0.95 & --- & --- & --- & CO3 \\
V1500~Cyg & $-0.28$ & 1.23 & --- & --- & --- & Ne2 \\
V2362~Cyg & $+0.25$ & 0.85 & --- & --- & --- & interp.\tablenotemark{e}\\
V2468~Cyg & $-0.06$ & 1.0 & --- & --- & --- & CO3 \\
V2491~Cyg & $-0.40$ & 1.35 & --- & 1.35 & 1.35 & Ne2 \\
V2659~Cyg & $+0.37$ & 0.75 & --- & --- & --- & CO4 \\
PR~Lup & $+0.20$ & 0.90 & --- & --- & --- & CO3 \\
V959~Mon & $+0.18$ & 0.95 & --- & 0.95 & --- & CO3 \\
V959~Mon & $+0.18$ & 1.05 & --- & 1.05 & --- & Ne2 \\
V959~Mon & $+0.18$ & 1.1 & --- & 1.10 & --- & Ne3 \\
V390~Nor & $+0.14$ & 0.90 & --- & --- & --- & CO3 \\
V2575~Oph  & $+0.18$ & 0.90 & --- & --- & --- & CO3 \\
V2615~Oph  & $+0.04$ & 0.90 & --- & --- & --- & CO3 \\
V2670~Oph  & $+0.59$ & 0.70 & --- & --- & --- & CO3 \\
V2677~Oph  & $-0.17$ & 1.15 & --- & --- & --- & Ne2 \\
V2944~Oph  & $+0.14$ & 0.85 & --- & --- & --- & CO3 \\
V574~Pup  & $+0.0$ & 1.05 & --- & 1.05 & 1.05 & Ne2 \\
V597~Pup  & $-0.18$ & 1.2 & --- & 1.2 & 1.2 & Ne3 \\
T~Pyx    & $-0.10$ & 1.15 & --- & 1.15 & 1.15 & CO4 \\
V5114~Sgr & $-0.12$ & 1.15 & --- & --- & --- & Ne2 \\
V5116~Sgr & $-0.14$ & 1.1 & --- & --- & --- & Ne3 \\
V5117~Sgr & $+0.10$ & 0.90 & --- & --- & --- & CO3 \\
V5579~Sgr & $+0.24$ & 0.85 & --- & --- & --- & CO3 \\
V5584~Sgr & $+0.10$ & 0.90 & --- & --- & --- & CO3 \\
V5585~Sgr & $+0.0$ & 0.98 & --- & --- & --- & CO3 \\
V5589~Sgr & $-0.67$ & 1.33 & --- & --- & --- & Ne2 \\
V5666~Sgr & $+0.20$ & 0.85 & --- & --- & --- & CO3 \\
V5667~Sgr & $+0.36$ & 0.78 & --- & --- & --- & CO4 \\
V5668~Sgr & $+0.27$ & 0.85 & --- & --- & --- & CO3 \\
U~Sco    & $-1.22$ & 1.37 & --- & 1.37 & 1.37 & evol.\tablenotemark{f} \\
V1281~Sco & $-0.07$ & 1.13 & --- & --- & --- & Ne2 \\
V1313~Sco & $-0.22$ & 1.20 & --- & --- & --- & Ne2 \\
V1324~Sco & $+0.32$ & 0.80 & --- & --- & --- & CO2 \\
V1535~Sco & $-0.26$ & 1.20 & --- & --- & --- & Ne2 \\
V496~Sct & $+0.30$ & 0.85 & --- & --- & --- & CO3 \\
FH~Ser & $+0.40$ & 0.75 & --- & --- & --- & CO3 \\
NR~TrA & $+0.55$ & 0.75 & --- & --- & --- & CO3 \\
V382~Vel & $-0.29$ & 1.23 & --- & --- & 1.23 & Ne2 \\
PU~Vul & $+0.40$ & 0.70 & 0.70 & --- & --- & $(0.7,0.29,0.01)$ \\
PW~Vul & $+0.30$ & 0.85 & 0.85 & --- & --- & CO4 \\
QU~Vul & $+0.59$ & 0.70 & --- & --- & --- & CO3 \\
V459~Vul & $-0.04$ & 1.15 & --- & --- & --- & Ne2 \\
\enddata
\tablenotetext{a}{WD mass estimated from the $f_{\rm s}$ timescale.}
\tablenotetext{b}{WD mass estimated from the UV~1455~\AA\  fit.}
\tablenotetext{c}{WD mass estimated from the $t_{\rm SSS-on}$ fit.}
\tablenotetext{d}{WD mass estimated from the $t_{\rm SSS-off}$ fit.}
\tablenotetext{e}{WD mass estimated from the linear interpolation
of the $\log f_{\rm s}$ versus WD mass relation \citep[see][]{hac18kb}.}
\tablenotetext{f}{WD mass estimated from the time-evolution
calculation with a Henyey type code \citep[see, e.g.,][]{hac18kb}.}
\end{deluxetable*}


\begin{figure*}
\plotone{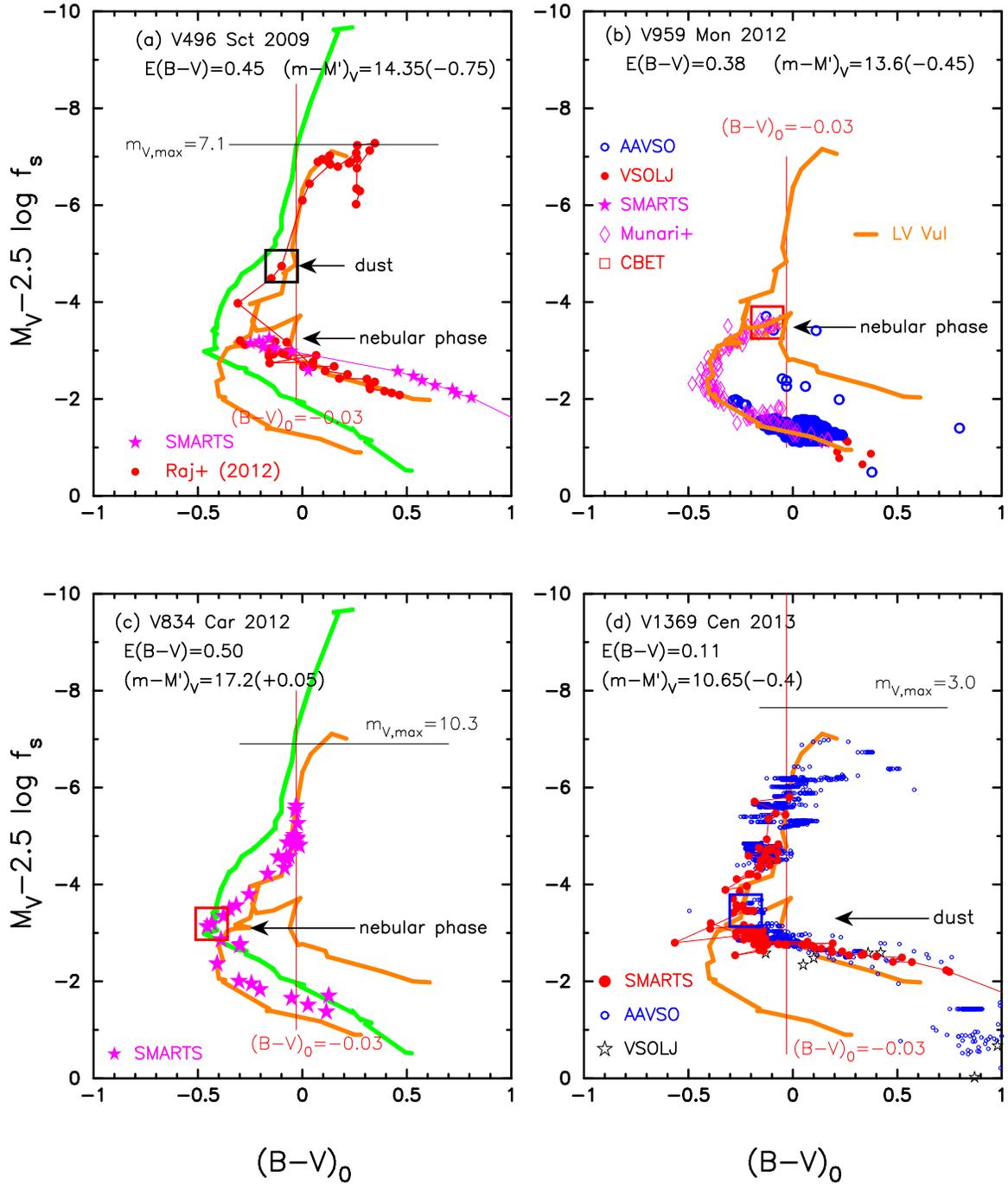}
\caption{
Time-stretched $(B-V)_0$-$(M_V-2.5\log f_{\rm s})$ 
color-magnitude diagram for (a) V496~Sct, (b) V959~Mon,
(c) V834~Car, and (d) V1369~Cen.  
The vertical solid red line labeled ``$(B-V)_0=-0.03$'' is the intrinsic
$B-V$ color of optically thick free-free emission \citep{hac14k}.
The thick solid orange lines represent the template track of LV~Vul while
the thick solid green lines denote that of V1500~Cyg.
}
\label{hr_diagram_v496_sct_v959_mon_v834_car_v1369_cen_outburst}
\end{figure*}


\begin{figure*}
\plotone{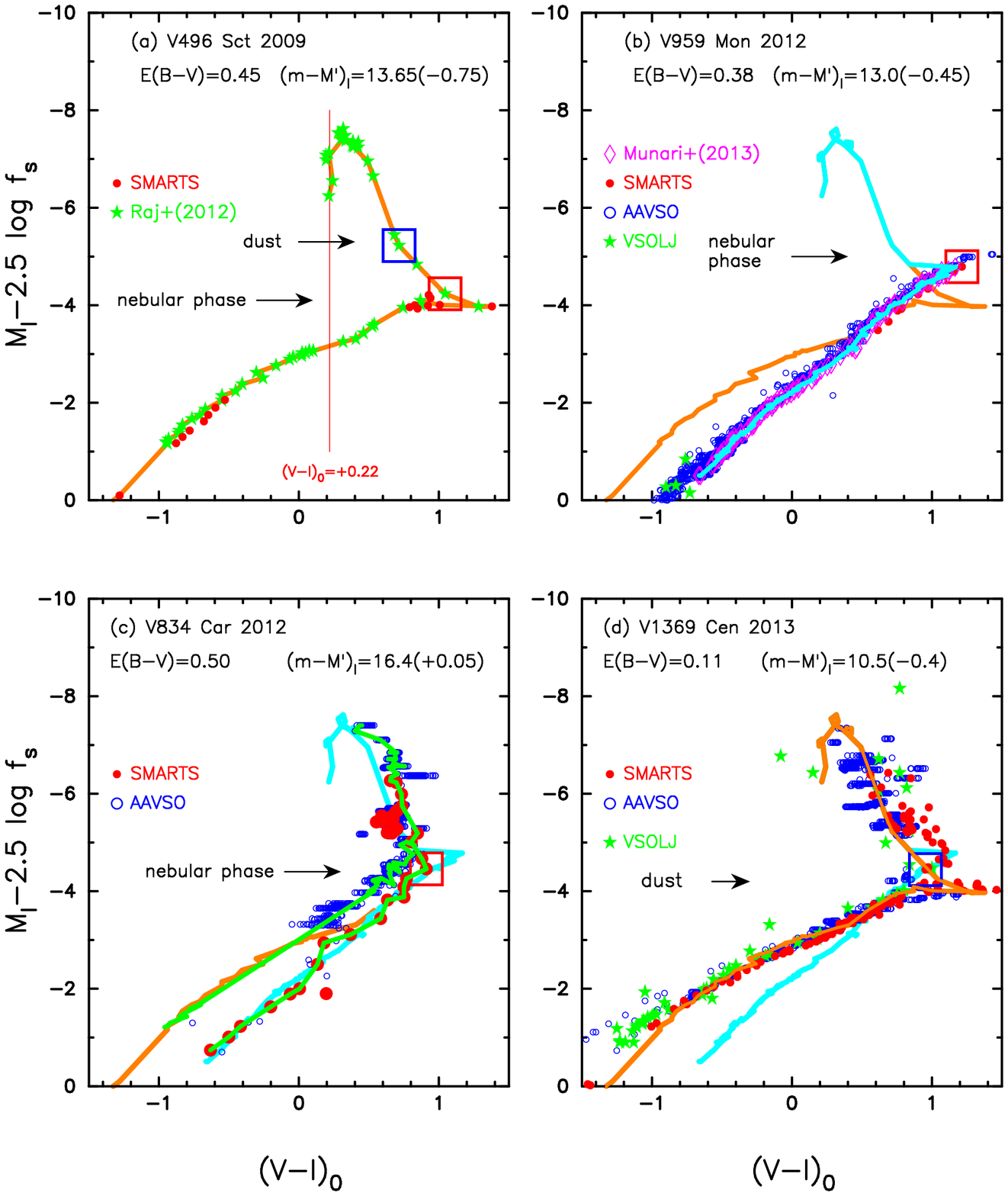}
\caption{
Time-stretched $(V-I)_0$-$(M_I-2.5\log f_{\rm s})$ 
color-magnitude diagram for (a) V496~Sct, (b) V959~Mon,
(c) V834~Car, and (d) V1369~Cen.  In panel (a), the thick solid orange
lines represent the template track of V496~Sct.
The vertical solid red line of $(V-I)_0=+0.22$ denotes the intrinsic
$V-I$ color of optically thick free-free emission.  In panel (b), 
the lower thick solid cyan line represents the template track of V959~Mon.
In panel (c), the track of V834~Car almost overlaps with the template track
of V496~Sct (orange line) in the early phase but overlaps with the track of
V959~Mon (cyan line) in the later phase.  In panel (d), the track of 
V1369~Cen almost overlaps with the track of V496~Sct (orange line).
\label{hr_diagram_v496_sct_v959_mon_v834_car_v1369_cen_outburst_vi}}
\end{figure*}

\section{Time-Stretched $(V-I)_0$-$(M_I-2.5\log \lowercase{f}_{\rm 
\lowercase{s}})$ Color-Magnitude Diagram of the LV~Vul Type}
\label{vi_color_magnitude_diagram_lv_vul}

In this paper, we analyze 52 novae in the 
$(V-I)_0$-$(M_I-2.5\log f_{\rm s})$ diagram, where
$(V-I)_0$ is the intrinsic $V-I$ color, $M_I$ is the absolute
$I$ magnitude, and $f_{\rm s}$ is the timescaling factor of a target
nova against the timescale of LV~Vul.
We obtain the $(V-I)_0$ via
\begin{equation}
(V-I)_0 = (V-I) - 1.6 E(B-V),
\label{dereddening_eq_vi}
\end{equation}
where the factor of $1.6$ is taken from \citet{rie85}.
We adopt the relation between the distance modulus in $I$ band,
distance, and color excess to a nova, i.e.,
\begin{equation}
(m-M)_I = 1.5 E(B-V) + 5 \log (d/10~{\rm pc}),
\label{distance_modulus_ri}
\end{equation}
where the factor $A_I/E(B-V)=1.5$ is taken from \citet{rie85}. 
In this series of papers \citep[e.g.,][]{hac19kb}, 
we have already divided many novae into two types, LV~Vul type 
and V1500~Cyg type, depending on their shape and position of tracks in the
$(B-V)_0$-$(M_V-2.5\log f_{\rm s})$ diagram.
In this section, we analyze 12 novae of the LV~Vul
type in the order of V496~Sct, V959~Mon, V834~Car, V1369~Cen, V1663~Aql,
V2615~Oph, V5666~Sgr, V2659~Cyg, PW~Vul, V382~Vel, V5117~Sgr, and V2362~Cyg.

Unfortunately, no $I$/$I_{\rm C}$ data are available for LV~Vul, which
is a template nova in this series of papers.  Therefore,
we adopt the tracks of V496~Sct and V959~Mon as a template
in the $(V-I)_0$-$(M_I-2.5\log f_{\rm s})$ diagram.  This is because
the tracks of V496~Sct and V959~Mon almost overlap with that of
LV~Vul in the $(B-V)_0$-$(M_V-2.5\log f_{\rm s})$ diagram.
To confirm the similarity, we first show the track in the
$(B-V)_0$-$(M_V-2.5\log f_{\rm s})$ diagram, and then show the track
in the $(V-I)_0$-$(M_I-2.5\log f_{\rm s})$ diagram.

\subsection{V496~Sct 2009}
\label{v496_sct_vi}
\citet{hac19ka} obtained $E(B-V)=0.45$, $(m-M)_V=13.7$, $d=2.9$~kpc,
and $\log f_{\rm s}= +0.30$ for V496~Sct.
We reanalyzed the $BVI_{\rm C}$ light/color curves of V496~Sct in Appendix
\ref{v496_sct_bvi} and obtained similar parameters as those of 
\citet{hac19ka} except for $(m-M)_V=13.6$ and $d=2.76$~kpc.
Then, we have $(m-M')_V=13.6+0.75=14.35$ and plot the
$(B-V)_0$-$(M_V-2.5\log f_{\rm s})$ diagram in Figure
\ref{hr_diagram_v496_sct_v959_mon_v834_car_v1369_cen_outburst}(a).
This time-stretched color-magnitude diagram is essentially the same
as Figure 15(d) of \citet{hac19ka}.
V496~Sct almost overlaps with the upper branch of LV~Vul.
Therefore, we expect that these two nova tracks overlap each other
also in the $(V-I)_0$-$(M_I-2.5\log f_{\rm s})$ diagram.

The distance modulus in $I_{\rm C}$ band, $(m-M)_I=12.9$, is taken from
Appendix \ref{v496_sct_bvi}.
Then, its time-stretched value of
\begin{equation}
(m-M')_I \equiv (m - (M - 2.5 \log f_{\rm s}))_I
\label{time-stretched_distance_modulus_i}
\end{equation}
is $(m-M')_I=12.9+0.75=13.65$.
The peak $I_{\rm C}$ brightness is $M'_I= M_I-2.5\log f_{\rm s}
= -6.9 - 0.75 = -7.65$ from the data of \citet{raj12}.

We plot the $(V-I)_0$-$(M_I-2.5\log f_{\rm s})$ diagram in Figure 
\ref{hr_diagram_v496_sct_v959_mon_v834_car_v1369_cen_outburst_vi}(a).
Here, we adopt the $BVI_{\rm C}$ data from 
the Small and Medium Aperture Telescope System \citep[SMARTS,][]{wal12}
and \citet{raj12}.
We define the template track of V496~Sct by the thick solid orange line
mainly from the data of \citet{raj12}.

In the rising phase of the $I_{\rm C}$ magnitude,
the track is close to the vertical solid red line of
$(V-I)_0= +0.22$, which is the intrinsic $V-I$ color of optically thick
free-free emission, i.e., $F_\nu\propto \nu^{2/3}$ \citep{wri75}.
Then the track goes to the redder side after the $I_{\rm C}$ peak
because the emission lines such as \ion{O}{1} $\lambda\lambda 7774$,
8446, and \ion{Ca}{2} $\lambda\lambda 8498$, 8542 contribute to the
$I_{\rm C}$ band and make the $V-I$ color redder.
After the nova entered the nebular phase, strong emission lines such as
[\ion{O}{3}] $\lambda\lambda 4959$, 5007, and [\ion{N}{2}] $\lambda 5755$
contribute much to the $V$ band and make $V-I$ color bluer
\citep[see Figures 6 and 7 of][for the spectral evolution of V496~Sct]{raj12}.

\subsection{V959~Mon 2012}
\label{v959_mon_vi}
\citet{hac18k} examined this nova in detail and obtained $E(B-V)=0.38$,
$(m-M)_V=13.15$, $d=2.5$~kpc, and $\log f_{\rm s}= +0.14$.
We reanalyze the $BVI_{\rm C}$ light/color curves of V959~Mon in Appendix 
\ref{v959_mon_bvi} and obtained the similar results of $E(B-V)=0.38$,
$(m-M)_V=13.15$, $d=2.5$~kpc, and $\log f_{\rm s}= +0.18$.  Then, 
we have $(m-M')_V=13.15+0.45=13.6$ and plot the
$(B-V)_0$-$(M_V-2.5\log f_{\rm s})$ diagram in Figure
\ref{hr_diagram_v496_sct_v959_mon_v834_car_v1369_cen_outburst}(b).
This time-stretched color-magnitude diagram is similar to
Figure 5(b) of \citet{hac18k}.  The track of V959~Mon almost overlaps
with the lower branch of LV~Vul (orange line).
No early phase light curve of V959~Mon is available because of the
proximity to the Sun.  
We suppose that these two nova tracks overlap each other
also in the $(V-I)_0$-$(M_I-2.5\log f_{\rm s})$ diagram, although
no $I$/$I_{\rm C}$ band data are available for LV~Vul.

The distance modulus in $I_{\rm C}$ band, $(m-M)_I=12.55$, is 
taken from the result in Appendix \ref{v959_mon_bvi}.
Then, we have $(m-M')_I=12.55+0.45=13.0$.
We plot the $(V-I)_0$-$(M_I-2.5\log f_{\rm s})$ diagram in Figure 
\ref{hr_diagram_v496_sct_v959_mon_v834_car_v1369_cen_outburst_vi}(b).
Here, we adopt the $BVI_{\rm C}$ data from \citet{mun13b}, the
archives of the Variable Star Observers League of Japan (VSOLJ), 
the American Association of Variable Star Observers (AAVSO), and SMARTS.
We define the template track of V959~Mon by the thick solid cyan line
mainly from the data of \citet{mun13b}.  Assuming that the early phase
data are similar to those of V496~Sct, we construct the data of
V959~Mon (solid cyan line) in the early phase.  Here, we further 
assumed that the template track of V496~Sct should be
located on or above the V959~Mon track because a part of the V496~Sct
track is below the line of V959~Mon and this part is owing to the dust 
blackout effect.  If no dust blackout occurs, the original V496~Sct track 
should be located on or above the V959~Mon track.

In the $(B-V)_0$-$(M_V-2.5\log f_{\rm s})$ diagram of Figure 
\ref{hr_diagram_v496_sct_v959_mon_v834_car_v1369_cen_outburst}(b),
V959~Mon follows the lower branch of LV~Vul track.
Thus, we expect that the LV~Vul type novae have two branches also in the
$(V-I)_0$-$(M_I-2.5\log f_{\rm s})$ diagram and V959~Mon follows
the lower branch as shown in Figure 
\ref{hr_diagram_v496_sct_v959_mon_v834_car_v1369_cen_outburst_vi}(b).
This expectation is further examined in the next subsections.

\subsection{V834~Car 2012}
\label{v834_car_vi}
\citet{hac19kb} obtained $E(B-V)=0.50$, $(m-M)_V=17.25$, $d=14$~kpc,
and $\log f_{\rm s}= -0.19$.  We reanalyzed the $BVI_{\rm C}$ light/color
curves of V834~Car and obtained $E(B-V)=0.50$, $(m-M)_V=17.25$, $d=14$~kpc,
and $\log f_{\rm s}= -0.02$ in Appendix \ref{v834_car_bvi}.
The main difference is the timescaling factor of $f_{\rm s}$.  Then,
we have $(m-M')_V=17.25-0.05=17.2$ and plot the
$(B-V)_0$-$(M_V-2.5\log f_{\rm s})$ diagram in Figure
\ref{hr_diagram_v496_sct_v959_mon_v834_car_v1369_cen_outburst}(c).
This time-stretched color-magnitude diagram is slightly shifted up compared
with the previous result \citep{hac19kb},
but still consistently follows the lower branch of LV~Vul (orange line).

The distance modulus in $I_{\rm C}$ band, $(m-M)_I=16.45$, is 
taken from Appendix \ref{v834_car_bvi}.
Then, we have $(m-M')_I=16.45-0.05=16.4$.
We plot the $(V-I)_0$-$(M_I-2.5\log f_{\rm s})$ diagram in Figure 
\ref{hr_diagram_v496_sct_v959_mon_v834_car_v1369_cen_outburst_vi}(c).
Here, we adopt the $BVI_{\rm C}$ data from AAVSO and SMARTS.
In this diagram, the SMARTS data of V834~Car (filled red circles)
almost follows the track of V959~Mon (cyan line).
The overlapping of V834~Car and V959~Mon in both the $V$ and $I_{\rm C}$
color-magnitude diagrams strongly suggests the presence of the lower
branch after the nebular phase started.  The data of AAVSO (unfilled blue
circles) seems to correspond to the (upper) branch (upper orange line)
of V496~Sct.   In what follows, we assume that LV~Vul has two branches
in the $(V-I)_0$-$(M_I-2.5\log f_{\rm s})$ diagram after the start of
nebular phase.  One branch is depicted by V496~Sct and the other is
represented by V959~Mon.  We denote the both branches by the same
orange line as shown in Figures
\ref{hr_diagram_v1663_aql_v2615_oph_v5666_sgr_v2659_cyg_outburst_vi}
and \ref{hr_diagram_lv_vul_4type_4fig_vi}.

\subsection{V1369~Cen 2013}
\label{v1369_cen_vi}
\citet{hac19ka} obtained $E(B-V)=0.11$, $(m-M)_V=10.25$, $d=0.96$~kpc,
and $\log f_{\rm s}= +0.17$.
We reanalyzed the $BVI_{\rm C}$ light/color
curves of V1369~Cen in Appendix \ref{v1369_cen_bvi}
and confirmed the same results.  Then, we have $(m-M')_V=10.25+0.4=10.65$
and plot the $(B-V)_0$-$(M_V-2.5\log f_{\rm s})$ diagram in Figure
\ref{hr_diagram_v496_sct_v959_mon_v834_car_v1369_cen_outburst}(d).
This time-stretched color-magnitude diagram is essentially the same
as Figure 9(c) of \citet{hac19ka}.  The track of V1369~Cen overlaps 
with the upper branch of LV~Vul (orange line). 
Therefore, we expect that these two novae overlap 
in the $(V-I)_0$-$(M_I-2.5\log f_{\rm s})$ diagram, too.

The distance modulus in $I_{\rm C}$ band, $(m-M)_I=10.1$, is 
taken from Appendix \ref{v1369_cen_bvi}.  Then, we have 
$(m-M')_I=10.1+0.4=10.5$.  The peak $I_{\rm C}$ brightness is 
$M'_I= M_I-2.5\log f_{\rm s}= -6.95 - 0.4 = -7.35$.
We plot the $(V-I)_0$-$(M_I-2.5\log f_{\rm s})$ diagram in Figure 
\ref{hr_diagram_v496_sct_v959_mon_v834_car_v1369_cen_outburst_vi}(d).
Here, we adopt the $BVI_{\rm C}$ data from AAVSO, VSOLJ, and SMARTS.
We confirm that the track of V1369~Cen almost overlaps with the template track
of V496~Sct (or the upper branch of LV~Vul).

Thus, the overlapping of V1369~Cen and V496~Sct
in the $(V-I)_0$-$(M_I-2.5\log f_{\rm s})$ diagram supports the results of
$E(B-V)=0.11$, $(m-M)_I=10.1$, $d=0.96$~kpc,
and $\log f_{\rm s}= +0.17$ for V1369~Cen. 

To summarize the results of the four novae, 
(1) we confirm that the track splits into the two branches 
also in the $(V-I)_0$-$(M_I-2.5\log f_{\rm s})$ diagram and 
(2) the overlapping in both the $(B-V)_0$-$(M_V-2.5\log f_{\rm s})$
and $(V-I)_0$-$(M_I-2.5\log f_{\rm s})$
diagrams strongly supports that the adopted values
of $E(B-V)$, $(m-M)_V$, $(m-M)_I$, $d$, and $\log f_{\rm s}$
are reasonable.


\begin{figure*}
\plotone{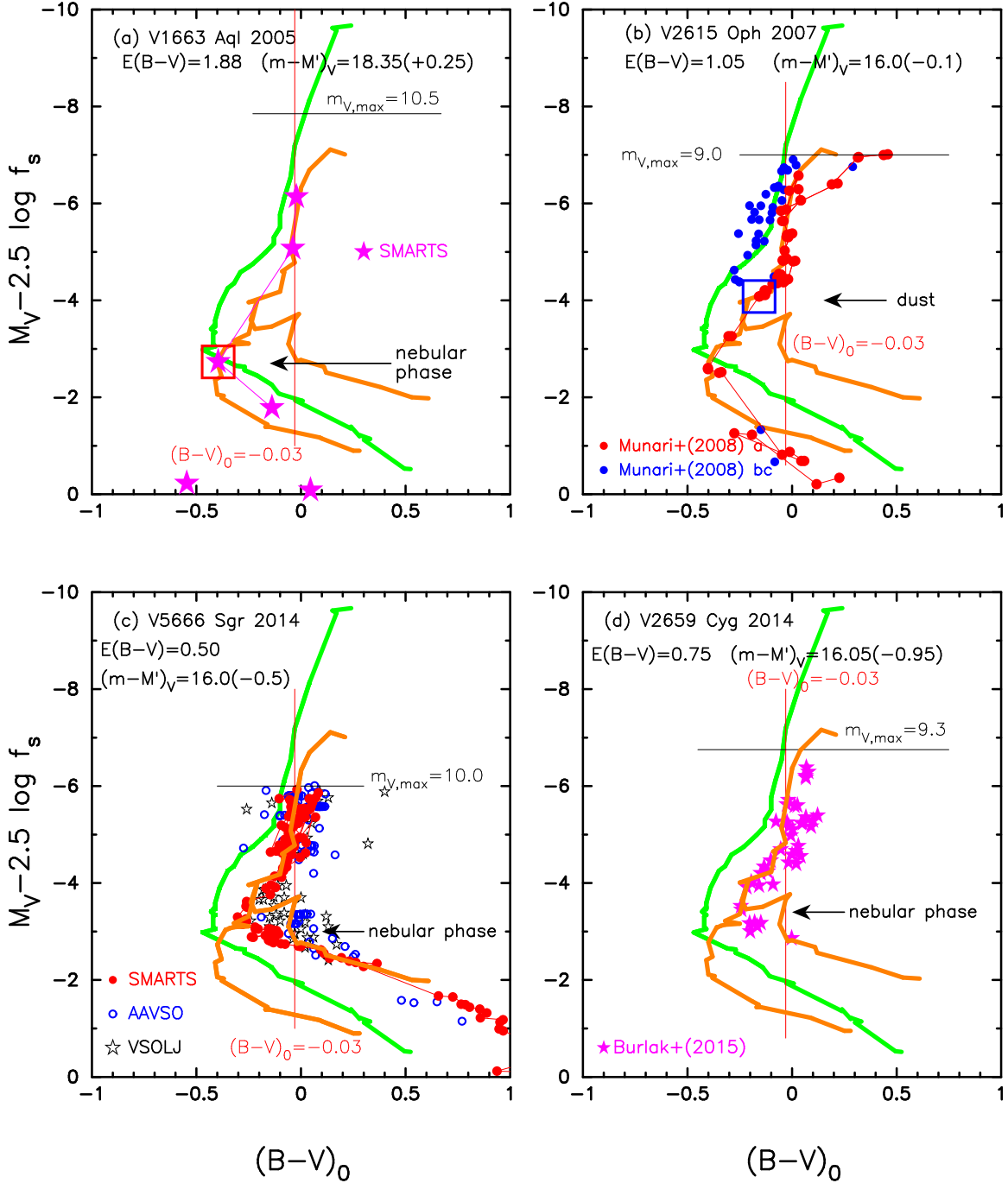}
\caption{
Same as Figure
\ref{hr_diagram_v496_sct_v959_mon_v834_car_v1369_cen_outburst}, but 
for (a) V1663~Aql, (b) V2615~Oph, (c) V5666~Sgr, and (d) V2659~Cyg.
\label{hr_diagram_v1663_aql_v2615_oph_v5666_sgr_v2659_cyg_outburst}}
\end{figure*}


\begin{figure*}
\plotone{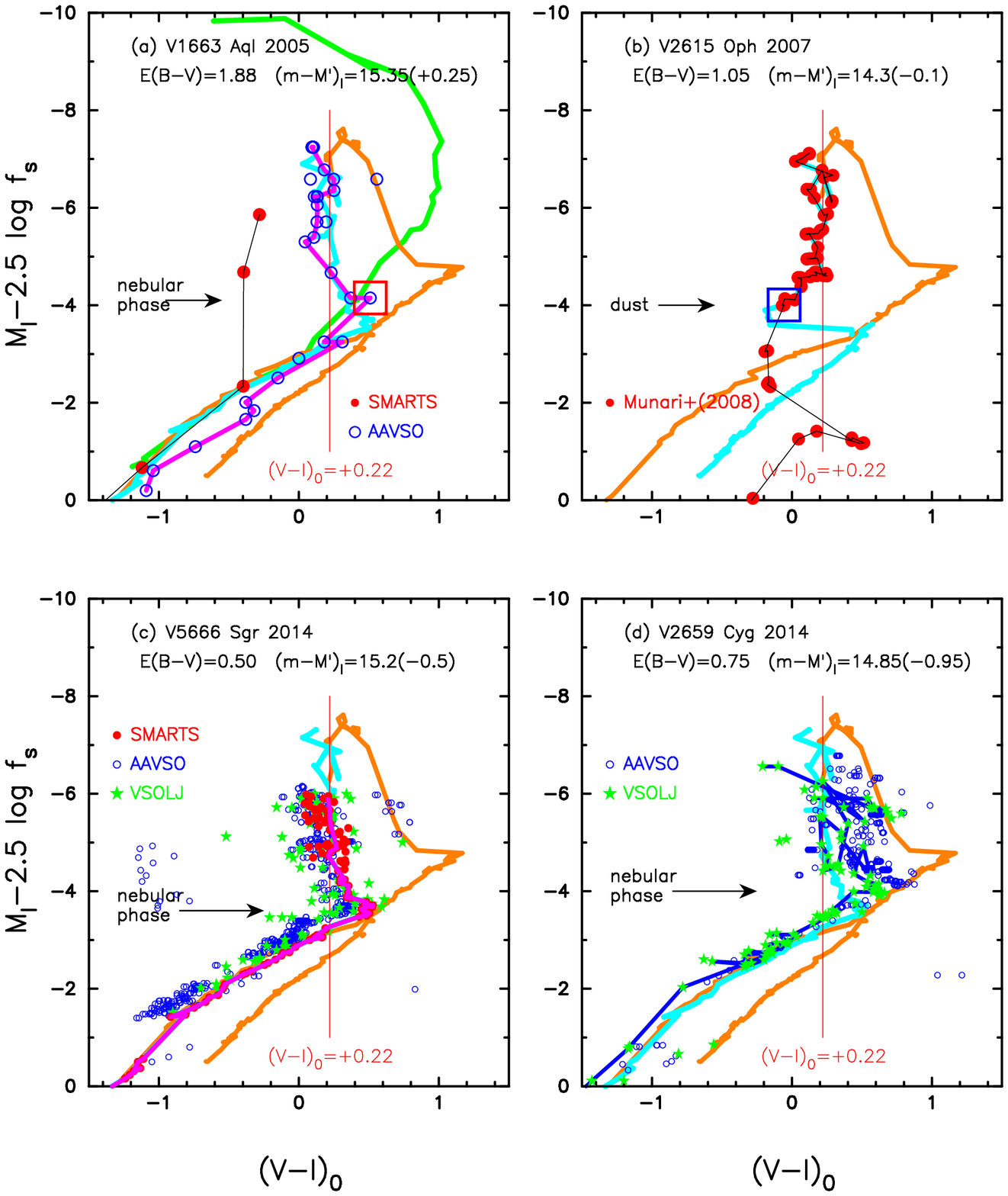}
\caption{
Same as Figure 
\ref{hr_diagram_v496_sct_v959_mon_v834_car_v1369_cen_outburst_vi},
but for (a) V1663~Aql, (b) V2615~Oph, (c) V5666~Sgr, and (d) V2659~Cyg.
The thick solid magenta lines in panels (a) and (c) represent the tracks
of V1663~Aql and V5666~Sgr, respectively. 
In panel (d), the thick solid blue line indicates the track of V2659~Cyg.
The thick solid cyan lines in panels (a), (c), and (d) depict
the reconstructed template track of V1663~Aql/V5666~Sgr while
the thick solid cyan line in panel (b) denotes the reconstructed
template track of V2615~Oph.  
The thick solid green line in panel (a) represents the template track
of V1500~Cyg in Section \ref{v1500_cyg_vi}.
See the text for more details.
\label{hr_diagram_v1663_aql_v2615_oph_v5666_sgr_v2659_cyg_outburst_vi}}
\end{figure*}

\subsection{V1663~Aql 2005}
\label{v1663_aql_vi}
\citet{hac19kb} obtained $E(B-V)=1.88$, $(m-M)_V=18.15$, $d=2.9$~kpc,
and $\log f_{\rm s}= -0.08$.
We have reanalyzed the $BVI_{\rm C}$ light/color curves of V1663~Aql
in Appendix \ref{v1663_aql_bvi} and obtained
$E(B-V)=1.88$, $(m-M)_V=18.6$, $d=3.6$~kpc, and $\log f_{\rm s}= -0.10$.
Then, we have $(m-M')_V=18.6-0.25=18.35$ and plot the
$(B-V)_0$-$(M_V-2.5\log f_{\rm s})$ diagram in Figure
\ref{hr_diagram_v1663_aql_v2615_oph_v5666_sgr_v2659_cyg_outburst}(a).
The track is slightly shifted up than the previous result.
The several data points of SMARTS
(filled magenta stars) almost follows the LV~Vul track (orange line)
in the early and middle phases and then transfers from the LV~Vul 
to the V1500~Cyg track (green line) in the later phase.

The distance modulus in $I_{\rm C}$ band is $(m-M)_I=15.6$ in 
Appendix \ref{v1663_aql_bvi}.
Then, we have $(m-M')_I=15.6-0.25=15.35$.
We plot the $(V-I)_0$-$(M_I-2.5\log f_{\rm s})$ diagram in Figure 
\ref{hr_diagram_v1663_aql_v2615_oph_v5666_sgr_v2659_cyg_outburst_vi}(a).
Here, we adopt the $BVI_{\rm C}$ data from AAVSO and SMARTS.
In the early phase, the track of V1663~Aql (blue open circles connected
by the thick magenta line: AAVSO data) goes down along
the vertical solid red line of $(V-I)_0=+0.22$. 
In the later phase, it transfers from $(V-I)_0=+0.22$ to 
the track of V1500~Cyg (green line).

We define the template track of V1663~Aql by the thick solid magenta
line from the data of AAVSO.  
In the early phase, i.e., before the nebular phase started,
the track follows $(V-I)_0=+0.22$, 
whereas the track of V496~Sct (orange line) goes toward the red. 
This is because
strong emission lines such as \ion{O}{1} and \ion{Ca}{2} triplet
contribute to the $I_{\rm C}$ band and make the $V-I$ color redder 
in V496~Sct \citep{raj12}.  
We also plot the template track of V5666~Sgr by the thick cyan line
(see Section \ref{v5666_sgr_vi}).  The track of V1663~Aql almost 
follows the track of V5666~Sgr.

\subsection{V2615~Oph 2007}
\label{v2615_oph_vi}
\citet{hac19ka} obtained $E(B-V)=0.90$, $(m-M)_V=15.95$, $d=4.3$~kpc,
and $\log f_{\rm s}= +0.20$.
We have reanalyzed the $BVI_{\rm C}$ multi-band light/color curves
of V2615~Oph in Appendix \ref{v2615_oph_bvi} and obtained
$E(B-V)=1.05$, $(m-M)_V=15.9$, $d=3.4$~kpc, and $\log f_{\rm s}= +0.04$.
The main differences are the color excess of $E(B-V)=1.05$ and the 
timescaling factor of $\log f_{\rm s}= +0.04$.  Then, we have $(m-M')_V=
15.9+0.1=16.0$ and plot the $(B-V)_0$-$(M_V-2.5\log f_{\rm s})$ diagram
in Figure \ref{hr_diagram_v1663_aql_v2615_oph_v5666_sgr_v2659_cyg_outburst}(b).
This time-stretched color-magnitude diagram is slightly shifted down
than that in Figure 15(a) of \citet{hac19ka}.  The data of \citet{mun08a}
almost follows the lower branch of LV~Vul (orange line), although
an optically-thick dust shell forms at
$M'_V\equiv M_V-2.5\log f_{\rm s}= -4.0 - 0.1= -4.1$.
Here, we plot separately the $B-V$ data of group ``a'' (filled red circles)
and groups ``b'' and ``c'' (filled blue circles) of \citet{mun08a}.
The group ``a'' data follows the LV~Vul track (orange line) while
the groups ``bc'' data are located on the V1500~Cyg track
(green line).

The distance modulus in $I_{\rm C}$ band, $(m-M)_I=14.2$, is 
taken from Appendix \ref{v2615_oph_bvi}.
Then, we have $(m-M')_I=14.2+0.1=14.3$.
The peak $I_{\rm C}$ brightness is $M'_I\equiv M_I-2.5\log f_{\rm s}= 
-7.0 - 0.1 = -7.1$.
We plot the $(V-I)_0$-$(M_I-2.5\log f_{\rm s})$ diagram in Figure 
\ref{hr_diagram_v1663_aql_v2615_oph_v5666_sgr_v2659_cyg_outburst_vi}(b).
Here, we adopt the $BVI_{\rm C}$ data from the group ``a'' of \citet{mun08a}.
In the early phase, V2615~Oph goes down along the line of $(V-I)_0=+0.22$. 
In the middle and later phases, the brightness drops due to the dust blackout.

For later use, we recover the brightness from the dust blackout
and locate it on the intrinsic position (thick cyan line)
of the V2615~Oph track in Figure
\ref{hr_diagram_v1663_aql_v2615_oph_v5666_sgr_v2659_cyg_outburst_vi}(b),
assuming
that the $(V-I_{\rm C})_0$ color is not affected by the dust shell and
the time-stretched $I_{\rm C}$ light curve follows the other novae
in Figure
\ref{v2615_oph_v5114_sgr_v1369_cen_v496_sct_i_vi_color_logscale}(a) of
Appendix \ref{v2615_oph_bvi}.
In what follows, we adopt this cyan line as a template nova track
of V2615~Oph. We plot this template track of V2615~Oph also in Figure
\ref{hr_diagram_lv_vul_4type_4fig_vi}(c).

\subsection{V5666~Sgr 2014}
\label{v5666_sgr_vi}
\citet{hac19ka} obtained $E(B-V)=0.50$, $(m-M)_V=15.4$, $d=5.8$~kpc,
and $\log f_{\rm s}= +0.25$.
We have reanalyzed the $BVI_{\rm C}$ multi-band light/color curves
of V5666~Sgr in Appendix \ref{v5666_sgr_bvi} and obtained
$E(B-V)=0.50$, $(m-M)_V=15.5$, $d=6.2$~kpc, and $\log f_{\rm s}= +0.20$.
The main difference is the timescaling factor of $\log f_{\rm s}= +0.20$.
Then, we have $(m-M')_V=15.5+0.5= 16.0$, which is almost the same as our
previous value of $(m-M')_V= 15.4 + 0.625 = 16.05$, and plot the 
$(B-V)_0$-$(M_V-2.5\log f_{\rm s})$ diagram in Figure
\ref{hr_diagram_v1663_aql_v2615_oph_v5666_sgr_v2659_cyg_outburst}(c).
This time-stretched color-magnitude diagram is similar to
Figure 9(d) of \citet{hac19ka}.  The track of V5666~Sgr
almost follows the upper branch of LV~Vul (orange line).

The distance modulus in $I_{\rm C}$ band, $(m-M)_I=14.7$, is 
taken from Appendix \ref{v5666_sgr_bvi}.
Then, we have $(m-M')_I= 14.7+0.5= 15.2$.
The peak $I_{\rm C}$ brightness is 
$M'_I\equiv M_I-2.5\log f_{\rm s}= -5.5 - 0.5 = -6.0$.
We plot the $(V-I)_0$-$(M_I-2.5\log f_{\rm s})$ diagram in Figure 
\ref{hr_diagram_v1663_aql_v2615_oph_v5666_sgr_v2659_cyg_outburst_vi}(c).
Here, we adopt the $BVI_{\rm C}$ data from AAVSO, VSOLJ, and SMARTS.
In the early phase, the track of V5666~Sgr almost follows 
the track of V1663~Aql/V2615~Oph  (cyan line), i.e., $(V-I)_0=+0.22$.
In the later phase, it follows the track of V496~Sct (upper orange line). 

We define the track of V5666~Sgr by the thick solid magenta
line from the data of SMARTS as shown in Figure 
\ref{hr_diagram_v1663_aql_v2615_oph_v5666_sgr_v2659_cyg_outburst_vi}(c).
We can see two overlappings: one is in the 
$(B-V)_0$-$(M_V-2.5\log f_{\rm s})$ and the other is in the 
$(V-I)_0$-$(M_I-2.5\log f_{\rm s})$.
Both the overlappings strongly support the results of
$E(B-V)=0.50$, $(m-M)_I=14.7$, $d=6.2$~kpc,
and $\log f_{\rm s}= +0.20$ for V5666~Sgr. 

We make a template track of V5666~Sgr both from the early phases of
V1663~Aql and V2615~Oph and from the middle and later phases of
V5666~Sgr and plot the template track (cyan line), as shown in Figure
\ref{hr_diagram_lv_vul_4type_4fig_vi}(d).


\begin{figure*}
\plotone{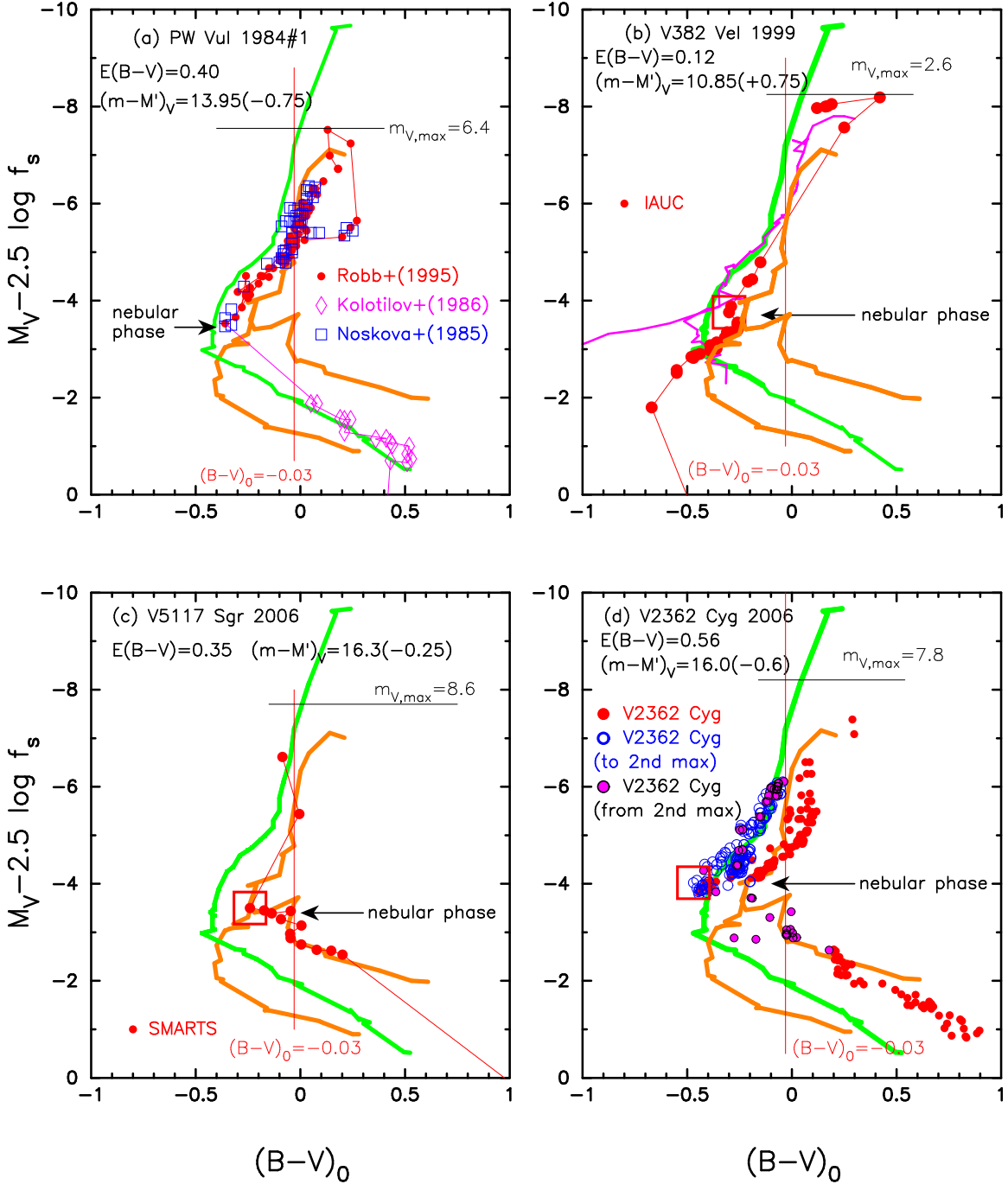}
\caption{
Same as Figure
\ref{hr_diagram_v496_sct_v959_mon_v834_car_v1369_cen_outburst}, but 
for (a) PW~Vul, (b) V382~Vel, (c) V5117~Sgr, and (d) V2362~Cyg.
\label{hr_diagram_pw_vul_v382_vel_v5117_sgr_v2362_cyg_outburst}}
\end{figure*}


\begin{figure*}
\plotone{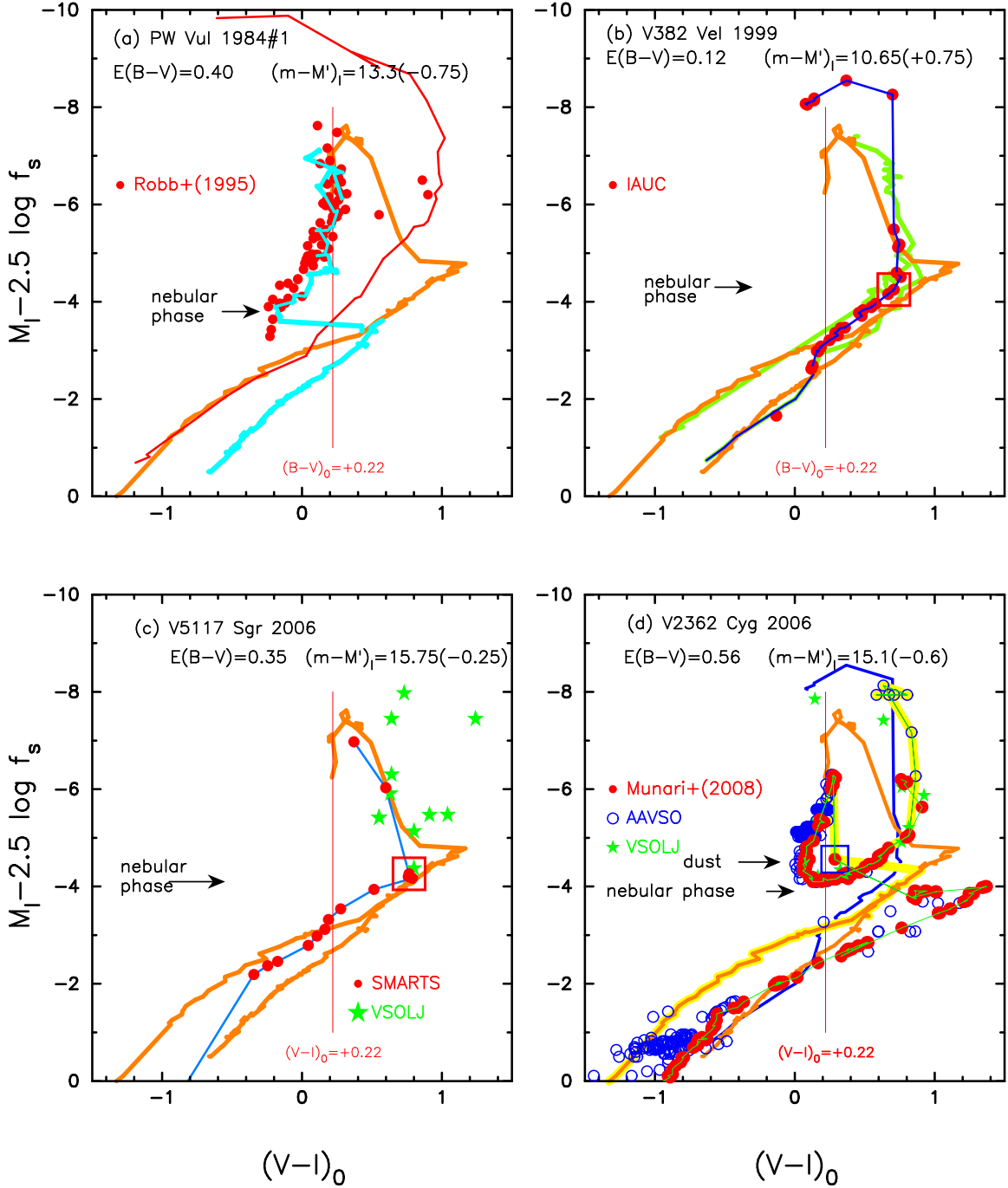}
\caption{
Same as Figure 
\ref{hr_diagram_v496_sct_v959_mon_v834_car_v1369_cen_outburst_vi},
but for (a) PW~Vul, (b) V382~Vel, (c) V5117~Sgr, and (d) V2362~Cyg.
The thick solid orange lines depict the reconstructed
template track of V496~Sct/V959~Mon.  In panel (a), the thick solid
cyan line denotes the reconstructed template track of
V2615~Oph and the solid red line represents the template track
of V1500~Cyg in Section \ref{v1500_cyg_vi}.  In panel (b), the thick
solid light-green line depicts the template track of V834~Car 
while the solid blue line denotes the track of V382~Vel.  
In panel (c), the cyan-blue line connects the V5117~Sgr data obtained by
SMARTS.  
In panel (d), the solid blue line indicates the template
track of V382~Vel while the thick solid yellow line represents the 
reconstructed template track of V2362~Cyg.  The thin light-green line
connects the original V2362~Cyg data obtained by \citet{mun08b} and AAVSO.
\label{hr_diagram_pw_vul_v382_vel_v5117_sgr_v2362_cyg_outburst_vi}}
\end{figure*}


\begin{figure*}
\plotone{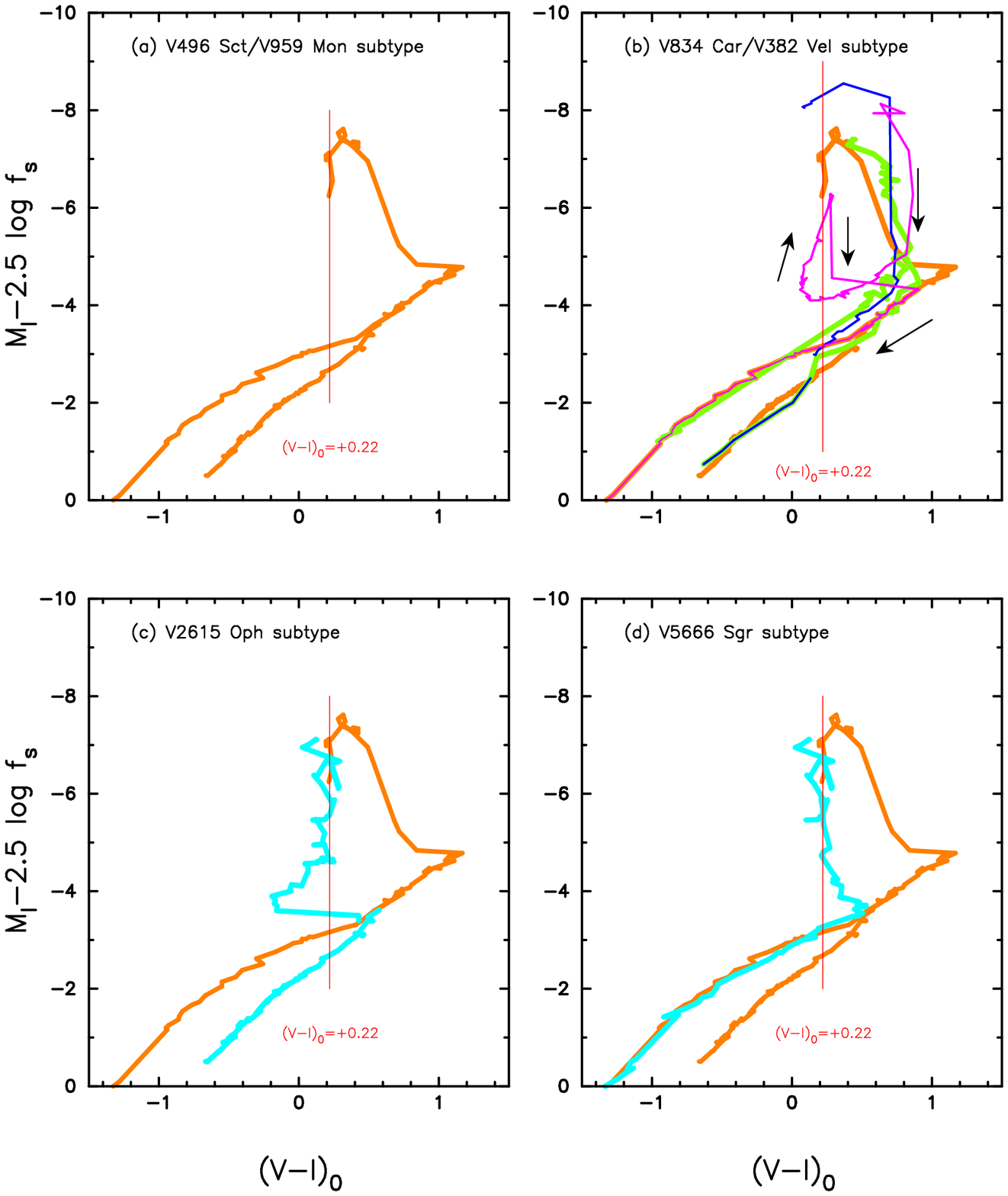}
\caption{
Template tracks of the LV~Vul type in the $(V-I)_0$-$(M_I-2.5\log f_{\rm s})$
diagram for (a) V496~Sct/V959~Mon subtype (thick solid orange line), 
(b) V834~Car/V382~Vel subtype (thick solid light-green line and blue line),
and (c) V2615~Oph subtype (thick solid cyan line),
(d) V5666~Sgr subtype (thick solid cyan line).
In panel (a), the track split into two branches in the later
phase after the nebular phase starts.
In panel (b), we add the V2362~Cyg track (thin magenta line), which
shows a prominent secondary maximum.  The evolution direction is
indicated by arrows.  
The vertical solid red line of $(V-I)_0=+0.22$ denotes the intrinsic
$V-I$ color of optically thick free-free emission.
\label{hr_diagram_lv_vul_4type_4fig_vi}}
\end{figure*}

\subsection{V2659~Cyg 2014}
\label{v2659_cyg_vi}
\citet{hac19kb} obtained $E(B-V)=0.80$, $(m-M)_V=15.7$, $d=4.4$~kpc,
and $\log f_{\rm s}= +0.52$ based on the $BV$ data.
We have reanalyzed the multi-band $UBVI_{\rm C}$ light curves in
Appendix \ref{v2659_cyg_ubvi} and obtained the new parameters of
$E(B-V)=0.75$, $(m-M)_V=15.1$, $d=3.6$~kpc,
and $\log f_{\rm s}= +0.37$.  The main differences are the distance
modulus in $V$ band and the timescaling factor.
We have $(m-M')_V=15.1 + 0.925= 16.05$ and plot the
$(B-V)_0$-$(M_V-2.5\log f_{\rm s})$ diagram in Figure
\ref{hr_diagram_v1663_aql_v2615_oph_v5666_sgr_v2659_cyg_outburst}(d).
We adopt only the data of \citet{bur15}.  The track of V2659~Cyg
almost follows the upper branch of LV~Vul (orange line).

The distance modulus in $I_{\rm C}$ band, $(m-M)_I=13.9$, is 
taken from Appendix \ref{v2659_cyg_ubvi}.
Then, we have $(m-M')_I=13.9 + 0.925= 14.85$.
The peak $I_{\rm C}$ brightness is 
$M'_I\equiv M_I-2.5\log f_{\rm s}= -5.98 - 0.925 = -6.9$ form the VSOLJ data.
We plot the $(V-I)_0$-$(M_I-2.5\log f_{\rm s})$ diagram in Figure 
\ref{hr_diagram_v1663_aql_v2615_oph_v5666_sgr_v2659_cyg_outburst_vi}(d).
Here, we adopt the $BVI_{\rm C}$ data from AAVSO and VSOLJ.
The track of V2659~Cyg (solid blue line) is between the track of 
V1663~Aql/V5666~Sgr (thick solid cyan line) and V496~Sct (upper 
thick solid orange line).  
The overlapping of V2659~Cyg and V1663~Aql/V5666~Sgr (or V496~Sct)
in the $(V-I)_0$-$(M_I-2.5\log f_{\rm s})$ diagram as well as
the overlappings of V2659~Cyg and LV~Vul both in the 
$(B-V)_0$-$(M_V-2.5\log f_{\rm s})$ and $(U-B)_0$-$(M_B-2.5\log f_{\rm s})$ 
diagrams may support the results of $E(B-V)=0.75$, $(m-M)_I=13.9$,
$d=3.6$~kpc, and $\log f_{\rm s}= +0.37$ for V2659~Cyg.

\subsection{PW~Vul 1984\#1}
\label{pw_vul_vi}
We have already reanalyzed the $UBVI$ data of PW~Vul
and obtained a new parameter set of $E(B-V)=0.40$, $(m-M)_U=11.7$,
$(m-M)_B=11.6$, $(m-M)_V=13.2$, $(m-M)_I=11.3$,
$d=2.5$~kpc, and $\log f_{\rm s}= +0.30$ in Section \ref{pw_vul_ub} 
and Appendix \ref{pw_vul_ubvi}.  
We plot the $(B-V)_0$-$(M_V-2.5\log f_{\rm s})$ diagram in Figure 
\ref{hr_diagram_pw_vul_v382_vel_v5117_sgr_v2362_cyg_outburst}(a) and
the $(V-I)_0$-$(M_I-2.5\log f_{\rm s})$ diagram in Figure 
\ref{hr_diagram_pw_vul_v382_vel_v5117_sgr_v2362_cyg_outburst_vi}(a).
The $(B-V)_0$-$(M_V-2.5\log f_{\rm s})$ track almost follows 
the LV~Vul track (orange lines) in the early and middle phases.
After the nebular phase started, the track goes along the track of
V1500~Cyg (green line).

In the $(V-I)_0$-$(M_I-2.5\log f_{\rm s})$ diagram of Figure 
\ref{hr_diagram_pw_vul_v382_vel_v5117_sgr_v2362_cyg_outburst_vi}(a),
we add the track of V1500~Cyg (red line) in Section \ref{v1500_cyg_vi}.
The $(V-I)_0$-$(M_I-2.5\log f_{\rm s})$ track of PW~Vul follows 
the reconstructed template track of V2615~Oph (cyan line) until the
nebular phase started.  We cannot obtain the track after the nebular
phase started because no $I$ data are available.  However, it could
follow the track of V1500~Cyg (solid red line) because the track of
PW~Vul follows the V1500~Cyg track in the
$(B-V)_0$-$(M_V-2.5\log f_{\rm s})$ diagram.   
The overlapping of PW~Vul and V2615~Oph in the
$(V-I)_0$-$(M_I-2.5\log f_{\rm s})$ diagram supports the values of
$E(B-V)=0.40$, $(m-M)_I=11.3$, $d=2.5$~kpc, and $\log f_{\rm s}= +0.30$.

\subsection{V382~Vel 1999}
\label{v382_vel_vi}
We have already reanalyzed the $UBVI$ data of V382~Vel
and obtained a new parameter set of $E(B-V)=0.12$, $(m-M)_U=11.78$,
$(m-M)_B=11.71$, $(m-M)_V=11.6$, $(m-M)_I=11.41$,
$d=1.76$~kpc, and $\log f_{\rm s}= -0.29$ in Section \ref{v382_vel_ub} 
and Appendix \ref{v382_vel_ubvi}.
We plot the $(B-V)_0$-$(M_V-2.5\log f_{\rm s})$ diagram in Figure 
\ref{hr_diagram_pw_vul_v382_vel_v5117_sgr_v2362_cyg_outburst}(b) and
the $(V-I)_0$-$(M_I-2.5\log f_{\rm s})$ diagram in Figure 
\ref{hr_diagram_pw_vul_v382_vel_v5117_sgr_v2362_cyg_outburst_vi}(b).
The $(B-V)_0$-$(M_V-2.5\log f_{\rm s})$ track almost follows 
the LV~Vul track (orange lines) in the middle phase.
The $(V-I)_0$-$(M_I-2.5\log f_{\rm s})$ track follows the template track 
of V834~Car (light-green line) in the middle and later phases.
The overlappings of the V382~Vel track with the LV~Vul track
in the $(B-V)_0$-$(M_V-2.5\log f_{\rm s})$ and with the V834~Car
track in the $(V-I)_0$-$(M_I-2.5\log f_{\rm s})$
suggests that the new values of $E(B-V)= 0.12$,
$d=1.76$~kpc, and $\log f_{\rm s}= -0.29$ are reasonable.

\subsection{V5117~Sgr 2006}
\label{v5117_sgr_vi}
\citet{hac19kb} obtained $E(B-V)=0.53$, $(m-M)_V=16.0$, $d=7.5$~kpc,
and $\log f_{\rm s}= +0.05$.
We have reanalyzed the $BVI_{\rm C}$ data of V5117~Sgr in Appendix 
\ref{v5117_sgr_bvi} and obtained a new parameter set of $E(B-V)=0.35$,
$(m-M)_B=16.4$,  $(m-M)_V=16.05$,  $(m-M)_I=15.5$, $d=9.8$~kpc,
and $\log f_{\rm s}= +0.10$.
Then, we have $(m-M')_V= 16.05+0.25= 16.3$ and plot the
$(B-V)_0$-$(M_V-2.5\log f_{\rm s})$ diagram in Figure
\ref{hr_diagram_pw_vul_v382_vel_v5117_sgr_v2362_cyg_outburst}(c).
\citet{hac19kb} concluded that V5117~Sgr belongs to the V1500~Cyg type
because the track is located on the template track of V1500~Cyg (green line).
However, the new track of V5117~Sgr in the $(B-V)_0$-$(M_V-2.5\log f_{\rm s})$ 
diagram is located on the template track of LV~Vul (orange line).
Therefore, we recategorized V5117~Sgr as the LV~Vul type in the
$(B-V)_0$-$(M_V-2.5\log f_{\rm s})$ diagram.

The distance modulus in $I_{\rm C}$ band, $(m-M)_I=15.5$, is 
taken from Appendix \ref{v5117_sgr_bvi}.
Then, we have $(m-M')_I= 15.5+0.25= 15.75$.
The peak $I_{\rm C}$ brightness is 
$M'_I= M_I-2.5\log f_{\rm s}= -7.72 - 0.25 = -7.97$.
We plot the $(V-I)_0$-$(M_I-2.5\log f_{\rm s})$ diagram in Figure 
\ref{hr_diagram_pw_vul_v382_vel_v5117_sgr_v2362_cyg_outburst_vi}(c).
Here, we adopt the $BVI_{\rm C}$ data from VSOLJ and SMARTS. 
The track of V5117~Sgr closely follows the template track of
V496~Sct (upper orange line).

We define the template track of V5117~Sgr by the cyan-blue line
from the data of SMARTS as shown in Figure 
\ref{hr_diagram_pw_vul_v382_vel_v5117_sgr_v2362_cyg_outburst_vi}(c).
The overlapping of V5117~Sgr and V496~Sct
in the $(V-I)_0$-$(M_I-2.5\log f_{\rm s})$ diagram may support the results of
$E(B-V)=0.35$, $(m-M)_I=15.5$, $d=9.8$~kpc,
and $\log f_{\rm s}= +0.10$ for V5117~Sgr.

\subsection{V2362~Cyg 2006}
\label{v2362_cyg_vi}
\citet{hac19ka} obtained $E(B-V)=0.60$, $(m-M)_V=15.4$, $d=5.1$~kpc,
and $\log f_{\rm s}= +0.25$.
We have reanalyzed the $BVI_{\rm C}$ multi-band light/color curves
of V2362~Cyg in Appendix \ref{v2362_cyg_bvi} and obtained a new set of
parameters, $E(B-V)=0.56$, $(m-M)_V=15.4$, $d=5.4$~kpc,
and $\log f_{\rm s}= +0.25$.  The main difference is the color excess.
We have $(m-M')_V=15.4+0.625=16.0$ and plot the
$(B-V)_0$-$(M_V-2.5\log f_{\rm s})$ diagram in Figure
\ref{hr_diagram_pw_vul_v382_vel_v5117_sgr_v2362_cyg_outburst}(d).
This diagram is essentially the same as Figure 13(d) of \citet{hac19ka}, 
but is shifted slightly toward the red.  
This nova shows a prominent secondary peak and makes a large loop
in the color-magnitude diagram.  
The track of V2362~Cyg almost follows the track of LV~Vul
(orange line) in the early decline phase.  The nova brightens up 
(secondary maximum) and follows the track of V1500~Cyg in the middle
phase.  In the final decline phase from the secondary maximum, 
it follows again the track of LV~Vul (upper orange line).
Thus, we conclude that V2362~Cyg belongs to the LV~Vul type
except for the secondary maximum phase.

The distance modulus in $I_{\rm C}$ band, $(m-M)_I=14.5$, is 
taken from Appendix \ref{v2362_cyg_bvi}.
Then, we have $(m-M')_I=14.5+0.625=15.1$.
The peak $I_{\rm C}$ brightness is 
$M'_I= M_I-2.5\log f_{\rm s}= -7.53 - 0.625 = -8.15$ from the data of AAVSO.
We plot the $(V-I)_0$-$(M_I-2.5\log f_{\rm s})$ diagram in Figure 
\ref{hr_diagram_pw_vul_v382_vel_v5117_sgr_v2362_cyg_outburst_vi}(d).
Here, we adopt the $BVI_{\rm C}$ data from \cite{mun08b}, AAVSO, and VSOLJ.
In the early phase, the track of V2362~Cyg roughly follows the track of
V382~Vel (solid blue line).  Then, it brightens up, reaches a secondary 
maximum, and then goes down along the thin solid red line of $(V-I)_0=+0.22$.
The nova formed an optically thin dust shell 
\citep[e.g.,][]{ara10, lyn08a, mun08b}.  The brightness
slightly drops and the $V-I$ color becomes redder as shown in Figure
\ref{hr_diagram_pw_vul_v382_vel_v5117_sgr_v2362_cyg_outburst_vi}(d).
In the later phase, it approaches the tracks of V496~Sct/V959~Mon
(lower orange line).  The data line of \citet{mun08b} is parallel to
the upper branch of V496~Sct/M959~Mon track after an optically thin
dust shell formed.  Therefore, our reconstructed template track follows
the upper branch of orange lines.  This is because the V2362~Cyg track almost
follow the upper branch of LV~Vul track in the 
$(B-V)$-$(M_V-2.5\log f_{\rm s})$ diagram and the $I_{\rm C}$ brightness
could be affected by the dust shell even in the later phase.

To summarize, the V2362~Cyg track first follows the track of
V382~Vel (blue line), then undergoes a secondary maximum,
and finally approaches the track of V496~Sct/V959~Mon (lower orange line).
We have reconstructed the template track of V2362~Cyg (yellow line)
by recovering the $I_{\rm C}$ brightness and $V-I_{\rm C}$ color
from the optically thin dust blackout
(see also Figure \ref{hr_diagram_lv_vul_4type_4fig_vi}(b)).

The path is very complicated but these paths roughly overlap with 
the other template novae in the $(V-I)_0$-$(M_I-2.5\log f_{\rm s})$ diagram.  
This may support the results of $E(B-V)=0.56$,
$(m-M)_I=14.5$, $d=5.4$~kpc, and $\log f_{\rm s}= +0.25$ for V2362~Cyg.

\subsection{Summary of the LV~Vul type}
\label{summary_lv_vul_type_vi}
We have classified nova tracks into two types on the  
$(B-V)_0$-$(M_V-2.5\log f_{\rm s})$ diagram,
that is, LV~Vul type and V1500~Cyg type \citep{hac19ka, hac19kb}.
We further divide the LV~Vul type into four subtypes on the 
$(V-I)_0$-$(M_I-2.5\log f_{\rm s})$ diagram,
that is, V496~Sct/V959~Mon subtype, V834~Car/V382~Vel subtype,
V2615~Oph subtype, and V5666~Sgr subtype, as shown in Figure
\ref{hr_diagram_lv_vul_4type_4fig_vi}.

No $I$ or $I_{\rm C}$ data are available for LV~Vul.
Therefore, we first define the template track of the LV~Vul type novae
with the combination of V496~Sct and V959~Mon as shown in Figure
\ref{hr_diagram_lv_vul_4type_4fig_vi}(a). 
The upper branch of the orange lines corresponds to the V496~Sct
and the lower branch represents the V959~Mon 
in the $(V-I)_0$-$(M_I-2.5\log f_{\rm s})$ diagram.
In the template track of V496~Sct/V959~Mon, we omit the part
of the V496~Sct track below the line (track) of V959~Mon,
because this part is owing to the dust blackout effect.  
If no dust blackout occurs, the ordinary V496~Sct track
should be located on or above the V959~Mon track.

The track splits into two branches just after the nebular phase starts.
This is because there are small differences between the $V$ filters
at their blue edges and strong emission lines such as [\ion{O}{3}]
contribute differently to their $V$ magnitudes as already discussed
in \citet{hac14k, hac15k, hac16k, hac16kb}. 
The V1369~Cen track almost overlaps with that of V496~Sct
in the $(V-I)_0$-$(M_I-2.5\log f_{\rm s})$ diagram as shown in Figure
\ref{hr_diagram_v496_sct_v959_mon_v834_car_v1369_cen_outburst_vi}(d).
This overlapping supports the template track of V496~Sct.

We also define the template track of V834~Car and V382~Vel by the 
light-green and blue lines, respectively, as shown in Figure 
\ref{hr_diagram_lv_vul_4type_4fig_vi}(b).
This track shows a smooth turn from toward the red to toward the
blue compared with the sharp cusp turn of V496~Sct/V959~Mon subtypes.
It is interesting that the V2362~Cyg track almost follows the track of
V382~Vel (blue line) before the secondary maximum phase
and then it follows the V496~Sct/V959~Mon track after the optically thin
dust blackout started.
We have reconstructed the template track of V2362~Cyg (magenta line)
by recovering the $I_{\rm C}$ brightness from the optically thin
dust blackout.

The intrinsic $V-I$ color of optically thick free-free emission is
$(V-I)_0= +0.22$.  In the early phase of nova outbursts, if no strong
emission lines such as \ion{O}{1} and \ion{Ca}{2} triplet contribute
to the $I_{\rm C}$ band and free-free emission dominates the spectra
of the novae, its $V-I$ color should be $(V-I)_0= +0.22$.
The V1663~Aql, V2615~Oph, V5666~Sgr, and V2659~Cyg tracks almost follow
the vertical red line of $(V-I)_0= +0.22$ in the early phase
of the nova outbursts, as shown in Figure
\ref{hr_diagram_v1663_aql_v2615_oph_v5666_sgr_v2659_cyg_outburst_vi}.
We define the template track of V2615~Oph by the cyan line
as shown in Figure \ref{hr_diagram_lv_vul_4type_4fig_vi}(c). 
This track has a sharp cusp at/near the turning points of the start of 
nebular phase.  We finally define the template track of V5666~Sgr by the 
cyan line as shown in Figure \ref{hr_diagram_lv_vul_4type_4fig_vi}(d).


\begin{figure*}
\plotone{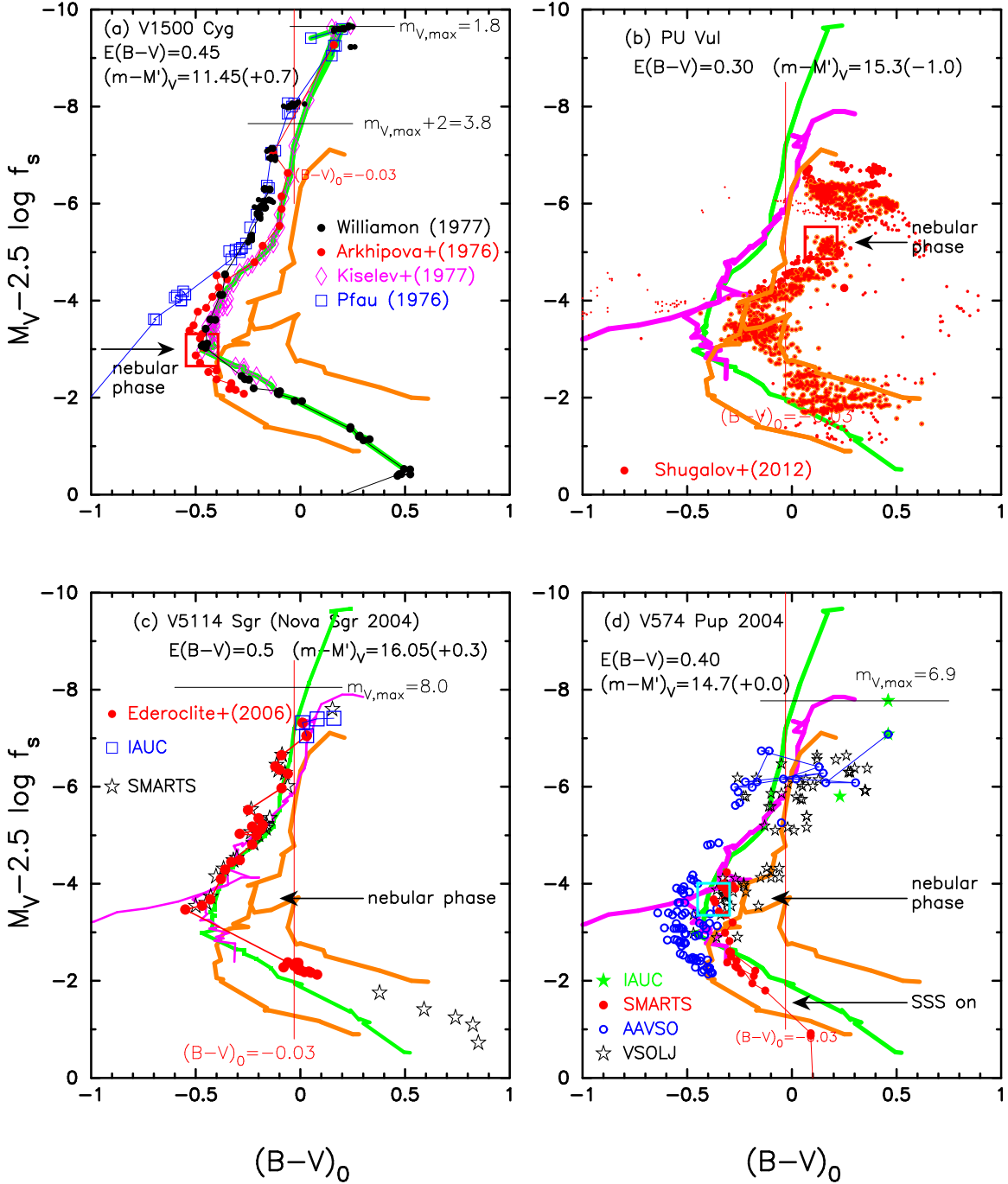}
\caption{
Same as Figure
\ref{hr_diagram_v496_sct_v959_mon_v834_car_v1369_cen_outburst}, but 
for (a) V1500~Cyg, (b) PU~Vul, (c) V5114~Sgr, and (d) V574~Pup.
See the text for the sources of observational data.
\label{hr_diagram_v1500_cyg_pu_vul_v5114_sgr_v574_pup_outburst}}
\end{figure*}


\begin{figure*}
\plotone{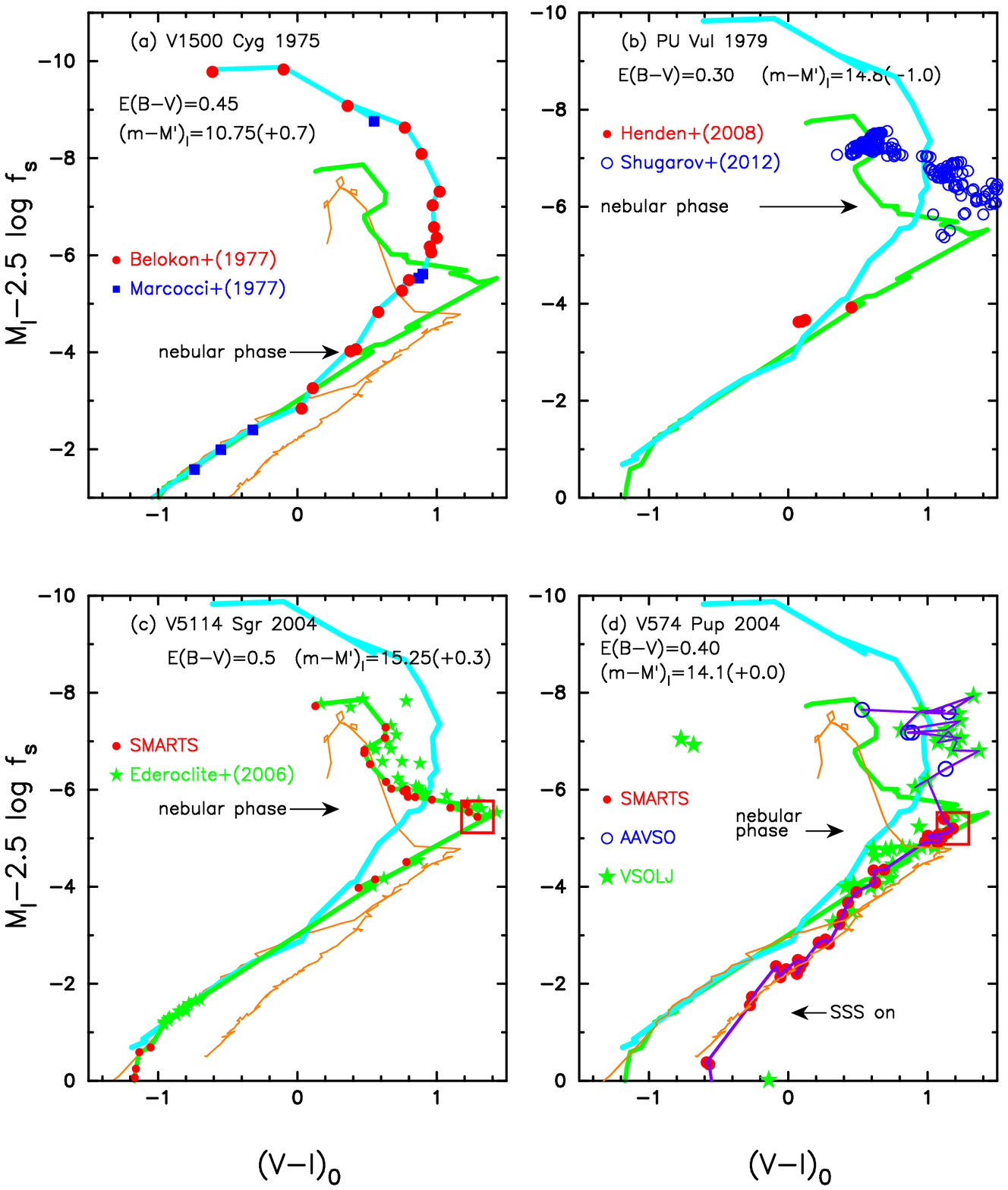}
\caption{
Same as Figure 
\ref{hr_diagram_v496_sct_v959_mon_v834_car_v1369_cen_outburst_vi},
but for (a) V1500~Cyg, (b) PU~Vul, (c) V5114~Sgr, and (d) V574~Pup.
The thick solid cyan lines depict the template track of V1500~Cyg.
The thin solid orange lines represent the template track of
V496~Sct/V959~Mon.  The thick solid green lines denote the template
track of V5114~Sgr.  In panel (d), the blue-magenta line shows 
the template track of V574~Pup.
See the text for the sources of observational data.
\label{hr_diagram_v1500_cyg_pu_vul_v5114_sgr_v574_pup_outburst_vi}}
\end{figure*}

\section{Time-Stretched $(V-I)_0$-$(M_I-2.5\log \lowercase{f}_{\rm 
\lowercase{s}})$ Color-Magnitude Diagram of the V1500~Cyg Type}
\label{vi_color_magnitude_diagram_v1500_cyg}

\subsection{V1500~Cyg 1975}
\label{v1500_cyg_vi}
We have revised the parameters of V1500~Cyg in 
Section \ref{v1500_cyg_ub} and Appendix \ref{v1500_cyg_ubvi}
and discussed the $(U-B)_0$-$(M_B-2.5\log f_{\rm s})$ diagram
in Section \ref{v1500_cyg_ub}.  Using the new set of $E(B-V)=0.45$,
$(m-M)_U=12.9$, $(m-M)_B=12.6$, $(m-M)_V=12.15$, $(m-M)_I=11.42$, 
$d=1.41$~kpc, and $\log f_{\rm s}= -0.28$,
we plot the $(B-V)_0$-$(M_V-2.5 \log f_{\rm s})$ diagram in Figure 
\ref{hr_diagram_v1500_cyg_pu_vul_v5114_sgr_v574_pup_outburst}(a).
Here, we adopt the $UBV$ data from \citet{ark76}, \citet{kis77}, 
\citet{pfa76}, and \citet{wil77}.  
We have already defined the template track of V1500~Cyg
by the thick solid green line.
It moves redward before the peak and comes back toward the blue.
After the peak it goes down almost straight.
After the nebular phase starts, the track turns to the red.
This is because strong [\ion{O}{3}] lines contribute to the $V$ band
in the nebular phase.

Using $(m-M')_I=11.45 - 0.7 = 10.75$ and $E(V-I)= 1.6 E(B-V)= 0.72$,
we also plot the $(V-I)_0$-$(M_I - 2.5\log f_{\rm s})$ diagram in Figure 
\ref{hr_diagram_v1500_cyg_pu_vul_v5114_sgr_v574_pup_outburst_vi}(a).
Here, we adopt the $VI$ data from \citet{bel77} and \citet{mar77}.  
We define the template track of V1500~Cyg by the thick solid cyan line
both from the data of \citet{bel77} and \citet{mar77}. 
Here, we shift Belokon \& Larionov's data toward blue by 0.75 mag 
and overlap them with those of \citet{mar77}, 
because the $V-I$ colors of \citet{bel77} are systematically
redder by 0.75 mag than those of \citet{mar77}.  
We plot two other template tracks, one is the V5114~Sgr (green line)
in Section \ref{v5114_sgr_vi} and the other is the V496~Sct/V959~Mon
(thin orange line) in Sections \ref{v496_sct_vi} and \ref{v959_mon_vi}. 
V5114~Sgr belongs to the V1500~Cyg type while V496~Sct/V959~Mon
belong to the LV~Vul type in the $(B-V)_0$-$(M_V-2.5 \log f_{\rm s})$ diagram.
The V1500~Cyg track deviates largely from the V496~Sct/V959~Mon 
template track (orange line) in the early and middle phases in Figure 
\ref{hr_diagram_v1500_cyg_pu_vul_v5114_sgr_v574_pup_outburst_vi}(a).
However, these three template tracks merge into one in the later phase.

\subsection{PU~Vul 1979}
\label{pu_vul_vi}
We have reanalyzed the $UBVI_{\rm C}$ data of PU~Vul 
in the $(U-B)_0$-$(M_B-2.5\log f_{\rm s})$ diagram in Figure 
\ref{hr_diagram_u_sco_t_pyx_v2659_cyg_pu_vul_outburst_ub}(d),
$(B-V)_0$-$(M_V-2.5\log f_{\rm s})$ diagram in Figure 
\ref{hr_diagram_v1500_cyg_pu_vul_v5114_sgr_v574_pup_outburst}(b), and
$(V-I)_0$-$(M_I-2.5\log f_{\rm s})$ diagram in Figure 
\ref{hr_diagram_v1500_cyg_pu_vul_v5114_sgr_v574_pup_outburst_vi}(b).
The $UBVI_{\rm C}$ data of PU~Vul are taken from \citet{hen08} and
\citet{shu12}.
As already explained in Section \ref{pu_vul_ub}, we have obtained
$\log f_{\rm s}= +0.40$ common in all the three color-magnitude diagrams,
taking $d=4.7$~kpc and $E(B-V)=0.30$ from \citet{kat12mh}.
The $(B-V)_0$-$(M_V-2.5\log f_{\rm s})$ track almost follows 
the LV~Vul track (orange lines) while
the $(V-I)_0$-$(M_I-2.5\log f_{\rm s})$ track almost follows 
the V5114~Sgr track (green line) in Section \ref{v5114_sgr_vi}.
The overlappings of PU~Vul with the V5114~Sgr
tracks in the $(V-I)_0$-$(M_I-2.5\log f_{\rm s})$ diagram
suggests that the set of $E(B-V)= 0.30$, $d=4.7$~kpc,
and $\log f_{\rm s}= +0.40$ are reasonable. 

The PU~Vul track is close to the LV~Vul track in the
$(B-V)_0$-$(M_V-2.5\log f_{\rm s})$ diagram. 
On the other hand, it is close to the V1974~Cyg track in the 
$(U-B)_0$-$(M_B-2.5\log f_{\rm s})$ diagram (Figure
\ref{hr_diagram_u_sco_t_pyx_v2659_cyg_pu_vul_outburst_ub}(d))
and is close to the
V5114~Sgr track in the $(V-I)_0$-$(M_I-2.5\log f_{\rm s})$ diagram.  
Because V1974~Cyg and V5114~Cyg belong to the V1500~Cyg type in the
$(B-V)_0$-$(M_V-2.5\log f_{\rm s})$ diagram, we may 
conclude that PU~Vul belongs to the V1500~Cyg type in the
$(V-I)_0$-$(M_I-2.5\log f_{\rm s})$ diagram.

\subsection{V5114~Sgr 2004}
\label{v5114_sgr_vi}
\citet{hac19ka} obtained $E(B-V)=0.47$, $(m-M)_V=16.65$, $d=10.9$~kpc,
and $\log f_{\rm s}= -0.12$ based on the $UBV$ light/color curves.
We have reanalyzed the $UBVI_{\rm C}$ data of V5114~Sgr in Appendix
\ref{v5114_sgr_ubvi} and obtain $E(B-V)=0.50$, $(m-M)_U=17.17$, 
$(m-M)_B=16.85$, $(m-M)_V=16.35$, $(m-M)_I=15.55$, 
$d=9.1$~kpc, and $\log f_{\rm s}= -0.12$.
Then, we have $(m-M')_V=16.35-0.3=16.05$ and plot the
$(B-V)_0$-$(M_V-2.5\log f_{\rm s})$ diagram in Figure
\ref{hr_diagram_v1500_cyg_pu_vul_v5114_sgr_v574_pup_outburst}(c).
The track of V5114~Sgr almost follows the track of V1974~Cyg (magenta line).
Therefore, V5114~Sgr belongs to the V1500~Cyg type
because V1974~Cyg is a member of the V1500~Cyg type \citep{hac19ka, hac19kb}. 
We expect that the V5114~Sgr track overlaps with the V1974~Cyg track
in the $(V-I)_0$-$(M_I-2.5\log f_{\rm s})$ diagram.
However, no sufficient $I/I_{\rm C}$ data of V1974~Cyg are available.

The distance modulus in $I_{\rm C}$ band, $(m-M)_I=15.55$, is taken
from Appendix \ref{v5114_sgr_ubvi}.
Then, we have $(m-M')_I=15.55-0.3=15.25$.
The peak $I_{\rm C}$ brightness is 
$M'_I= M_I-2.5\log f_{\rm s}= -8.19 + 0.3 = -7.9$.
We plot the $(V-I)_0$-$(M_I-2.5\log f_{\rm s})$ diagram in Figure 
\ref{hr_diagram_v1500_cyg_pu_vul_v5114_sgr_v574_pup_outburst_vi}(c).
Here, we adopt the $UBVI_{\rm C}$ data from \citet{ede06} and SMARTS.
In the early phase, the track of V5114~Sgr is located on a slightly redder
and brighter position than the V496~Sct/V959~Mon template track (orange
line).  In the later phase, it approaches and almost overlaps with
the tracks of V496~Sct (upper orange line) and V1500~Cyg (cyan line). 

We define the template track of V5114~Sgr by the thick solid green
line from the data of \citet{ede06} and SMARTS as shown in Figure 
\ref{hr_diagram_v1500_cyg_pu_vul_v5114_sgr_v574_pup_outburst_vi}(c).
We adopt this green line as one of the template tracks for the 
V1500~Cyg type in the $(V-I)_0$-$(M_I-2.5\log f_{\rm s})$ diagram.

\subsection{V574~Pup 2004}
\label{v574_pup_vi}
\citet{hac19ka} obtained $E(B-V)=0.45$, $(m-M)_V=15.0$, $d=5.3$~kpc,
and $\log f_{\rm s}= +0.10$.
We have reanalyzed the $BVI_{\rm C}K_{\rm s}$ multi-band light/color
curves in Appendix \ref{v574_pup_bvik} and obtained the new
parameters of $E(B-V)=0.40$, $(m-M)_V=14.7$, $d=5.7$~kpc,
and $\log f_{\rm s}= +0.0$.  The main differences are the color excess
and the timescaling factor.
We have $(m-M')_V=14.7+0.0=14.7$ and plot the
$(B-V)_0$-$(M_V-2.5\log f_{\rm s})$ diagram in Figure
\ref{hr_diagram_v1500_cyg_pu_vul_v5114_sgr_v574_pup_outburst}(d).
The AAVSO data (unfilled open blue circles) of V574~Pup almost follows
the track of V1974~Cyg (magenta line) while the SMARTS data follows
the LV Vul track (orange line).
The $(B-V)_0$ colors of SMARTS stays at $(B-V)_0\sim +0.2$
after the SSS phase starts \citep{nes07}.
This could be the color of an irradiated accretion disk.

The distance modulus in $I_{\rm C}$ band, $(m-M)_I=14.1$, is 
obtained in Appendix \ref{v574_pup_bvik}.
Then, we have $(m-M')_I=14.1+0.0=14.1$.
The peak $I_{\rm C}$ brightness is 
$M'_I= M_I-2.5\log f_{\rm s}= -8.0 - 0.0 = -8.0$.
We plot the $(V-I)_0$-$(M_I-2.5\log f_{\rm s})$ diagram in Figure 
\ref{hr_diagram_v1500_cyg_pu_vul_v5114_sgr_v574_pup_outburst_vi}(d).
Here, we adopt the $BVI_{\rm C}$ data from AAVSO, VSOLJ, and SMARTS.
In the early phase, the track of V574~Pup is located on a slightly redder
position than those of V5114~Sgr (green line) and V1500~Cyg (cyan line).  
In the middle phase, it overlaps with the track of V5114~Sgr 
and then slightly drops than that of V5114~Sgr.
In the later phase, it overlaps with the lower branch of V496~Sct/V959~Mon
and then slightly drops than that of V496~Sct/V959~Mon after the SSS phase
started.  This is also an effect of the accretion disk irradiated by
a hot WD.

Therefore, we define the template track of V574~Pup by the thick solid
blue-magenta line as shown in Figure 
\ref{hr_diagram_v1500_cyg_pu_vul_v5114_sgr_v574_pup_outburst_vi}(d).


\begin{figure*}
\plotone{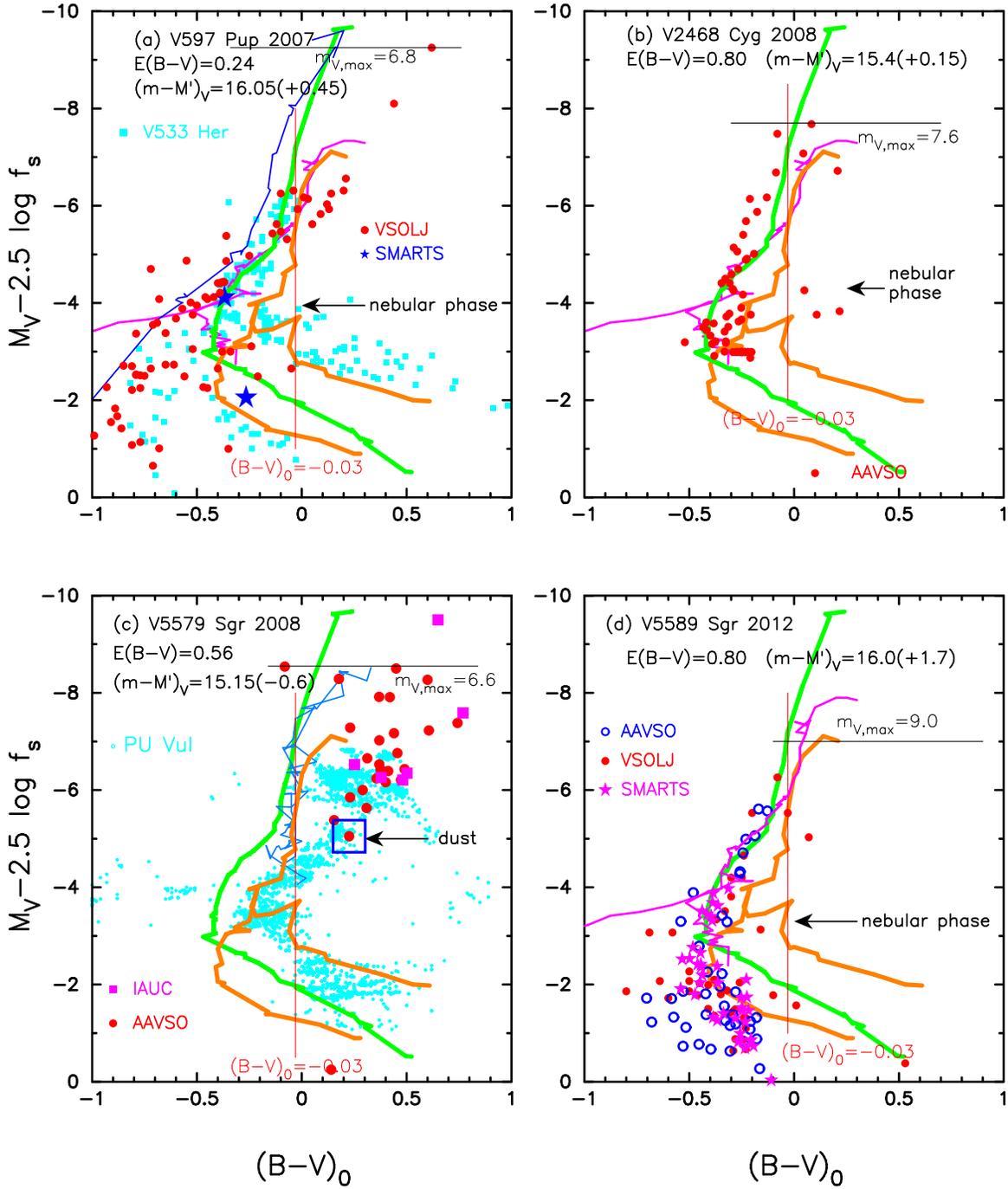}
\caption{
Same as Figure 
\ref{hr_diagram_v496_sct_v959_mon_v834_car_v1369_cen_outburst}, but 
for (a) V597~Pup, (b) V2468~Cyg, (c) V5579~Sgr, and (d) V5589~Sgr.
In panel (a), we add the track of V533~Her (filled cyan squares)
and another track of V1500~Cyg (thin solid blue line) obtained by
\citet{pfa76}.
In panel (c), we add the tracks of V1668~Cyg (thin solid cyan-blue lines)
and PU~Vul (small unfilled cyan circles).
In panel (d), we add the track of V1974~Cyg (solid magenta lines).
\label{hr_diagram_v597_pup_v2468_cyg_v5579_sgr_v5589_sgr_outburst}}
\end{figure*}


\begin{figure*}
\plotone{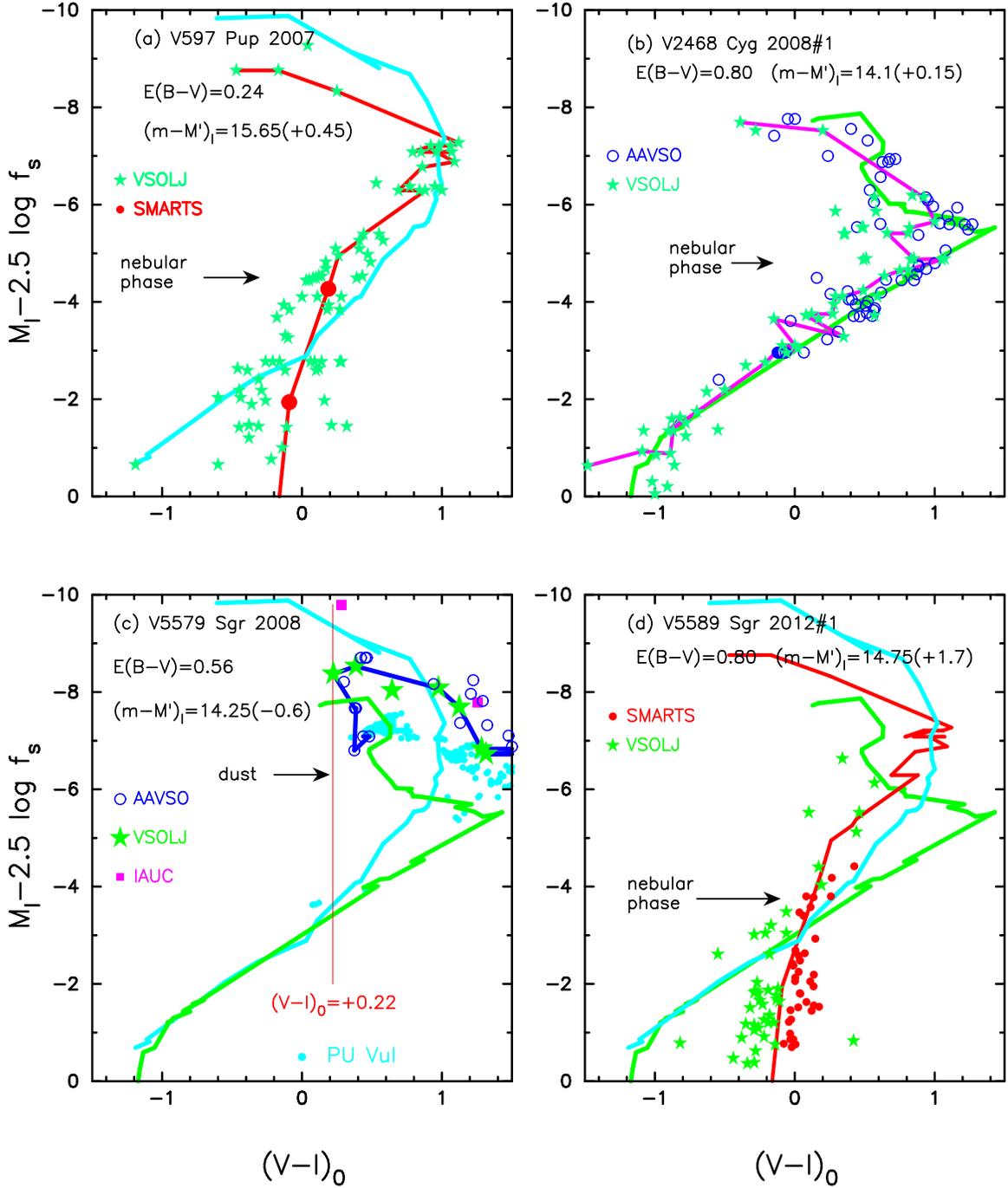}
\caption{
Same as Figure 
\ref{hr_diagram_v496_sct_v959_mon_v834_car_v1369_cen_outburst_vi},
but for (a) V597~Pup, (b) V2468~Cyg, (c) V5579~Sgr, and (d) V5589~Sgr.
The thick solid cyan, green, magenta, and red lines denote the template
tracks of V1500~Cyg, V5114~Sgr, V2468~Cyg, and V597~Pup, respectively.
\label{hr_diagram_v597_pup_v2468_cyg_v5579_sgr_v5589_sgr_outburst_vi}}
\end{figure*}


\begin{figure*}
\plotone{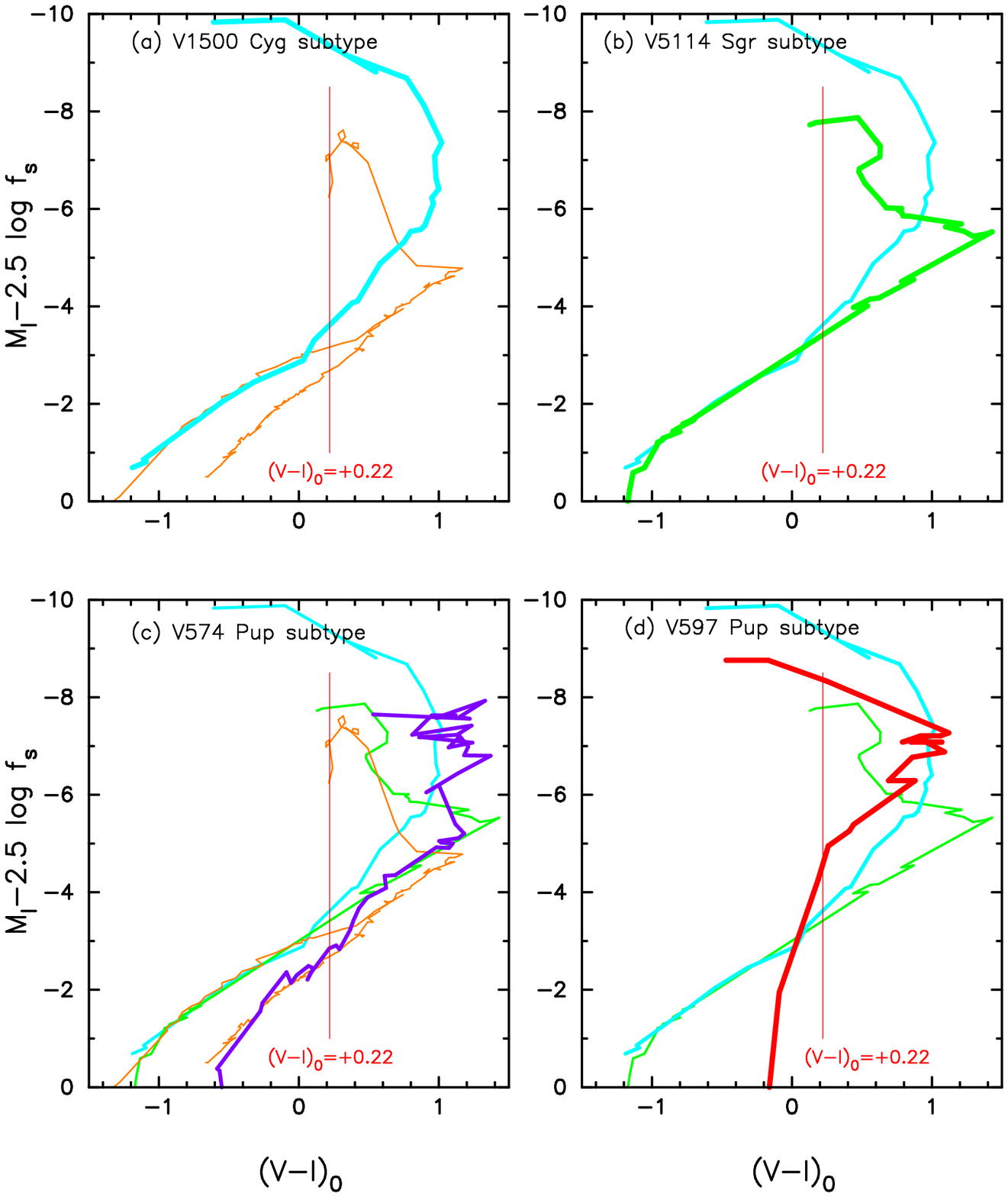}
\caption{
Template tracks of the V1500~Cyg type novae 
in the $(V-I)_0$-$(M_I-2.5\log f_{\rm s})$ diagram;
(a) V1500~Cyg subtype (thick solid cyan line), 
(b) V5114~Sgr subtype (thick solid green line),
(c) V574~Pup subtype (thick solid blue-magenta line), and 
(d) V597~Pup subtype (thick solid red line).
We add the vertical solid red line of $(V-I)_0=+0.22$. 
In panel (a), we add the reconstructed template track of V496~Sct/V959~Mon
subtype (thin solid orange lines) for comparison, which is 
an LV~Vul type nova.   The redder $(V-I)_0$ colors of V574~Pup
and V597~Pup in the later phase ($M'_I \equiv M_I -2.5 \log f_{\rm s} 
> -2$) originate from a large contribution 
of an irradiated accretion disk or companion star.   
\label{hr_diagram_v1500_cyg_type_4fig_vi}}
\end{figure*}

\subsection{V597~Pup 2007}
\label{v597_pup_vi}
\citet{hac19kb} obtained $E(B-V)=0.24$, $(m-M)_V=16.4$, $d=13.5$~kpc,
and $\log f_{\rm s}= -0.18$ for V597~Pup.
We have reanalyzed the multi-band light/color curves of V597~Pup in
Appendix \ref{v597_pup_bvi} and obtained a similar set of $E(B-V)=0.24$, 
$(m-M)_I=16.1$, $(m-M)_V=16.5$, $(m-M)_B=16.75$, $d=14.2$~kpc, 
and $\log f_{\rm s}= -0.18$.
Then, we have $(m-M')_V=16.5-0.45=16.05$ and plot the
$(B-V)_0$-$(M_V-2.5\log f_{\rm s})$ diagram in Figure
\ref{hr_diagram_v597_pup_v2468_cyg_v5579_sgr_v5589_sgr_outburst}(a).
This nova shows a super-bright peak like V1500~Cyg as already discussed
in Figure 45 of \citet{hac19kb}.  The peak $V$ brightness reaches
$M_V\sim -9.7$.  The data of VSOLJ are so scattered that we cannot
precisely specify a track.  We adopt the SMARTS data (filled blue
stars) that almost follows the tracks of V1500~Cyg (green line) and
regard that V597~Pup belongs to the V1500~Cyg type.

We add another V1500 Cyg track (thin blue line) from the data of 
\citet{pfa76} and the V533~Her data (filled cyan squares) for comparison. 
We think that the data of VSOLJ (filled red circles)
almost follow this blue line and the V533~Her track.  These two also 
suggest that the V597~Pup belongs to the V1500~Cyg type,
because V533~Her also belongs to the V1500~Cyg type.  

The distance modulus in $I_{\rm C}$ band, $(m-M)_I=16.1$, is 
taken from Appendix \ref{v597_pup_bvi}.
Then, we have $(m-M')_I=16.1-0.45=15.65$.
The peak $I_{\rm C}$ brightness is 
$M'_I= M_I-2.5\log f_{\rm s}= -9.7 + 0.45 = -9.25$.
We plot the $(V-I)_0$-$(M_I-2.5\log f_{\rm s})$ diagram in Figure 
\ref{hr_diagram_v597_pup_v2468_cyg_v5579_sgr_v5589_sgr_outburst_vi}(a).
Here, we adopt the $BVI_{\rm C}$ data from VSOLJ and SMARTS. 
In the early phase, the track of V597~Pup is close to the track of
V1500~Cyg (solid cyan line).  Then, it deviates from and is located
slightly above the V1500~Cyg track in the middle phase.
Finally, it drops sharply near $(V-I)_0\sim 0.0$.
We suppose that this almost constant color of $(V-I)_0\sim 0.0$
is due to the effect of an irradiated accretion disk (or a companion star).
We construct the template track of V597~Pup (solid red line)
from the data of VSOLJ in the early phase and then from the data of
SMARTS in the later phase, as shown in Figures 
\ref{hr_diagram_v597_pup_v2468_cyg_v5579_sgr_v5589_sgr_outburst_vi}(a)
and \ref{hr_diagram_v1500_cyg_type_4fig_vi}(d).

\subsection{V2468~Cyg 2008}
\label{v2468_cyg_vi}
\citet{hac19ka} obtained $E(B-V)=0.65$, $(m-M)_V=16.2$, $d=6.9$~kpc,
and $\log f_{\rm s}= +0.38$.
We have reanalyzed the $BVI_{\rm C}$ multi-band light/color curves
of V2468~Cyg in Appendix \ref{v2468_cyg_bvi} and obtained a new set
of $E(B-V)=0.80$, $(m-M)_V=15.55$, $d=4.1$~kpc,
and $\log f_{\rm s}= -0.06$ for V2468~Cyg.
The main difference is the timescaling factor of
$\log f_{\rm s}= -0.06$.  This is because we overlap the $V-I_{\rm C}$ 
and $B-V$ color curves of V2468~Cyg with other novae as much as possible.

Then, we have $(m-M')_V=15.55 - 0.15= 15.4$ and plot the
$(B-V)_0$-$(M_V-2.5\log f_{\rm s})$ diagram in Figure
\ref{hr_diagram_v597_pup_v2468_cyg_v5579_sgr_v5589_sgr_outburst}(b).
The peak $V$ brightness reaches $M'_V= M_V-2.5\log f_{\rm s} 
\sim -7.95 + 0.15 = -7.8$ from the data of AAVSO.
The track almost follow the track of V1500~Cyg (green line) or
V1974~Cyg (magenta line), although the AAVSO data are rather scattered. 

The distance modulus in $I_{\rm C}$ band, $(m-M)_I=14.25$, is 
taken from Appendix \ref{v2468_cyg_bvi}.
Then, we have $(m-M')_I=14.25 - 0.15 = 14.1$.
The peak $I_{\rm C}$ brightness is 
$M'_I= M_I-2.5\log f_{\rm s}= -7.9 + 0.15 = -7.75$.
We plot the $(V-I)_0$-$(M_I-2.5\log f_{\rm s})$ diagram in Figure 
\ref{hr_diagram_v597_pup_v2468_cyg_v5579_sgr_v5589_sgr_outburst_vi}(b).
Here, we adopt the $BVI_{\rm C}$ data from AAVSO and VSOLJ.
The track of V2468~Cyg almost follows 
the template track of V5114~Sgr (solid green line).
We define the template track of V2468~Cyg by the thick solid
magenta line from the data of VSOLJ as shown in Figure 
\ref{hr_diagram_v597_pup_v2468_cyg_v5579_sgr_v5589_sgr_outburst_vi}(b).
The overlapping of V2468~Cyg and V5114~Sgr tracks may support 
our new set of parameters, i.e., $E(B-V)=0.80$, $(m-M)_I=14.25$,
$d=4.1$~kpc, and $\log f_{\rm s}= -0.06$.

\subsection{V5579~Sgr 2008}
\label{v5579_sgr_vi}
This nova shows a deep dust blackout about 20 days after the outburst.
\citet{hac19kb} obtained $E(B-V)=0.82$, $(m-M)_V=15.95$, $d=4.8$~kpc,
and $\log f_{\rm s}= +0.28$.
We have reanalyzed the $BVI_{\rm C}$ light/color curves in Appendix
\ref{v5579_sgr_bvi} and obtained a new set of parameters, i.e., 
$E(B-V)=0.56$, $(m-M)_V=14.55$, $d=3.6$~kpc, and $\log f_{\rm s}= +0.24$.
Then, we have $(m-M')_V=14.55 + 0.6 = 15.15$ and plot the
$(B-V)_0$-$(M_V-2.5\log f_{\rm s})$ diagram in Figure
\ref{hr_diagram_v597_pup_v2468_cyg_v5579_sgr_v5589_sgr_outburst}(c).
The peak $V$ brightness reaches $M_V= -7.94$ from the data of AAVSO.
Then, we have $M'_V= M_V-2.5\log f_{\rm s}= -8.54$.
Soon after the $V$ peak, the nova entered the dust blackout phase.
Therefore, we do not correctly determine the type of the track.  

The distance modulus in $I_{\rm C}$ band, $(m-M)_I=13.65$, is 
taken from Appendix \ref{v5579_sgr_bvi}.
Then, we have $(m-M')_I=13.65+0.6=14.25$.
The peak $I_{\rm C}$ brightness is 
$M'_I= M_I-2.5\log f_{\rm s}= -7.9 - 0.6 = -8.5$ from the data of VSOLJ.
We plot the $(V-I)_0$-$(M_I-2.5\log f_{\rm s})$ diagram in Figure 
\ref{hr_diagram_v597_pup_v2468_cyg_v5579_sgr_v5589_sgr_outburst_vi}(c).
Here, we adopt the $BVI_{\rm C}$ data from AAVSO, VSOLJ, and IAU Circular.
We define the track of V5579~Sgr by the thick solid blue line connecting
the data of AAVSO and VSOLJ.
We also plot the track of PU~Vul by the small filled cyan circles. 
In the very early phase, the track of V5579~Sgr starts near the position
of PU~Vul and goes up near the line of $(V-I)_0= +0.22$ 
(vertical solid red line).  Then it almost follows the track of
V5114~Sgr (green line) and then V1500~Cyg (cyan line) and
PU~Vul (small filled cyan circles) until the dust blackout phase started.

The rough overlapping of V5579~Sgr and V5114~Sgr
in the early phase on the $(V-I)_0$-$(M_I-2.5\log f_{\rm s})$ diagram
may support the results of $E(B-V)=0.56$, $(m-M)_I=13.65$, $d=3.6$~kpc,
and $\log f_{\rm s}= +0.24$ for V5579~Sgr.

\subsection{V5589~Sgr 2012\#1}
\label{v5589_sgr_vi}
\citet{hac19kb} obtained $E(B-V)=0.84$, $(m-M)_V=17.6$, $d=10.0$~kpc,
and $\log f_{\rm s}= -0.67$.  We have reanalyzed the $BVI_{\rm C}$
light/color curves of V5589~Sgr in Appendix \ref{v5589_sgr_bvi} and
obtained a new set of parameters, i.e., $E(B-V)=0.80$, $(m-M)_V=17.7$,
$d=11.0$~kpc, and $\log f_{\rm s}= -0.67$. 
Then, we have $(m-M')_V=17.7 - 1.675 = 16.0$ and plot the
$(B-V)_0$-$(M_V-2.5\log f_{\rm s})$ diagram in Figure
\ref{hr_diagram_v597_pup_v2468_cyg_v5579_sgr_v5589_sgr_outburst}(d).
This nova belongs to the very fast nova \citep{pay57},
because $t_2=5$ days and $t_3=10.9$ days \citep{thomp17}.
\citet{mro15} obtained the orbital period of 1.5923 days.
Therefore, the nova has a donor star that has already evolved off the 
main-sequence like the recurrent nova U~Sco.  The nova has a large
accretion disk, which is irradiated by the central WD, during the SSS phase.
This effect makes the $B-V$ color bluer as shown in Figure 
\ref{hr_diagram_v597_pup_v2468_cyg_v5579_sgr_v5589_sgr_outburst}(d).
U~Sco also shows this kind of bluer nature in the color-magnitude diagram
as already shown in Figure 94 of \citet{hac19kb}.
The peak $V$ brightness reaches $M_V= -8.7$ from the data of \citet{mro15}.
Then, we have $M'_V= M_V-2.5\log f_{\rm s}= -8.7 + 1.675= -7.0$.
The track of V5589~Sgr almost follows the track of V1500~Cyg (green line).
Therefore, we may conclude that V5589~Sgr belongs to the V1500~Cyg type.

The distance modulus in $I_{\rm C}$ band, $(m-M)_I=16.42$, is 
taken from Appendix \ref{v5589_sgr_bvi}.
Then, we have $(m-M')_I=16.42 - 1.675 = 14.75$.
The peak $I_{\rm C}$ brightness is 
$M'_I= M_I-2.5\log f_{\rm s}= -8.3 + 1.675 = -6.65$ from the data of VSOLJ.
We plot the $(V-I)_0$-$(M_I-2.5\log f_{\rm s})$ diagram in Figure 
\ref{hr_diagram_v597_pup_v2468_cyg_v5579_sgr_v5589_sgr_outburst_vi}(d).
Here, we adopt the $BVI_{\rm C}$ data from AAVSO, VSOLJ, and SMARTS. 
We plot the template tracks of V597~Pup (red line: V1500 Cyg type).
The track of V5589~Sgr almost follows the track of V597~Pup although
the data of VSOLJ are rather scattered. 

Thus, the redder $(V-I)_0$ colors of V597~Pup and V5589~Sgr than
those of V1500~Cyg and V5114~Sgr in the
later phase ($M'_I\equiv M_I - 2.5 \log f_{\rm s} > -2$) originate
from a large contribution of an irradiated accretion disk or 
companion star.
The rough overlapping of V5589~Sgr and the template track of V597~Pup
may support the results of $E(B-V)=0.80$, $(m-M)_I=16.42$, $d=11.0$~kpc,
and $\log f_{\rm s}= -0.67$ for V5589~Sgr.

\subsection{Summary of the V1500~Cyg type}
\label{summary_v1500_cyg_type_vi}
In this section, we examined the V1500~Cyg type novae (Figure
\ref{hr_diagram_v1500_cyg_type_4fig_vi}).  Comparing with the LV~Vul type
novae (Figure \ref{hr_diagram_lv_vul_4type_4fig_vi}),
the tracks in the $(V-I)_0$-$(M_I-2.5\log f_{\rm s})$
diagram show a brighter and redder excursion 
at $M'_I= M_I-2.5 \log f_{\rm s} < -6$.
Figure \ref{hr_diagram_v1500_cyg_type_4fig_vi}(a) compares the track
of V1500~Cyg with the track of V496~Sct (LV~Vul type).
The $(V-I)_0$ color of V1500~Cyg (cyan line) is rather redder than
that of V496~Sct (thin solid orange line) because strong emission lines 
such as \ion{O}{1} $\lambda\lambda 7774$,
8446, and \ion{Ca}{2} $\lambda\lambda 8498$, 8542 contribute much 
to the $I_{\rm C}$ band and make the $V-I$ color redder. 

We divide the V1500~Cyg type into four subtypes
on the $(V-I)_0$-$(M_I-2.5\log f_{\rm s})$ diagram, that is, 
V1500~Cyg subtype, V5114~Sgr subtype, V574~Pup subtype, and V597~Pup subtype,
as summarized in Figure \ref{hr_diagram_v1500_cyg_type_4fig_vi}.
The $(V-I)_0$ color of V5114~Sgr (green line) is also redder than
that of V496~Sct (thin orange line).
The V574~Pup track is located much redder than those of V1500~Cyg and
V5114~Sgr in the early phase, but in the middle and later phase it 
overlaps with that of V5114~Sgr and the lower branch of V496~Sct/V959~Mon
as shown in Figure \ref{hr_diagram_v1500_cyg_type_4fig_vi}(c).  Then,
it deviates from them in the very later phase
($M'_I\equiv M_I-2.5 \log f_{\rm s} > -2$).
This is partly because an irradiated accretion disk or companion star
contributes to the $V-I_{\rm C}$ color of V574~Pup.
The V597~Pup track (red line) is closely located to that of V1500~Cyg
(cyan line) in the early phase as shown in Figure 
\ref{hr_diagram_v1500_cyg_type_4fig_vi}(d),
but it drops and stays at  
$(V-I_{\rm C})_0\sim -0.2$ in the later phase of
$M'_I\equiv M_I-2.5 \log f_{\rm s} > -2$. 
This is again because an irradiated accretion disk and companion star
contribute much to the $V-I_{\rm C}$ color than those of the V574~Pup type.

It should be noted that the intrinsic $V-I$ color of optically 
thick free-free emission is $(V-I)_0= +0.22$.  
If no strong emission lines such as \ion{O}{1} and \ion{Ca}{2} triplet
contribute to the $I_{\rm C}$ band and free-free emission dominates
spectra of novae in the early phase of nova outbursts, 
its $(V-I)_0$ color should be $(V-I)_0= +0.22$.


\begin{figure*}
\plotone{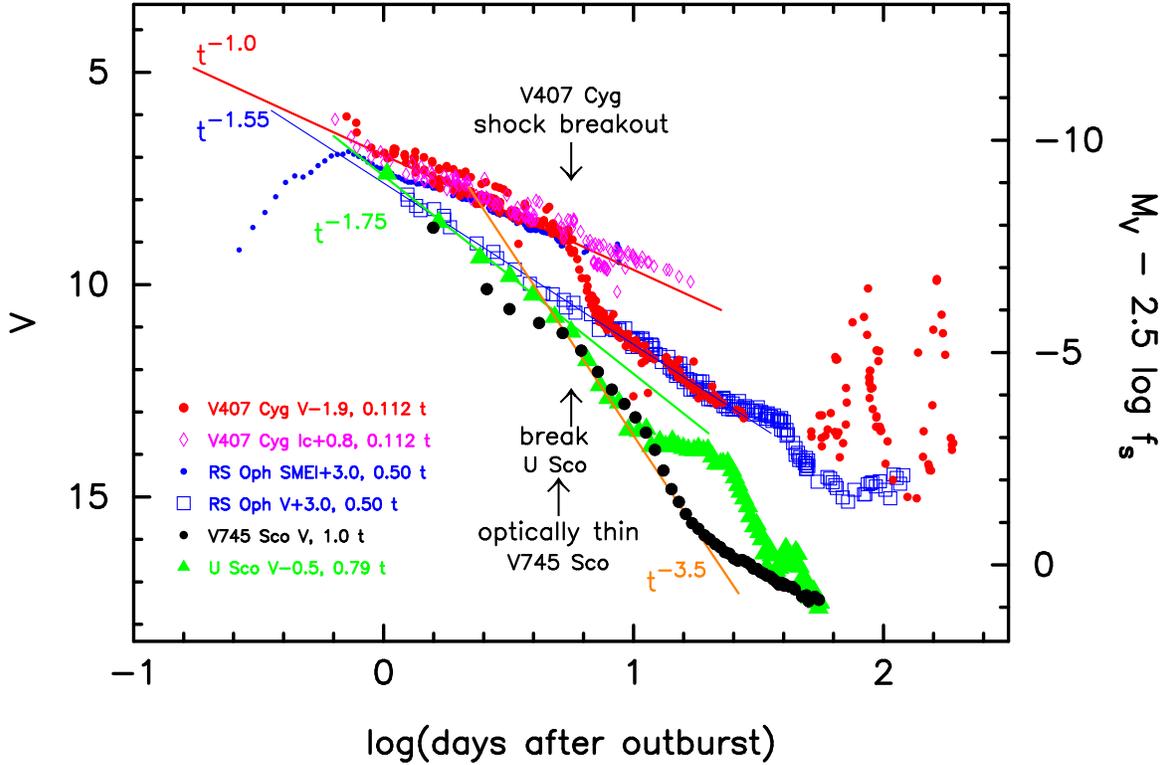}
\caption{
Nova $V$ light curves for V407~Cyg, RS~Oph, V745~Sco, and U~Sco on 
logarithmic timescales.  We add the $I_{\rm C}$ light curves of V407~Cyg and
the {\it SMEI} magnitudes \citep{hou10} of RS~Oph for comparison.
We also add the time-stretched absolute magnitude, $M_V - 2.5 \log f_{\rm s}$,
against V745~Sco ($f_{\rm s}=1$ for V745~Sco), in the right ordinate.
We depict the four decline trends of $F_\nu \propto t^{-1.0}$, $t^{-1.55}$,
$t^{-1.75}$, and $t^{-3.5}$ by the solid red, blue, green, and orange
lines, respectively.  The sources of the light curve data are all the same 
as those in \citet{hac18kb}.
\label{v745_sco_u_sco_v407_cyg_rs_oph_v_template_no2}}
\end{figure*}


\begin{figure*}
\plotone{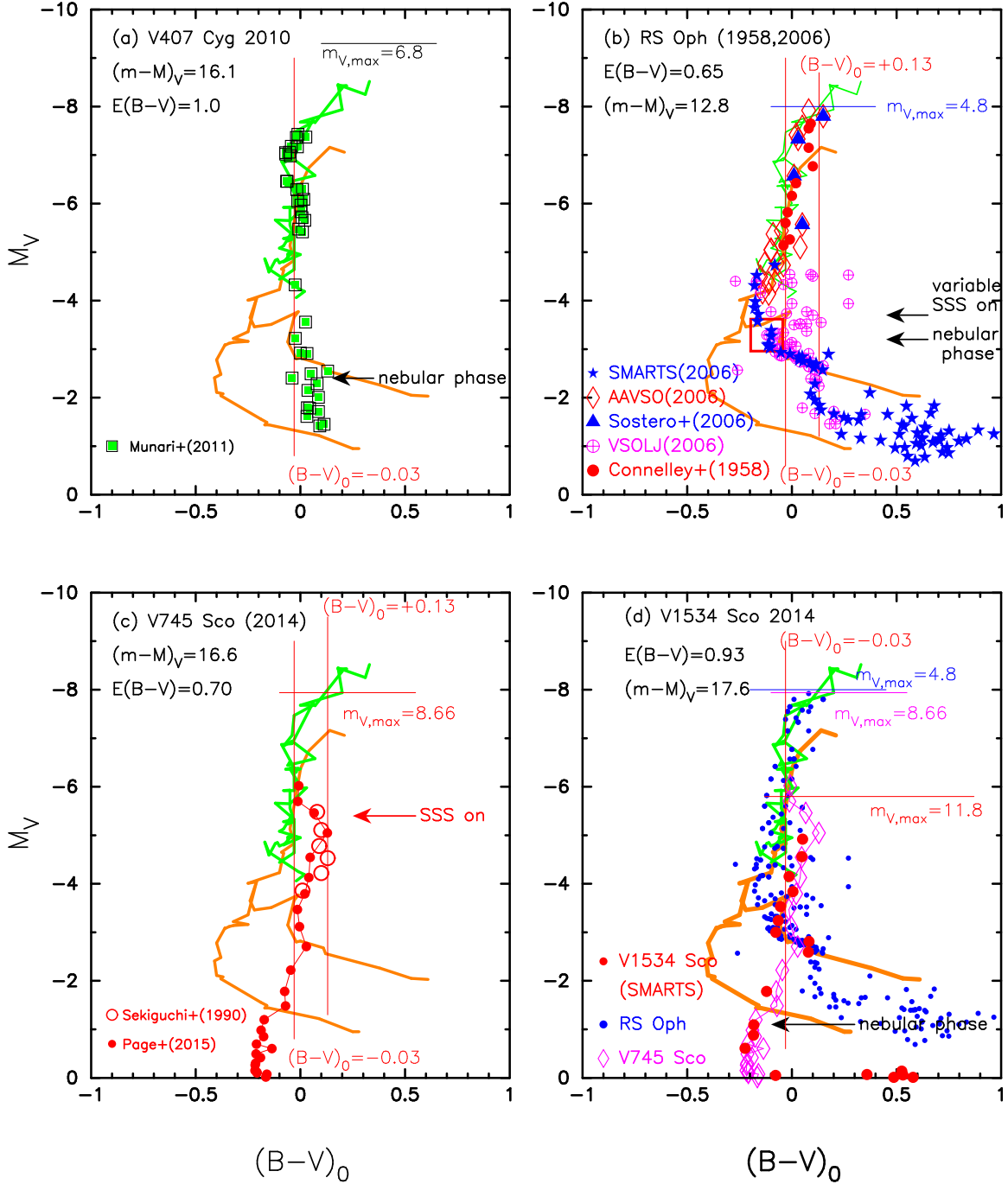}
\caption{
Color-magnitude diagram in outburst
for (a) V407~Cyg, (b) RS~Oph, (c) V745~Sco, and (d) V1534~Sco.
\label{hr_diagram_v407_cyg_rs_oph__v745_sco_v1534_sco_outburst_mv}}
\end{figure*}


\begin{figure*}
\plotone{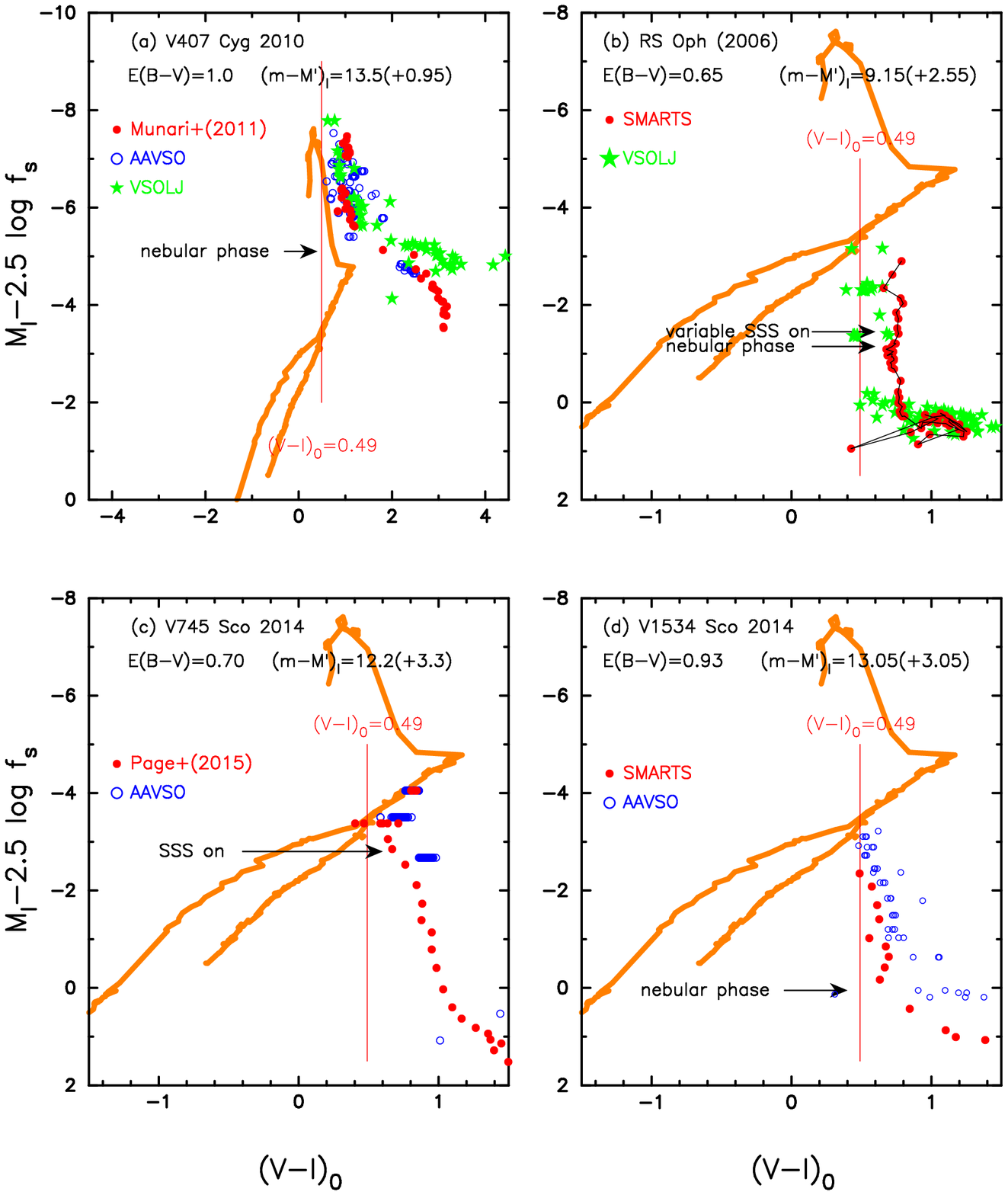}
\caption{
Same as Figure 
\ref{hr_diagram_v496_sct_v959_mon_v834_car_v1369_cen_outburst_vi},
but for (a) V407~Cyg, (b) RS~Oph, (c) V745~Sco, and (d) V1534~Sco.
The vertical thin solid lines show the intrinsic color of
$(V-I)_0= +0.49$ for optically thin free-free emission.
\label{hr_diagram_v407_cyg_rs_oph_v745_sco_v1534_sco_outburst_vi}}
\end{figure*}

\section{Exceptional Type of Novae}
\label{exceptional_type_novae_vi}
\citet{hac19ka, hac19kb} divided many classical novae into two types
in the $(B-V)_0$-$(M_V-2.5 \log f_{\rm s})$ diagram, that is, LV~Vul and 
V1500~Cyg types.  However, several exceptional novae cannot be clearly 
classified into the two classes.  In this section, we study the reasons
for their exceptionalities.

\citet{hac18kb} analyzed light curves of many fast and recurrent novae.
Some of them deviate from the universal decline law.  We show such examples
in Figure \ref{v745_sco_u_sco_v407_cyg_rs_oph_v_template_no2}.
The decay trend of V407~Cyg is $F_\nu \propto t^{-1.0}$ (solid red line)
in the early phase, being different from that of the universal decline law,
$F_\nu \propto t^{-1.75}$, where $t$ is the time after the outburst.
V745~Sco barely shows the slope of $F_\nu \propto t^{-1.75}$ (solid green
line), which is soon followed by a much steeper decline of 
$F_\nu \propto t^{-3.5}$ (solid orange line).

We first explain the reason for $F_\nu \propto t^{-1.0}$.
V407~Cyg is a symbiotic star consisting of a WD and a Mira \citep{mun90}.
The Mira companion emits massive cool winds that form circumstellar
matter (CSM) around the binary.  Just after the nova explosion, ejecta 
plunge into the CSM and create a strong shock.  Such a CSM shock, which is 
frequently observed in Type IIn supernovae, shows a slope of 
$F_\nu \propto t^{-1}$ \citep[see, e.g.,][for SN~2005ip]{mor13}.
Therefore, we regard the $F_\nu \propto t^{-1}$ slope to be a result of a 
strong CSM shock.  The shock had broken out of the CSM on day $\sim 45$
in Figure \ref{v745_sco_u_sco_v407_cyg_rs_oph_v_template_no2}.
Then the continuum flux drops and decays as $F_\nu \propto t^{-1.55}$
(solid blue line) like RS~Oph. 

RS~Oph decays as $F_\nu \propto t^{-1.55}$ in the $V$ band
but as $F_\nu \propto t^{-1.0}$ in the {\it Solar Mass Ejection Imager (SMEI)} 
light curve \citep{hou10}.  The {\it SMEI} band has a peak quantum 
efficiency at 7000\AA\  with an FWHM of 3000\AA.
RS~Oph is a binary consisting of a WD and a red giant (not a Mira).
Cool winds from the red giant are much less massive than in V407~Cyg.
Even if ejecta created a CSM shock, it was much weaker than in V407~Cyg.  
As a result, the continuum flux by the CSM shock is too small to emit 
a slope of $F_\nu \propto t^{-1.0}$.  On the other hand, the CSM shock
increases the H$\alpha$ flux.  It contributes to the {\it SMEI} band and
could make a slope of $F_\nu \propto t^{-1.0}$ in the {\it SMEI} band.

V745~Sco, T~CrB, and V1534~Sco show no evidence of a CSM shock
in their $V$ light curves although they have a red-giant companion.
Their decay trends in the $V$ band almost overlap with the decay of
V838~Her $V$ \citep[see, e.g.,][]{hac18kb}.  
This clearly shows that their CSM is extremely
less massive because V838~Her has a main-sequence companion \citep[e.g.,
][]{ing92, lei92}.

Next, we explain the reason for the rapid decline of 
$F_\nu \propto t^{-3.5}$ in U~Sco and V745~Sco.  In general,
the nova decay trend transfers from $F_\nu \propto t^{-1.75}$ 
(universal decline) to $F_\nu \propto t^{-3.5}$ (rapid decline)
in model light curves of novae calculated by \citet{hac06kb}.  
This change corresponds to the quick drop in the wind mass-loss rate 
together with the rapid shrinking of the photosphere.  
\citet{hac06kb} called this transition ``the break.''
We suppose that this occurs on day $\sim 8$ in U~Sco
(upward black arrow labeled ``break U~Sco''
in Figure \ref{v745_sco_u_sco_v407_cyg_rs_oph_v_template_no2}).

The break in V745~Sco comes slightly earlier than, but roughly 
coincides with, that of U~Sco.  This is also close to the epoch 
when the ejecta of V745~Sco became optically thin (upward black 
arrow labeled ``optically thin V745~Sco'').
We suppose that the break occurs when the ejecta become optically 
thick to thin.  The break comes very early in U~Sco and
V745~Sco because are very little the ejecta masses 
in these very high mass WDs \citep[$1.37-1.38~M_\sun$, see][]{hac18kb}.

In what follows, we examine the 
$(V-I)_0$-$(M_I-2.5\log f_{\rm s})$ diagrams for exceptional novae,
in the order of more massive CSM interaction, that is, 
in the order of V407~Cyg, RS~Oph, V745~Sco, and V1534~Sco.

\subsection{V407~Cyg 2010}
\label{v407_cyg_vi}
\citet{hac18kb} obtained $E(B-V)=1.0$, $(m-M)_V=16.1$, $d=3.9$~kpc,
and $\log f_{\rm s}= -0.37$ for V407~Cyg.
We have $(m-M')_V=16.1 - 0.925= 15.2$.  
However, we plot the 
usual $(B-V)_0$-$M_V$ diagram for V407~Cyg in Figure
\ref{hr_diagram_v407_cyg_rs_oph__v745_sco_v1534_sco_outburst_mv}(a).
This diagram is essentially the same as Figure 6(d) of \citet{hac18kb}.
The track almost goes down along the vertical line of $(B-V)_0= -0.03$.
The color of $(B-V)_0= -0.03$ is the intrinsic color of optically
thick free-free emission \citep{hac14k}.
We cannot compare the track with the V1500~Cyg or LV~Vul type
because the track goes almost straight down.

The distance modulus in $I_{\rm C}$ band of $(m-M)_I=14.45$ 
is calculated from Equation (\ref{distance_modulus_ri}) and
the results of \citet{hac18kb}.
Then, we have $(m-M')_I=14.45 - 0.925=13.5$.
The peak $I_{\rm C}$ brightness is 
$M'_I= M_I-2.5\log f_{\rm s}= -9.15 + 0.925 = -8.2$ from the data of VSOLJ.
We plot the $(V-I)_0$-$(M_I-2.5\log f_{\rm s})$ diagram in Figure 
\ref{hr_diagram_v407_cyg_rs_oph_v745_sco_v1534_sco_outburst_vi}(a).
Here, we adopt the $BVI_{\rm C}$ data from \citet{mun11c}, AAVSO,
and VSOLJ.  
The track of V407~Cyg is located by $\Delta (V-I)= 1-2$ mag redder
than that of V496~Sct/V959~Mon (LV~Vul) subtype.
This is because the CSM shock interaction contributes much to 
the $I_{\rm C}$ magnitude and makes the $V-I$ color redder.

\subsection{RS~Oph 2006}
\label{rs_oph_vi}
\citet{hac18kb} obtained $E(B-V)=0.65$, $(m-M)_V=12.8$, $d=1.4$~kpc,
and $\log f_{\rm s}= -1.02$ for RS~Oph.
We have $(m-M')_V=12.8 - 2.55= 10.25$.  We plot the 
usual $(B-V)_0$-$M_V$ diagram for RS~Oph in Figure
\ref{hr_diagram_v407_cyg_rs_oph__v745_sco_v1534_sco_outburst_mv}(b).
This diagram is essentially the same as Figure 6(c) of \citet{hac18kb}.
The track almost follows the upper branch of LV~Vul 
(upper thick solid orange line).  Then, it goes down along the vertical
line of $(B-V)_0= +0.13$.  The early decline along the color of 
$(B-V)_0= -0.03$ is consistent with the intrinsic color of optically
thick free-free emission \citep{hac14k}.  The later change to
$(B-V)_0= +0.13$ is also consistent with the intrinsic color of 
optically thin free-free emission \citep{hac14k}.
This change might occur from the ejecta being optically thick to
being optically thin.

The distance modulus in $I_{\rm C}$ band of $(m-M)_I=11.7$ 
is calculated from Equation (\ref{distance_modulus_ri}) and
the results of \citet{hac18kb}.
Then, we have $(m-M')_I=11.7 - 2.55=9.15$.
The peak $I_{\rm C}$ brightness is 
$M'_I= M_I-2.5\log f_{\rm s}= -7.96 + 2.55 = -5.4$ from the data of VSOLJ.
We plot the $(V-I)_0$-$(M_I-2.5\log f_{\rm s})$ diagram in Figure 
\ref{hr_diagram_v407_cyg_rs_oph_v745_sco_v1534_sco_outburst_vi}(b).
Here, we adopt the $BVI_{\rm C}$ data from VSOLJ and SMARTS.  
The track of RS~Oph is located at $(V-I)_0 \sim +0.7$ until 
$M'_I= M_I -2.5 \log f_{\rm s} \lesssim +0.0$.
This color is about $+0.2$ mag redder than the vertical thin solid
red line of $(V-I)_0= +0.49$, which is the intrinsic $V-I$ color of
optically thin free-free emission.

\subsection{V745~Sco 2014}
\label{v745_sco_vi}
\citet{hac18kb} obtained $E(B-V)=0.70$, $(m-M)_V=16.6$, $d=7.8$~kpc,
and $\log f_{\rm s}= -1.32$ for V745~Sco.
We have $(m-M')_V=16.6 - 3.3= 13.3$.  We plot the 
usual $(B-V)_0$-$M_V$ diagram for V745~Sco in Figure
\ref{hr_diagram_v407_cyg_rs_oph__v745_sco_v1534_sco_outburst_mv}(c).
This diagram is essentially the same as Figure 6(a) of \citet{hac18kb}.
The track almost goes down along the vertical red line of $(B-V)_0= -0.03$ 
until $M_V \sim -5.5$.  It moves to the redder line of $(B-V)_0= +0.13$
and goes down along this line.  Then, the track gradually moves to 
the bluer region over the line of $(B-V)_0= -0.03$. 
The jump from $(B-V)_0= -0.03$ to $(B-V)_0= +0.13$ at $M_V=-5.5$
corresponds to the transition of the ejecta from being optically thick
to thin as denoted by the red arrow labeled ``SSS on.'' 

The distance modulus in $I_{\rm C}$ band of $(m-M)_I=15.5$ 
is calculated from Equation (\ref{distance_modulus_ri}) and
the results of \citet{hac18kb}.
Then, we have $(m-M')_I=15.5 - 3.3=12.2$.
The peak $I_{\rm C}$ brightness is 
$M'_I= M_I-2.5\log f_{\rm s}= -7.35 + 3.3 = -4.05$ from \citet{pag15}.
We plot the $(V-I)_0$-$(M_I-2.5\log f_{\rm s})$ diagram in Figure 
\ref{hr_diagram_v407_cyg_rs_oph_v745_sco_v1534_sco_outburst_vi}(c).
Here, we adopt the $BVI_{\rm C}$ data from AAVSO and \citet{pag15}.
The track of V745~Sco in the early phase is located 
close to the vertical thin solid red line of $(V-I)_0= +0.49$.


\subsection{V1534~Sco 2014}
\label{v1534_sco_vi}
\citet{hac18kb} obtained $E(B-V)=0.93$, $(m-M)_V=17.6$, $d=8.8$~kpc,
and $\log f_{\rm s}= -1.22$ for V1534~Sco.
We have $(m-M')_V=17.6 - 3.05= 14.55$.  We plot the 
usual $(B-V)_0$-$M_V$ diagram for V1534~Sco in Figure
\ref{hr_diagram_v407_cyg_rs_oph__v745_sco_v1534_sco_outburst_mv}(d).
This diagram is essentially the same as Figure 6(b) of \citet{hac18kb}.
The track almost overlaps with that of V745~Sco.

The distance modulus in $I_{\rm C}$ band of $(m-M)_I=16.1$ 
taken from the results of \citet{hac18kb}.
Then, we have $(m-M')_I=16.1 - 3.05=13.05$.
The peak $I_{\rm C}$ brightness is 
$M'_I= M_I-2.5\log f_{\rm s}= -6.27 + 3.05 = -3.22$ from the data of AAVSO.
We plot the $(V-I)_0$-$(M_I-2.5\log f_{\rm s})$ diagram in Figure 
\ref{hr_diagram_v407_cyg_rs_oph_v745_sco_v1534_sco_outburst_vi}(d).
Here, we adopt the $BVI_{\rm C}$ data from AAVSO and SMARTS.   
The track of V1534~Sco in the early phase is located close to
the vertical thin solid red line of $(V-I)_0= +0.49$.



\begin{figure*}
\plotone{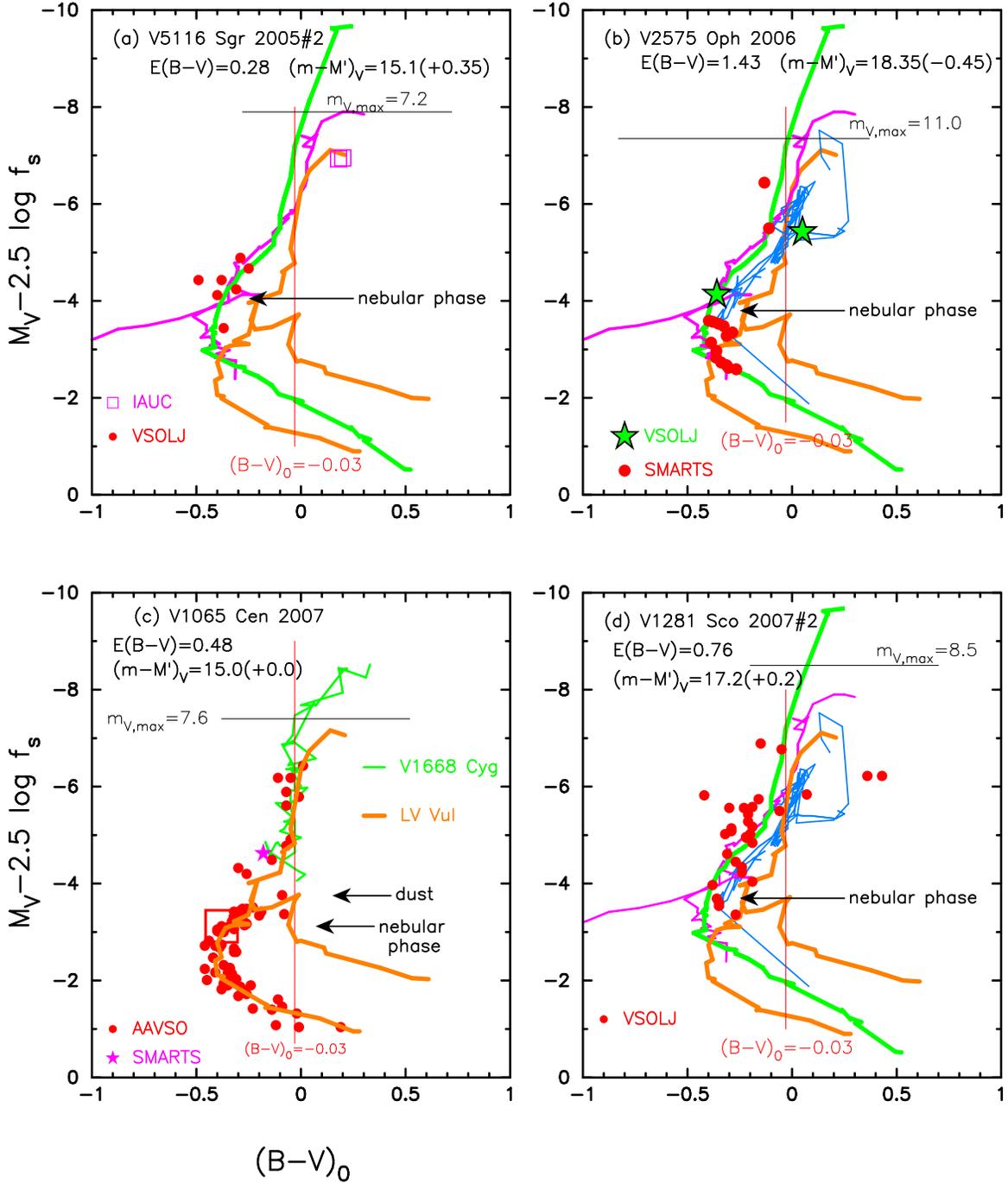}
\caption{
Same as Figure 
\ref{hr_diagram_v496_sct_v959_mon_v834_car_v1369_cen_outburst}, 
but for (a) V5116~Sgr, (b) V2575~Oph, (c) V1065~Cen, and (d) V1281~Sco.
In panels (a), (b), and (d), we add the track of V1974~Cyg (magenta lines).
In panels (b) and (d), we add the track of PW~Vul (thin solid cyan-blue line).
In panel (c), we add the track of V1668~Cyg (thin solid green lines).
\label{hr_diagram_v5116_sgr_v2575_oph_v1065_cen_v1281_sco_outburst}}
\end{figure*}


\begin{figure*}
\plotone{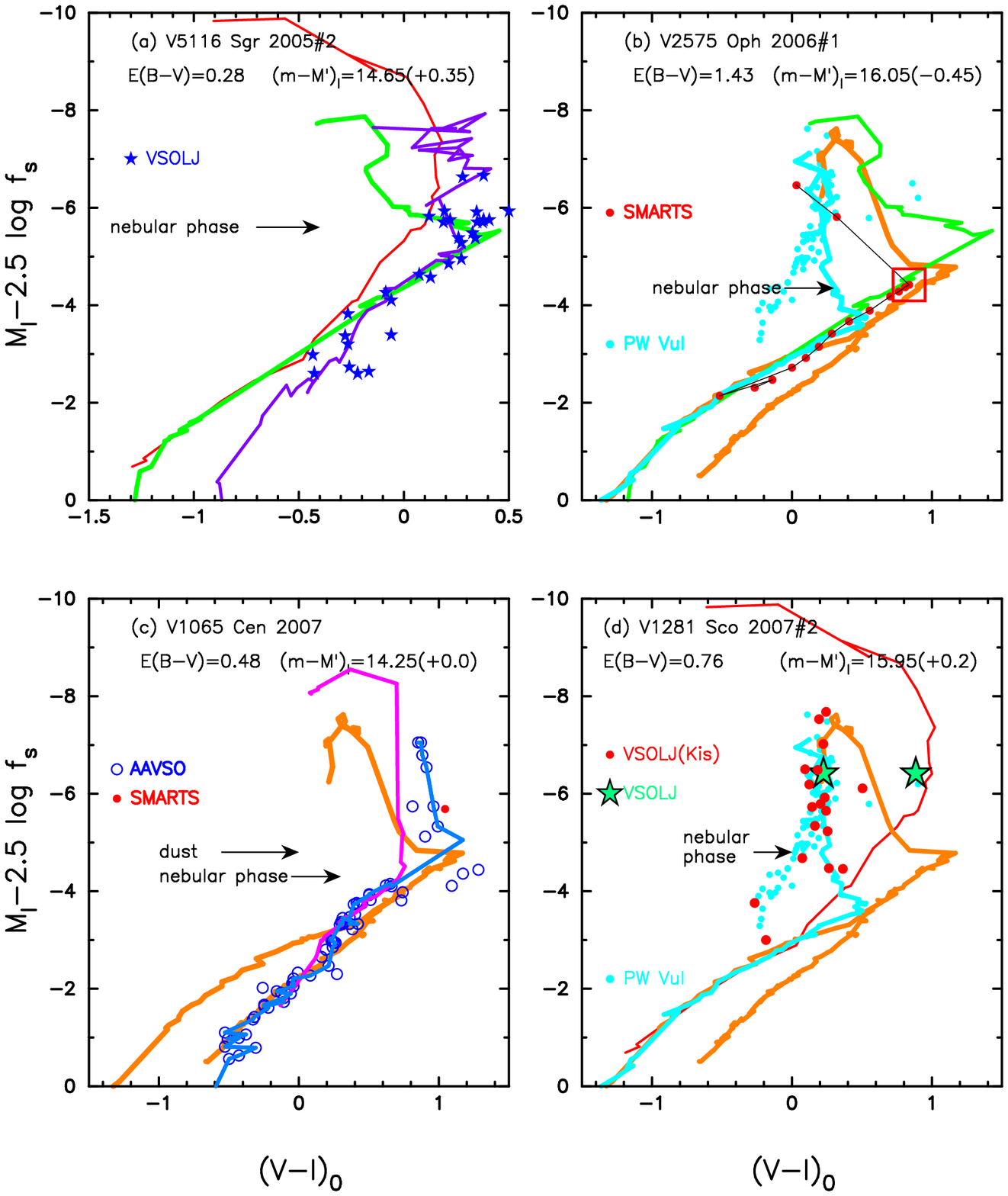}
\caption{
Same as Figure 
\ref{hr_diagram_v496_sct_v959_mon_v834_car_v1369_cen_outburst_vi},
but for (a) V5116~Sgr, (b) V2575~Oph, (c) V1065~Cen, and (d) V1281~Sco.
In panel (a), we add the track of V574~Pup (blue-magenta line)
in Section \ref{v574_pup_vi}.  In panels (a) and (d), we add the track
of V1500~Cyg (solid red line) in Section \ref{v1500_cyg_vi}.
In panels (b) and (d), we add the track of PW~Vul (filled cyan circles)
in Section \ref{pw_vul_vi} and the track of V5666~Sgr (cyan line)
in Section \ref{v5666_sgr_vi}.  In panel (c), we add the track 
of V382~Vel (magenta line) in Section \ref{v382_vel_vi}.
\label{hr_diagram_v5116_sgr_v2575_oph_v1065_cen_v1281_sco_outburst_vi}}
\end{figure*}

\section{Time-Stretched $(V-I)_0$-$(M_I-2.5\log \lowercase{f}_{\rm 
\lowercase{s}})$ Color-Magnitude Diagram of Other Novae}
\label{other_novae_vi}
We have examined 20 novae that belong to the typical LV~Vul
or V1500~Cyg type.  We have also examined four exceptional novae
that do not show a universal decline law.  In what follows,
we select other 28 novae that have enough $BVI_{\rm C}$ data and 
analyze their light and color curves based on the
$(V-I)_0$-$(M_I-2.5\log f_{\rm s})$ diagram method
in the order of discovery date.

\subsection{V5116~Sgr 2005\#2}
\label{v5116_sgr_vi}
\citet{hac19kb} obtained $E(B-V)=0.23$, $(m-M)_V=16.05$, $d=12$~kpc,
and $\log f_{\rm s}= +0.20$.
We have reanalyzed the $BVI_{\rm C}$ multi-band light/color curves
of V5116~Sgr in Appendix \ref{v5116_sgr_bvi} and obtained the new
parameter set of $E(B-V)=0.28$, $(m-M)_V=15.45$, $d=8.2$~kpc,
and $\log f_{\rm s}= -0.14$.  The essential difference is the
reddening of $E(B-V)=0.28$, distance modulus in $V$ band of 
$(m-M)_V=15.45$, and timescaling factor of $\log f_{\rm s}= -0.14$.
Therefore, we have $(m-M')_V=15.45 - 0.35=15.1$ and plot 
the $(B-V)_0$-$(M_V-2.5\log f_{\rm s})$ diagram in Figure 
\ref{hr_diagram_v5116_sgr_v2575_oph_v1065_cen_v1281_sco_outburst}(a).
The track almost follows the V1500~Cyg/V1974~Cyg tracks 
(green/magenta lines, respectively).
V5116~Sgr belongs to the V1500~Cyg type.

The distance modulus in $I_{\rm C}$ band, $(m-M)_I=15.0$, is 
taken from Appendix \ref{v5116_sgr_bvi}.
Then, we have $(m-M')_I=15.0 - 0.35=14.65$.
The peak $I_{\rm C}$ brightness is 
$M'_I= M_I-2.5\log f_{\rm s}= -7.6 + 0.35 = -7.25$ from 
IAU Circular No. 8559.
We plot the $(V-I)_0$-$(M_I-2.5\log f_{\rm s})$ diagram in Figure 
\ref{hr_diagram_v5116_sgr_v2575_oph_v1065_cen_v1281_sco_outburst_vi}(a).
Here, we adopt the $BVI_{\rm C}$ data from VSOLJ.
The track of V5116~Sgr broadly follows the track of V5114~Sgr (green line)
or V574~Pup (blue-magenta line).

This is consistent with the previous result that V5116~Sgr belongs to
the V1500~Cyg type in the $(B-V)_0$-$(M_V-2.5\log f_{\rm s})$ diagram.
The overlapping of V5116~Sgr with the track of V5114~Sgr or V574~Pup
on the $(V-I)_0$-$(M_I-2.5\log f_{\rm s})$ diagram
supports our new results of $E(B-V)=0.28$, $(m-M)_I=15.0$, $d=8.2$~kpc,
and $\log f_{\rm s}= -0.14$ for V5116~Sgr.

\subsection{V2575~Oph 2006\#1}
\label{v2575_oph_vi}
\citet{hac19kb} obtained $E(B-V)=1.43$, $(m-M)_V=17.85$, $d=4.9$~kpc,
and $\log f_{\rm s}= +0.11$.
We have reanalyzed the $BVI_{\rm C}$ multi-band light/color curves
of V2575~Oph in Appendix \ref{v2575_oph_bvi} and obtained 
a new parameter set of $E(B-V)=1.43$, $(m-M)_V=17.9$, $d=4.9$~kpc,
and $\log f_{\rm s}= +0.18$.  
Then, we have $(m-M')_V=17.9 + 0.45=18.35$ and plot 
the $(B-V)_0$-$(M_V-2.5\log f_{\rm s})$ diagram in Figure 
\ref{hr_diagram_v5116_sgr_v2575_oph_v1065_cen_v1281_sco_outburst}(b).
The track is close to the track of PW~Vul (cyan-blue line).

The distance modulus in $I_{\rm C}$ band, $(m-M)_I=15.6$, is 
taken from Appendix \ref{v2575_oph_bvi}.
Then, we have $(m-M')_I=15.6 + 0.45=16.05$.
The peak $I_{\rm C}$ brightness is 
$M'_I= M_I-2.5\log f_{\rm s}= -6.7 - 0.45 = -7.15$ from the data of VSOLJ. 
We plot the $(V-I)_0$-$(M_I-2.5\log f_{\rm s})$ diagram in Figure 
\ref{hr_diagram_v5116_sgr_v2575_oph_v1065_cen_v1281_sco_outburst_vi}(b).
Here, we adopt the $BVI_{\rm C}$ data from SMARTS.
We add the track of PW~Vul (filled cyan circles).
The track of V2575~Oph broadly follows the tracks of
PW~Vul in the early phase, V496~Sct (upper orange line) in the middle
and later phases, also V5666~Sgr (cyan line) in the later phase.

The rough overlapping of V2575~Oph with the tracks of V496~Sct and V5666~Sgr
on the $(V-I)_0$-$(M_I-2.5\log f_{\rm s})$ diagram
supports the results of $E(B-V)=1.43$, $(m-M)_I=15.6$, $d=4.9$~kpc,
and $\log f_{\rm s}= +0.18$ for V2575~Oph.

\subsection{V1065~Cen 2007}
\label{v1065_cen_vi}
\citet{hac18k} obtained $E(B-V)=0.45$, $(m-M)_V=15.0$, $d=5.3$~kpc,
and $\log f_{\rm s}= +0.0$.  We have reanalyzed the $BVI_{\rm C}$
multi-band light/color curves of V1065~Cen in Appendix 
\ref{v1065_cen_bvi} and obtained a similar set of parameters except
$E(B-V)=0.48$ and $d=5.0$~kpc. 
Then, we have $(m-M')_V= 15.0 + 0.0 = 15.0$.  We plot 
the $(B-V)_0$-$(M_V-2.5\log f_{\rm s})$ diagram in Figure
\ref{hr_diagram_v5116_sgr_v2575_oph_v1065_cen_v1281_sco_outburst}(c).
The track almost overlaps with the lower branch of LV~Vul (orange line).
Therefore, V1065~Cen belongs to the LV~Vul type.
This is consistent with the previous result that V1065~Cen belongs to
the LV~Vul type and overlaps with the lower branch. 

The distance modulus in $I_{\rm C}$ band,
$(m-M)_I=14.25$, is taken from Appendix \ref{v1065_cen_bvi}.
Then, we have $(m-M')_I=14.25 + 0.0 = 14.25$.
The peak $I_{\rm C}$ brightness is 
$M'_I= M_I-2.5\log f_{\rm s}= -7.05 - 0.0 = -7.05$ from the data of AAVSO. 
We plot the $(V-I)_0$-$(M_I-2.5\log f_{\rm s})$ diagram in Figure 
\ref{hr_diagram_v5116_sgr_v2575_oph_v1065_cen_v1281_sco_outburst_vi}(c).
Here, we adopt the $BVI_{\rm C}$ data from AAVSO and SMARTS.
The track of V1065~Cen broadly follows the lower branch of
V496~Sct/V959~Mon subtype (orange lines).
We have constructed a template track of V1065~Cen (solid cyan-blue line)
by recovering the brightness from the optically thin dust blackout.
For comparison, we also plot the V382~Vel (magenta line).
The overlapping of V1065~Cen with the track of V496~Sct/V959~Mon
on the $(V-I)_0$-$(M_I-2.5\log f_{\rm s})$ diagram
supports the results of $E(B-V)=0.48$, $(m-M)_I=14.25$, $d=5.0$~kpc,
and $\log f_{\rm s}= +0.0$ for V1065~Cen.

\subsection{V1281~Sco 2007\#2}
\label{v1281_sco_vi}
\citet{hac19kb} obtained $E(B-V)=0.82$, $(m-M)_V=17.4$, $d=9.4$~kpc,
and $\log f_{\rm s}= -0.07$.  We have reanalyzed the $BVI_{\rm C}$ 
multi-band light/color curves of V1281~Sco in Appendix \ref{v1281_sco_bvi}
and obtained a new set of parameters, i.e.,
$E(B-V)=0.76$, $(m-M)_V=17.4$, $d=10.1$~kpc, and $\log f_{\rm s}= -0.07$.
Then, we have $(m-M')_V=17.4 - 0.175=17.2$ and plot
the $(B-V)_0$-$(M_V-2.5\log f_{\rm s})$ diagram in Figure 
\ref{hr_diagram_v5116_sgr_v2575_oph_v1065_cen_v1281_sco_outburst}(d).
V1281~Sco shows multiple secondary peaks and this feature is similar to
that of PW~Vul.  The track broadly overlaps with that of PW~Vul 
(cyan-blue line) or V1500~Cyg (green line).

The distance modulus in $I_{\rm C}$ band, $(m-M)_I=16.15$, is 
taken from Appendix \ref{v1281_sco_bvi}.
Then, we have $(m-M')_I=16.15 - 0.175=15.95$.
The peak $I_{\rm C}$ brightness is 
$M'_I= M_I-2.5\log f_{\rm s}= -7.88 + 0.175 = -7.7$ from the data of VSOLJ. 
We plot the $(V-I)_0$-$(M_I-2.5\log f_{\rm s})$ diagram in Figure 
\ref{hr_diagram_v5116_sgr_v2575_oph_v1065_cen_v1281_sco_outburst_vi}(d).
Here, we adopt the $BVI_{\rm C}$ data from VSOLJ.
We separately plot the VSOLJ data observed by S. Kiyota (labeled ``Kis'':
red filled circles) and by the other people (filled stars).
We add the tracks of PW~Vul (filled cyan circles) and V1500~Cyg (red line).
The track of V1281~Sco broadly follows the tracks of
PW~Vul (filled cyan circles).
It could follow the V1500~Cyg (red line) or V5666~Sgr (cyan line)
track in the later phase.
The overlapping of V1281~Sco with PW~Vul
on the $(V-I)_0$-$(M_I-2.5\log f_{\rm s})$ diagram
supports the results of $E(B-V)=0.76$, $(m-M)_I=16.15$, $d=10$~kpc,
and $\log f_{\rm s}= -0.07$ for V1281~Sco.


\begin{figure*}
\plotone{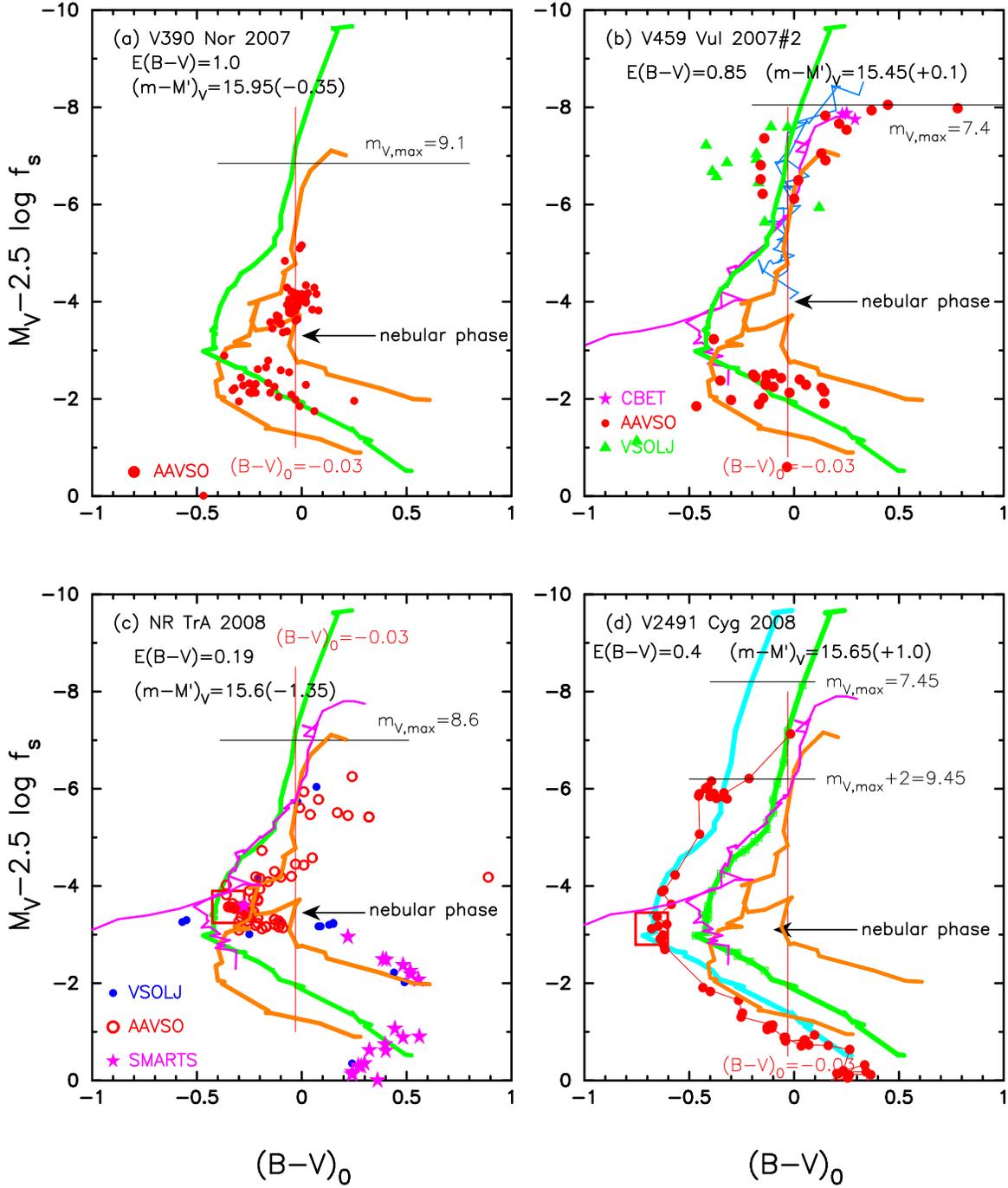}
\caption{
Same as Figure 
\ref{hr_diagram_v496_sct_v959_mon_v834_car_v1369_cen_outburst}, 
but for (a) V390~Nor, (b) V459~Vul, (c) NR~TrA, and (d) V2491~Cyg.
In panel (b), we add the track of V1668~Cyg (thin solid cyan-blue lines).
In panels (c)(d), we add the track of V1974~Cyg (magenta lines).
In panel (d), we add the track of V1500~Cyg shifted by $\Delta (B-V)= 
-0.25$ (thick solid cyan line).
\label{hr_diagram_v390_nor_v459_vul_nr_tra_v2491_cyg_outburst}}
\end{figure*}


\begin{figure*}
\plotone{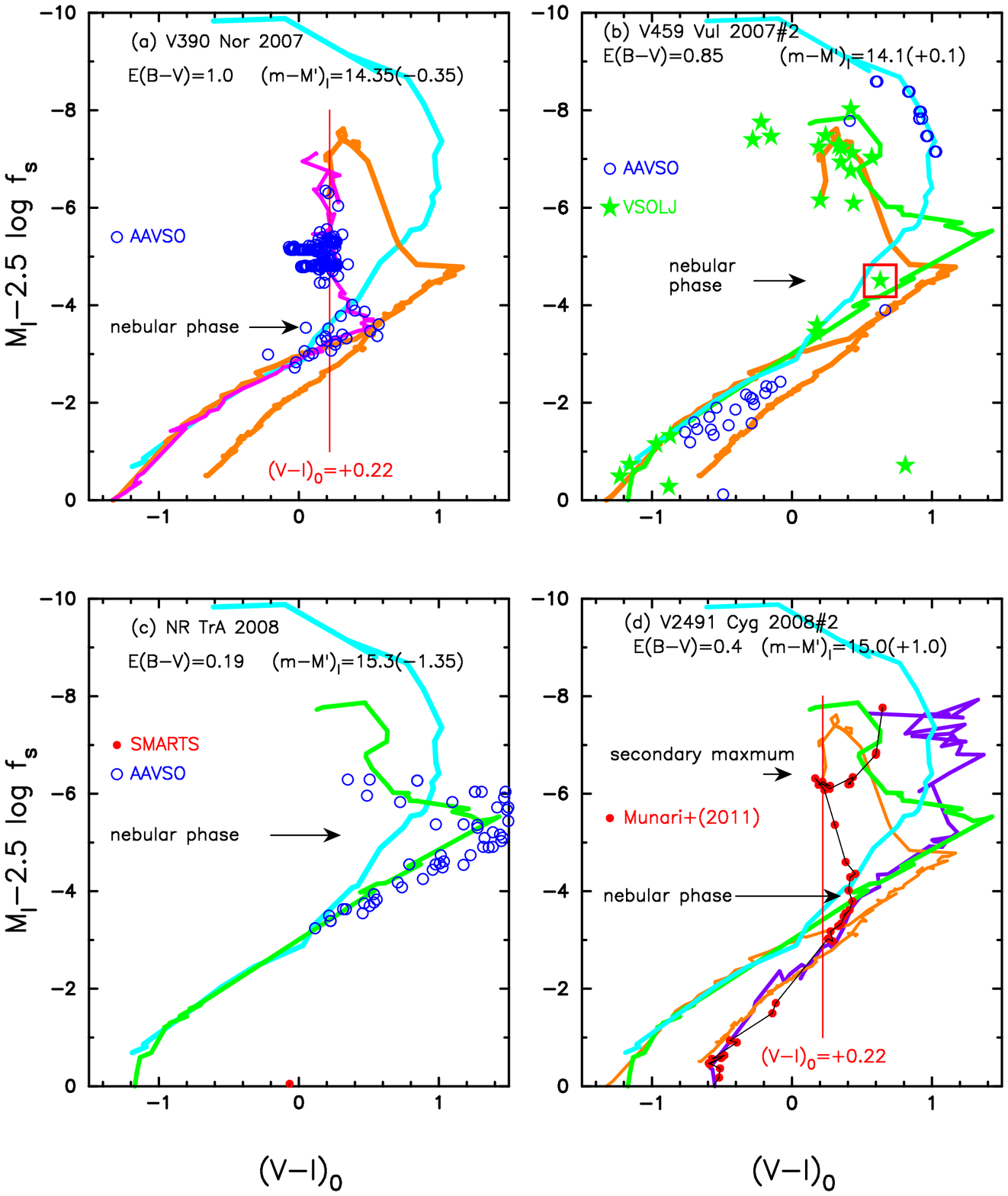}
\caption{
Same as Figure 
\ref{hr_diagram_v496_sct_v959_mon_v834_car_v1369_cen_outburst_vi},
but for (a) V390~Nor, (b) V459~Vul, (c) NR~TrA,
and (d) V2491~Cyg.
The thick solid cyan lines correspond to the template track of V1500~Cyg.
In panel (a), we add the template track of V5666~Sgr (magenta line).
In panel (d), we add the template track of V574~Pup (thick magenta-blue line).
\label{hr_diagram_v390_nor_v459_vul_nr_tra_v2491_cyg_outburst_vi}}
\end{figure*}

\subsection{V390~Nor 2007}
\label{v390_nor_vi}
\citet{hac19kb} obtained $E(B-V)=0.89$, $(m-M)_V=16.6$, $d=5.8$~kpc,
and $\log f_{\rm s}= +0.45$.  We have reanalyzed the $BVI_{\rm C}$
multi-band light/color curves of V390~Nor in Appendix \ref{v390_nor_bvi}
and obtained a new set of parameters, i.e.,
$E(B-V)=1.0$, $(m-M)_V=15.6$, $d=3.2$~kpc, and $\log f_{\rm s}= +0.14$.
Then, we have $(m-M')_V=15.6 + 0.35=15.95$ and plot the
$(B-V)_0$-$(M_V-2.5\log f_{\rm s})$ diagram in Figure
\ref{hr_diagram_v390_nor_v459_vul_nr_tra_v2491_cyg_outburst}(a).
The track of V390~Nor broadly follows the LV~Vul track in the early
and middle phase, then it transfers to the track of V1500~Cyg
(green line) in the later phase.

The distance modulus in $I_{\rm C}$ band, $(m-M)_I=14.0$, is 
taken from Appendix \ref{v390_nor_bvi}.
Then, we have $(m-M')_I=14.0+0.35=14.35$.
The peak $I_{\rm C}$ brightness is 
$M'_I= M_I-2.5\log f_{\rm s}= -6.0 - 0.35 = -6.35$ from the data of AAVSO.
We plot the $(V-I)_0$-$(M_I-2.5\log f_{\rm s})$ diagram in Figure 
\ref{hr_diagram_v390_nor_v459_vul_nr_tra_v2491_cyg_outburst_vi}(a).
Here, we adopt the $BVI_{\rm C}$ data from AAVSO.
We add the tracks of V5666~Sgr (magenta line; V5666~Sgr subtype of
LV~Vul type) and V1500~Cyg (cyan line; V1500~Cyg subtype of V1500~Cyg type).
The track of V390~Nor almost overlaps with the track of V5666~Sgr 
(magenta line) in the early phase and then transfers from that of V5666~Sgr
to V1500~Cyg (cyan line) in the middle and later phases.  
The overlapping of V390~Nor with V5666~Sgr and V1500~Cyg
in the $(V-I)_0$-$(M_I-2.5\log f_{\rm s})$ diagram supports our new 
results of $E(B-V)=1.0$, $(m-M)_I=15.6$, $d=3.2$~kpc,
and $\log f_{\rm s}= +0.14$ for V390~Nor.

\subsection{V459~Vul 2007\#2}
\label{v459_vul_vi}
\citet{hac19kb} obtained $E(B-V)=0.90$, $(m-M)_V=15.45$, $d=3.4$~kpc,
and $\log f_{\rm s}= -0.15$.
We have reanalyzed the $BVI_{\rm C}$ multi-band light/color curves of
V459~Vul in Appendix \ref{v459_vul_bvi} and obtained a new parameter set
of $E(B-V)=0.85$, $(m-M)_V=15.55$, $d=3.8$~kpc, and $\log f_{\rm s}= -0.04$.
Then, we have $(m-M')_V=15.55-0.1 = 15.45$ and plot the 
$(B-V)_0$-$(M_V-2.5\log f_{\rm s})$ diagram in Figure
\ref{hr_diagram_v390_nor_v459_vul_nr_tra_v2491_cyg_outburst}(b). 
The track of V459~Vul broadly follows the template track of V1500~Cyg 
(green line).

The distance modulus in $I_{\rm C}$ band, $(m-M)_I=14.2$, is 
taken from Appendix \ref{v459_vul_bvi}.
Then, we have $(m-M')_I=14.2-0.1 = 14.1$.
The peak $I_{\rm C}$ brightness is 
$M'_I= M_I-2.5\log f_{\rm s}= -8.7 + 0.1 = -8.6$ from the data of AAVSO.
We plot the $(V-I)_0$-$(M_I-2.5\log f_{\rm s})$ diagram in Figure 
\ref{hr_diagram_v390_nor_v459_vul_nr_tra_v2491_cyg_outburst_vi}(b).
Here, we adopt the $BVI_{\rm C}$ data from AAVSO and VSOLJ.
The track of V459~Vul almost overlaps with the V1500~Cyg subtype
(cyan line) or V5114~Sgr subtype (green line).
The rough overlapping of V459~Vul with the V1500~Cyg subtype
in the $(V-I)_0$-$(M_I-2.5\log f_{\rm s})$ diagram supports our new
results of $E(B-V)=0.85$, $(m-M)_I=14.2$, $d=3.8$~kpc,
and $\log f_{\rm s}= -0.04$ for V459~Vul.

\subsection{NR~TrA 2008}
\label{nr_tra_vi}
\citet{hac19kb} obtained $E(B-V)=0.24$, $(m-M)_V=15.35$, $d=8.3$~kpc,
and $\log f_{\rm s}= +0.43$.  We have reanalyzed the $BVI_{\rm C}$ 
multi-band light/color curves in Appendix \ref{nr_tra_bvi} and obtained
a new set of parameters, i.e., $E(B-V)=0.19$, $(m-M)_V=14.25$, 
$d=5.4$~kpc, and $\log f_{\rm s}= +0.55$ for NR~TrA.
Then, we have $(m-M')_V=14.25 + 1.375 = 15.6$ and plot
the $(B-V)_0$-$(M_V-2.5\log f_{\rm s})$ diagram in Figure
\ref{hr_diagram_v390_nor_v459_vul_nr_tra_v2491_cyg_outburst}(c).
In the early phase, the track of NR~TrA broadly follows
the upper branch of LV~Vul (orange line).

The distance modulus in $I_{\rm C}$ band, $(m-M)_I=13.95$, is 
taken from Appendix \ref{nr_tra_bvi}.
Then, we have $(m-M')_I=13.95+1.375=15.3$.
The peak $I_{\rm C}$ brightness is 
$M'_I= M_I-2.5\log f_{\rm s}= -5.07 - 1.375 = -6.45$ from the data of VSOLJ.
We plot the $(V-I)_0$-$(M_I-2.5\log f_{\rm s})$ diagram in Figure 
\ref{hr_diagram_v390_nor_v459_vul_nr_tra_v2491_cyg_outburst_vi}(c).
Here, we adopt the $BVI_{\rm C}$ data from AAVSO and SMARTS.
The track of NR~TrA almost overlaps with the V5114~Sgr track (green line).
The rough overlapping of NR~TrA and V5114~Sgr 
on the $(V-I)_0$-$(M_I-2.5\log f_{\rm s})$ diagram supports our new
results of $E(B-V)=0.19$, $(m-M)_I=13.95$, $d=5.4$~kpc,
and $\log f_{\rm s}= +0.55$ for NR~TrA.

\subsection{V2491~Cyg 2008\#2}
\label{v2491_cyg_vi}
V2491~Cyg is characterized by a secondary maximum
\citep[see, e.g., Figure 1 of][]{hac09ka}.
\citet{hac19ka} obtained $E(B-V)=0.45$, $(m-M)_V=17.4$, $d=15.9$~kpc,
and $\log f_{\rm s}= -0.34$.  We have reanalyzed the $BVI_{\rm C}$
multi-band light/color curves of V2491~Cyg in Appendix \ref{v2491_cyg_bvi}
and obtained a new set of parameters, i.e., $E(B-V)=0.40$, 
$(m-M)_V=16.65$, $d=12.1$~kpc, and $\log f_{\rm s}= -0.40$.
Then, we have $(m-M')_V=16.65-1.0=15.65$ and plot
the $(B-V)_0$-$(M_V-2.5\log f_{\rm s})$ diagram in Figure 
\ref{hr_diagram_v390_nor_v459_vul_nr_tra_v2491_cyg_outburst}(d).
The $BVI_{\rm C}$ data are taken from \citet{mun11}.
The V2491~Cyg track almost overlaps with the track (cyan line)
of V1500~Cyg, which is blue-shifted by $\Delta (B-V)= -0.25$.
This is because the metallicity of
V2491~Cyg is subsolar, i.e., [Fe/H]$=-0.25$ \citep{mun11}.
\citet{hac19ka} concluded that V2491~Cyg belongs to the V1500~Cyg type.

The distance modulus in $I_{\rm C}$ band, $(m-M)_I=16.0$, is 
taken from Appendix \ref{v2491_cyg_bvi}.
Then, we have $(m-M')_I=16.0-1.0=15.0$.
The peak $I_{\rm C}$ brightness is 
$M'_I= M_I-2.5\log f_{\rm s}= -9.28 + 1.0 = -8.3$ from the data of VSOLJ.
We plot the $(V-I)_0$-$(M_I-2.5\log f_{\rm s})$ diagram in Figure 
\ref{hr_diagram_v390_nor_v459_vul_nr_tra_v2491_cyg_outburst_vi}(d).
Here, we adopt the $BVI_{\rm C}$ data from \citet{mun11}.
The secondary maximum may cloud the intrinsic color and brightness in the
$(V-I)_0$-$(M_I-2.5\log f_{\rm s})$ diagram.
However, we think that the track of V2491~Cyg broadly follows
the V574~Pup track (magenta-blue line).
The rough overlapping of V2491~Cyg with V574~Pup in the 
$(V-I)_0$-$(M_I-2.5\log f_{\rm s})$ diagram
supports our results of $E(B-V)=0.40$, $(m-M)_I=16.0$, $d=12.1$~kpc,
and $\log f_{\rm s}= -0.40$ for V2491~Cyg.


\begin{figure*}
\plotone{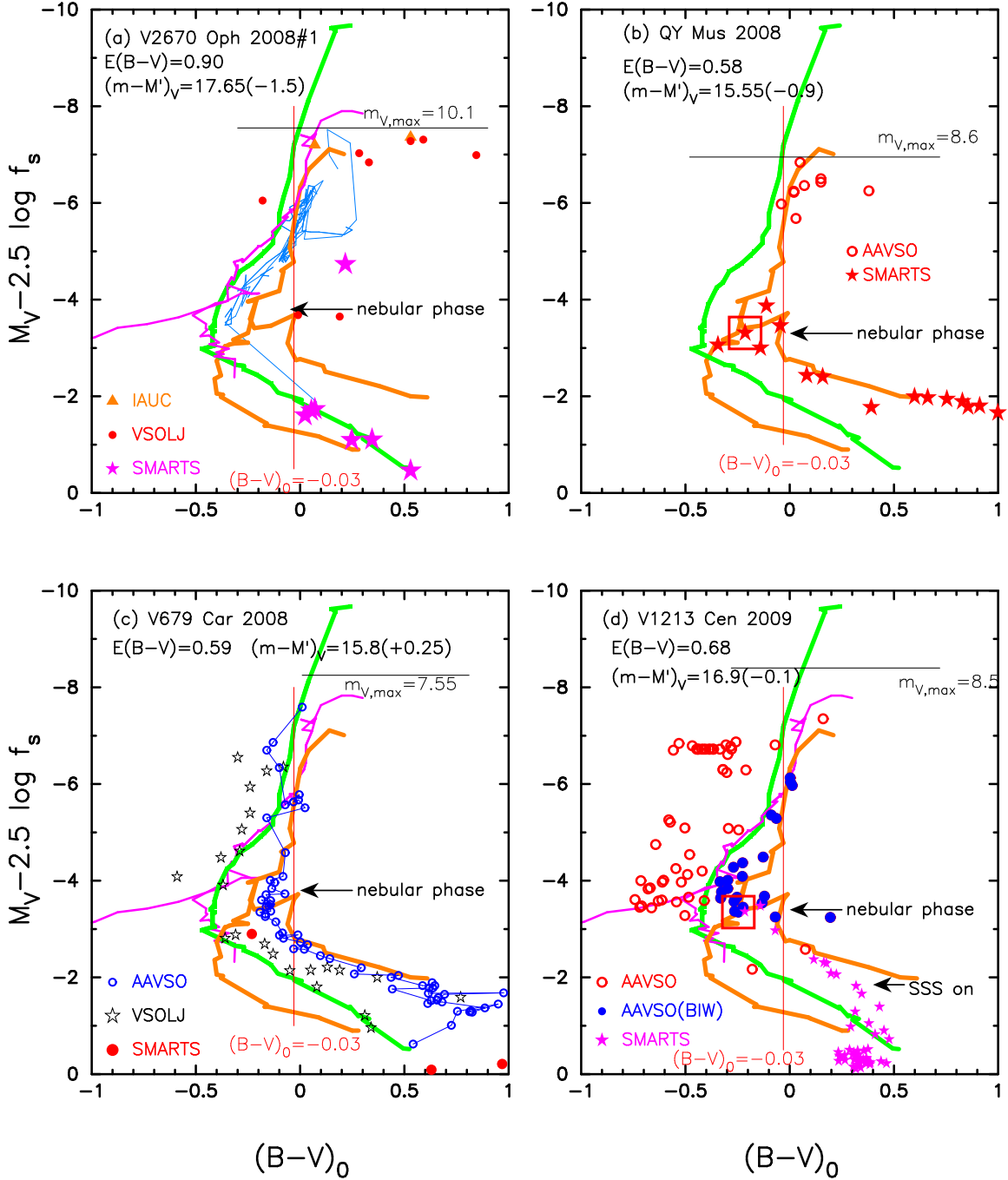}
\caption{
Same as Figure 
\ref{hr_diagram_v496_sct_v959_mon_v834_car_v1369_cen_outburst},
but for (a) V2670~Oph, (b) QY~Mus, (c) V679~Car, and (d) V1213~Cen.
In panel (a), we add the track of PW~Vul (thin solid cyan-blue line).
In panels (a), (c), and (d), we add the track of V1974~Cyg (magenta lines).
\label{hr_diagram_v2670_oph_qy_mus_v679_car_v1213_cen_outburst}}
\end{figure*}


\begin{figure*}
\plotone{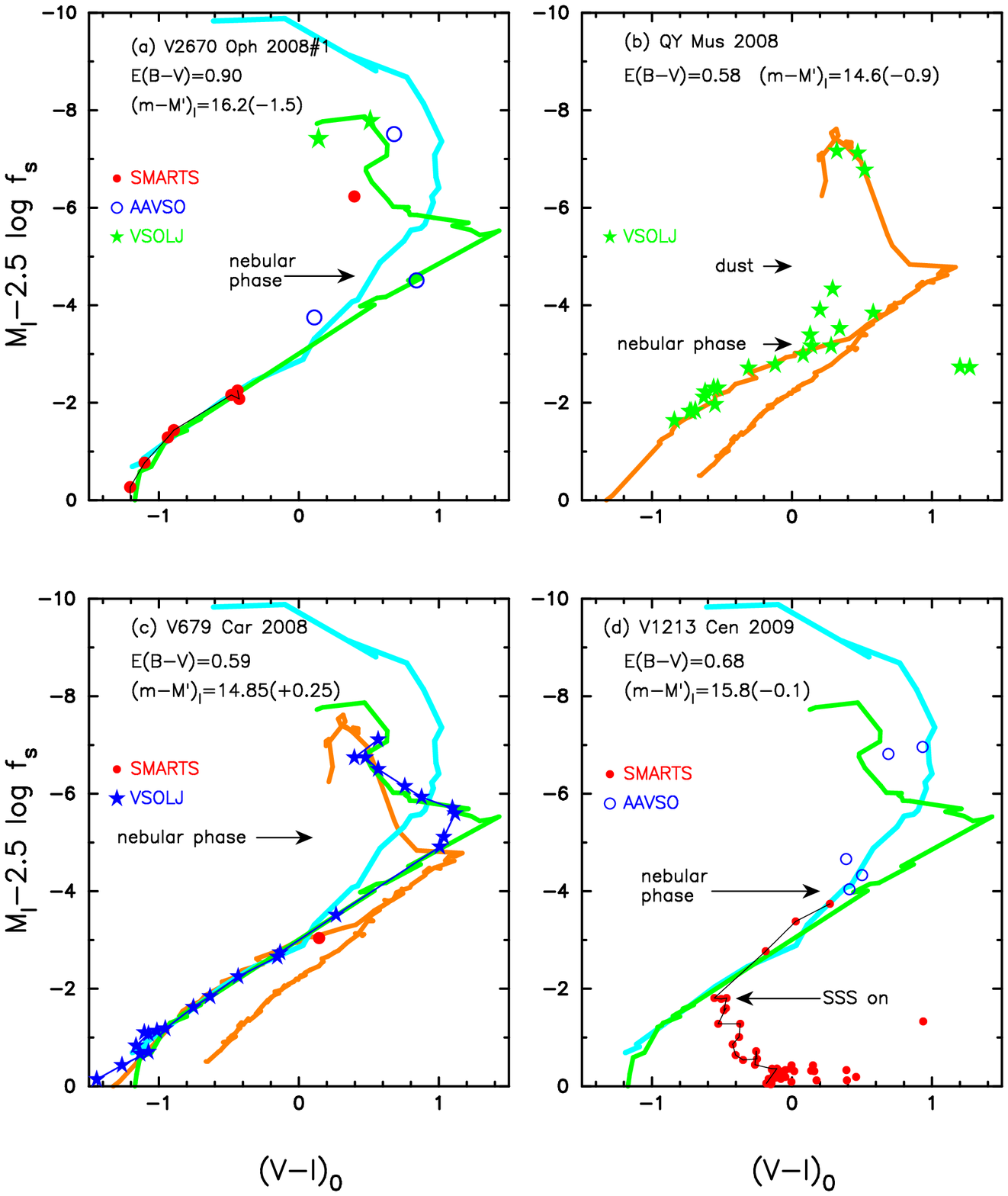}
\caption{
Same as Figure 
\ref{hr_diagram_v496_sct_v959_mon_v834_car_v1369_cen_outburst_vi},
but for (a) V2670~Oph, (b) QY~Mus, (c) V679~Car, and (d) V1213~Cen.
\label{hr_diagram_v2670_oph_qy_mus_v679_car_v1213_cen_outburst_vi}}
\end{figure*}

\subsection{V2670~Oph 2008\#1}
\label{v2670_oph_vi}
\citet{hac19kb} obtained $E(B-V)=1.05$, $(m-M)_V=17.6$, $d=7.4$~kpc,
and $\log f_{\rm s}= +0.33$.
We have reanalyzed the $BVI_{\rm C}$ light/color curves of V2670~Oph
in Appendix \ref{v2670_oph_bvi} and obtained a new set of
$E(B-V)=0.90$, $(m-M)_V=16.15$, $d=4.7$~kpc,
and $\log f_{\rm s}= +0.59$ for V2670~Oph.
Then, we have $(m-M')_V=16.15+1.475=17.65$ and plot
the $(B-V)_0$-$(M_V-2.5\log f_{\rm s})$ diagram 
in Figure \ref{hr_diagram_v2670_oph_qy_mus_v679_car_v1213_cen_outburst}(a).
The $BVI_{\rm C}$ light curves of V2670~Oph show a few to several 
secondary peaks like those of NR~TrA in Section \ref{nr_tra_vi}.
This kind of secondary peaks are also observed in PW~Vul.
We add the track of PW~Vul (thin solid cyan-blue line) 
in Figure \ref{hr_diagram_v2670_oph_qy_mus_v679_car_v1213_cen_outburst}(a). 
The PW~Vul track shows a loop in the early phase.  
The data of V2670~Oph are so scattered in the early and middle phases.
In the later phase, V2670~Oph follows the track of V1500~Cyg (green line). 
We regard that V2670~Oph belongs to the V1500~Cyg type.   

The distance modulus in $I_{\rm C}$ band, $(m-M)_I=14.7$, is 
taken from Appendix \ref{v2670_oph_bvi}.
Then, we have $(m-M')_I=14.7+1.475=16.2$.
The peak $I_{\rm C}$ brightness is 
$M'_I= M_I-2.5\log f_{\rm s}= -6.28 - 1.475 = -7.75$ from the data of VSOLJ.
We plot the $(V-I)_0$-$(M_I-2.5\log f_{\rm s})$ diagram in Figure 
\ref{hr_diagram_v2670_oph_qy_mus_v679_car_v1213_cen_outburst_vi}(a).
Here, we adopt the $BVI_{\rm C}$ data from AAVSO, VSOLJ, and SMARTS.
The track of V2670~Oph almost overlaps with the V5114~Sgr track 
(green line).  Thus, V2670~Oph belongs to the V1500~Cyg type.  
The overlapping of V2670~Oph and V5114~Sgr in the 
$(V-I)_0$-$(M_I-2.5\log f_{\rm s})$ diagram supports the results of
$E(B-V)=0.90$, $(m-M)_I=14.7$, $d=4.7$~kpc,
and $\log f_{\rm s}= +0.59$ for V2670~Oph.

\subsection{QY~Mus 2008}
\label{qy_mus_vi}
\citet{hac19kb} obtained $E(B-V)=0.58$, $(m-M)_V=14.65$, $d=3.7$~kpc,
and $\log f_{\rm s}= +0.35$.  We have reanalyzed the $BVI_{\rm C}$
light/color curves of QY~Mus in Appendix \ref{qy_mus_bvi} and
obtained a similar set of parameters for QY~Mus.
Then, we have $(m-M')_V=14.65+0.875=15.55$ and plot 
the $(B-V)_0$-$(M_V-2.5\log f_{\rm s})$ diagram 
in Figure \ref{hr_diagram_v2670_oph_qy_mus_v679_car_v1213_cen_outburst}(b).
The track of QY~Mus follows well the upper branch of LV~Vul (orange line).

The distance modulus in $I_{\rm C}$ band, $(m-M)_I=13.7$, is 
taken from Appendix \ref{qy_mus_bvi}.
Then, we have $(m-M')_I=13.7+0.875=14.6$.
The peak $I_{\rm C}$ brightness is 
$M'_I= M_I-2.5\log f_{\rm s}= -6.26 - 0.875 = -7.1$ from the data of VSOLJ.
We plot the $(V-I)_0$-$(M_I-2.5\log f_{\rm s})$ diagram in Figure 
\ref{hr_diagram_v2670_oph_qy_mus_v679_car_v1213_cen_outburst_vi}(b).
Here, we adopt the $BVI_{\rm C}$ data from VSOLJ.
The VSOLJ data (filled green stars) of QY~Mus almost overlaps
with the upper branch of V496~Sct/V959~Mon subtype (orange line).
The overlapping of QY~Mus and V496~Sct/V959~Mon 
on the $(V-I)_0$-$(M_I-2.5\log f_{\rm s})$ diagram supports the results of
$E(B-V)=0.58$, $(m-M)_I=13.7$, $d=3.7$~kpc,
and $\log f_{\rm s}= +0.35$ for QY~Mus.

\subsection{V679~Car 2008}
\label{v679_car_vi}
\citet{hac19ka} obtained $E(B-V)=0.69$, $(m-M)_V=16.1$, $d=6.2$~kpc,
and $\log f_{\rm s}= +0.0$.  We have reanalyzed the $BVI_{\rm C}$
light/color curves of V679~Car in Appendix \ref{v679_car_bvi} and
obtained a new set of parameters, i.e., $E(B-V)=0.59$, $(m-M)_V=16.05$,
$d=7.0$~kpc, and $\log f_{\rm s}= -0.10$ for V679~Car.
Then, we have $(m-M')_V=16.05-0.25=15.8$ and plot 
the $(B-V)_0$-$(M_V-2.5\log f_{\rm s})$ diagram in Figure
\ref{hr_diagram_v2670_oph_qy_mus_v679_car_v1213_cen_outburst}(c).
The track of V679~Car (AAVSO, unfilled blue circles) 
slightly deviates from, in the early phase,
but broadly follows, in the middle and later phases, that of LV~Vul 
while the data of VSOLJ (unfilled black stars) broadly follows
the V1500~Cyg track (green line).  Therefore, we regard V679~Car
to belong to the V1500~Cyg type from the data of VSOLJ.

The distance modulus in $I_{\rm C}$ band, $(m-M)_I=15.1$, is 
taken from Appendix \ref{v679_car_bvi}.
Then, we have $(m-M')_I=15.1-0.25=14.85$.
The peak $I_{\rm C}$ brightness is 
$M'_I= M_I-2.5\log f_{\rm s}= -7.72 + 0.25 = -7.5$ from the data of VSOLJ.
We plot the $(V-I)_0$-$(M_I-2.5\log f_{\rm s})$ diagram in Figure 
\ref{hr_diagram_v2670_oph_qy_mus_v679_car_v1213_cen_outburst_vi}(c).
Here, we adopt the $BVI_{\rm C}$ data from VSOLJ and SMARTS.
The track of V679~Car overlaps well with the V5114~Sgr (green line) track.
The overlapping of V679~Car with the V5114~Sgr track
on the $(V-I)_0$-$(M_I-2.5\log f_{\rm s})$ diagram supports the results
of $E(B-V)=0.59$, $(m-M)_I=15.1$, $d=7.0$~kpc,
and $\log f_{\rm s}= -0.10$ for V679~Car.

\subsection{V1213~Cen 2009}
\label{v1213_cen_vi}
\citet{hac19kb} obtained $E(B-V)=0.78$, $(m-M)_V=16.95$, $d=8.1$~kpc,
and $\log f_{\rm s}= +0.05$.  We have reanalyzed the $BVI_{\rm C}$
light/color curves of V1213~Cen in Appendix \ref{v1213_cen_bvi} and
obtained a similar set of parameters, i.e., $E(B-V)=0.68$, 
$(m-M)_V=16.8$, $d=8.6$~kpc, and $\log f_{\rm s}= +0.05$.
Then, we have $(m-M')_V=16.8 + 0.125= 16.9$ and plot 
the $(B-V)_0$-$(M_V-2.5\log f_{\rm s})$ diagram in Figure
\ref{hr_diagram_v2670_oph_qy_mus_v679_car_v1213_cen_outburst}(d).
The data of AAVSO are scattered among the observers.  We separately 
plot the data for the observer code ``BIW'' (filled blue circles).
This V1213~Cen track almost follows the LV~Vul track (orange line)
until the nebular phase started.
The nova entered the SSS phase at $M'_V= M_V-2.5\log f_{\rm s}= -2$ and
$(B-V)_0\sim +0.3$.  Then the $(B-V)_0$ color stays at $\sim +0.3$. 
This could be a color of an accretion disk irradiated by a hot WD
like in V574~Pup in Section \ref{v574_pup_vi}.
The supersoft X-ray light curve was well reproduced with a $1.0~M_\sun$
WD (Ne2) model as shown in Figure
\ref{v1213_cen_lv_vul_v4743_sgr_v_bv_ub_color_curve_logscale}
of Appendix \ref{v1213_cen_bvi}.
\citep[See][for X-ray observation.]{schw10, schw11} 

The distance modulus in $I_{\rm C}$ band, $(m-M)_I=15.7$, is 
taken from Appendix \ref{v1213_cen_bvi}.
Then, we have $(m-M')_I=15.7+0.125=15.8$.
The peak $I_{\rm C}$ brightness is 
$M'_I= M_I-2.5\log f_{\rm s}= -7.8 - 0.125 = -7.9$ from the data of AAVSO.
We plot the $(V-I)_0$-$(M_I-2.5\log f_{\rm s})$ diagram in Figure 
\ref{hr_diagram_v2670_oph_qy_mus_v679_car_v1213_cen_outburst_vi}(d).
Here, we adopt the $BVI_{\rm C}$ data from AAVSO and SMARTS.
The track of V1213~Cen almost follows the track of V1500~Cyg subtype
(cyan line) until the SSS phase started at 
$M'_I= M_I-2.5\log f_{\rm s}= -2$ and $(V-I)_0= -0.6$.
The overlapping of V1213~Cen and V1500~Cyg 
on the $(V-I)_0$-$(M_I-2.5\log f_{\rm s})$ diagram supports the results of
$E(B-V)=0.68$, $(m-M)_I=15.7$, $d=8.6$~kpc,
and $\log f_{\rm s}= +0.05$ for V1213~Cen.


\begin{figure*}
\plotone{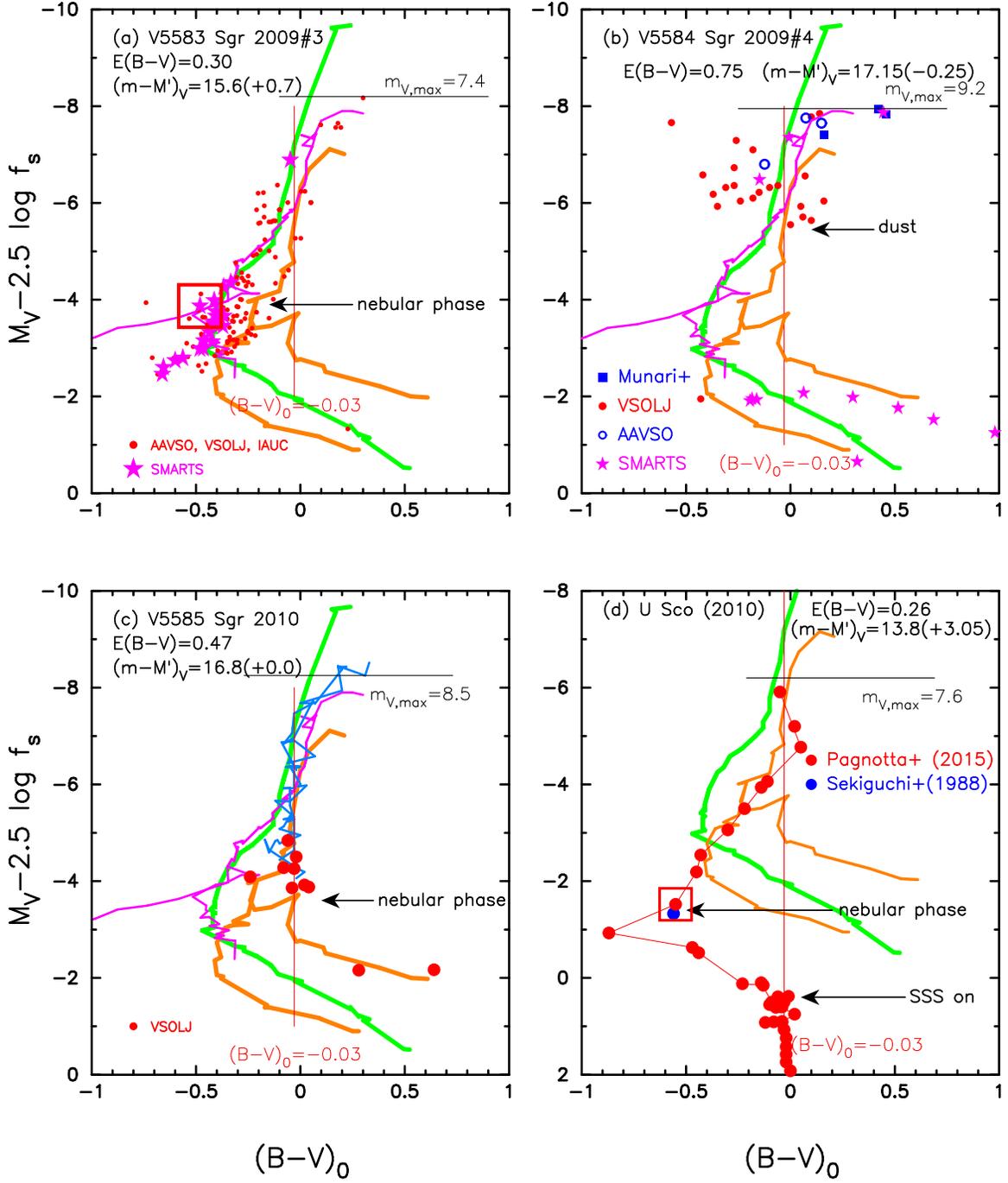}
\caption{
Same as Figure 
\ref{hr_diagram_v496_sct_v959_mon_v834_car_v1369_cen_outburst}, 
but for (a) V5583~Sgr, (b) V5584~Sgr, (c) V5585~Sgr, and (d) U~Sco.
In panels (a), (b), and (c), we add the track of V1974~Cyg (magenta lines).
In panel (c), we add the track of V1668~Cyg (thin solid cyan-blue lines).
\label{hr_diagram_v5583_sgr_v5584_sgr_v5585_sgr_u_sco_outburst}}
\end{figure*}


\begin{figure*}
\plotone{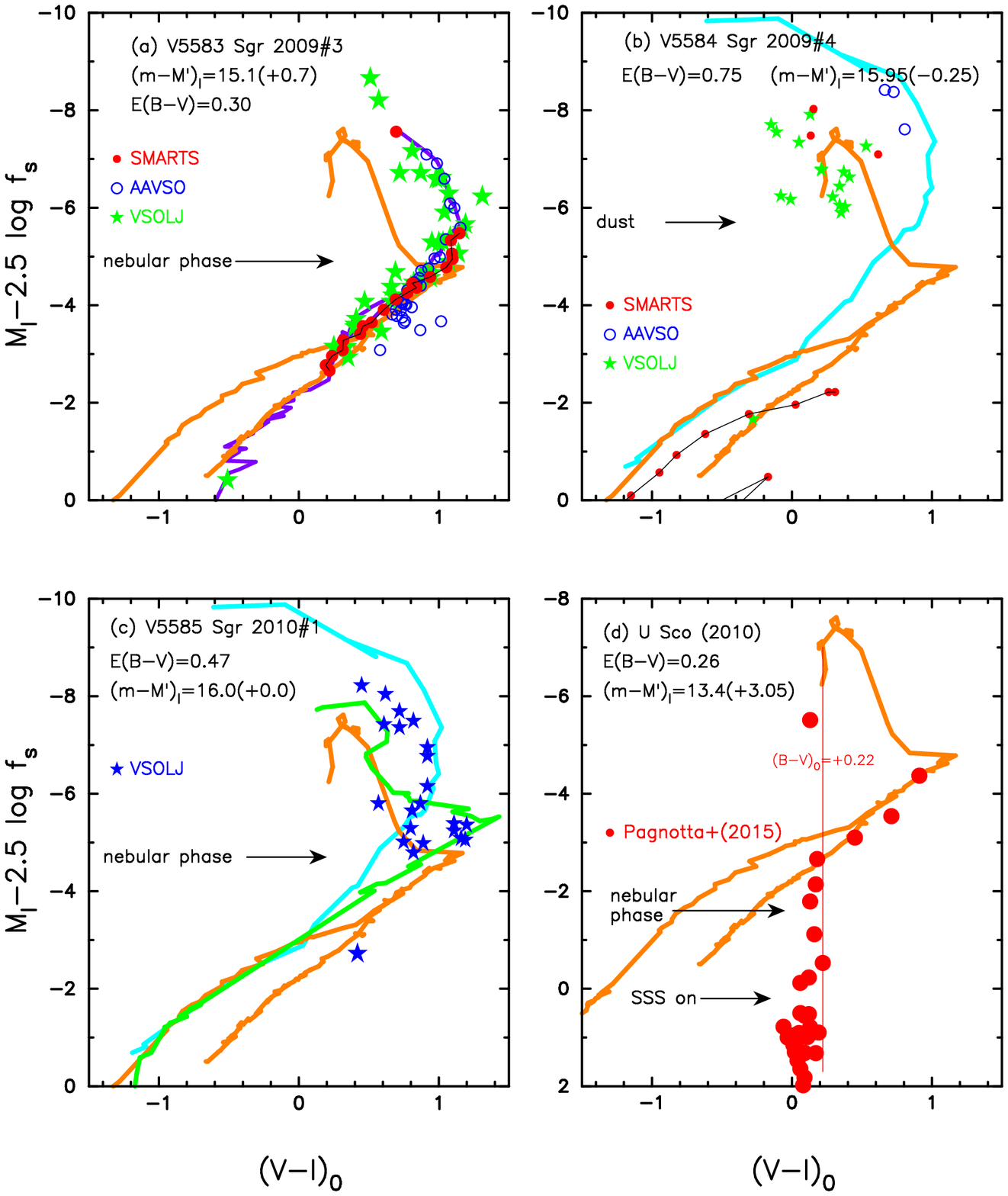}
\caption{
Same as Figure 
\ref{hr_diagram_v496_sct_v959_mon_v834_car_v1369_cen_outburst_vi},
but for (a) V5583~Sgr, (b) V5584~Sgr, (c) V5585~Sgr, and (d) U~Sco (2010).
In panel (a), we add the V1065~Cen track 
(thin solid blue-magenta line) in Section \ref{v1065_cen_vi}.
\label{hr_diagram_v5583_sgr_v5584_sgr_v5585_sgr_u_sco_outburst_vi}}
\end{figure*}

\subsection{V5583~Sgr 2009\#3}
\label{v5583_sgr_vi}
\citet{hac19kb} obtained $E(B-V)=0.30$, $(m-M)_V=16.3$, $d=12$~kpc,
and $\log f_{\rm s}= -0.29$.  We have reanalyzed the $BVI_{\rm C}$ 
light/color curves of V5583~Sgr in Appendix \ref{v5583_sgr_bvi}
and obtained a similar set of parameters for V5583~Sgr.  Then, 
we have $(m-M')_V=16.3 - 0.725 = 15.6$ and plot
the $(B-V)_0$-$(M_V-2.5\log f_{\rm s})$ diagram in Figure
\ref{hr_diagram_v5583_sgr_v5584_sgr_v5585_sgr_u_sco_outburst}(a).
The track of V5583~Sgr broadly follows the track of V1500~Cyg (green line)
or V1974~Cyg (magenta line).

The distance modulus in $I_{\rm C}$ band, $(m-M)_I=15.8$, is 
taken from Appendix \ref{v5583_sgr_bvi}.
Then, we have $(m-M')_I=15.8-0.725=15.1$.
The peak $I_{\rm C}$ brightness is 
$M'_I= M_I-2.5\log f_{\rm s}= -9.36 + 0.725 = -8.6$ from the data of VSOLJ.
We plot the $(V-I)_0$-$(M_I-2.5\log f_{\rm s})$ diagram in Figure 
\ref{hr_diagram_v5583_sgr_v5584_sgr_v5585_sgr_u_sco_outburst_vi}(a).
Here, we adopt the $BVI_{\rm C}$ data from AAVSO, VSOLJ, and SMARTS.
The track of V5583~Sgr almost follows the track of 
V1065~Cen (solid blue-magenta line).  The overlapping of V5583~Sgr with
V1065~Cen on the $(V-I)_0$-$(M_I-2.5\log f_{\rm s})$ diagram supports 
the results of $E(B-V)=0.30$, $(m-M)_I=15.8$, $d=12$~kpc,
and $\log f_{\rm s}= -0.29$ for V5583~Sgr.

\subsection{V5584~Sgr 2009\#4}
\label{v5584_sgr_vi}
\citet{hac19kb} obtained $E(B-V)=0.70$, $(m-M)_V=16.7$, $d=8.0$~kpc,
and $\log f_{\rm s}= +0.13$.  We have reanalyzed the $BVI_{\rm C}$
light/color curves of V5584~Sgr in Appendix \ref{v5584_sgr_bvi} and
obtained a new set of parameters, i.e., $E(B-V)=0.75$, $(m-M)_V=16.9$,
$d=8.2$~kpc, and $\log f_{\rm s}= +0.10$.  The main differences are
the reddening of $E(B-V)=0.75$ and
the timescaling factor of $\log f_{\rm s}= +0.10$.
Then, we have $(m-M')_V=16.9 + 0.25=17.15$ and plot
the $(B-V)_0$-$(M_V-2.5\log f_{\rm s})$ diagram in Figure 
\ref{hr_diagram_v5583_sgr_v5584_sgr_v5585_sgr_u_sco_outburst}(b).
The track of V5584~Sgr broadly follows the track of V1974~Cyg
until the dust blackout occurs.

The distance modulus in $I_{\rm C}$ band, $(m-M)_I=15.7$, is 
taken from Appendix \ref{v5584_sgr_bvi}.
Then, we have $(m-M')_I=15.7+0.25=15.95$.
The peak $I_{\rm C}$ brightness is 
$M'_I= M_I-2.5\log f_{\rm s}= -8.17 - 0.25 = -8.42$ from the data of AAVSO.
We plot the $(V-I)_0$-$(M_I-2.5\log f_{\rm s})$ diagram in Figure 
\ref{hr_diagram_v5583_sgr_v5584_sgr_v5585_sgr_u_sco_outburst_vi}(b).
Here, we adopt the $BVI_{\rm C}$ data from AAVSO, VSOLJ, and SMARTS.
The track of V5584~Sgr almost follows the track of V496~Sct/V959~Mon
(orange line) until the dust blackout occurs.
The broad overlapping of V5584~Sgr and V496~Sct/V959~Mon on the 
$(V-I)_0$-$(M_I-2.5\log f_{\rm s})$ diagram may support the results of
$E(B-V)=0.75$, $(m-M)_I=15.7$, $d=8.2$~kpc,
and $\log f_{\rm s}= +0.10$ for V5584~Sgr.

\subsection{V5585~Sgr 2010}
\label{v5585_sgr_vi}
\citet{hac19kb} obtained $E(B-V)=0.47$, $(m-M)_V=16.7$, $d=11$~kpc,
and $\log f_{\rm s}= +0.10$.  We have reanalyzed the $BVI_{\rm C}$
light/color curves of V5585~Sgr in Appendix \ref{v5585_sgr_bvi}
and obtained a new set of parameters,
i.e., $E(B-V)=0.47$, $(m-M)_V=16.8$, $d=11.6$~kpc, 
and $\log f_{\rm s}= +0.0$.  Then, we have $(m-M')_V=16.8+0.0=16.8$
and plot the $(B-V)_0$-$(M_V-2.5\log f_{\rm s})$ diagram in Figure
\ref{hr_diagram_v5583_sgr_v5584_sgr_v5585_sgr_u_sco_outburst}(c).
The track of V5585~Sgr broadly follows the track of LV~Vul (orange line)
or V1668~Cyg (cyan-blue line) in the early and middle phases.

The distance modulus in $I_{\rm C}$ band, $(m-M)_I=16.0$, is 
taken from Appendix \ref{v5585_sgr_bvi}.
Then, we have $(m-M')_I=16.0+0.0=16.0$.
The peak $I_{\rm C}$ brightness is 
$M'_I= M_I-2.5\log f_{\rm s}= -8.2 - 0.0 = -8.2$ from the data of VSOLJ.
We plot the $(V-I)_0$-$(M_I-2.5\log f_{\rm s})$ diagram in Figure 
\ref{hr_diagram_v5583_sgr_v5584_sgr_v5585_sgr_u_sco_outburst_vi}(c).
Here, we adopt the $BVI_{\rm C}$ data from VSOLJ.
The track of V5585~Sgr almost follows the track of V5114~Sgr (green line)
and then V1500~Cyg (cyan line).
The rough overlapping of V5585~Sgr and V5114~Sgr
on the $(V-I)_0$-$(M_I-2.5\log f_{\rm s})$ diagram may support the results
of $E(B-V)=0.47$, $(m-M)_I=16.0$, $d=11.6$~kpc,
and $\log f_{\rm s}= +0.0$ for V5585~Sgr.

\subsection{U~Sco 2010}
\label{u_sco_vi}
U~Sco is a recurrent nova with the recurrence time of $\sim 10$~yr
\citep[e.g.,][]{sch10b}.
As already discussed in Section
\ref{exceptional_type_novae_vi}, U~Sco has a very short
period of the universal decline trend of $F_\nu\propto t^{-1.75}$
and quickly shifts to a steeper decline of $F_\nu\propto t^{-3.5}$. 
\citet{hac18kb} obtained $E(B-V)=0.26$, $(m-M)_V=16.3$, $d=12.6$~kpc,
and $\log f_{\rm s}= -1.32$ for the U~Sco 2010 outburst.
We have reanalyzed the $UBVI_{\rm C}$ light/color curves of U~Sco
in Appendix \ref{u_sco_ubvik} and obtained a new set of parameters,
i.e., $E(B-V)=0.26$, $(m-M)_V=16.85$, $d=16.2$~kpc,
and $\log f_{\rm s}= -1.22$.  Then, we have $(m-M')_V=16.85 - 3.05= 13.8$
and plot the $(B-V)_0$-$(M_V-2.5\log f_{\rm s})$ diagram in Figure
\ref{hr_diagram_v5583_sgr_v5584_sgr_v5585_sgr_u_sco_outburst}(d).
The track of U~Sco broadly follows the track of LV~Vul in the early phase
on the $(B-V)_0$-$(M_V-2.5\log f_{\rm s})$ diagram.

The distance modulus in $I_{\rm C}$ band, $(m-M)_I=16.45$, is 
taken from Appendix \ref{u_sco_ubvik}.
Then, we have $(m-M')_I=16.45-3.05=13.4$.
The peak $I_{\rm C}$ brightness is 
$M'_I= M_I-2.5\log f_{\rm s}= -9.75 + 3.05 = -6.7$ from the data of AAVSO.
We plot the $(V-I)_0$-$(M_I-2.5\log f_{\rm s})$ diagram in Figure 
\ref{hr_diagram_v5583_sgr_v5584_sgr_v5585_sgr_u_sco_outburst_vi}(d).
Here, we adopt the $BVI_{\rm C}$ data from \citet{pagnotta15}.
The track of U~Sco is located on the track of V496~Sct/V959~Mon
(orange line) until $M'_I= M_I-2.5\log f_{\rm s}= -2.5$ and
$(V-I)_0 = +0.2$ and then goes down along
almost the line of $(V-I)_0 \sim +0.2$. 
\citet{hac18kb} discussed that the constancy in the $(B-V)_0$ color
originates from an accretion disk irradiated by a central
hot WD during the SSS phase \citep[see Figure 27(a) of][]{hac18kb}.
The constancy in the $(V-I)_0$ color should be the same
origin as that in the $(B-V)_0$ color \citep[see, e.g.,][]{kat20sh}.
The rough overlappings between U~Sco and LV~Vul on the
$(U-B)_0$-$(M_B-2.5\log f_{\rm s})$ and
$(B-V)_0$-$(M_V-2.5\log f_{\rm s})$ diagrams and between
U~Sco and V496~Sct/V959~Mon on the $(V-I)_0$-$(M_I-2.5\log f_{\rm s})$
diagram support our new values of $E(B-V)=0.26$, $(m-M)_V=16.85$,
$(m-M)_I=16.45$, $d=16.2$~kpc, and $\log f_{\rm s}= -1.22$ for U~Sco.


\begin{figure*}
\plotone{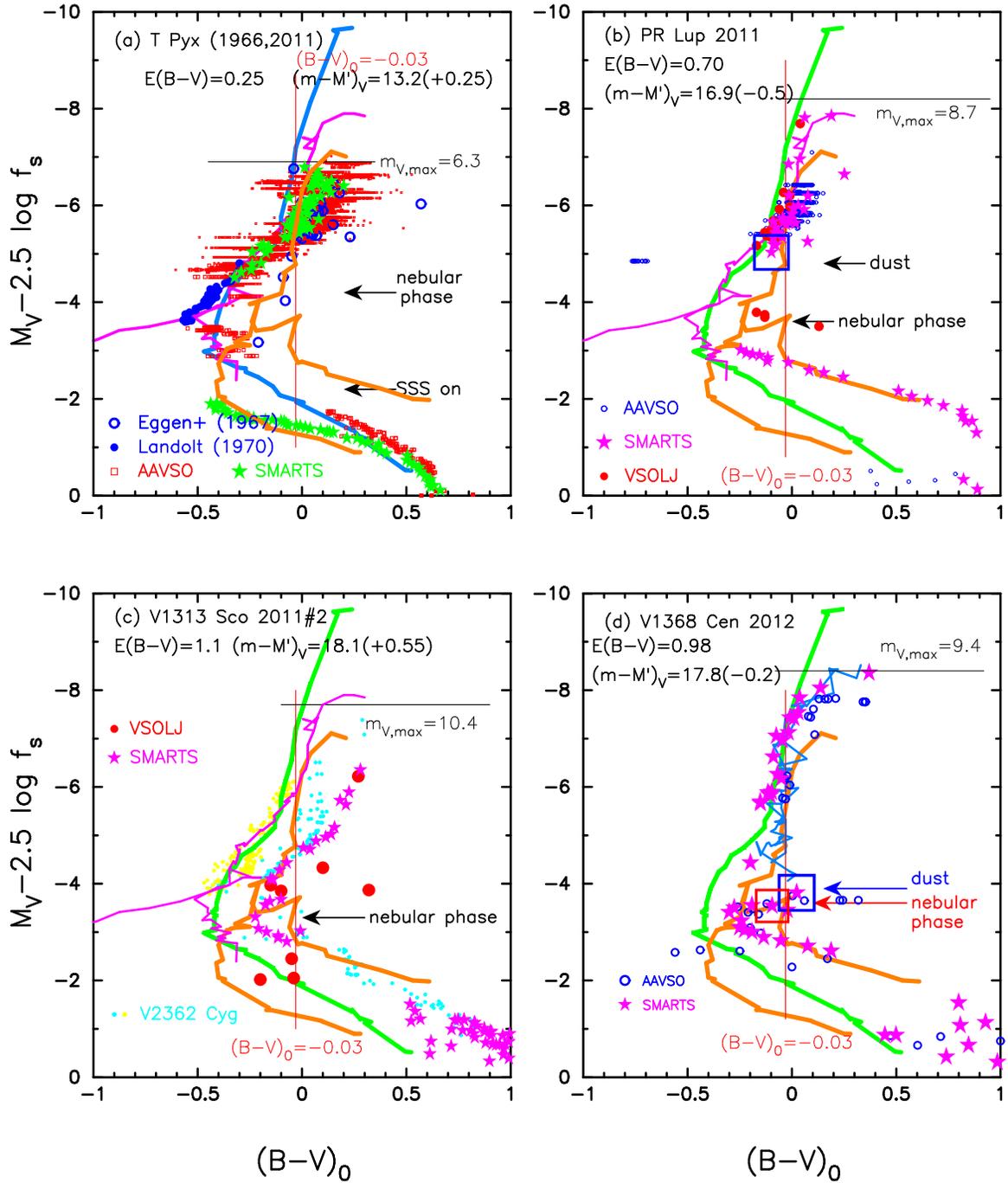}
\caption{
Same as Figure 
\ref{hr_diagram_v496_sct_v959_mon_v834_car_v1369_cen_outburst}, 
but for (a) T~Pyx, (b) PR~Lup, (c) V1313~Sco, and (d) V1368~Cen.
In panel (a), the thick solid cyan-blue line is the track of V1500~Cyg.
In panels (a), (b), and (c), we add the track of V1974~Cyg (magenta lines).
In panel (c), we add the track of V2362~Cyg (cyan dots except for the 
secondary maximum and yellow dots for the secondary maximum duration).
In panel (d), we add the track of V1668~Cyg (thin solid cyan-blue lines).
\label{hr_diagram_t_pyx_pr_lup_v1313_sco_v1368_cen_outburst_bv}}
\end{figure*}


\begin{figure*}
\plotone{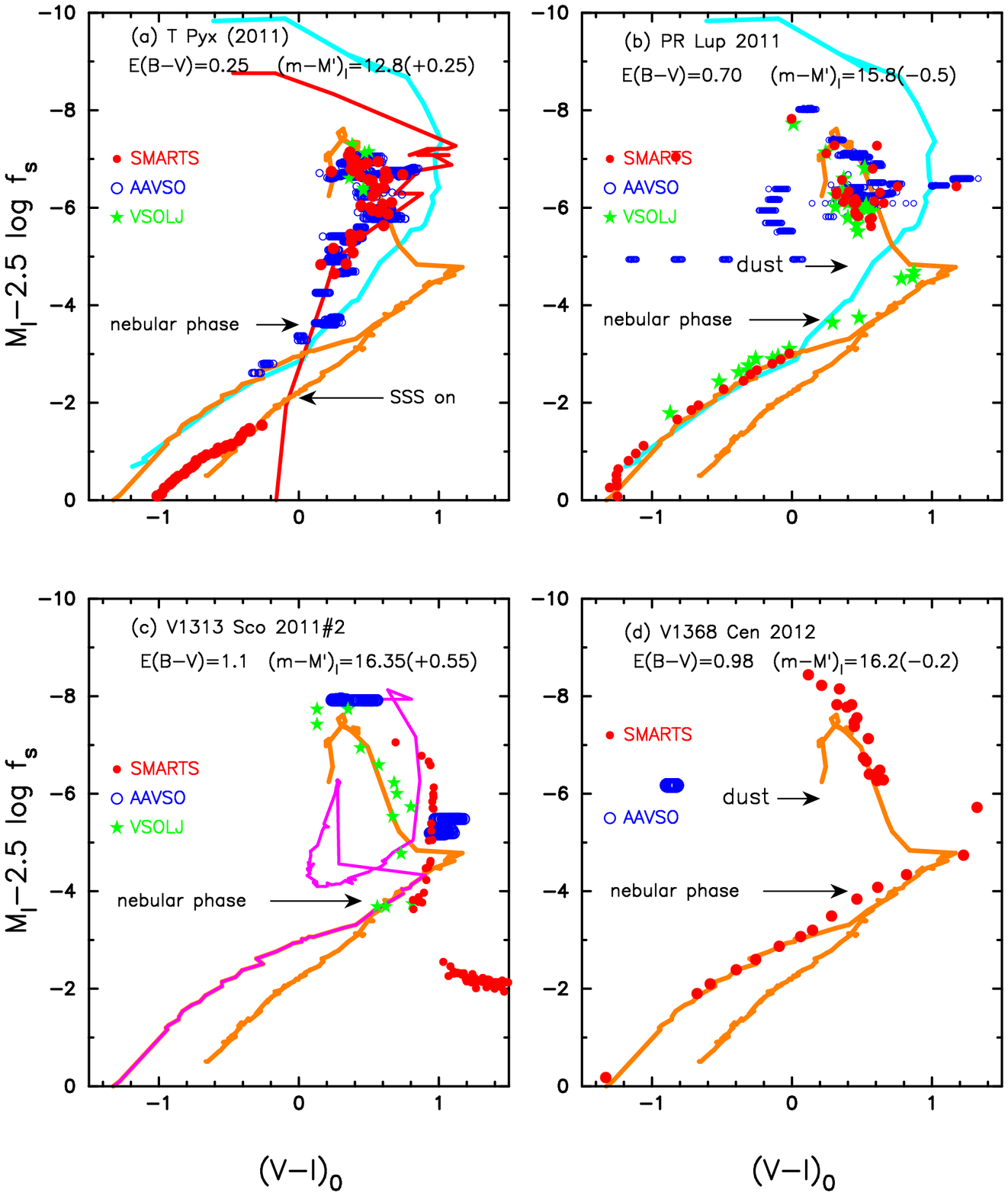}
\caption{
Same as Figure 
\ref{hr_diagram_v496_sct_v959_mon_v834_car_v1369_cen_outburst_vi},
but for (a) T~Pyx, (b) PR~Lup, (c) V1313~Sco, and (d) V1368~Cen.
In panel (a), we add the template track of V597~Pup (solid red line)
in Section \ref{v597_pup_vi}.
In panel (c), we add the track of V2362~Cyg (magenta line) in Section
\ref{v2362_cyg_vi}.
\label{hr_diagram_t_pyx_pr_lup_v1313_sco_v1368_cen_outburst_vi}}
\end{figure*}

\subsection{T~Pyx 2011}
\label{t_pyx_vi}
\citet{hac14k} obtained $E(B-V)=0.25$, $(m-M)_V=13.8$, and $d=4.0$~kpc.
We have reanalyzed the $UBVI_{\rm C}$ light/color curves of the T~Pyx
2011 outburst in Appendix \ref{t_pyx_ubvi} and obtained a new
parameter set of $E(B-V)=0.25$, $(m-M)_V=13.45$, $d=3.4$~kpc, and
$\log f_{\rm s}= -0.10$.  We have already discussed the distance and
reddening of T~Pyx on the 
$(U-B)_0$-$(M_B-2.5\log f_{\rm s})$ diagram in Section \ref{t_pyx_ub}.
Then, we have $(m-M')_V=13.45 - 0.25=13.2$
and plot the $(B-V)_0$-$(M_V-2.5\log f_{\rm s})$ diagram in Figure
\ref{hr_diagram_t_pyx_pr_lup_v1313_sco_v1368_cen_outburst_bv}(a).
The track of T~Pyx follows the track of LV~Vul in the early phase,
then goes down along the tracks of V1500~Cyg/V1974~Cyg in the 
middle phase, and finally follows again the lower branch of
LV~Vul in the later phase.  Thus, it transfers from the LV~Vul to 
V1500~Cyg, and then from V1500~Cyg to LV~Vul type 
in the $(B-V)_0$-$(M_V-2.5\log f_{\rm s})$ diagram. 

The distance modulus in $I_{\rm C}$ band, $(m-M)_I=13.05$, is 
taken from Appendix \ref{t_pyx_ubvi}.
Then, we have $(m-M')_I=13.05-0.25=12.8$.
The peak $I_{\rm C}$ brightness is 
$M'_I= M_I-2.5\log f_{\rm s}= -7.45 + 0.25 = -7.2$ from the data of AAVSO.
We plot the $(V-I)_0$-$(M_I-2.5\log f_{\rm s})$ diagram in Figure 
\ref{hr_diagram_t_pyx_pr_lup_v1313_sco_v1368_cen_outburst_vi}(a).
Here, we adopt the $BVI_{\rm C}$ data from AAVSO, VSOLJ, and SMARTS.
We add the track of V597~Pup (solid red line), which belongs to
the V1500~Cyg type.  The track of T~Pyx first follows the track of 
V496~Sct/V959~Mon (orange line), then that of V597~Pup, and finally 
that of V496~Sct/V959~Mon again.  The data of AAVSO (unfilled blue circles)
follows the track of V1500~Cyg (cyan line) in the later phase.
This behavior is consistent with that of AAVSO in the
$(B-V)_0$-$(M_V-2.5\log f_{\rm s})$ diagram of Figure 
\ref{hr_diagram_t_pyx_pr_lup_v1313_sco_v1368_cen_outburst_bv}(a). 
The overlapping of T~Pyx with V496~Sct/V959~Mon and V597~Pup
on the $(V-I)_0$-$(M_I-2.5\log f_{\rm s})$ diagram may support the results of
$E(B-V)=0.25$, $(m-M)_I=13.0$, $d=3.4$~kpc,
and $\log f_{\rm s}= -0.10$ for T~Pyx.

\subsection{PR~Lup 2011}
\label{pr_lup_vi}
\citet{hac19kb} obtained $E(B-V)=0.74$, $(m-M)_V=16.1$, $d=5.8$~kpc,
and $\log f_{\rm s}= +0.23$.  We have reanalyzed the $BVI_{\rm C}$ 
multi-band light/color curves of PR~Lup in Appendix \ref{pr_lup_bvi}
and obtained a new parameter set of $E(B-V)=0.70$, $(m-M)_V=16.4$,
$d=7.0$~kpc, and $\log f_{\rm s}= +0.20$.  Then, we have 
$(m-M')_V=16.4 + 0.5=16.9$ and plot the $(B-V)_0$-$(M_V-2.5\log f_{\rm s})$
diagram in Figure 
\ref{hr_diagram_t_pyx_pr_lup_v1313_sco_v1368_cen_outburst_bv}(b).
The track of PR~Lup almost follows the upper branch of LV~Vul
(orange line).

The distance modulus in $I_{\rm C}$ band, $(m-M)_I=15.3$, is 
taken from Appendix \ref{pr_lup_bvi}.
Then, we have $(m-M')_I=15.3 + 0.5 = 15.8$.
The peak $I_{\rm C}$ brightness is 
$M'_I= M_I-2.5\log f_{\rm s}= -7.6 - 0.5 = -8.1$ from the data of AAVSO.
We plot the $(V-I)_0$-$(M_I-2.5\log f_{\rm s})$ diagram in Figure 
\ref{hr_diagram_t_pyx_pr_lup_v1313_sco_v1368_cen_outburst_vi}(b).
Here, we adopt the $BVI_{\rm C}$ data from AAVSO, VSOLJ, and SMARTS.
The track of PR~Lup almost follows the track of V496~Sct/V959~Mon
(upper branch of orange line).
This is consistent with the above result that the track of PR~Lup
follows the upper branch of LV~Vul in the 
$(B-V)_0$-$(M_V-2.5\log f_{\rm s})$ diagram.
The overlapping of PR~Lup and V496~Sct
on the $(V-I)_0$-$(M_I-2.5\log f_{\rm s})$ diagram supports the results of
$E(B-V)=0.70$, $(m-M)_I=15.3$, $d=7.0$~kpc,
and $\log f_{\rm s}= +0.20$ for PR~Lup.

\subsection{V1313~Sco 2011\#2}
\label{v1313_sco_vi}
\citet{hac19kb} obtained $E(B-V)=1.30$, $(m-M)_V=19.0$, $d=9.9$~kpc,
and $\log f_{\rm s}= -0.22$.  We have reanalyzed the $BVI_{\rm C}$
light/color curves of V1313~Sco in Appendix \ref{v1313_sco_bvi}
and obtained a new set of parameters, i.e.,
$E(B-V)=1.1$, $(m-M)_V=18.65$, $d=11.2$~kpc,
and $\log f_{\rm s}= -0.22$.  The main difference is the color excess
of $E(B-V)=1.1$.
Then, we have $(m-M')_V=18.65 - 0.55 = 18.1$ and plot the 
$(B-V)_0$-$(M_V-2.5\log f_{\rm s})$ diagram in Figure 
\ref{hr_diagram_t_pyx_pr_lup_v1313_sco_v1368_cen_outburst_bv}(c).
We add the track of V2362~Cyg (cyan dots except for
the secondary maximum and yellow dots during the secondary maximum phase).
The track of V1313~Sco follows well the track of V2362~Cyg 
except for the duration of secondary maximum.  We regard V1313~Sco to
belong to the LV~Vul type because V2362~Cyg is a member of the LV~Vul type.

The distance modulus in $I_{\rm C}$ band, $(m-M)_I=16.9$, is 
taken from Appendix \ref{v1313_sco_bvi}.  Then, we have 
$(m-M')_I=16.9 - 0.55 = 16.35$.
The peak $I_{\rm C}$ brightness is 
$M'_I= M_I-2.5\log f_{\rm s}= -8.7 + 0.55 = -8.15$ from the data of AAVSO.
We plot the $(V-I)_0$-$(M_I-2.5\log f_{\rm s})$ diagram in Figure 
\ref{hr_diagram_t_pyx_pr_lup_v1313_sco_v1368_cen_outburst_vi}(c).
Here, we adopt the $BVI_{\rm C}$ data from AAVSO, VSOLJ, and SMARTS.
The VSOLJ data (filled green stars) of V1313~Sco almost follows
the track of V496~Sct/V959~Mon (orange line) while the AAVSO and SMARTS
data (unfilled blue circles and filled red circles) broadly follow 
the V2362~Cyg track (magenta line) until the nebular phase started.
After the nebular phase started, the $(V-I_{\rm C})_0$ color turned
to the red (rightward).  \citet{wal12} suggested that V1313~Sco is
a symbiotic nova, of which the companion star is a red giant. 
The redward excursion in the later phase is owing to the red color of
the companion star both in the $(B-V)_0$-$(M_V-2.5\log f_{\rm s})$ and
$(V-I)_0$-$(M_I-2.5\log f_{\rm s})$ diagrams.
The rough overlapping of V1313~Sco and V496~Sct or V2362~Cyg
on the $(V-I)_0$-$(M_I-2.5\log f_{\rm s})$ diagram supports the new
results of $E(B-V)=1.1$, $(m-M)_I=16.9$, $d=11.2$~kpc,
and $\log f_{\rm s}= -0.22$ for V1313~Sco.

\subsection{V1368~Cen 2012}
\label{v1368_cen_vi}
\citet{hac19kb} obtained $E(B-V)=0.93$, $(m-M)_V=17.6$, $d=8.8$~kpc,
and $\log f_{\rm s}= +0.10$.  We have reanalyzed the $BVI_{\rm C}$
multi-band light/color curves of V1368~Cen in Appendix \ref{v1368_cen_bvi}
and obtained a slightly different results from those in \citet{hac19kb},
that is, $E(B-V)=0.98$, $(m-M)_B=18.6$, $(m-M)_V=17.6$, $(m-M)_I=16.04$, 
$d=8.2$~kpc, and $\log f_{\rm s}= +0.07$.  Then,
we have $(m-M')_V=17.6 + 0.175 = 17.8$ and plot the 
$(B-V)_0$-$(M_V-2.5\log f_{\rm s})$ diagram in Figure
\ref{hr_diagram_t_pyx_pr_lup_v1313_sco_v1368_cen_outburst_bv}(d).
The track of V1368~Cen follows the upper branch of LV~Vul except for
the dust blackout phase.
This is consistent with the previous result that the track of V1368~Cen
follows the upper branch of LV~Vul in the 
$(B-V)_0$-$(M_V-2.5\log f_{\rm s})$ diagram \citep{hac19kb}.

The distance modulus in $I_{\rm C}$ band, $(m-M)_I=16.04$, is 
taken from Appendix \ref{v1368_cen_bvi}.
Then, we have $(m-M')_I = 16.04 + 0.175 = 16.2$.
The peak $I_{\rm C}$ brightness is 
$M'_I= M_I-2.5\log f_{\rm s}= -8.28 - 0.2 = -8.5$ from the data of SMARTS.
We plot the $(V-I)_0$-$(M_I-2.5\log f_{\rm s})$ diagram in Figure 
\ref{hr_diagram_t_pyx_pr_lup_v1313_sco_v1368_cen_outburst_vi}(d).
Here, we adopt the $BVI_{\rm C}$ data from AAVSO and SMARTS.
The track of V1368~Cen almost follows the track of V496~Sct/V959~Mon
(upper branch of orange line).  The overlapping of V1368~Cen and V496~Sct 
on the $(V-I)_0$-$(M_I-2.5\log f_{\rm s})$ diagram supports the results of
$E(B-V)=0.98$, $(m-M)_I=16.04$, $d=8.2$~kpc,
and $\log f_{\rm s}= +0.07$ for V1368~Cen.


\begin{figure*}
\plotone{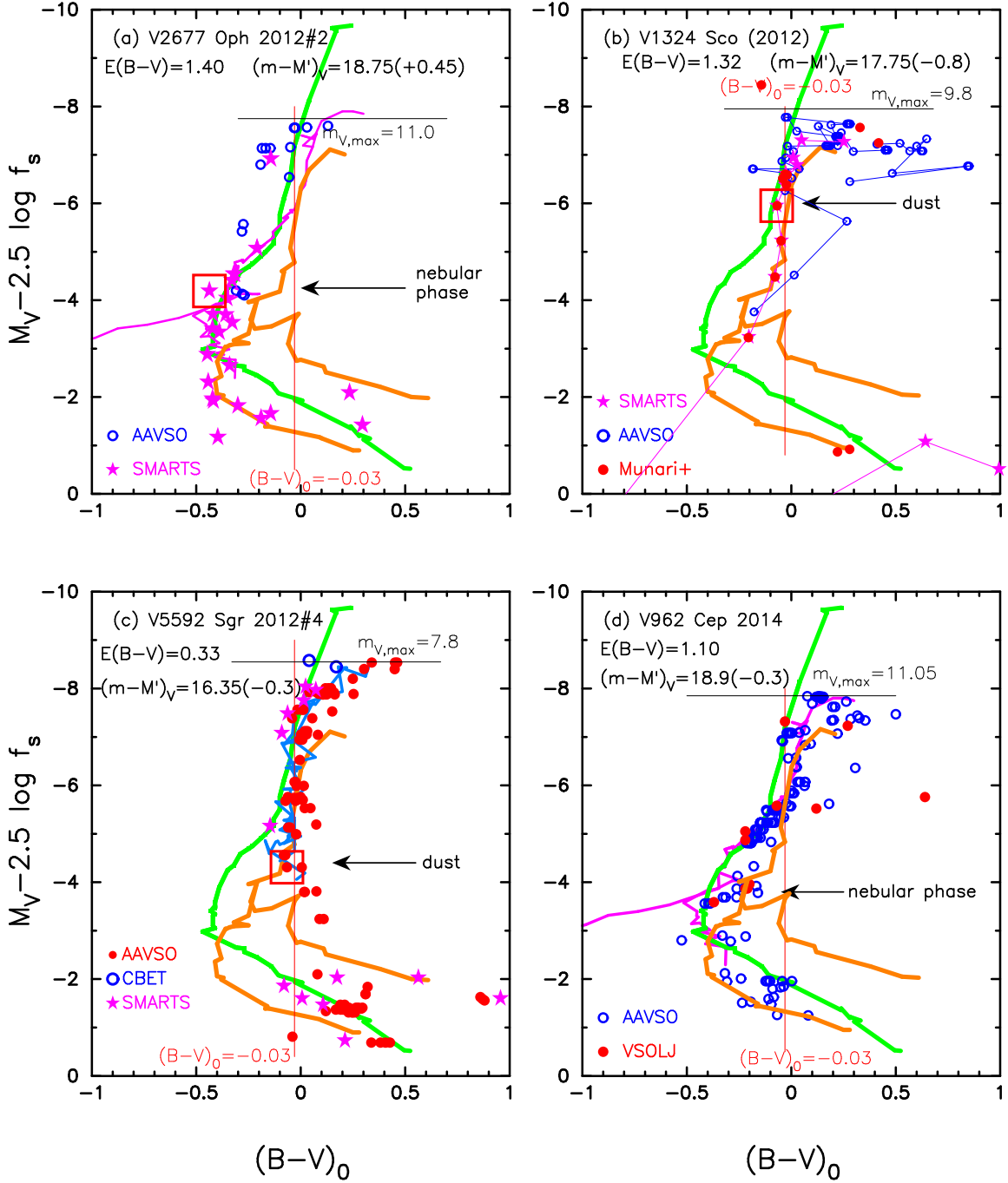}
\caption{
Same as Figure 
\ref{hr_diagram_v496_sct_v959_mon_v834_car_v1369_cen_outburst}, 
but for (a) V2677~Oph, (b) V1324~Sco, (c) V5592~Sgr, and (d) V962~Cep.
In panels (a) and (d), we add the track of V1974~Cyg 
(thin solid magenta lines).
In panel (c), we add the track of V1668~Cyg (thin solid cyan-blue lines).
\label{hr_diagram_v2677_oph_v1324_sco_v5592_sgr_v962_cep_outburst_bv}}
\end{figure*}


\begin{figure*}
\plotone{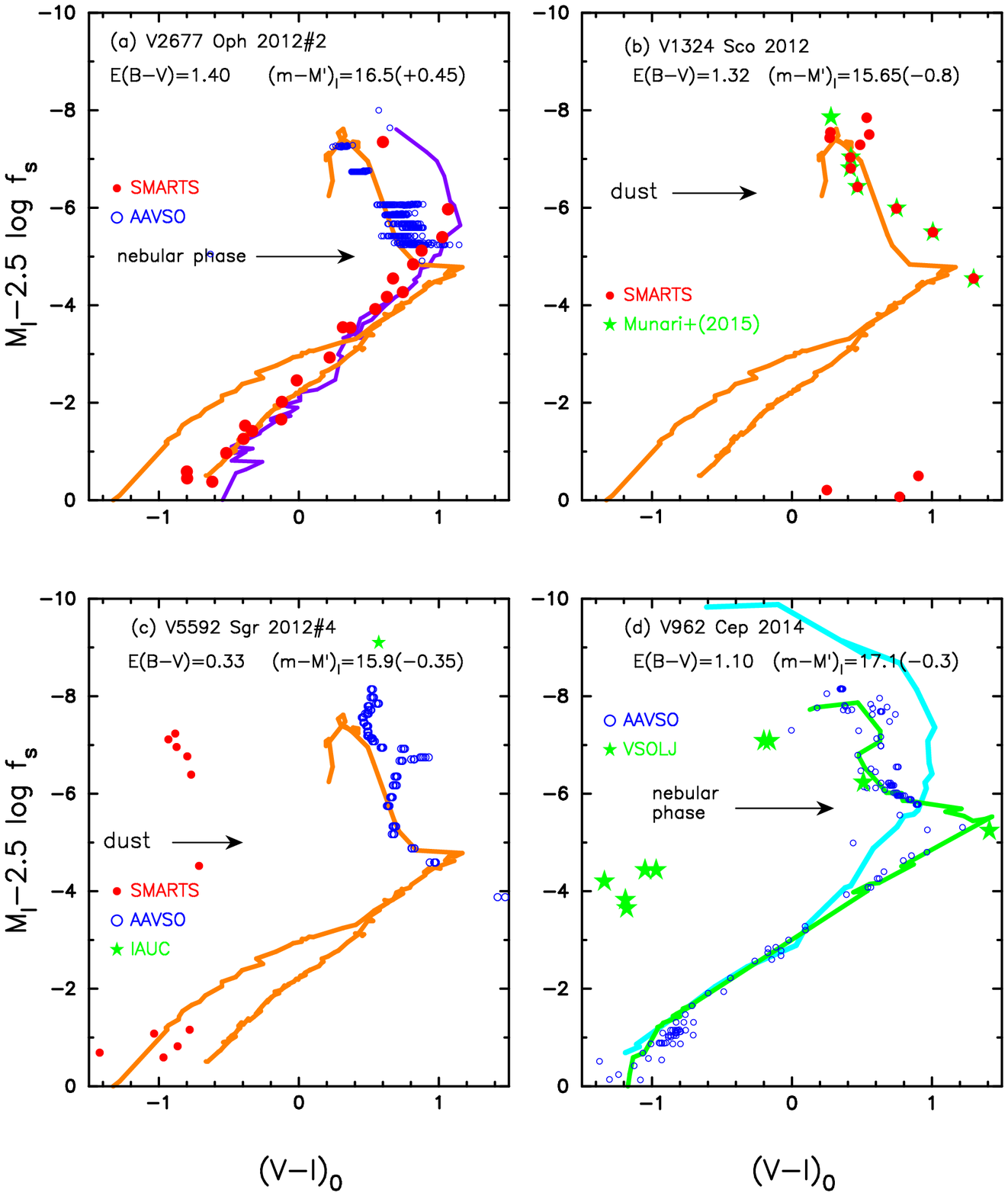}
\caption{
Same as Figure 
\ref{hr_diagram_v496_sct_v959_mon_v834_car_v1369_cen_outburst_vi},
but for (a) V2677~Oph, (b) V1324~Sco, (c) V5592~Sgr, and (d) V962~Cep.
In panel (a), we add the template track of V1065~Cen (solid blue-magenta line).
\label{hr_diagram_v2677_oph_v1324_sco_v5592_sgr_v962_cep_outburst_vi}}
\end{figure*}

\subsection{V2677~Oph 2012\#2}
\label{v2677_oph_vi}
\citet{hac19kb} obtained $E(B-V)=1.30$, $(m-M)_V=19.2$, $d=10.7$~kpc,
and $\log f_{\rm s}= -0.17$.  We have reanalyzed the $BVI_{\rm C}$
light/color curves of V2677~Oph in Appendix \ref{v2677_oph_bvi} and
obtained a new set of parameters, i.e., $E(B-V)=1.40$, $(m-M)_V=19.2$,
$d=9.4$~kpc, and $\log f_{\rm s}= -0.17$.  The main difference is 
the reddening of $E(B-V)=1.40$.
Then, we have $(m-M')_V= 19.2 - 0.425 = 18.75$ and plot the
$(B-V)_0$-$(M_V-2.5\log f_{\rm s})$ diagram in Figure
\ref{hr_diagram_v2677_oph_v1324_sco_v5592_sgr_v962_cep_outburst_bv}(a).
The track of V2677~Oph broadly follows the V1500~Cyg/V1974~Cyg 
(green/magenta lines) until the nebular phase started.
Then, it transfers from the track of V1500~Cyg
to the lower branch of LV~Vul (orange line).  

The distance modulus in $I_{\rm C}$ band, $(m-M)_I=16.95$, is 
taken from Appendix \ref{v2677_oph_bvi}.
Then, we have $(m-M')_I=16.95 - 0.425 = 16.5$.
The peak $I_{\rm C}$ brightness is 
$M'_I= M_I-2.5\log f_{\rm s}= -8.45 + 0.425 = -8.0$ from the data of AAVSO.
We plot the $(V-I)_0$-$(M_I-2.5\log f_{\rm s})$ diagram in Figure 
\ref{hr_diagram_v2677_oph_v1324_sco_v5592_sgr_v962_cep_outburst_vi}(a).
Here, we adopt the $BVI_{\rm C}$ data from AAVSO and SMARTS.
The track of V2677~Oph almost follows the track of V1065~Cen (blue-magenta
line) in the early phase and then goes along the lower branch of 
V496~Sct/V959~Mon (orange line).  
The rough overlapping of V2677~Oph and V1065~Cen and V496~Sct/V959~Mon 
on the $(V-I)_0$-$(M_I-2.5\log f_{\rm s})$ diagram
may support the new results of
$E(B-V)=1.40$, $(m-M)_I=16.95$, $d=9.4$~kpc,
and $\log f_{\rm s}= -0.17$ for V2677~Oph.

\subsection{V1324~Sco 2012}
\label{v1324_sco_vi}
\citet{hac19kb} obtained $E(B-V)=1.32$, $(m-M)_V=16.95$, $d=3.7$~kpc,
and $\log f_{\rm s}= +0.32$.  We have reanalyzed the $BVI_{\rm C}$ 
light/color curves of V1324~Sco in Appendix \ref{v1324_sco_bvi} and
obtained the same parameters as those obtained by \citet{hac19kb};
note that $\log f_{\rm s}= +0.28$ in their Table 1 is a typographical 
error of  $\log f_{\rm s}= \log 2.1 = +0.32$. 
We have $(m-M')_V=16.95 + 0.8 = 17.75$ and plot the
$(B-V)_0$-$(M_V-2.5\log f_{\rm s})$ diagram in Figure 
\ref{hr_diagram_v2677_oph_v1324_sco_v5592_sgr_v962_cep_outburst_bv}(b).
The track of V1324~Sco broadly follows the track of LV~Vul (orange
line) until the dust blackout starts.

The distance modulus in $I_{\rm C}$ band, $(m-M)_I=14.85$, is 
taken from Appendix \ref{v1324_sco_bvi}.
Then, we have $(m-M')_I= 14.85 + 0.8 = 15.65$.
The peak $I_{\rm C}$ brightness is $M'_I= M_I-2.5\log f_{\rm s}= 
-7.05 - 0.8 = -7.85$ from the data of \citet{mun15wh}.
We plot the $(V-I)_0$-$(M_I-2.5\log f_{\rm s})$ diagram in Figure 
\ref{hr_diagram_v2677_oph_v1324_sco_v5592_sgr_v962_cep_outburst_vi}(b).
Here, we adopt the $BVI_{\rm C}$ data from SMARTS and \citet{mun15wh}.
The track of V1324~Sco almost follows the track of V496~Sct/V959~Mon
(orange line) until the dust blackout starts.
The rough overlapping of V1324~Sco and V496~Sct/V959~Mon
on the $(V-I)_0$-$(M_I-2.5\log f_{\rm s})$ diagram may support the results of
$E(B-V)=1.32$, $(m-M)_I=14.85$, $d=3.7$~kpc,
and $\log f_{\rm s}= +0.32$ for V1324~Sco.

\subsection{V5592~Sgr 2012\#4}
\label{v5592_sgr_vi}
\citet{hac19kb} obtained $E(B-V)=0.33$, $(m-M)_V=16.05$, $d=10$~kpc,
and $\log f_{\rm s}= +0.13$.  We have reanalyzed the $BVI_{\rm C}$
light/color curves of V5592~Sgr in Appendix \ref{v5592_sgr_bvi} and
obtained the same parameters as those obtained by \citet{hac19kb}. 
Then, we have $(m-M')_V= 16.05 + 0.325 = 16.35$ and plot the
$(B-V)_0$-$(M_V-2.5\log f_{\rm s})$ diagram in Figure 
\ref{hr_diagram_v2677_oph_v1324_sco_v5592_sgr_v962_cep_outburst_bv}(c).
The track of V5592~Sgr follows the track of LV~Vul (orange line)
until the dust blackout starts.

The distance modulus in $I_{\rm C}$ band, $(m-M)_I=15.55$, is 
taken from Appendix \ref{v5592_sgr_bvi}.
Then, we have $(m-M')_I= 15.55 + 0.325 = 15.9$.
The peak $I_{\rm C}$ brightness is $M'_I= M_I-2.5\log f_{\rm s}= 
-8.75 - 0.325 = -9.1$ from the data of CBET No.3177.
We plot the $(V-I)_0$-$(M_I-2.5\log f_{\rm s})$ diagram in Figure 
\ref{hr_diagram_v2677_oph_v1324_sco_v5592_sgr_v962_cep_outburst_vi}(c).
Here, we adopt the $BVI_{\rm C}$ data from CBET No.3177, AAVSO, and SMARTS.
The track of V5592~Sgr (unfilled blue circles; AAVSO data)
almost follows the track of V496~Sct/V959~Mon (orange line)
until the dust blackout starts.
The rough overlapping of V5592~Sgr and V496~Sct/V959~Mon until the dust
blackout on the $(V-I)_0$-$(M_I-2.5\log f_{\rm s})$ diagram
may support the results of $E(B-V)=0.33$, $(m-M)_I=15.55$, $d=10$~kpc,
and $\log f_{\rm s}= +0.13$ for V5592~Sgr.

\subsection{V962~Cep 2014}
\label{v962_cep_vi}
\citet{hac19kb} obtained $E(B-V)=1.10$, $(m-M)_V=18.45$, $d=10.2$~kpc,
and $\log f_{\rm s}= +0.12$.  We have reanalyzed the $BVI_{\rm C}$
light/color curves of V962~Cep in Appendix \ref{v962_cep_bvi}
and obtained a new parameter set of $E(B-V)=1.10$, $(m-M)_V=18.6$, 
$d=10.9$~kpc, and $\log f_{\rm s}= +0.12$.
Then, we have $(m-M')_V= 18.6 + 0.3 = 18.9$ and plot the
$(B-V)_0$-$(M_V-2.5\log f_{\rm s})$ diagram in Figure 
\ref{hr_diagram_v2677_oph_v1324_sco_v5592_sgr_v962_cep_outburst_bv}(d).
\citet{hac19kb} concluded that V962~Cep belongs to the LV~Vul type because
the track of V962~Cep almost follows the lower branch of LV~Vul.  In Figure 
\ref{hr_diagram_v2677_oph_v1324_sco_v5592_sgr_v962_cep_outburst_bv}(d),
however, V962~Cep follows the V1500~Cyg or V1974~Cyg track rather than
the LV~Vul track.  Therefore, we regard that V962~Cep belongs to
the V1500~Cyg type.

The distance modulus in $I_{\rm C}$ band, $(m-M)_I=16.8$, is 
taken from Appendix \ref{v962_cep_bvi}.
Then, we have $(m-M')_I= 16.8 + 0.3 = 17.1$.
The peak $I_{\rm C}$ brightness is 
$M'_I= M_I-2.5\log f_{\rm s}= -7.8 - 0.3 = -8.1$ from the data of AAVSO.
We plot the $(V-I)_0$-$(M_I-2.5\log f_{\rm s})$ diagram in Figure 
\ref{hr_diagram_v2677_oph_v1324_sco_v5592_sgr_v962_cep_outburst_vi}(d).
Here, we adopt the $BVI_{\rm C}$ data from AAVSO and VSOLJ.
The track of V962~Cep follows the track of V5114~Sgr (green line).
The overlapping of V962~Cep with V5114~Sgr (V1500~Cyg type) 
on the $(V-I)_0$-$(M_I-2.5\log f_{\rm s})$ diagram supports the results of
$E(B-V)=1.10$, $(m-M)_I=16.8$, $d=10.9$~kpc,
and $\log f_{\rm s}= +0.12$ for V962~Cep.


\begin{figure*}
\plotone{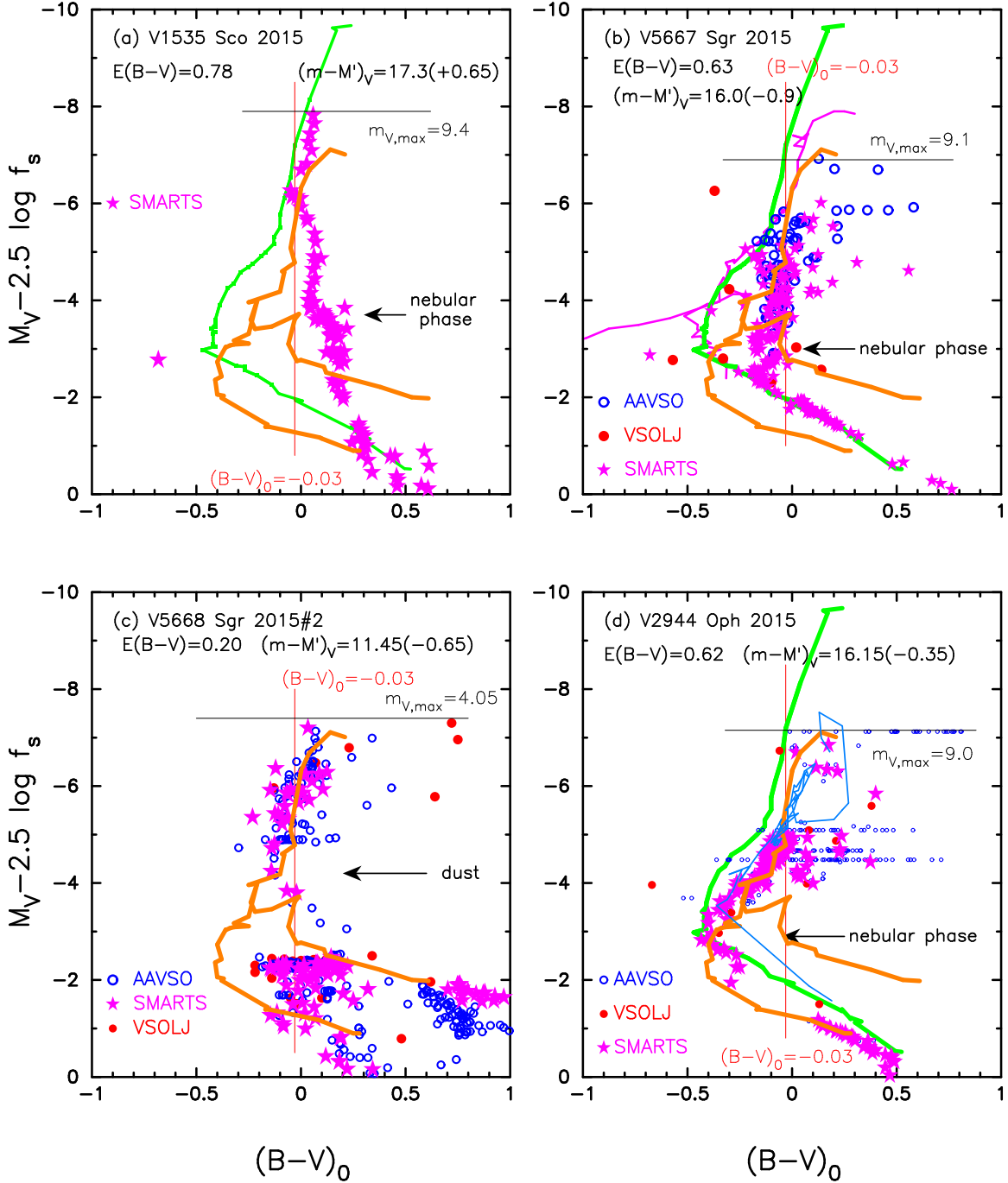}
\caption{
Same as Figure 
\ref{hr_diagram_v496_sct_v959_mon_v834_car_v1369_cen_outburst}, but for 
(a) V1535~Sco, (b) V5667~Sgr, (c) V5668~Sgr, and (d) V2944~Oph.
In panel (b), we add the track of V1974~Cyg (thin solid magenta lines).
In panel (d), we add the track of PW~Vul (thin solid cyan-blue line).
\label{hr_diagram_v1535_sco_v5667_sgr_v5668_sgr_v2944_oph_outburst_bv}}
\end{figure*}


\begin{figure*}
\plotone{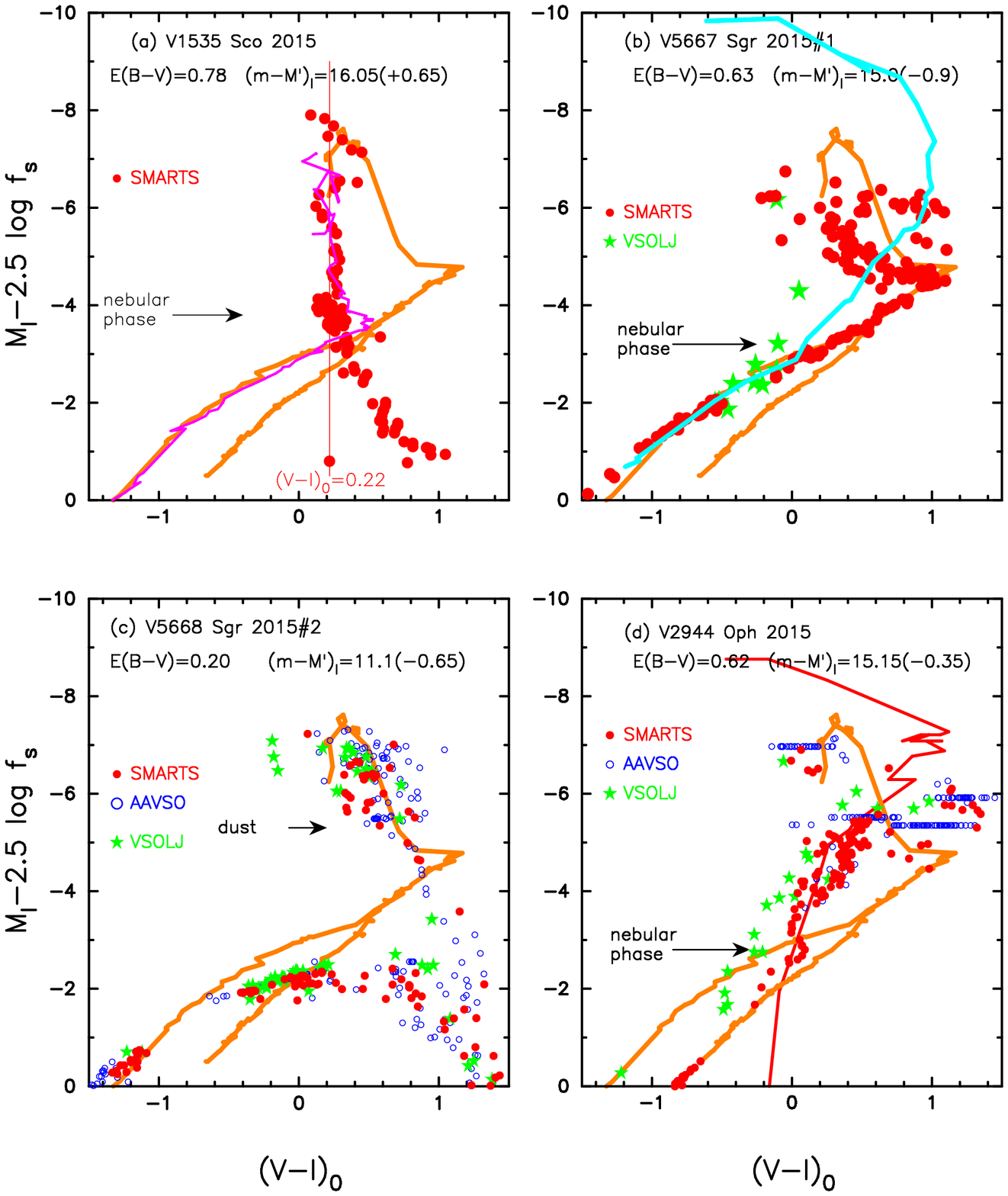}
\caption{
Same as Figure 
\ref{hr_diagram_v496_sct_v959_mon_v834_car_v1369_cen_outburst_vi},
but for (a) V1535~Sco, (b) V5667~Sgr, (c) V5668~Sgr, and (d) V2944~Oph.
In panel (a), we add the template track of V5666~Sgr (thin solid magenta line).
In panel (b), we add the template track of V1500~Cyg (thick solid cyan line).
In panel (d), we add the template track of V597~Pup (thin solid red line).
\label{hr_diagram_v1535_sco_v5667_sgr_v5668_sgr_v2944_oph_outburst_vi}}
\end{figure*}

\subsection{V1535~Sco 2015}
\label{v1535_sco_vi}
\citet{hac19kb} obtained $E(B-V)=0.78$, $(m-M)_V=18.3$, $d=15$~kpc,
and $\log f_{\rm s}= +0.38$.  We have reanalyzed the $BVI_{\rm C}$
light/color curves of V1535~Sco in Appendix \ref{v1535_sco_bvi}
and obtained a new set of parameters, i.e., $E(B-V)=0.78$,
$(m-M)_V=17.95$, $d=12.8$~kpc, and $\log f_{\rm s}= -0.26$.
The main difference is the timescaling factor of $\log f_{\rm s}= -0.26$.
Then, we have $(m-M')_V=17.95 - 0.65 = 17.3$
and plot the $(B-V)_0$-$(M_V-2.5\log f_{\rm s})$ diagram in Figure 
\ref{hr_diagram_v1535_sco_v5667_sgr_v5668_sgr_v2944_oph_outburst_bv}(a).
The companion star to the WD of V1535~Sco is a K3III or K4III giant
\citep[e.g.,][]{sri15, mun17}.  The shape of the V1535~Sco track
is almost straight and similar to that of V407~Cyg in Figure
\ref{hr_diagram_v407_cyg_rs_oph__v745_sco_v1534_sco_outburst_mv}(a).
The companion star could substantially contribute to the $B-V$ color in 
the later phase.

The distance modulus in $I_{\rm C}$ band, $(m-M)_I=16.7$, is 
taken from Appendix \ref{v1535_sco_bvi}.
Then, we have $(m-M')_I=16.7-0.65=16.05$.
The peak $I_{\rm C}$ brightness is 
$M'_I= M_I-2.5\log f_{\rm s}= -8.55 + 0.65 = -7.9$ from the data of SMARTS.
We plot the $(V-I)_0$-$(M_I-2.5\log f_{\rm s})$ diagram in Figure 
\ref{hr_diagram_v1535_sco_v5667_sgr_v5668_sgr_v2944_oph_outburst_vi}(a).
Here, we adopt the $BVI_{\rm C}$ data from SMARTS.
The track of V1535~Sco goes down straight almost along the track of 
V5666~Sgr (magenta lines) in the early and middle phases.
Then it turns to the red (rightward).  This is because the companion
star contributes to the $(V-I_{\rm C})_0$ color
as shown in V1313~Sco of Figure
\ref{hr_diagram_t_pyx_pr_lup_v1313_sco_v1368_cen_outburst_vi}(c).
The overlapping of V1535~Sco and V5666~Sgr 
on the $(V-I)_0$-$(M_I-2.5\log f_{\rm s})$ diagram may support the results of
$E(B-V)=0.78$, $(m-M)_I=16.7$, $d=12.8$~kpc,
and $\log f_{\rm s}= -0.26$ for V1535~Sco.

\subsection{V5667~Sgr 2015\#1}
\label{v5667_sgr_vi}
\citet{hac19kb} obtained $E(B-V)=0.63$, $(m-M)_V=15.4$, $d=4.9$~kpc,
and $\log f_{\rm s}= +0.57$.  We have reanalyzed the $BVI_{\rm C}$
light/color curves of V5667~Sgr in Appendix \ref{v5667_sgr_bvi}
and obtained a new parameter set of $E(B-V)=0.63$, $(m-M)_V=15.1$,
$d=4.3$~kpc, and $\log f_{\rm s}= +0.36$.
Then, we have $(m-M')_V=15.1+0.9=16.0$ and plot the 
$(B-V)_0$-$(M_V-2.5\log f_{\rm s})$ diagram in Figure 
\ref{hr_diagram_v1535_sco_v5667_sgr_v5668_sgr_v2944_oph_outburst_bv}(b).
The track of V5667~Sgr follows LV~Vul (orange line) until the nebular
phase started.  After that, it seems to follow the track of V1500~Cyg
(green line).

The distance modulus in $I_{\rm C}$ band, $(m-M)_I=14.1$, is 
taken from Appendix \ref{v5667_sgr_bvi}.
Then, we have $(m-M')_I=14.1+0.9=15.0$.
The peak $I_{\rm C}$ brightness is 
$M'_I= M_I-2.5\log f_{\rm s}= -5.85 - 0.9 = -6.75$ from the data of SMARTS.
We plot the $(V-I)_0$-$(M_I-2.5\log f_{\rm s})$ diagram in Figure 
\ref{hr_diagram_v1535_sco_v5667_sgr_v5668_sgr_v2944_oph_outburst_vi}(b).
Here, we adopt the $BVI_{\rm C}$ data from VSOLJ and SMARTS.
V5667~Sgr shows multiple peaks and the $(V-I_{\rm C})_0$ color
goes back and forth at/near $M'_I\equiv M_I - 2.5 \log f_{\rm s} \sim -5$. 
The track of V5667~Sgr broadly follows the track of V496~Sct/V959~Mon
(orange line) in the early and middle phases.  Then it goes along the 
track of V1500~Cyg (cyan line).  Thus, the rough overlapping of V5667~Sgr 
with V496~Sct in the early and middle phases and with V1500~Cyg in the later
phase on the $(V-I)_0$-$(M_I-2.5\log f_{\rm s})$ diagram supports 
the results of $E(B-V)=0.63$, $(m-M)_I=14.1$, $d=4.3$~kpc,
and $\log f_{\rm s}= +0.36$ for V5667~Sgr.

\subsection{V5668~Sgr 2015\#2}
\label{v5668_sgr_vi}
\citet{hac19kb} obtained $E(B-V)=0.20$, $(m-M)_V=11.0$, $d=1.2$~kpc,
and $\log f_{\rm s}= +0.27$.  We have reanalyzed the $BVI_{\rm C}$ 
light/color curves of V5668~Sgr in Appendix \ref{v5668_sgr_bvi}
and obtained a new parameter set of $E(B-V)=0.20$, $(m-M)_V=10.8$,
$d=1.1$~kpc, and $\log f_{\rm s}= +0.27$.
Then, we have $(m-M')_V=10.8+0.675=11.45$ and plot the 
$(B-V)_0$-$(M_V-2.5\log f_{\rm s})$ diagram in Figure 
\ref{hr_diagram_v1535_sco_v5667_sgr_v5668_sgr_v2944_oph_outburst_bv}(c).
The track of V5668~Sgr almost follows the upper branch of LV~Vul
except for the dust blackout phase.

The distance modulus in $I_{\rm C}$ band, $(m-M)_I=10.45$, is 
taken from Appendix \ref{v5668_sgr_bvi}.
Then, we have $(m-M')_I=10.45+0.675=11.1$.
The peak $I_{\rm C}$ brightness is 
$M'_I= M_I-2.5\log f_{\rm s}= -6.8 - 0.675 = -7.5$ from the data of VSOLJ.
We plot the $(V-I)_0$-$(M_I-2.5\log f_{\rm s})$ diagram in Figure 
\ref{hr_diagram_v1535_sco_v5667_sgr_v5668_sgr_v2944_oph_outburst_vi}(c).
Here, we adopt the $BVI_{\rm C}$ data from AAVSO, VSOLJ, and SMARTS.
The track of V5668~Sgr overlaps with the track of V496~Sct
(upper branch of orange line) except for the dust blackout phase.
The overlapping of V5668~Sgr and V496~Sct except for the dust blackout phase
on the $(V-I)_0$-$(M_I-2.5\log f_{\rm s})$ diagram supports the results of
$E(B-V)=0.20$, $(m-M)_I=10.45$, $d=1.1$~kpc,
and $\log f_{\rm s}= +0.27$ for V5668~Sgr.

\subsection{V2944~Oph 2015}
\label{v2944_oph_vi}
\citet{hac19kb} obtained $E(B-V)=0.62$, $(m-M)_V=16.5$, $d=8.2$~kpc,
and $\log f_{\rm s}= +0.25$.  We have reanalyzed the $BVI_{\rm C}$
light/color curves of V2944~Oph in Appendix \ref{v2944_oph_bvi}
and obtained a new parameter set of $E(B-V)=0.62$, $(m-M)_V=15.8$,
$d=6.0$~kpc, and $\log f_{\rm s}= +0.14$.
Then, we have $(m-M')_V=15.8+0.35=16.15$ and plot the 
$(B-V)_0$-$(M_V-2.5\log f_{\rm s})$ diagram in Figure 
\ref{hr_diagram_v1535_sco_v5667_sgr_v5668_sgr_v2944_oph_outburst_bv}(d).
The track of V2944~Oph overlaps with the track of LV~Vul in the
early phase,  then follows the track of V1500~Cyg in the middle phase,
and finally again goes along the lower branch of LV~Vul (orange line)
in the very later phase.
This behavior is similar to T~Pyx.

The distance modulus in $I_{\rm C}$ band, $(m-M)_I=14.8$, is 
taken from Appendix \ref{v2944_oph_bvi}.
Then, we have $(m-M')_I=14.8+0.35=15.15$.
The peak $I_{\rm C}$ brightness is 
$M'_I= M_I-2.5\log f_{\rm s}= -7.3 - 0.35 = -7.65$ from the data of AAVSO.
We plot the $(V-I)_0$-$(M_I-2.5\log f_{\rm s})$ diagram in Figure 
\ref{hr_diagram_v1535_sco_v5667_sgr_v5668_sgr_v2944_oph_outburst_vi}(d).
Here, we adopt the $BVI_{\rm C}$ data from AAVSO, VSOLJ, and SMARTS.
We add the track of V597~Pup (solid red line).
The track of V2944~Oph is close to the track of V496~Sct/V959~Mon
(orange line) in the early phase.  Then, it goes along the track of 
V597~Pup (red line) in the middle phase, and then finally follows the track
of V959~Mon (SMARTS: lower branch of orange line).
The overlapping of V2944~Oph with V496~Sct/V959~Mon in the early phase,
with V597~Pup in the middle phase, and with V959~Mon in the later phase
on the $(V-I)_0$-$(M_I-2.5\log f_{\rm s})$ diagram
may support the results of
$E(B-V)=0.62$, $(m-M)_I=14.8$, $d=6.0$~kpc,
and $\log f_{\rm s}= +0.14$ for V2944~Oph.




\section{Conclusions}
\label{conclusions}
Based on the time-stretching method of nova light curves together 
with our experience of the $(B-V)_0$-$(M_V - 2.5 \log f_{\rm s})$ diagram 
\citep{hac18k, hac18kb, hac19ka, hac19kb},
we newly construct $(U-B)_0$-$(M_B - 2.5 \log f_{\rm s})$ diagrams
for 16 classical novae and $(V-I)_0$-$(M_I - 2.5 \log f_{\rm s})$ 
diagrams for 52 classical novae. 
Our results are summarized as follows:\\

\noindent
{\bf 1.} We made the $(U-B)_0$-$(M_B - 2.5 \log f_{\rm s})$
diagrams of 16 novae and classified them into two types, LV~Vul and 
V1500~Cyg types, which are consistent with the classification in the  
$(B-V)_0$-$(M_V - 2.5 \log f_{\rm s})$ diagram.
The overlapping of the tracks between the target and template novae 
substantially refined the distance, reddening, distance modulus,
and timescaling factor of each nova, which are already derived
from the time-stretching method of nova light curves together with the 
$(B-V)_0$-$(M_V - 2.5 \log f_{\rm s})$ diagram.\\

\noindent
{\bf 2.} 
Many novae basically follow the universal decline law \citep{hac06kb}.
We classified the $(V-I)_0$-$(M_I - 2.5 \log f_{\rm s})$
diagrams of such 52 novae into two types, LV~Vul and V1500~Cyg types,
depending on the classification in the  
$(B-V)_0$-$(M_V - 2.5 \log f_{\rm s})$ diagram.  We further
subclassify the LV~Vul type into four subtypes,
V496~Sct/V959~Mon subtype, V834~Car/V382~Vel subtype,
V2615~Oph subtype, and V5666~Sgr subtype, 
and the V1500~Cyg type into four subtypes, V1500~Cyg subtype,
V5114~Sgr subtype, V574~Pup subtype, and V597~Pup subtype,
on the $(V-I)_0$-$(M_I - 2.5 \log f_{\rm s})$ diagram.
Each track roughly overlaps with the template track.
This overlapping of the tracks between the target and template novae 
refined the distance, reddening, distance modulus,
and timescaling factor of each nova, which are already derived
from the time-stretching method of nova light curves together with the 
$(B-V)_0$-$(M_V - 2.5 \log f_{\rm s})$ diagram.\\

\noindent
{\bf 3.} There are exceptional novae that do not exactly follow 
the universal decline law.
We construct the $(V-I)_0$-$(M_I - 2.5 \log f_{\rm s})$ diagrams for 
such five novae, i.e., V407~Cyg, RS~Oph, V745~Sco, V1534~Sco,
and U~Sco.  Their tracks are different
from those of the normal novae that follow the universal decline law.    
We discussed the reasons of the deviations in each nova.\\

\noindent
{\bf 4.} After the $M_B$ peak, many novae typically go down
in the $(U-B)_0$-$(M_B - 2.5 \log f_{\rm s})$ diagram along 
the vertical line of $(U-B)_0 = -0.97$, which is the intrinsic 
$U-B$ color of optically thick free-free emission.  We showed that 
many tracks of novae overlap with that of LV~Vul or V1500~Cyg between 
$M'_B= M_B-2.5\log f_{\rm s}\sim -6$ and $-4$.
After the nebular phase starts near $M'_B= M_B-2.5\log f_{\rm s}\sim -4$,
the tracks of LV~Vul type turn to the red. 
This is because strong emission lines contribute much more to the $B$ band 
than to the $U$ band in the nebular phase.
The tracks of V1500~Cyg type depart from that of LV~Vul.  
This is because strong [\ion{Ne}{3}] and [\ion{Ne}{5}] 
lines contribute to the $U$ band in the nebular phase.
\\

\noindent
{\bf 5.} In the early phase of the V2615~Oph and V5666~Sgr subtypes 
(LV~Vul type), the tracks go up or down along the vertical line of 
$(V-I)_0 = +0.22$ in the $(V-I)_0$-$(M_I - 2.5 \log f_{\rm s})$ diagram, 
where $(V-I)_0 = +0.22$ is the intrinsic $V-I$ color of optically thick 
free-free emission.   After the nova entered the nebular phase, strong 
emission lines such as [\ion{O}{3}] $\lambda\lambda 4959$, 5007, and 
[\ion{N}{2}] $\lambda 5755$ contribute much to the $V$ band and make 
$V-I$ color bluer.  Then the track turns to the blue.\\

\noindent
{\bf 6.} For the other subtypes, the track usually goes toward the red,
much redder than $(V-I)_0 = +0.22$ in the 
$(V-I)_0$-$(M_I - 2.5 \log f_{\rm s})$ diagram.
This is because the emission lines of \ion{O}{1} $\lambda\lambda 7774$,
8446, and \ion{Ca}{2} $\lambda\lambda 8498$, 8542 contribute to the
$I_{\rm C}$ band and make the $V-I$ color redder.
After the nova entered the nebular phase, strong emission lines such as
[\ion{O}{3}] and [\ion{N}{2}] contribute much to the $V$ band and make 
$V-I$ color bluer.  Then the track turns to the blue.\\

\noindent
{\bf 7.} We determined the color excesses, distances, and timescaling
factors of total 60 novae.  We also estimate their WD masses.
The results are summarized 
in Tables \ref{extinction_various_novae} and \ref{wd_mass_novae}.
We obtained a more accurate parameter set with
(two or) three time-stretched color-magnitude diagrams
rather than only one $(B-V)_0$-$(M_V - 2.5 \log f_{\rm s})$ 
diagram.  Finally, we emphasize that the three
$(U-B)_0$-$(M_B - 2.5 \log f_{\rm s})$,
$(B-V)_0$-$(M_V - 2.5 \log f_{\rm s})$, and
$(V-I)_0$-$(M_I - 2.5 \log f_{\rm s})$ diagrams are useful to understand
the physical nature of each nova as well as to confirm
the distance, reddening, distance moduli, and timescaling factor
of each target nova. \\

\acknowledgments
     We are grateful to
the American Association of Variable Star Observers
(AAVSO) and the Variable Star Observers League of Japan (VSOLJ)
for the archival data of various novae.
  We also thank the anonymous referee for useful comments
that improved the manuscript.




\appendix

\section{Time-Stretched Multi-band Light Curves of Novae}
\label{light_curves}
Many classical novae follow the universal decline law \citep{hac06kb}.
\citet{hac10k, hac14k, hac15k, hac16k, hac16kb} established the
time-stretching method based on the universal decline law.
Using this time-stretching method,
\citet{hac19ka, hac19kb} obtained the distance moduli in
$UBVI_{\rm C}$ bands for many classical novae.
However, they did not examine simultaneous overlaps both in the
$\log (t/f_{\rm s})$-$(M_I - 2.5 \log f_{\rm s})$ light curves and
$\log (t/f_{\rm s})$-$(V-I)_0$ color curves.
In this Appendix, using simultaneous overlaps 
both in the time-stretched $I$ light curves and $V-I$ color curves,
we obtain more precisely the distance moduli in $UBVI_{\rm C}$
bands for selected classical novae that are not well studied yet in our 
previous papers.  We reanalyzed four novae, in the order of V5114~Sgr,
V2362~Cyg, V1065~Cen, and V959~Mon.


\begin{figure}
\plotone{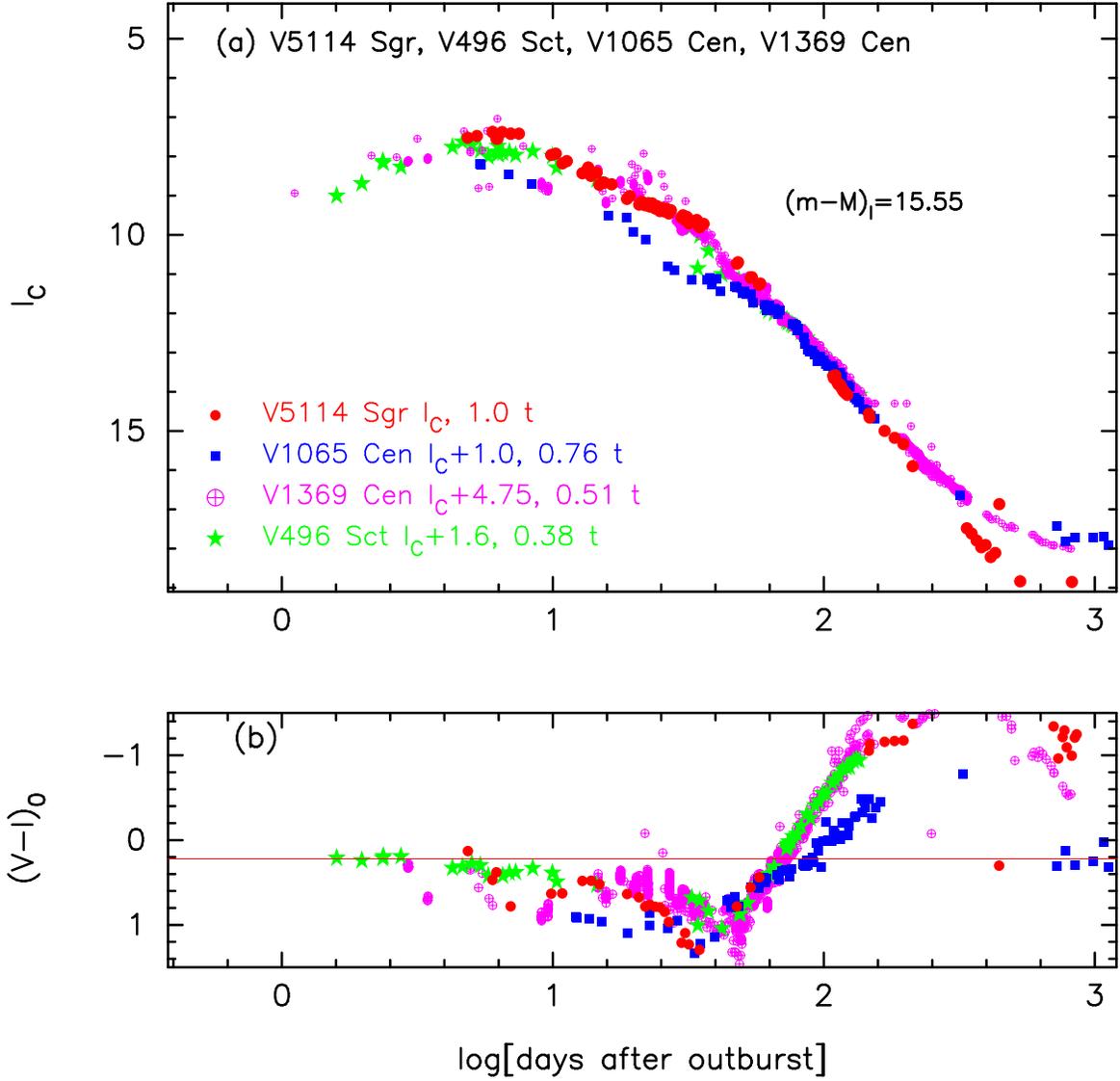}
\caption{
The (a) $I_{\rm C}$ light curve and (b) $(V-I_{\rm C})_0$ color curve
of V5114~Sgr on a logarithmic timescale.  We add the $I_{\rm C}$ light and
$(V-I_{\rm C})_0$ color curves of V1065~Cen, V1369~Cen, and V496~Sct.
Each light curve is horizontally moved by $\Delta \log t = \log f_{\rm s}$
and vertically shifted by $\Delta I_{\rm C}$ 
with respect to that of V5114~Sgr,
as indicated in the figure by, for example, ``V1369~Cen I$_C +4.75$,
0.51 t,'' where $\Delta I_{\rm C}= +4.75$ and $f_{\rm s}= 0.51$.
In panel (b), the horizontal thin solid red line shows 
the intrinsic $V-I$ color of optically thick free-free emission,
that is, $(V-I)_0= +0.22$.
\label{v5114_sgr_v496_sct_v1065_cen_v1369_cen_i_vi_logscale}}
\end{figure}


\begin{figure}
\epsscale{0.9}
\plotone{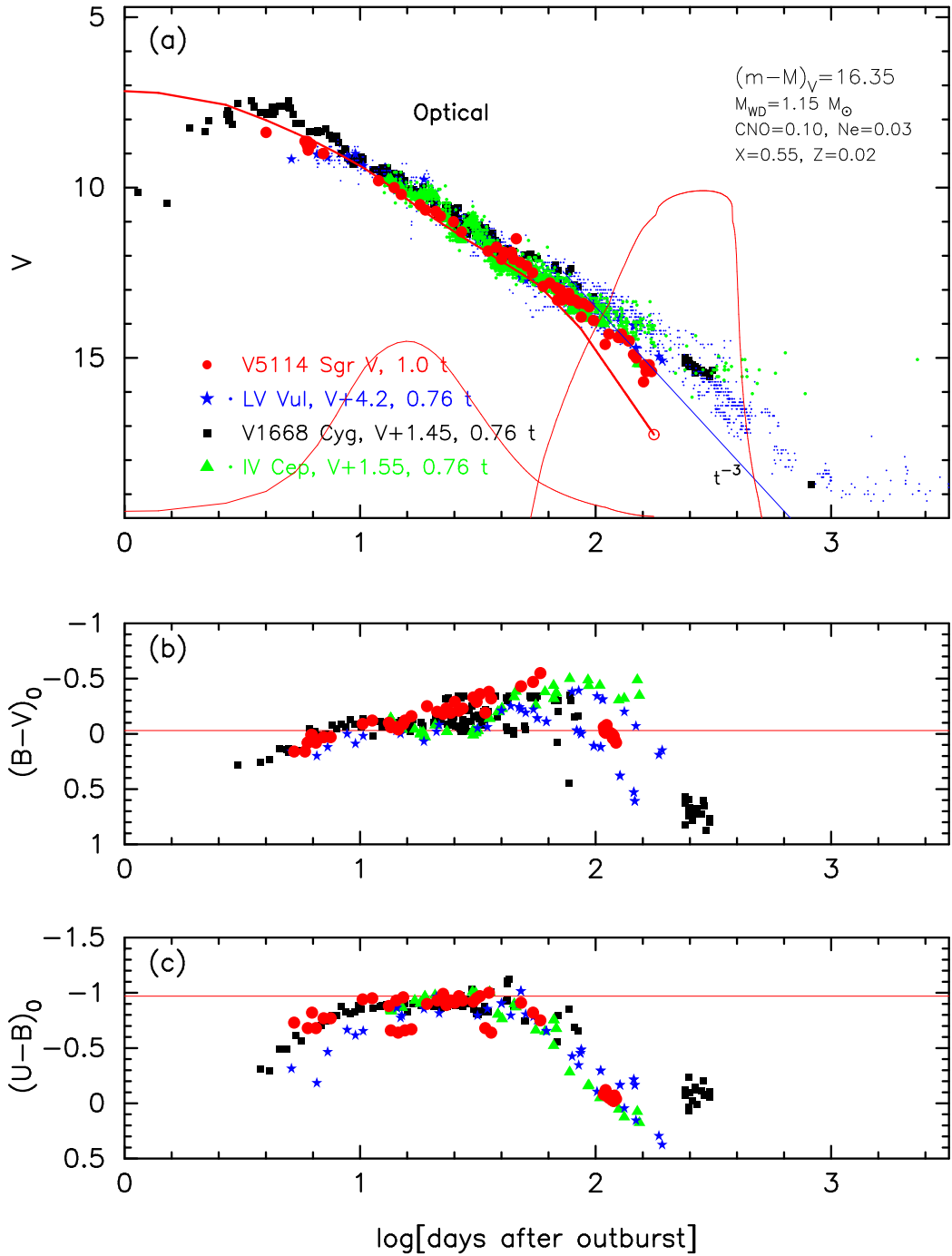}
\caption{
The (a) $V$ light curve, (b) $(B-V)_0$, and (c) $(U-B)_0$ color curves
of V5114~Sgr as well as those of LV~Vul, V1668~Cyg, and IV~Cep.
In panel (a), small blue and green dots represent the visual magnitudes
of LV~Vul and IV~Cep, respectively. 
The straight solid blue line labeled ``$t^{-3}$''
indicates the flux from homologously expanding ejecta
\citep[see, e.g.,][]{woo97, hac06kb}.
We add a $1.15~M_\sun$ WD model $V$ light curve (solid red lines) with the
chemical composition of Ne2 \citep{hac10k} for V5114~Sgr.  
The left lower solid red line denotes the UV~1455\AA\ 
flux and the right solid red line corresponds to the soft X-ray flux
($0.2 - 1.0$~keV). 
In panel (b), the horizontal thin solid red line shows the intrinsic
$B-V$ color of optically thick free-free emission, that is, 
$(B-V)_0= -0.03$ \citep{hac14k}.
In panel (c), the thin solid red line shows the intrinsic $U-B$ color
of optically thick free-free emission, that is, $(U-B)_0= -0.97$
\citep{hac14k}.
\label{v5114_sgr_lv_vul_v1668_cyg_iv_cep_v_bv_ub_logscale}}
\end{figure}


\begin{figure}
\plotone{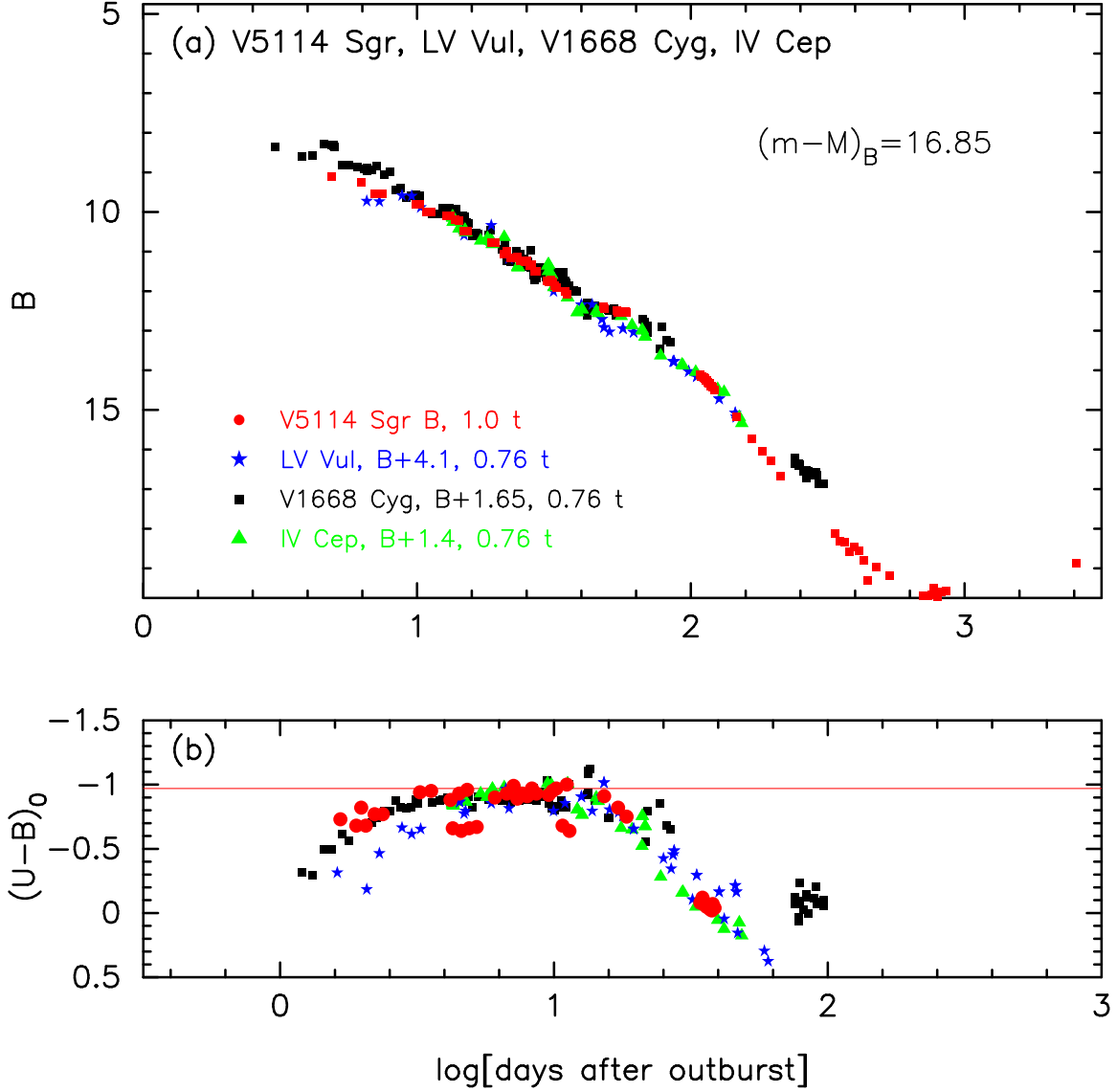}
\caption{
The (a) $B$ light and (b) $(U-B)_0$ color curves of V5114~Sgr
as well as those of LV~Vul, V1668~Cyg, and IV~Cep.
The $UBVI_{\rm C}$ data of V5114~Sgr are taken from 
\citet{ede06} and SMARTS \citep{wal12}.
The $UBV$ data of LV~Vul, V1668~Cyg, and IV~Cep are the same as
those in \citet{hac19ka}.  
In panel (b), the thin solid red line shows the intrinsic $U-B$ color
of optically thick free-free emission, that is, $(U-B)_0= -0.97$
\citep{hac14k}.
\label{v5114_sgr_lv_vul_v1668_cyg_iv_cep_b_ub_logscale}}
\end{figure}


\begin{figure}
\epsscale{0.55}
\plotone{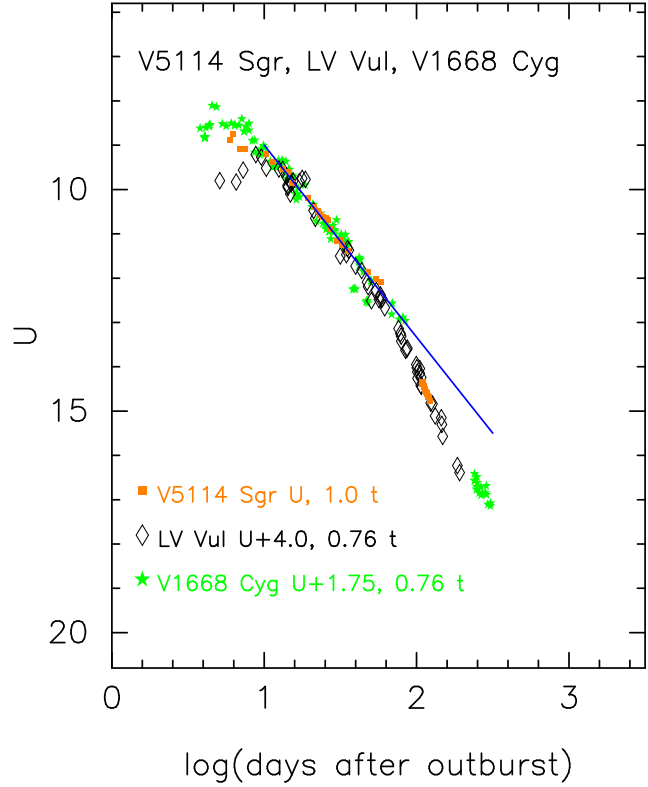}
\caption{
The time-stretched $U$ light curves of V5114~Sgr as well as LV~Vul 
and V1668~Cyg.  The data of V5114~Sgr are taken from \citet{ede06} and
SMARTS \citep{wal12}.  The solid blue line shows the trend of 
$F_\nu \propto t^{-1.75}$ for the universal decline law \citep{hac06kb}.
\label{v5114_sgr_lv_vul_v1668_cyg_u_only_logscale}}
\end{figure}


\begin{figure*}
\plotone{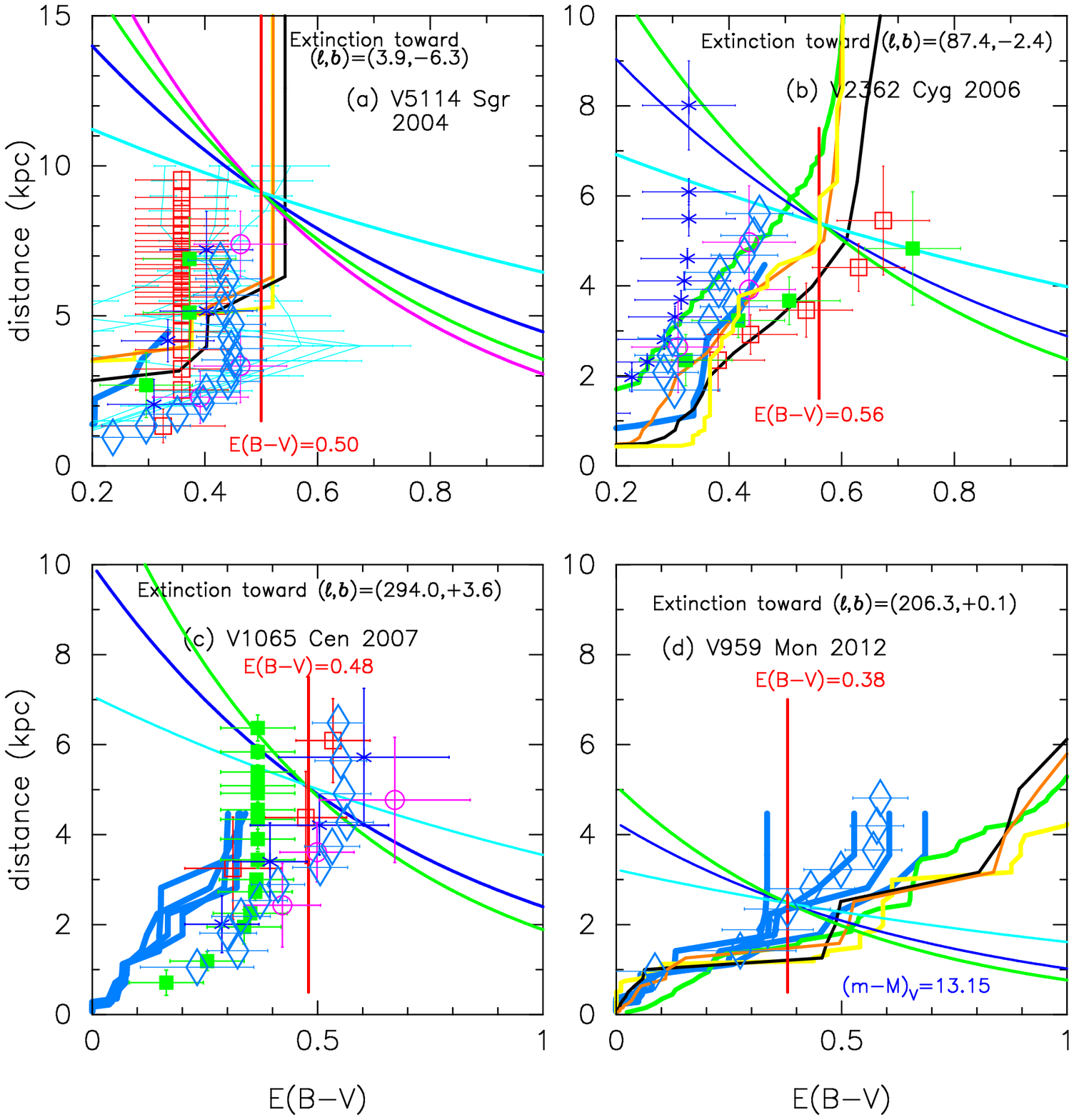}
\caption{
Various distance-reddening relations toward
(a) V5114~Sgr, (b) V2362~Cyg, (c) V1065~Cen, and (d) V959~Mon.
In panel (a), we plot $(m-M)_U= 17.15$ and Equation 
(\ref{distance_modulus_ru}), $(m-M)_B= 16.85$ and Equation 
(\ref{distance_modulus_rb}), $(m-M)_V= 16.35$ and Equation 
(\ref{distance_modulus_rv}), and $(m-M)_I= 15.55$ and
Equation (\ref{distance_modulus_ri}) with the magenta, green, blue,
and cyan lines, respectively.  
These four lines cross at $d=9.1$~kpc and $E(B-V)=0.50$.
The vertical solid red line indicates the reddening of $E(B-V)=0.50$.
The unfilled red squares, filled green squares, blue asterisks, 
and unfilled magenta circles, each with error bars,
represent the relation given by \citet{mar06}.
The thick solid black line depicts the relation given by \citet{gre15} 
while the orange one denotes their revised version \citep{gre18}.
They have again revised their distance-reddening relations
\citep[][thick solid yellow lines]{gre19}.
The unfilled cyan-blue diamonds with error bars 
represent the relation of \citet{ozd18}.
We also add the thick solid cyan-blue line given by \citet{chen19}.
We plot four relations given by \citet{schu14} by the very thin cyan lines.
In panels (b) and (d), the thick solid green lines denote 
the distance-reddening relation given by \citet{sal14}.
In panels (c) and (d), we add four thick cyan-blue lines of \citet{chen19},
which correspond to four nearby directions toward V1065~Cen, the galactic
coordinates of $(\ell, b)= (293\fdg95, +3\fdg55)$, $(293\fdg95, +3\fdg65)$,
$(294\fdg05, +3\fdg55)$, and $(294\fdg05, +3\fdg65)$, and toward V959~Mon, 
the galactic coordinates of
$(\ell, b)= (206\fdg25, +0\fdg05)$, $(206\fdg35, +0\fdg05)$,
$(206\fdg25, +0\fdg15)$, and $(206\fdg35, +0\fdg15)$.
See Section 2.2 of \citet{hac19ka} for more details
of various distance-reddening relations.
\label{distance_reddening_v5114_sgr_v2362_cyg_v1065_cen_v959_mon}}
\end{figure*}

\subsection{V5114~Sgr 2004}
\label{v5114_sgr_ubvi}
We obtain the reddening and distance toward V5114~Sgr 
with the time-stretching method.
\citet{hac19ka} analyzed the $V$ light and $B-V$ color curves and obtained
the color excess, distance modulus in $V$ band, distance,
and timescaling factor based on the $UBV$ data.
Here, we reanalyze the $UBVI_{\rm C}$ light curves
and examine the above various parameters.
Figure \ref{v5114_sgr_v496_sct_v1065_cen_v1369_cen_i_vi_logscale}
shows the (a) $I_{\rm C}$ light and (b) $(V-I)_0$ color curves of
V5114~Sgr, V1065~Cen, V1369~Cen, and V496~Sct.  
Each light curve is horizontally moved by $\Delta \log t = \log f_{\rm s}$
and vertically shifted by $\Delta V$ with respect to that of V5114~Sgr,
as indicated in the figure.  For example, ``V496~Sct I$_{\rm C}+1.6$, 0.38 t''
means that $\Delta I_{\rm C}= +1.6$ and $f_{\rm s}= 0.38$.
These four $I_{\rm C}$ light and $(V-I_{\rm C})_0$ color curves 
broadly overlap with each other.
Applying Equation (8) of \citet{hac19ka} for the $I$ band to Figure
\ref{v5114_sgr_v496_sct_v1065_cen_v1369_cen_i_vi_logscale}(a),
we obtain
\begin{eqnarray}
(m&-&M)_{I, \rm V5114~Sgr} \cr
&=& ((m - M)_I + \Delta I_{\rm C})
_{\rm V1065~Cen} - 2.5 \log 0.76 \cr
&=& 14.26 + 1.0\pm0.2 + 0.3 = 15.56\pm0.2 \cr
&=& ((m - M)_I + \Delta I_{\rm C})
_{\rm V1369~Cen} - 2.5 \log 0.51 \cr
&=& 10.11 + 4.75\pm0.2 + 0.725 = 15.58\pm0.2 \cr
&=& ((m - M)_I + \Delta I_{\rm C})
_{\rm V496~Sct} - 2.5 \log 0.38 \cr
&=& 12.9 + 1.6\pm0.2 + 1.05 = 15.55\pm0.2,
\label{distance_modulus_i_vi_v5114_sgr}
\end{eqnarray}
where we adopt
$(m-M)_{I, \rm V1065~Cen}= 14.26$ in Appendix \ref{v1065_cen_bvi}, 
$(m-M)_{I, \rm V1369~Cen}=10.11$ from \citet{hac19ka}, and
$(m-M)_{I, \rm V496~Sct}=12.9$ in Appendix \ref{v496_sct_bvi}. 
Thus, we obtain $(m-M)_{I, \rm V5114~Sgr}= 15.56\pm0.2$

Figure \ref{v5114_sgr_lv_vul_v1668_cyg_iv_cep_v_bv_ub_logscale} 
shows the (a) $V$ light, (b) $(B-V)_0$, and (c) $(U-B)_0$ color curves
of V5114~Sgr as well as those of LV~Vul, V1668~Cyg, and IV~Cep. 
Applying Equation (4) of \citet{hac19ka} for the $V$ band to Figure
\ref{v5114_sgr_lv_vul_v1668_cyg_iv_cep_v_bv_ub_logscale}(a), we obtain
\begin{eqnarray}
(m&-&M)_{V, \rm V5114~Sgr} \cr
&=& ((m - M)_V + \Delta V)_{\rm LV~Vul} - 2.5 \log 0.76 \cr
&=& 11.85 + 4.2\pm0.2 + 0.30 = 16.35\pm0.2 \cr
&=& ((m - M)_V + \Delta V)_{\rm V1668~Cyg} - 2.5 \log 0.76 \cr
&=& 14.6 + 1.45\pm0.2 + 0.30 = 16.35\pm0.2 \cr
&=& ((m - M)_V + \Delta V)_{\rm IV~Cep} - 2.5 \log 0.76 \cr
&=& 14.5 + 1.55\pm0.2 + 0.30 = 16.35\pm0.2.
\label{distance_modulus_v_bv_ub_v5114_sgr}
\end{eqnarray}
Thus, we obtain $(m-M)_{V, \rm V5114~Sgr}= 16.35\pm0.2$.

Figure \ref{v5114_sgr_lv_vul_v1668_cyg_iv_cep_b_ub_logscale} shows
the (a) $B$ light and (b) $(U-B)_0$ color curves of V5114~Sgr as well as
those of LV~Vul, V1668~Cyg, and IV~Cep.
Applying Equation (7) of \citet{hac19ka} for the $B$ band to
Figure \ref{v5114_sgr_lv_vul_v1668_cyg_iv_cep_b_ub_logscale}(a),
we obtain
\begin{eqnarray}
(m&-&M)_{B, \rm V5114~Sgr} \cr
&=& ((m - M)_B + \Delta B)_{\rm LV~Vul} - 2.5 \log 0.76 \cr
&=& 12.45 + 4.1\pm0.2 + 0.30 = 16.85\pm0.2 \cr
&=& ((m - M)_B + \Delta B)_{\rm V1668~Cyg} - 2.5 \log 0.76 \cr
&=& 14.9 + 1.65\pm0.2 + 0.30 = 16.85\pm0.2 \cr
&=& ((m - M)_B + \Delta B)_{\rm IV~Cep} - 2.5 \log 0.76 \cr
&=& 15.15 + 1.4\pm0.2 + 0.30 = 16.85\pm0.2,
\label{distance_modulus_b_ub_v5114_sgr}
\end{eqnarray}
where we adopt $(m-M)_{B, \rm LV~Vul}= 11.85 + 1.0\times 0.6 =12.45$,
$(m-M)_{B, \rm V1668~Cyg}= 14.6 + 1.0\times 0.3 =14.9$.
and $(m-M)_{B, \rm IV~Cep}= 14.5 + 1.0\times 0.65 =15.15$ all from
\citet{hac19ka, hac19kb}.
Thus, we obtain $(m-M)_{B, \rm V5114~Sgr}= 16.85\pm0.2$.

Figure \ref{v5114_sgr_lv_vul_v1668_cyg_u_only_logscale} shows
the $U$ light curves of V5114~Sgr together
with those of LV~Vul and V1668~Cyg.
It should be noted that the timescaling factor $f_{\rm s}$ of each nova 
is well determined by the three overlaps of $(V-I)_0$, $(B-V)_0$, 
and $(U-B)_0$ color curves.  
Applying Equation (6) of \citet{hac19ka} for the $U$ band to
Figure \ref{v5114_sgr_lv_vul_v1668_cyg_u_only_logscale}, we obtain
\begin{eqnarray}
(m&-&M)_{U, \rm V5114~Sgr} \cr
&=& ((m - M)_U + \Delta U)_{\rm LV~Vul} - 2.5 \log 0.76 \cr
&=& 12.85 + 4.0\pm0.2 + 0.30 = 17.15\pm0.2 \cr
&=& ((m - M)_U + \Delta U)_{\rm V1668~Cyg} - 2.5 \log 0.76 \cr
&=& 15.1 + 1.75\pm0.2 + 0.30 = 17.15\pm0.2,
\label{distance_modulus_u_v5114_sgr_lv_vul_v1668_cyg}
\end{eqnarray}
where we adopt $(m-M)_{U, \rm LV~Vul}=11.85 + (4.75-3.1) \times 0.60
=12.85$, and $(m-M)_{U, \rm V1668~Cyg}=14.6 + (4.75-3.1) \times 0.30
=15.10$.
Thus, we obtain $(m-M)_{U, \rm V5114~Sgr}= 17.15\pm0.2$.

We plot the distance moduli in $U$, $B$, $V$, and $I_{\rm C}$ bands
in Figure \ref{distance_reddening_v5114_sgr_v2362_cyg_v1065_cen_v959_mon}(a)
by the magenta, green, blue, and cyan lines, that is, $(m-M)_U= 17.15$, 
$(m-M)_B= 16.85$, $(m-M)_V= 16.35$, and $(m-M)_I= 15.55$ together with
Equations (\ref{distance_modulus_ru}), 
(\ref{distance_modulus_rb}), (\ref{distance_modulus_rv}), 
and (\ref{distance_modulus_ri}), respectively.
These four lines cross at $d=9.1$~kpc and $E(B-V)=0.50$.
The reddening of $E(B-V)=0.50$ is between those of \citet{ozd18} 
and \citet{mar06} and those of \citet{gre15, gre18, gre19}.  
See Section 2.2 of \citet{hac19ka} for more details
of various distance-reddening relations in Figure
\ref{distance_reddening_v5114_sgr_v2362_cyg_v1065_cen_v959_mon}.
We also obtained the timescaling factor of 
$\log f_{\rm s}= \log 0.76 = -0.12$ against that of LV~Vul.  
The new results are summarized in Tables 
\ref{extinction_various_novae} and \ref{wd_mass_novae}.


\begin{figure}
\plotone{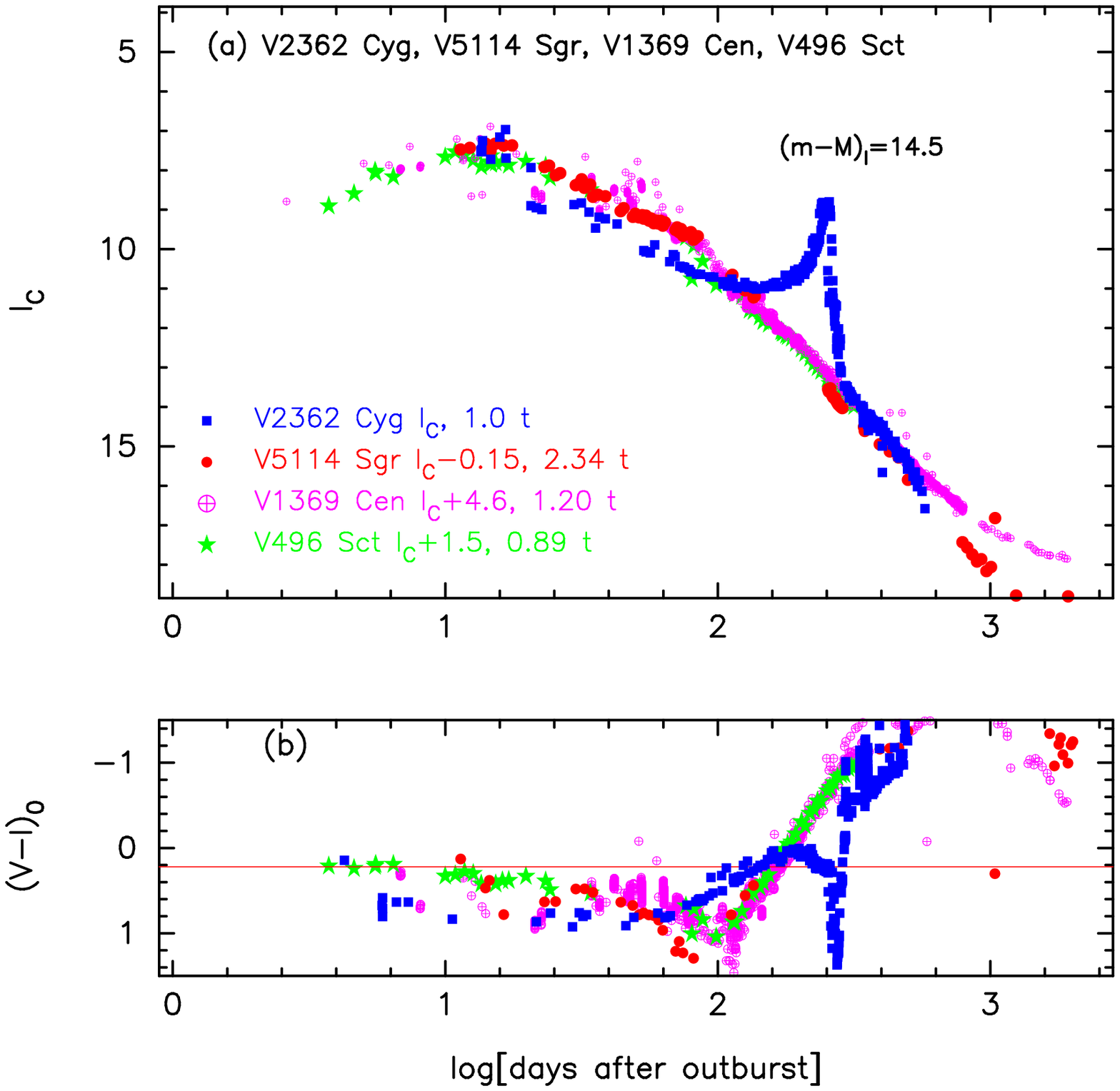}
\caption{
The (a) $I_{\rm C}$ light curve and (b) $(V-I_{\rm C})_0$ color curve
of V2362~Cyg as well as those of V5114~Sgr, V1369~Cen, and V496~Sct.
\label{v2362_cyg_v5114_sgr_v1369_cen_v496_sct_i_vi_color_logscale}}
\end{figure}


\begin{figure}
\plotone{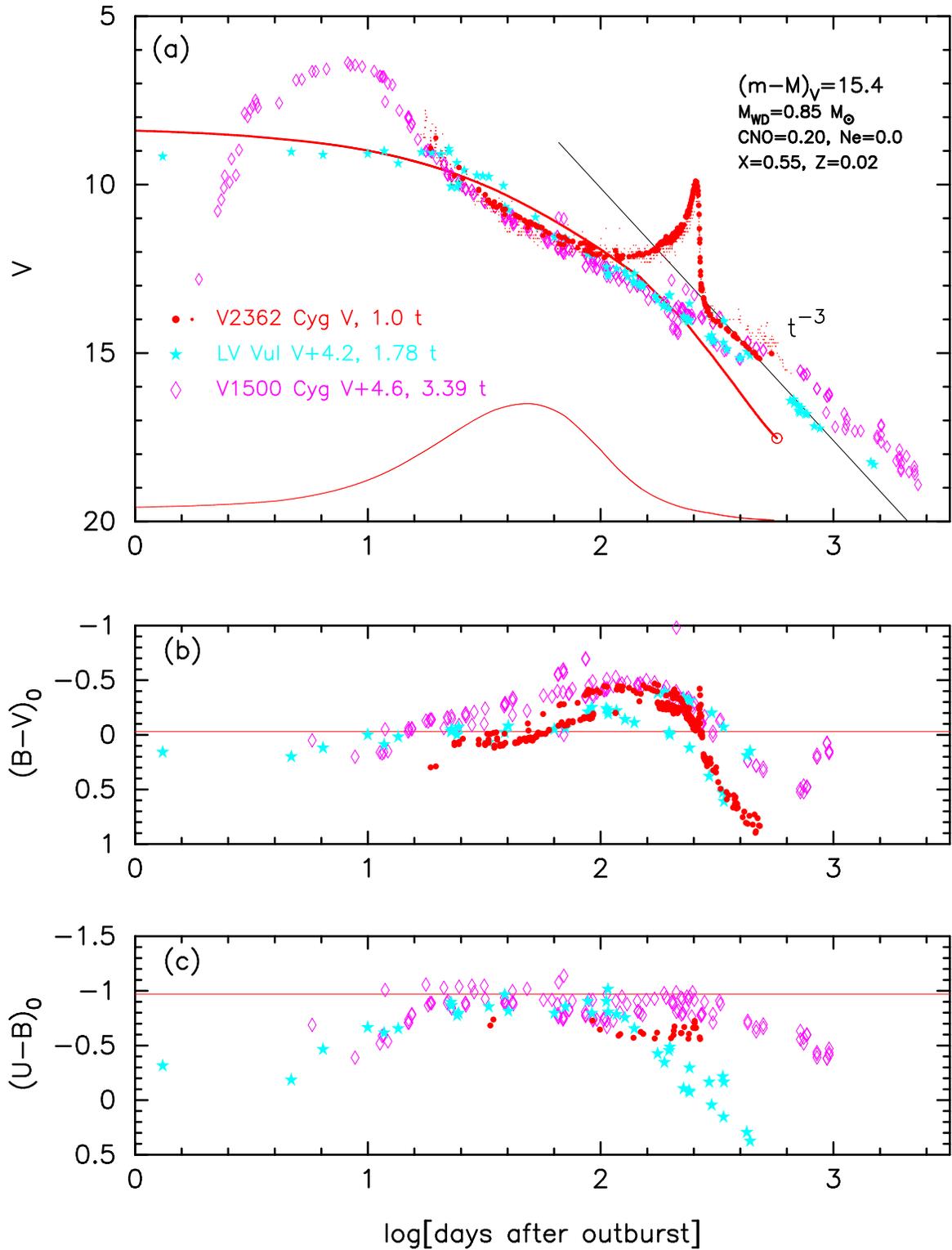}
\caption{
The (a) $V$ light, (b) $(B-V)_0$, and (c) $(U-B)_0$ color curves
of V2362~Cyg as well as those of LV~Vul and V1500~Cyg.
\label{v2362_cyg_lv_vul_v1500_cyg_v_bv_ub_color_logscale_no2}}
\end{figure}


\begin{figure}
\epsscale{0.55}
\plotone{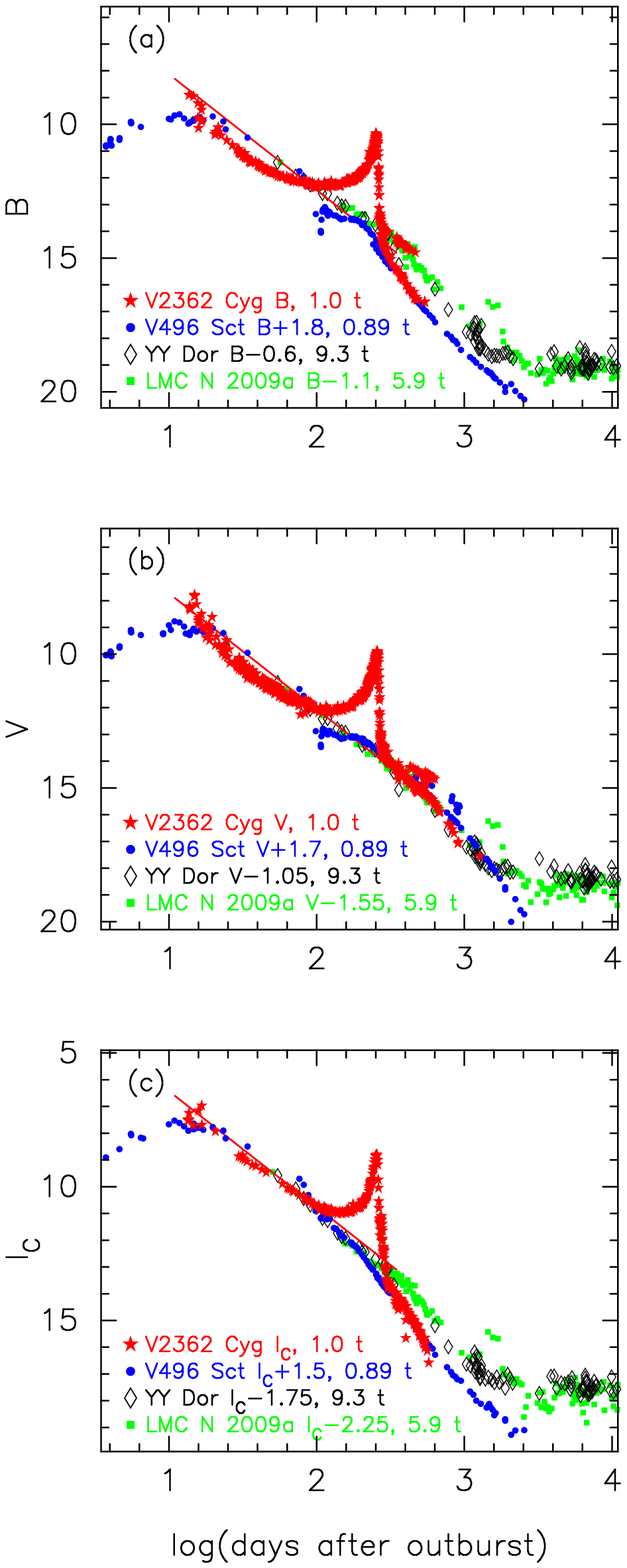}
\caption{
The (a) $B$, (b) $V$, and (c) $I_{\rm C}$ light curves of 
V2362~Cyg together with those of V496~Sct, YY~Dor, and LMC~N~2009a.
The data of V2362~Cyg are taken from \citet{mun08b}, AAVSO, and VSOLJ.
The $BVI_{\rm C}$ data of YY~Dor and LMC~N~2009a are taken from SMARTS.
\label{v2362_cyg_v496_sct_v382_vel_yy_dor_lmcn_2009a_b_v_i_logscale_3fig}}
\end{figure}

\subsection{V2362~Cyg 2006}
\label{v2362_cyg_bvi}
\citet{hac19ka} analyzed the $V$ light and $B-V$ color curves and obtained
the color excess, distance modulus in $V$ band, distance,
and timescaling factor.  Here, we reanalyze the $BVI_{\rm C}$ multi-band
light/color curves and examine the various parameters.
Figure \ref{v2362_cyg_v5114_sgr_v1369_cen_v496_sct_i_vi_color_logscale}
shows the $I_{\rm C}$ light and $(V-I_{\rm C})_0$ color curves of V2362~Cyg
as well as V5114~Sgr, V1369~Cen, and V496~Sct.
The $UBVI_{\rm C}$ data of V2362~Cyg are taken from AAVSO, VSOLJ, 
and \citet{mun08b}.
Here, we adopt the color excess of $E(B-V)= 0.56$ as mentioned below.
We adopt the timescaling factor $\log f_{\rm s}= +0.25$ after \citet{hac19ka}.
We apply Equation (8) of \citet{hac19ka} for the $I$ band to Figure
\ref{v2362_cyg_v5114_sgr_v1369_cen_v496_sct_i_vi_color_logscale}(a)
and obtain
\begin{eqnarray}
(m&-&M)_{I, \rm V2362~Cyg} \cr
&=& ((m - M)_I + \Delta I_{\rm C})
_{\rm V5114~Sgr} - 2.5 \log 2.34 \cr
&=& 15.55 - 0.15\pm0.2 - 0.925 = 14.48\pm0.2 \cr
&=& ((m - M)_I + \Delta I_{\rm C})
_{\rm V1369~Cen} - 2.5 \log 1.20 \cr
&=& 10.11 + 4.6\pm0.2 - 0.2 = 14.51\pm0.2 \cr
&=& ((m - M)_I + \Delta I_{\rm C})
_{\rm V496~Sct} - 2.5 \log 0.89 \cr
&=& 12.9 + 1.5\pm0.2 + 0.125 = 14.52\pm0.2,
\label{distance_modulus_i_vi_v2362_cyg}
\end{eqnarray}
where we adopt
$(m-M)_{I, \rm V5114~Sgr}=15.55$ from Appendix \ref{v5114_sgr_ubvi},
$(m-M)_{I, \rm V1369~Cen}=10.11$ from \citet{hac19ka}, and
$(m-M)_{I, \rm V496~Sct}=12.9$ in Appendix \ref{v496_sct_bvi}.
Thus, we obtain $(m-M)_{I, \rm V2362~Cyg}= 14.5\pm0.2$.

Figure \ref{v2362_cyg_lv_vul_v1500_cyg_v_bv_ub_color_logscale_no2}
shows the (a) $V$, (b) $(B-V)_0$, and (c) $(U-B)_0$ evolutions of
V2362~Cyg as well as LV~Vul and V1500~Cyg.  
Applying Equation (4) of \citet{hac19ka} for the $V$ band to them,
we have the relation
\begin{eqnarray}
(m&-&M)_{V, \rm V2362~Cyg} \cr
&=& ((m - M)_V + \Delta V)_{\rm LV~Vul} - 2.5 \log 1.78 \cr
&=& 11.85 + 4.2\pm0.2 - 0.625 = 15.42\pm0.2 \cr
&=& ((m - M)_V + \Delta V)_{\rm V1500~Cyg} - 2.5 \log 3.39 \cr
&=& 12.15 + 4.6\pm0.2 - 1.325 = 15.42\pm0.2,
\label{distance_modulus_v_bv_v2362_cyg}
\end{eqnarray}
where we adopt $(m-M)_{V, \rm LV~Vul}=11.85$ from \citet{hac19ka},
and $(m-M)_{V, \rm V1500~Cyg}=12.15$ in Appendix \ref{v1500_cyg_ubvi}.
Thus, we obtain $(m-M)_V=15.42\pm0.1$ for V2362~Cyg, being consistent with
Hachisu \& Kato's (2019a) results.

Figure 
\ref{v2362_cyg_v496_sct_v382_vel_yy_dor_lmcn_2009a_b_v_i_logscale_3fig}
shows the three band light curves of V2362~Cyg as well as V496~Sct and
the Large Magellanic Cloud (LMC) novae, YY~Dor and LMC~N~2009a.
For the $B$ band, we apply Equation (7) of \citet{hac19ka} to Figure 
\ref{v2362_cyg_v496_sct_v382_vel_yy_dor_lmcn_2009a_b_v_i_logscale_3fig}(a)
and obtain
\begin{eqnarray}
(m&-&M)_{B, \rm V2362~Cyg} \cr
&=& ((m - M)_B + \Delta B)_{\rm V496~Sct} - 2.5 \log 0.89 \cr
&=& 14.05 + 1.8\pm0.2 + 0.125 = 15.98\pm0.2 \cr
&=& ((m - M)_B + \Delta B)_{\rm YY~Dor} - 2.5 \log 9.3 \cr
&=& 18.98 - 0.6\pm0.2 - 2.425 = 15.96\pm0.2 \cr
&=& ((m - M)_B + \Delta B)_{\rm LMC~N~2009a} - 2.5 \log 5.9 \cr
&=& 18.98 - 1.1\pm0.2 - 1.925 = 15.96\pm0.2,
\label{distance_modulus_b_v2362_cyg_v496_sct_yy_dor_lmcn2009a}
\end{eqnarray}
where we adopt
$(m-M)_{B, \rm V496~Sct}=14.05$ in Appendix \ref{v496_sct_bvi}, and
$(m-M)_{B, \rm YY~Dor}=18.49 + 4.1\times 0.12=18.98$ and
$(m-M)_{B, \rm LMC~N~2009a}=18.49 + 4.1\times 0.12=18.98$ both from
\citet{hac18kb}.
\citet{hac18kb} adopted the distance modulus of $\mu_0\equiv 
(m-M)_0 = 18.49$ and reddening of $E(B-V)= 0.12$ toward the LMC novae.
Thus, we obtain $(m-M)_{B, \rm V2362~Cyg}= 15.97\pm0.2$.

For the $V$ light curves, we apply Equation (4) of \citet{hac19ka} to Figure
\ref{v2362_cyg_v496_sct_v382_vel_yy_dor_lmcn_2009a_b_v_i_logscale_3fig}(b)
and obtain
\begin{eqnarray}
(m&-&M)_{V, \rm V2362~Cyg} \cr
&=& ((m - M)_V + \Delta V)_{\rm V496~Sct} - 2.5 \log 0.89 \cr
&=& 13.6 + 1.7\pm0.2 + 0.125 = 15.42\pm0.2 \cr
&=& ((m - M)_V + \Delta V)_{\rm YY~Dor} - 2.5 \log 9.3 \cr
&=& 18.86 - 1.05\pm0.2 - 2.425 = 15.39\pm0.2 \cr
&=& ((m - M)_V + \Delta V)_{\rm LMC~N~2009a} - 2.5 \log 5.9 \cr
&=& 18.86 - 1.55\pm0.2 - 1.925 = 15.39\pm0.2,
\label{distance_modulus_v_v2362_cyg_v496_sct_yy_dor_lmcn2009a}
\end{eqnarray}
where we adopt
$(m-M)_{V, \rm V496~Sct}=13.6$ in Appendix \ref{v496_sct_bvi}, and
$(m-M)_{V, \rm YY~Dor}=18.49 + 3.1\times 0.12= 18.86$ and
$(m-M)_{V, \rm LMC~N~2009a}=18.49 + 3.1\times 0.12=18.86$ both from
\citet{hac18kb}.
Thus, we obtain $(m-M)_{V, \rm V2362~Cyg}= 15.4\pm0.1$.

We apply Equation (8) of \citet{hac19ka} for the $I$ band to Figure
\ref{v2362_cyg_v496_sct_v382_vel_yy_dor_lmcn_2009a_b_v_i_logscale_3fig}(c)
and obtain
\begin{eqnarray}
(m&-&M)_{I, \rm V2362~Cyg} \cr
&=& ((m - M)_I + \Delta I_C)_{\rm V496~Sct} - 2.5 \log 0.89 \cr
&=& 12.9 + 1.5\pm0.3 + 0.125 = 14.52\pm0.3 \cr
&=& ((m - M)_I + \Delta I_C)_{\rm YY~Dor} - 2.5 \log 9.3 \cr
&=& 18.67 - 1.75\pm0.3 - 2.425 = 14.49\pm0.3 \cr
&=& ((m - M)_I + \Delta I_C)_{\rm LMC~N~2009a} - 2.5 \log 5.9 \cr
&=& 18.67 - 2.25\pm0.3 - 1.925 = 14.49\pm0.3,
\label{distance_modulus_i_v2362_cyg_v496_sct_yy_dor_lmcn2009a}
\end{eqnarray}
where we adopt
$(m-M)_{I, \rm V496~Sct}=12.9$ in Appendix \ref{v496_sct_bvi}, and
$(m-M)_{I, \rm YY~Dor}=18.49 + 1.5\times 0.12=18.67$ and
$(m-M)_{I, \rm LMC~N~2009a}=18.49 + 1.5\times 0.12=18.67$ both from
\citet{hac18kb}.
Thus, we obtain $(m-M)_{I, \rm V2362~Cyg}= 14.5\pm0.2$.

We plot these three distance moduli in Figure 
\ref{distance_reddening_v5114_sgr_v2362_cyg_v1065_cen_v959_mon}(b).
The three lines cross at $d=5.4$~kpc and $E(B-V)=0.56$.
The crossing point is consistent with the distance-reddening relation
given by \citet[][thick solid yellow line]{gre19}.
Thus, we obtain a parameter set of $E(B-V)=0.56\pm0.05$, 
$(m-M)_V= 15.4\pm0.1$, $(m-M)_I= 14.5\pm0.2$, $d=5.4\pm0.5$~kpc,
and $\log f_{\rm s}= +0.25$ against that of LV~Vul.


\begin{figure}
\plotone{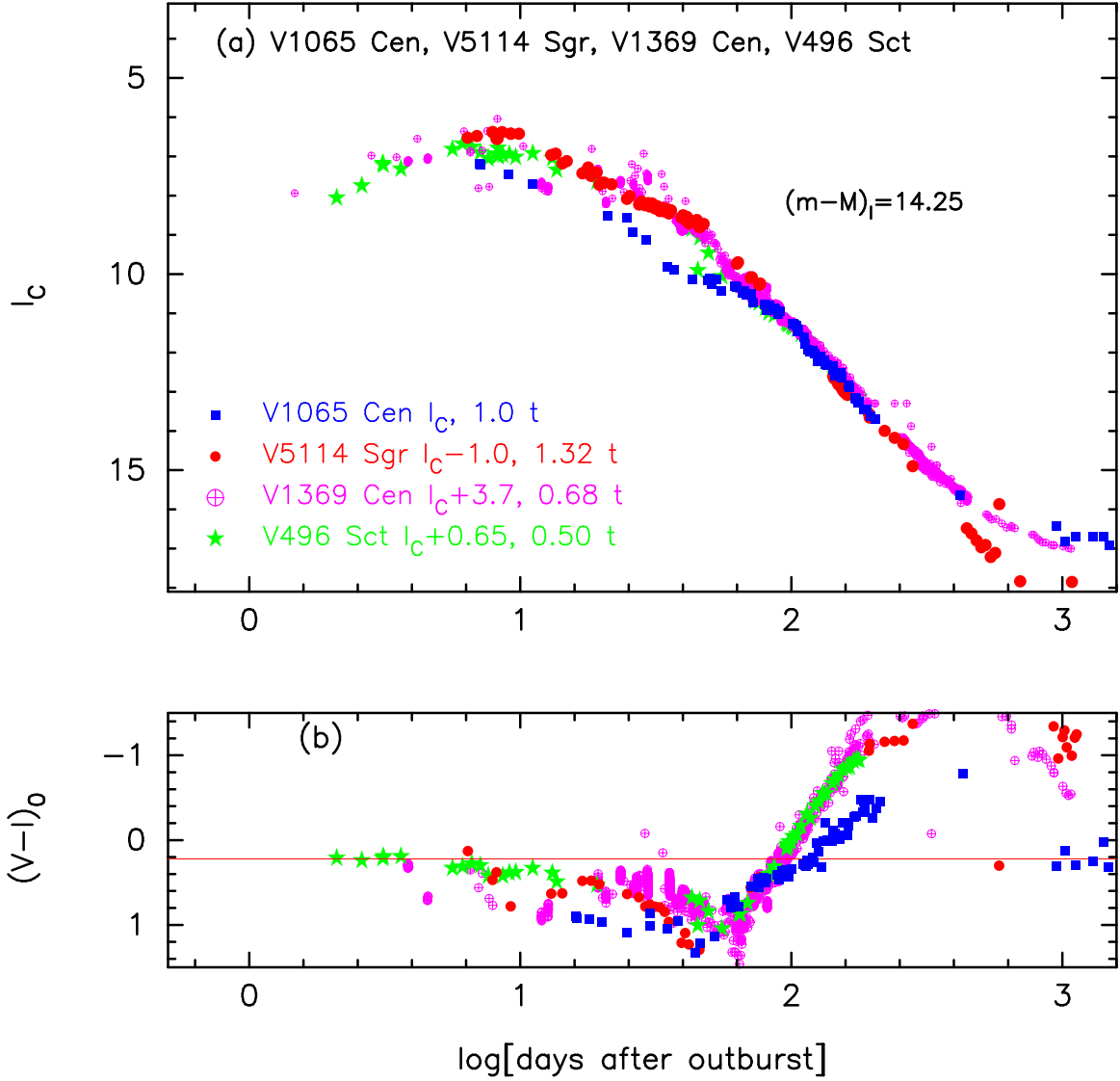}
\caption{
The (a) $I_{\rm C}$ light curve and (b) $(V-I_{\rm C})_0$ color curve
of V1065~Cen as well as those of V5114~Sgr, V1369~Cen, and V496~Sct.
\label{v1065_cen_v5114_sgr_v1369_cen_v496_sct_i_vi_color_logscale}}
\end{figure}


\begin{figure}
\plotone{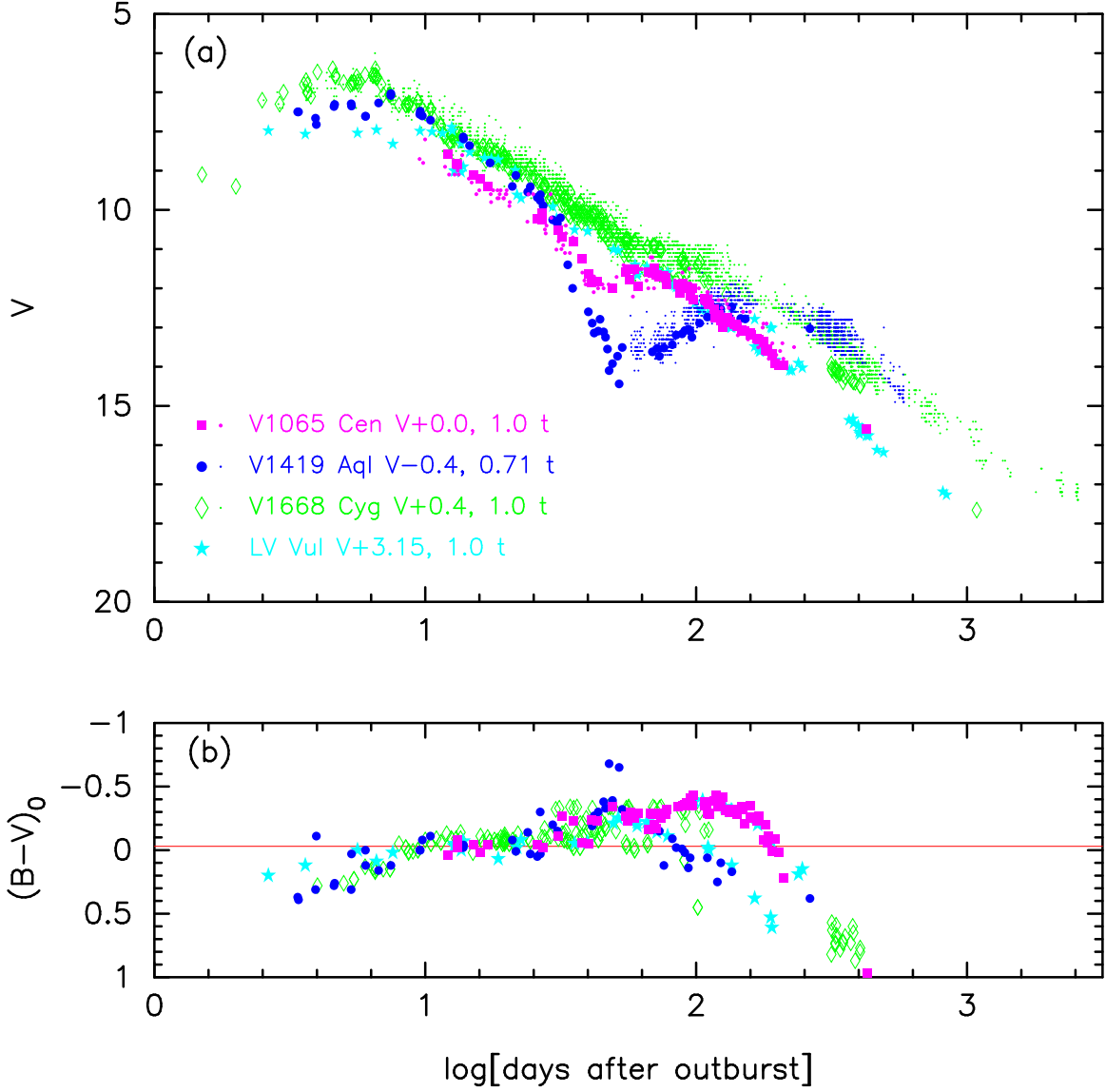}
\caption{
The (a) $V$ light curve and (b) $(B-V)_0$ color curve
of V1065~Cen as well as those of V1419~Aql, V1668~Cyg, and LV~Vul.
\label{v1065_cen_v1419_aql_v1668_cyg_v1974_cyg_v_bv_ub_color_logscale_no2}}
\end{figure}

\subsection{V1065~Cen 2007}
\label{v1065_cen_bvi}
We have reanalyzed the $BVI_{\rm C}$ multi-band light/color curves
of V1065~Cen based on the time-stretching method.  
We obtained a new parameter set of $E(B-V)=0.48$, $(m-M)_V=15.0$,
$d=5.0$~kpc, and $\log f_{\rm s}= +0.0$ against that of LV~Vul,
as mentioned below.
Figure \ref{v1065_cen_v5114_sgr_v1369_cen_v496_sct_i_vi_color_logscale}
shows the $I_{\rm C}$ light and $(V-I_{\rm C})_0$ color curves of V1065~Cen
as well as V5114~Sgr, V1369~Cen, and V496~Sct.
Here, we adopt $\log f_{\rm s}= +0.0$ after \citet{hac18k}.
The $BVI_{\rm C}$ data of V1065~Cen are taken from AAVSO, VSOLJ and SMARTS.
We apply Equation (8) of \citet{hac19ka} for the $I$ band to Figure
\ref{v1065_cen_v5114_sgr_v1369_cen_v496_sct_i_vi_color_logscale}(a)
and obtain
\begin{eqnarray}
(m&-&M)_{I, \rm V1065~Cen} \cr
&=& ((m - M)_I + \Delta I_{\rm C})
_{\rm V5114~Sgr} - 2.5 \log 1.32 \cr
&=& 15.55 - 1.0\pm0.2 - 0.3  = 14.25\pm0.2 \cr
&=& ((m - M)_I + \Delta I_{\rm C})
_{\rm V1369~Cen} - 2.5 \log 0.68 \cr
&=& 10.11 + 3.7\pm0.2 + 0.425 = 14.24\pm0.2 \cr
&=& ((m - M)_I + \Delta I_{\rm C})
_{\rm V496~Sct} - 2.5 \log 0.5 \cr
&=& 12.9 + 0.65\pm0.2 + 0.75  = 14.3\pm0.2,
\label{distance_modulus_i_vi_v1065_cen}
\end{eqnarray}
where we adopt
$(m-M)_{I, \rm V5114~Sgr}=15.55$ from Appendix \ref{v5114_sgr_ubvi},
$(m-M)_{I, \rm V1369~Cen}=10.11$ from \citet{hac19ka}, and
$(m-M)_{I, \rm V496~Sct}=12.9$ in Appendix \ref{v496_sct_bvi}.
Thus, we obtain $(m-M)_{I, \rm V1065~Cen}= 14.26\pm0.2$.

Figure \ref{v1065_cen_v1419_aql_v1668_cyg_v1974_cyg_v_bv_ub_color_logscale_no2}
shows the (a) visual and $V$, and (b) $(B-V)_0$ evolutions of V1065~Cen
as well as V1419~Aql, V1668~Cyg, and  LV~Vul.  
Applying Equation (4) of \citet{hac19ka} for the $V$ band to them,
we have the relation
\begin{eqnarray}
(m&-&M)_{V, \rm V1065~Cen} \cr
&=& ((m - M)_V + \Delta V)_{\rm LV~Vul} - 2.5 \log 1.0 \cr
&=& 11.85 + 3.15\pm0.2 - 0.0 = 15.0\pm0.2 \cr
&=& ((m - M)_V + \Delta V)_{\rm V1668~Cyg} - 2.5 \log 1.0 \cr
&=& 14.6 + 0.4\pm0.2 - 0.0 = 15.0\pm0.2 \cr
&=& ((m - M)_V + \Delta V)_{\rm V1419~Aql} - 2.5 \log 0.71 \cr
&=& 15.0 - 0.4\pm0.2 + 0.375 = 15.0\pm0.2,
\label{distance_modulus_v_bv_v1065_cen}
\end{eqnarray}
where we adopt $(m-M)_{V, \rm LV~Vul}=11.85$,
$(m-M)_{V, \rm V1668~Cyg}=14.6$, and
$(m-M)_{V, \rm V1419~Aql}=15.0$, all from \citet{hac19ka}.
Thus, we obtain $(m-M)_V=15.0\pm0.1$ for V1065~Cen, which is consistent with
Hachisu \& Kato's (2018a) results.


\begin{figure*}
\epsscale{0.55}
\plotone{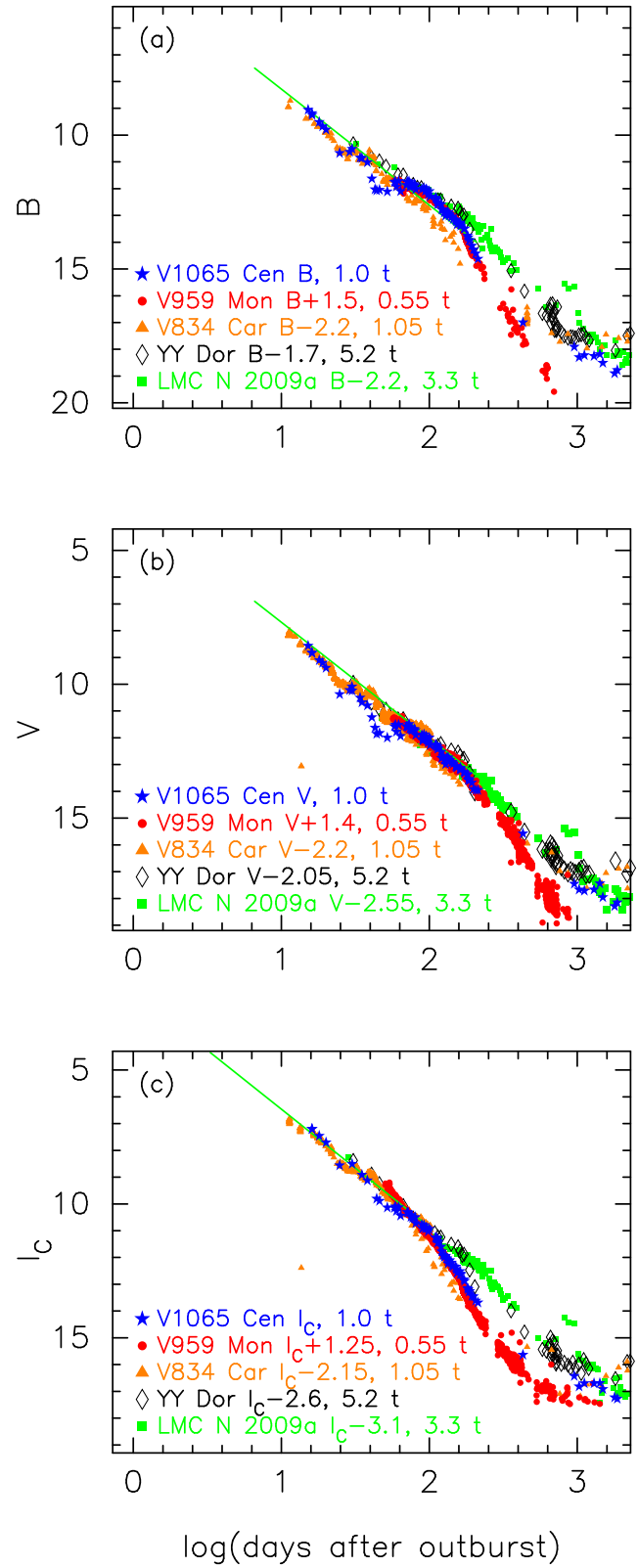}
\caption{
Same as Figure 
\ref{v2362_cyg_v496_sct_v382_vel_yy_dor_lmcn_2009a_b_v_i_logscale_3fig},
but for V1065~Cen (filled blue stars).
The data of V1065~Cen are taken from AAVSO and SMARTS.
\label{v1065_cen_v959_mon_v834_car_yy_dor_lmcn_2009a_b_v_i_logscale_3fig}}
\end{figure*}

Figure \ref{v1065_cen_v959_mon_v834_car_yy_dor_lmcn_2009a_b_v_i_logscale_3fig}
shows the $BVI_{\rm C}$ light curves of V1065~Cen as well as V959~Mon,
V834~Car, YY~Dor, and LMC~N~2009a.
We apply Equation (7) of \citet{hac19ka} for the $B$ band to Figure
\ref{v1065_cen_v959_mon_v834_car_yy_dor_lmcn_2009a_b_v_i_logscale_3fig}(a)
and obtain
\begin{eqnarray}
(m&-&M)_{B, \rm V1065~Cen} \cr
&=& ((m - M)_B + \Delta B)_{\rm V959~Mon} - 2.5 \log 0.55 \cr
&=& 13.53 + 1.5\pm0.2 + 0.45 = 15.48\pm0.2 \cr
&=& ((m - M)_B + \Delta B)_{\rm V834~Car} - 2.5 \log 1.05 \cr
&=& 17.75 - 2.2\pm0.2 - 0.05 = 15.5\pm0.2 \cr
&=& ((m - M)_B + \Delta B)_{\rm YY~Dor} - 2.5 \log 5.2 \cr
&=& 18.98 - 1.7\pm0.2 - 1.8 = 15.48\pm0.2 \cr
&=& ((m - M)_B + \Delta B)_{\rm LMC~N~2009a} - 2.5 \log 3.3 \cr
&=& 18.98 - 2.2\pm0.2 - 1.3 = 15.48\pm0.2.
\label{distance_modulus_b_v1065_cen_v959_mon_v834_car_yy_dor_lmcn2009a}
\end{eqnarray}
Thus, we have $(m-M)_{B, \rm V1065~Cen}= 15.48\pm0.1$.

For the $V$ light curves in Figure
\ref{v1065_cen_v959_mon_v834_car_yy_dor_lmcn_2009a_b_v_i_logscale_3fig}(b),
we similarly obtain
\begin{eqnarray}
(m&-&M)_{V, \rm V1065~Cen} \cr   
&=& ((m - M)_V + \Delta V)_{\rm V959~Mon} - 2.5 \log 0.55 \cr
&=& 13.15 + 1.4\pm0.2 + 0.45 = 15.0\pm0.2 \cr
&=& ((m - M)_V + \Delta V)_{\rm V834~Car} - 2.5 \log 1.05 \cr
&=& 17.25 - 2.2\pm0.2 - 0.05 = 15.0\pm0.2 \cr
&=& ((m - M)_V + \Delta V)_{\rm YY~Dor} - 2.5 \log 5.2 \cr
&=& 18.86 - 2.05\pm0.2 - 1.8 = 15.01\pm0.2 \cr
&=& ((m - M)_V + \Delta V)_{\rm LMC~N~2009a} - 2.5 \log 3.3 \cr
&=& 18.86 - 2.55\pm0.2 -1.3 = 15.01\pm0.2.
\label{distance_modulus_v_v1065_cen_v959_mon_v834_car_yy_dor_lmcn2009a}
\end{eqnarray}
We have $(m-M)_{V, \rm V1065~Cen}= 15.0\pm0.1$.

We apply Equation (8) of \citet{hac19ka} for the $I$ band to Figure
\ref{v1065_cen_v959_mon_v834_car_yy_dor_lmcn_2009a_b_v_i_logscale_3fig}(c)
and obtain
\begin{eqnarray}
(m&-&M)_{I, \rm V1065~Cen} \cr
&=& ((m - M)_I + \Delta I_C)_{\rm V959~Mon} - 2.5 \log 0.55 \cr
&=& 12.54 + 1.25\pm0.2 + 0.45 = 14.24\pm 0.2 \cr
&=& ((m - M)_I + \Delta I_C)_{\rm V834~Car} - 2.5 \log 1.05 \cr
&=& 16.44 - 2.15\pm0.2 - 0.05 = 14.24\pm 0.2 \cr
&=& ((m - M)_I + \Delta I_C)_{\rm YY~Dor} - 2.5 \log 5.2 \cr
&=& 18.67 - 2.6\pm0.2 - 1.8 = 14.27\pm 0.2 \cr
&=& ((m - M)_I + \Delta I_C)_{\rm LMC~N~2009a} - 2.5 \log 3.3 \cr
&=& 18.67 - 3.1\pm0.2 - 1.3 = 14.27\pm 0.2.
\label{distance_modulus_i_v1065_cen_v959_mon_v834_car_yy_dor_lmcn2009a}
\end{eqnarray}
Thus, we have $(m-M)_{I, \rm V1065~Cen}= 14.25\pm0.1$,

We plot the three distance moduli
in Figure \ref{distance_reddening_v5114_sgr_v2362_cyg_v1065_cen_v959_mon}(c).
The three lines cross at $d= 5.0$~kpc and $E(B-V)= 0.48$.  Here,
we add the four thick cyan-blue lines of \citet{chen19}, which correspond
to four nearby directions toward V1065~Cen, i.e., the galactic coordinates of
$(\ell, b)= (293\fdg95, +3\fdg55)$, $(293\fdg95, +3\fdg65)$,
$(294\fdg05, +3\fdg55)$, and $(294\fdg05, +3\fdg65)$.
The crossing point is broadly consistent with the distance-reddening
relations given by \citet{mar06}.  The reddening value of $E(B-V)= 0.48$
at $d=5$~kpc is larger than those of \citet{chen19} but slightly smaller
than that of \citet{ozd18}.


\begin{figure}
\plotone{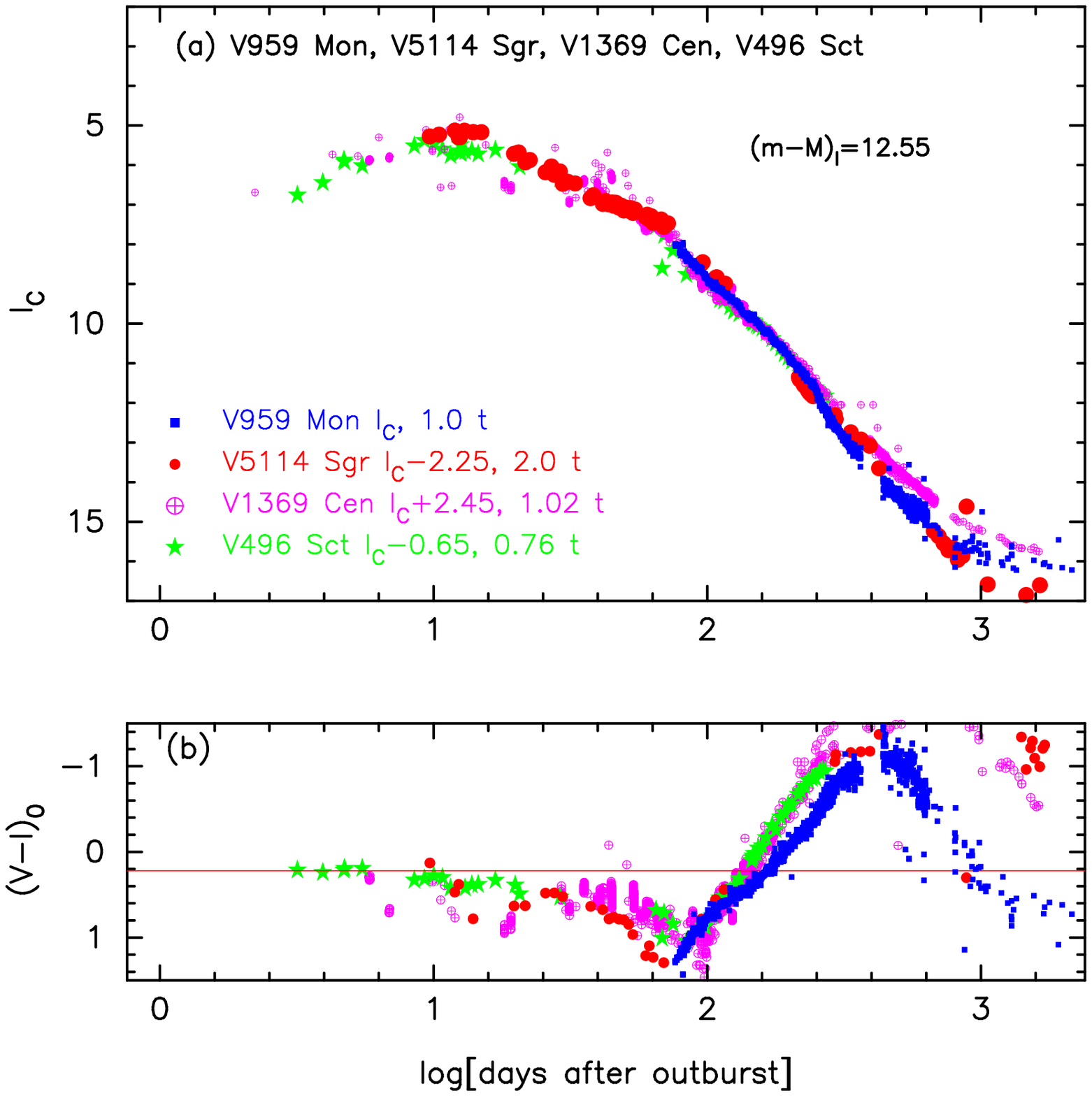}
\caption{
The (a) $I_{\rm C}$ light curve and (b) $(V-I_{\rm C})_0$ color curve
of V959~Mon as well as those of V5114~Sgr, V1369~Cen, and V496~Sct.
\label{v959_mon_v5114_sgr_v1369_cen_v496_sct_i_vi_logscale}}
\end{figure}


\begin{figure}
\plotone{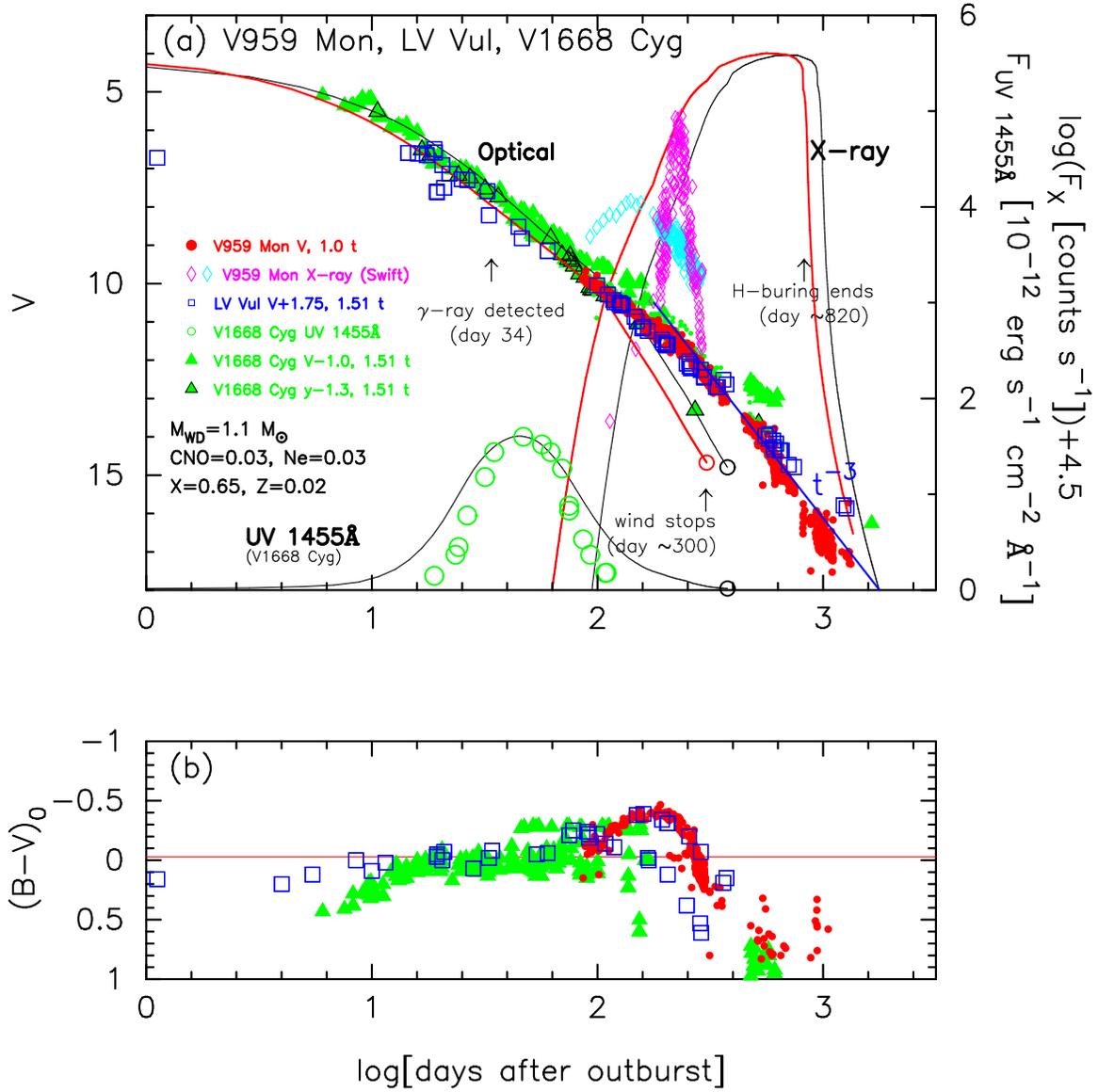}
\caption{
The (a) $V$ light curve and (b) $(B-V)_0$ color curve
of V959~Mon as well as those of LV~Vul and V1668~Cyg.
In panel (a), we add a $1.1~M_\sun$ WD model \citep[Ne3, 
solid red lines,][]{hac16k} for V959~Mon.  The solid black lines denote
a $0.98~M_\sun$ WD model \citep[CO3,][]{hac16k} for V1668~Cyg.   
\label{v959_mon_v1668_cyg_lv_vul_v_bv_logscale_no4}}
\end{figure}


\begin{figure}
\epsscale{0.55}
\plotone{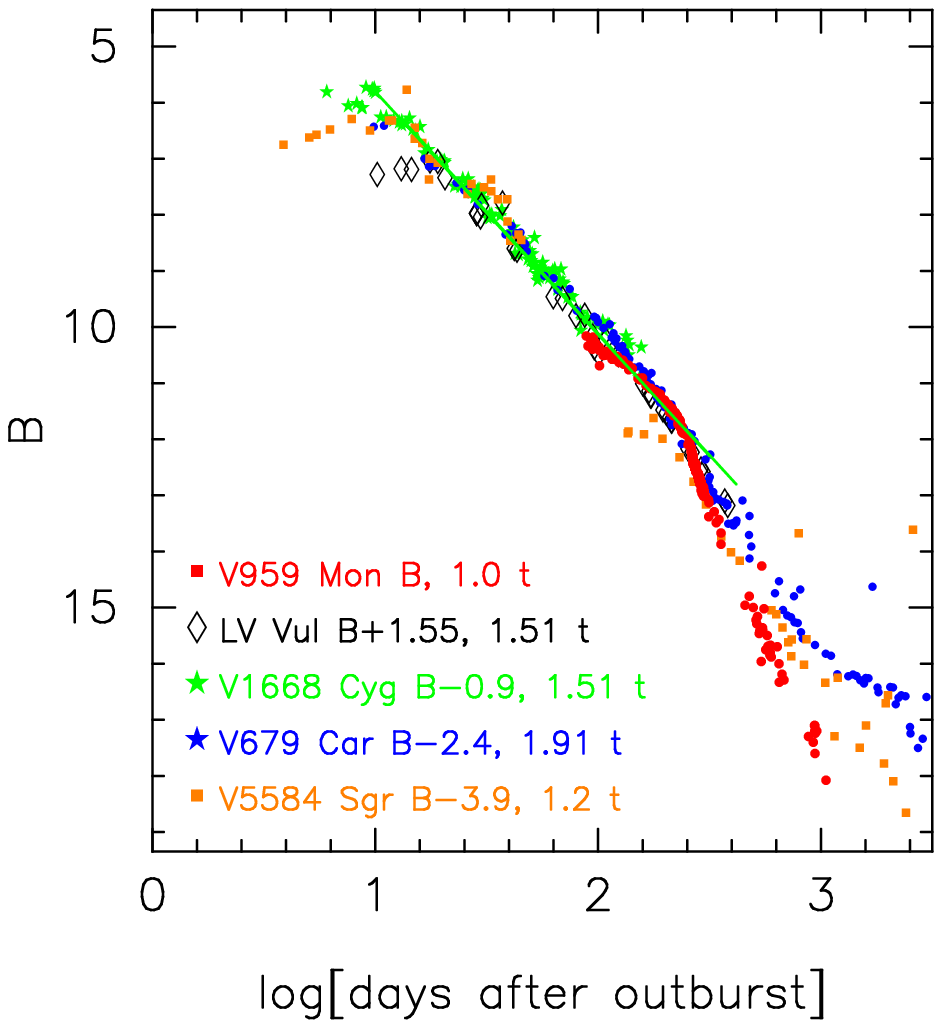}
\caption{
Same as Figure 
\ref{v2362_cyg_v496_sct_v382_vel_yy_dor_lmcn_2009a_b_v_i_logscale_3fig},
but for V959~Mon (filled red circles).
The data of V959~Mon are taken from \citet{mun13b}, AAVSO, VSOLJ,
and SMARTS.  
\label{v959_mon_v5584_sgr_lv_vul_v1668_cyg_b_only_logscale}}
\end{figure}

\subsection{V959~Mon 2012}
\label{v959_mon_bvi}
We have reanalyzed the $BVI_{\rm C}$ multi-band 
light/color curves of V959~Mon based on the time-stretching method.  
The main differences are the timescaling factor of $\log f_{\rm s}=
+0.18$ (the previous value of $+0.14$) against that of LV~Vul, although
the distance modulus in $V$ band of $(m-M)_V= 13.15$ is the same as
the previous value of $13.15$.
Figure \ref{v959_mon_v5114_sgr_v1369_cen_v496_sct_i_vi_logscale}
shows the $I_{\rm C}$ light and $(V-I_{\rm C})_0$ color curves of
V959~Mon as well as V5114~Sgr, V1369~Cen, and V496~Sct.
The $BVI_{\rm C}$ data of V959~Mon are taken from \citet{mun13b}.
Using the color excess of $E(B-V)= 0.38$ after \citet{mun13b},
we adopt the timescaling factor of $\log f_{\rm s}= +0.18$
in order to overlap the $(V-I_{\rm C})_0$ color curve of V959~Mon
with the other novae, as shown in
Figure \ref{v959_mon_v5114_sgr_v1369_cen_v496_sct_i_vi_logscale}(b).
We apply Equation (8) of \citet{hac19ka} for the $I$ band to Figure
\ref{v959_mon_v5114_sgr_v1369_cen_v496_sct_i_vi_logscale}(a)
and obtain
\begin{eqnarray}
(m&-&M)_{I, \rm V959~Mon} \cr
&=& ((m - M)_I + \Delta I_{\rm C})
_{\rm V5114~Sgr} - 2.5 \log 2.0 \cr
&=& 15.55 -2.25\pm0.2 - 0.75 = 12.55\pm0.2 \cr
&=& ((m - M)_I + \Delta I_{\rm C})
_{\rm V1369~Cen} - 2.5 \log 1.02 \cr
&=& 10.11 + 2.45\pm0.2 - 0.025 = 12.54\pm0.2 \cr
&=& ((m - M)_I + \Delta I_{\rm C})
_{\rm V496~Sct} - 2.5 \log 0.76 \cr
&=& 12.9 - 0.65\pm0.2 + 0.3 = 12.55\pm0.2,
\label{distance_modulus_i_vi_v959_mon}
\end{eqnarray}
where we adopt
$(m-M)_{I, \rm V5114~Sgr}=15.55$ from Appendix \ref{v5114_sgr_ubvi},
$(m-M)_{I, \rm V1369~Cen}=10.11$ from \citet{hac19ka}, and
$(m-M)_{I, \rm V496~Sct}=12.9$ in Appendix \ref{v496_sct_bvi}.
Thus, we obtain $(m-M)_{I, \rm V959~Mon}= 12.55\pm0.2$.

We plot the $V$ and $B-V$ data of V959~Mon in
Figure \ref{v959_mon_v1668_cyg_lv_vul_v_bv_logscale_no4} 
as well as LV~Vul and V1668~Cyg.
Applying Equation (4) of \citet{hac19ka} for the $V$ band to 
Figure \ref{v959_mon_v1668_cyg_lv_vul_v_bv_logscale_no4}(a),
we have the relation of
\begin{eqnarray}
(m&-&M)_{V, \rm V959~Mon} \cr
&=& ((m - M)_V + \Delta V)_{\rm LV~Vul} - 2.5 \log 1.51 \cr
&=& 11.85 + 1.75\pm0.2 - 0.45 = 13.15\pm0.2 \cr
&=& ((m - M)_V + \Delta V)_{\rm V1668~Cyg} - 2.5 \log 1.51 \cr
&=& 14.6 - 1.0\pm0.2 - 0.45 = 13.15\pm0.2.
\label{distance_modulus_v_bv_v959_mon_lv_vul_v1668_cyg}
\end{eqnarray}
Thus, we obtain $(m-M)_V=13.15\pm0.2$ and $\log f_{\rm s} = \log 1.51
= +0.18$ against the template nova LV~Vul.

Figure \ref{v959_mon_v5584_sgr_lv_vul_v1668_cyg_b_only_logscale} shows
the $B$ light curves of V959~Mon
together with those of LV~Vul, V1668~Cyg, V679~Car, and V5584~Sgr.
We apply Equation (7) of \citet{hac19ka} for the $B$ band to Figure
\ref{v959_mon_v5584_sgr_lv_vul_v1668_cyg_b_only_logscale}
and obtain
\begin{eqnarray}
(m&-&M)_{B, \rm V959~Mon} \cr
&=& ((m - M)_B + \Delta B)_{\rm LV~Vul} - 2.5 \log 1.51 \cr
&=& 12.45 + 1.55\pm0.2 - 0.45 = 13.55\pm0.2 \cr
&=& ((m - M)_B + \Delta B)_{\rm V1668~Cyg} - 2.5 \log 1.51 \cr
&=& 14.9 - 0.9\pm0.2 - 0.45 = 13.55\pm0.2 \cr
&=& ((m - M)_B + \Delta B)_{\rm V679~Car} - 2.5 \log 1.91 \cr
&=& 16.64 - 2.4\pm0.2 - 0.7 = 13.54\pm0.2 \cr
&=& ((m - M)_B + \Delta B)_{\rm V5584~Sgr} - 2.5 \log 1.2 \cr
&=& 17.65 - 3.9\pm0.2 - 0.20 = 13.55\pm0.2,
\label{distance_modulus_b_v959_mon_lv_vul_v1668_cyg}
\end{eqnarray}
where we adopt $(m-M)_{B, \rm LV~Vul}= 12.45$,
$(m-M)_{B, \rm V1668~Cyg}= 14.9$, both from \citet{hac19ka},
$(m-M)_{B, \rm V679~Car}= 16.05 + 0.59 = 16.64$
from Appendix \ref{v679_car_bvi}, and
$(m-M)_{B, \rm V5584~Sgr}= 16.9 + 0.75 = 17.65$
from Appendix \ref{v5584_sgr_bvi}.
We have $(m-M)_{B, \rm V959~Mon}= 13.55\pm0.2$.

The three distance moduli of $(m-M)_B=13.55$, $(m-M)_V=13.15$, 
and $(m-M)_I=12.55$, are similar to those obtained by \citet{hac18k}.
We plot these three relations in Figure 
\ref{distance_reddening_v5114_sgr_v2362_cyg_v1065_cen_v959_mon}(d),
which cross at the distance of $d= 2.5$~kpc
and the extinction of $E(B-V)=0.38$. 
The crossing point is consistent with the distance-reddening relation
(unfilled cyan-blue diamonds) given by \citet{ozd18} and 
\citet{chen19}.  Here, we add the four thick cyan-blue lines of 
\citet{chen19}, which correspond to four nearby directions toward V959~Mon, 
i.e., the galactic coordinates of
$(\ell, b)= (206\fdg25, +0\fdg05)$, $(206\fdg35, +0\fdg05)$,
$(206\fdg25, +0\fdg15)$, and $(206\fdg35, +0\fdg15)$.
The new results are summarized in Tables 
\ref{extinction_various_novae} and \ref{wd_mass_novae}.

\section{Revised Analysis of Nova Light Curves}
\label{revised_analysis}

We are able to obtain a more accurate parameter set of color excess
$E(B-V)$, distance moduli $(m-M)_U$, $(m-M)_B$, $(m-M)_V$, 
$(m-M)_I$, distance $d$, and timescaling factor $\log f_{\rm s}$
by using multi-band $UBVI$ light/color curves and
$(U-B)_0$-$(M_B-2.5 \log f_{\rm s})$,
$(B-V)_0$-$(M_V-2.5 \log f_{\rm s})$,
$(V-I)_0$-$(M_I-2.5 \log f_{\rm s})$ diagrams
rather than by using only $BV$ light/color curves and
$(B-V)_0$-$(M_V-2.5 \log f_{\rm s})$ diagram in the previous
analysis \citep{hac19kb}.
Here, we reanalyze 44 novae that were studied in our previous works
\citep{hac19ka, hac19kb} and update their parameter sets by using
$UBVI$ multi-band light/color curves.  The new results are much more
consistent with the other observations like the distances of {\it Gaia} DR2.
We present 44 novae in the order of discovery date of the outburst.


\begin{figure}
\plotone{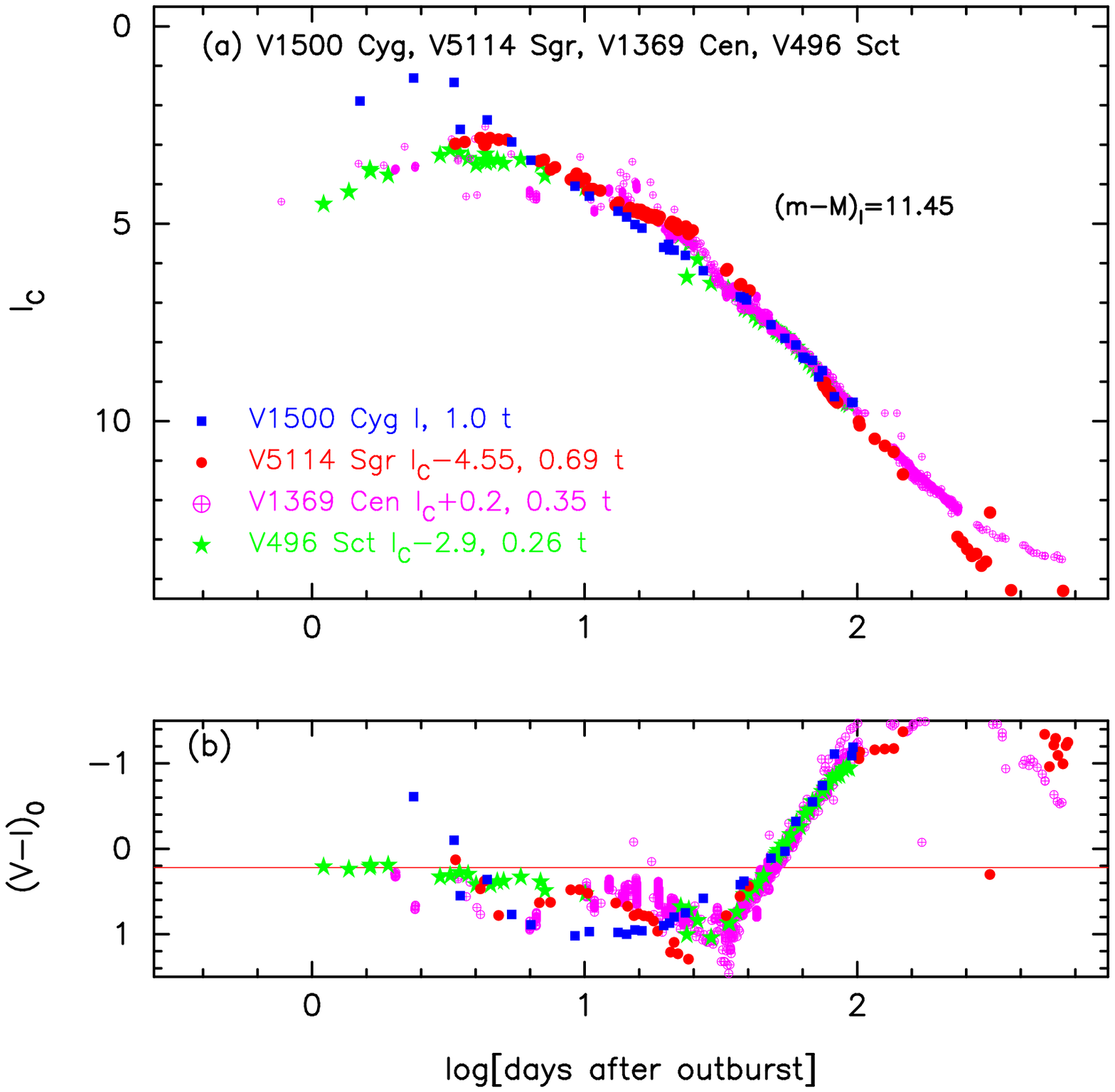}
\caption{
The (a) $I$ light curve and (b) $(V-I)_0$ color curve of V1500~Cyg
as well as those of V5114~Sgr, V1369~Cen, and V496~Sct.
The $UBVI$ data of V1500~Cyg are taken from 
\citet{mar77}, \citet{bel77}, \citet{wil77}, \citet{tem79},
\citet{pfa76}, and \citet{ark76}.
\label{v1500_cyg_v5114_sgr_v1369_cen_v496_sct_i_vi_color_logscale}}
\end{figure}


\begin{figure}
\plotone{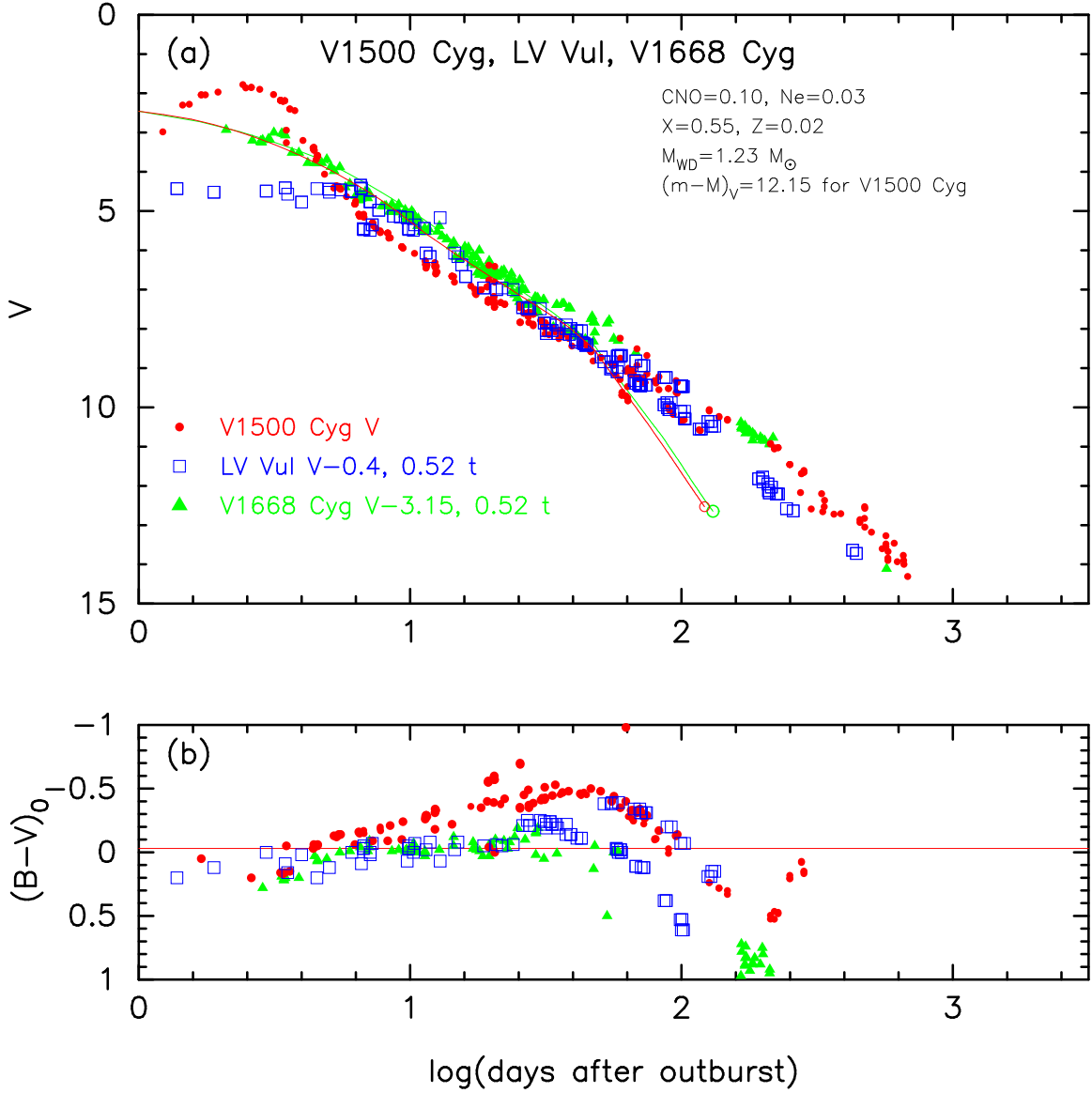}
\caption{
The (a) $V$ light and (b) $(B-V)_0$  color curves
of V1500~Cyg as well as those of LV~Vul and V1668~Cyg.
In panel (a), we show the V1500~Cyg model light curve (thin solid red lines)
of a $1.23~M_\sun$ WD \citep[Ne2,][]{hac10k}.
We also add a $0.98~M_\sun$ WD model (CO3, solid green lines) for V1668~Cyg.
\label{v1500_cyg_lv_vul_v1668_cyg_v_bv_logscale}}
\end{figure}


\begin{figure}
\plotone{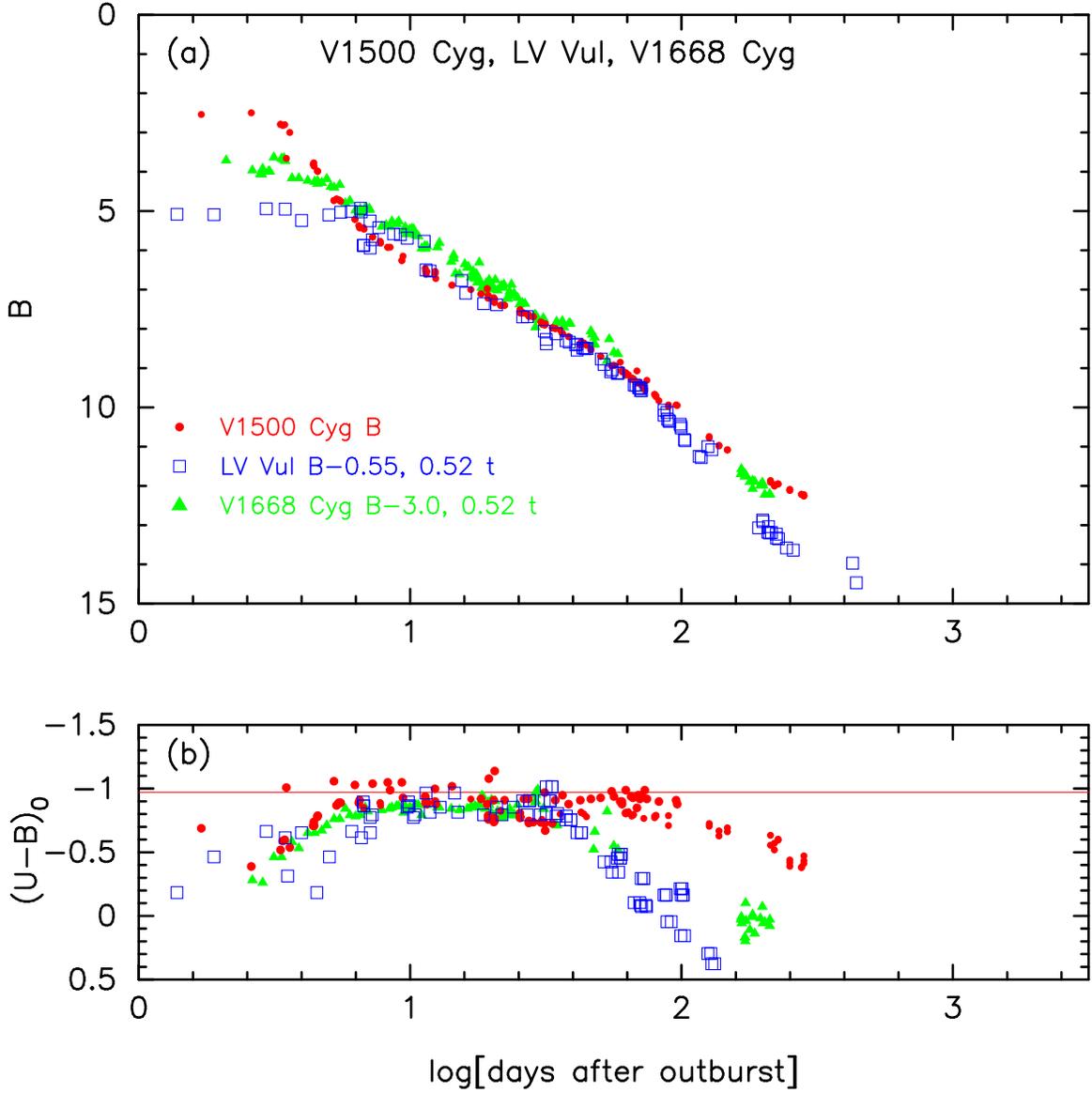}
\caption{
The (a) $B$ light and (b) $(U-B)_0$ color curves of V1500~Cyg as well as
those of LV~Vul and V1668~Cyg.
\label{v1500_cyg_lv_vul_v1668_cyg_b_ub_color_logscale}}
\end{figure}


\begin{figure*}
\plottwo{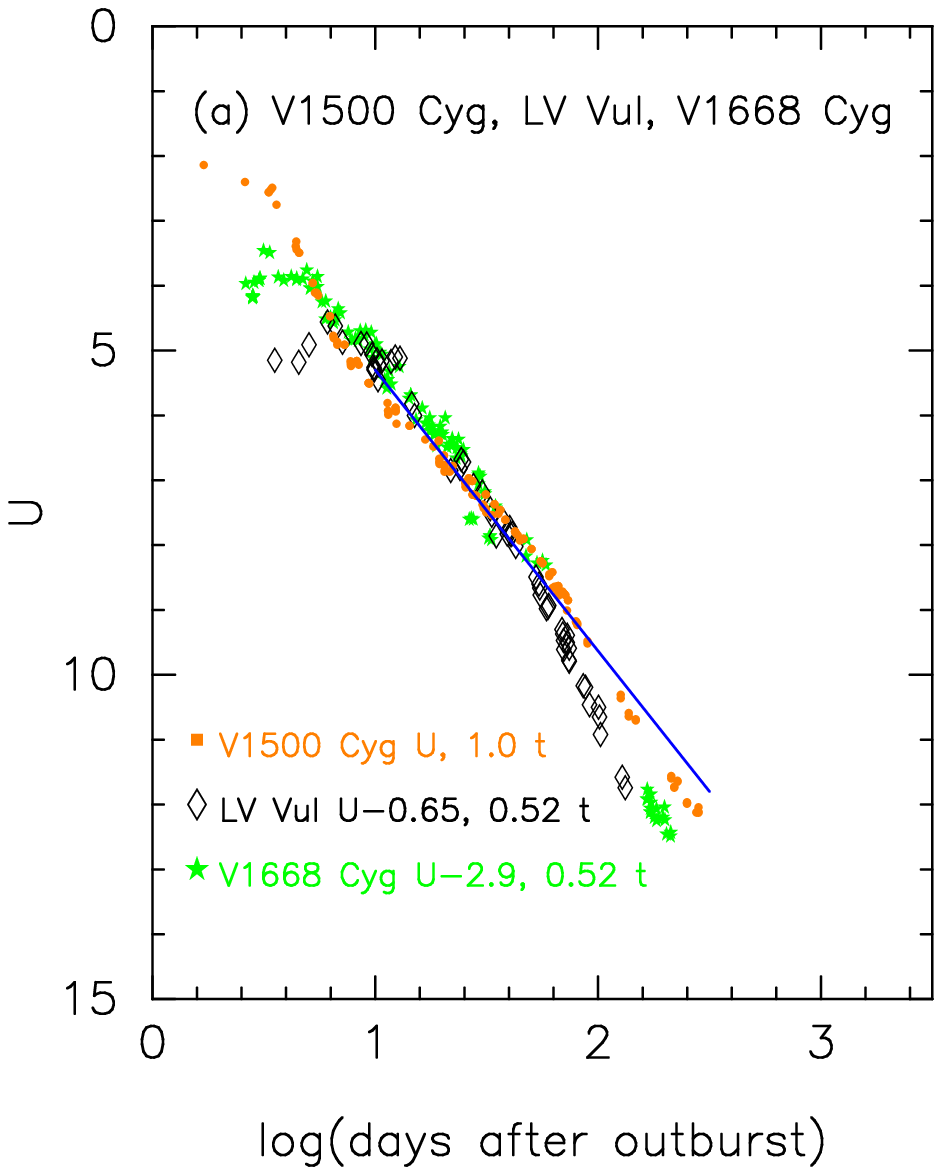}{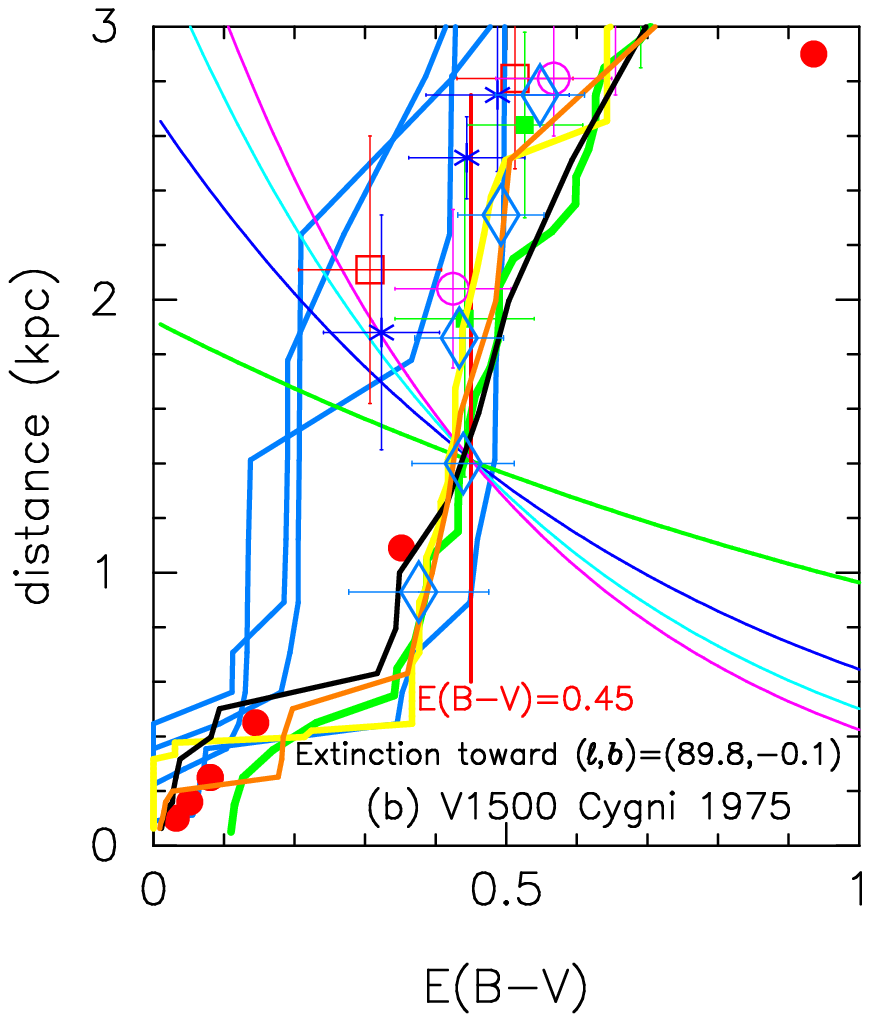}
\caption{
(a) The $U$ light curves of V1500~Cyg as well as
those of LV~Vul and V1668~Cyg.  
(b) Various distance-reddening relations toward V1500~Cyg.
The four thin lines of magenta, cyan, blue, and green 
denote the distance-reddening relations given by 
$(m-M)_U= 12.9$, $(m-M)_B= 12.6$, $(m-M)_V= 12.15$, 
and $(m-M)_I= 11.42$, respectively.
The filled red circles denote the distance
and reddening of nearby stars given by \citet{you76}.
Other symbols and lines are the same as those in
Figure \ref{distance_reddening_v5114_sgr_v2362_cyg_v1065_cen_v959_mon}.
\label{distance_reddening_v1500_cyg_xxxxxx}}
\end{figure*}

\subsection{V1500~Cyg 1975}
\label{v1500_cyg_ubvi}
V1500~Cyg is an important template nova, so we first reanalyze
the $UBVI$ multi-band light/color curves of V1500~Cyg based on the
time-stretching method.  We obtain the distance moduli
in $UBVI$ bands and examine the timescaling factor, distance,
and reddening toward V1500~Cyg.   Figure
\ref{v1500_cyg_v5114_sgr_v1369_cen_v496_sct_i_vi_color_logscale}
shows the (a) $I$ light and (b) $(V-I)_0$ color curves of V1500~Cyg 
as well as V5114~Sgr, V1369~Cen, and V496~Sct.
The $VI$ data of V1500~Cyg are taken from \citet{mar77} and
\citet{bel77}.   The $V-I$ colors of \citet{bel77} are systematically
redder by 0.75 mag than those of \citet{mar77}.  We shift
them up by 0.75 mag and overlap them with those of \citet{mar77}.  
Adopting the color excess of $E(B-V)= 0.45$, we redefine the timescaling
factor $\log f_{\rm s}= -0.28$ of V1500~Cyg against that of LV~Vul.
This is because the $(V-I)_0$ color evolution of V1500~Cyg overlaps with
the other novae as much as possible, as shown in Figure
\ref{v1500_cyg_v5114_sgr_v1369_cen_v496_sct_i_vi_color_logscale}(b).  
Then, we apply Equation (8) of \citet{hac19ka} for the $I$ band to Figure
\ref{v1500_cyg_v5114_sgr_v1369_cen_v496_sct_i_vi_color_logscale}(a)
and obtain
\begin{eqnarray}
(m&-&M)_{I, \rm V1500~Cyg} \cr
&=& ((m - M)_I + \Delta I_{\rm C})
_{\rm V5114~Sgr} - 2.5 \log 0.69 \cr
&=& 15.55 - 4.55\pm0.2 + 0.4 = 11.4\pm0.2 \cr
&=& ((m - M)_I + \Delta I_{\rm C})
_{\rm V1369~Cen} - 2.5 \log 0.35 \cr
&=& 10.11 + 0.2\pm0.2 + 1.125 = 11.43\pm0.2 \cr
&=& ((m - M)_I + \Delta I_{\rm C})
_{\rm V496~Sct} - 2.5 \log 0.26 \cr
&=& 12.9 - 2.9\pm0.2 + 1.45 = 11.45\pm0.2,
\label{distance_modulus_i_vi_v1500_cyg}
\end{eqnarray}
where we adopt
$(m-M)_{I, \rm V5114~Sgr}=15.55$ from Appendix \ref{v5114_sgr_ubvi},
$(m-M)_{I, \rm V1369~Cen}=10.11$ from \citet{hac19ka}, and
$(m-M)_{I, \rm V496~Sct}=12.9$ in Appendix \ref{v496_sct_bvi}. 
Thus, we obtain $(m-M)_{I, \rm V1500~Cyg}= 11.43\pm0.2$.

Figure 
\ref{v1500_cyg_lv_vul_v1668_cyg_v_bv_logscale}
shows the (a) $V$ light and (b) $(B-V)_0$ color curves of V1500~Cyg
as well as LV~Vul and V1668~Cyg.  Here, we adopt
the set of $E(B-V)= 0.45$ and $\log f_{\rm s} = -0.28$ after
the $I$ light and $(V-I)_0$ color curves analysis mentioned above.
We apply Equation (4) of \citet{hac19ka} to Figure 
\ref{v1500_cyg_lv_vul_v1668_cyg_v_bv_logscale}(a)
and obtain
\begin{eqnarray}
(m&-&M)_{V, \rm V1500~Cyg} \cr
&=& ((m - M)_V + \Delta V)_{\rm LV~Vul} - 2.5 \log 0.52 \cr
&=& 11.85 - 0.4\pm0.2 + 0.7 = 12.15\pm0.2 \cr
&=& ((m - M)_V + \Delta V)_{\rm V1668~Cyg} - 2.5 \log 0.52 \cr
&=& 14.6 - 3.15\pm0.2 + 0.7 = 12.15\pm0.2,
\label{distance_modulus_v_bv_v1500_cyg}
\end{eqnarray}
where we adopt $(m-M)_{V, \rm LV~Vul}=11.85$
and $(m-M)_{V, \rm V1668~Cyg}=14.6$ both from \citet{hac19ka}.
Thus, we obtain $(m-M)_{V, \rm V1500~Cyg}= 12.15\pm0.1$.

Figure 
\ref{v1500_cyg_lv_vul_v1668_cyg_b_ub_color_logscale}
shows the (a) $B$ light and (b) $(U-B)_0$ color curves of V1500~Cyg
as well as V1668~Cyg and LV~Vul.
For the $B$ band, we apply Equation (7) of \citet{hac19ka} to Figure 
\ref{v1500_cyg_lv_vul_v1668_cyg_b_ub_color_logscale}(a)
and obtain
\begin{eqnarray}
(m&-&M)_{B, \rm V1500~Cyg} \cr
&=& ((m - M)_B + \Delta B)_{\rm LV~Vul} - 2.5 \log 0.52 \cr
&=& 12.45 - 0.55\pm0.2 + 0.7 = 12.6\pm0.2 \cr
&=& ((m - M)_B + \Delta B)_{\rm V1668~Cyg} - 2.5 \log 0.52 \cr
&=& 14.9 - 3.0\pm0.2 + 0.7 = 12.6\pm0.2,
\label{distance_modulus_b_ub_v1500_cyg}
\end{eqnarray}
where we adopt and $(m-M)_{B, \rm LV~Vul}= 12.45$ 
and $(m-M)_{B, \rm V1668~Cyg}= 14.9$ both from \citet{hac19ka}.
Thus, we obtain $(m-M)_{B, \rm V1500~Cyg}= 12.6\pm0.2$.

Using the timescaling factor of $\log f_{\rm s}= -0.28$, we plot
the $U$ band light curve of V1500~Cyg as well as
LV~Vul and V1668~Cyg in Figure
\ref{distance_reddening_v1500_cyg_xxxxxx}(a).
We apply Equation (6) of \citet{hac19ka} for the $U$ band to Figure
\ref{distance_reddening_v1500_cyg_xxxxxx}(a)
and obtain
\begin{eqnarray}
(m&-&M)_{U, \rm V1500~Cyg} \cr
&=& ((m - M)_U + \Delta U)_{\rm LV~Vul} - 2.5 \log 0.52 \cr
&=& 12.85 - 0.65\pm0.2 + 0.7 = 12.9\pm0.2 \cr
&=& ((m - M)_U + \Delta U)_{\rm V1668~Cyg} - 2.5 \log 0.52 \cr
&=& 15.1 - 2.9\pm0.2 + 0.7 = 12.9\pm0.2, 
\label{distance_modulus_u_v1500_cyg}
\end{eqnarray}
where we adopt 
$(m-M)_{U, \rm LV~Vul}= 12.85$, and
$(m-M)_{U, \rm V1668~Cyg}= 15.10$ both from \citet{hac19ka}.
Thus, we obtain $(m-M)_{U, \rm V1500~Cyg}= 12.9\pm0.2$.

We plot the four distance moduli
in Figure \ref{distance_reddening_v1500_cyg_xxxxxx}(b).
These four lines cross at $d=1.41\pm0.2$~kpc and $E(B-V)=0.45\pm0.05$.
The crossing point is close to the distance-reddening relations
given by \citet[][thick solid green line]{sal14}, 
\citet[][thick solid black and orange lines]{gre15, gre18}, 
and \citet[][unfilled cyan-blue diamonds with error bars]{ozd16}.


\begin{figure}
\plotone{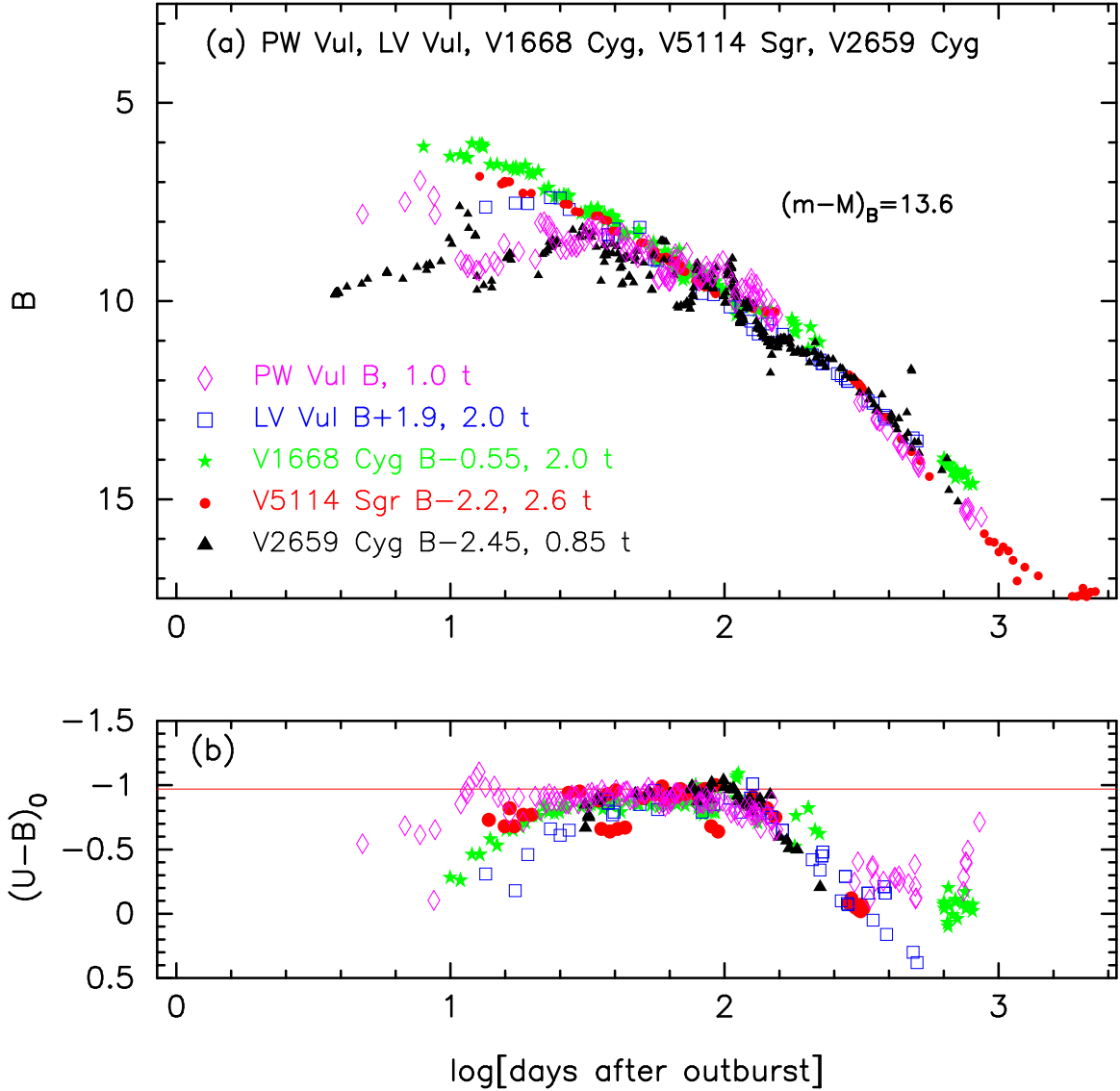}
\caption{
The (a) $B$ light curve and (b) $(U-B)_0$ color curve of PW~Vul
as well as those of LV~Vul, V1668~Cyg, V5114~Sgr, and V2659~Cyg.
The $UBV$ data of PW~Vul are taken
from \citet{nos85}, \citet{kol86}, and \citet{rob95}.
\label{pw_vul_v2659_cyg_v5114_sgr_v1668_cyg_lv_vul_b_ub_color_logscale}}
\end{figure}


\begin{figure}
\plotone{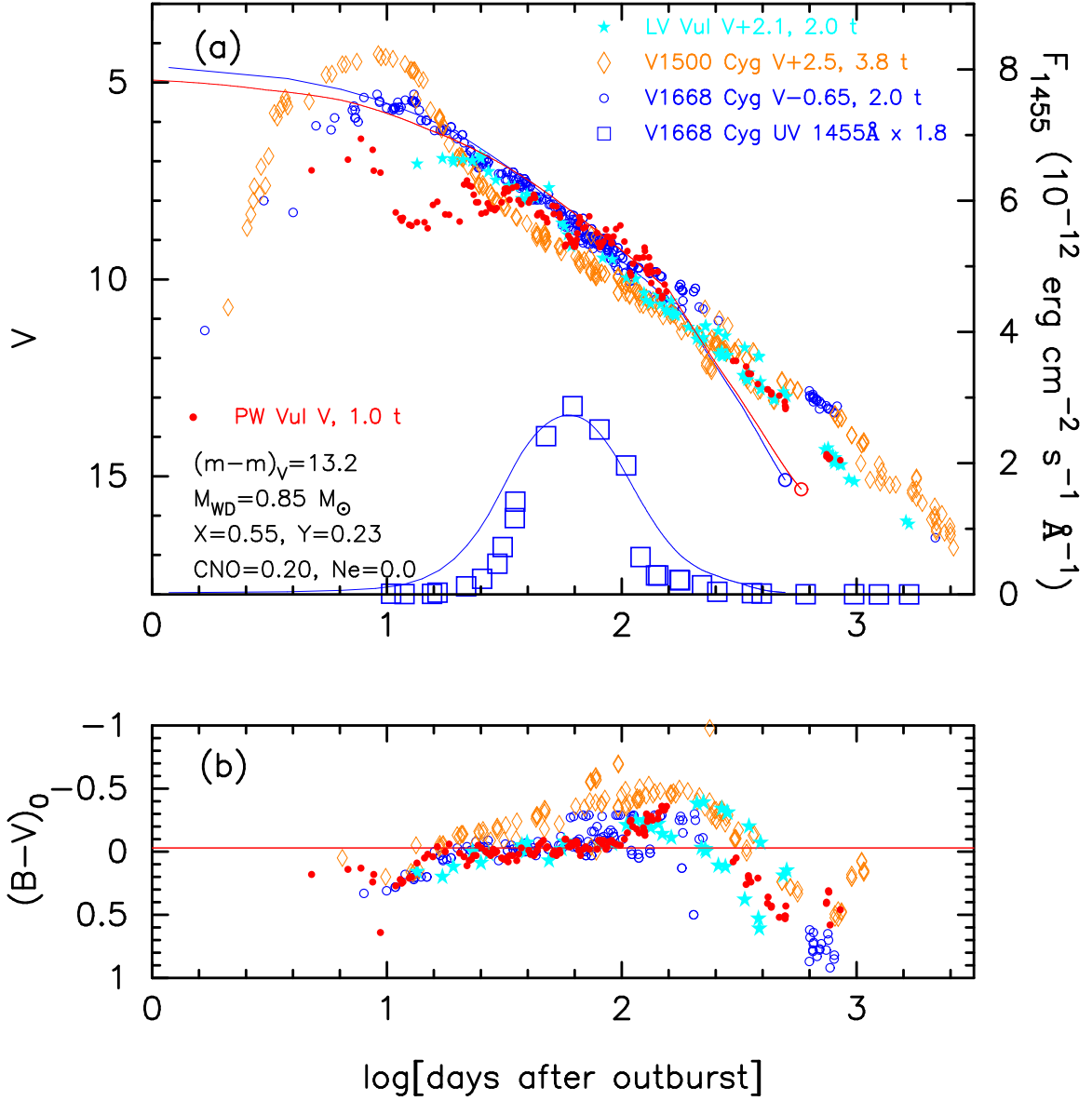}
\caption{
The (a) $V$ light curve and (b) $(B-V)_0$ color curves
of PW~Vul as well as those of LV~Vul, V1500~Cyg, and V1668~Cyg.
In panel (a), we show the PW~Vul model light curve (thin solid red lines)
of a $0.85~M_\sun$ WD \citep[CO4,][]{hac15k}.
We also add a $0.98~M_\sun$ WD model (CO3, thin solid blue lines)
for V1668~Cyg.
\label{pw_vul_lv_vul_v1668_cyg_v1500_cyg_v_bv_ub_color_logscale_no2}}
\end{figure}


\begin{figure}
\plotone{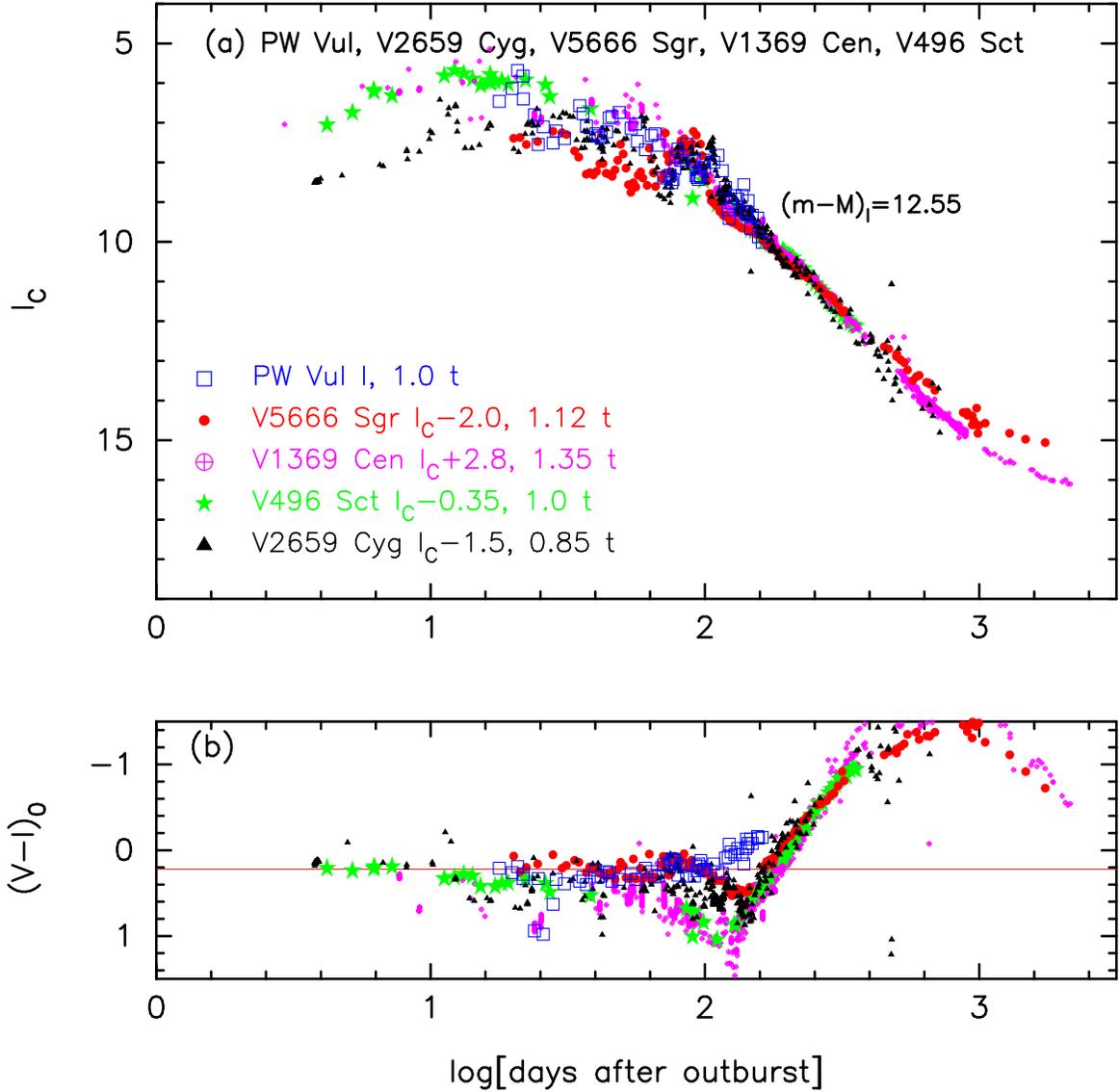}
\caption{
The (a) $I$ light curve and (b) $(V-I)_0$ color curve of PW~Vul
as well as those of V5666~Sgr, V1369~Cen, V496~Sct, and V2659~Cyg.
The $UBVI$ data of PW~Vul are taken from \citet{rob95}.
\label{pw_vul_v2659_cyg_v5666_sgr_v1369_cen_v496_sct_lv_vul_i_vi_color_logscale}}
\end{figure}


\begin{figure*}
\plottwo{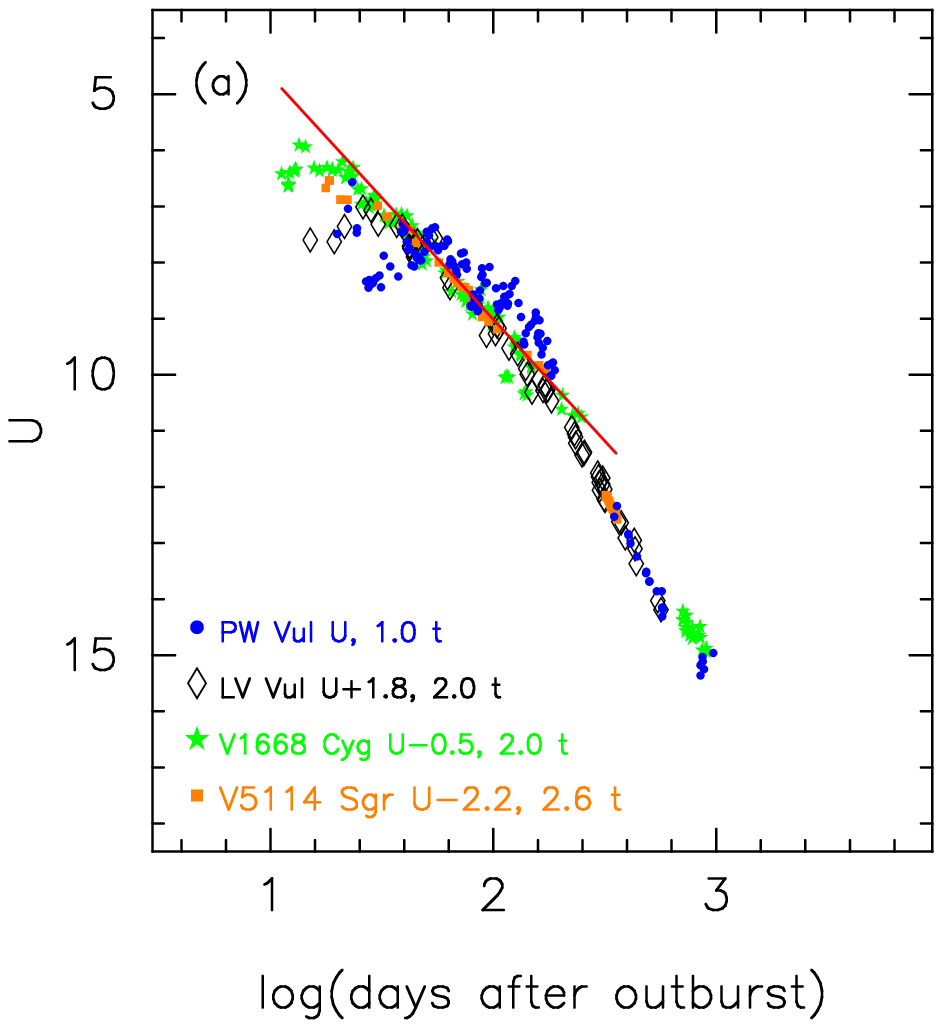}{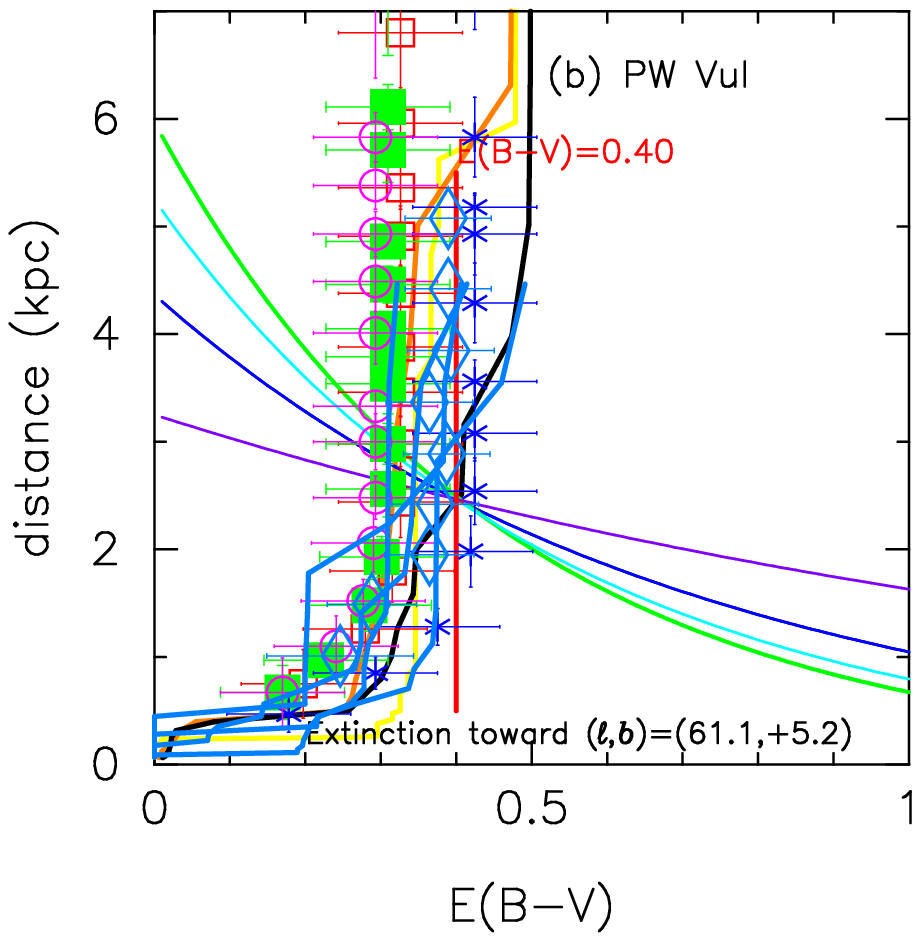}
\caption{
(a) The $U$ light curve of PW~Vul is plotted together with
those of LV~Vul, V1668~Cyg, and V5114~Sgr.  
(b) Various distance-reddening relations toward PW~Vul.  The four thin lines
of green, cyan, blue, and blue-magenta denote the distance-reddening
relations given by $(m-M)_U= 13.88$, $(m-M)_B= 13.58$, $(m-M)_V= 13.2$, 
and $(m-M)_I= 12.56$, respectively.
\label{distance_reddening_pw_vul_xxxxxx}}
\end{figure*}

\subsection{PW~Vul 1984\#1}
\label{pw_vul_ubvi}
We have reanalyzed the $UBVI$ multi-band light/color curves of PW~Vul
based on the time-stretching method.  
We first obtain the distance moduli in $BVI$ bands.
Figure 
\ref{pw_vul_v2659_cyg_v5114_sgr_v1668_cyg_lv_vul_b_ub_color_logscale}
shows the (a) $B$ light and (b) $(U-B)_0$ color curves of PW~Vul
as well as LV~Vul, V1668~Cyg, V5114~Sgr, and V2659~Cyg. 
For the $B$ band, we apply Equation (7) of \citet{hac19ka} to Figure 
\ref{pw_vul_v2659_cyg_v5114_sgr_v1668_cyg_lv_vul_b_ub_color_logscale}(a)
and obtain
\begin{eqnarray}
(m&-&M)_{B, \rm PW~Vul} \cr
&=& ((m - M)_B + \Delta B)_{\rm LV~Vul} - 2.5 \log 2.0 \cr
&=& 12.45 + 1.9\pm0.2 - 0.75 = 13.6\pm0.2 \cr
&=& ((m - M)_B + \Delta B)_{\rm V1668~Cyg} - 2.5 \log 2.0 \cr
&=& 14.9 - 0.55\pm0.2 - 0.75 = 13.6\pm0.2 \cr
&=& ((m - M)_B + \Delta B)_{\rm V5114~Sgr} - 2.5 \log 2.6 \cr
&=& 16.85 - 2.2\pm0.2 - 1.05 = 13.6\pm0.2 \cr
&=& ((m - M)_B + \Delta B)_{\rm V2659~Cyg} - 2.5 \log 0.85 \cr
&=& 15.85 - 2.45\pm0.2 + 0.175 = 13.58\pm0.2, 
\label{distance_modulus_b_ub_pw_vul}
\end{eqnarray}
where we adopt $(m-M)_{B, \rm LV~Vul}= 12.45$ 
and $(m-M)_{B, \rm V1668~Cyg}= 14.9$ both from \citet{hac19ka},
$(m-M)_{B, \rm V5114~Sgr}=16.85$ in Appendix \ref{v5114_sgr_ubvi},
$(m-M)_{B, \rm V2659~Cyg}=15.85$ in Appendix \ref{v2659_cyg_ubvi}.
Here, we adopt $E(B-V)= 0.40$ in order to overlap the $(U-B)_0$ color
of PW~Vul with those of LV~Vul, V1668~Cyg, V5114~Sgr, and V2659~Cyg, 
and obtain $\log f_{\rm s} = +0.30$ against the timescale of
LV~Vul in oder to overlap the $B$ light and $U-B$ color curves
with these five novae as much as possible by eye.  We will check
the set of $E(B-V)= 0.40$ and $\log f_{\rm s} = +0.30$ below.
Thus, we obtain $(m-M)_{B, \rm PW~Vul}= 13.60\pm0.2$.

Figure 
\ref{pw_vul_lv_vul_v1668_cyg_v1500_cyg_v_bv_ub_color_logscale_no2}
shows the (a) $V$ light and (b) $(B-V)_0$ color curves of PW~Vul
as well as LV~Vul, V1500~Cyg, and V1668~Cyg.  Here, we adopt
the set of $E(B-V)= 0.40$ and $\log f_{\rm s} = +0.30$ after
the $B$ light and $(U-B)_0$ color curve analysis mentioned above.
We apply Equation (4) of \citet{hac19ka} to Figure 
\ref{pw_vul_lv_vul_v1668_cyg_v1500_cyg_v_bv_ub_color_logscale_no2}(a)
and obtain
\begin{eqnarray}
(m&-&M)_{V, \rm PW~Vul} \cr
&=& ((m - M)_V + \Delta V)_{\rm LV~Vul} - 2.5 \log 2.0 \cr
&=& 11.85 + 2.1\pm0.2 - 0.75 = 13.2\pm0.2 \cr
&=& ((m - M)_V + \Delta V)_{\rm V1500~Cyg} - 2.5 \log 3.8 \cr
&=& 12.15 + 2.5\pm0.2 - 1.45 = 13.2\pm0.2 \cr
&=& ((m - M)_V + \Delta V)_{\rm V1668~Cyg} - 2.5 \log 2.0 \cr
&=& 14.6 - 0.65\pm0.2 - 0.75 = 13.2\pm0.2, 
\label{distance_modulus_v_pw_vul_v1500_cyg_v1668_cyg_lv_vul}
\end{eqnarray}
where we adopt $(m-M)_{V, \rm LV~Vul}=11.85$ and
$(m-M)_{V, \rm V1668~Cyg}=14.6$ both from \citet{hac19ka},
and $(m-M)_{V, \rm V1500~Cyg}=12.15$ in Appendix \ref{v1500_cyg_ubvi}.
Thus, we obtain $(m-M)_{V, \rm PW~Vul}= 13.2\pm0.1$.

Figure
\ref{pw_vul_v2659_cyg_v5666_sgr_v1369_cen_v496_sct_lv_vul_i_vi_color_logscale}
shows the (a) $I$ light and (b) $(V-I)_0$ color curves of PW~Vul as well as
V5666~Sgr, V1369~Cen, V496~Sct, and V2659~Cyg.
We apply Equation (8) of \citet{hac19ka} for the $I$ band to Figure
\ref{pw_vul_v2659_cyg_v5666_sgr_v1369_cen_v496_sct_lv_vul_i_vi_color_logscale}(a)
and obtain
\begin{eqnarray}
(m&-&M)_{I, \rm PW~Vul} \cr
&=& ((m - M)_I + \Delta I_{\rm C})
_{\rm V5666~Sgr} - 2.5 \log 1.12 \cr
&=& 14.7 - 2.0\pm0.3 - 0.13 = 12.57\pm0.3 \cr
&=& ((m - M)_I + \Delta I_{\rm C})
_{\rm V1369~Cen} - 2.5 \log 1.35 \cr
&=& 10.11 + 2.8\pm0.3 - 0.33 = 12.58\pm0.3 \cr
&=& ((m - M)_I + \Delta I_{\rm C})
_{\rm V496~Sct} - 2.5 \log 1.0 \cr
&=& 12.9 - 0.35\pm0.3 - 0.0 = 12.55\pm0.3 \cr
&=& ((m - M)_I + \Delta I_{\rm C})
_{\rm V2659~Cyg} - 2.5 \log 0.85 \cr
&=& 13.9 - 1.5\pm0.3 + 0.18 = 12.58\pm0.3, 
\label{distance_modulus_i_v2659_cyg_v5666_sgr_v1369_cen_v496_sct}
\end{eqnarray}
where we adopt
$(m-M)_{I, \rm V5666~Sgr}=14.7$ in Appendix \ref{v5666_sgr_bvi}, 
$(m-M)_{I, \rm V1369~Cen}=10.11$ from \citet{hac19ka} and
$(m-M)_{I, \rm V496~Sct}=12.9$ in Appendix \ref{v496_sct_bvi}.
$(m-M)_{I, \rm V2659~Cyg}=13.9$ in Appendix \ref{v2659_cyg_ubvi}.
Thus, we obtain $(m-M)_{I, \rm PW~Vul}= 12.57\pm0.2$.

Using the timescaling factor of $\log f_{\rm s}= +0.30$, we plot
the $U$ band light curves of PW~Vul as well as LV~Vul, V1668~Cyg, 
and V5114~Sgr in Figure \ref{distance_reddening_pw_vul_xxxxxx}(a).
We apply Equation (6) of \citet{hac19ka} for the $U$ band to Figure
\ref{distance_reddening_pw_vul_xxxxxx}(a) and obtain
\begin{eqnarray}
(m&-&M)_{U, \rm PW~Vul} \cr
&=& ((m - M)_U + \Delta U)_{\rm LV~Vul} - 2.5 \log 2.0 \cr
&=& 12.85 + 1.8\pm0.2 - 0.75 = 13.9\pm0.2 \cr
&=& ((m - M)_U + \Delta U)_{\rm V1668~Cyg} - 2.5 \log 2.0 \cr
&=& 15.1 - 0.5\pm0.2 - 0.75 = 13.85\pm0.2 \cr 
&=& ((m - M)_U + \Delta U)_{\rm V5114~Sgr} - 2.5 \log 2.6 \cr
&=& 17.15 - 2.2\pm0.2 - 1.05 = 13.9\pm0.2, 
\label{distance_modulus_u_pw_vul}
\end{eqnarray}
where we adopt $(m-M)_{U, \rm LV~Vul}= 12.85$ and
$(m-M)_{U, \rm V1668~Cyg}= 15.10$ both from \citet{hac19ka}, and
$(m-M)_{U, \rm V5114~Sgr}= 17.15$ in Appendix \ref{v5114_sgr_ubvi}. 
Thus, we obtain $(m-M)_{U, \rm PW~Vul}= 13.88\pm0.2$.   

Figure \ref{distance_reddening_pw_vul_xxxxxx}(b) depicts various
distance-reddening relations. 
We plot the four distance moduli in $U$, $B$, $V$, and 
$I$ bands by the thin solid lines of green, cyan, blue, and blue-magenta,
respectively.
These four lines cross at $d=2.46\pm0.2$~kpc and $E(B-V)=0.40\pm0.05$.
The crossing point is close to the distance-reddening relation
(thick black line) of \citet{gre15}, that (blue asterisks) of 
\citet{mar06}, and those (thick solid cyan-blue lines) 
given by \citet{chen19}.  Here,
we add the four thick cyan-blue lines of \citet{chen19}, which correspond
to four nearby directions toward PW~Vul, i.e., the galactic coordinates of
$(\ell, b)= (61\fdg05,  +5\fdg15)$, $(61\fdg05,  +5\fdg25)$,
$(61\fdg15,  +5\fdg15)$, and $(61\fdg15,  +5\fdg25)$.

We finally examine our set of $E(B-V)= 0.40$, $(m-M)_B= 13.6$,
$(m-M)_V= 13.2$, $(m-M)_I= 12.55$, and $\log f_{\rm s}= +0.30$ in the three 
time-stretched color-magnitude diagrams of Figures
\ref{hr_diagram_pw_vul_v1419_aql_v382_vel_v5114_sgr_outburst_ub}(a),
\ref{hr_diagram_pw_vul_v382_vel_v5117_sgr_v2362_cyg_outburst}(a) and 
\ref{hr_diagram_pw_vul_v382_vel_v5117_sgr_v2362_cyg_outburst_vi}(a).
The track of PW~Vul overlaps well with that of the LV~Vul template track
(orange line) in the $(U-B)_0$-$(M_B-2.5 \log f_{\rm s})$ diagram,
that is, Figures 
\ref{hr_diagram_pw_vul_v1419_aql_v382_vel_v5114_sgr_outburst_ub}(a),
in the $(B-V)_0$-$(M_V-2.5 \log f_{\rm s})$ diagram, that is,
Figure \ref{hr_diagram_pw_vul_v382_vel_v5117_sgr_v2362_cyg_outburst}(a). 
This supports the above set of values.
The $(V-I)_0$-$(M_I-2.5 \log f_{\rm s})$ diagram of PW~Vul is already
discussed in Section \ref{pw_vul_vi}.
The track of PW~Vul almost follows the reconstructed template track of
V2615~Oph (cyan line) in Figure
\ref{hr_diagram_pw_vul_v382_vel_v5117_sgr_v2362_cyg_outburst_vi}(a), that is,
in the $(V-I)_0$-$(M_I-2.5 \log f_{\rm s})$ diagram.
This may also support the reddening value of
$E(B-V)= 0.40$, the distance modulus of $(m-M)_I= 12.55$, and 
the timescaling factor of $\log f_{\rm s}= +0.30$.



\begin{figure}
\plotone{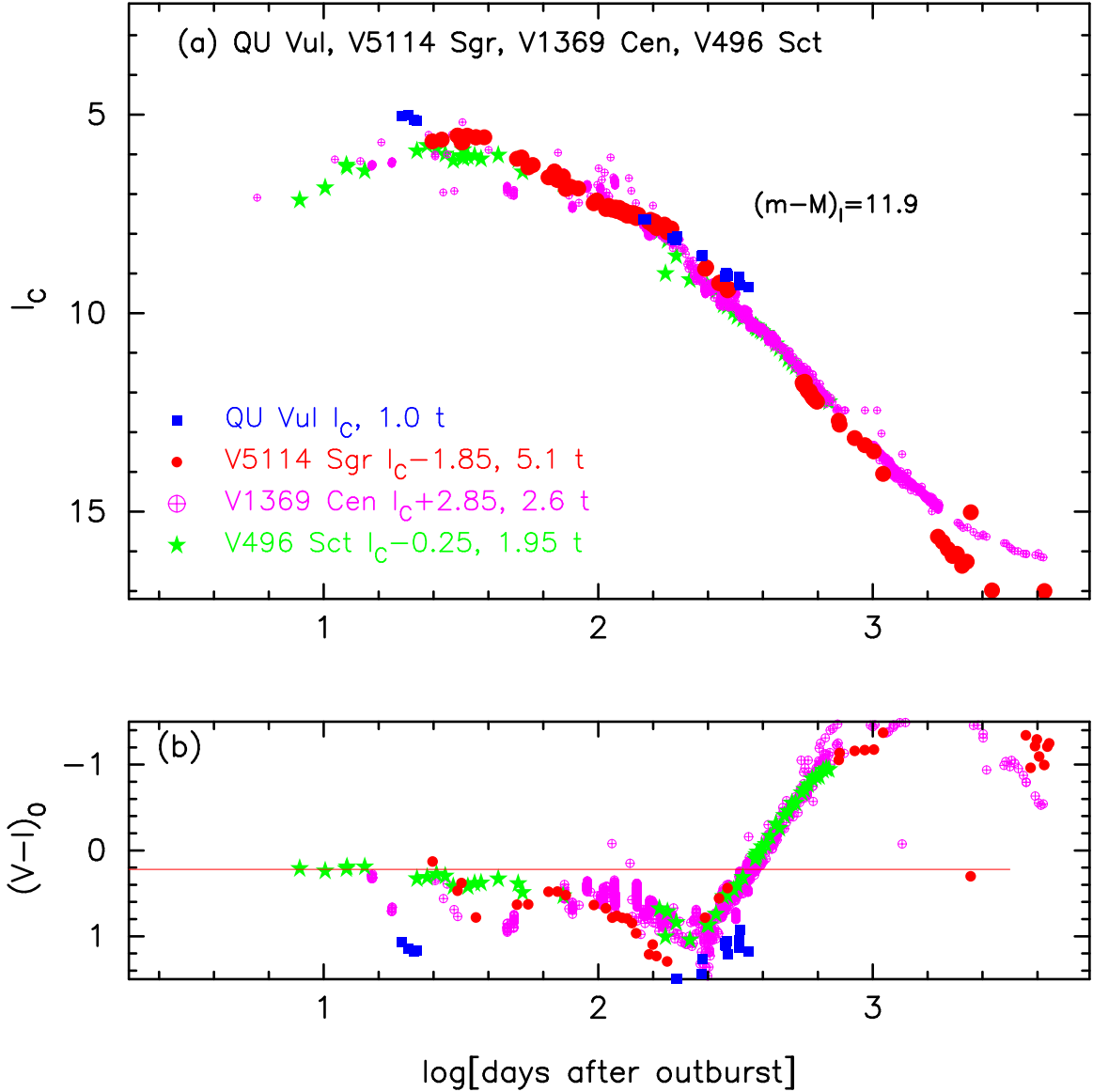}
\caption{
The (a) $I_{\rm C}$ light curve and (b) $(V-I_{\rm C})_0$ color curve
of QU~Vul as well as those of V5114~Sgr, V1369~Cen, and V496~Sct.
The $UBVI_{\rm C}$ data of QU~Vul are taken from \citet{ber88}.
\label{qu_vul_v5114_sgr_v1369_cen_v496_sct_i_vi_color_logscale}}
\end{figure}


\begin{figure*}
\gridline{\fig{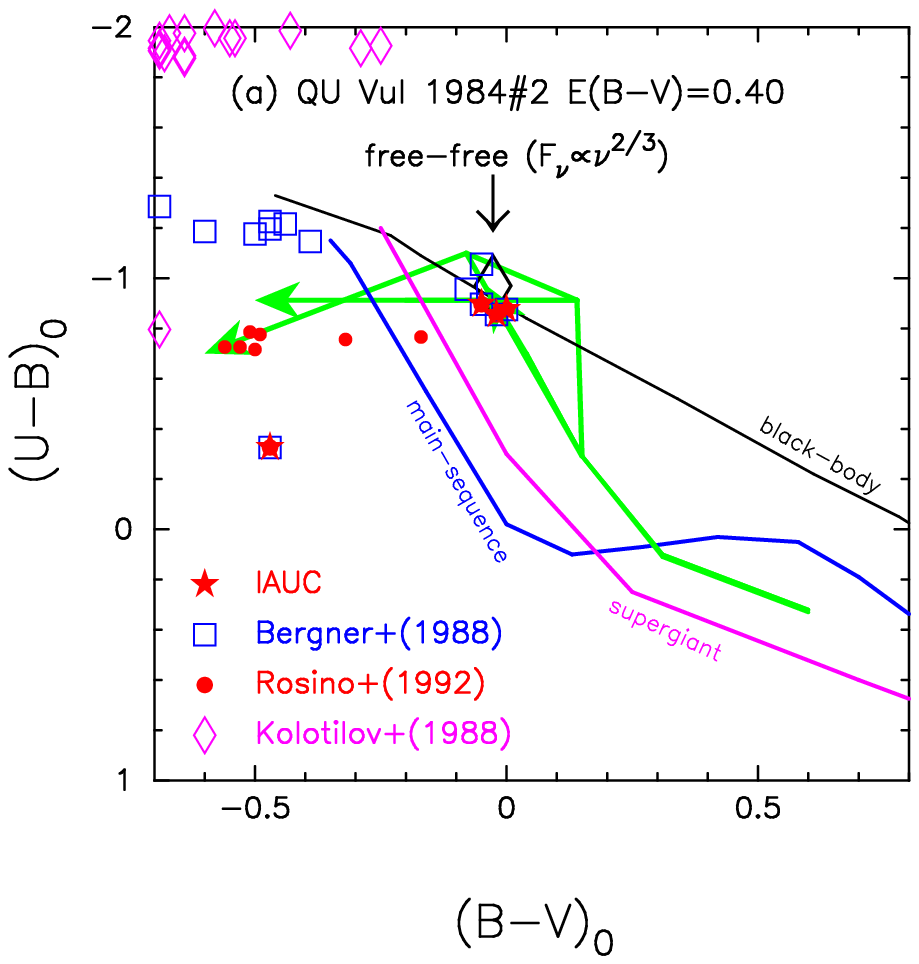}{0.4\textwidth}{(a)}
          \fig{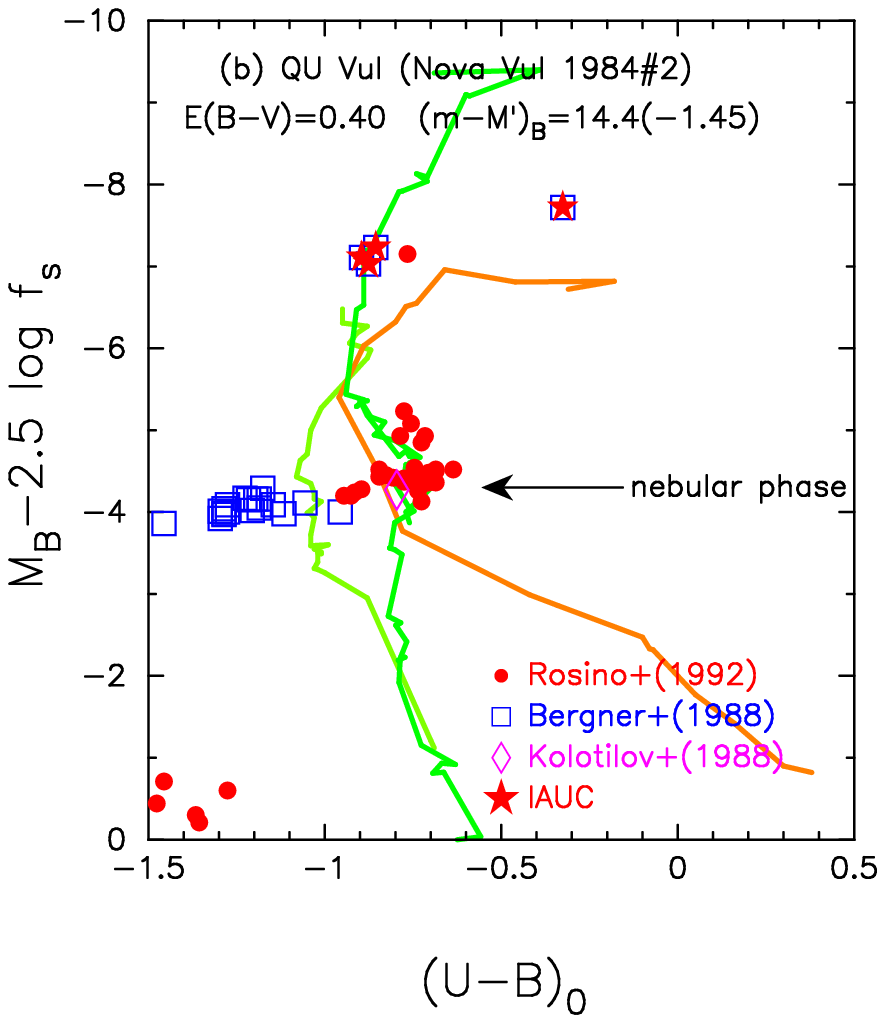}{0.4\textwidth}{(b)}
          }
\gridline{\fig{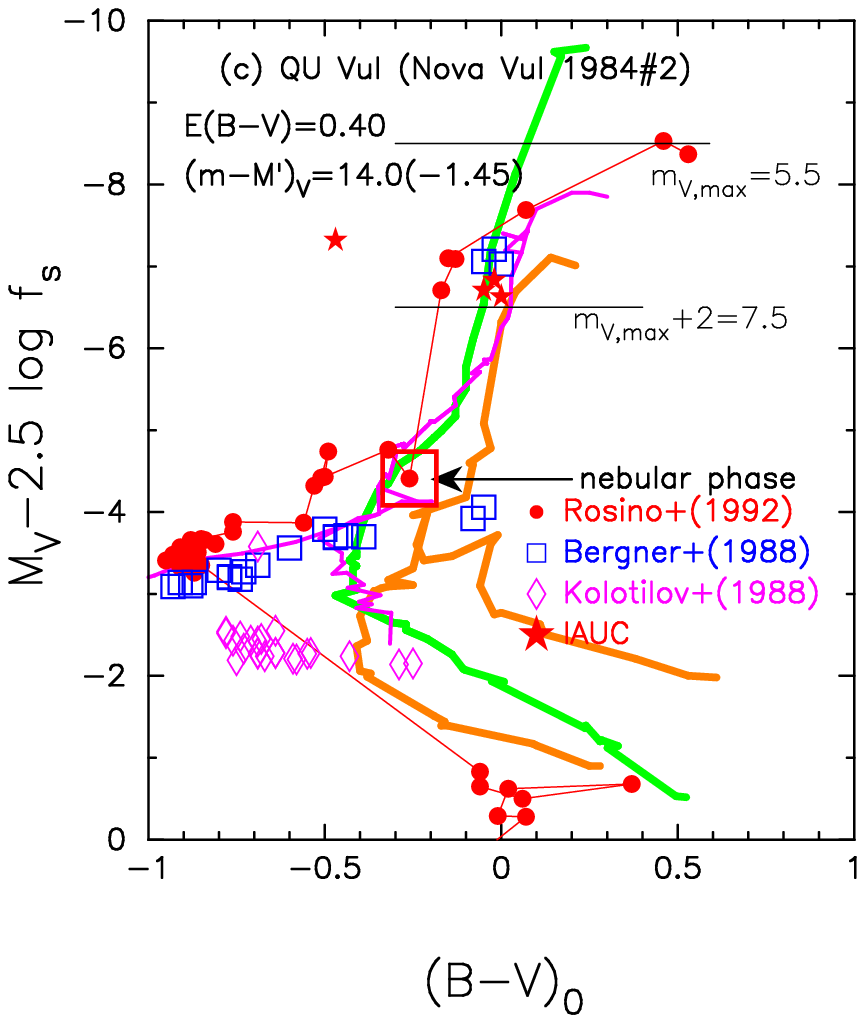}{0.4\textwidth}{(c)}
          \fig{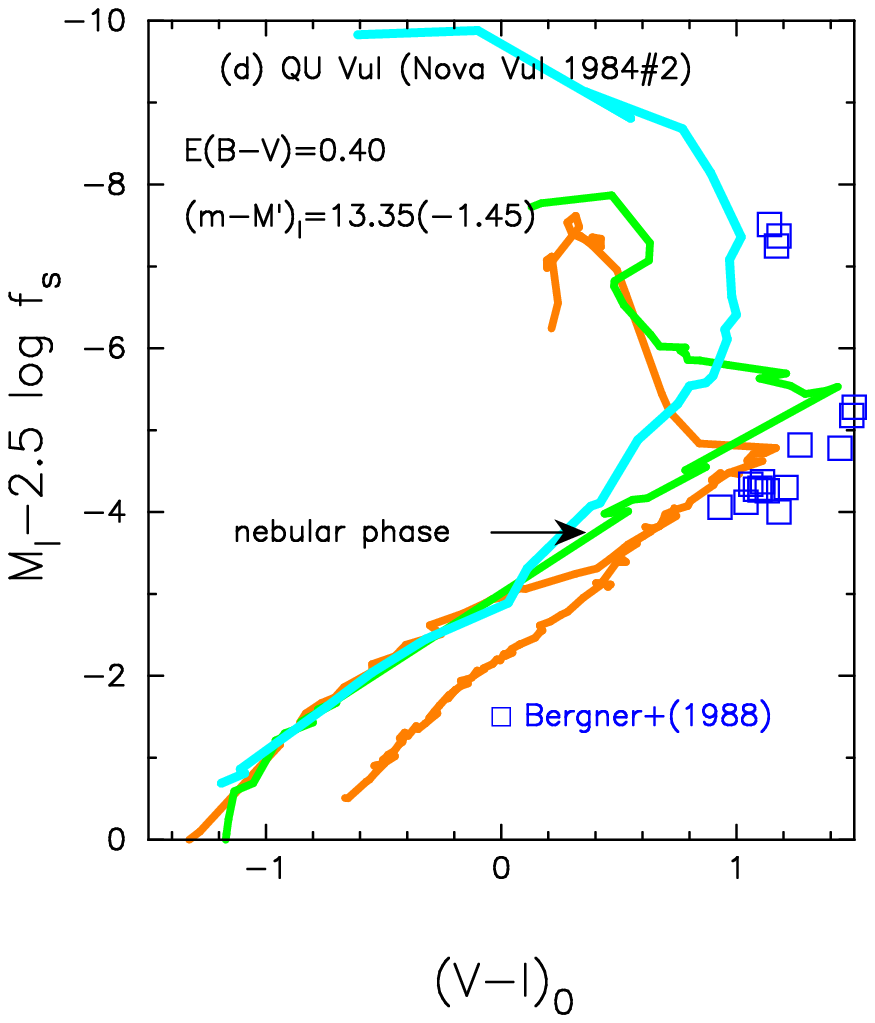}{0.4\textwidth}{(d)}
          }
\caption{The
(a) $(B-V)_0$-$(U-B)_0$ color-color diagram, 
(b) time-stretched $(U-B)_0$-$(M_B-2.5 \log f_{\rm s})$
color-magnitude diagram,
(c) $(B-V)_0$-$(M_V-2.5 \log f_{\rm s})$ diagram,
and (d) $(V-I)_0$-$(M_I-2.5 \log f_{\rm s})$ diagram of QU~Vul.
\label{color_magnitude_diagram_qu_vul}}
\end{figure*}

\subsection{QU~Vul 1984\#2}
\label{qu_vul_ubvi}
The $V$ light and $B-V$ color curves of QU~Vul were studied in \citet{hac16k},
but we do not discuss the time-stretched color-magnitude diagrams 
in the main text of the present work.  This nova shows a peculiar behavior 
in the color-magnitude diagram so that we analyze the 
$UBVI_{\rm C}$ multi-band light/color curves of QU~Vul in this Appendix.
Figure \ref{qu_vul_v5114_sgr_v1369_cen_v496_sct_i_vi_color_logscale}
shows the $I_{\rm C}$ light and $V-I_{\rm C}$ color curves of QU~Vul
as well as V5114~Sgr, V1369~Cen, and V496~Sct.
The $UBVI_{\rm C}$ data of QU~Vul are taken from \citet{ber88}.
Adopting the color excess of $E(B-V)= 0.40$ as mentioned below,
we confirm the timescaling factor $\log f_{\rm s}= +0.59$ for QU~Vul
as shown in Figure 
\ref{qu_vul_v5114_sgr_v1369_cen_v496_sct_i_vi_color_logscale}(b).
We apply Equation (8) of \citet{hac19ka} for the $I$ band to Figure
\ref{qu_vul_v5114_sgr_v1369_cen_v496_sct_i_vi_color_logscale}(a)
and obtain
\begin{eqnarray}
(m&-&M)_{I, \rm QU~Vul} \cr
&=& ((m - M)_I + \Delta I_{\rm C})
_{\rm V5114~Sgr} - 2.5 \log 5.1 \cr
&=& 15.55 - 1.85\pm0.2 - 1.78 = 11.92\pm0.2 \cr
&=& ((m - M)_I + \Delta I_{\rm C})
_{\rm V1369~Cen} - 2.5 \log 2.6 \cr
&=& 10.11 + 2.85\pm0.2 - 1.05 = 11.91\pm0.2 \cr
&=& ((m - M)_I + \Delta I_{\rm C})
_{\rm V496~Sct} - 2.5 \log 1.95 \cr
&=& 12.9 - 0.25\pm0.2 - 0.725 = 11.92\pm0.2,
\label{distance_modulus_i_vi_qu_vul}
\end{eqnarray}
where we adopt
$(m-M)_{I, \rm V5114~Sgr}=15.55$ from Appendix \ref{v5114_sgr_ubvi},
$(m-M)_{I, \rm V1369~Cen}=10.11$ from \citet{hac19ka}, and
$(m-M)_{I, \rm V496~Sct}=12.9$ in Appendix \ref{v496_sct_bvi}.
Thus, we obtain $(m-M)_{I, \rm QU~Vul}= 11.92\pm0.2$.


\citet{hac16k} obtained $E(B-V)= 0.55$, $(m-M)_V= 13.6$, $d= 2.4$~kpc,
and $\log f_{\rm s}= +0.33$.  We reanalyzed the $V$ light and $B-V$ color
curves and obtained a new set of parameters, that is,
$E(B-V)= 0.40$, $(m-M)_V= 12.55$, $d= 1.83$~kpc, and $\log f_{\rm s}= +0.59$. 
We plot the $(B-V)_0$-$(U-B)_0$ color-color diagram in Figure
\ref{color_magnitude_diagram_qu_vul}(a).
We have $(m-M')_B= 12.55 + 0.40 + 2.5\times 0.59 = 14.4$,
$(m-M')_V= 12.55 + 2.5\times 0.59 = 14.0$, and
$(m-M')_I= 11.9 + 2.5\times 0.59 = 13.35$, and plot the
$(U-B)_0$-$(M_B-2.5\log f_{\rm s})$,
$(B-V)_0$-$(M_V-2.5\log f_{\rm s})$,
$(V-I)_0$-$(M_I-2.5\log f_{\rm s})$ diagrams in Figure
\ref{color_magnitude_diagram_qu_vul}(b)(c)(d).
The $UBV$ data of QU~Vul are taken from IAU Circular No.4033,
\citet{ber88}, \citet{kol88}, and \citet{ros92}.
In the $(B-V)_0$-$(U-B)_0$ diagram of Figure
\ref{color_magnitude_diagram_qu_vul}(a), color-color data
are rather different among various observers.  
Only the data of \citet{ros92} almost follow 
the template track of nova-giant sequence 
\citep[green lines with an arrow;][]{hac14k}.
In the $(U-B)_0$-$(M_B-2.5\log f_{\rm s})$ diagram of Figure
\ref{color_magnitude_diagram_qu_vul}(b), various data
broadly follow the template tracks of LV~Vul (orange line)
and V1500~Cyg (green line).  
Also in the $(B-V)_0$-$(M_V-2.5\log f_{\rm s})$ diagram of Figure
\ref{color_magnitude_diagram_qu_vul}(c),
the data points broadly follow the template tracks of V1500~Cyg (green line)
or V1974~Cyg (magenta line).  
In the $(V-I)_0$-$(M_I-2.5\log f_{\rm s})$ diagram of Figure
\ref{color_magnitude_diagram_qu_vul}(d), the data of \citet{ber88}
are located slightly below but broadly follow the template track of
V496~Sct/V959~Mon (orange line) or V5114~Sgr (green line). 

The three distance-reddening lines
of $(m-M)_B=12.95$, $(m-M)_V=12.55$, and $(m-M)_I=11.91$ consistently 
cross at $E(B-V)=0.40$ and $d=1.83$~kpc, although they are not shown. 
The Gaia DR2 distance determination suggests a distance of
$1.786^{+3.495}_{-0.196}$~kpc in the ``Bronze sample'' of \citet{schaefer18}.
This is consistent with our distance estimate of $d= 1.83\pm0.2$~kpc.
Thus, we consistently obtain $E(B-V)= 0.40$, $(m-M)_V= 12.55$,
$d= 1.83$~kpc, $(m-M)_I= 11.9$, and $\log f_{\rm s}= +0.59$ for QU~Vul.


\begin{figure}
\plotone{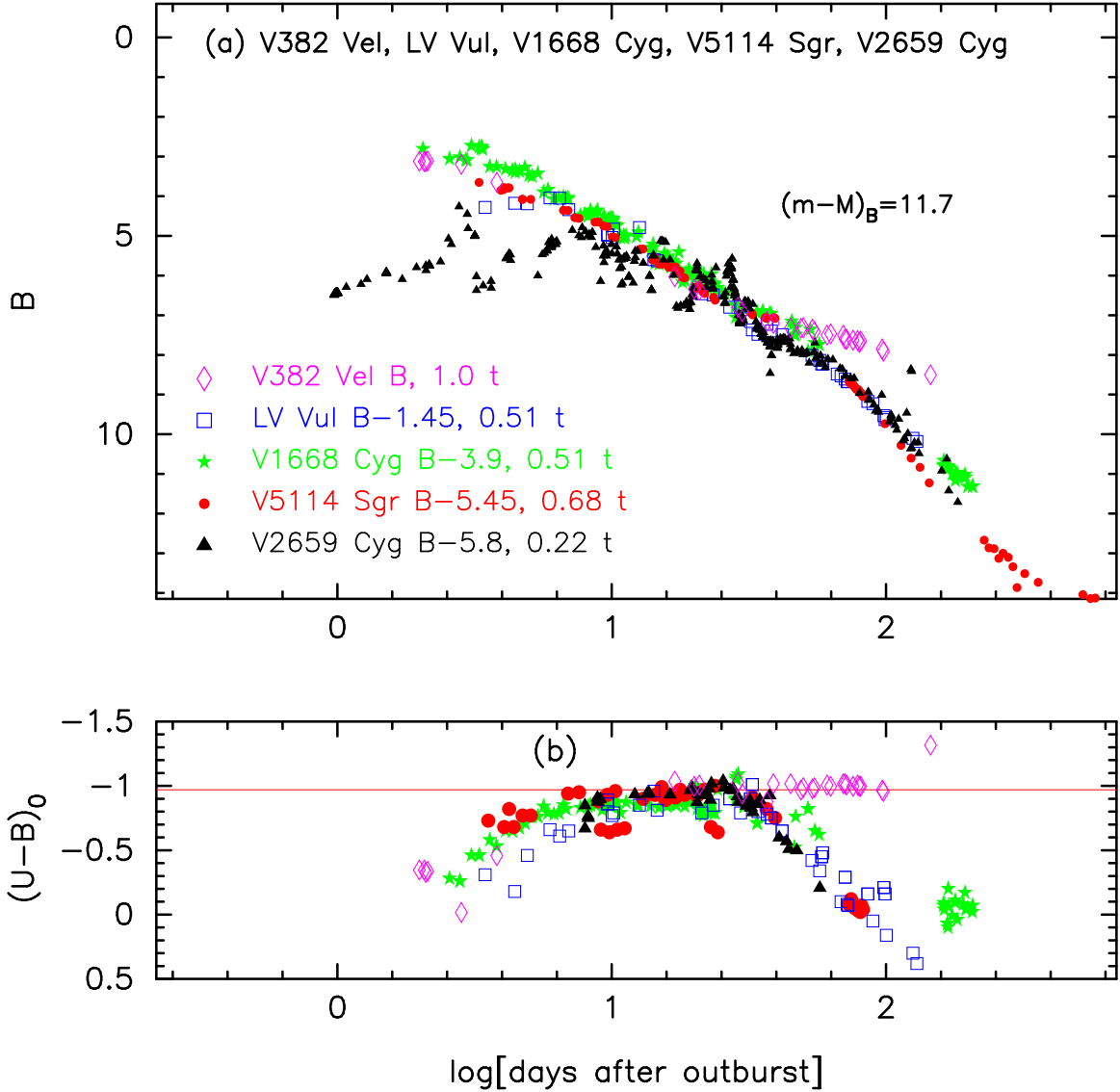}
\caption{
The (a) $B$ light curve and (b) $(U-B)_0$ color curve of V382~Vel 
as well as those of LV~Vul, V1668~Cyg, V5114~Sgr, and V2659~Cyg.
The $UBVI$ data of V382~Vel are taken from 
IAU Circular No. 7176, 7179, 7196, 7209, 7216, 7226, 7232, 7238, and 7277.
\label{v382_vel_v2659_cyg_v5114_sgr_v1668_cyg_lv_vul_b_ub_color_logscale}}
\end{figure}


\begin{figure}
\plotone{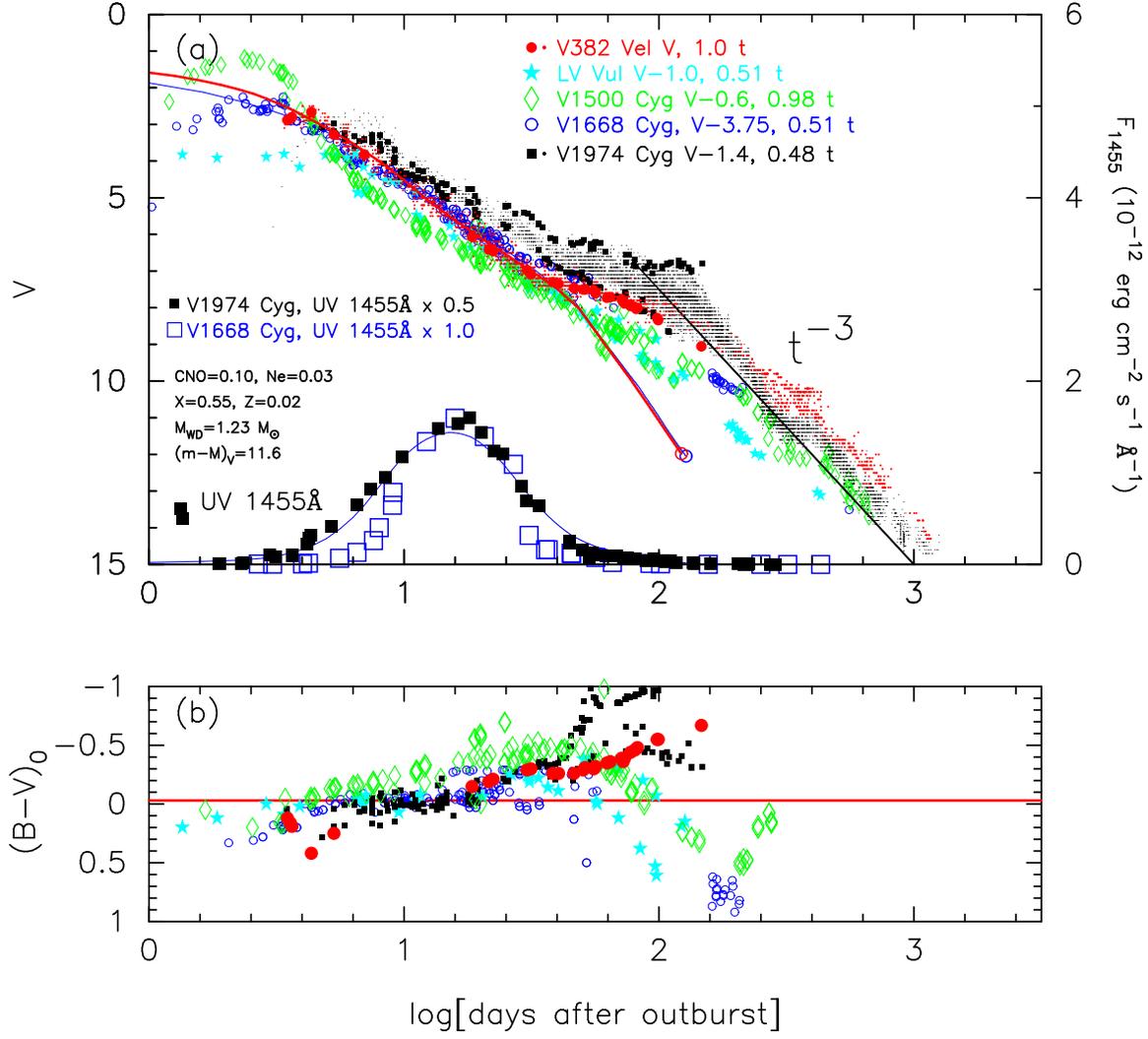}
\caption{
The (a) $V$ light and (b) $(B-V)_0$ color curves of V382~Vel as well as
those of LV~Vul, V1500~Cyg, V1668~Cyg, and V1974~Cyg.
In panel (a), we show the V382~Vel model light curve (thin solid red lines)
of a $1.23~M_\sun$ WD (Ne2).  We also add a $0.98~M_\sun$ WD model 
(CO3, thin solid blue lines) for V1668~Cyg.
\label{v382_vel_v1500_cyg_lv_vul_v1668_cyg_v1974_cyg_v_bv_logscale_no2}}
\end{figure}


\begin{figure}
\plotone{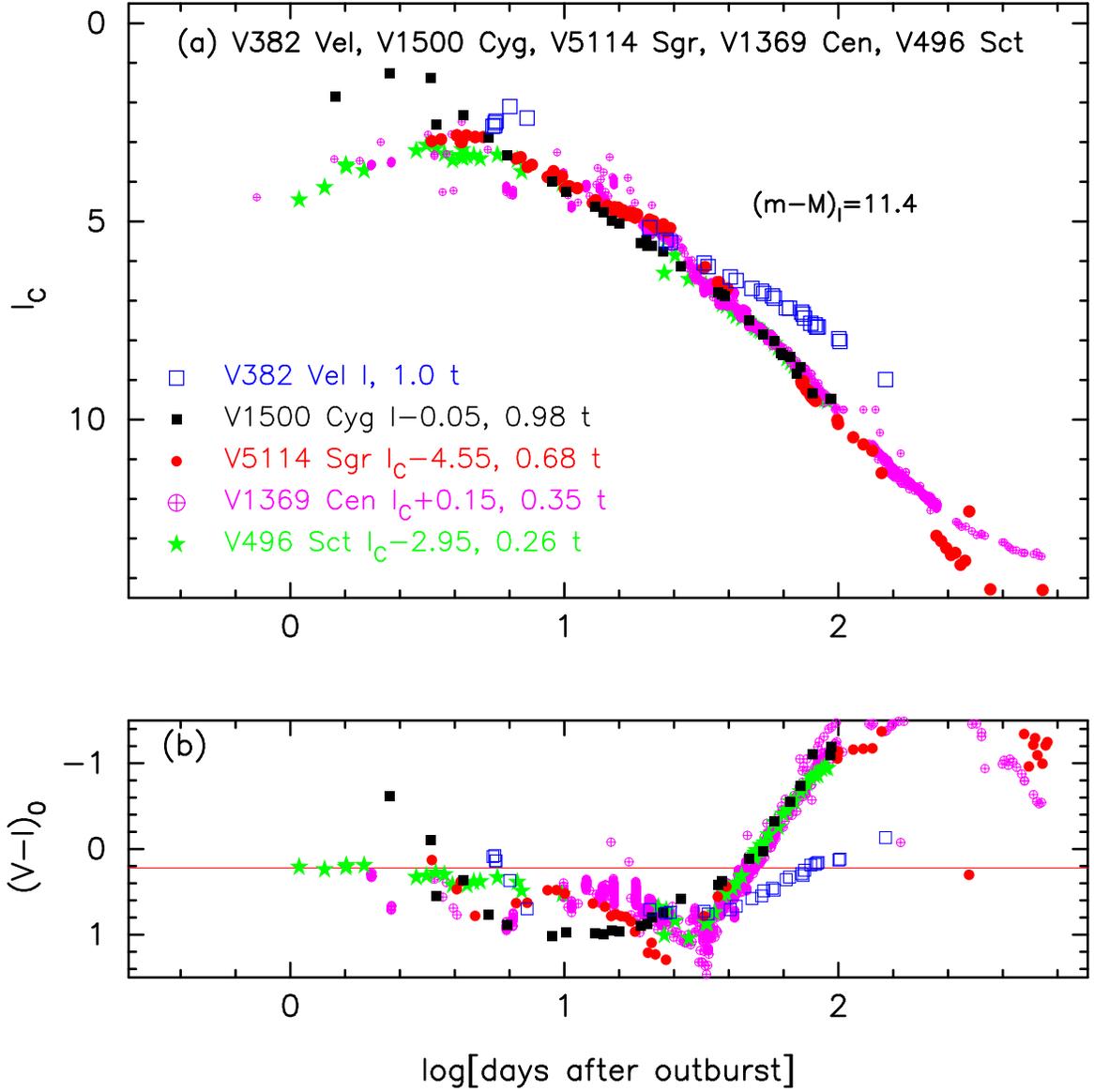}
\caption{
The (a) $I$ light curve and (b) $(V-I)_0$ color curve of V382~Vel
as well as those of V1500~Cyg, V5114~Sgr, V1369~Cen, and V496~Sct.
\label{v382_vel_v5114_sgr_v1369_cen_v496_sct_v1500_cyg_i_vi_color_logscale}}
\end{figure}


\begin{figure*}
\plottwo{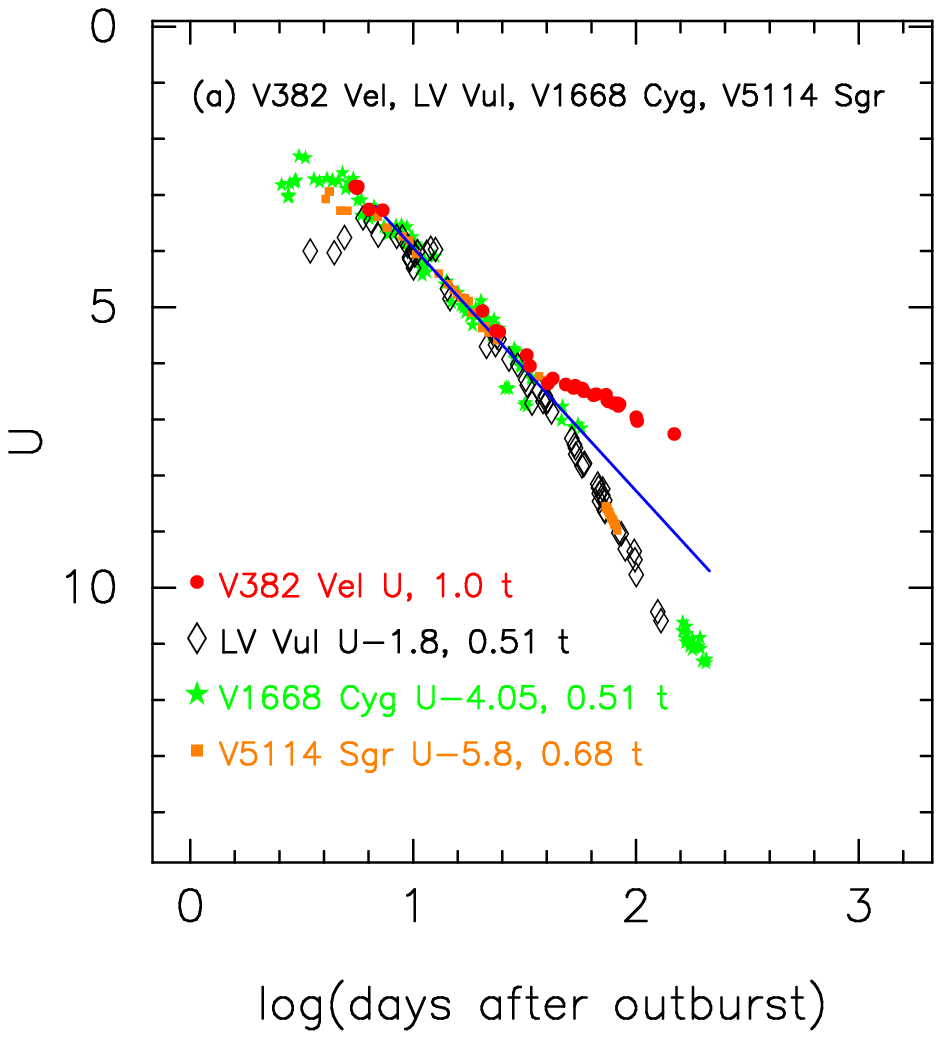}{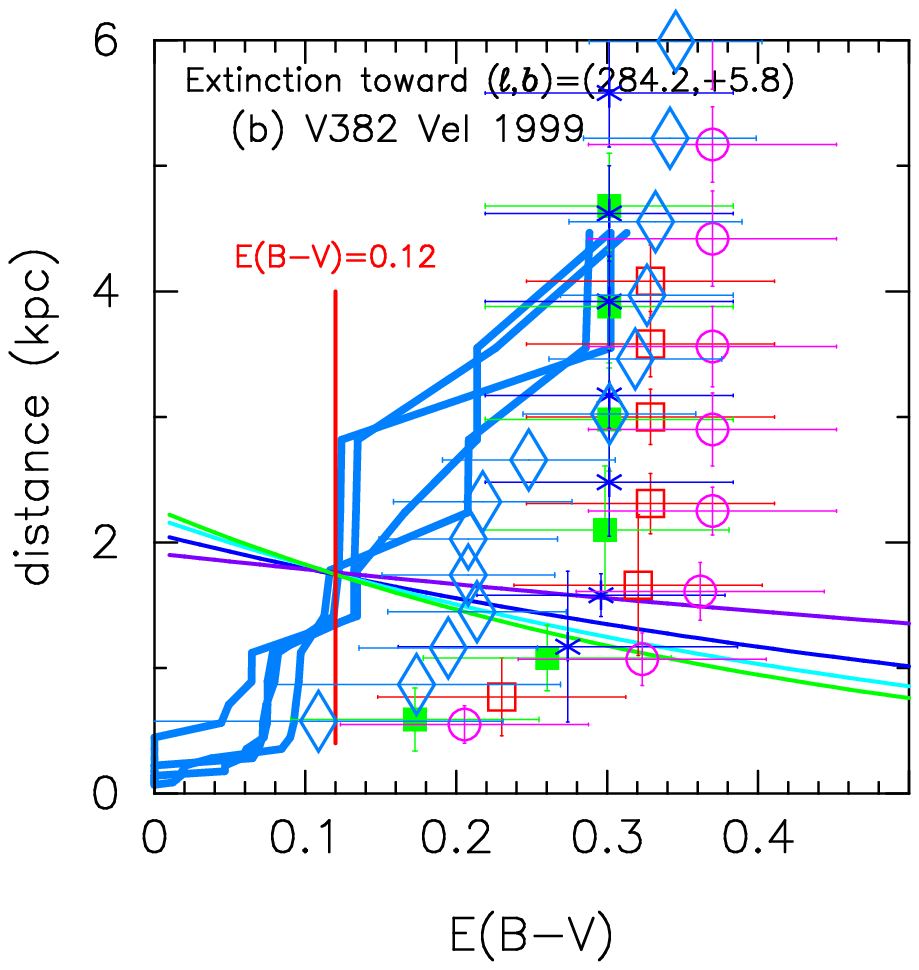}
\caption{
(a) The $U$ light curves of V382~Vel as well as
those of LV~Vul, V1668~Cyg, and V5114~Sgr.  
(b) Various distance-reddening relations toward V382~Vel.
The thin solid lines of green, cyan, blue, and blue-magenta 
denote the distance-reddening relations given by 
$(m-M)_U= 11.78$, $(m-M)_B= 11.71$, $(m-M)_V= 11.58$, 
and $(m-M)_I= 11.41$, respectively.
\label{color_color_distance_reddening_v382_vel_xxxxxx}}
\end{figure*}

\subsection{V382~Vel 1999}
\label{v382_vel_ubvi}
We reanalyze the $UBVI$ multi-band light/color curves of V382~Vel based
on the time-stretching method.  We adopt $\log f_{\rm s}= -0.29$ after
\citet{hac16k}.  We first obtain the distance moduli in $UBVI$ bands.
Figure 
\ref{v382_vel_v2659_cyg_v5114_sgr_v1668_cyg_lv_vul_b_ub_color_logscale}
shows the (a) $B$ light and (b) $(U-B)_0$ color curves of V382~Vel
as well as LV~Vul, V1668~Cyg, V5114~Sgr, and V2659~Cyg, 
where we use $E(B-V)= 0.12$ to deredden the $U-B$ color as mentioned below.  
For the $B$ band, we apply Equation (7) of \citet{hac19ka} to Figure 
\ref{v382_vel_v2659_cyg_v5114_sgr_v1668_cyg_lv_vul_b_ub_color_logscale}(a)
and obtain
\begin{eqnarray}
(m&-&M)_{B, \rm V382~Vel} \cr
&=& ((m - M)_B + \Delta B)_{\rm LV~Vul} - 2.5 \log 0.51 \cr
&=& 12.45 - 1.45\pm0.2 + 0.725 = 11.72\pm0.2 \cr
&=& ((m - M)_B + \Delta B)_{\rm V1668~Cyg} - 2.5 \log 0.51 \cr
&=& 14.9 - 3.9\pm0.2 + 0.725 = 11.72\pm0.2 \cr
&=& ((m - M)_B + \Delta B)_{\rm V5114~Sgr} - 2.5 \log 0.68 \cr
&=& 16.85 - 5.45\pm0.2 + 0.3 = 11.7\pm0.2 \cr
&=& ((m - M)_B + \Delta B)_{\rm V2659~Cyg} - 2.5 \log 0.22 \cr
&=& 15.85 - 5.8\pm0.2 + 1.65 = 11.7\pm0.2,
\label{distance_modulus_b_ub_v382_vel}
\end{eqnarray}
where we adopt
$(m-M)_{B, \rm LV~Vul}= 12.45$ 
and $(m-M)_{B, \rm V1668~Cyg}= 14.9$ both from \citet{hac19ka},
$(m-M)_{B, \rm V5114~Sgr}=16.85$ from Appendix \ref{v5114_sgr_ubvi}, and
$(m-M)_{B, \rm V2659~Cyg}=15.85$ from Appendix \ref{v2659_cyg_ubvi}.
Here, we adopt $E(B-V)= 0.12$ in order to overlap the $(U-B)_0$ color
of V382~Vel to those of LV~Vul, V1668~Cyg, V5114~Sgr, and V2659~Cyg, 
and $\log f_{\rm s} = -0.29$ against the timescale of LV~Vul
after \citet{hac19ka}.  We will check the set of $E(B-V)= 0.12$ 
and $\log f_{\rm s} = -0.29$ below.
Thus, we obtain $(m-M)_{B, \rm V382~Vel}= 11.71\pm0.2$.

Figure 
\ref{v382_vel_v1500_cyg_lv_vul_v1668_cyg_v1974_cyg_v_bv_logscale_no2}
shows the (a) $V$ light and (b) $(B-V)_0$ color curves of V382~Vel
as well as LV~Vul, V1500~Cyg, V1668~Cyg, and V1974~Cyg.
Here, we adopt the set of $E(B-V)= 0.12$ and $\log f_{\rm s} = -0.29$
after the $B$ light and $U-B$ color curves analysis mentioned above.
We apply Equation (4) of \citet{hac19ka} to Figure 
\ref{v382_vel_v1500_cyg_lv_vul_v1668_cyg_v1974_cyg_v_bv_logscale_no2}(a)
and obtain
\begin{eqnarray}
(m&-&M)_{V, \rm V382~Vel} \cr
&=& ((m - M)_V + \Delta V)_{\rm LV~Vul} - 2.5 \log 0.51 \cr
&=& 11.85 - 1.0\pm0.2 + 0.725 = 11.58\pm0.2 \cr
&=& ((m - M)_V + \Delta V)_{\rm V1500~Cyg} - 2.5 \log 0.98 \cr
&=& 12.15 - 0.6\pm0.2 + 0.025 = 11.58\pm0.2 \cr
&=& ((m - M)_V + \Delta V)_{\rm V1668~Cyg} - 2.5 \log 0.51 \cr
&=& 14.6 - 3.75\pm0.2 + 0.725 = 11.58\pm0.2 \cr
&=& ((m - M)_V + \Delta V)_{\rm V1974~Cyg} - 2.5 \log 0.48 \cr
&=& 12.2 - 1.4\pm0.2 + 0.8 = 11.6\pm0.2, 
\label{distance_modulus_v_v382_vel_pw_vul_v1500_cyg_v1668_cyg_lv_vul}
\end{eqnarray}
where we adopt
$(m-M)_{V, \rm V1500~Cyg}=12.15$ in Section \ref{v1500_cyg_ubvi},
and $(m-M)_{V, \rm LV~Vul}=11.85$,
$(m-M)_{V, \rm V1668~Cyg}=14.6$,
$(m-M)_{V, \rm V1974~Cyg}=12.2$ from \citet{hac19ka}.
Thus, we obtain $(m-M)_{V, \rm V382~Vel}= 11.58\pm0.1$.

Figure
\ref{v382_vel_v5114_sgr_v1369_cen_v496_sct_v1500_cyg_i_vi_color_logscale}
shows the $I$ light and $(V-I)_0$ color curves of V382~Vel as well as
V1500~Cyg, V5114~Sgr, V1369~Cen, and V496~Sct.
We apply Equation (8) of \citet{hac19ka} for the $I$ band to Figure
\ref{v382_vel_v5114_sgr_v1369_cen_v496_sct_v1500_cyg_i_vi_color_logscale}(a)
and obtain
\begin{eqnarray}
(m&-&M)_{I, \rm V382~Vel} \cr
&=& ((m - M)_I + \Delta I)_{\rm V1500~Cyg} - 2.5 \log 0.98 \cr
&=& 11.42 - 0.05\pm0.2 + 0.025 = 11.4\pm0.2 \cr
&=& ((m - M)_I + \Delta I_C)_{\rm V5114~Sgr} - 2.5 \log 0.68 \cr
&=& 15.55 - 4.55\pm0.2 + 0.425 = 11.42\pm0.2 \cr
&=& ((m - M)_I + \Delta I_C)_{\rm V1369~Cen} - 2.5 \log 0.35 \cr
&=& 10.11 + 0.15\pm0.2 + 1.15 = 11.41\pm0.2 \cr
&=& ((m - M)_I + \Delta I_C)_{\rm V496~Sct} - 2.5 \log 0.26 \cr
&=& 12.9 - 2.95\pm0.2 + 1.475 = 11.42\pm0.2,
\label{distance_modulus_i_v382_vel_v2659_cyg_v5666_sgr_v1369_cen_v496_sct}
\end{eqnarray}
where we adopt
$(m-M)_{I, \rm V1500~Cyg}=11.42$ from Appendix \ref{v1500_cyg_ubvi},
$(m-M)_{I, \rm V5114~Sgr}=15.55$ from Appendix \ref{v5114_sgr_ubvi},
$(m-M)_{I, \rm V1369~Cen}=10.11$ from \citet{hac19ka}, and
$(m-M)_{I, \rm V496~Sct}=12.9$ in Appendix \ref{v496_sct_bvi}.
Thus, we obtain $(m-M)_{I, \rm V382~Vel}= 11.41\pm0.2$.

Using the timescaling factor of $\log f_{\rm s}= -0.29$, we plot
the $U$ band light curves of V382~Vel as well as
LV~Vul, V1668~Cyg, and V5114~Sgr in Figure
\ref{color_color_distance_reddening_v382_vel_xxxxxx}(a).
We apply Equation (6) of \citet{hac19ka} for the $U$ band to Figure
\ref{color_color_distance_reddening_v382_vel_xxxxxx}(a)
and obtain
\begin{eqnarray}
(m&-&M)_{U, \rm V382~Vel} \cr
&=& ((m - M)_U + \Delta U)_{\rm LV~Vul} - 2.5 \log 0.51 \cr
&=& 12.85 - 1.8\pm0.2 + 0.725 = 11.78\pm0.2 \cr
&=& ((m - M)_U + \Delta U)_{\rm V1668~Cyg} - 2.5 \log 0.51 \cr
&=& 15.1 - 4.05\pm0.2 + 0.725 = 11.78\pm0.2 \cr 
&=& ((m - M)_U + \Delta U)_{\rm V5114~Sgr} - 2.5 \log 0.68 \cr
&=& 17.15 - 5.8\pm0.2 + 0.425 = 11.78\pm0.2, 
\label{distance_modulus_u_v382_vel}
\end{eqnarray}
where we adopt $(m-M)_{U, \rm LV~Vul}= 12.85$ and 
$(m-M)_{U, \rm V1668~Cyg}= 15.10$ from \citet{hac19ka}, 
and $(m-M)_{U, \rm V5114~Sgr}= 17.15$ from Appendix \ref{v5114_sgr_ubvi}. 
Thus, we obtain $(m-M)_{U, \rm V382~Vel}= 11.78\pm0.2$.

Figure \ref{color_color_distance_reddening_v382_vel_xxxxxx}(b)
depicts various distance-reddening relations toward V382~Vel. 
We plot the four distance moduli in $U$, $B$, $V$, and 
$I$ bands by the four thin solid lines of green, cyan, blue, 
and blue-magenta, respectively.
These four lines cross at $d=1.76\pm0.2$~kpc and $E(B-V)=0.12\pm0.05$.
The crossing point is close to the distance-reddening relation
(thick solid cyan-blue line) given by \citet{chen19}.  Here,
we add the four thick cyan-blue lines of \citet{chen19}, which correspond
to four nearby directions toward V382~Vel, i.e., the galactic coordinates of
$(\ell, b)= (284\fdg15, +5\fdg75)$, $(284\fdg15, +5\fdg85)$,
$(284\fdg25, +5\fdg75)$, and $(284\fdg25, +5\fdg85)$.


\begin{figure}
\plotone{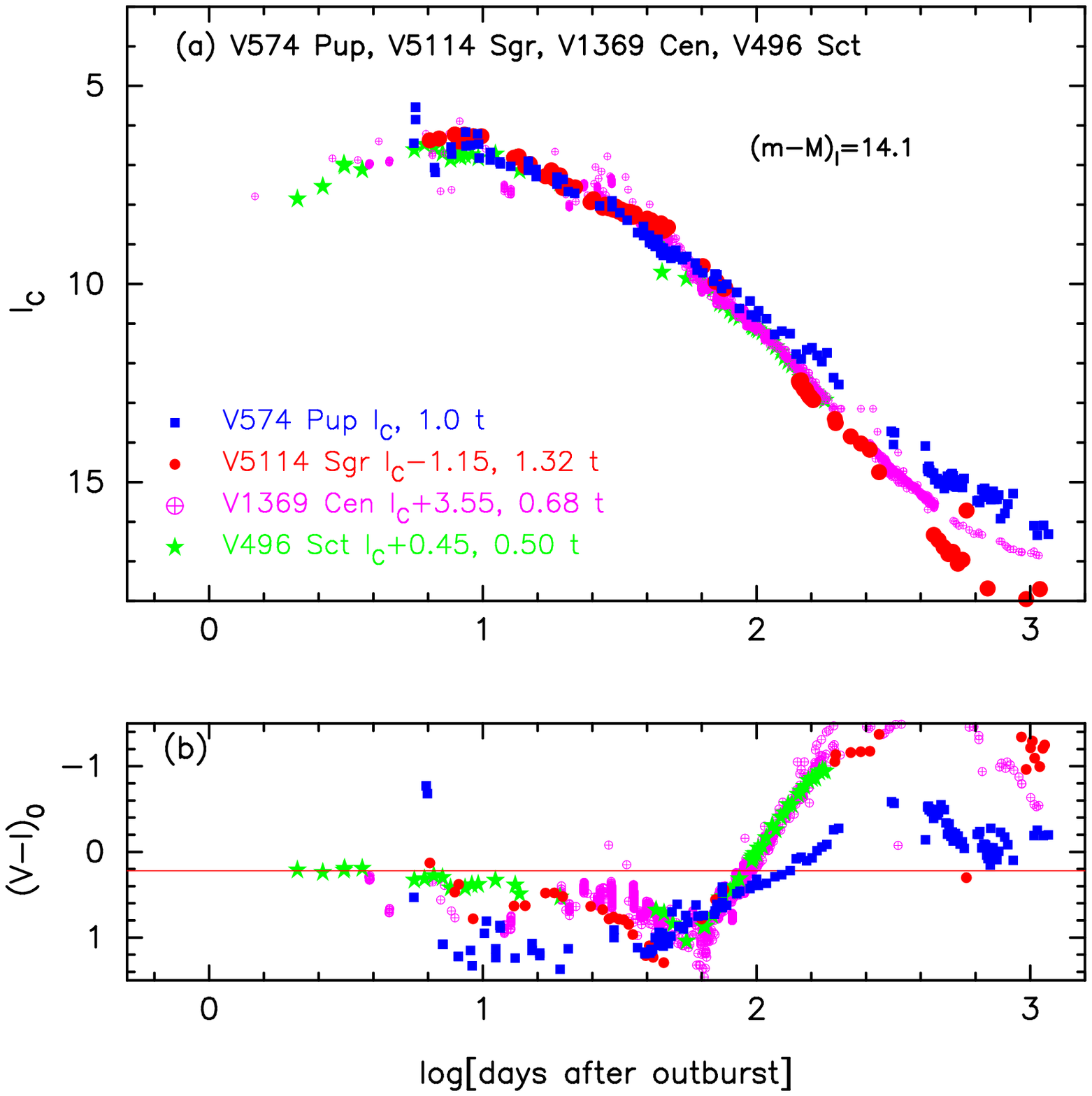}
\caption{
The (a) $I_{\rm C}$ light curve and (b) $(V-I_{\rm C})_0$ color curve
of V574~Pup as well as those of V5114~Sgr, V1369~Cen, and V496~Sct.
The $BVI_{\rm C}$ data of V574~Pup are taken from AAVSO, VSOLJ, and SMARTS. 
\label{v574_pup_v5114_sgr_v1369_cen_v496_sct_i_vi_color_logscale}}
\end{figure}


\begin{figure}
\plotone{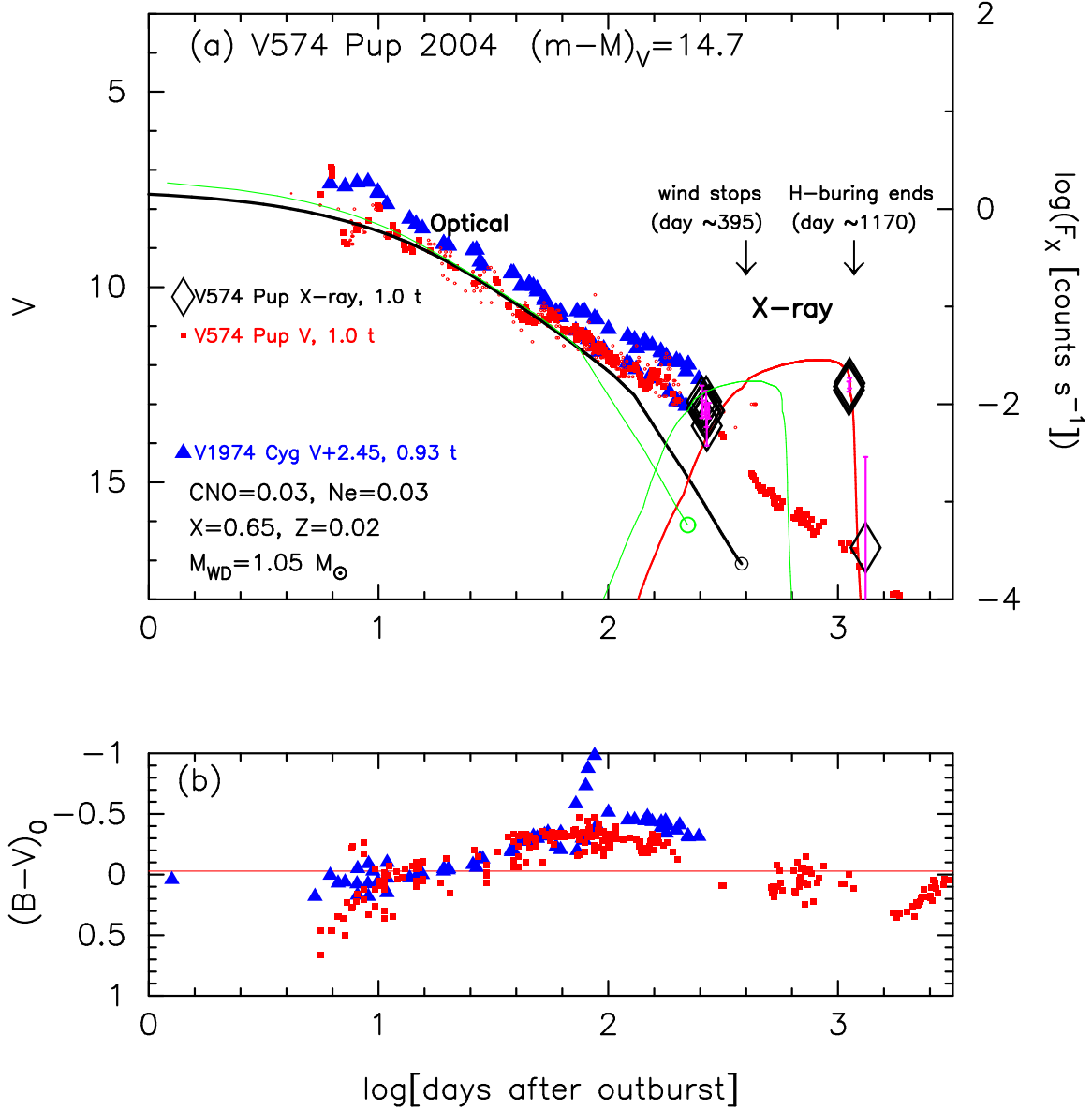}
\caption{
The (a) $V$ light and (b) $(B-V)_0$ color curves of V574~Pup
together with those of V1974~Cyg.
We also add a $1.05~M_\sun$ WD (Ne3, solid black/red lines) model
for V574~Pup as well as a $0.98~M_\sun$ WD (CO3, solid green lines)
model for V1974~Cyg.
\label{v574_pup_v1974_cyg_v_bv_logscale_no2}}
\end{figure}


\begin{figure}
\plottwo{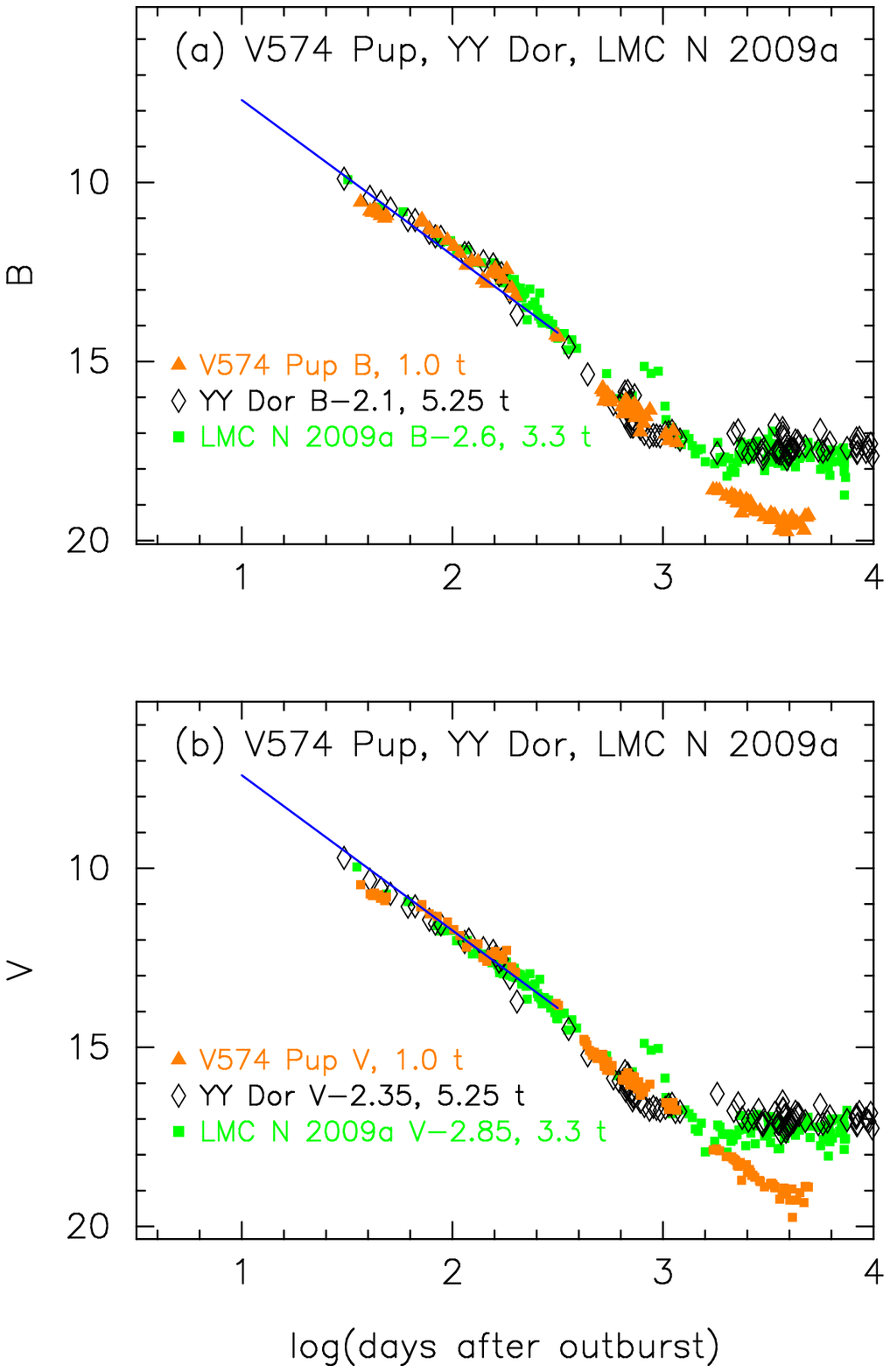}{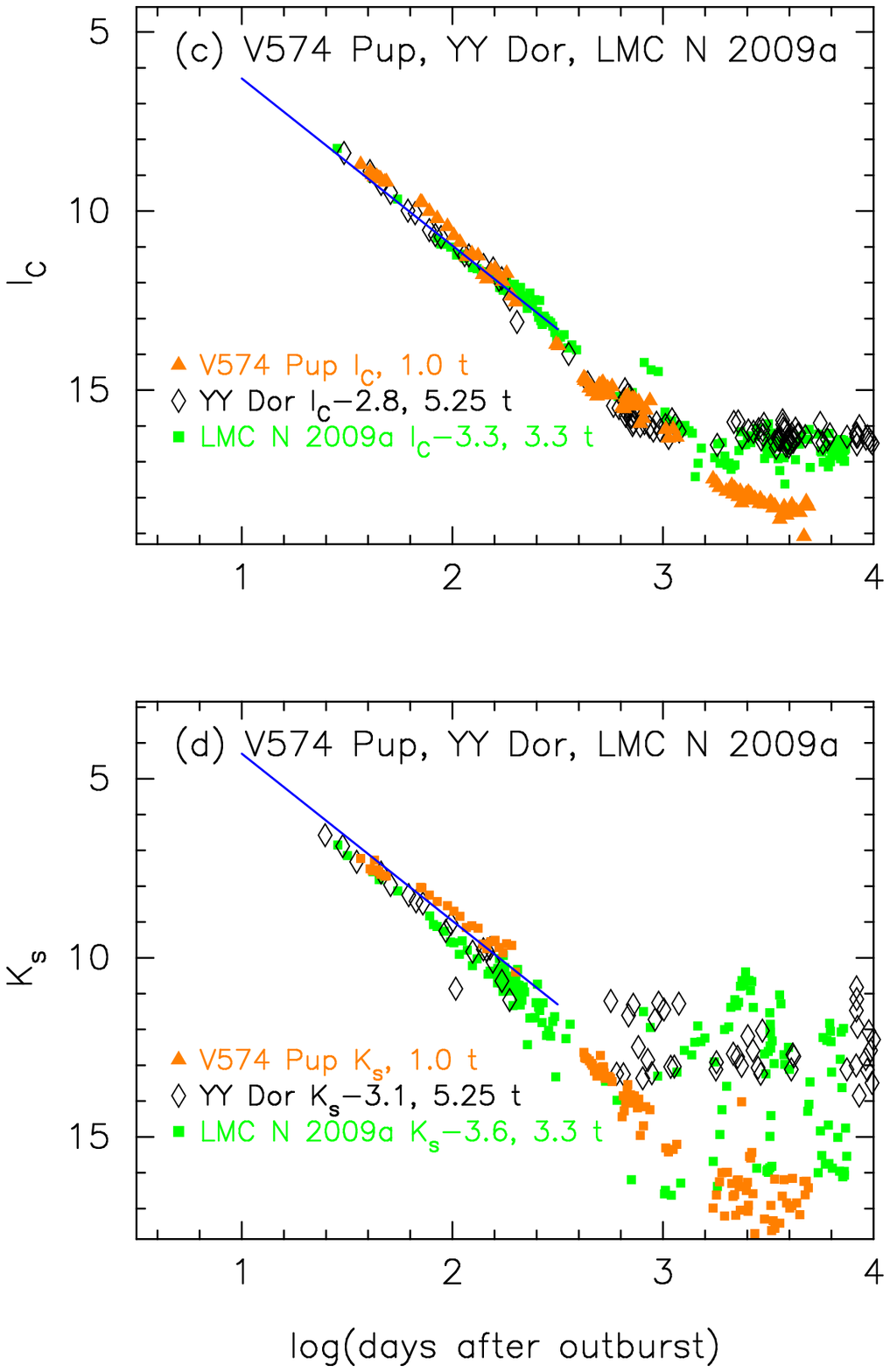}
\caption{
The (a) $B$, (b) $V$, (c) $I_{\rm C}$, and (d) $K_{\rm s}$
light curves of V574~Pup, YY~Dor, and LMC~N~2009a.
The $BVI_{\rm C}K_{\rm s}$ data of V574~Pup are taken from AAVSO, 
VSOLJ, and SMARTS.  The $BVI_{\rm C}K_{\rm s}$ data of YY~Dor and 
LMC~N~2009a are taken from SMARTS. 
\label{v574_pup_yy_dor_lmcn2009a_b_v_i_k_logscale}}
\end{figure}


\begin{figure*}
\epsscale{0.65}
\plotone{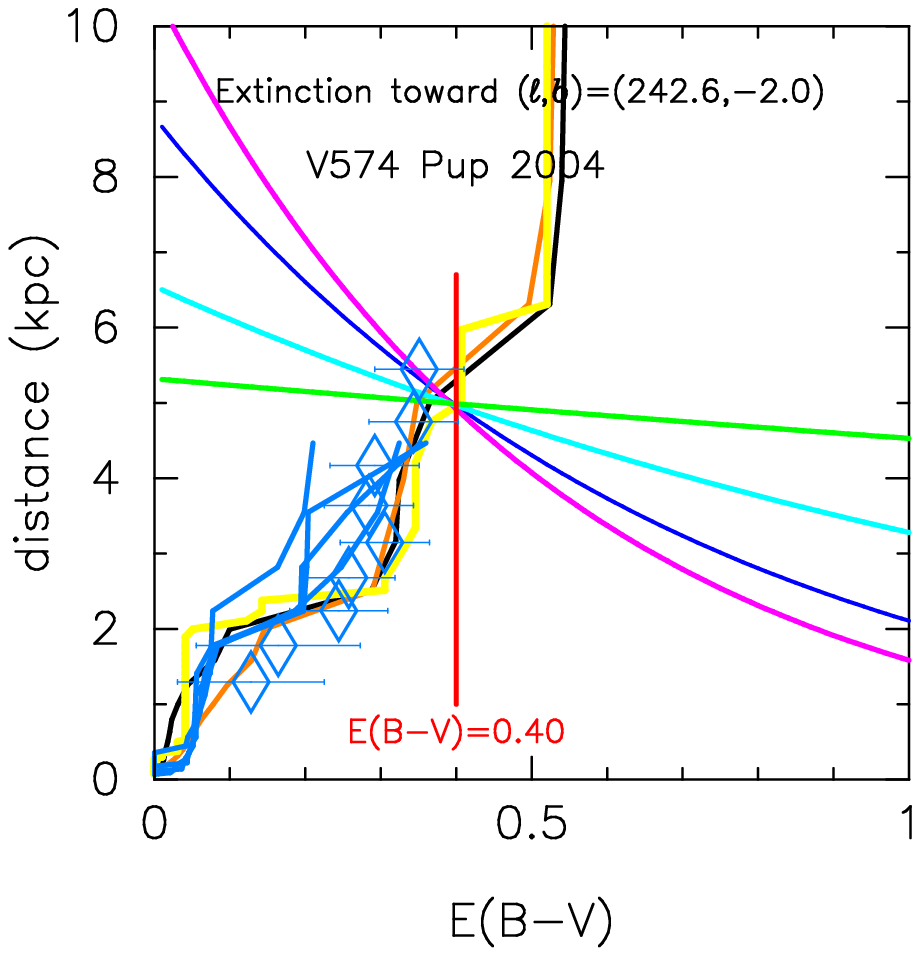}
\caption{
(a) Various distance-reddening relations toward V574~Pup.
The four thin lines of magenta, blue, cyan, and green denote 
the distance-reddening relations given by 
$(m-M)_B= 15.1$, $(m-M)_V= 14.7$, $(m-M)_I= 14.08$, and $(m-M)_K= 13.63$,
respectively.
\label{distance_reddening_v574_pup_bvik_xxxxxx}}
\end{figure*}

\subsection{V574~Pup 2004}
\label{v574_pup_bvik}
We have reanalyzed the $BVI_{\rm C}K_{\rm s}$ multi-band 
light/color curves of V574~Pup based on the time-stretching method.  
Figure \ref{v574_pup_v5114_sgr_v1369_cen_v496_sct_i_vi_color_logscale}
shows the (a) $I_{\rm C}$ light and (b) $(V-I_{\rm C})_0$ color curves of
V574~Pup as well as V5114~Sgr, V1369~Cen, and V496~Sct.
The $BVI_{\rm C}$ data of V574~Pup are taken from AAVSO, VSOLJ, and SMARTS.
We adopt the color excess of $E(B-V)= 0.30$ as mentioned below.
We apply Equation (8) of \citet{hac19ka} for the $I$ band to Figure
\ref{v574_pup_v5114_sgr_v1369_cen_v496_sct_i_vi_color_logscale}(a)
and obtain
\begin{eqnarray}
(m&-&M)_{I, \rm V574~Pup} \cr
&=& ((m - M)_I + \Delta I_{\rm C})
_{\rm V5114~Sgr} - 2.5 \log 1.32 \cr
&=& 15.55 - 1.15\pm0.2 - 0.3 = 14.1\pm0.2 \cr
&=& ((m - M)_I + \Delta I_{\rm C})
_{\rm V1369~Cen} - 2.5 \log 0.68 \cr
&=& 10.11 + 3.55\pm0.2 + 0.425 = 14.08\pm0.2 \cr
&=& ((m - M)_I + \Delta I_{\rm C})
_{\rm V496~Sct} - 2.5 \log 0.50 \cr
&=& 12.9 + 0.45\pm0.2 + 0.75 = 14.1\pm0.2,
\label{distance_modulus_i_vi_v574_pup}
\end{eqnarray}
where we adopt
$(m-M)_{I, \rm V5114~Sgr}=15.55$ from Appendix \ref{v5114_sgr_ubvi},
$(m-M)_{I, \rm V1369~Cen}=10.11$ from \citet{hac19ka}, and
$(m-M)_{I, \rm V496~Sct}=12.9$ in Appendix \ref{v496_sct_bvi}.
Thus, we obtain $(m-M)_{I, \rm V574~Pup}= 14.1\pm0.2$.

Figure \ref{v574_pup_v1974_cyg_v_bv_logscale_no2} shows
the (a) $V$ light and (b) $(B-V)_0$ color curves 
together with those of V1974~Cyg.
Applying Equation (4) of \citet{hac19ka} to them,
we have the relation
\begin{eqnarray}
(m&-&M)_{V, \rm V574~Pup} \cr
&=& ((m - M)_V + \Delta V)_{\rm V1974~Cyg} - 2.5 \log 0.93 \cr
&=& 12.2 + 2.45\pm0.2 + 0.075 = 14.72\pm0.2,
\label{distance_modulus_v_bv_v574_pup}
\end{eqnarray}
where we adopt $(m-M)_{V, \rm V1974~Cyg}=12.2$ from \citet{hac19ka}.
Thus, we obtain $(m-M)_V=14.7\pm0.2$ for V574~Pup and
$\log f_{\rm s}= \log 1.0 = 0.0$ against that of LV~Vul.

We further study the distance and reddening with a different set of novae.
Figure \ref{v574_pup_yy_dor_lmcn2009a_b_v_i_k_logscale} shows
the (a) $B$, (b) $V$, (c) $I_{\rm C}$, and (d) $K_{\rm s}$
light curves of V574~Pup together with those of YY~Dor and LMC~N~2009a.
The $K_{\rm s}$ data of V574~Pup are taken from SMARTS.
We apply Equation (7) of \citet{hac19ka} for the $B$ band to Figure
\ref{v574_pup_yy_dor_lmcn2009a_b_v_i_k_logscale}(a)
and obtain
\begin{eqnarray}
(m&-&M)_{B, \rm V574~Pup} \cr
&=& ((m - M)_B + \Delta B)_{\rm YY~Dor} - 2.5 \log 5.25 \cr
&=& 18.98 - 2.1\pm0.2 - 1.8 = 15.08\pm0.2 \cr
&=& ((m - M)_B + \Delta B)_{\rm LMC~N~2009a} - 2.5 \log 3.3 \cr
&=& 18.98 - 2.6\pm0.2 - 1.3 = 15.08\pm0.2.
\label{distance_modulus_b_v574_pup_yy_dor_lmcn2009a}
\end{eqnarray}
Thus, we obtain $(m-M)_{B, \rm V574~Pup}= 15.1\pm0.1$.

For the $V$ light curves in Figure
\ref{v574_pup_yy_dor_lmcn2009a_b_v_i_k_logscale}(b),
we similarly obtain
\begin{eqnarray}
(m&-&M)_{V, \rm V574~Pup} \cr
&=& ((m - M)_V + \Delta V)_{\rm YY~Dor} - 2.5 \log 5.25 \cr
&=& 18.86 - 2.35\pm0.2 - 1.8 = 14.71\pm0.2 \cr
&=& ((m - M)_V + \Delta V)_{\rm LMC~N~2009a} - 2.5 \log 3.3 \cr
&=& 18.86 - 2.85\pm0.2 - 1.3 = 14.71\pm0.2.
\label{distance_modulus_v_v574_pup_yy_dor_lmcn2009a}
\end{eqnarray}
Thus, we obtain $(m-M)_{V, \rm V574~Pup}= 14.71\pm0.1$, which is
consistent with Equation (\ref{distance_modulus_v_bv_v574_pup}).

We apply Equation (8) of \citet{hac19ka} for the $I$-band to Figure
\ref{v574_pup_yy_dor_lmcn2009a_b_v_i_k_logscale}(c) and obtain
\begin{eqnarray}
(m&-&M)_{I, \rm V574~Pup} \cr
&=& ((m - M)_I + \Delta I_{\rm C})_{\rm YY~Dor} - 2.5 \log 5.25 \cr
&=& 18.67 - 2.8\pm0.3 - 1.8 = 14.07\pm 0.3 \cr
&=& ((m - M)_I + \Delta I_{\rm C})_{\rm LMC~N~2009a} - 2.5 \log 3.3 \cr
&=& 18.67 - 3.3\pm0.3 - 1.3 = 14.07\pm 0.3.
\label{distance_modulus_i_v574_pup_yy_dor_lmcn2009a}
\end{eqnarray}
Thus, we obtain $(m-M)_{I, \rm V574~Pup}= 14.07\pm0.2$.

We apply Equation (9) of \citet{hac19ka} for the $K$-band to Figure
\ref{v574_pup_yy_dor_lmcn2009a_b_v_i_k_logscale}(d) and obtain
\begin{eqnarray}
(m&-&M)_{K, \rm V574~Pup} \cr
&=& ((m - M)_K + \Delta K_{\rm s})_{\rm YY~Dor} - 2.5 \log 5.25 \cr
&=& 18.53 - 3.1\pm0.3 - 1.8 = 13.63\pm 0.3 \cr
&=& ((m - M)_K + \Delta K_{\rm s})_{\rm LMC~N~2009a} - 2.5 \log 3.3 \cr
&=& 18.53 - 3.6\pm0.3 - 1.3 = 13.63\pm 0.3.
\label{distance_modulus_k_v574_pup_yy_dor_lmcn2009a}
\end{eqnarray}
Thus, we obtain $(m-M)_{K, \rm V574~Pup}= 13.63\pm0.2$.

Figure \ref{distance_reddening_v574_pup_bvik_xxxxxx} shows various
distance-reddening relations toward V574~Pup.  The four thin lines of
$(m-M)_B= 15.1$, $(m-M)_V= 14.7$, $(m-M)_I= 14.08$, and $(m-M)_K= 13.63$,
cross at $d=5.0$~kpc and $E(B-V)=0.40$.
The crossing point is broadly consistent with the distance-reddening
relation (cyan-blue diamonds) given by \citet{ozd18}.
Thus, we obtain $E(B-V)=0.40\pm0.05$ and $d=5.0\pm0.5$~kpc.


\begin{figure}
\plotone{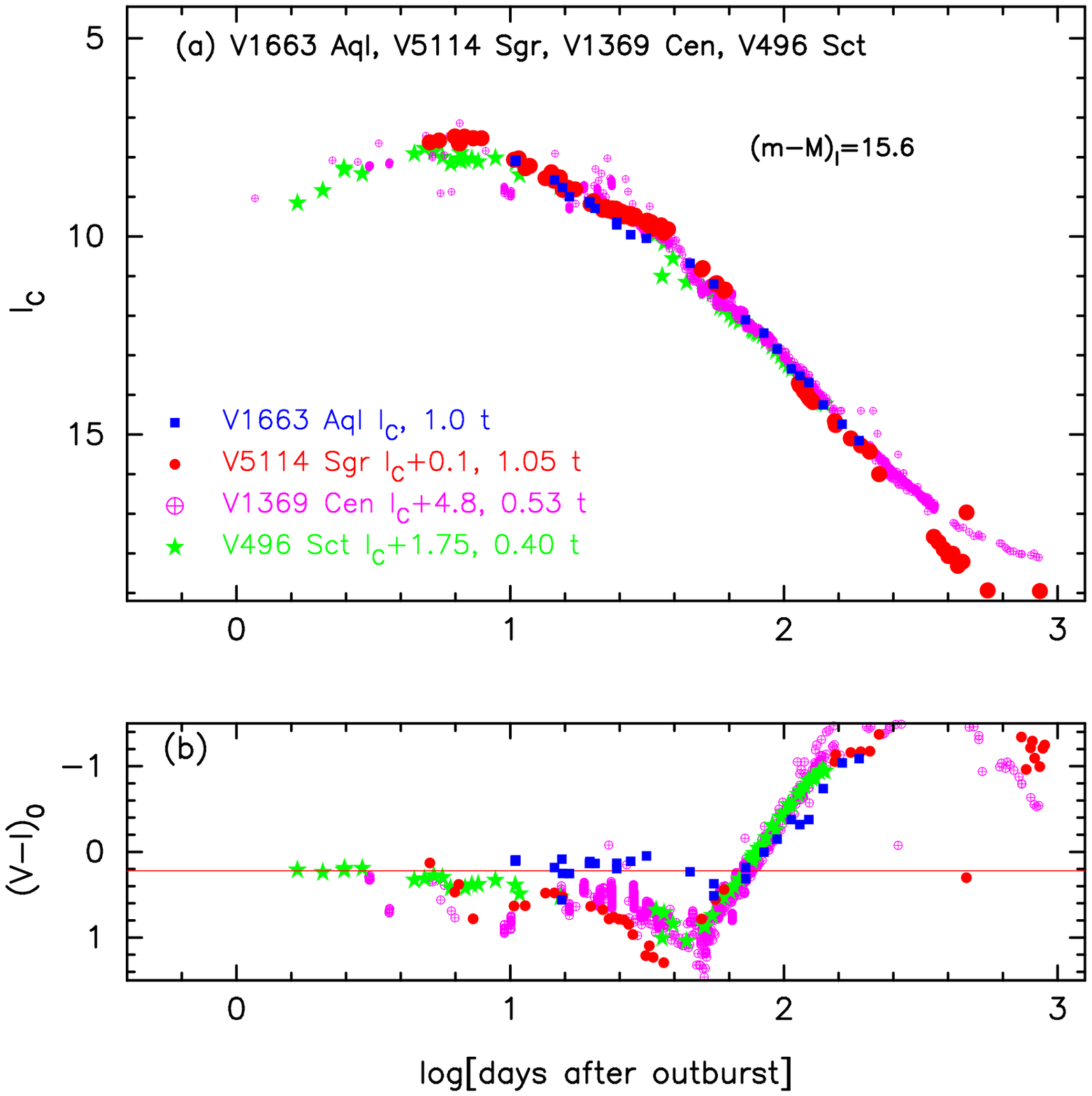}
\caption{
The (a) $I_{\rm C}$ light curve and (b) $(V-I_{\rm C})_0$ color curve
of V1663~Aql as well as those of V5114~Sgr, V1369~Cen, and V496~Sct.
The $BVI_{\rm C}$ data of V1663~Aql are taken from AAVSO, VSOLJ, and SMARTS. 
\label{v1663_aql_v5114_sgr_v1369_cen_v496_sct_i_vi_color_logscale}}
\end{figure}


\begin{figure}
\plotone{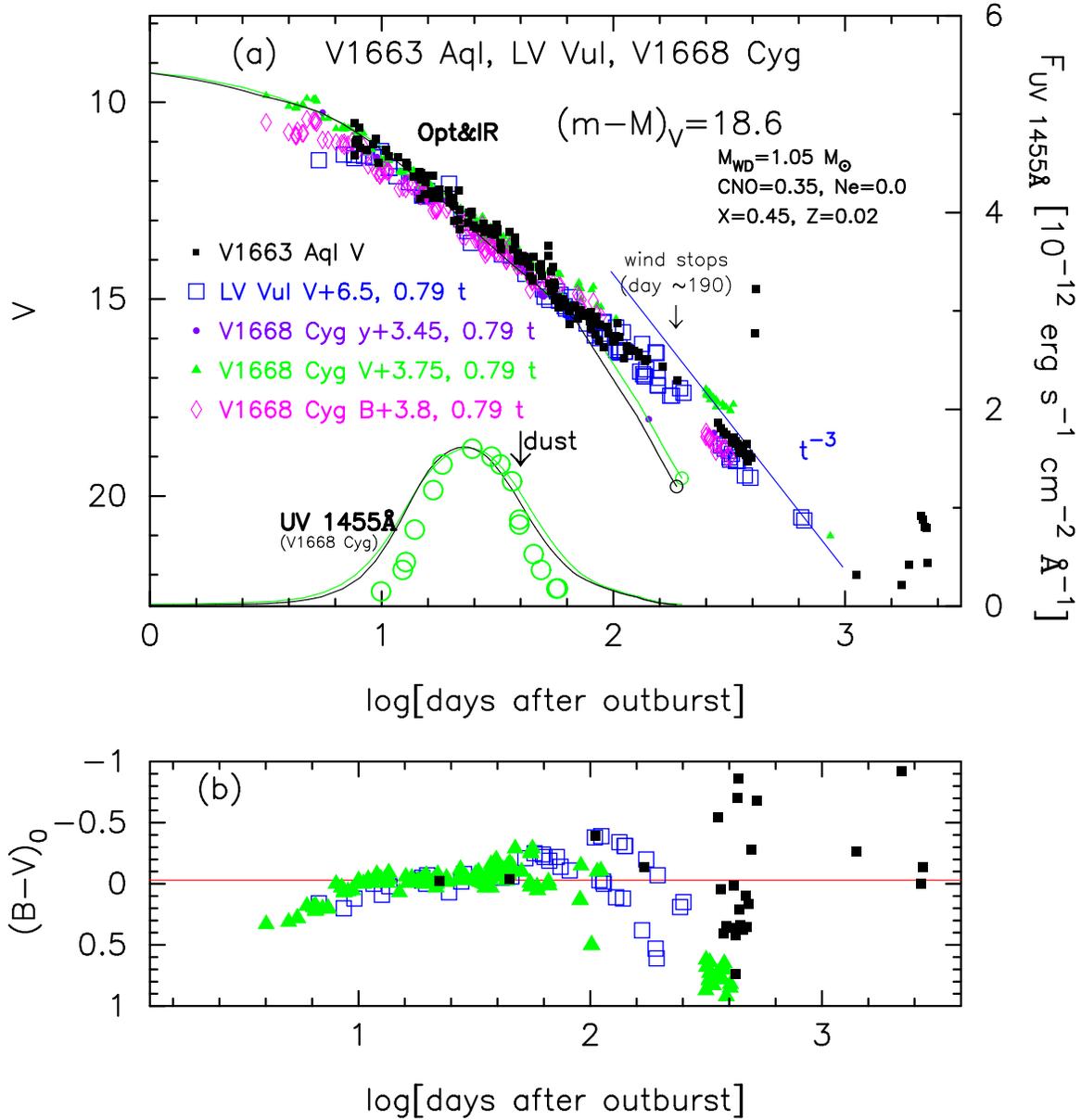}
\caption{
The (a) $V$ light and (b) $(B-V)_0$ color curves of V1663~Aql
as well as those of LV~Vul and V1668~Cyg.
In panel (a), we add a $1.05~M_\sun$ WD model (CO3, solid black lines)
for V1663~Aql as well as a $0.98~M_\sun$ WD model (CO3, solid green lines)
for V1668~Cyg.
\label{v1663_aql_lv_vul_v1668_cyg_v_bv_logscale_no2}}
\end{figure}


\begin{figure}
\plottwo{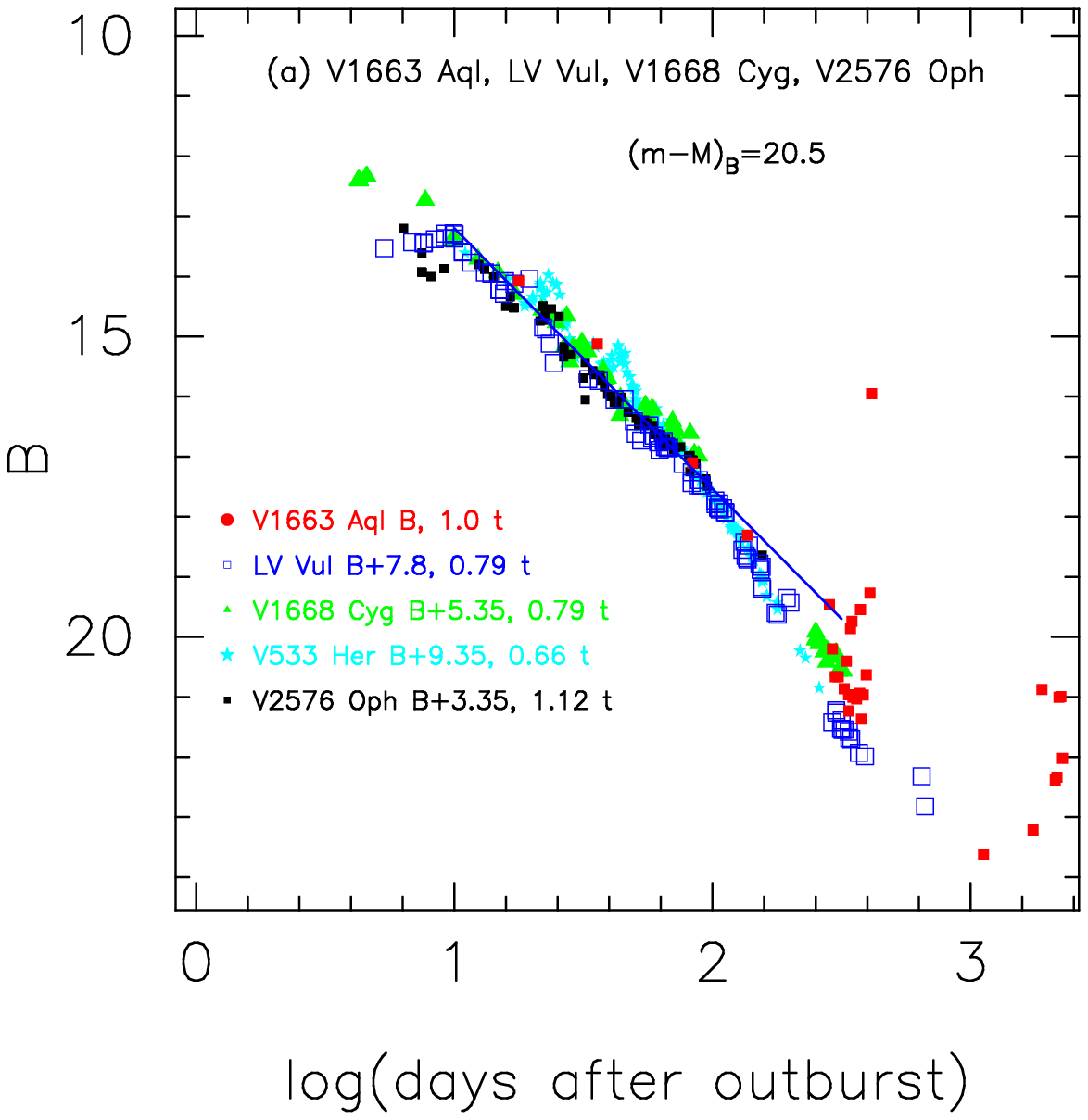}{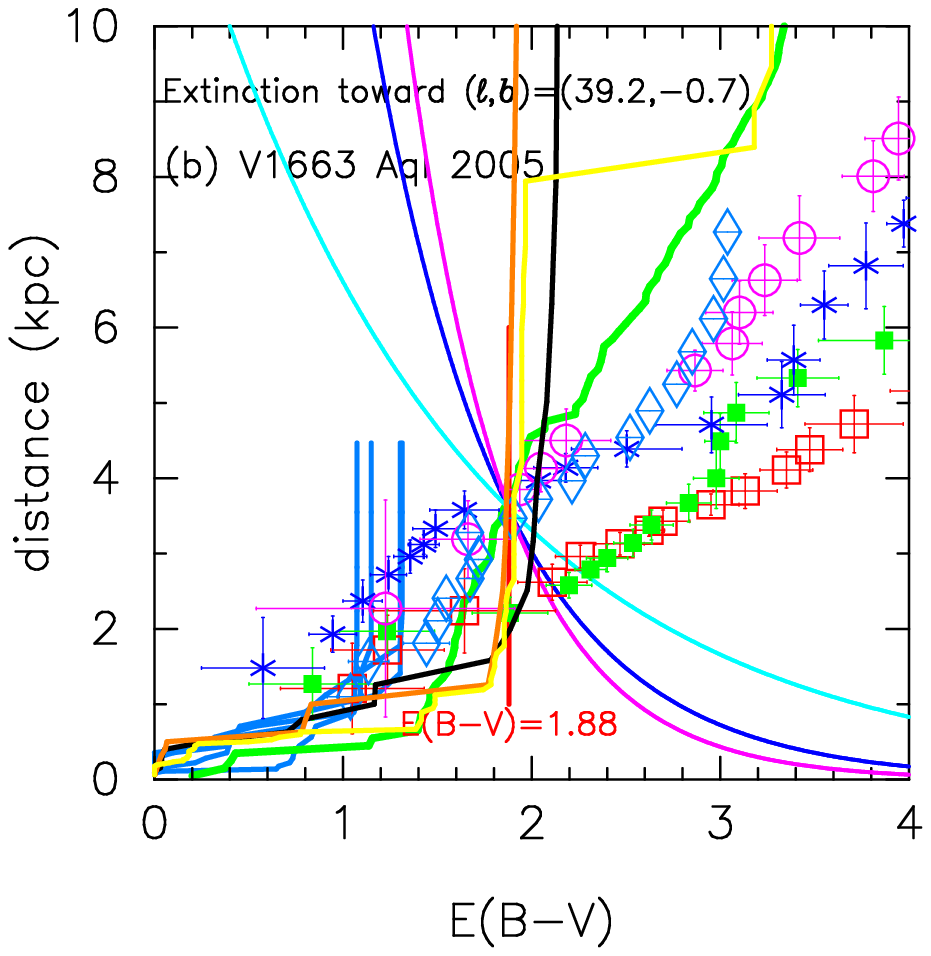}
\caption{
(a) The $B$ light curves of V1663~Aql, LV~Vul, V1668~Cyg, V533~Her,
and V2576~Oph.
The $B$ data of V1663~Aql are taken from AAVSO, VSOLJ, and SMARTS.
(b) Various distance-reddening relations toward V1663~Aql.
The three thin solid lines of magenta, blue, and cyan denote 
the distance-reddening relations given by  $(m-M)_B= 20.49$, 
$(m-M)_V= 18.6$, and $(m-M)_I= 15.6$, respectively.  
\label{v1663_aql_v2576_oph_v1668_cyg_lv_vul_b_only_logscale}}
\end{figure}

\subsection{V1663~Aql 2005}
\label{v1663_aql_bvi}
We have reanalyzed the $BVI_{\rm C}$ multi-band 
light/color curves of V1663~Aql based on the time-stretching method.  
Figure \ref{v1663_aql_v5114_sgr_v1369_cen_v496_sct_i_vi_color_logscale}
shows the (a) $I_{\rm C}$ light and (b) $(V-I_{\rm C})_0$ color curves of
V1663~Aql as well as V5114~Sgr, V1369~Cen, and V496~Sct.
The $BVI_{\rm C}$ data of V1663~Aql are taken from AAVSO, VSOLJ, and SMARTS.
We adopt the color excess of $E(B-V)= 1.88$ in order to overlap
the $(V-I)_0$ color curve of V1663~Aql with the other novae, as shown in
Figure \ref{v1663_aql_v5114_sgr_v1369_cen_v496_sct_i_vi_color_logscale}(b).
We apply Equation (8) of \citet{hac19ka} for the $I$ band to Figure
\ref{v1663_aql_v5114_sgr_v1369_cen_v496_sct_i_vi_color_logscale}(a)
and obtain
\begin{eqnarray}
(m&-&M)_{I, \rm V1663~Aql} \cr
&=& ((m - M)_I + \Delta I_{\rm C})
_{\rm V5114~Sgr} - 2.5 \log 1.05 \cr
&=& 15.55 + 0.1\pm0.2 - 0.05 = 15.6\pm0.2 \cr
&=& ((m - M)_I + \Delta I_{\rm C})
_{\rm V1369~Cen} - 2.5 \log 0.53 \cr
&=& 10.11 + 4.8\pm0.2 + 0.675 = 15.58\pm0.2 \cr
&=& ((m - M)_I + \Delta I_{\rm C})
_{\rm V496~Sct} - 2.5 \log 0.40 \cr
&=& 12.9 + 1.75\pm0.2 + 1.0 = 15.65\pm0.2,
\label{distance_modulus_i_vi_v1663_aql}
\end{eqnarray}
where we adopt
$(m-M)_{I, \rm V5114~Sgr}=15.55$ from Appendix \ref{v5114_sgr_ubvi},
$(m-M)_{I, \rm V1369~Cen}=10.11$ from \citet{hac19ka}, and
$(m-M)_{I, \rm V496~Sct}=12.9$ in Appendix \ref{v496_sct_bvi}.
Thus, we obtain $(m-M)_{I, \rm V1663~Aql}= 15.61\pm0.2$.

Figure \ref{v1663_aql_lv_vul_v1668_cyg_v_bv_logscale_no2} shows
the (a) $V$ light and (b) $(B-V)_0$ color curves 
as well as those of LV~Vul and V1668~Cyg.
Applying Equation (4) of \citet{hac19ka} to them,
we have the relation
\begin{eqnarray}
(m&-&M)_{V, \rm V1663~Aql} \cr
&=& ((m - M)_V + \Delta V)_{\rm LV~Vul} - 2.5 \log 0.79 \cr
&=& 11.85 + 6.5\pm0.2 + 0.25 = 18.6\pm0.2 \cr
&=& ((m - M)_V + \Delta V)_{\rm V1668~Cyg} - 2.5 \log 0.79 \cr
&=& 14.6 + 3.75\pm0.2 + 0.25 = 18.6\pm0.2,
\label{distance_modulus_v_bv_v1663_aql}
\end{eqnarray}
where we adopt $(m-M)_{V, \rm LV~Vul}=11.85$ and
$(m-M)_{V, \rm V1668~Cyg}=14.6$, both from \citet{hac19ka}.
Thus, we obtain $(m-M)_V=18.6\pm0.1$ for V1663~Aql and
$\log f_{\rm s}= \log 0.79 = -0.10$ against that of LV~Vul.

Figure \ref{v1663_aql_v2576_oph_v1668_cyg_lv_vul_b_only_logscale}(a)
shows the $B$ light curves of V1663~Aql
together with those of LV~Vul, V1668~Cyg, V533~Her, and V2576~Oph.
We apply Equation (7) of \citet{hac19ka}
for the $B$ band to Figure
\ref{v1663_aql_v2576_oph_v1668_cyg_lv_vul_b_only_logscale}(a)
and obtain
\begin{eqnarray}
(m&-&M)_{B, \rm V1663~Aql} \cr
&=& ((m - M)_B + \Delta B)_{\rm LV~Vul} - 2.5 \log 0.79 \cr
&=& 12.45 + 7.8\pm0.2 + 0.25 = 20.5\pm0.2 \cr
&=& ((m - M)_B + \Delta B)_{\rm V1668~Cyg} - 2.5 \log 0.79 \cr
&=& 14.9 + 5.35\pm0.2 + 0.25 = 20.5\pm0.2 \cr
&=& ((m - M)_B + \Delta B)_{\rm V533~Her} - 2.5 \log 0.66 \cr
&=& 10.69 + 9.35\pm0.2 + 0.45 = 20.49\pm0.2 \cr
&=& ((m - M)_B + \Delta B)_{\rm V2576~Oph} - 2.5 \log 1.12 \cr
&=& 17.25 + 3.35\pm0.2 - 0.125 = 20.48\pm0.2.
\label{distance_modulus_b_v1663_aql_lv_vul_v1668_cyg}
\end{eqnarray}
We have $(m-M)_{B, \rm V1663~Aql}= 20.49\pm0.2$.

We plot the distance-reddening relations of
$(m-M)_B=20.49$, $(m-M)_V=18.6$, and $(m-M)_I=15.6$ in Figure
\ref{v1663_aql_v2576_oph_v1668_cyg_lv_vul_b_only_logscale}(b).
The three thin solid lines of magenta, blue, and cyan 
consistently cross at $d=3.6$~kpc and $E(B-V)=1.88$.  
This point is also consistent with the distance-reddening relations
(orange and yellow lines) given by \citet{gre18, gre19} as well as
those of \citet[][green line]{sal14} 
and \citet[][unfilled cyan-blue diamonds]{ozd18}.


\begin{figure}
\plotone{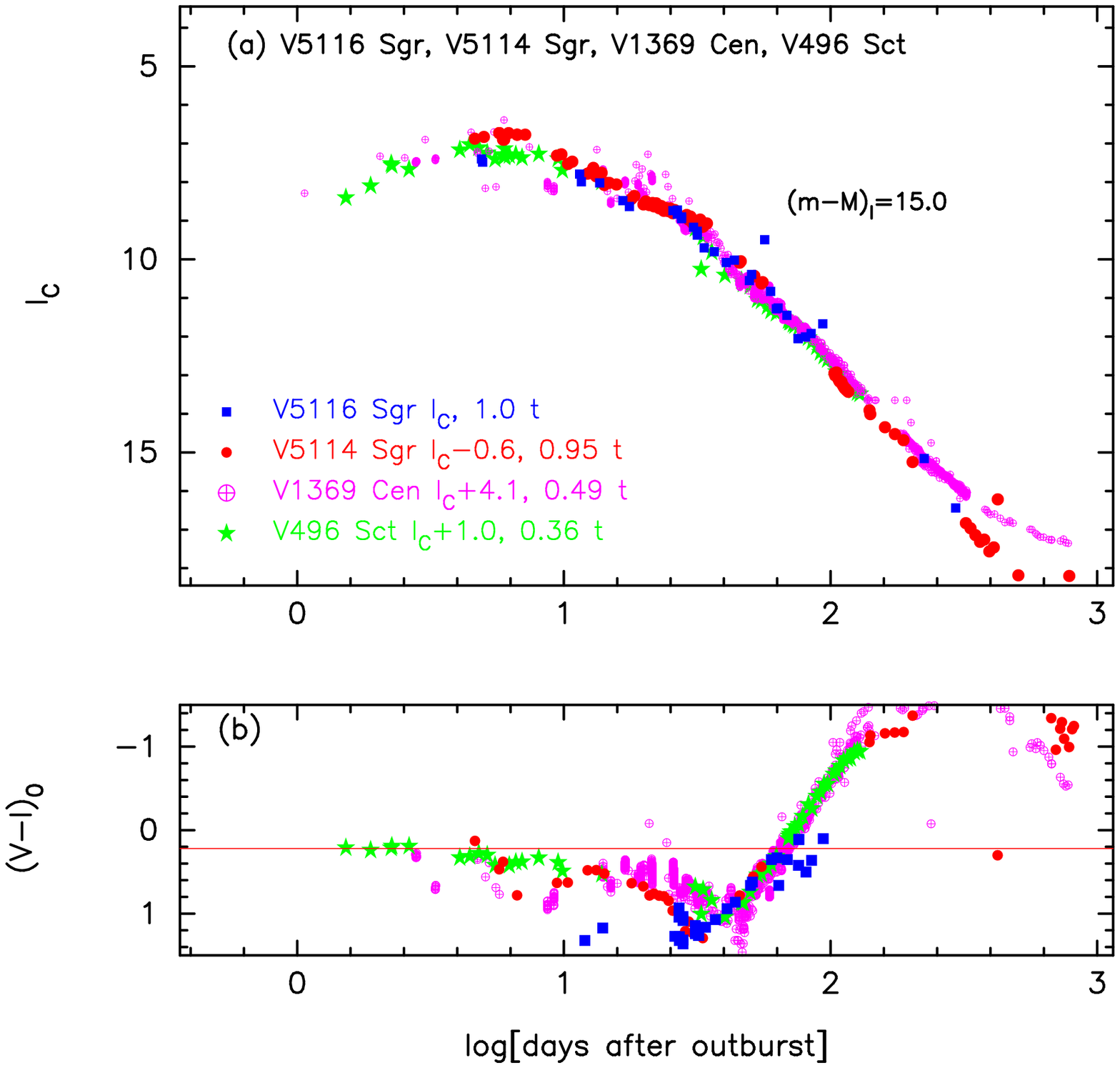}
\caption{
The (a) $I_{\rm C}$ light curve and (b) $(V-I_{\rm C})_0$ color curve
of V5116~Sgr as well as those of V5114~Sgr, V1369~Cen, and V496~Sct.
\label{v5116_sgr_v5114_sgr_v1369_cen_v496_sct_i_vi_color_logscale}}
\end{figure}


\begin{figure}
\plotone{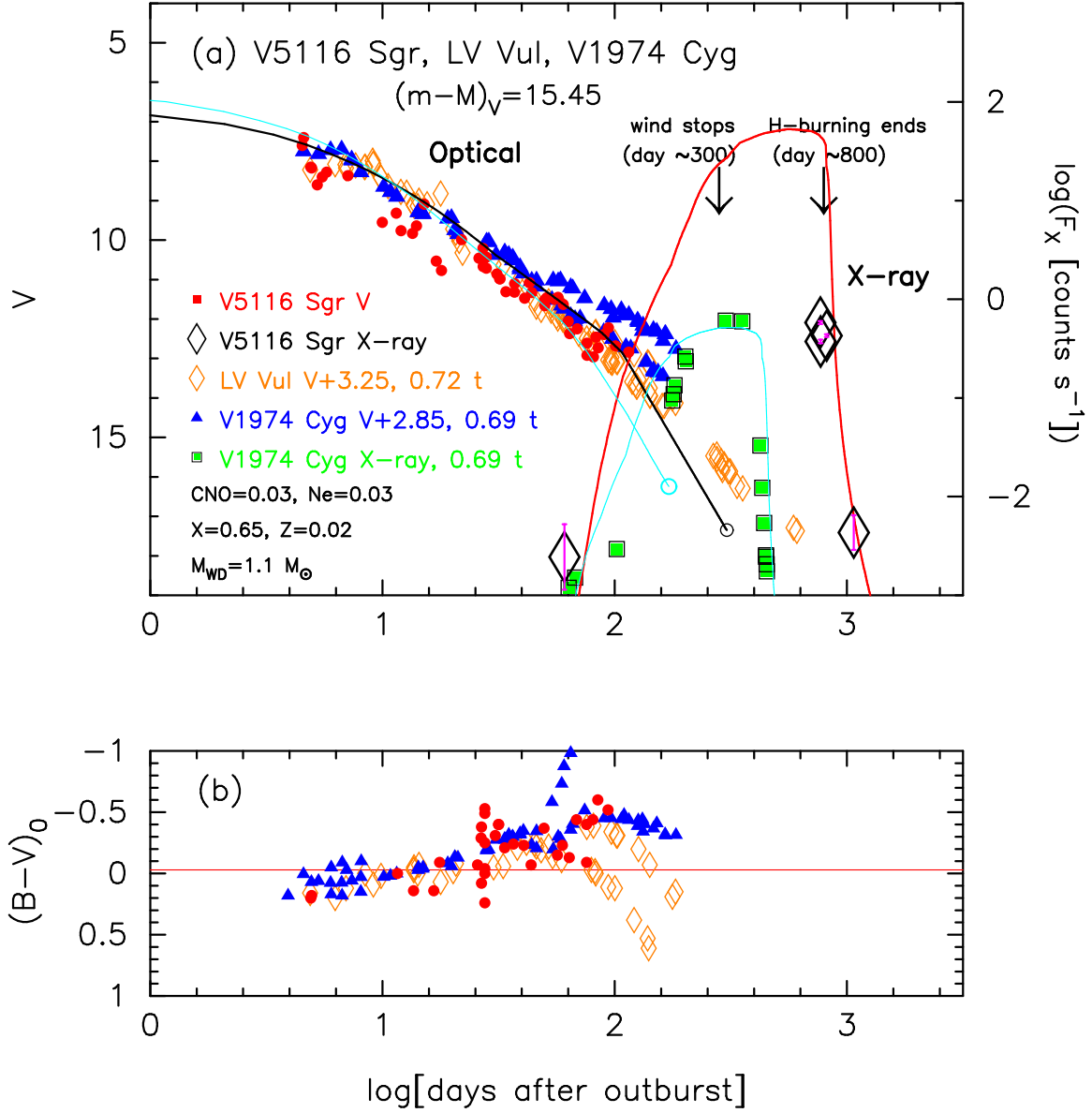}
\caption{
The (a) $V$ light curve and (b) $(B-V)_0$ color curve
of V5116~Sgr as well as those of LV~Vul and V1974~Cyg.
In panel (a), we show a $1.1~M_\sun$ WD model (Ne3, solid black line 
for $V$ and solid red line for X-ray) for V5116~Sgr as well as 
a $0.98~M_\sun$ WD model (CO3, solid cyan line) for V1974~Cyg. 
\label{v5116_sgr_v1974_cyg_x65z02o03ne03_v_bv_color_logscale_no2}}
\end{figure}


\begin{figure*}
\plottwo{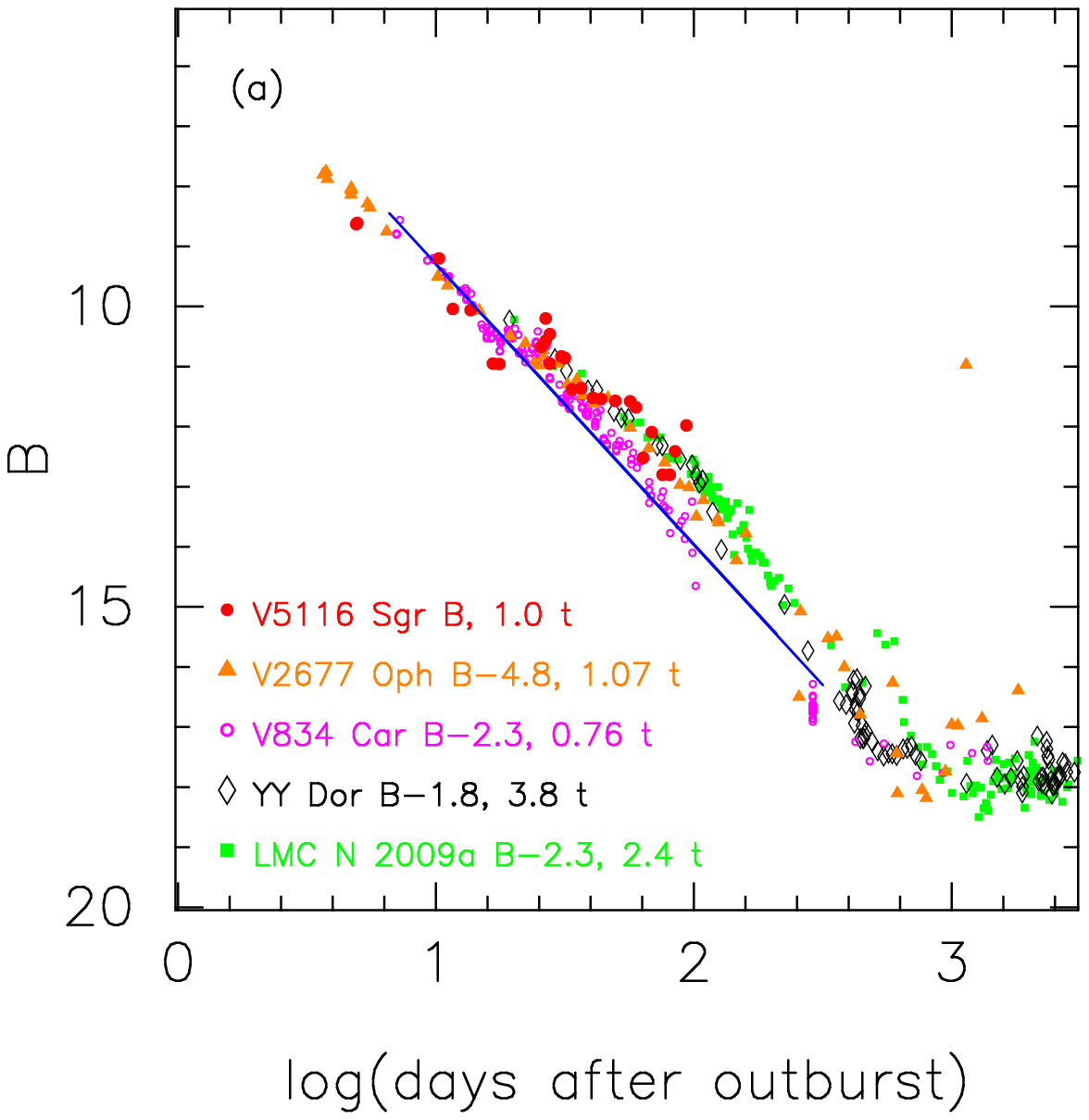}{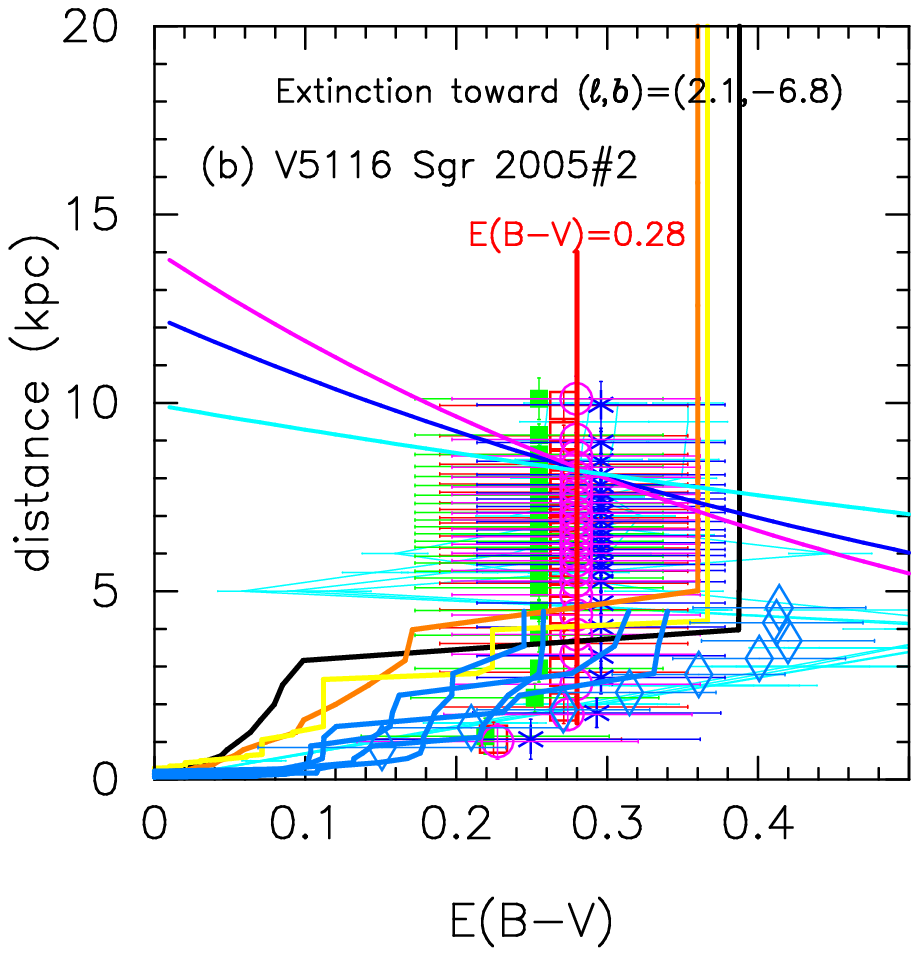}
\caption{
(a) The $B$ light curves of V5116~Sgr
as well as V2677~Oph, V834~Car, YY~Dor, and LMC~N~2009a.
The $BVI_{\rm C}$ data of V5116~Sgr are taken from AAVSO, VSOLJ, SMARTS,
and IAU Circular.
(b) Various distance-reddening relations toward V5116~Sgr.
The thin solid lines of magenta, blue, and cyan denote 
the distance-reddening relations given by $(m-M)_B= 15.74$, 
$(m-M)_V= 15.45$, and $(m-M)_I= 14.99$, respectively.
\label{distance_reddening_v5116_sgr_bvi_xxxxxx}}
\end{figure*}

\subsection{V5116~Sgr 2005}
\label{v5116_sgr_bvi}
We have reanalyzed the $BVI_{\rm C}$ multi-band light/color curves
of V5116~Sgr based on the time-stretching method.  
The important revised point is the timescaling factor of $f_{\rm s}$,
which is changed from the previous $\log f_{\rm s}= +0.20$ to
the present $f_{\rm s}= -0.14$ in order to overlap the $V-I_{\rm C}$
color curve of V5116~Sgr with other novae as shown in Figure
\ref{v5116_sgr_v5114_sgr_v1369_cen_v496_sct_i_vi_color_logscale}(b).
Figure \ref{v5116_sgr_v5114_sgr_v1369_cen_v496_sct_i_vi_color_logscale}
shows the (a) $I_{\rm C}$ light and (b) $(V-I_{\rm C})_0$ color curves of
V5116~Sgr as well as V5114~Sgr, V1369~Cen, and V496~Sct.
The $BVI_{\rm C}$ data of V5116~Sgr are taken from AAVSO, VSOLJ, SMARTS,
and IAU Circular No.8559.
Adopting the color excess of $E(B-V)= 0.28$ mentioned below,
we redetermine the timescaling factor $\log f_{\rm s}= -0.14$ for V5116~Sgr.
This is because the $(V-I)_0$ color evolution of V5116~Sgr overlaps with
the other novae as much as possible, as shown in Figure
\ref{v5116_sgr_v5114_sgr_v1369_cen_v496_sct_i_vi_color_logscale}(b).  
Then, we apply Equation (8) of \citet{hac19ka} for the $I$ band to Figure
\ref{v5116_sgr_v5114_sgr_v1369_cen_v496_sct_i_vi_color_logscale}(a)
and obtain
\begin{eqnarray}
(m&-&M)_{I, \rm V5116~Sgr} \cr
&=& ((m - M)_I + \Delta I_{\rm C})
_{\rm V5114~Sgr} - 2.5 \log 0.95 \cr
&=& 15.55 - 0.6\pm0.2 + 0.05 = 15.0\pm0.2 \cr
&=& ((m - M)_I + \Delta I_{\rm C})
_{\rm V1369~Cen} - 2.5 \log 0.49 \cr
&=& 10.11 + 4.1\pm0.2 + 0.775 = 14.99\pm0.2 \cr
&=& ((m - M)_I + \Delta I_{\rm C})
_{\rm V496~Sct} - 2.5 \log 0.36 \cr
&=& 12.9 + 1.0\pm0.2 + 1.1 = 15.0\pm0.2,
\label{distance_modulus_i_vi_v5116_sgr}
\end{eqnarray}
where we adopt
$(m-M)_{I, \rm V5114~Sgr}=15.55$ from Appendix \ref{v5114_sgr_ubvi},
$(m-M)_{I, \rm V1369~Cen}=10.11$ from \citet{hac19ka}, and
$(m-M)_{I, \rm V496~Sct}=12.9$ in Appendix \ref{v496_sct_bvi}.
Thus, we obtain $(m-M)_{I, \rm V5116~Sgr}= 15.0\pm0.2$.

Figure \ref{v5116_sgr_v1974_cyg_x65z02o03ne03_v_bv_color_logscale_no2}
shows the (a) $V$ and (b) $(B-V)_0$ evolutions of V5116~Sgr
as well as LV~Vul and V1974~Cyg.  
Applying Equation (4) of \citet{hac19ka} for the $V$ band to them,
we have the relation
\begin{eqnarray}
(m&-&M)_{V, \rm V5116~Sgr} \cr
&=& ((m - M)_V + \Delta V)_{\rm LV~Vul} - 2.5 \log 0.72 \cr
&=& 11.85 + 3.25\pm0.2 + 0.35 = 15.45\pm0.2 \cr
&=& ((m - M)_V + \Delta V)_{\rm V1974~Cyg} - 2.5 \log 0.69 \cr
&=& 12.2 + 2.85\pm0.2 + 0.4 = 15.45\pm0.2,
\label{distance_modulus_v_bv_v5116_sgr}
\end{eqnarray}
where we adopt $(m-M)_{V, \rm LV~Vul}=11.85$ and
$(m-M)_{V, \rm V1974~Cyg}=12.2$, both from \citet{hac19ka}.
Thus, we obtain $(m-M)_V=15.45\pm0.1$ for V5116~Sgr.

Figure \ref{distance_reddening_v5116_sgr_bvi_xxxxxx}(a)
shows the $B$ light curve of V5116~Sgr
together with those of V2677~Oph, V834~Car, and the LMC novae YY~Dor
and LMC~N~2009a.
We apply Equation (7) of \citet{hac19ka} for the $B$ band to Figure
\ref{distance_reddening_v5116_sgr_bvi_xxxxxx}(a)
and obtain
\begin{eqnarray}
(m&-&M)_{B, \rm V5116~Sgr} \cr
&=& ((m - M)_B + \Delta B)_{\rm V2677~Oph} - 2.5 \log 1.07 \cr
&=& 20.6 - 4.8\pm0.2 - 0.075 = 15.73\pm0.2 \cr
&=& ((m - M)_B + \Delta B)_{\rm V834~Car} - 2.5 \log 0.76 \cr
&=& 17.75 - 2.3\pm0.2 + 0.3 = 15.75\pm0.2 \cr
&=& ((m - M)_B + \Delta B)_{\rm YY~Dor} - 2.5 \log 3.8 \cr
&=& 18.98 - 1.8\pm0.2 - 1.45 = 15.73\pm0.2 \cr
&=& ((m - M)_B + \Delta B)_{\rm LMC~N~2009a} - 2.5 \log 2.4 \cr
&=& 18.98 - 2.3\pm0.2 - 0.95 = 15.73\pm0.2,
\label{distance_modulus_b_v5116_sgr_v2677_oph_v834_car_yy_dor_lmcn2009a}
\end{eqnarray}
where we adopt $(m-M)_{B, \rm V2677~Oph}= 20.6$ in Appendix \ref{v2677_oph_bvi}
and $(m-M)_{B, \rm V834~Car}= 17.75$ from \citet{hac19kb}.
Thus, we have $(m-M)_{B, \rm V5116~Sgr}= 15.74\pm0.1$.

We plot $(m-M)_B= 15.74$, $(m-M)_V= 15.45$, and $(m-M)_I= 14.99$,
in Figure \ref{distance_reddening_v5116_sgr_bvi_xxxxxx}(b),
which cross at $d=8.2$~kpc and $E(B-V)=0.28$.
The crossing point is broadly consistent with the distance-reddening
relations given by \citet{mar06} and \citet{chen19}.
Thus, we have $E(B-V)=0.28\pm0.05$ and $d=8.2\pm1$~kpc.


\begin{figure}
\plotone{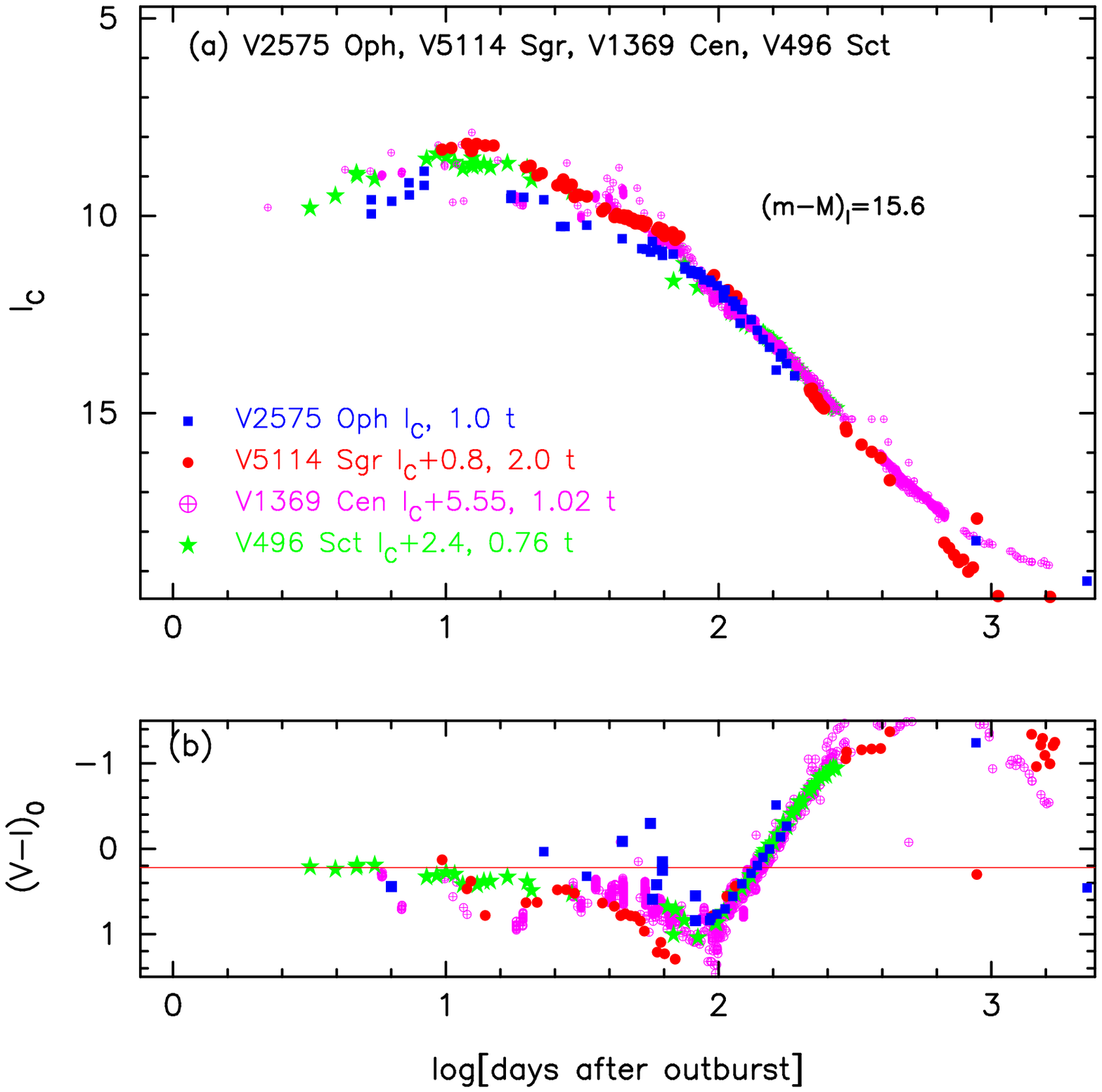}
\caption{
The (a) $I_{\rm C}$ light curve and (b) $(V-I_{\rm C})_0$ color curve
of V2575~Oph as well as those of V5114~Sgr, V1369~Cen, and V496~Sct.
\label{v2575_oph_v5114_sgr_v1369_cen_v496_sct_i_vi_color_logscale}}
\end{figure}


\begin{figure}
\plotone{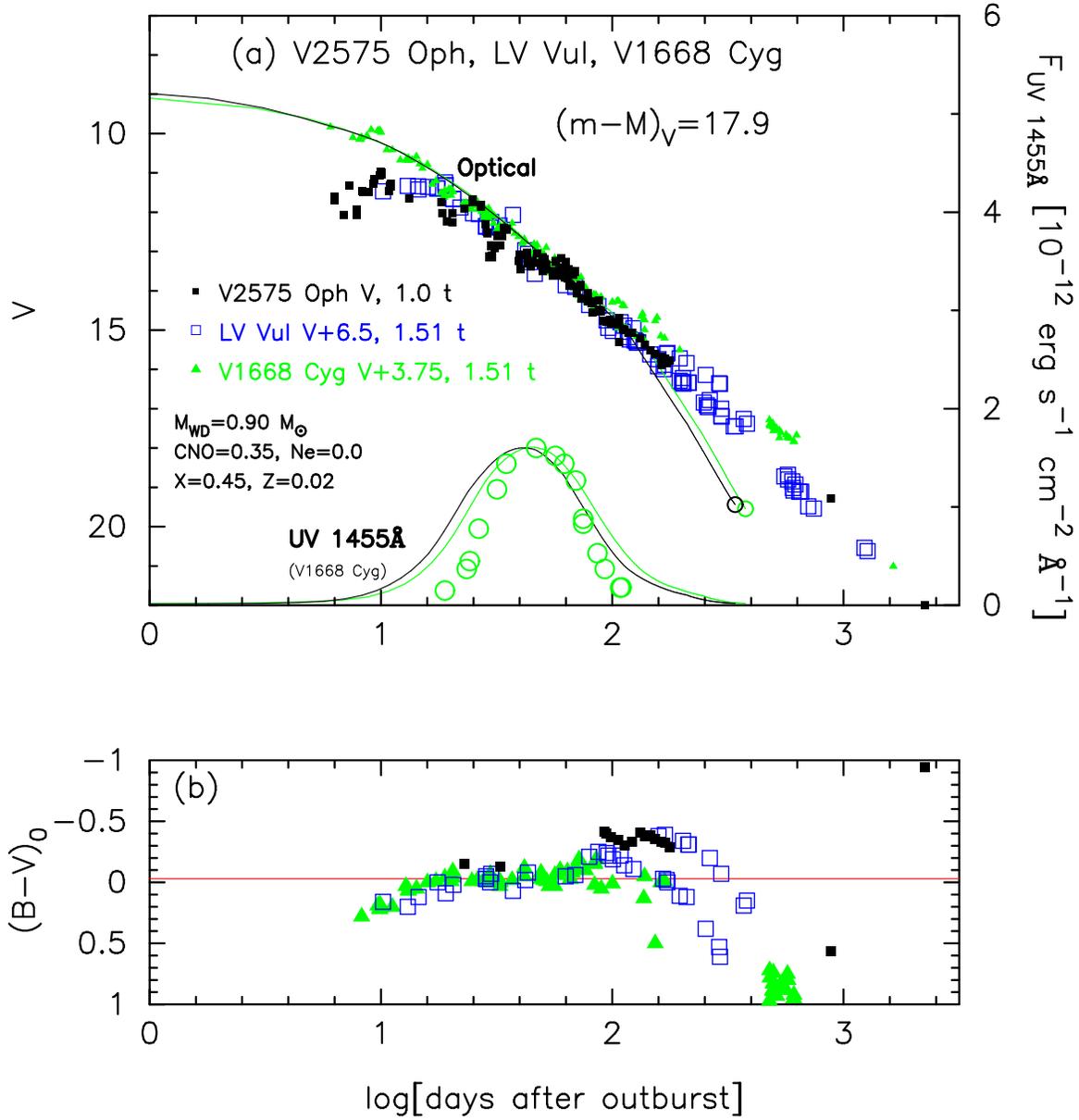}
\caption{
The (a) $V$ light curve and (b) $(B-V)_0$ color curve
of V2575~Oph as well as those of LV~Vul and V1668~Cyg.
In panel (a), we plot a $0.9~M_\sun$ WD model (CO3, solid black lines)
for V2575~Oph as well as a $0.98~M_\sun$ WD model (CO3, solid green lines)
for V1668~Cyg. 
\label{v2575_oph_v1668_cyg_lv_vul_v_bv_logscale_no2}}
\end{figure}


\begin{figure*}
\plottwo{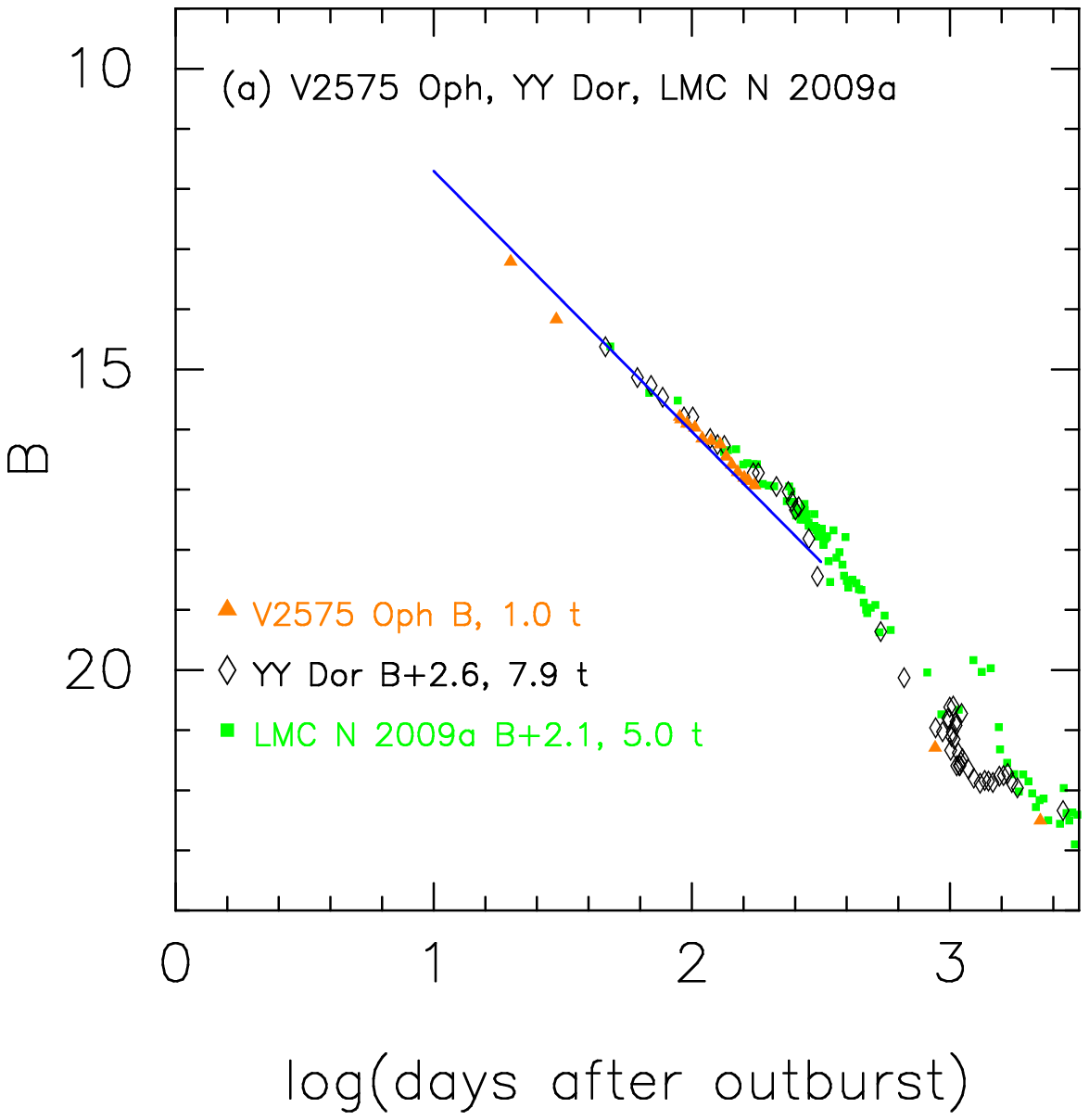}{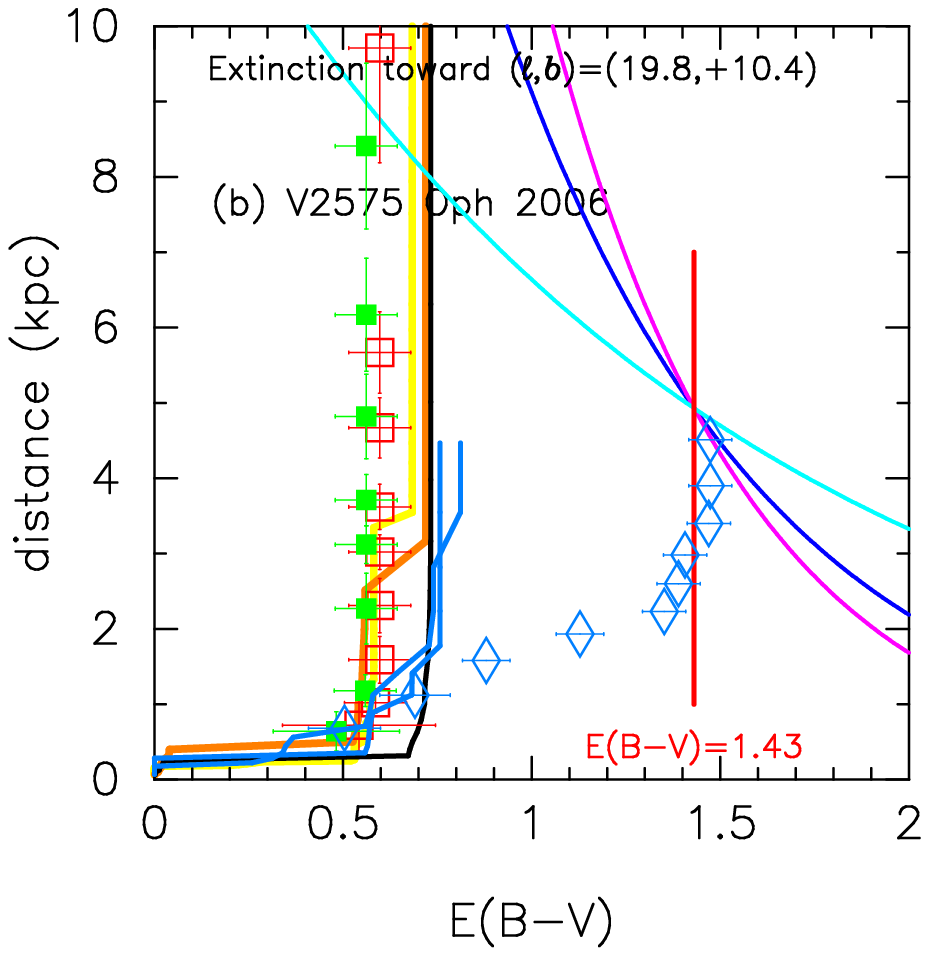}
\caption{
(a) The $B$ light curves of V2575~Oph
as well as the LMC novae YY~Dor and LMC~N~2009a.
The $BVI_{\rm C}$ data of V2575~Oph are taken from AAVSO, VSOLJ and SMARTS.
(b) Various distance-reddening relations toward V2575~Oph.
The thin solid lines of magenta, blue, and cyan denote the distance-reddening
relations given by $(m-M)_B= 19.33$, $(m-M)_V= 17.9$, 
and $(m-M)_I= 15.62$, respectively.
\label{distance_reddening_v2575_oph_bvi_xxxxxx}}
\end{figure*}

\subsection{V2575~Oph 2006\#1}
\label{v2575_oph_bvi}
We have reanalyzed the $BVI_{\rm C}$ multi-band light/color curves
of V2575~Oph based on the time-stretching method.  
The important revised point is the timescaling factor of $f_{\rm s}$,
which is changed from the previous $\log f_{\rm s}= +0.11$ to
the present $\log f_{\rm s}= +0.18$ in order to overlap the $V-I_{\rm C}$
color curves of V2575~Oph and other novae as shown in Figure
\ref{v2575_oph_v5114_sgr_v1369_cen_v496_sct_i_vi_color_logscale}(b).
Figure \ref{v2575_oph_v5114_sgr_v1369_cen_v496_sct_i_vi_color_logscale}
shows the (a) $I_{\rm C}$ light and (b) $V-I_{\rm C}$ color curves
of V2575~Oph as well as V5114~Sgr, V1369~Cen, and V496~Sct.
The $BVI_{\rm C}$ data of V2575~Oph are taken from VSOLJ and SMARTS.
Adopting the color excess of $E(B-V)= 1.43$ after \citet{hac19kb},
we obtain the timescaling factor $\log f_{\rm s}= +0.18$ of V2575~Oph.
We apply Equation (8) of \citet{hac19ka} for the $I$ band to Figure
\ref{v2575_oph_v5114_sgr_v1369_cen_v496_sct_i_vi_color_logscale}(a)
and obtain
\begin{eqnarray}
(m&-&M)_{I, \rm V2575~Oph} \cr
&=& ((m - M)_I + \Delta I_{\rm C})
_{\rm V5114~Sgr} - 2.5 \log 2.0 \cr
&=& 15.55 + 0.8\pm0.2 - 0.75 = 15.6\pm0.2 \cr
&=& ((m - M)_I + \Delta I_{\rm C})
_{\rm V1369~Cen} - 2.5 \log 1.02 \cr
&=& 10.11 + 5.55\pm0.2 - 0.025 = 15.64\pm0.2 \cr
&=& ((m - M)_I + \Delta I_{\rm C})
_{\rm V496~Sct} - 2.5 \log 0.76 \cr
&=& 12.9 + 2.4\pm0.2 + 0.3 = 15.65\pm0.2,
\label{distance_modulus_i_vi_v2575_oph}
\end{eqnarray}
where we adopt
$(m-M)_{I, \rm V5114~Sgr}=15.55$ from Appendix \ref{v5114_sgr_ubvi},
$(m-M)_{I, \rm V1369~Cen}=10.11$ from \citet{hac19ka}, and
$(m-M)_{I, \rm V496~Sct}=12.9$ in Appendix \ref{v496_sct_bvi}. 
Thus, we obtain $(m-M)_{I, \rm V2575~Oph}= 15.61\pm0.2$.

Figure \ref{v2575_oph_v1668_cyg_lv_vul_v_bv_logscale_no2}
shows the (a) $V$ and (b) $(B-V)_0$ evolutions of V2575~Oph
as well as LV~Vul and V1668~Cyg.  
Applying Equation (4) of \citet{hac19ka} for the $V$ band to them,
we have the relation
\begin{eqnarray}
(m&-&M)_{V, \rm V2575~Oph} \cr
&=& ((m - M)_V + \Delta V)_{\rm LV~Vul} - 2.5 \log 1.51 \cr
&=& 11.85 + 6.5\pm0.2 - 0.45 = 17.9\pm0.2 \cr
&=& ((m - M)_V + \Delta V)_{\rm V1668~Cyg} - 2.5 \log 1.51 \cr
&=& 14.6 + 3.75\pm0.2 - 0.45 = 17.9\pm0.2,
\label{distance_modulus_v_bv_v2575_oph}
\end{eqnarray}
where we adopt $(m-M)_{V, \rm LV~Vul}=11.85$ and
$(m-M)_{V, \rm V1668~Cyg}=14.6$, both from \citet{hac19ka}.
Thus, we obtain $(m-M)_V=17.9\pm0.1$ for V2575~Oph.

Figure \ref{distance_reddening_v2575_oph_bvi_xxxxxx}(a) shows the $B$ light
curves of V2575~Oph together with those of YY~Dor and LMC~N~2009a.
We apply Equation (7) of \citet{hac19ka} for the $B$ band to 
Figure \ref{distance_reddening_v2575_oph_bvi_xxxxxx}(a) and obtain
\begin{eqnarray}
(m&-&M)_{B, \rm V2575~Oph} \cr
&=& ((m - M)_B + \Delta B)_{\rm YY~Dor} - 2.5 \log 7.9 \cr
&=& 18.98 + 2.6\pm0.2 - 2.25 = 19.33\pm0.2 \cr
&=& ((m - M)_B + \Delta B)_{\rm LMC~N~2009a} - 2.5 \log 5.0 \cr
&=& 18.98 + 2.1\pm0.2 - 1.75 = 19.33\pm0.2.
\label{distance_modulus_b_v2575_oph_yy_dor_lmcn2009a}
\end{eqnarray}
Thus, we obtain $(m-M)_{B, \rm V2575~Oph}= 19.33\pm0.1$.

We plot the three distance moduli in Figure
\ref{distance_reddening_v2575_oph_bvi_xxxxxx}(b),
which cross at $d=4.9$~kpc and $E(B-V)=1.43$. 
The crossing point is consistent with the distance-reddening relation
(unfilled cyan-blue diamonds) given by \citet{ozd18}.
Thus, we obtain $E(B-V)=1.43\pm0.05$ and $d=4.9\pm0.5$~kpc.


\begin{figure}
\plotone{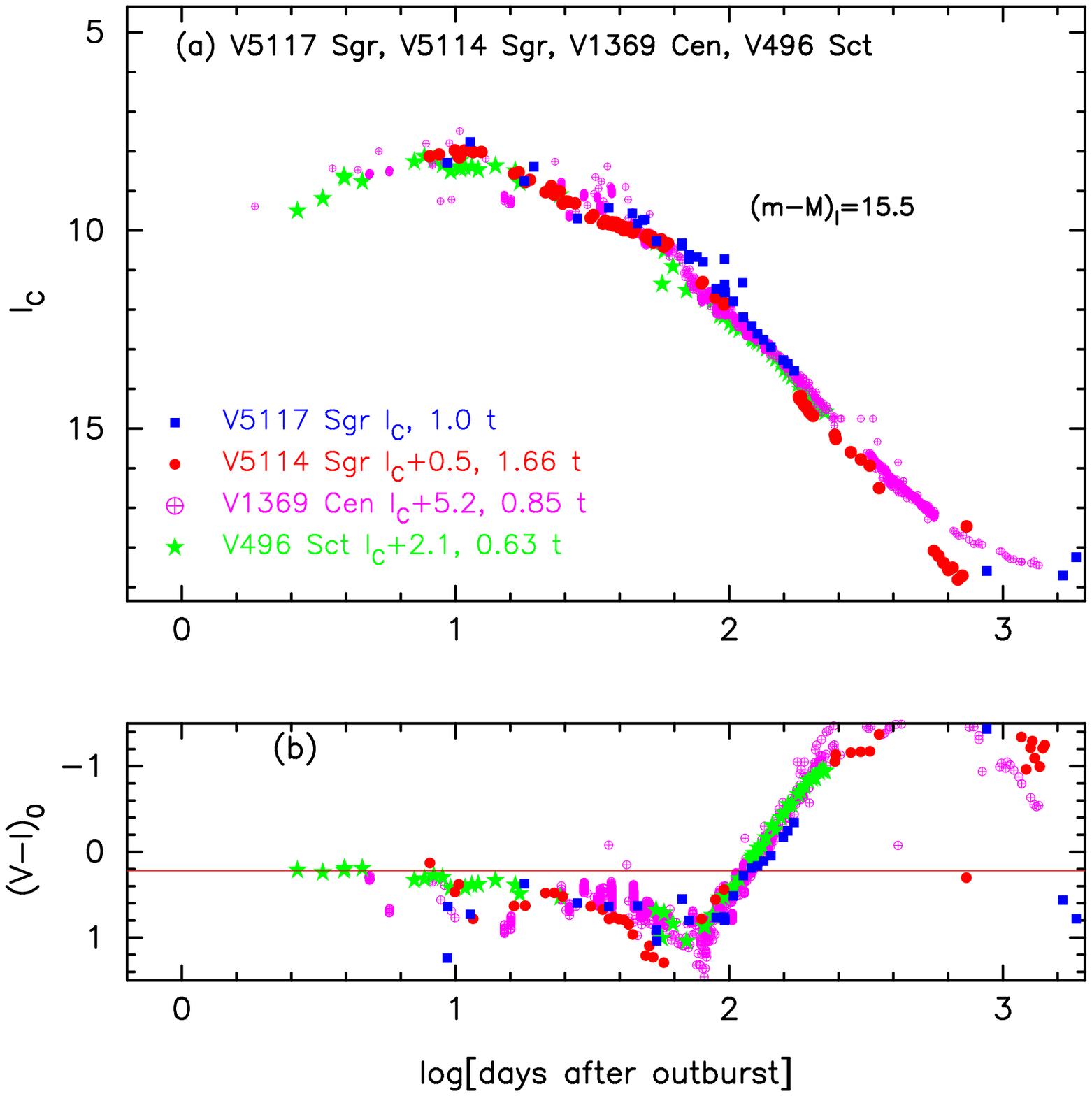}
\caption{
The (a) $I_{\rm C}$ light and (b) $(V-I_{\rm C})_0$ color curve of
V5117~Sgr as well as those of V5114~Sgr, V1369~Cen, and V496~Sct.
\label{v5117_sgr_v5114_sgr_v1369_cen_v496_sct_i_vi_color_logscale}}
\end{figure}


\begin{figure}
\plotone{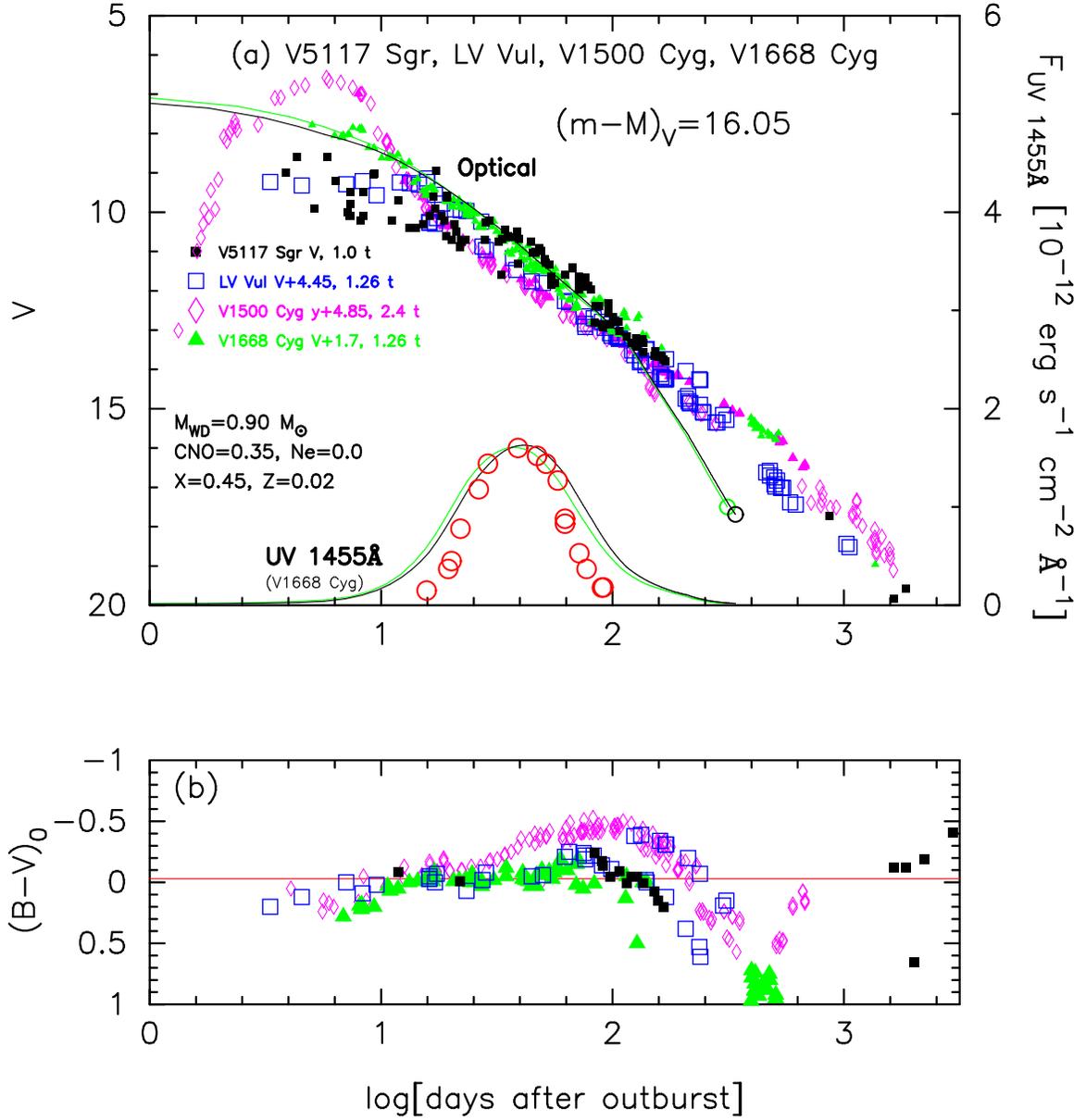}
\caption{
The (a) $V$ light and (b) $(B-V)_0$ color curves of V5117~Sgr
as well as those of LV~Vul, V1500~Cyg, and V1668~Cyg.
In panel (a), we show a $0.90~M_\sun$ WD model (CO3, thin solid black lines)
for V5117~Sgr as well as a $0.98~M_\sun$ WD model (CO3, 
thin solid green lines) for V1668~Cyg.
\label{v5117_sgr_v1668_cyg_v1500_cyg_lv_vul_v_bv_ub_logscale}}
\end{figure}


\begin{figure*}
\plottwo{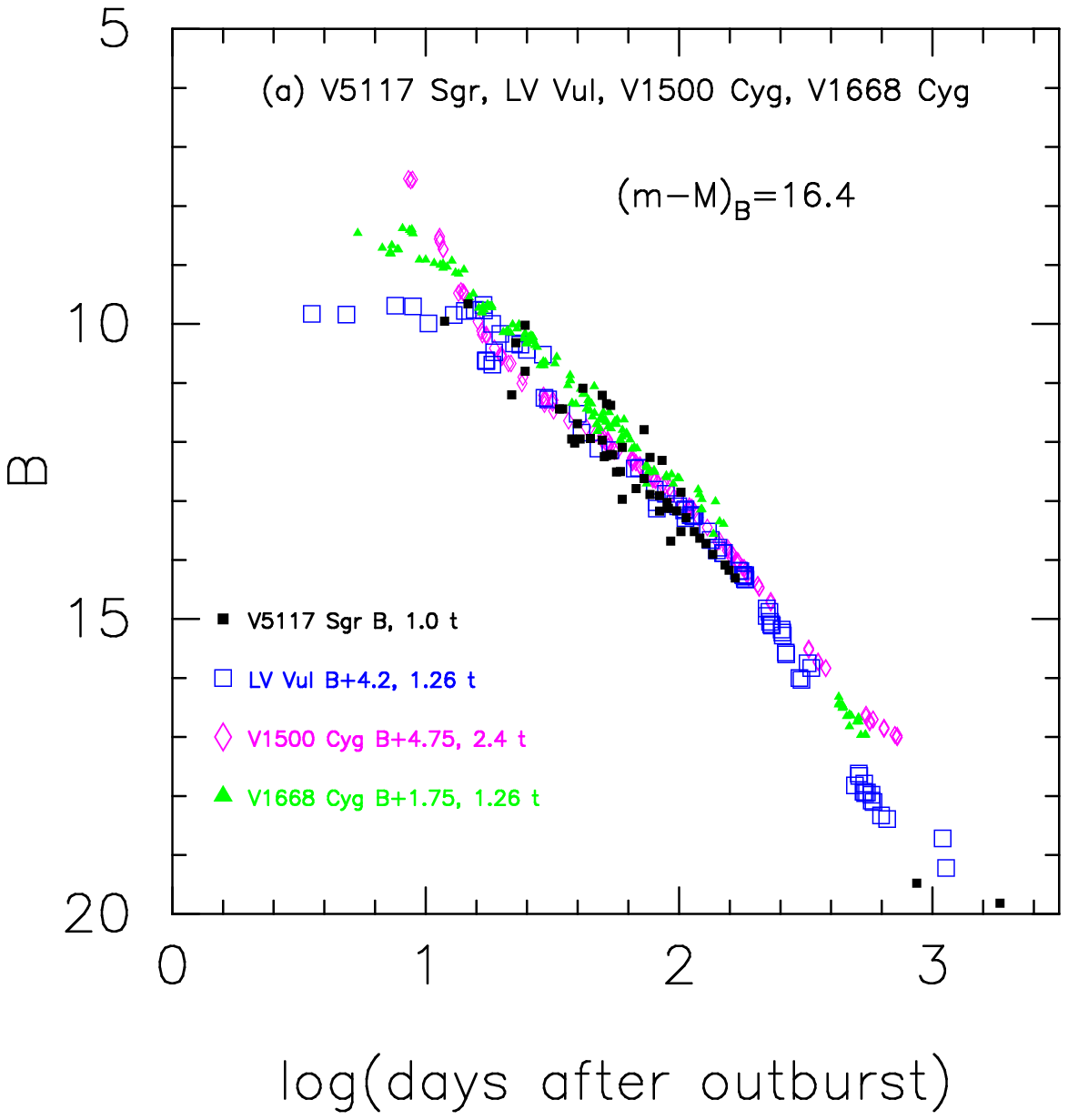}{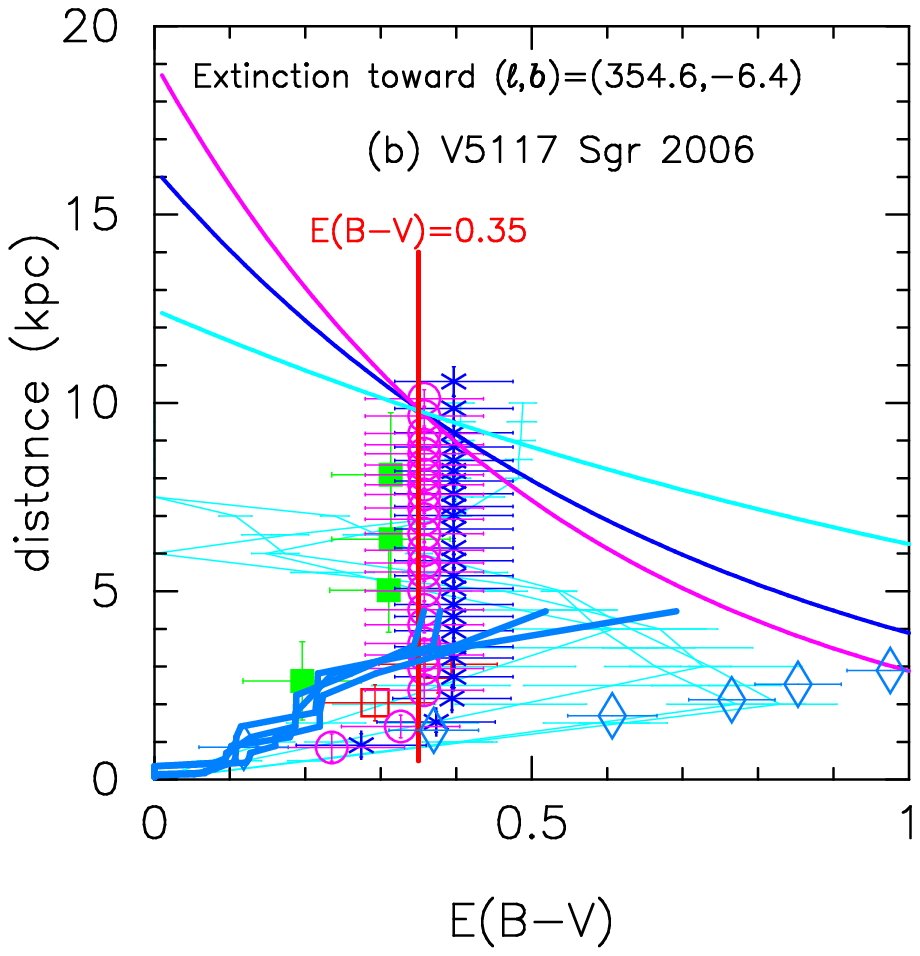}
\caption{
(a) The $B$ light curves of V5117~Sgr as well as LV~Vul, V1500~Cyg, 
and V1668~Cyg.  
(b) Various distance-reddening relations toward V5117~Sgr.
The thin solid lines of magenta, blue, and cyan denote the distance-reddening
relations given by $(m-M)_B= 16.4$, $(m-M)_V= 16.05$, and $(m-M)_I= 15.49$,
respectively.  
\label{distance_reddening_v5117_sgr_bvi_xxxxxx}}
\end{figure*}

\subsection{V5117~Sgr 2006}
\label{v5117_sgr_bvi}
We have reanalyzed the $BVI_{\rm C}$ multi-band light/color curves
of V5117~Sgr and obtained the new parameters.
Figure \ref{v5117_sgr_v5114_sgr_v1369_cen_v496_sct_i_vi_color_logscale}
shows the (a) $I_{\rm C}$ light and (b) $(V-I_{\rm C})_0$ color curves
of V5117~Sgr as well as V5114~Sgr, V1369~Cen, and V496~Sct.
The $BVI_{\rm C}$ data of V5117~Sgr are taken from VSOLJ and SMARTS.
Adopting the color excess of $E(B-V)= 0.35$ as mentioned below,
we obtain the timescaling factor $\log f_{\rm s}= +0.10$ for V5117~Sgr.
This is because the $(V-I)_0$ color evolution of V5117~Sgr overlaps with
the other novae as much as possible, as shown in Figure
\ref{v5117_sgr_v5114_sgr_v1369_cen_v496_sct_i_vi_color_logscale}(b).  
Then, we apply Equation (8) of \citet{hac19ka} for the $I$ band to Figure
\ref{v5117_sgr_v5114_sgr_v1369_cen_v496_sct_i_vi_color_logscale}(a)
and obtain
\begin{eqnarray}
(m&-&M)_{I, \rm V5117~Sgr} \cr
&=& ((m - M)_I + \Delta I_{\rm C})
_{\rm V5114~Sgr} - 2.5 \log 1.66 \cr
&=& 15.55 + 0.5\pm0.2 - 0.55 = 15.5\pm0.2 \cr
&=& ((m - M)_I + \Delta I_{\rm C})
_{\rm V1369~Cen} - 2.5 \log 0.85 \cr
&=& 10.11 + 5.2\pm0.2 + 0.175 = 15.49\pm0.2 \cr
&=& ((m - M)_I + \Delta I_{\rm C})
_{\rm V496~Sct} - 2.5 \log 0.63 \cr
&=& 12.9 + 2.1\pm0.2 + 0.5 = 15.5\pm0.2,
\label{distance_modulus_i_vi_v5117_sgr}
\end{eqnarray}
where we adopt
$(m-M)_{I, \rm V5114~Sgr}=15.55$ from Appendix \ref{v5114_sgr_ubvi},
$(m-M)_{I, \rm V1369~Cen}=10.11$ from \citet{hac19ka}, and
$(m-M)_{I, \rm V496~Sct}=12.9$ in Appendix \ref{v496_sct_bvi}.
Thus, we obtain $(m-M)_{I, \rm V5117~Sgr}= 15.5\pm0.2$.

Figure \ref{v5117_sgr_v1668_cyg_v1500_cyg_lv_vul_v_bv_ub_logscale}
shows the (a) $V$ light and (b) $(B-V)_0$ color curves of V5117~Sgr
as well as LV~Vul, V1500~Cyg, and V1668~Cyg.  
We apply Equation (4) of \citet{hac19ka} to Figure 
\ref{v5117_sgr_v1668_cyg_v1500_cyg_lv_vul_v_bv_ub_logscale}(a)
and obtain
\begin{eqnarray}
(m&-&M)_{V, \rm V5117~Sgr} \cr
&=& ((m - M)_V + \Delta V)_{\rm LV~Vul} - 2.5 \log 1.26 \cr
&=& 11.85 + 4.45\pm0.2 - 0.25 = 16.05\pm0.2 \cr
&=& ((m - M)_V + \Delta V)_{\rm V1500~Cyg} - 2.5 \log 2.4 \cr
&=& 12.15 + 4.85\pm0.2 - 0.95 = 16.05\pm0.2 \cr
&=& ((m - M)_V + \Delta V)_{\rm V1668~Cyg} - 2.5 \log 1.26 \cr
&=& 14.6 + 1.7\pm0.2 - 0.25 = 16.05\pm0.2,
\label{distance_modulus_v_bv_v5117_sgr}
\end{eqnarray}
where we adopt
$(m-M)_{V, \rm LV~Vul}=11.85$ and $(m-M)_{V, \rm V1668~Cyg}=14.6$
both from \citet{hac19ka}, and 
$(m-M)_{V, \rm V1500~Cyg}=12.15$ from Appendix \ref{v1500_cyg_ubvi}. 
Thus, we obtain $(m-M)_{V, \rm V5117~Sgr}= 16.05\pm0.1$.

Figure \ref{distance_reddening_v5117_sgr_bvi_xxxxxx}(a)
shows the $B$ light curves of V5117~Sgr
as well as LV~Vul, V1500~Cyg, and V1668~Cyg.
We apply Equation (7) of \citet{hac19ka} to Figure 
\ref{distance_reddening_v5117_sgr_bvi_xxxxxx}(a)
and obtain
\begin{eqnarray}
(m&-&M)_{B, \rm V5117~Sgr} \cr
&=& ((m - M)_B + \Delta B)_{\rm LV~Vul} - 2.5 \log 1.26 \cr
&=& 12.45 + 4.2\pm0.2 - 0.25 = 16.4\pm0.2 \cr
&=& ((m - M)_B + \Delta B)_{\rm V1500~Cyg} - 2.5 \log 2.4 \cr
&=& 12.6 + 4.75\pm0.2 - 0.95 = 16.4\pm0.2 \cr
&=& ((m - M)_B + \Delta B)_{\rm V1668~Cyg} - 2.5 \log 1.26 \cr
&=& 14.9 + 1.75\pm0.2 - 0.25 = 16.4\pm0.2, 
\label{distance_modulus_b_v5117_sgr_lv_vul_v1500_cyg_v1668_cyg}
\end{eqnarray}
where we adopt
$(m-M)_{B, \rm LV~Vul}=12.45$ and $(m-M)_{B, \rm V1668~Cyg}=14.9$ both
from \citet{hac19ka}, and $(m-M)_{B, \rm V1500~Cyg}=12.15 + 0.45= 12.6$ 
from Appendix \ref{v1500_cyg_ubvi}.
Thus, we obtain $(m-M)_{B, \rm V5117~Sgr}= 16.4\pm0.1$.

We plot $(m-M)_B= 16.4$, $(m-M)_V= 16.05$, and $(m-M)_I= 15.49$,
which cross at $d=9.8$~kpc and $E(B-V)=0.35$, in Figure
\ref{distance_reddening_v5117_sgr_bvi_xxxxxx}(b).
The crossing point is consistent with the distance-reddening relations
given by \citet{mar06} and \citet{chen19}.
Thus, we obtain $E(B-V)=0.35\pm0.05$ and $d=9.8\pm1$~kpc.


\begin{figure}
\plotone{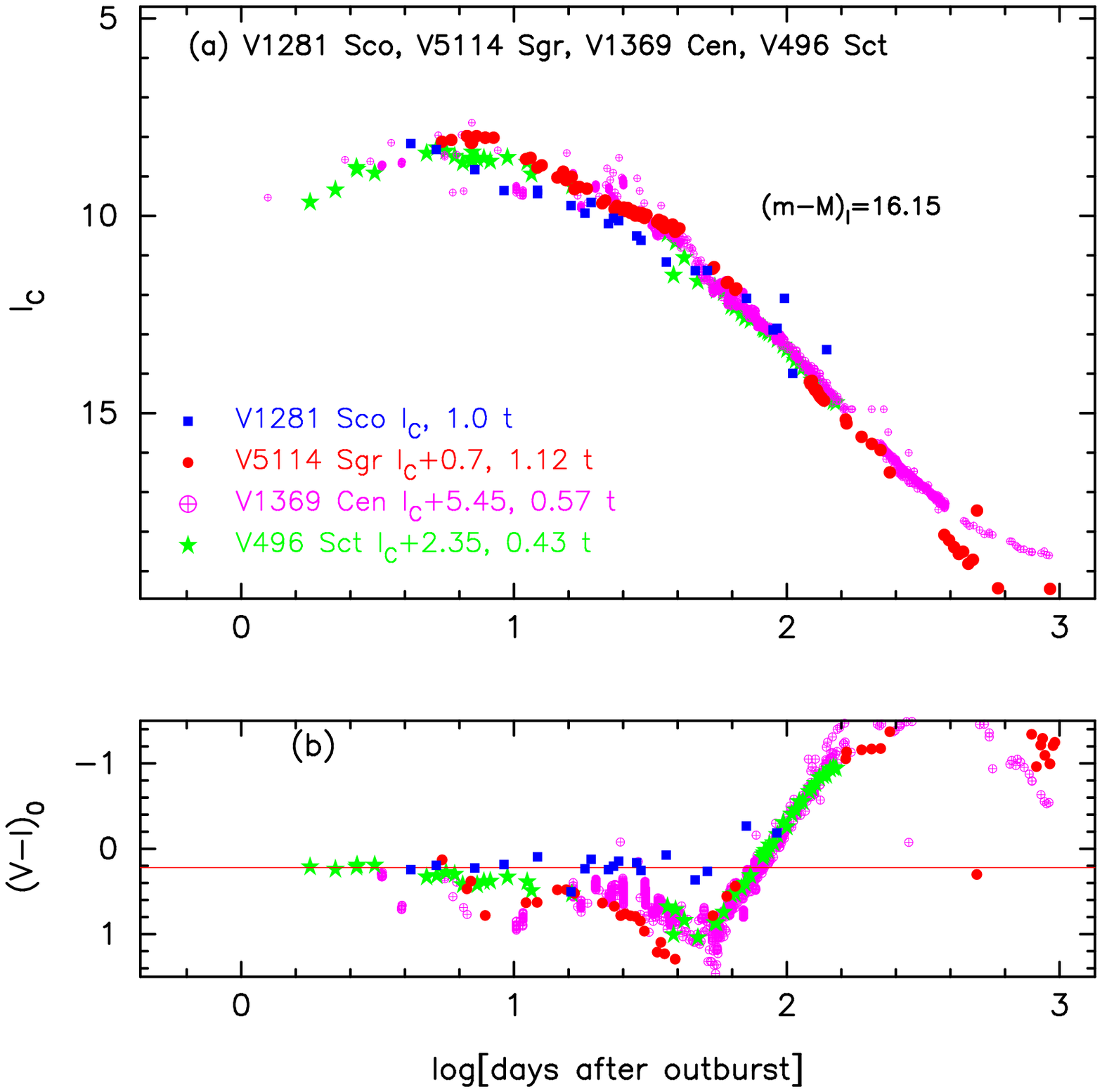}
\caption{
The (a) $I_{\rm C}$ light curve and (b) $(V-I_{\rm C})_0$ color curve
of V1281~Sco as well as those of V5114~Sgr, V1369~Cen, and V496~Sct.
\label{v1281_sco_v5114_sgr_v1369_cen_v496_sct_i_vi_color_logscale}}
\end{figure}


\begin{figure}
\plotone{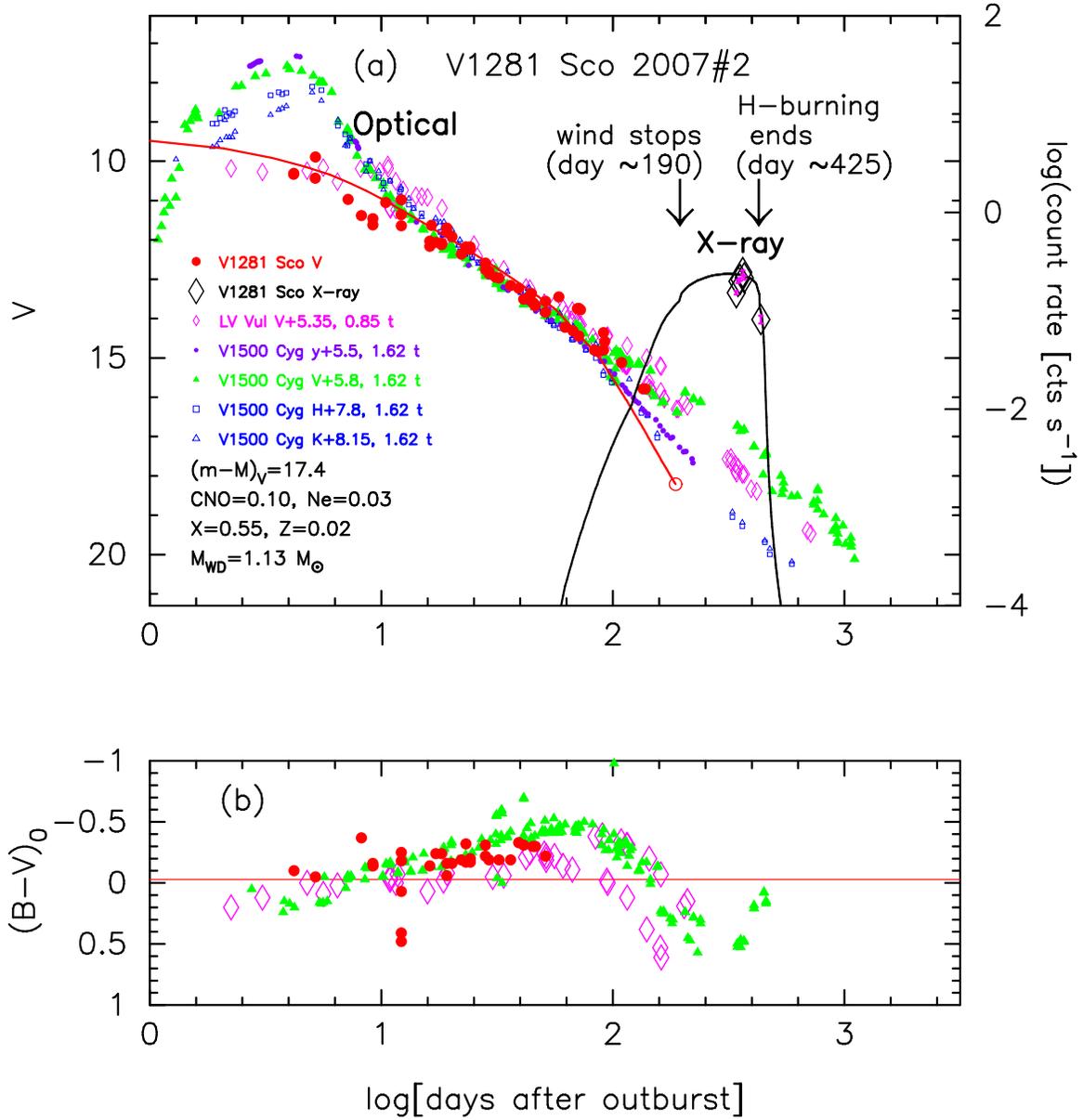}
\caption{
The (a) $V$ light curve and (b) $(B-V)_0$ color curve of V1281~Sco
as well as those of LV~Vul and V1500~Cyg.  In panel (a), 
we show a $1.13~M_\sun$ WD model (Ne2, solid red and black lines)
for V1281~Sco.
\label{v1281_sco_v1500_cyg_v_bv_m1130_x55z02o10ne03_logscale_no2}}
\end{figure}


\begin{figure}
\plottwo{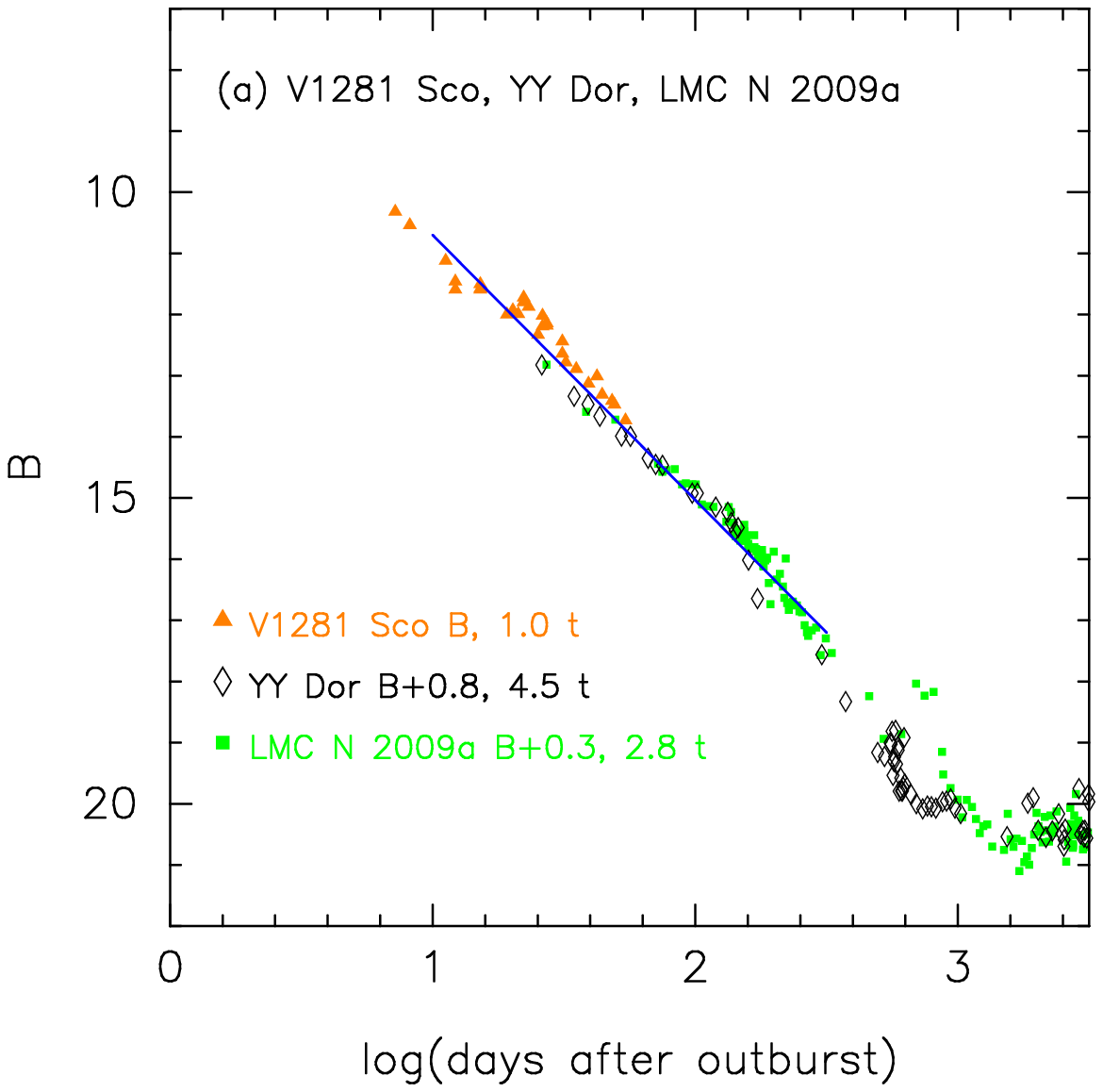}{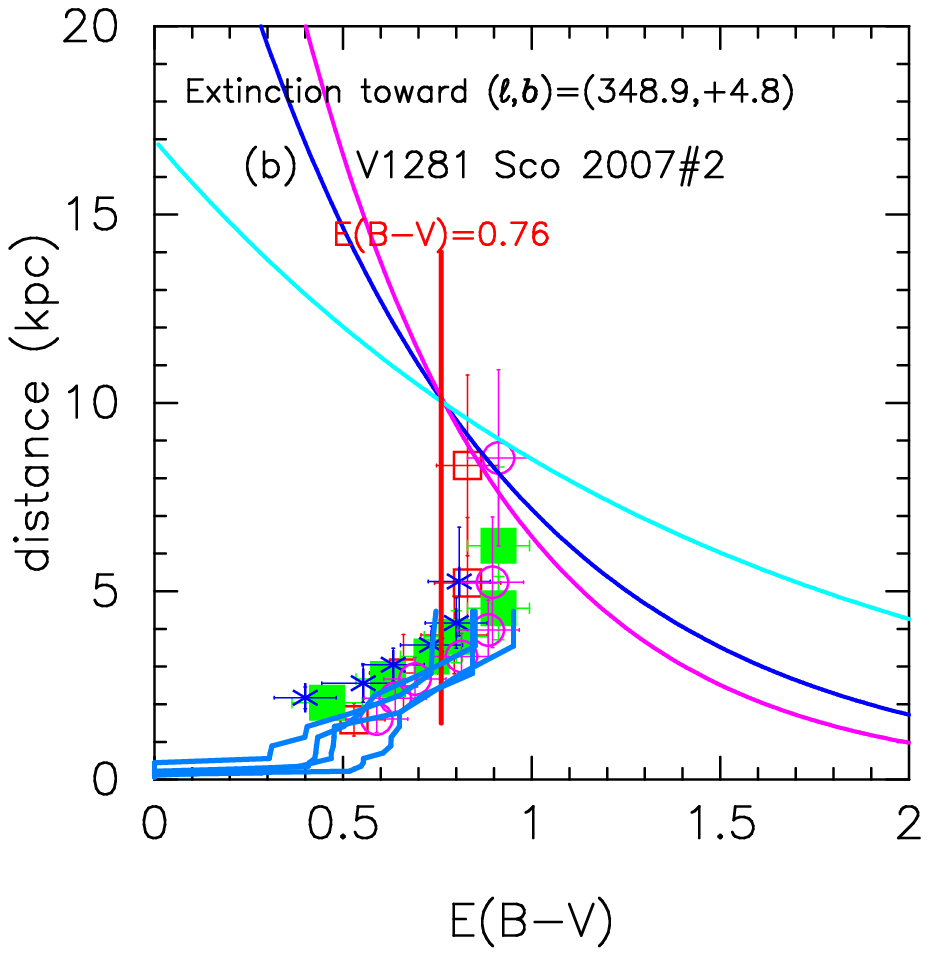}
\caption{
(a) The $B$ light curves of V1281~Sco, YY~Dor, and LMC~N~2009a.
The $B$ data of V1281~Sco are taken from VSOLJ.
(b) Various distance-reddening relations toward V1281~Sco.
The thin solid lines of magenta, blue, and cyan denote the distance-reddening
relations given by $(m-M)_B= 18.15$, $(m-M)_V= 17.38$, 
and $(m-M)_I= 16.15$, respectively.
\label{v1281_sco_yy_dor_lmcn_2009a_b_only_logscale}}
\end{figure}

\subsection{V1281~Sco 2007}
\label{v1281_sco_bvi}
We have reanalyzed the $BVI_{\rm C}$ multi-band light/color curves
of V1281~Sco based on the time-stretching method.  
Figure \ref{v1281_sco_v5114_sgr_v1369_cen_v496_sct_i_vi_color_logscale}
shows the (a) $I_{\rm C}$ light and (b) $(V-I_{\rm C})_0$ color curves
of V1281~Sco as well as V5114~Sgr, V1369~Cen, and V496~Sct.
The $BVI_{\rm C}$ data of V1281~Sco are taken from VSOLJ.
Adopting the color excess of $E(B-V)= 0.76$ as mentioned below,
we obtain the timescaling factor $\log f_{\rm s}= -0.07$ for V1281~Sco.
We apply Equation (8) of \citet{hac19ka} for the $I$ band to Figure
\ref{v1281_sco_v5114_sgr_v1369_cen_v496_sct_i_vi_color_logscale}(a)
and obtain
\begin{eqnarray}
(m&-&M)_{I, \rm V1281~Sco} \cr
&=& ((m - M)_I + \Delta I_{\rm C})
_{\rm V5114~Sgr} - 2.5 \log 1.12 \cr
&=& 15.55 + 0.7\pm0.2 - 0.125  = 16.13\pm0.2 \cr
&=& ((m - M)_I + \Delta I_{\rm C})
_{\rm V1369~Cen} - 2.5 \log 0.57 \cr
&=& 10.11 + 5.45\pm0.2 + 0.6 = 16.16\pm0.2 \cr
&=& ((m - M)_I + \Delta I_{\rm C})
_{\rm V496~Sct} - 2.5 \log 0.43 \cr
&=& 12.9 + 2.35\pm0.2 + 0.925 = 16.17\pm0.2,
\label{distance_modulus_i_vi_v1281_sco}
\end{eqnarray}
where we adopt
$(m-M)_{I, \rm V5114~Sgr}=15.55$ from Appendix \ref{v5114_sgr_ubvi},
$(m-M)_{I, \rm V1369~Cen}=10.11$ from \citet{hac19ka}, and
$(m-M)_{I, \rm V496~Sct}=12.9$ in Appendix \ref{v496_sct_bvi}.
Thus, we obtain $(m-M)_{I, \rm V1281~Sco}= 16.15\pm0.2$.

Figure \ref{v1281_sco_v1500_cyg_v_bv_m1130_x55z02o10ne03_logscale_no2}
shows the (a) $V$ and (b) $(B-V)_0$ evolutions of V1281~Sco
as well as LV~Vul and V1500~Cyg.  
Applying Equation (4) of \citet{hac19ka} for the $V$ band to them,
we have the relation
\begin{eqnarray}
(m&-&M)_{V, \rm V1281~Sco} \cr
&=& ((m - M)_V + \Delta V)_{\rm LV~Vul} - 2.5 \log 0.85 \cr
&=& 11.85 + 5.35\pm0.2 + 0.175 = 17.38\pm0.2 \cr
&=& ((m - M)_V + \Delta V)_{\rm V1500~Cyg} - 2.5 \log 1.62 \cr
&=& 12.15 + 5.8\pm0.2 - 0.525 = 17.38\pm0.2,
\label{distance_modulus_v_bv_v1281_sco}
\end{eqnarray}
where we adopt $(m-M)_{V, \rm LV~Vul}=11.85$ from \citet{hac19ka}
and $(m-M)_{V, \rm V1500~Cyg}=12.15$ in Section \ref{v1500_cyg_ubvi}.
Thus, we obtain $(m-M)_V=17.38\pm0.1$ for V1281~Sco, which is 
consistent with Hachisu \& Kato's (2019b) results.

Figure \ref{v1281_sco_yy_dor_lmcn_2009a_b_only_logscale}(a) shows
the $B$ light curves of V1281~Sco
together with those of YY~Dor and LMC~N~2009a.
We apply Equation (7) of \citet{hac19ka} for the $B$ band to Figure
\ref{v1281_sco_yy_dor_lmcn_2009a_b_only_logscale}(a)
and obtain
\begin{eqnarray}
(m&-&M)_{B, \rm V1281~Sco} \cr
&=& ((m - M)_B + \Delta B)_{\rm YY~Dor} - 2.5 \log 4.5 \cr
&=& 18.98 + 0.8\pm0.2 - 1.63 = 18.15\pm0.2 \cr
&=& ((m - M)_B + \Delta B)_{\rm LMC~N~2009a} - 2.5 \log 2.8 \cr
&=& 18.98 + 0.3\pm0.2 - 1.13 = 18.15\pm0.2.
\label{distance_modulus_b_v1281_sco_yy_dor_lmcn2009a}
\end{eqnarray}
Thus, we have $(m-M)_{B, \rm V1281~Sco}= 18.15\pm0.1$.

We plot $(m-M)_B= 18.15$, $(m-M)_V= 17.38$, and $(m-M)_I= 16.15$,
which broadly cross at $d=10$~kpc and $E(B-V)=0.76$, in Figure
\ref{v1281_sco_yy_dor_lmcn_2009a_b_only_logscale}(b).
The crossing point is consistent with the distance-reddening relations
given by \citet{mar06} and \citet{chen19}.
Thus, we have $E(B-V)=0.76\pm0.05$ and $d=10\pm1$~kpc.

\begin{figure}
\plotone{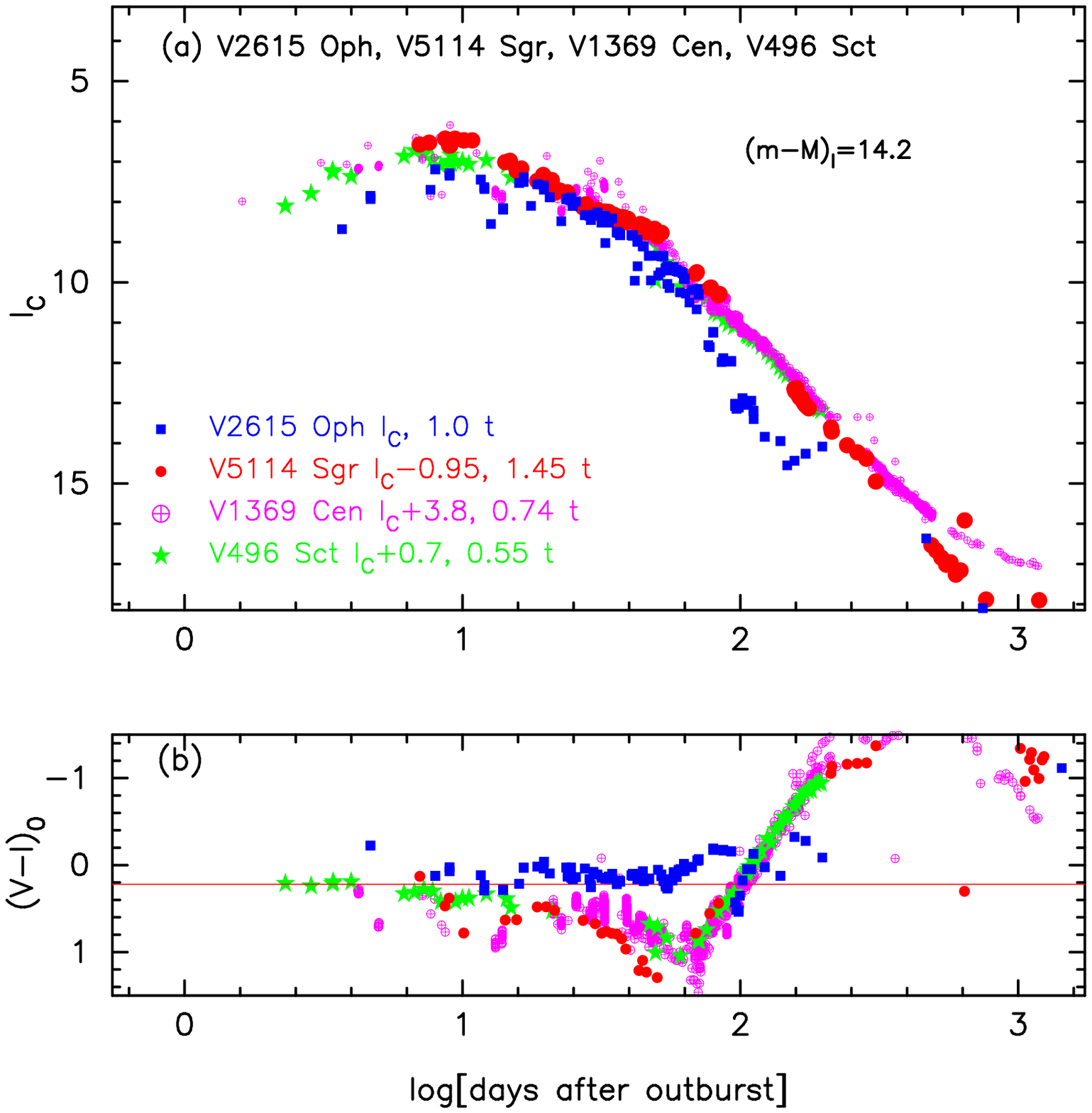}
\caption{
The (a) $I_{\rm C}$ light curve and (b) $(V-I_{\rm C})_0$ color curve
of V2615~Oph as well as those of V5114~Sgr, V1369~Cen, and V496~Sct.
The $BVI_{\rm C}$ data of V2615~Oph are taken from \citet{mun08a}.
\label{v2615_oph_v5114_sgr_v1369_cen_v496_sct_i_vi_color_logscale}}
\end{figure}


\begin{figure}
\plotone{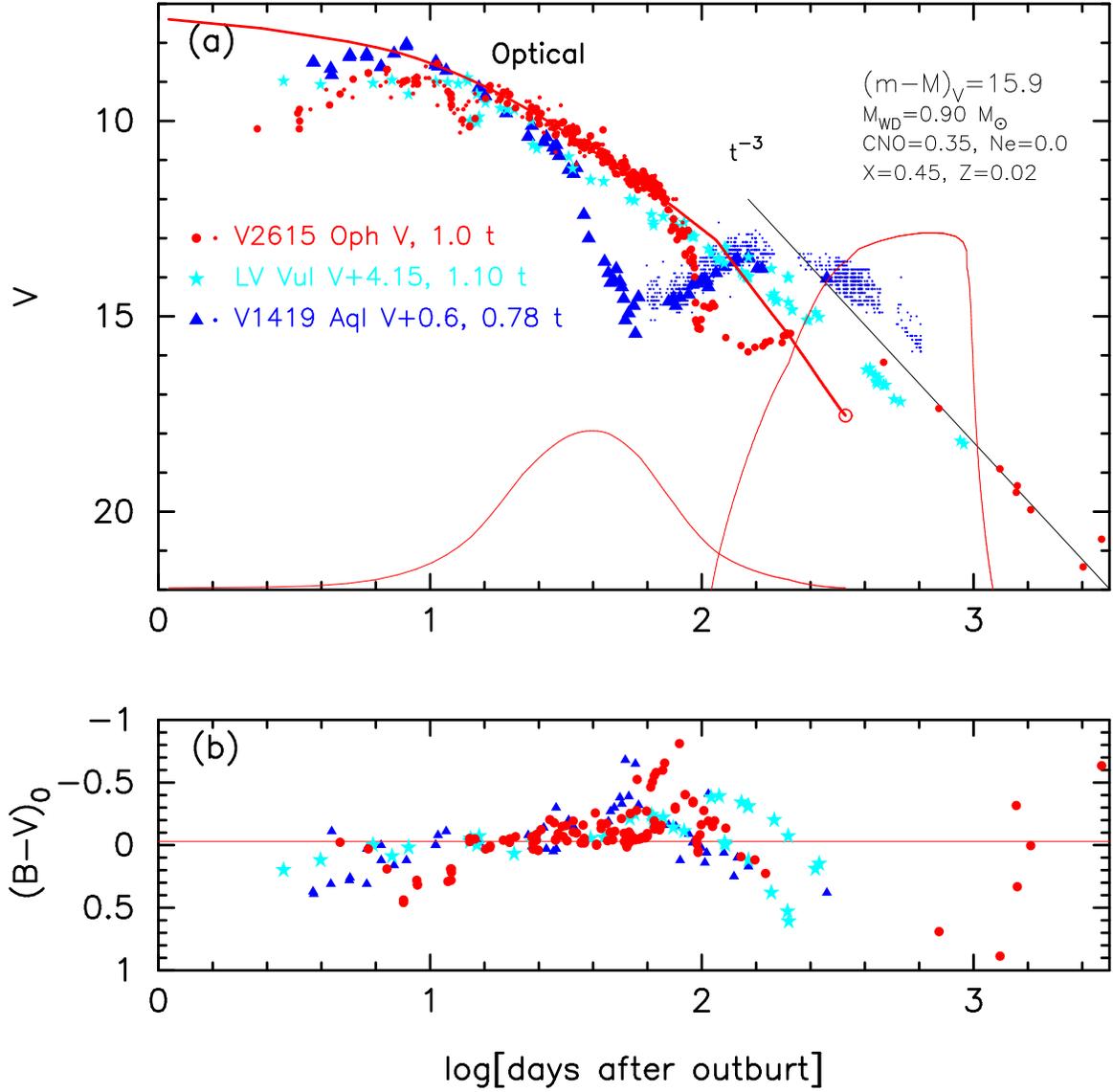}
\caption{
The (a) $V$ light and (b) $(B-V)_0$ color curves of V2615~Oph 
as well as those of LV~Vul and V1419~Aql.
In panel (a), we plot a $0.90~M_\sun$ WD model (CO3, 
solid red lines) for V2615~Oph. 
\label{v2615_oph_lv_vul_v1419_aql_v_bv_logscale_no2}}
\end{figure}


\begin{figure}
\plottwo{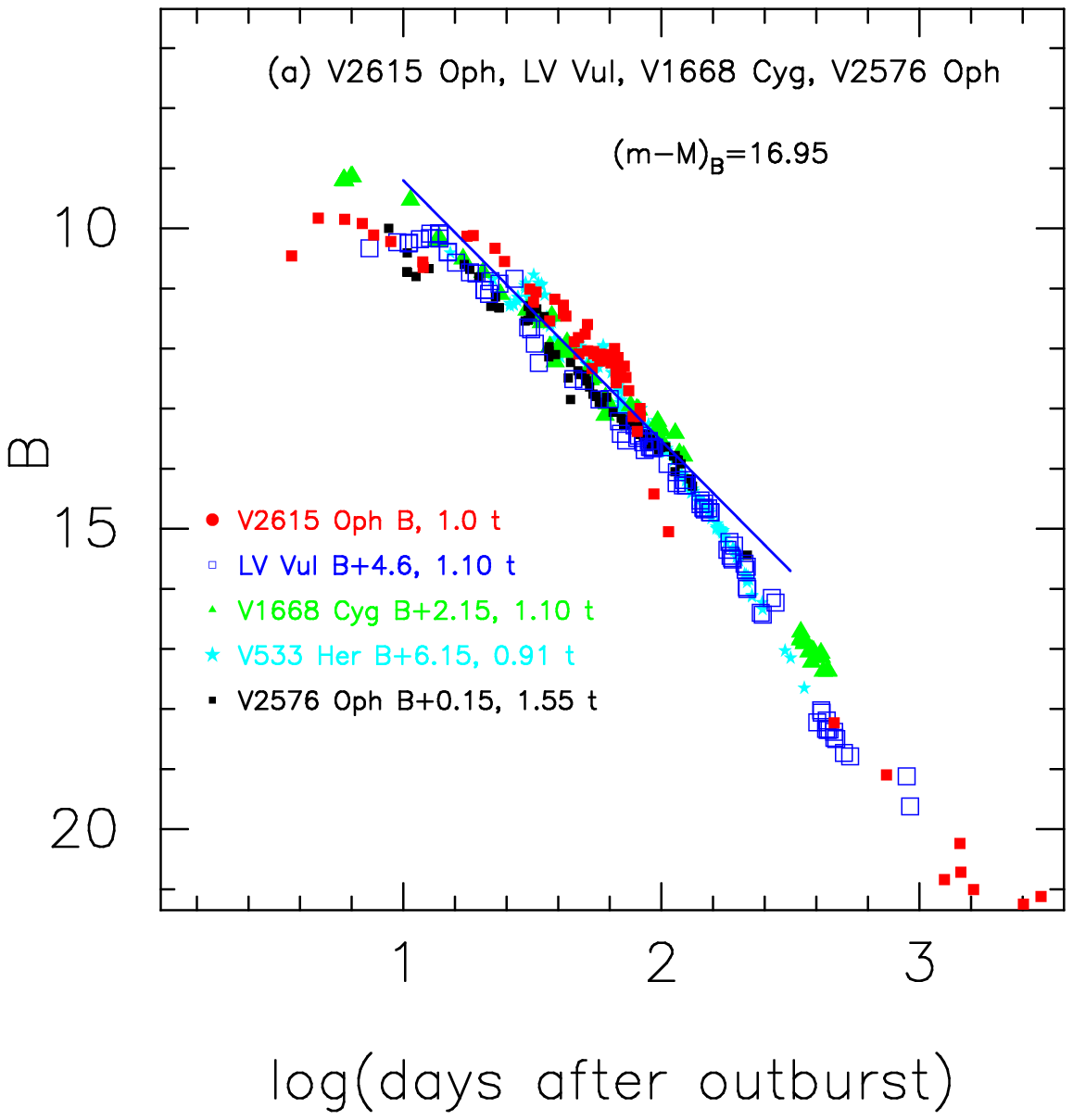}{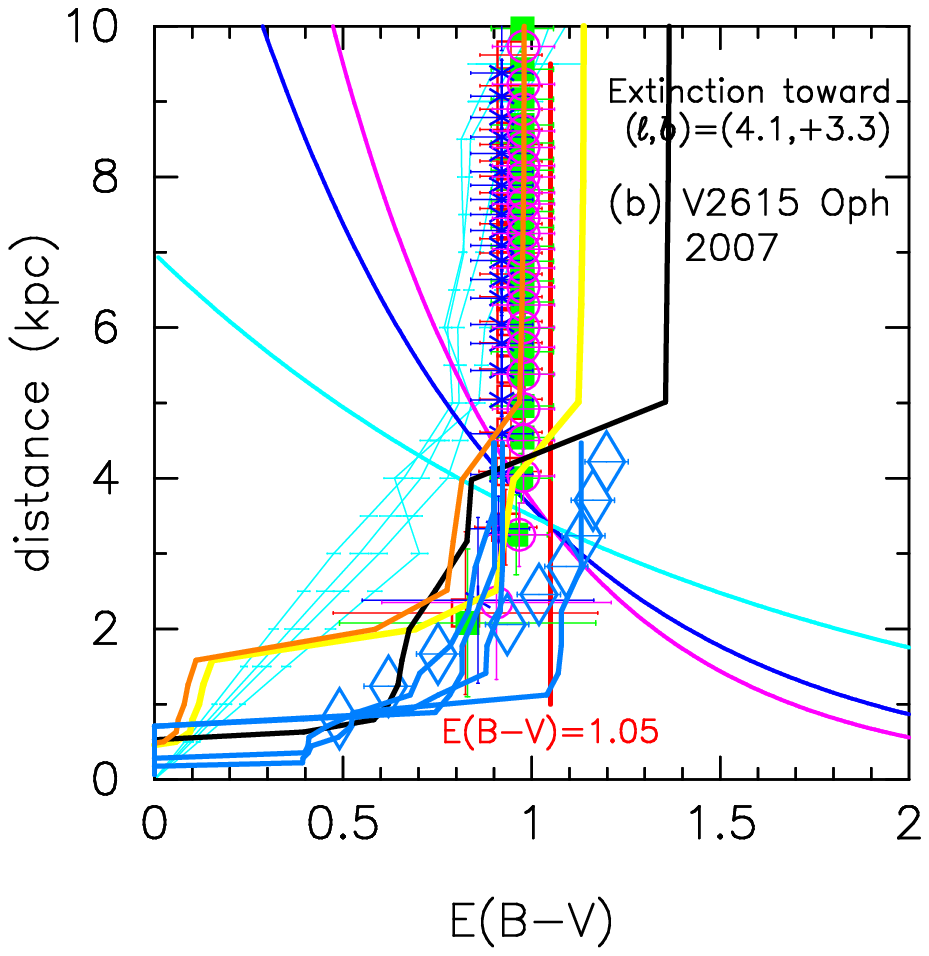}
\caption{
(a) The $B$ light curves of V2615~Oph, LV~Vul, V1668~Cyg, V533~Her,
and V2576~Oph.
The $B$ data of V2615~Oph are taken from AAVSO, VSOLJ, SMARTS, and
\citet{mun08a}.
(b) Various distance-reddening relations toward V2615~Oph.
The thin solid lines of magenta, blue, and cyan denote the distance-reddening
relations given by $(m-M)_B= 16.94$, $(m-M)_V= 15.89$, 
and $(m-M)_I= 14.22$, respectively.
\label{v2615_oph_v2576_oph_v1668_cyg_lv_vul_b_only_logscale}}
\end{figure}

\subsection{V2615~Oph 2007}
\label{v2615_oph_bvi}
We have reanalyzed the $BVI_{\rm C}$ multi-band 
light/color curves of V2615~Oph based on the time-stretching method.  
Figure \ref{v2615_oph_v5114_sgr_v1369_cen_v496_sct_i_vi_color_logscale}
shows the (a) $I_{\rm C}$ light and (b) $(V-I_{\rm C})_0$ color curves
of V2615~Oph as well as V5114~Sgr, V1369~Cen, and V496~Sct.
The $BVI_{\rm C}$ data of V2615~Oph are taken from \citet{mun08a}.
We adopt the color excess of $E(B-V)= 1.05$ as mentioned below
and overlap the $(V-I)_0$ color curve of V2615~Oph with the other novae
for the timescaling factor of $\log f_{\rm s}= +0.04$, as shown in
Figure \ref{v2615_oph_v5114_sgr_v1369_cen_v496_sct_i_vi_color_logscale}(b).
We apply Equation (8) of \citet{hac19ka} for the $I$ band to Figure
\ref{v2615_oph_v5114_sgr_v1369_cen_v496_sct_i_vi_color_logscale}(a)
and obtain
\begin{eqnarray}
(m&-&M)_{I, \rm V2615~Oph} \cr
&=& ((m - M)_I + \Delta I_{\rm C})
_{\rm V5114~Sgr} - 2.5 \log 1.45 \cr
&=& 15.55 - 0.95\pm0.2 - 0.4 = 14.2\pm0.2 \cr
&=& ((m - M)_I + \Delta I_{\rm C})
_{\rm V1369~Cen} - 2.5 \log 0.74 \cr
&=& 10.11 + 3.8\pm0.2 + 0.325 = 14.23\pm0.2 \cr
&=& ((m - M)_I + \Delta I_{\rm C})
_{\rm V496~Sct} - 2.5 \log 0.55 \cr
&=& 12.9 + 0.7\pm0.2 + 0.65 = 14.25\pm0.2,
\label{distance_modulus_i_vi_v2615_oph}
\end{eqnarray}
where we adopt
$(m-M)_{I, \rm V5114~Sgr}=15.55$ from Appendix \ref{v5114_sgr_ubvi},
$(m-M)_{I, \rm V1369~Cen}=10.11$ from \citet{hac19ka}, and
$(m-M)_{I, \rm V496~Sct}=12.9$ in Appendix \ref{v496_sct_bvi}.
Thus, we obtain $(m-M)_{I, \rm V2615~Oph}= 14.23\pm0.2$.

Figure \ref{v2615_oph_lv_vul_v1419_aql_v_bv_logscale_no2} shows 
the (a) $V$ light and (b) $(B-V)_0$ color curves of V2615~Oph
as well as LV~Vul and V1419~Aql.  
The $BV$ data are taken from \citet{mun08a}.
We also add the $V$ data taken from SMARTS \citep{wal12}.
Based on the time-stretching method, we have the relation of
\begin{eqnarray}
(m&-&M)_{V, \rm V2615~Oph} \cr
&=& ((m - M)_V + \Delta V)_{\rm LV~Vul} - 2.5 \log 1.10 \cr
&=& 11.85 + 4.15\pm0.2 - 0.10 = 15.9\pm0.2 \cr
&=& ((m - M)_V + \Delta V)_{\rm V1419~Aql} - 2.5 \log 0.78 \cr
&=& 15.0 + 0.6\pm0.2 + 0.275 = 15.88\pm0.2.
\label{distance_modulus_v_bv_v2615_oph_v1419_aql_lv_vul}
\end{eqnarray}
Thus, we obtain $\log f_{\rm s}= \log 1.10= +0.04$ against the template 
nova LV~Vul and $(m-M)_{V, \rm V2615~Oph}=15.89\pm0.2$.

Figure \ref{v2615_oph_v2576_oph_v1668_cyg_lv_vul_b_only_logscale}(a)
shows the $B$ light curves of V2615~Oph
together with those of LV~Vul, V1668~Cyg, V533~Her, and V2576~Oph.
We apply Equation (7) of \citet{hac19ka} for the $B$ band to
Figure \ref{v2615_oph_v2576_oph_v1668_cyg_lv_vul_b_only_logscale}(a)
and obtain
\begin{eqnarray}
(m&-&M)_{B, \rm V2615~Oph} \cr
&=& ((m - M)_B + \Delta B)_{\rm LV~Vul} - 2.5 \log 1.10 \cr
&=& 12.45 + 4.6\pm0.2 - 0.10 = 16.95\pm0.2 \cr
&=& ((m - M)_B + \Delta B)_{\rm V1668~Cyg} - 2.5 \log 1.10 \cr
&=& 14.9 + 2.15\pm0.2 - 0.10 = 16.95\pm0.2 \cr
&=& ((m - M)_B + \Delta B)_{\rm V533~Her} - 2.5 \log 0.91 \cr
&=& 10.69 + 6.15\pm0.2 + 0.10 = 16.94\pm0.2 \cr
&=& ((m - M)_B + \Delta B)_{\rm V2576~Oph} - 2.5 \log 1.55 \cr
&=& 17.25 + 0.15\pm0.2 - 0.475 = 16.93\pm0.2.
\label{distance_modulus_b_v2615_oph_lv_vul_v1668_cyg}
\end{eqnarray}
We have $(m-M)_{B, \rm V2615~Oph}= 16.94\pm0.2$.

The three distance moduli of $(m-M)_B=16.94$, $(m-M)_V=15.89$, 
and $(m-M)_I=14.22$, are plotted in Figure
\ref{v2615_oph_v2576_oph_v1668_cyg_lv_vul_b_only_logscale}(b).
These three relations cross at the distance of $d= 3.4$~kpc
and the extinction of $E(B-V)=1.05$.   The crossing point is
consistent with the distance-reddening relations given by
\citet{mar06} and \citet{chen19}.  Thus, we confirm again
that the color excess is $E(B-V)=1.05$ and the distance is $d= 3.4$~kpc.


\begin{figure}
\plotone{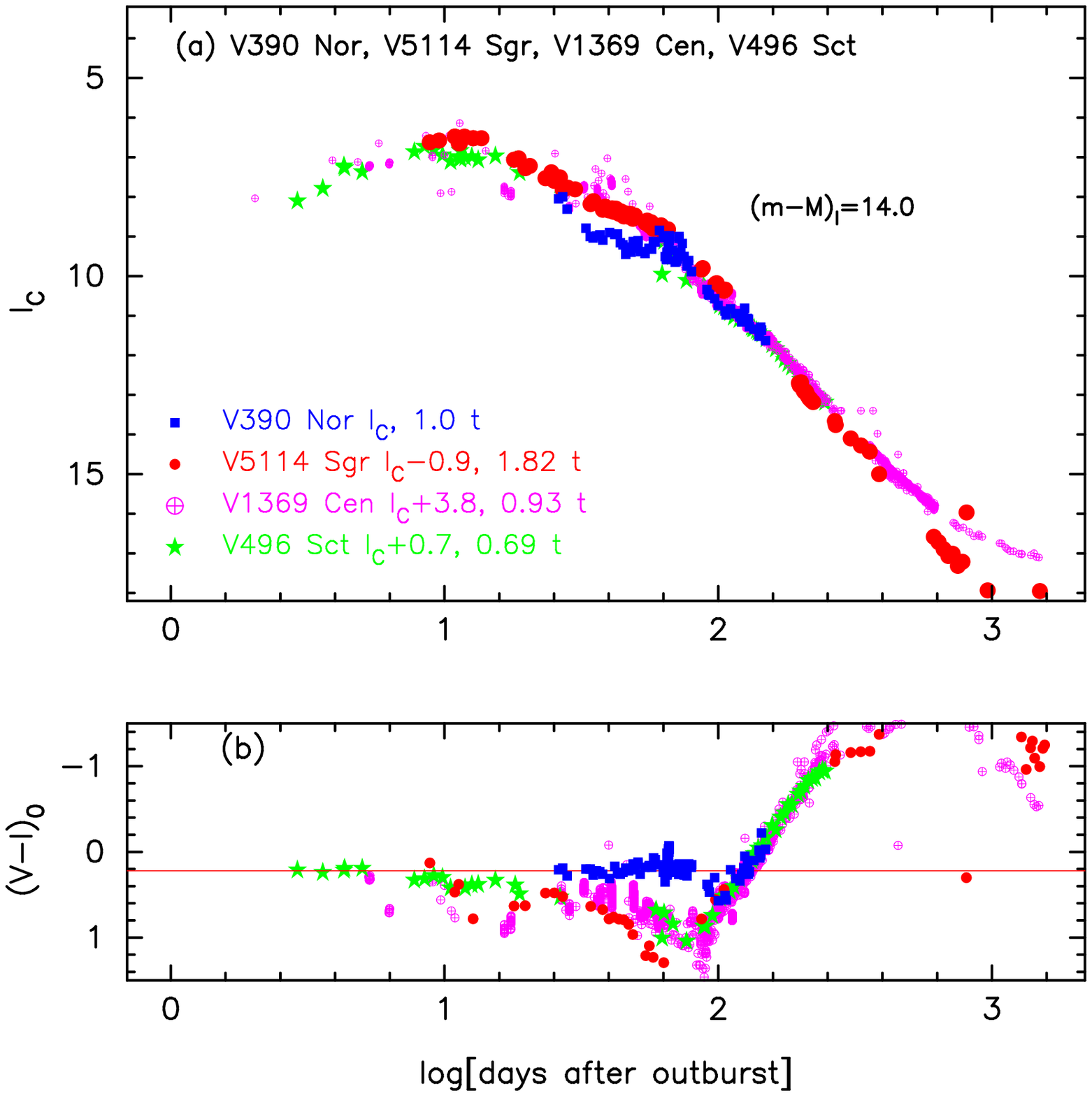}
\caption{
The (a) $I_{\rm C}$ light curve and (b) $(V-I_{\rm C})_0$ color curve
of V390~Nor as well as those of V5114~Sgr, V1369~Cen, and V496~Sct.
\label{v390_nor_v5114_sgr_v1369_cen_v496_sct_i_vi_color_logscale}}
\end{figure}


\begin{figure}
\plotone{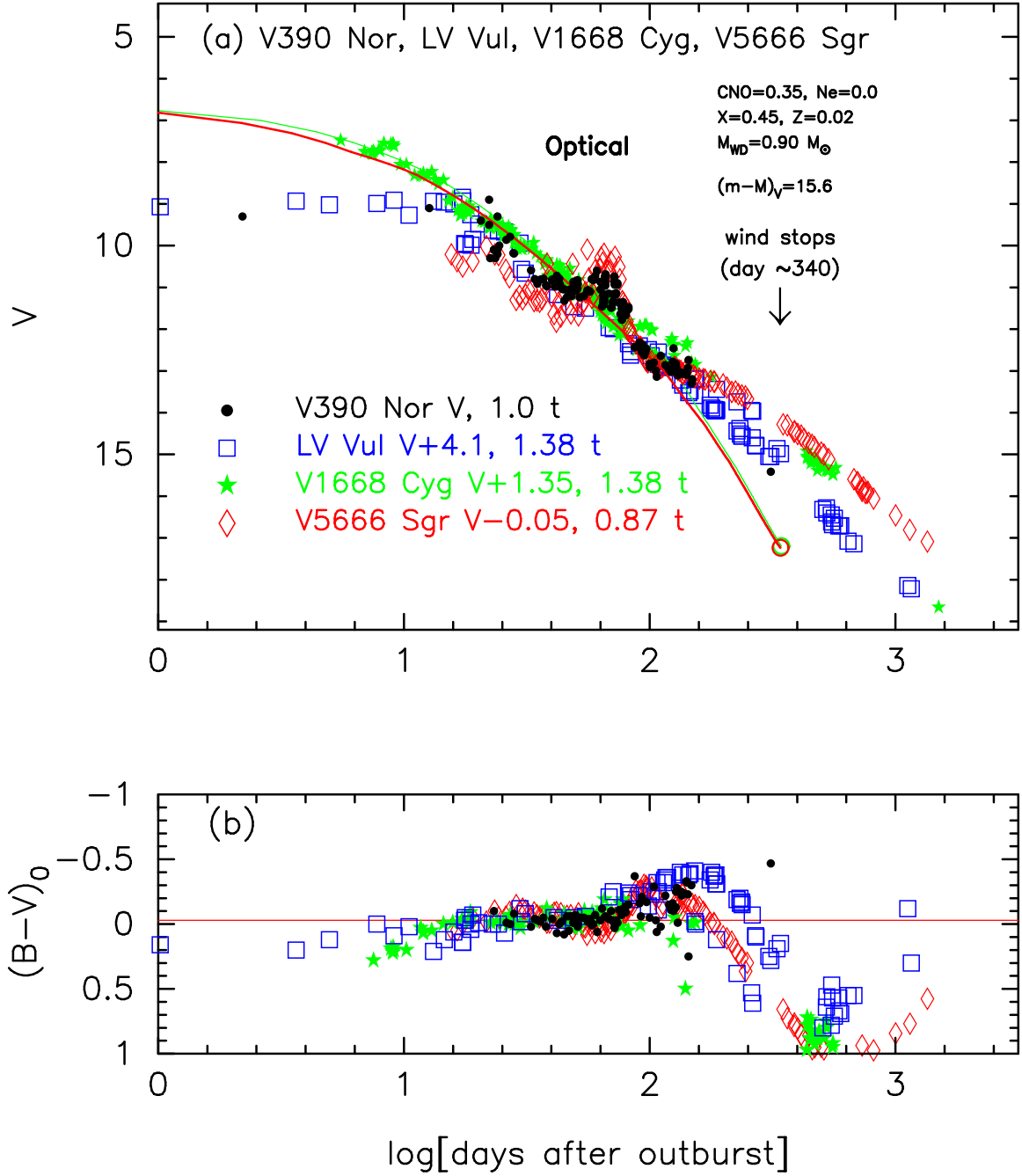}
\caption{
The (a) $V$ light curve and (b) $(B-V)_0$ color curve
of V390~Nor as well as those of LV~Vul, V1668~Cyg, and V5666~Sgr.
In panel (a), we plot a $0.90~M_\sun$ WD model (CO3, solid red line)
for V390~Nor as well as a $0.98~M_\sun$ WD model (CO3, solid green line)
for V1668~Cyg. 
\label{v390_nor_v5666_sgr_lv_vul_v1668_cyg_x55z02c10o10_logscale_no2}}
\end{figure}


\begin{figure*}
\plottwo{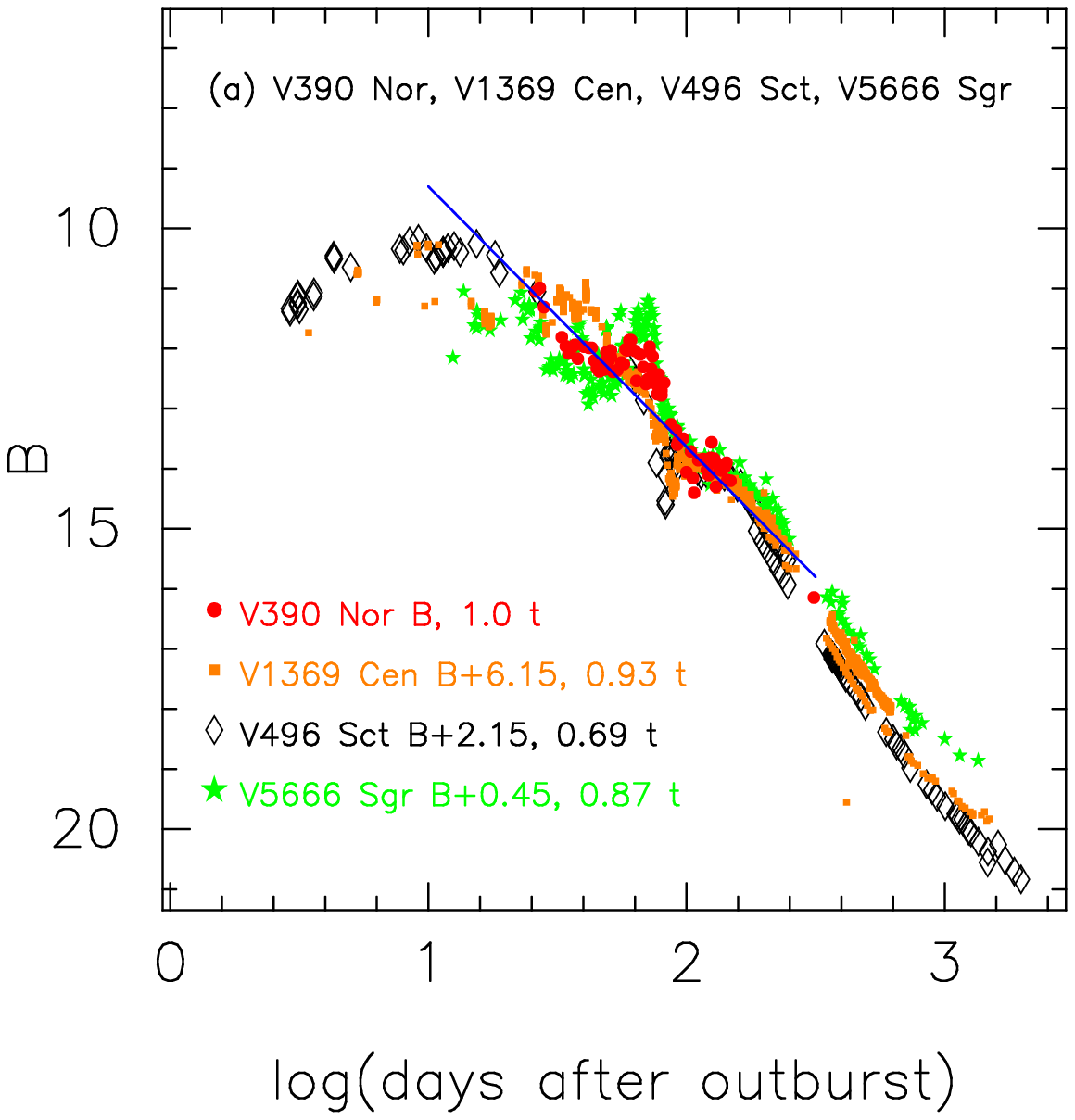}{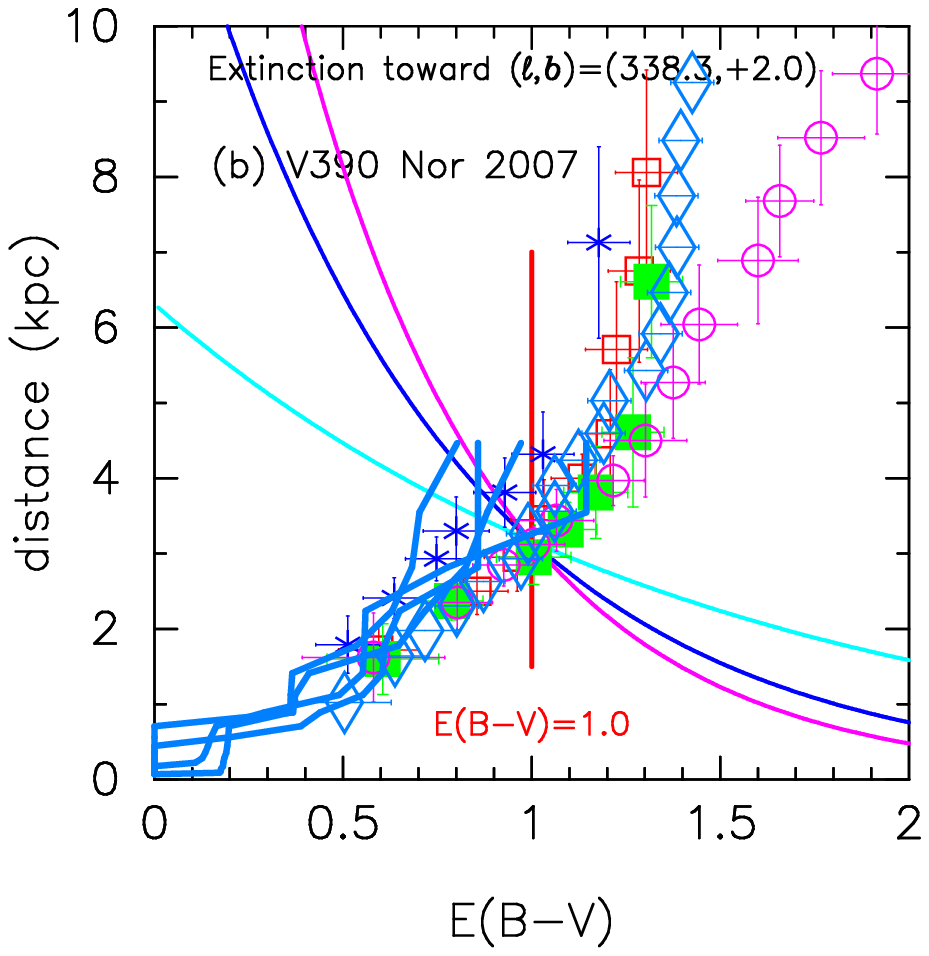}
\caption{
(a) The $B$ light curves of V390~Nor as well as V1369~Cen, V496~Sct,
and V5666~Sgr.
The $BVI_{\rm C}$ data of V390~Nor are taken from AAVSO.
(b) Various distance-reddening relations toward V390~Nor.
The thin solid lines of magenta, blue, and cyan denote the distance-reddening
relations given by $(m-M)_B=16.6$, $(m-M)_V=15.6$, and $(m-M)_I=14.0$,
respectively.
\label{distance_reddening_v390_nor_bvi_xxxxxx}}
\end{figure*}

\subsection{V390~Nor 2007}
\label{v390_nor_bvi}
We have reanalyzed the $BVI_{\rm C}$ multi-band light/color curves
of V390~Nor based on the time-stretching method.  
The important revised point is the timescaling factor of $f_{\rm s}$,
which is changed from the previous $\log f_{\rm s}= +0.45$ \citep{hac19kb}
to the present $f_{\rm s}= +0.14$ in order to overlap the $(V-I_{\rm C})_0$
color curve of V390~Nor and other novae as shown in Figure
\ref{v390_nor_v5114_sgr_v1369_cen_v496_sct_i_vi_color_logscale}(b).
Figure \ref{v390_nor_v5114_sgr_v1369_cen_v496_sct_i_vi_color_logscale}
shows the (a) $I_{\rm C}$ light and (b) $(V-I_{\rm C})_0$ color curves
of V390~Nor as well as V5114~Sgr, V1369~Cen, and V496~Sct.
The $BVI_{\rm C}$ data of V390~Nor are taken from AAVSO.
Adopting the color excess of $E(B-V)= 1.0$ as mentioned below,
we have obtained the timescaling factor $\log f_{\rm s}= +0.14$ for V390~Nor.
We apply Equation (8) of \citet{hac19ka} for the $I$ band to Figure
\ref{v390_nor_v5114_sgr_v1369_cen_v496_sct_i_vi_color_logscale}(a)
and obtain
\begin{eqnarray}
(m&-&M)_{I, \rm V390~Nor} \cr
&=& ((m - M)_I + \Delta I_{\rm C})_{\rm V5114~Sgr} - 2.5 \log 1.82 \cr
&=& 15.55 - 0.9\pm0.2 - 0.65  = 14.0\pm0.2 \cr
&=& ((m - M)_I + \Delta I_{\rm C})_{\rm V1369~Cen} - 2.5 \log 0.93 \cr
&=& 10.11 + 3.8\pm0.2 + 0.075 = 13.99\pm0.2 \cr
&=& ((m - M)_I + \Delta I_{\rm C})_{\rm V496~Sct} - 2.5 \log 0.69 \cr
&=& 12.9 + 0.7\pm0.2 + 0.4 = 14.0\pm0.2,
\label{distance_modulus_i_vi_v390_nor}
\end{eqnarray}
where we adopt
$(m-M)_{I, \rm V5114~Sgr}=15.55$ from Appendix \ref{v5114_sgr_ubvi},
$(m-M)_{I, \rm V1369~Cen}=10.11$ from \citet{hac19ka}, and
$(m-M)_{I, \rm V496~Sct}=12.9$ in Appendix \ref{v496_sct_bvi}.
Thus, we obtain $(m-M)_{I, \rm V390~Nor}= 14.0\pm0.2$.

Figure \ref{v390_nor_v5666_sgr_lv_vul_v1668_cyg_x55z02c10o10_logscale_no2}
shows the (a) $V$ and (b) $(B-V)_0$ evolutions of V390~Nor
as well as LV~Vul, V1668~Cyg, and V5666~Sgr.  
Applying Equation (4) of \citet{hac19ka} for the $V$ band to them,
we have the relation
\begin{eqnarray}
(m&-&M)_{V, \rm V390~Nor} \cr
&=& ((m - M)_V + \Delta V)_{\rm LV~Vul} - 2.5 \log 1.38 \cr
&=& 11.85 + 4.1\pm0.2 - 0.35 = 15.6\pm0.2 \cr
&=& ((m - M)_V + \Delta V)_{\rm V1668~Cyg} - 2.5 \log 1.38 \cr
&=& 14.6 + 1.35\pm0.2 - 0.35 = 15.6\pm0.2 \cr
&=& ((m - M)_V + \Delta V)_{\rm V5666~Sgr} - 2.5 \log 0.87 \cr
&=& 15.5 - 0.05\pm0.2 + 0.15 = 15.6\pm0.2,
\label{distance_modulus_v_bv_v390_nor}
\end{eqnarray}
where we adopt $(m-M)_{V, \rm LV~Vul}=11.85$ and
$(m-M)_{V, \rm V1668~Cyg}=14.6$ both from \citet{hac19ka}, 
and $(m-M)_{V, \rm V5666~Sgr}=15.5$ in Appendix \ref{v5666_sgr_bvi}. 
Thus, we obtain $(m-M)_{V, \rm V390~Nor}=15.6\pm0.1$.

Figure \ref{distance_reddening_v390_nor_bvi_xxxxxx}(a)
shows the $B$ light curves of V390~Nor
together with those of V1369~Cen, V496~Sct, and V5666~Sgr.
Applying Equation (7) for the $B$ band to Figure
\ref{distance_reddening_v390_nor_bvi_xxxxxx}(a),
we have the relation
\begin{eqnarray}
(m&-&M)_{B, \rm V390~Nor} \cr
&=& \left( (m-M)_B + \Delta B\right)_{\rm V1369~Cen} - 2.5 \log 0.93 \cr
&=& 10.36 + 6.15\pm0.3 + 0.075 = 16.59\pm0.3 \cr
&=& \left( (m-M)_B + \Delta B\right)_{\rm V496~Sct} - 2.5 \log 0.69 \cr
&=& 14.05 + 2.15\pm0.3 + 0.4 = 16.6\pm0.3 \cr
&=& \left( (m-M)_B + \Delta B\right)_{\rm V5666~Sgr} - 2.5 \log 0.87 \cr
&=& 16.0 + 0.45\pm0.3 + 0.15 = 16.6\pm0.3,
\label{distance_modulus_v396_nor_v1369_cen_v496_sct_v5666_sgr_b}
\end{eqnarray}
where we adopt $(m-M)_{B, \rm V1369~Cen}= 10.36$ from \citet{hac19ka},
$(m-M)_{B, \rm V496~Sct}= 14.05$ in Appendix \ref{v496_sct_bvi}, and
$(m-M)_{B, \rm V5666~Sgr}= 16.0$ in Appendix \ref{v5666_sgr_bvi}. 
Thus, we obtain $(m-M)_{B, \rm V390~Nor}=16.6\pm0.2$.

We plot $(m-M)_B=16.6$, $(m-M)_V=15.6$, and $(m-M)_I=14.0$,
which cross at $d= 3.2$~kpc and $E(B-V)=1.0$, in Figure
\ref{distance_reddening_v390_nor_bvi_xxxxxx}(b).
The crossing point is consistent with the distance-reddening relations
given by \citet{mar06}, \citet{ozd18}, and \citet{chen19}.
Thus, we obtain $E(B-V)=1.0\pm0.05$ and $d= 3.2\pm0.4$~kpc.


\begin{figure}
\plotone{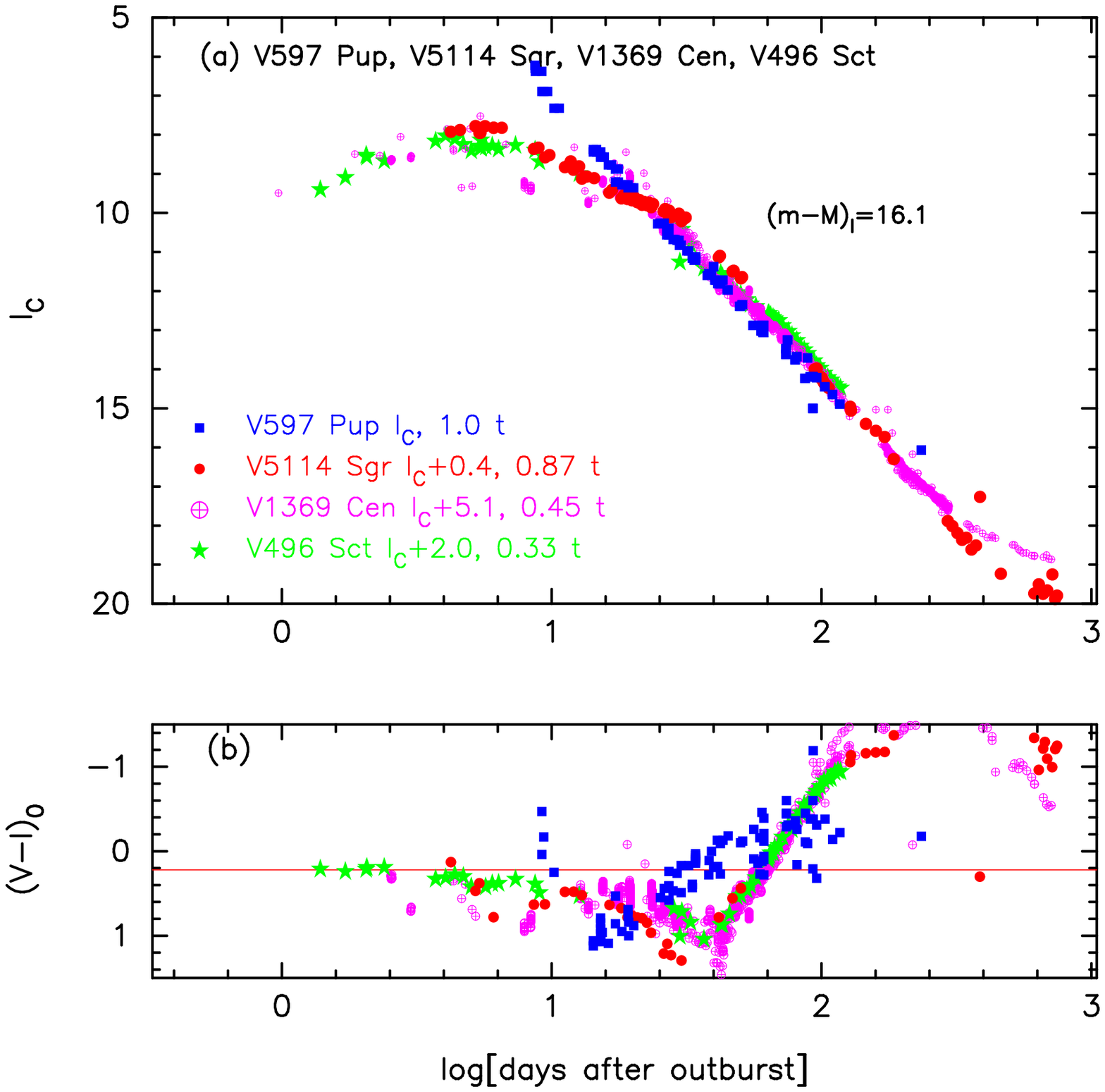}
\caption{
The (a) $I_{\rm C}$ light curve and (b) $(V-I_{\rm C})_0$ color curve 
of V597~Pup as well as those of V5114~Sgr, V1369~Cen, and V496~Sct.
\label{v597_pup_v5114_sgr_v1369_cen_v496_sct_i_vi_color_logscale}}
\end{figure}


\begin{figure}
\plotone{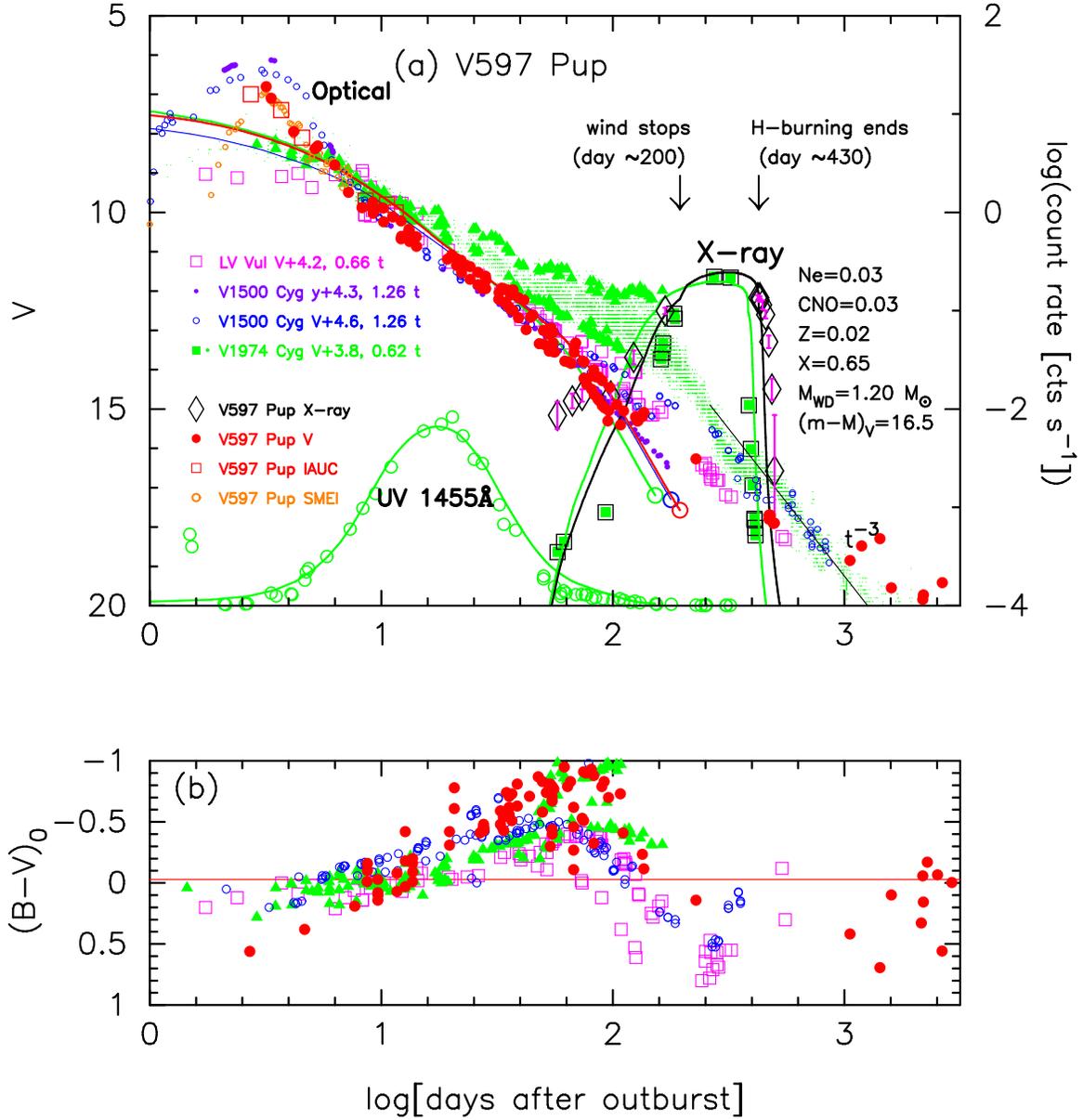}
\caption{
The (a) $V$ light curve and (b) $(B-V)_0$ color curve
of V597~Pup as well as those of LV~Vul, V1500~Cyg, and V1974~Cyg.
In panel (a), we show a $1.2~M_\sun$ WD model (Ne3, solid red and black lines)
for V597~Pup as well as a $0.98~M_\sun$ WD model (CO3, green lines)
for V1974~Cyg and a $1.2~M_\sun$ WD model (Ne2, blue line) for V1500~Cyg.  
\label{v597_pup_v1500_cyg_v1974_cyg_lv_vul_v_bv_x65z02o03ne03_logscale_no2}}
\end{figure}


\begin{figure*}
\plottwo{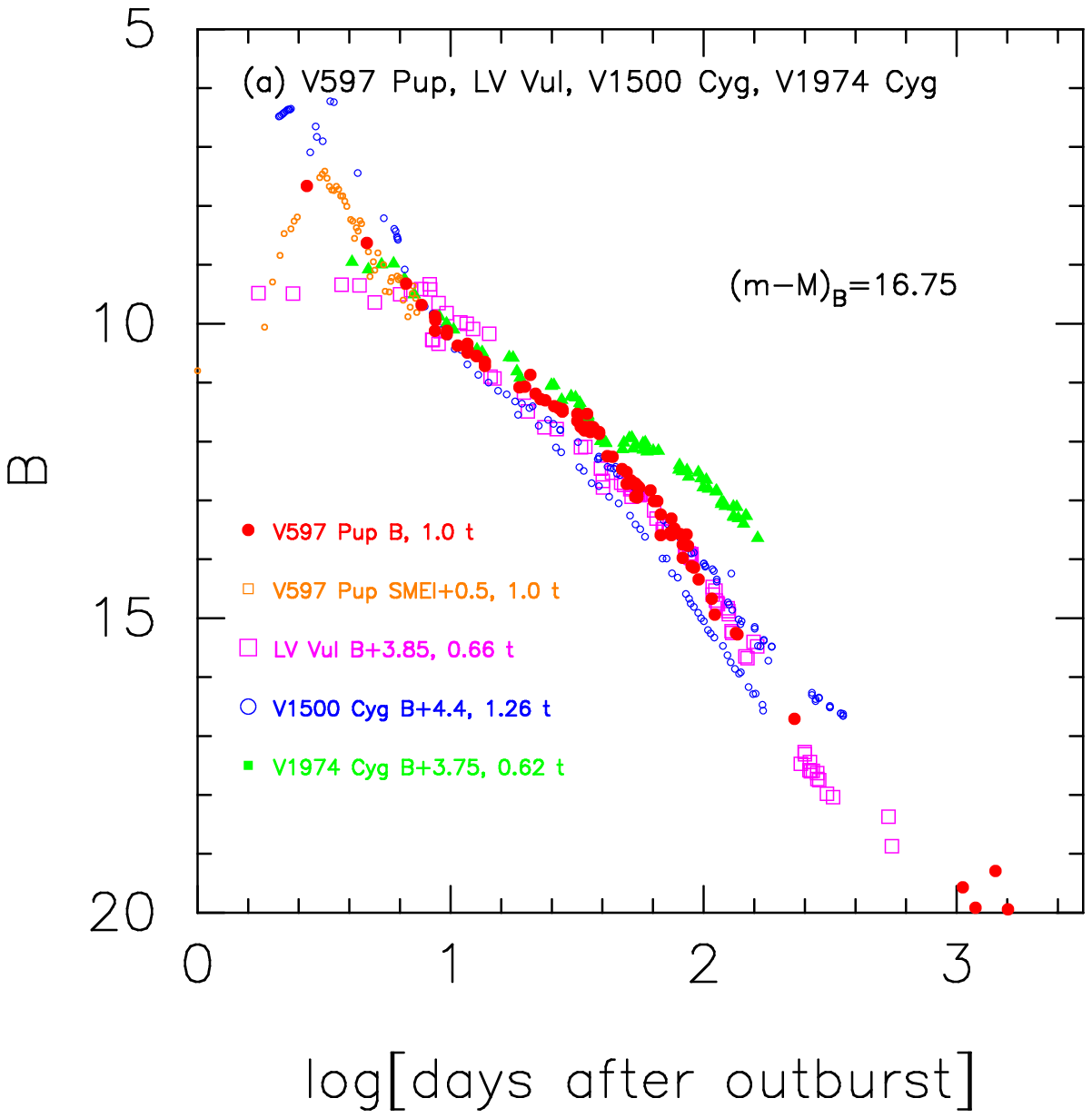}{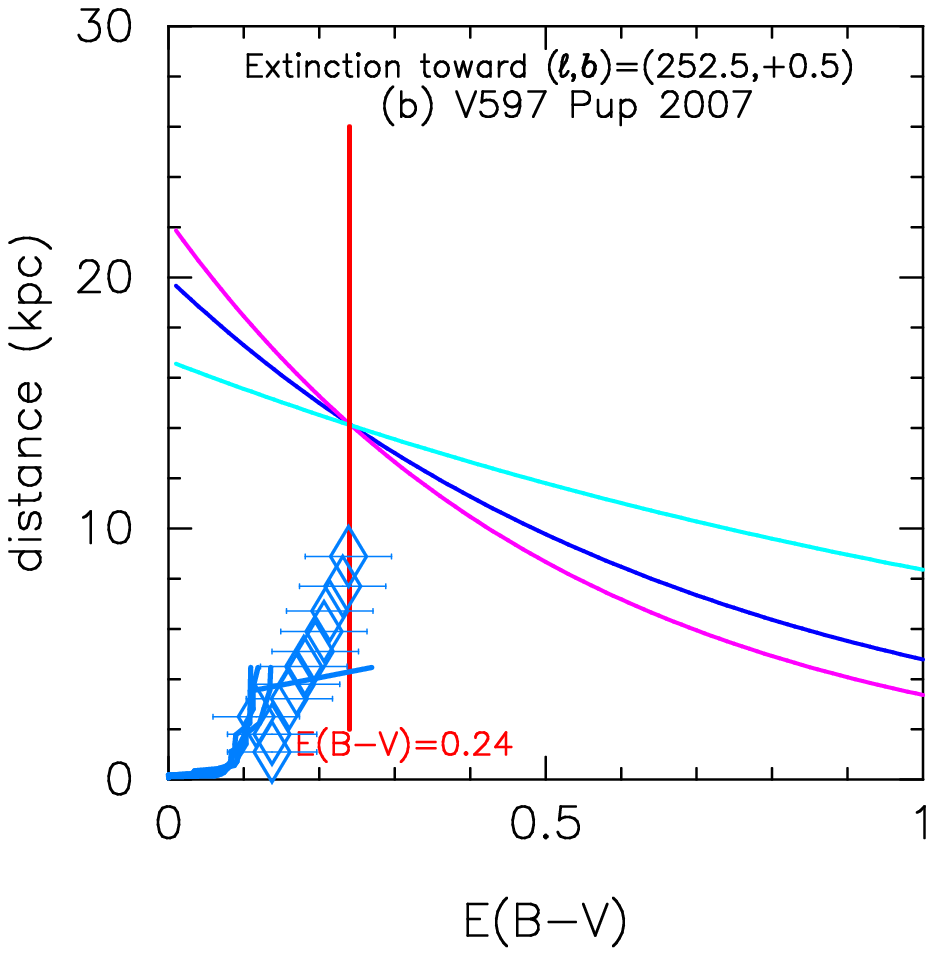}
\caption{
(a) The $B$ light curves of V597~Pup as well as LV~Vul, V1500~Cyg, 
and V1974~Cyg.  We add the {\it SMEI}
magnitudes of V597~Pup, which are taken from \citet{hou16}.
The $B$ data of V597~Pup are taken from AAVSO, VSOLJ, and SMARTS.
(b) Various distance-reddening relations toward V597~Pup.
The magenta, blue, and cyan lines denote the distance-reddening
relations given by $(m-M)_B= 16.75$, $(m-M)_V= 16.5$, 
and $(m-M)_I= 16.11$, respectively.
\label{distance_reddening_v597_pup_bvi_xxxxxx}}
\end{figure*}

\subsection{V597~Pup 2007}
\label{v597_pup_bvi}
We have reanalyzed the $BVI_{\rm C}$ multi-band light/color curves
of V597~Pup based on the time-stretching method.  
Figure \ref{v597_pup_v5114_sgr_v1369_cen_v496_sct_i_vi_color_logscale}
shows the (a) $I_{\rm C}$ light and (b) $(V-I_{\rm C})_0$ color curves
of V597~Pup as well as V5114~Sgr, V1369~Cen, and V496~Sct.  
The $BVI_{\rm C}$ data of V597~Pup are taken from VSOLJ and SMARTS.
Then, we apply Equation (8) of \citet{hac19ka} for the $I$ band to Figure
\ref{v597_pup_v5114_sgr_v1369_cen_v496_sct_i_vi_color_logscale}(a)
and obtain
\begin{eqnarray}
(m&-&M)_{I, \rm V597~Pup} \cr
&=& ((m - M)_I + \Delta I_{\rm C})_{\rm V5114~Sgr} - 2.5 \log 0.87 \cr
&=& 15.55 + 0.4\pm0.2 + 0.15= 16.1\pm0.2 \cr
&=& ((m - M)_I + \Delta I_{\rm C})_{\rm V1369~Cen} - 2.5 \log 0.45 \cr
&=& 10.11 + 5.1\pm0.2 + 0.875= 16.09\pm0.2 \cr
&=& ((m - M)_I + \Delta I_{\rm C})_{\rm V496~Sct} - 2.5 \log 0.33 \cr
&=& 12.9 + 2.0\pm0.2 + 1.2 = 16.1\pm0.2,
\label{distance_modulus_i_vi_v597_pup}
\end{eqnarray}
where we adopt
$(m-M)_{I, \rm V5114~Sgr}=15.55$ from Appendix \ref{v5114_sgr_ubvi},
$(m-M)_{I, \rm V1369~Cen}=10.11$ from \citet{hac19ka}, and
$(m-M)_{I, \rm V496~Sct}=12.9$ in Appendix \ref{v496_sct_bvi}.
Thus, we obtain $(m-M)_{I, \rm V597~Pup}= 16.1\pm0.2$.

Figure 
\ref{v597_pup_v1500_cyg_v1974_cyg_lv_vul_v_bv_x65z02o03ne03_logscale_no2}
shows the (a) $V$ light and (b) $(B-V)_0$ color curves of V597~Pup
as well as LV~Vul, V1500~Cyg, and V1974~Cyg.  
The timescaling factor of $\log f_{\rm s}= -0.18$ is mainly determined
from the supersoft X-ray light curve fitting.
We apply Equation (4) of \citet{hac19ka} to Figure 
\ref{v597_pup_v1500_cyg_v1974_cyg_lv_vul_v_bv_x65z02o03ne03_logscale_no2}(a)
and obtain
\begin{eqnarray}
(m&-&M)_{V, \rm V597~Pup} \cr
&=& ((m - M)_V + \Delta V)_{\rm LV~Vul} - 2.5 \log 0.66 \cr
&=& 11.85 + 4.2\pm0.2 + 0.45 = 16.5\pm0.2 \cr
&=& ((m - M)_V + \Delta V)_{\rm V1500~Cyg} - 2.5 \log 1.26 \cr
&=& 12.15 + 4.6\pm0.2 - 0.25 = 16.5\pm0.2 \cr
&=& ((m - M)_V + \Delta V)_{\rm V1974~Cyg} - 2.5 \log 0.62 \cr
&=& 12.2 + 3.8\pm0.2 + 0.525 = 16.52\pm0.2,
\label{distance_modulus_v_bv_v597_pup}
\end{eqnarray}
where we adopt $(m-M)_{V, \rm LV~Vul}=11.85$ and 
$(m-M)_{V, \rm V1974~Cyg}=12.2$ both from \citet{hac19ka}, and 
$(m-M)_{V, \rm V1500~Cyg}=12.15$ from Appendix \ref{v1500_cyg_ubvi}. 
Thus, we obtain $(m-M)_{V, \rm V597~Pup}= 16.5\pm0.1$.

We also plot the $B$ light curves of V597~Pup
together with LV~Vul, V1500~Cyg, and V1974~Cyg, in Figure
\ref{distance_reddening_v597_pup_bvi_xxxxxx}(a).
We apply Equation (7) of \citet{hac19ka} for the $B$ band to Figure
\ref{distance_reddening_v597_pup_bvi_xxxxxx}(a) and obtain
\begin{eqnarray}
(m&-&M)_{B, \rm V597~Pup} \cr
&=& ((m - M)_B + \Delta B)_{\rm LV~Vul} - 2.5 \log 0.66 \cr
&=& 12.45 + 3.85\pm0.2 + 0.45 = 16.75\pm0.2 \cr
&=& ((m - M)_B + \Delta B)_{\rm V1500~Cyg} - 2.5 \log 1.26 \cr
&=& 12.6 + 4.4\pm0.2 - 0.25 = 16.75\pm0.2 \cr
&=& ((m - M)_B + \Delta B)_{\rm V1974~Cyg} - 2.5 \log 0.62 \cr
&=& 12.5 + 3.75\pm0.2 + 0.525 = 16.77\pm0.2,
\label{distance_modulus_b_v597_pup_lv_vul_v1500_cyg_v1974_cyg}
\end{eqnarray}
where we adopt 
$(m-M)_{B, \rm LV~Vul}=11.85 + 0.6 = 12.45$ and
$(m-M)_{B, \rm V1974~Vul}=12.2 + 0.3 = 12.5$ both from \citet{hac19ka},
and $(m-M)_{B, \rm V1500~Cyg}=12.15 + 0.45 = 12.6$ from 
Appendix \ref{v1500_cyg_ubvi}. 
Thus, we obtain $(m-M)_{B, \rm V597~Pup}= 16.75\pm0.2$.

We obtain the three distance moduli in $B$, $V$, and $I_{\rm C}$ bands
and plot them in Figure \ref{distance_reddening_v597_pup_bvi_xxxxxx}(b).
These three lines cross at $d=14.2$~kpc and $E(B-V)=0.24$.
The crossing point is consistent with the distance-reddening relation
given by \citet{ozd18}.


\begin{figure}
\plotone{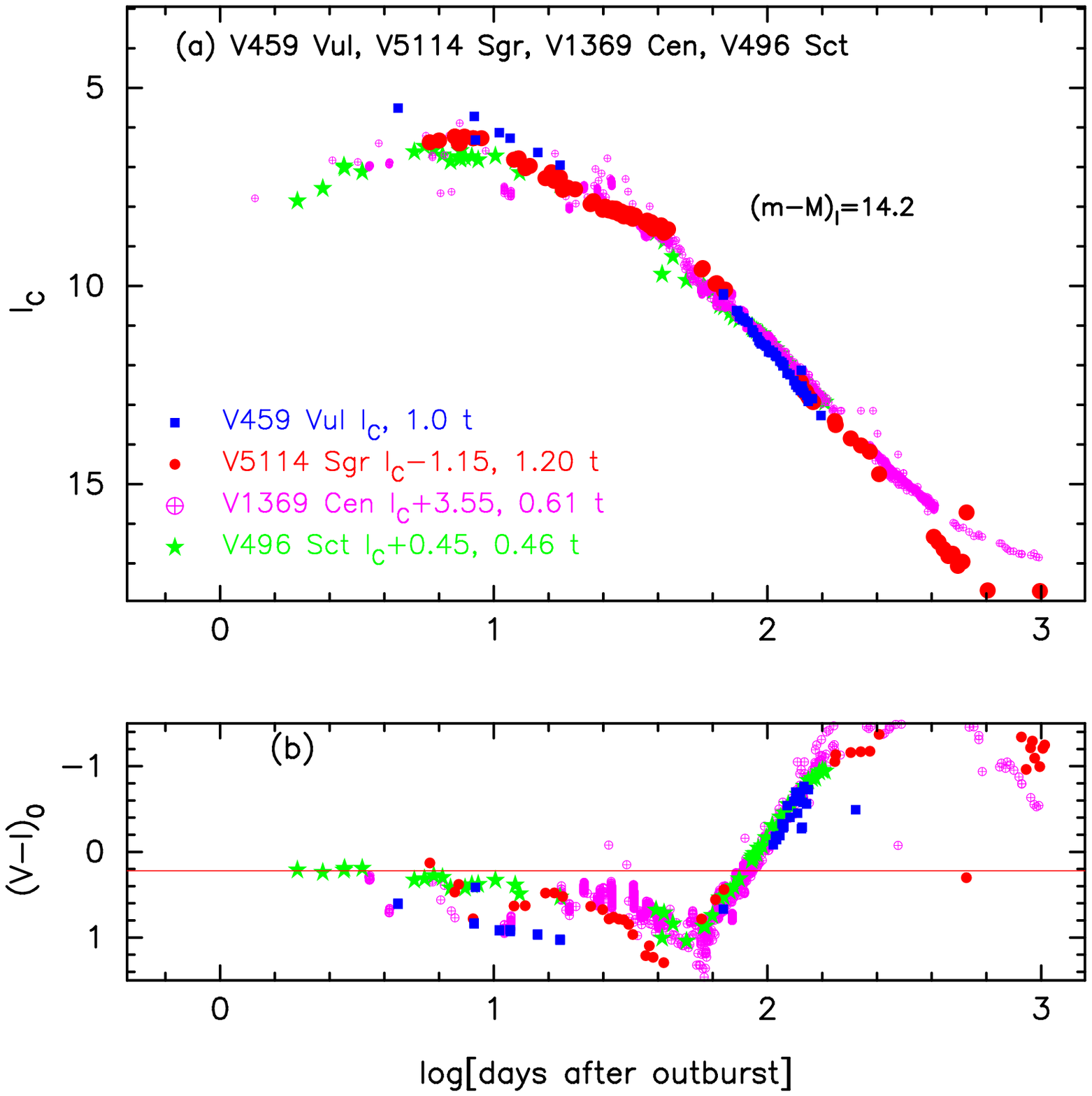}
\caption{
The (a) $I_{\rm C}$ light curve and (b) $(V-I_{\rm C})_0$ color curve
of V459~Vul as well as those of V5114~Sgr, V1369~Cen, and V496~Sct.
\label{v459_vul_v5114_sgr_v1369_cen_v496_sct_i_vi_color_logscale}}
\end{figure}


\begin{figure}
\plotone{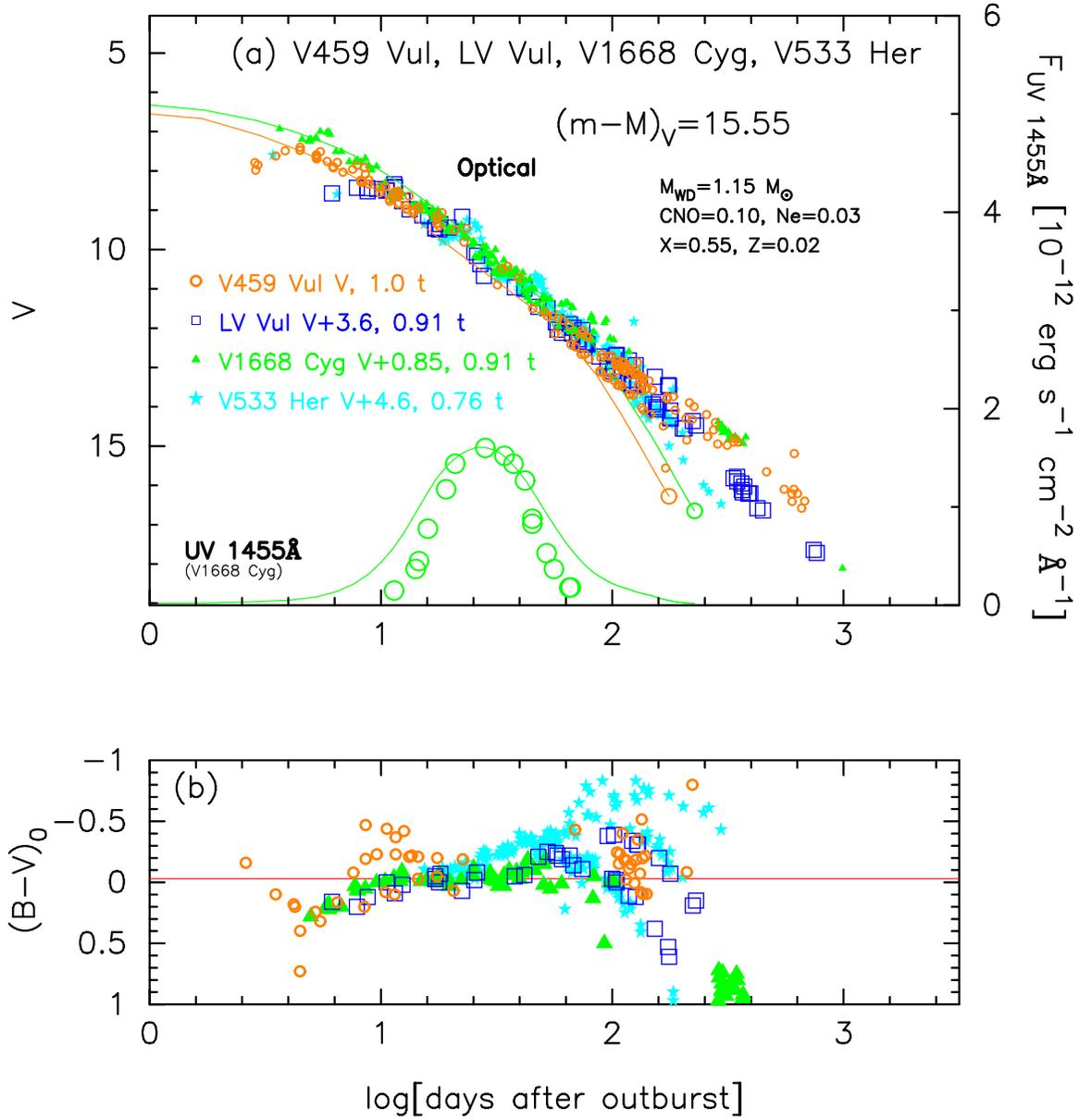}
\caption{
The (a) $V$ light curve and (b) $(B-V)_0$ color curve
of V459~Vul as well as those of LV~Vul, V1668~Cyg, and V533~Her.
In panel (a), we show a $1.15~M_\sun$ WD (Ne2, solid orange line) for
V459~Vul as well as a $0.98~M_\sun$ WD model (CO3, green lines) for V1668~Cyg.
\label{v459_vul_v533_her_v1668_cyg_lv_vul_v_bv_logscale_no2}}
\end{figure}


\begin{figure*}
\plottwo{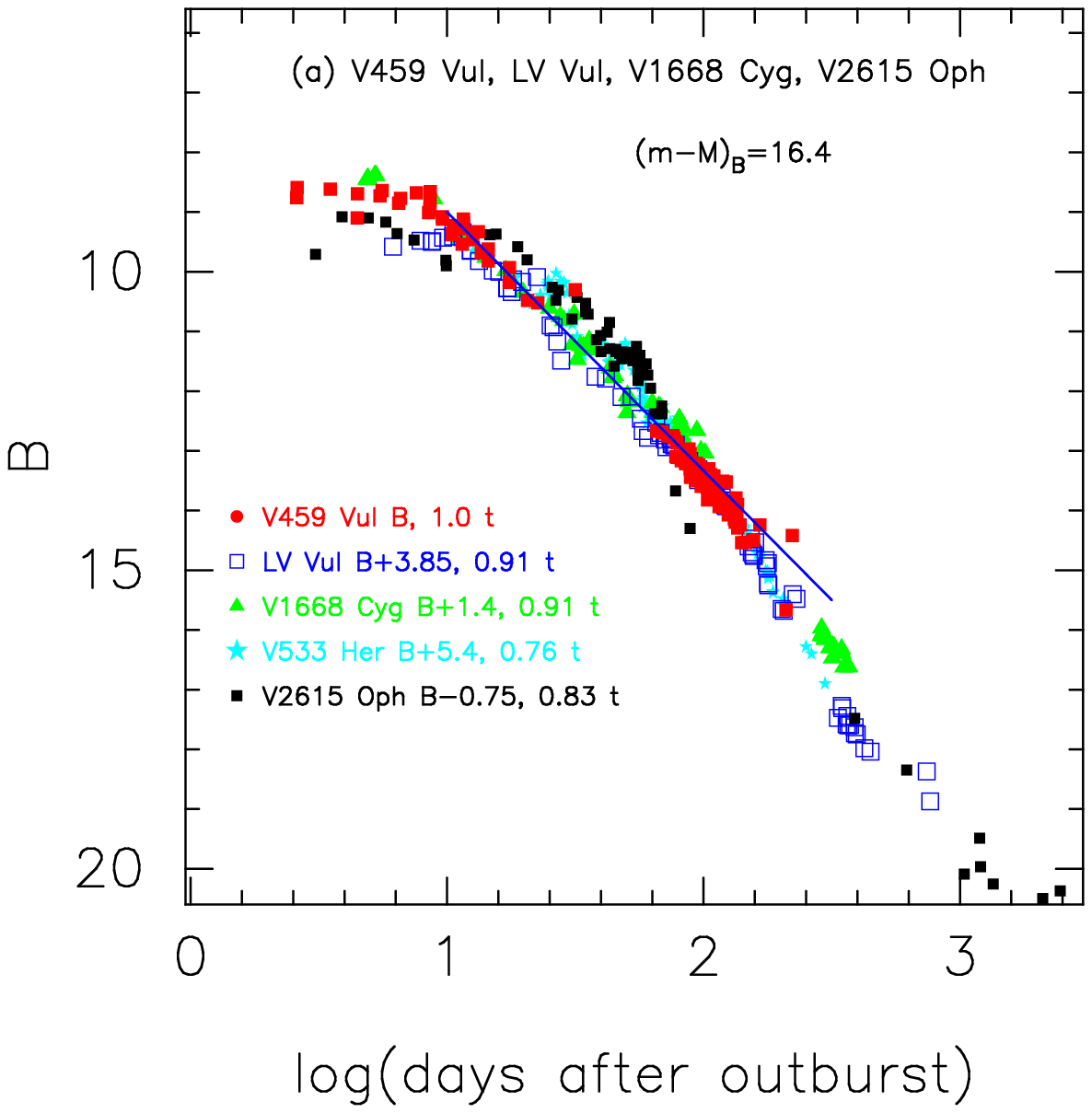}{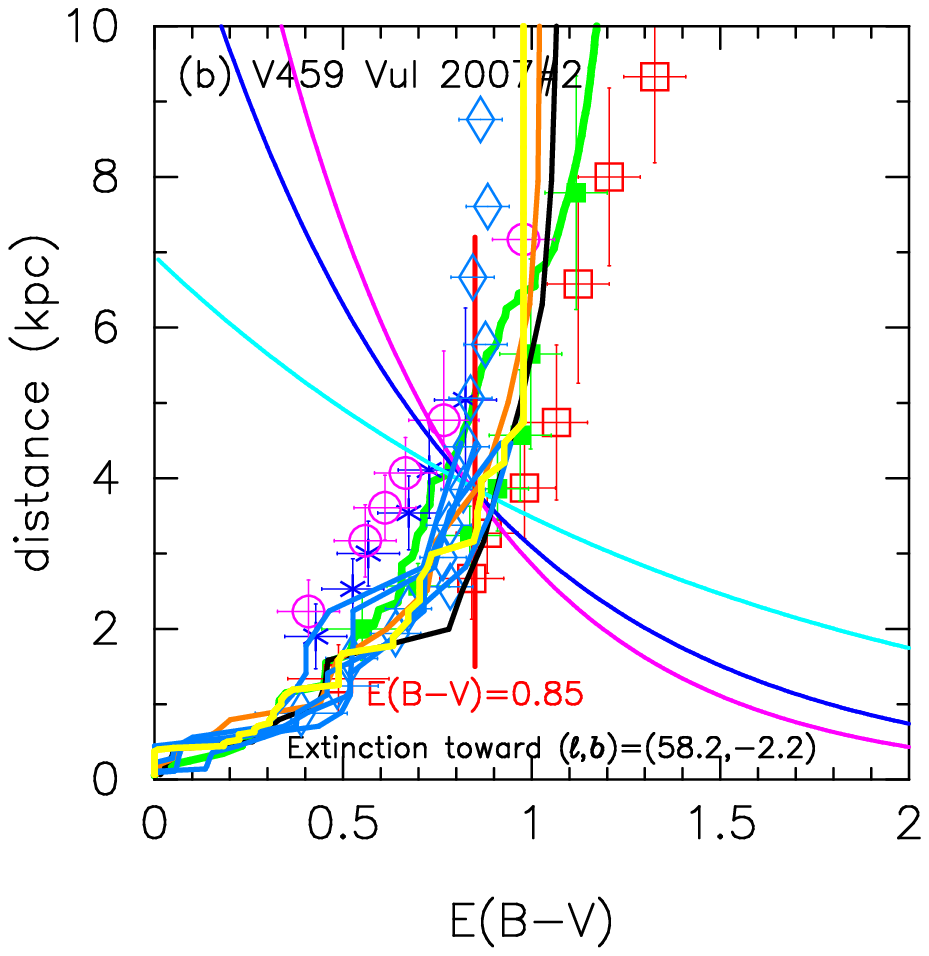}
\caption{
(a) The $B$ light curves of V459~Vul
as well as LV~Vul, V1668~Cyg, V533~Her, and V2615~Oph.
The $B$ data of V459~Vul are taken from AAVSO and VSOLJ.
(b) Various distance-reddening relations toward V459~Vul.
The thin solid lines of magenta, blue, and cyan denote the distance-reddening
relations given by $(m-M)_B= 16.4$, $(m-M)_V= 15.55$, 
and $(m-M)_I= 14.2$, respectively.
\label{distance_reddening_v459_vul_bvi_xxxxxx}}
\end{figure*}

\subsection{V459~Vul 2007\#2}
\label{v459_vul_bvi}
We have reanalyzed the $BVI_{\rm C}$ multi-band light/color curves
of V459~Vul based on the time-stretching method.  
The important revised point is the timescaling factor of $f_{\rm s}$,
which is changed from the previous $\log f_{\rm s}= -0.15$ to
the present $f_{\rm s}= -0.04$ in order to overlap the $(V-I_{\rm C})_0$
color curve of V459~Vul with other novae as shown in Figure
\ref{v459_vul_v5114_sgr_v1369_cen_v496_sct_i_vi_color_logscale}(b).
Figure \ref{v459_vul_v5114_sgr_v1369_cen_v496_sct_i_vi_color_logscale}
shows the (a) $I_{\rm C}$ light and (b) $(V-I_{\rm C})_0$ color curves
of V459~Vul as well as V5114~Sgr, V1369~Cen, and V496~Sct.
The $BVI_{\rm C}$ data of V459~Vul are taken from AAVSO and VSOLJ.
Adopting the color excess of $E(B-V)= 0.85$ mentioned below,
we obtained the timescaling factor $\log f_{\rm s}= -0.04$ for V459~Vul.
We apply Equation (8) of \citet{hac19ka} for the $I$ band to Figure
\ref{v459_vul_v5114_sgr_v1369_cen_v496_sct_i_vi_color_logscale}(a)
and obtain
\begin{eqnarray}
(m&-&M)_{I, \rm V459~Vul} \cr
&=& ((m - M)_I + \Delta I_{\rm C})
_{\rm V5114~Sgr} - 2.5 \log 1.20 \cr
&=& 15.55 - 1.15\pm0.2 - 0.2  = 14.2\pm0.2 \cr
&=& ((m - M)_I + \Delta I_{\rm C})
_{\rm V1369~Cen} - 2.5 \log 0.61 \cr
&=& 10.11 + 3.55\pm0.2 + 0.525 = 14.19\pm0.2 \cr
&=& ((m - M)_I + \Delta I_{\rm C})
_{\rm V496~Sct} - 2.5 \log 0.46 \cr
&=& 12.9 + 0.45\pm0.2 + 0.85 = 14.2\pm0.2,
\label{distance_modulus_i_vi_v459_vul}
\end{eqnarray}
where we adopt
$(m-M)_{I, \rm V5114~Sgr}=15.55$ from Appendix \ref{v5114_sgr_ubvi},
$(m-M)_{I, \rm V1369~Cen}=10.11$ from \citet{hac19ka}, and
$(m-M)_{I, \rm V496~Sct}=12.9$ in Appendix \ref{v496_sct_bvi}.
Thus, we obtain $(m-M)_{I, \rm V459~Vul}= 14.2\pm0.2$.

Figure \ref{v459_vul_v533_her_v1668_cyg_lv_vul_v_bv_logscale_no2} shows
the (a) $V$ light and (b) $(B-V)_0$ color curves of V459~Vul as well as
those of LV~Vul, V1668~Cyg, and V533~Her.  
Applying Equation (4) of \citet{hac19ka} to them,
we have the relation
\begin{eqnarray}
(m&-&M)_{V, \rm V459~Vul} \cr
&=& ((m - M)_V + \Delta V)_{\rm LV~Vul} - 2.5 \log 0.91 \cr
&=& 11.85 + 3.6\pm0.2 + 0.1 = 15.55\pm0.2 \cr
&=& ((m - M)_V + \Delta V)_{\rm V1668~Cyg} - 2.5 \log 0.91 \cr
&=& 14.6 + 0.85\pm0.2 + 0.1 = 15.55\pm0.2 \cr
&=& ((m - M)_V + \Delta V)_{\rm V533~Her} - 2.5 \log 0.76 \cr
&=& 10.65 + 4.6\pm0.2 + 0.3 = 15.55\pm0.2,
\label{distance_modulus_v_bv_v459_vul}
\end{eqnarray}
where we adopt $(m-M)_{V, \rm LV~Vul}=11.85$, 
$(m-M)_{V, \rm V1668~Cyg}=14.6$, and $(m-M)_{V, \rm V533~Her}=10.65$,
all from \citet{hac19ka}.  Thus, we obtain $(m-M)_{V, \rm V459~Vul}=
15.55\pm0.1$ and $f_{\rm s}=0.91$ against LV~Vul.

Figure \ref{distance_reddening_v459_vul_bvi_xxxxxx}(a) shows the $B$ light 
curves of V459~Vul together with those of LV~Vul, V1668~Cyg, V533~Her, 
and V2615~Oph.  We apply Equation (7) of \citet{hac19ka}
for the $B$ band to Figure \ref{distance_reddening_v459_vul_bvi_xxxxxx}(a)
and obtain
\begin{eqnarray}
(m&-&M)_{B, \rm V459~Vul} \cr
&=& ((m - M)_B + \Delta B)_{\rm LV~Vul} - 2.5 \log 0.91 \cr
&=& 12.45 + 3.85\pm0.2 + 0.1 = 16.4\pm0.2 \cr
&=& ((m - M)_B + \Delta B)_{\rm V1668~Cyg} - 2.5 \log 0.91 \cr
&=& 14.9 + 1.4\pm0.2 + 0.1 = 16.4\pm0.2 \cr
&=& ((m - M)_B + \Delta B)_{\rm V533~Her} - 2.5 \log 0.76 \cr
&=& 10.69 + 5.4\pm0.2 + 0.3 = 16.39\pm0.2 \cr
&=& ((m - M)_B + \Delta B)_{\rm V2615~Oph} - 2.5 \log 0.83 \cr
&=& 16.94 - 0.75\pm0.2 + 0.2 = 16.39\pm0.2.
\label{distance_modulus_b_v459_vul_lv_vul_v1668_cyg_v533_her}
\end{eqnarray}
We have $(m-M)_{B, \rm V459~Vul}= 16.4\pm0.2$.

We plot $(m-M)_B= 16.4$, $(m-M)_V= 15.55$, and $(m-M)_I= 14.21$,
which broadly cross at $d=3.8$~kpc and $E(B-V)=0.85$, as shown
in Figure \ref{distance_reddening_v459_vul_bvi_xxxxxx}(b).
The crossing point is consistent with the distance-reddening relations
(thick solid orange and yellow lines) given by \citet{gre18, gre19}.
Thus, we have obtained $E(B-V)=0.85\pm0.05$ and $d=3.8\pm0.4$~kpc
for V459~Vul.


\begin{figure}
\plotone{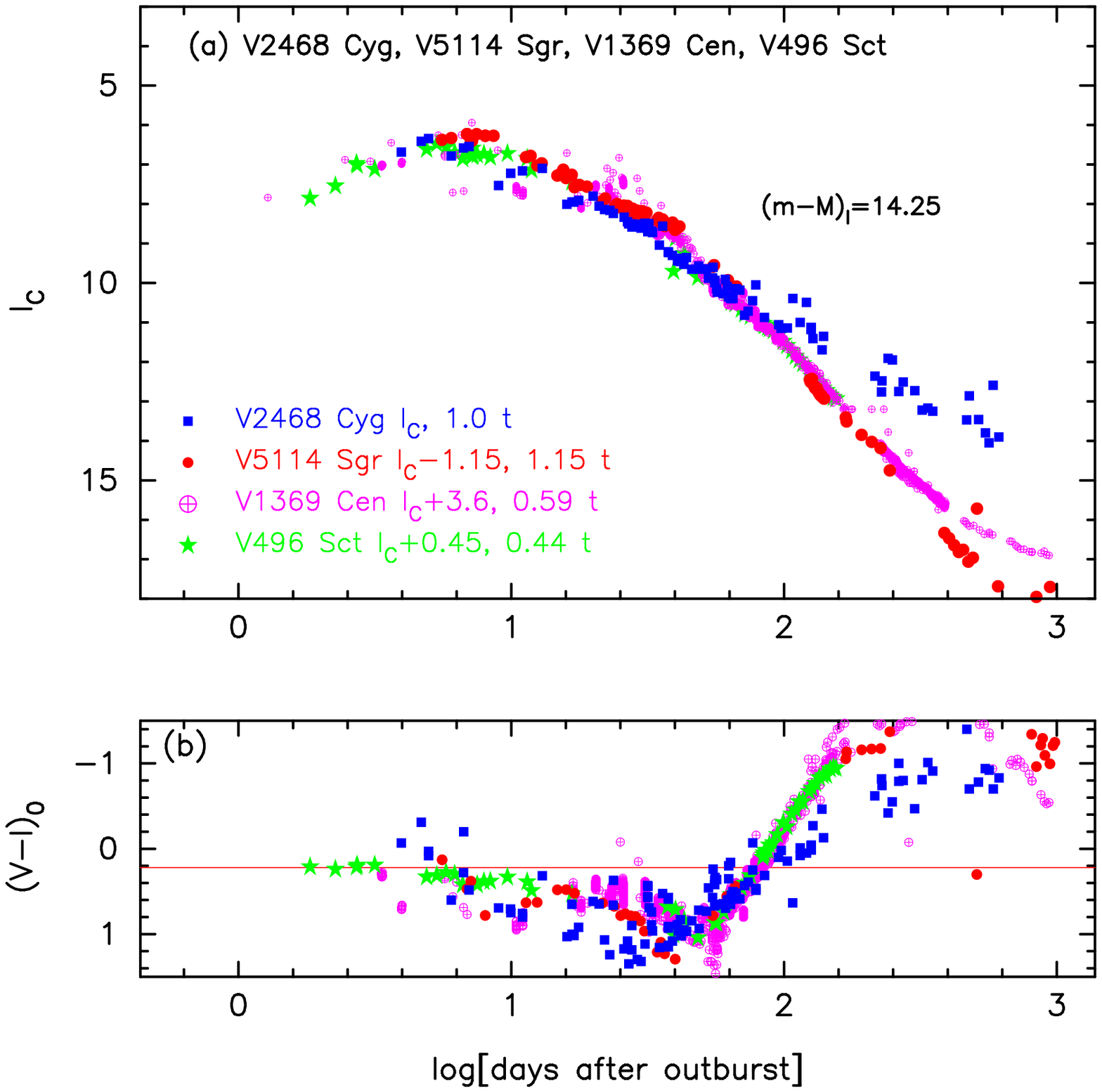}
\caption{
The (a) $I_{\rm C}$ light curve and (b) $(V-I_{\rm C})_0$ color curve 
of V2468~Cyg as well as those of V5114~Sgr, V1369~Cen, and V496~Sct.
\label{v2468_cyg_v5114_sgr_v1369_cen_v496_sct_i_vi_color_logscale}}
\end{figure}


\begin{figure}
\plotone{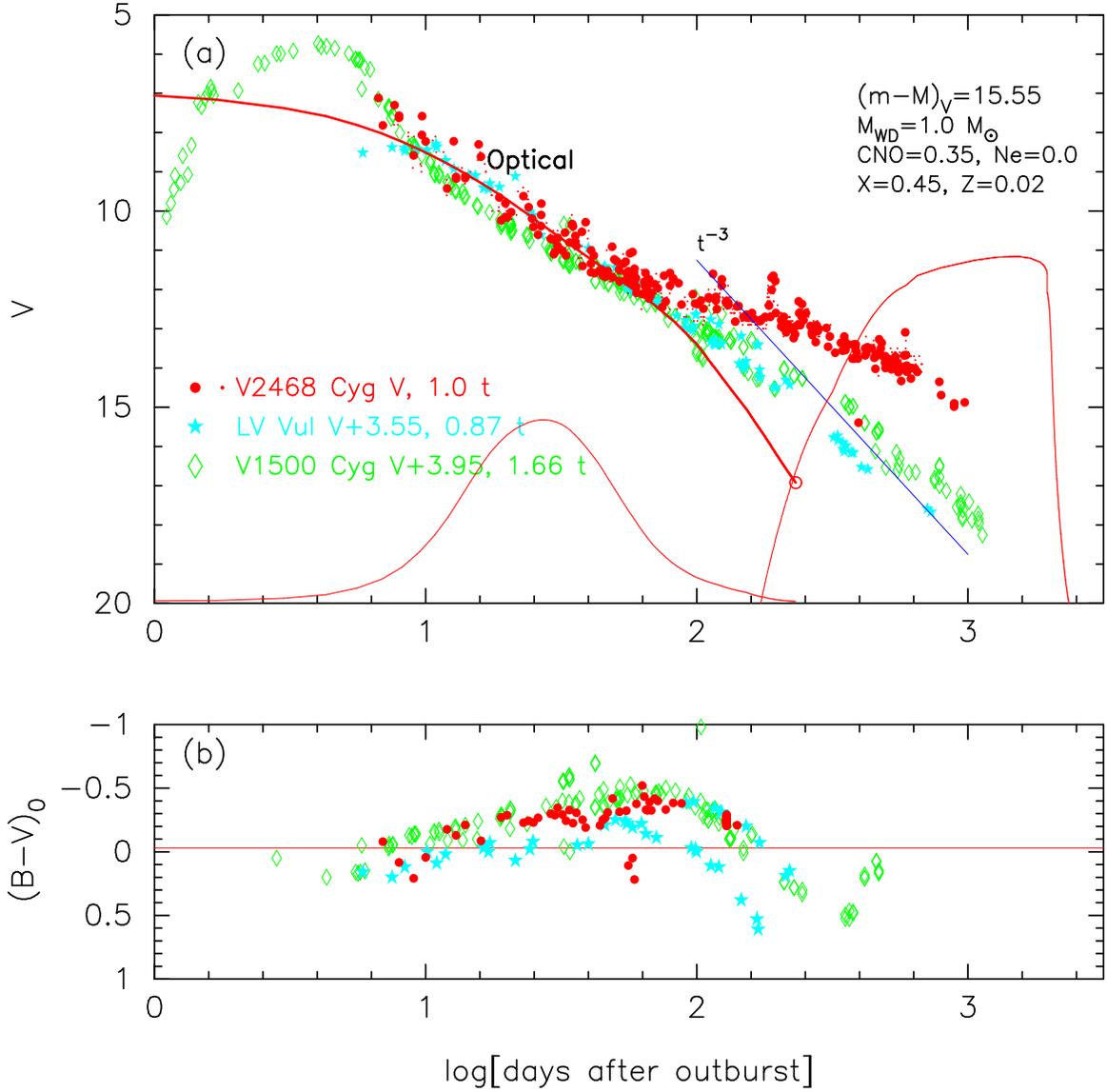}
\caption{
The (a) $V$ light curve and (b) $(B-V)_0$ color curve
of V2468~Cyg as well as those of LV~Vul and V1500~Cyg.
In panel (a), we show a $1.0~M_\sun$ WD model (CO3, solid red lines)
for V2468~Cyg.
\label{v2468_cyg_lv_vul_v1500_cyg_v_color_logscale_no2}}
\end{figure}


\begin{figure*}
\plottwo{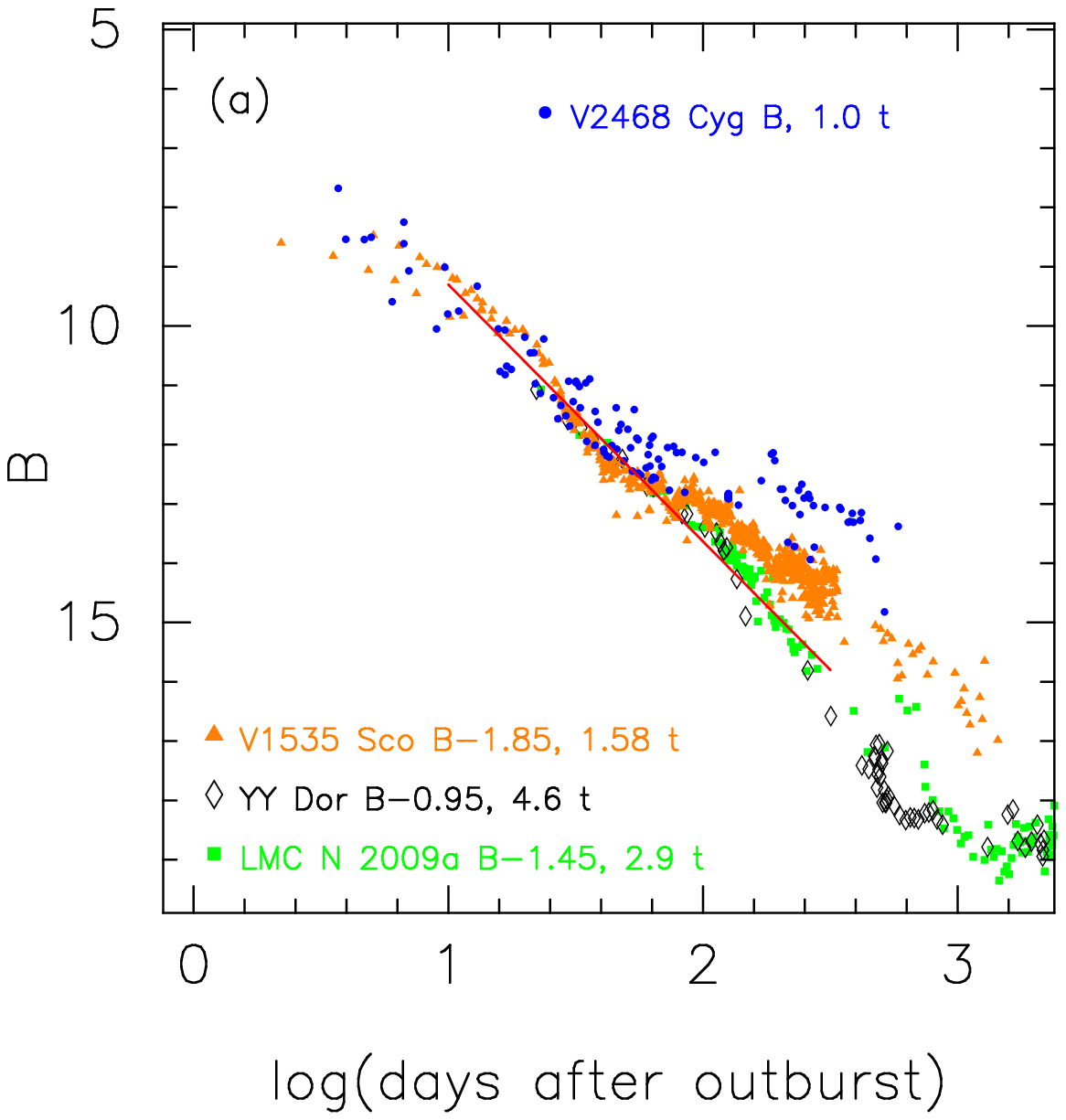}{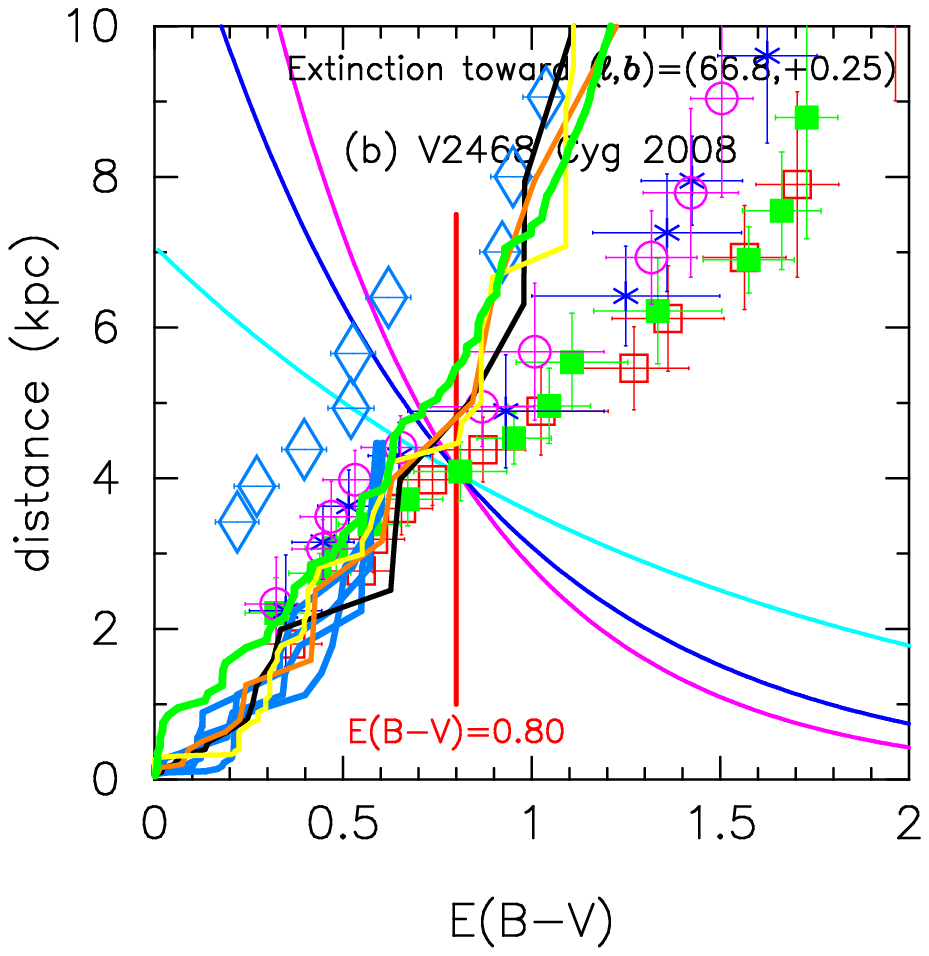}
\caption{
(a) The $B$ light curve of V2468~Cyg as well as V1535~Sco, 
YY~Dor, and LMC~N~2009a.
The $BVI_{\rm C}$ data of V2468~Cyg are taken from AAVSO and VSOLJ.
(b) Various distance-reddening relations toward V2468~Cyg.
The thin solid lines of magenta, blue, and cyan denote the distance-reddening
relations given by $(m-M)_B= 16.38$, $(m-M)_V= 15.55$, 
and $(m-M)_I= 14.25$, respectively.
\label{distance_reddening_v2468_cyg_bvi_xxxxxx}}
\end{figure*}

\subsection{V2468~Cyg 2008}
\label{v2468_cyg_bvi}
We have reanalyzed the $BVI_{\rm C}$ multi-band light/color curves
of V2468~Cyg based on the time-stretching method.  
The important revised point is the timescaling factor of $f_{\rm s}$,
which is changed from the previous $\log f_{\rm s}= +0.38$ to
the present $f_{\rm s}= -0.06$ in order to overlap the $V-I_{\rm C}$
and $B-V$ color curves of V2468~Cyg with other novae as shown in Figure
\ref{v2468_cyg_v5114_sgr_v1369_cen_v496_sct_i_vi_color_logscale}(b)
and \ref{v2468_cyg_lv_vul_v1500_cyg_v_color_logscale_no2}(b).
This large change in the timescaling factor does not affect much 
the $V$ light curve fitting partly because the $V$ data are scattered.
As a result, we have good overlapping in the $B-V$ color
curves as shown in Figure
\ref{v2468_cyg_lv_vul_v1500_cyg_v_color_logscale_no2}(b).  
This demonstrates an importance of multi-band light/color curves
analysis.

Figure \ref{v2468_cyg_v5114_sgr_v1369_cen_v496_sct_i_vi_color_logscale}
shows the (a) $I_{\rm C}$ light and (b) $(V-I_{\rm C})_0$ color curve
of V2468~Cyg as well as V5114~Sgr, V1369~Cen, and V496~Sct.
The $BVI_{\rm C}$ data of V2468~Cyg are taken from AAVSO and VSOLJ.
Adopting a new color excess of $E(B-V)= 0.80$ as mentioned below,
we redefine the timescaling factor $\log f_{\rm s}= -0.06$ of V2468~Cyg.
This is because the $(V-I)_0$ and $(B-V)_0$ color evolutions of V2468~Cyg
overlaps with the other novae as much as possible, as shown in Figure
\ref{v2468_cyg_v5114_sgr_v1369_cen_v496_sct_i_vi_color_logscale}(b)
and \ref{v2468_cyg_lv_vul_v1500_cyg_v_color_logscale_no2}(b).  
Then, we apply Equation (8) of \citet{hac19ka} for the $I$ band to Figure
\ref{v2468_cyg_v5114_sgr_v1369_cen_v496_sct_i_vi_color_logscale}(a)
and obtain
\begin{eqnarray}
(m&-&M)_{I, \rm V2468~Cyg} \cr
&=& ((m - M)_I + \Delta I_{\rm C})
_{\rm V5114~Sgr} - 2.5 \log 1.15 \cr
&=& 15.55 - 1.15\pm0.2 - 0.15 = 14.25\pm0.2 \cr
&=& ((m - M)_I + \Delta I_{\rm C})
_{\rm V1369~Cen} - 2.5 \log 0.59 \cr
&=& 10.11 + 3.6\pm0.2 + 0.575 = 14.28\pm0.2 \cr
&=& ((m - M)_I + \Delta I_{\rm C})
_{\rm V496~Sct} - 2.5 \log 0.44 \cr
&=& 12.9 + 0.45\pm0.2 + 0.9 = 14.25\pm0.2,
\label{distance_modulus_i_vi_v2468_cyg}
\end{eqnarray}
where we adopt
$(m-M)_{I, \rm V5114~Sgr}=15.55$ from Appendix \ref{v5114_sgr_ubvi},
and $(m-M)_{I, \rm V1369~Cen}=10.11$ from \citet{hac19ka}, and
$(m-M)_{I, \rm V496~Sct}=12.9$ in Appendix \ref{v496_sct_bvi}.
Thus, we obtain $(m-M)_{I, \rm V2468~Cyg}= 14.26\pm0.2$.

Figure 
\ref{v2468_cyg_lv_vul_v1500_cyg_v_color_logscale_no2}
shows the (a) $V$ light and (b) $(B-V)_0$ color curves of V2468~Cyg
as well as LV~Vul and V1500~Cyg.  Here, we adopt
$E(B-V)= 0.80$ and $\log f_{\rm s} = -0.06$ after the $I_{\rm C}$ light
and $(V-I_{\rm C})_0$ color curves analysis mentioned above.
We apply Equation (4) of \citet{hac19ka} to Figure 
\ref{v2468_cyg_lv_vul_v1500_cyg_v_color_logscale_no2}(a)
and obtain
\begin{eqnarray}
(m&-&M)_{V, \rm V2468~Cyg} \cr
&=& ((m - M)_V + \Delta V)_{\rm LV~Vul} - 2.5 \log 0.87 \cr
&=& 11.85 + 3.55\pm0.2 + 0.15 = 15.55\pm0.2 \cr
&=& ((m - M)_V + \Delta V)_{\rm V1500~Cyg} - 2.5 \log 1.66 \cr
&=& 12.15 + 3.95\pm0.2 - 0.55 = 15.55\pm0.2,
\label{distance_modulus_v_bv_v2468_cyg}
\end{eqnarray}
where we adopt
$(m-M)_{V, \rm LV~Vul}=11.85$ from \citet{hac19ka}, and
$(m-M)_{V, \rm V1500~Cyg}=12.15$ from Appendix \ref{v1500_cyg_ubvi}. 
Thus, we obtain $(m-M)_{V, \rm V2468~Cyg}= 15.55\pm0.1$.

We also plot the $B$ light curve of V2468~Cyg
together with V1535~Sco, YY~Dor, and LMC~N~2009a, in Figure
\ref{distance_reddening_v2468_cyg_bvi_xxxxxx}(a).
We apply Equation (7) of \citet{hac19ka} for the $B$ band  to Figure
\ref{distance_reddening_v2468_cyg_bvi_xxxxxx}(a)
and obtain
\begin{eqnarray}
(m&-&M)_{B, \rm V2468~Cyg} \cr
&=& ((m - M)_B + \Delta B)_{\rm V1535~Sco} - 2.5 \log 1.58 \cr
&=& 18.73 - 1.85\pm0.3 - 0.50 = 16.38\pm0.3 \cr
&=& ((m - M)_B + \Delta B)_{\rm YY~Dor} - 2.5 \log 4.6 \cr
&=& 18.98 - 0.95\pm0.3 - 1.65 = 16.38\pm0.3 \cr
&=& ((m - M)_B + \Delta B)_{\rm LMC~N~2009a} - 2.5 \log 2.9 \cr
&=& 18.98 - 1.45\pm0.3 - 1.15 = 16.38\pm0.3,
\label{distance_modulus_b_v2468_cyg_v1535_sco_v834_car_yy_dor_lmcn2009a}
\end{eqnarray}
where we adopt $(m-M)_{B, \rm V1535~Sco}=17.95 + 0.78 = 18.73$
in Appendix \ref{v1535_sco_bvi}. 
Thus, we obtain $(m-M)_{B, \rm V2468~Cyg}= 16.38\pm0.2$.

We obtain the three distance moduli in $B$, $V$, and $I_{\rm C}$ bands
and plot them in Figure
\ref{distance_reddening_v2468_cyg_bvi_xxxxxx}(b).
These three lines cross at $d=4.1$~kpc and $E(B-V)=0.80$.
The crossing point is consistent with the distance-reddening relations
given by \citet[][filled green squares]{mar06}
and \citet[][yellow line]{gre19}.


\begin{figure}
\plotone{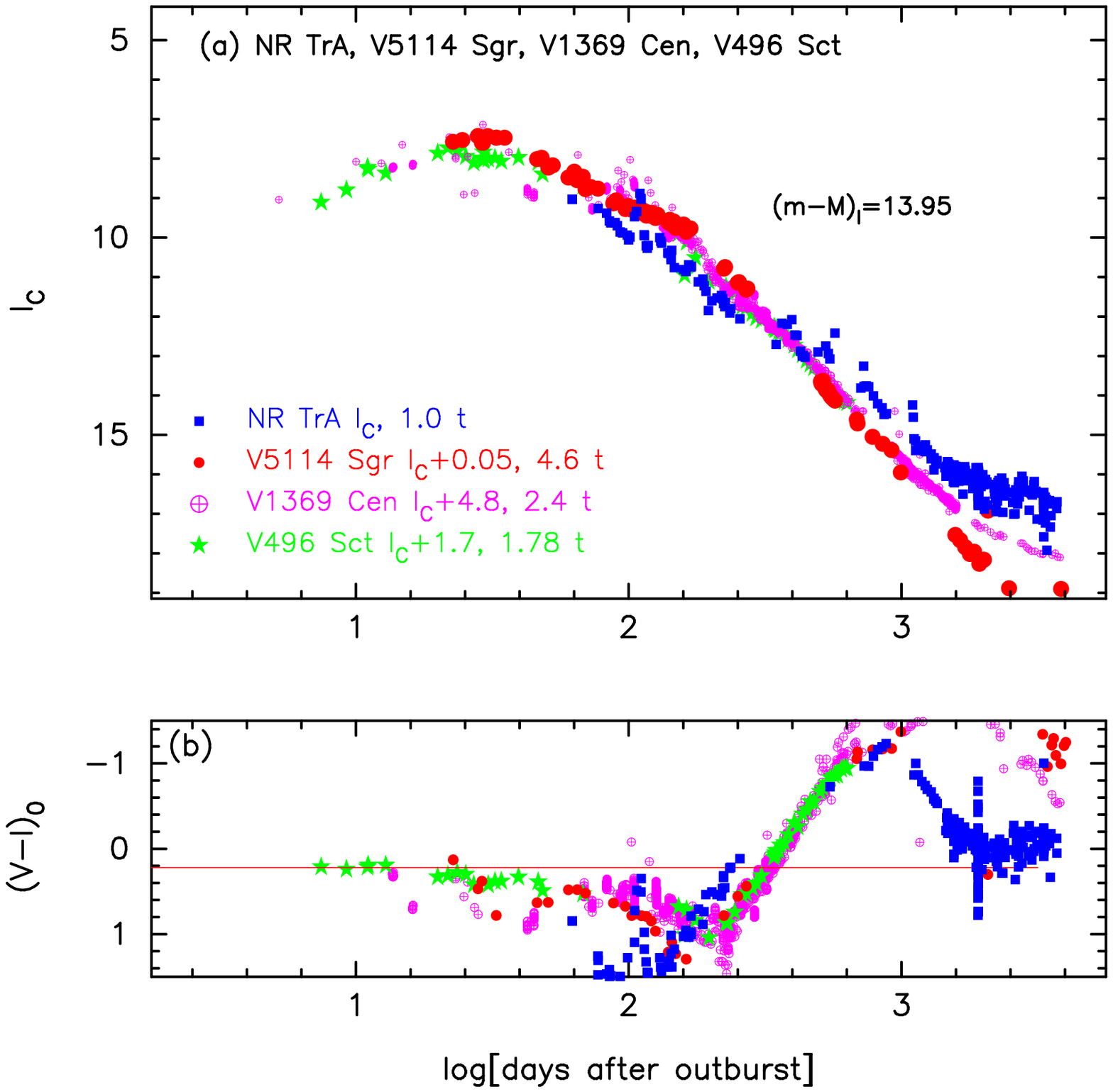}
\caption{
The (a) $I_{\rm C}$ light curve and (b) $(V-I_{\rm C})_0$ color curve
of NR~TrA as well as those of V5114~Sgr, V1369~Cen, and V496~Sct.
\label{nr_tra_v5114_sgr_v1369_cen_v496_sct_i_vi_color_logscale}}
\end{figure}


\begin{figure}
\plotone{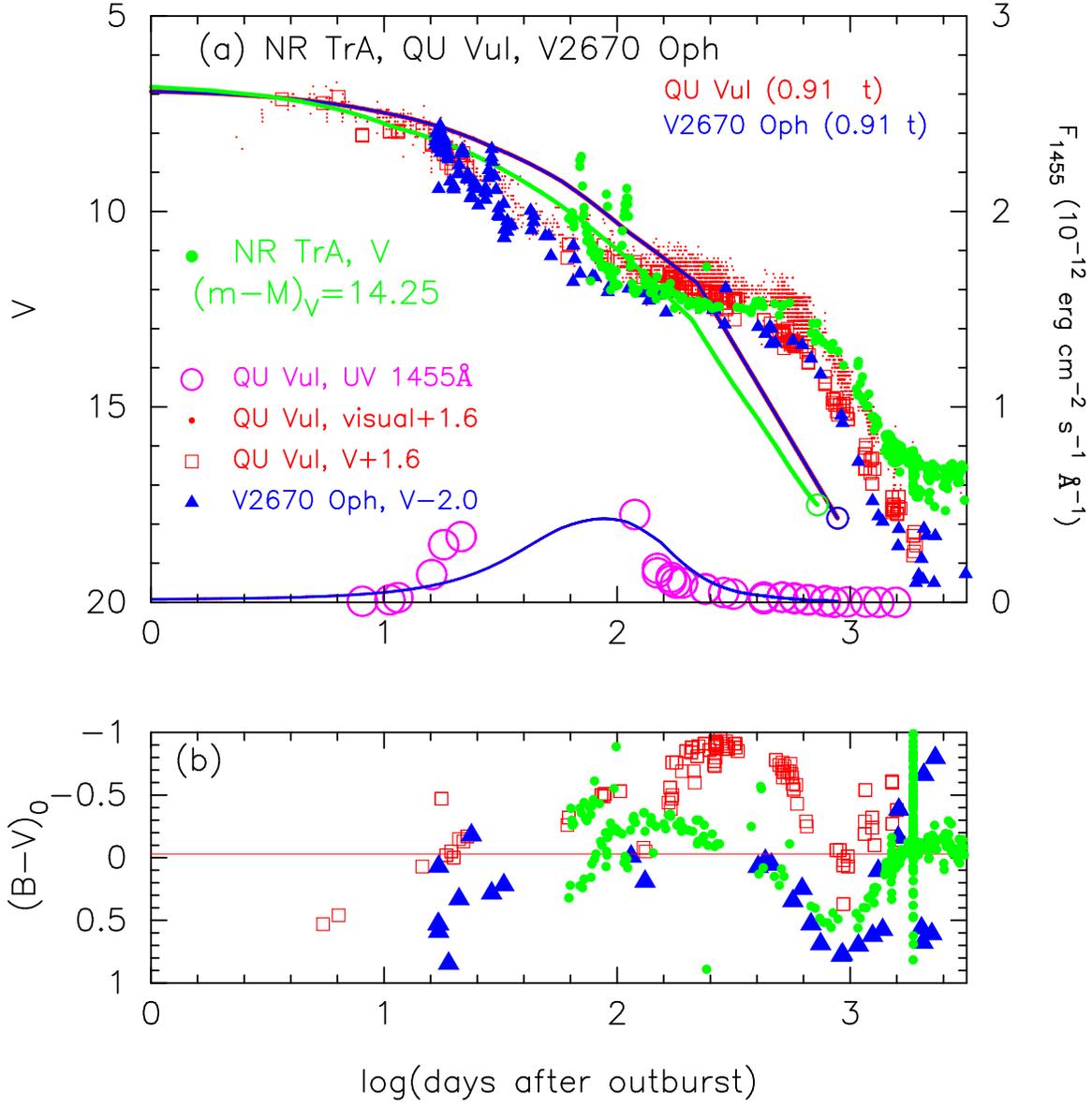}
\caption{
The (a) $V$ light curve and (b) $(B-V)_0$ color curve of NR~TrA 
together with QU~Vul and V2670~Oph.
The data of NR~TrA are taken from AAVSO, VSOLJ, and SMARTS.
In panel (a), we plot a $0.75~M_\sun$ WD model (CO3, solid green lines)
for NR~TrA as well as a $0.70~M_\sun$ WD model (CO3, solid blue lines)
both for V2670~Oph and QU~Vul.
\label{nr_tra_v2670_oph_qu_vul_v_bv_x55z02c10o10_logscale_no2}}
\end{figure}


\begin{figure*}
\plottwo{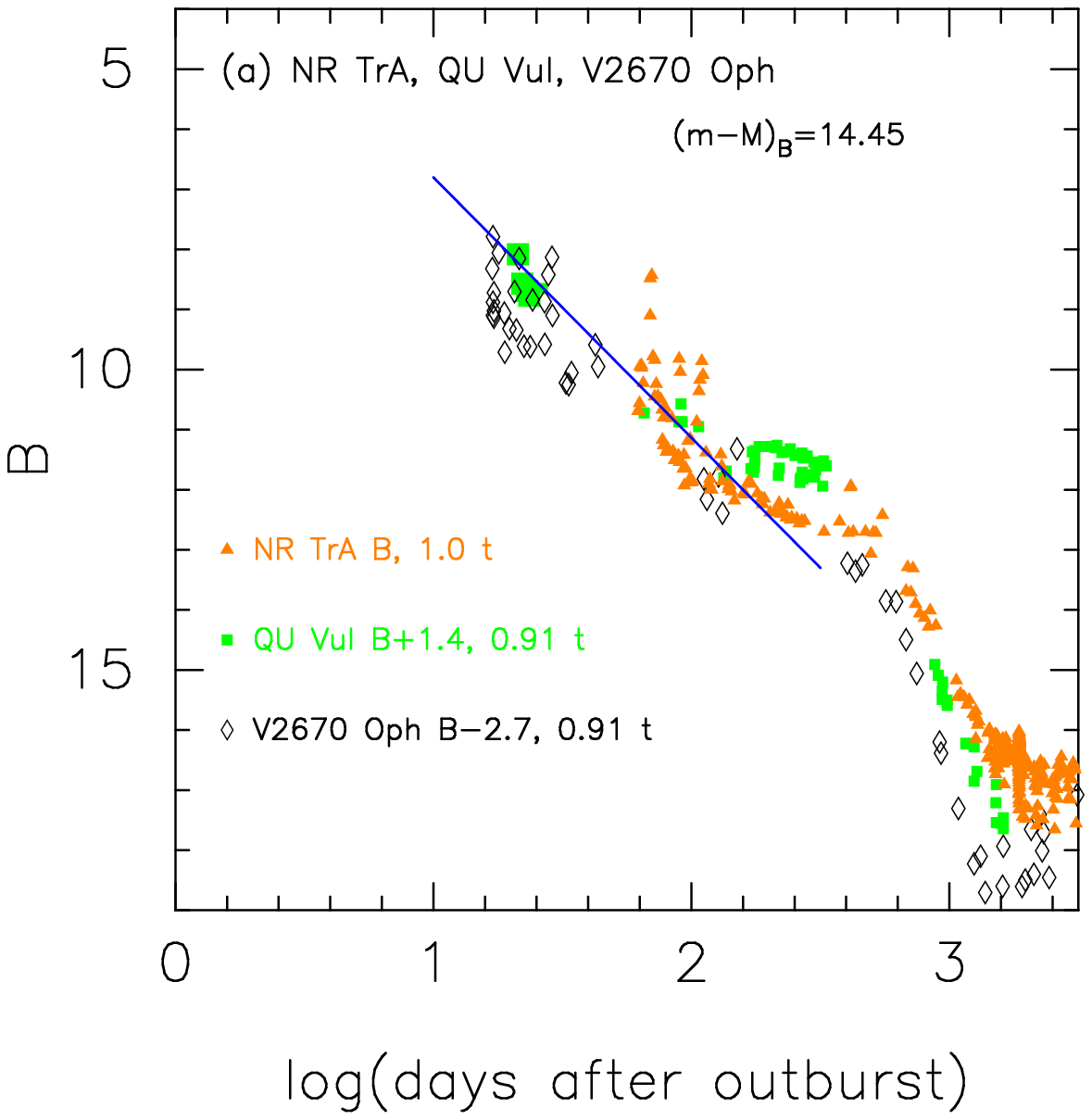}{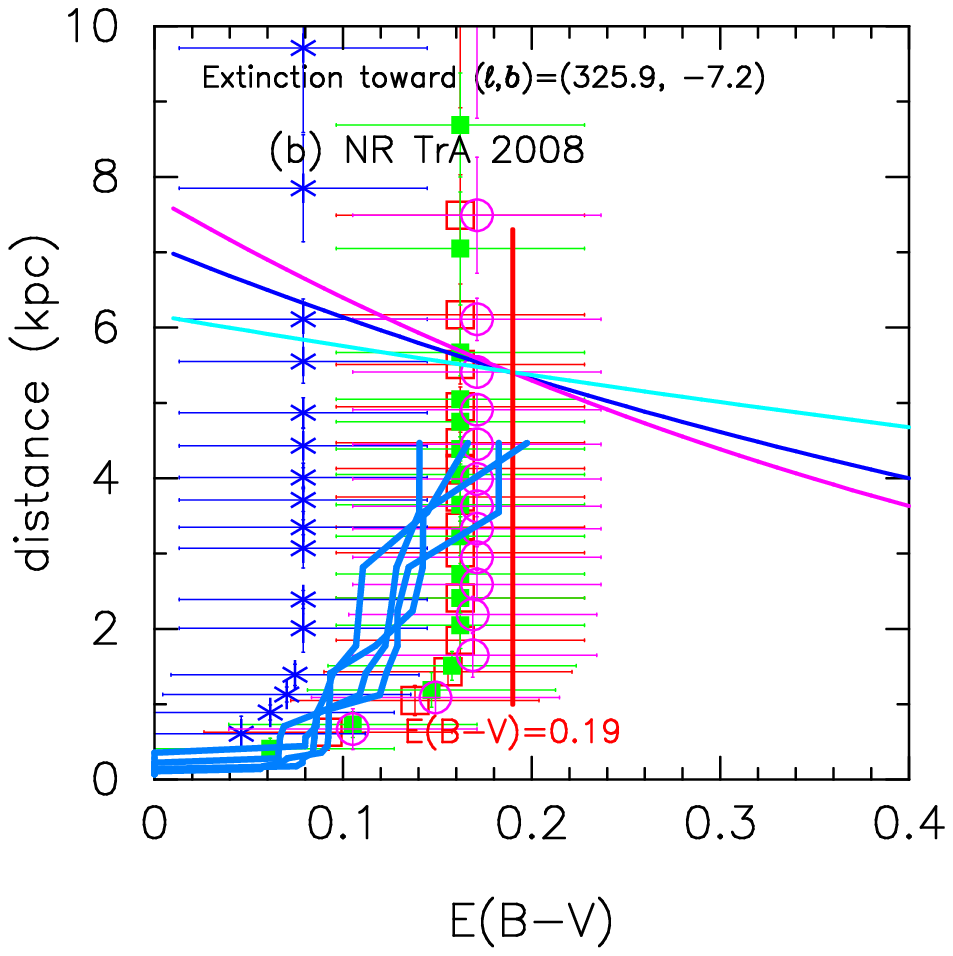}
\caption{
(a) The $B$ light curve of NR~TrA as well as QU~Vul and V2670~Oph. 
The $BVI_{\rm C}$ data of NR~TrA are taken from AAVSO, VSOLJ, and SMARTS.
(b) Various distance-reddening relations toward NR~TrA.
The thin solid lines of magenta, blue, and cyan denote the distance-reddening
relations given by $(m-M)_B=14.45$, $(m-M)_V=14.25$, and $(m-M)_I=13.95$,
respectively.
\label{distance_reddening_nr_tra_bvi_xxxxxx}}
\end{figure*}

\subsection{NR~TrA 2008}
\label{nr_tra_bvi}
We have reanalyzed the $BVI_{\rm C}$ multi-band light/color curves
of NR~TrA based on the time-stretching method.
Figure \ref{nr_tra_v5114_sgr_v1369_cen_v496_sct_i_vi_color_logscale}
shows the (a) $I_{\rm C}$ light and (b) $(V-I_{\rm C})_0$ color curves
of NR~TrA as well as V5114~Sgr, V1369~Cen, and V496~Sct.
The $BVI_{\rm C}$ data of NR~TrA are taken from AAVSO, VSOLJ, and SMARTS.
Adopting the color excess of $E(B-V)= 0.19$ as mentioned below,
we obtain a slightly larger factor of $\log f_{\rm s}= +0.55$ than that of
\citet{hac19kb}.
We apply Equation (8) of \citet{hac19ka} for the $I$ band to Figure
\ref{nr_tra_v5114_sgr_v1369_cen_v496_sct_i_vi_color_logscale}(a)
and obtain
\begin{eqnarray}
(m&-&M)_{I, \rm NR~TrA} \cr
&=& ((m - M)_I + \Delta I_{\rm C})
_{\rm V5114~Sgr} - 2.5 \log 4.6 \cr
&=& 15.55 + 0.05\pm0.2 - 1.675  = 13.93\pm0.2 \cr
&=& ((m - M)_I + \Delta I_{\rm C})
_{\rm V1369~Cen} - 2.5 \log 2.4 \cr
&=& 10.11 + 4.8\pm0.2 - 0.95 = 13.96\pm0.2 \cr
&=& ((m - M)_I + \Delta I_{\rm C})
_{\rm V496~Sct} - 2.5 \log 1.78 \cr
&=& 12.9 + 1.7\pm0.2 - 0.625 = 13.97\pm0.2,
\label{distance_modulus_i_vi_nr_tra}
\end{eqnarray}
where we adopt
$(m-M)_{I, \rm V5114~Sgr}=15.55$ from Appendix \ref{v5114_sgr_ubvi},
$(m-M)_{I, \rm V1369~Cen}=10.11$ from \citet{hac19ka}, and
$(m-M)_{I, \rm V496~Sct}=12.9$ in Appendix \ref{v496_sct_bvi}.
Thus, we obtain $(m-M)_{I, \rm NR~TrA}= 13.95\pm0.2$.


Figure \ref{nr_tra_v2670_oph_qu_vul_v_bv_x55z02c10o10_logscale_no2}
shows the (a) visual and $V$ light curves and (b) $(B-V)_0$ color
curves of NR~TrA together with QU~Vul and V2670~Oph.
Applying Equation (4) of \citet{hac19ka} to them,
we have the relation
\begin{eqnarray}
(m&-&M)_{V, \rm NR~TrA} \cr
&=& ((m - M)_V + \Delta V)_{\rm QU~Vul} - 2.5 \log 0.91 \cr
&=& 12.55 + 1.6\pm0.2 + 0.1 = 14.25\pm0.2 \cr
&=& ((m - M)_V + \Delta V)_{\rm V2670~Oph} - 2.5 \log 0.91 \cr
&=& 16.15 -2.0\pm0.2 + 0.1 = 14.25\pm0.2,
\label{distance_modulus_nr_tra_qu_vul}
\end{eqnarray}
where we adopt $(m-M)_{V, \rm QU~Vul}=12.55$ from Appendix \ref{qu_vul_ubvi}
and $(m-M)_{V, \rm V2670~Oph}=16.15$ from Appendix \ref{v2670_oph_bvi}.
Thus, we adopt $(m-M)_{V, \rm NR~TrA}= 14.25\pm0.2$.

Figure \ref{distance_reddening_nr_tra_bvi_xxxxxx}(a) shows
the $B$ light curves of NR~TrA
together with those of QU~Vul and V2670~Oph.
We apply Equation (7) for the $B$ band to
Figure \ref{distance_reddening_nr_tra_bvi_xxxxxx}(a) and obtain
\begin{eqnarray}
(m&-&M)_{B, \rm NR~TrA} \cr
&=& ((m - M)_B + \Delta B)_{\rm QU~Vul} - 2.5 \log 0.91 \cr
&=& 12.95 + 1.4\pm0.3 + 0.1 = 14.45\pm0.3 \cr
&=& ((m - M)_B + \Delta B)_{\rm V2670~Oph} - 2.5 \log 0.91 \cr
&=& 17.05 - 2.7\pm0.3 + 0.1 = 14.45\pm0.3,
\label{distance_modulus_b_nr_tra_v2670_oph_qu_vul}
\end{eqnarray}
where we adopt $(m - M)_{B, \rm QU~Vul}= 12.55+0.4= 12.95$ from Appendix
\ref{qu_vul_ubvi} and $(m - M)_{B, \rm V2670~Oph}= 16.15 + 0.90= 17.05$
from Appendix \ref{v2670_oph_bvi}.
We have $(m-M)_{B, \rm NR~TrA}= 14.45\pm0.3$.

We plot $(m-M)_B= 14.45$, $(m-M)_V= 14.25$, and $(m-M)_I=13.95$,
which cross at $d=5.4$~kpc and $E(B-V)=0.19$,
in Figure \ref{distance_reddening_nr_tra_bvi_xxxxxx}(b).
The crossing point is consistent with the distance-reddening relation
(unfilled magenta circles) given by \citet{mar06}.
Thus, we obtain $d=5.4\pm0.6$~kpc and $E(B-V)=0.19\pm0.05$ for NR~TrA.


\begin{figure}
\plotone{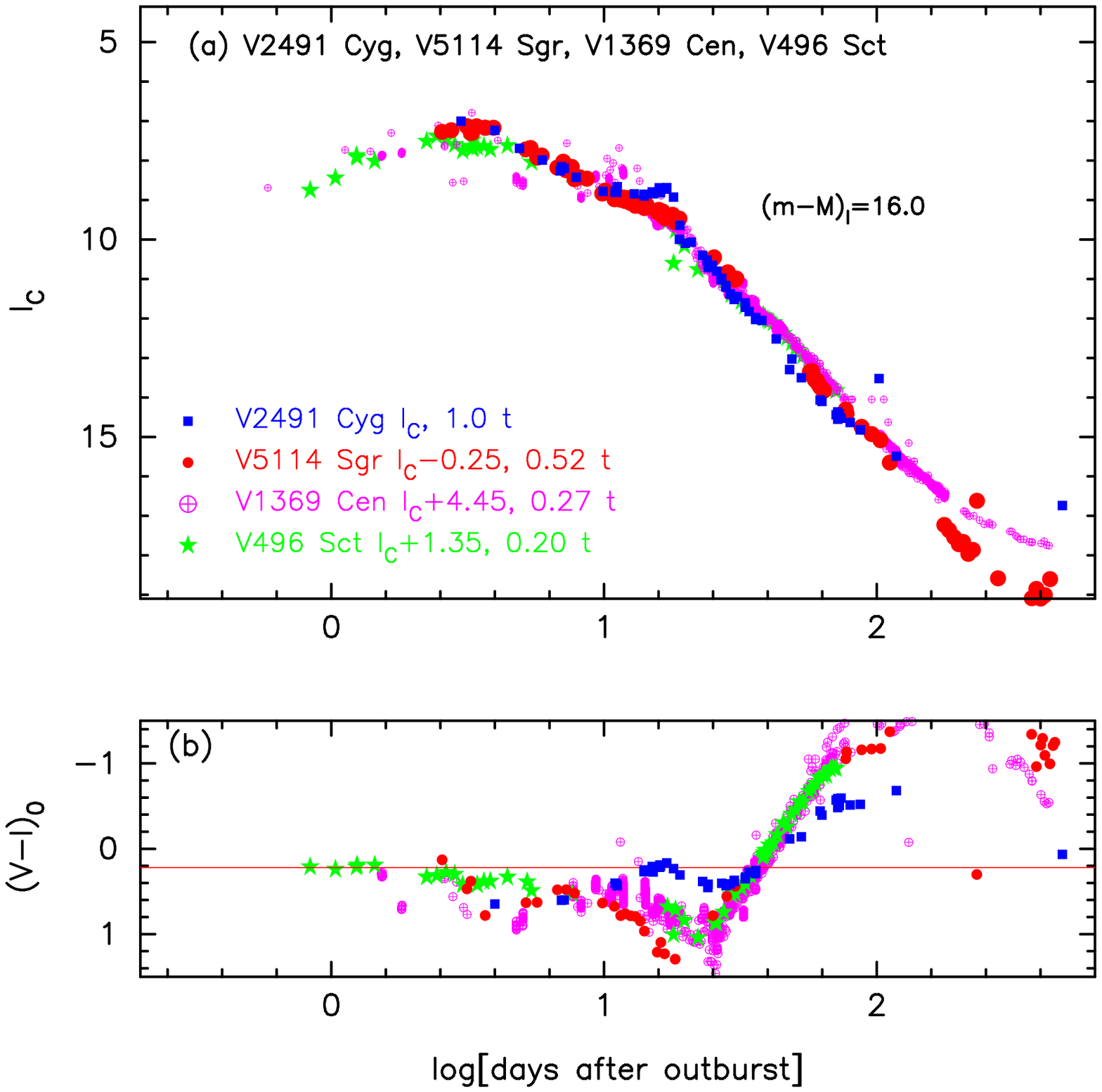}
\caption{
The (a) $I_{\rm C}$ light curve and (b) $(V-I_{\rm C})_0$ color curve
of V2491~Cyg as well as those of V5114~Sgr, V1369~Cen, and V496~Sct.
\label{v2491_cyg_v5114_sgr_v1369_cen_v496_sct_i_vi_color_logscale}}
\end{figure}


\begin{figure}
\plotone{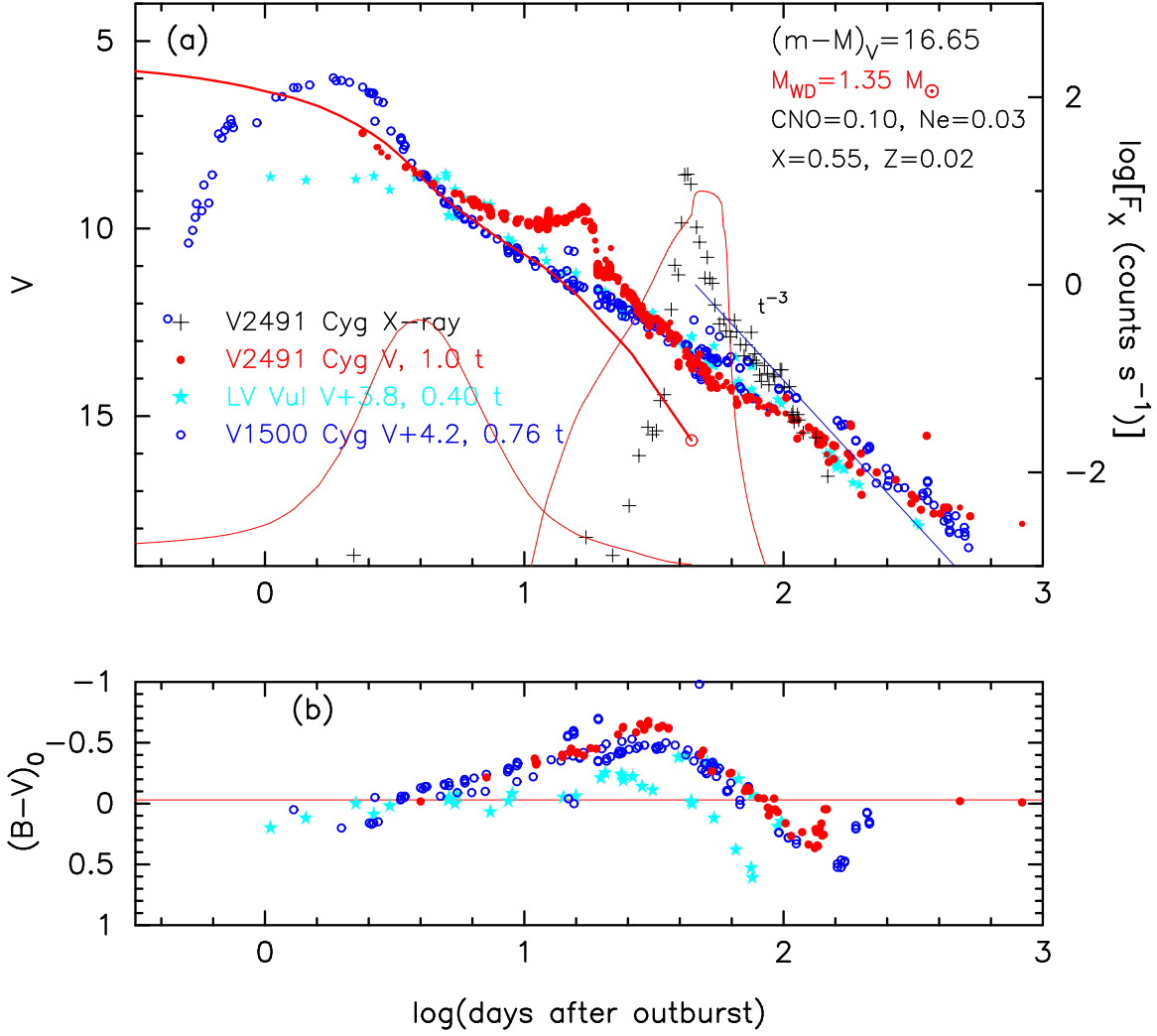}
\caption{
The (a) $V$ light and (b) $(B-V)_0$ color curves of V2491~Cyg 
as well as those of LV~Vul and V1500~Cyg.
The data of V2491~Cyg are the same as those in Figure 48 of \citet{hac19ka}.
We plot a $1.35~M_\sun$ WD model (Ne2, solid red lines) for V2491~Cyg.
\label{v2491_cyg_lv_vul_v1500cyg_cyg_v_color_logscale_no2}}
\end{figure}


\begin{figure*}
\plottwo{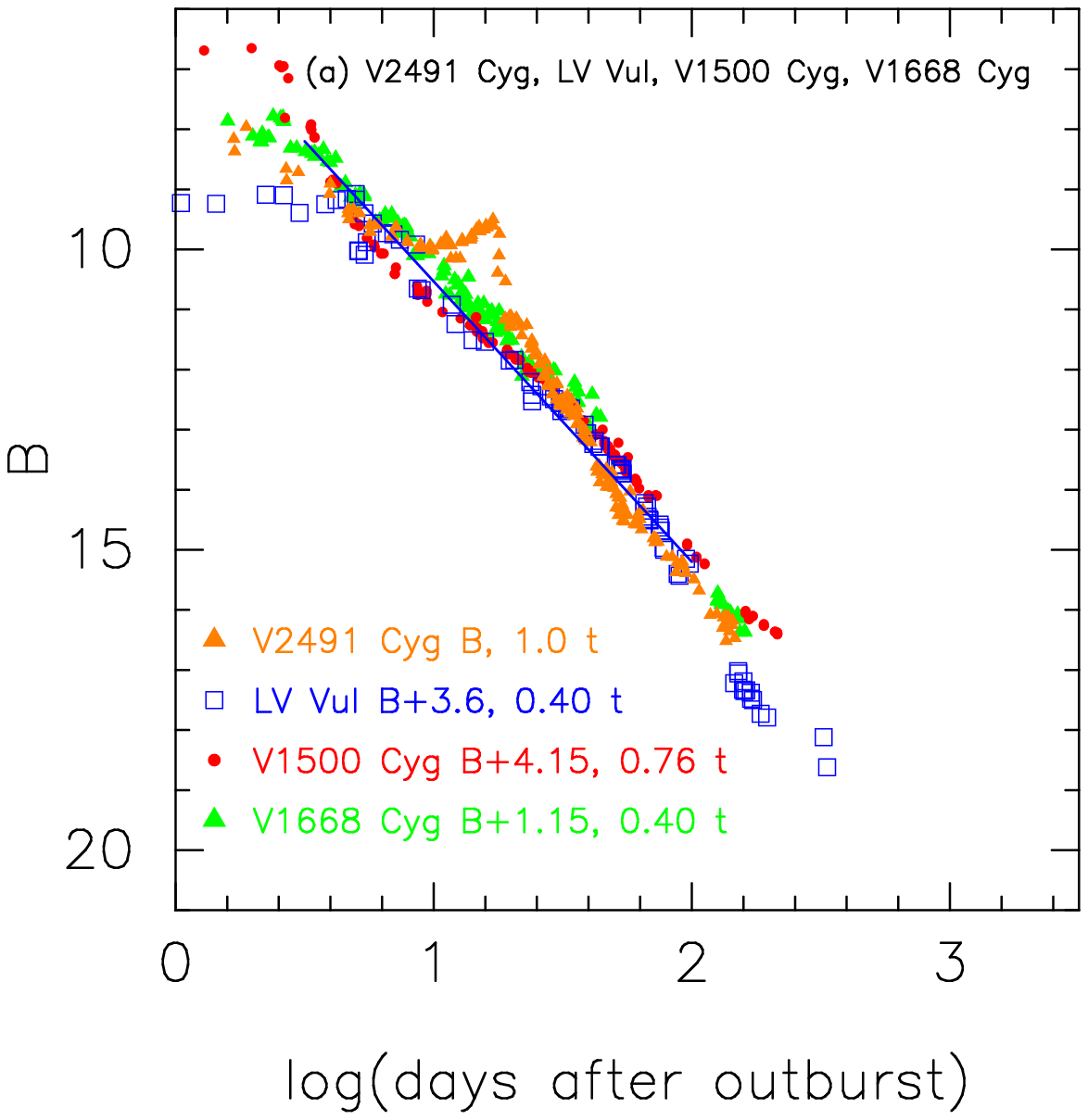}{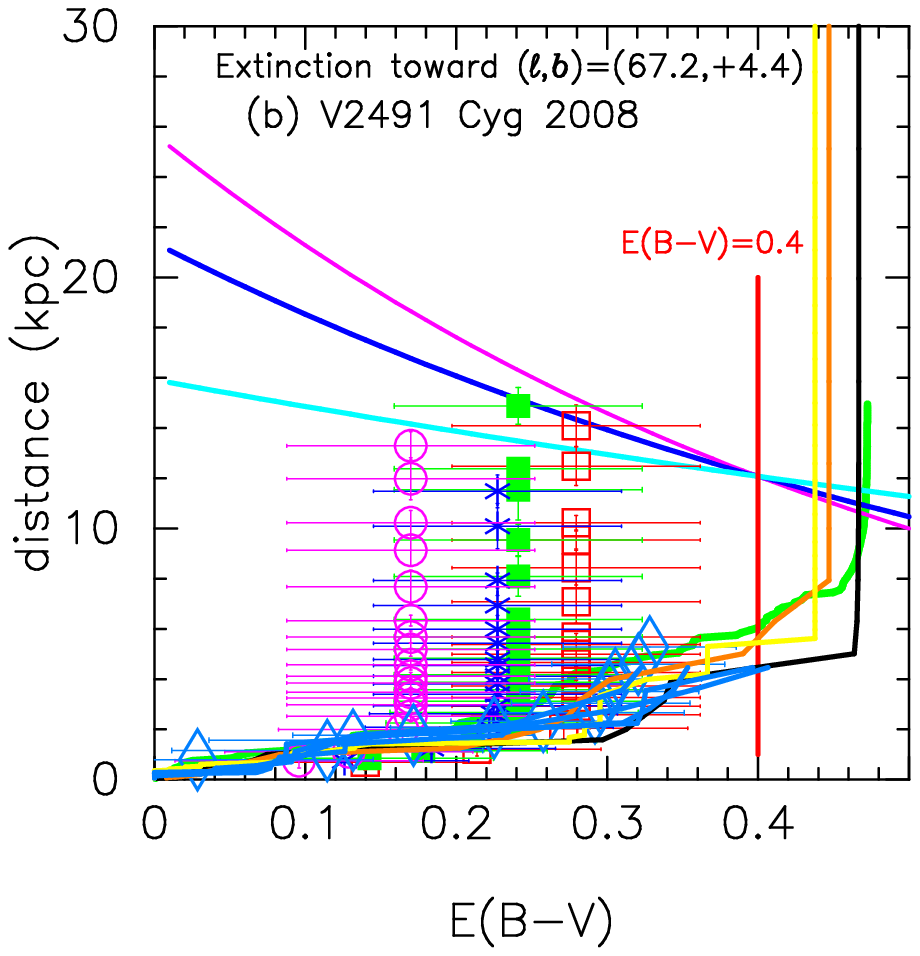}
\caption{
(a) The $B$ light curve of V2491~Cyg as well as LV~Vul, V1500~Cyg, 
and V1668~Cyg.
The $BVI_{\rm C}$ data of V2491~Cyg are taken from \citet{mun11}.
(b) Various distance-reddening relations toward V2491~Cyg.
The thin solid lines of magenta, blue, and cyan denote the distance-reddening
relations given by $(m-M)_B= 17.05$, $(m-M)_V= 16.65$, 
and $(m-M)_I= 16.0$, respectively.
\label{distance_reddening_v2491_cyg_bvi_xxxxxx}}
\end{figure*}

\subsection{V2491~Cyg 2008}
\label{v2491_cyg_bvi}
We have reanalyzed the $BVI_{\rm C}$ multi-band light/color curves
of V2491~Cyg based on the time-stretching method.  
Figure \ref{v2491_cyg_v5114_sgr_v1369_cen_v496_sct_i_vi_color_logscale}
shows the (a) $I_{\rm C}$ light and (b) $(V-I_{\rm C})_0$ color curves
of V2491~Cyg as well as V5114~Sgr, V1369~Cen, and V496~Sct.
The $BVI_{\rm C}$ data of V2491~Cyg are taken from \citet{mun11}.
Adopting the color excess of $E(B-V)= 0.40$ as mentioned below,
we redetermine the timescaling factor $\log f_{\rm s}= -0.40$
for V2491~Cyg from Figure 
\ref{v2491_cyg_v5114_sgr_v1369_cen_v496_sct_i_vi_color_logscale}(b)
and Figure \ref{v2491_cyg_lv_vul_v1500cyg_cyg_v_color_logscale_no2}(b).
We apply Equation (8) of \citet{hac19ka} for the $I$ band to Figure
\ref{v2491_cyg_v5114_sgr_v1369_cen_v496_sct_i_vi_color_logscale}(a)
and obtain
\begin{eqnarray}
(m&-&M)_{I, \rm V2491~Cyg} \cr
&=& ((m - M)_I + \Delta I_{\rm C})
_{\rm V5114~Sgr} - 2.5 \log 0.52 \cr
&=& 15.55 - 0.25\pm0.2 + 0.7 = 16.0\pm0.2 \cr
&=& ((m - M)_I + \Delta I_{\rm C})
_{\rm V1369~Cen} - 2.5 \log 0.27 \cr
&=& 10.11 + 4.45\pm0.2 + 1.425 = 15.99\pm0.2 \cr
&=& ((m - M)_I + \Delta I_{\rm C})
_{\rm V496~Sct} - 2.5 \log 0.20 \cr
&=& 12.9 + 1.35\pm0.2 + 1.75 = 16.0\pm0.2,
\label{distance_modulus_i_vi_v2491_cyg}
\end{eqnarray}
where we adopt
$(m-M)_{I, \rm V5114~Sgr}=15.55$ from Appendix \ref{v5114_sgr_ubvi},
$(m-M)_{I, \rm V1369~Cen}=10.11$ from \citet{hac19ka}, and
$(m-M)_{I, \rm V496~Sct}=12.9$ in Appendix \ref{v496_sct_bvi}.
Thus, we obtain $(m-M)_{I, \rm V2491~Cyg}= 16.0\pm0.2$.

Figure \ref{v2491_cyg_lv_vul_v1500cyg_cyg_v_color_logscale_no2} shows
the (a) $V$ light and (b) $(B-V)_0$ color curves of V2491~Cyg as well as 
LV~Vul and V1500~Cyg.  Applying Equation (4) of \citet{hac19ka} to Figure
\ref{v2491_cyg_lv_vul_v1500cyg_cyg_v_color_logscale_no2}(a),
we have the relation of
\begin{eqnarray}
(m&-&M)_{V, \rm V2491~Cyg} \cr
&=& ((m - M)_V + \Delta V)_{\rm LV~Vul} - 2.5 \log 0.40 \cr
&=& 11.85 + 3.8\pm0.3 + 1.0 = 16.65\pm0.3 \cr
&=& ((m - M)_V + \Delta V)_{\rm V1500~Cyg} - 2.5 \log 0.76 \cr
&=& 12.15 + 4.2\pm0.3 + 0.3 = 16.65\pm0.3.
\label{distance_modulus_v2491_cyg_lv_vul_iv_cep_v1668_cyg}
\end{eqnarray}
Thus, we obtain $\log f_{\rm s}= \log 0.40 = -0.40$ against 
the template nova LV~Vul and $(m-M)_{V, \rm V2491~Cyg}= 16.65\pm0.3$. 

We plot the $B$ light curves of V2491~Cyg together
with those of LV~Vul, V1500~Cyg, and V1668~Cyg in Figure
\ref{distance_reddening_v2491_cyg_bvi_xxxxxx}(a).
We apply Equation (7) of \citet{hac19ka}
for the $B$ band  to Figure
\ref{distance_reddening_v2491_cyg_bvi_xxxxxx}(a) and obtain
\begin{eqnarray}
(m&-&M)_{B, \rm V2491~Cyg} \cr
&=& ((m - M)_B + \Delta B)_{\rm LV~Vul} - 2.5 \log 0.40 \cr
&=& 12.45 + 3.6\pm0.3 + 1.0 = 17.05\pm0.3 \cr
&=& ((m - M)_B + \Delta B)_{\rm V1500~Cyg} - 2.5 \log 0.76 \cr
&=& 12.6 + 4.15\pm0.3 + 0.3 = 17.05\pm0.3 \cr
&=& ((m - M)_B + \Delta B)_{\rm V1668~Cyg} - 2.5 \log 0.40 \cr
&=& 14.9 + 1.15\pm0.3 + 1.0 = 17.05\pm0.3.
\label{distance_modulus_b_v2491_cyg_lv_vul_v1500_cyg_v1668_cyg}
\end{eqnarray}
Thus, we obtain $(m-M)_{B, \rm V2491~Cyg}= 17.05\pm0.2$.

We plot the three distance moduli in
Figure \ref{distance_reddening_v2491_cyg_bvi_xxxxxx}(b).
These three lines broadly cross at $d=12.1$~kpc and $E(B-V)=0.4$.
The crossing point is located between the distance-reddening relations
given by \citet[][open red squares]{mar06}
and by \citet[][solid yellow line]{gre19}.


\begin{figure}
\plotone{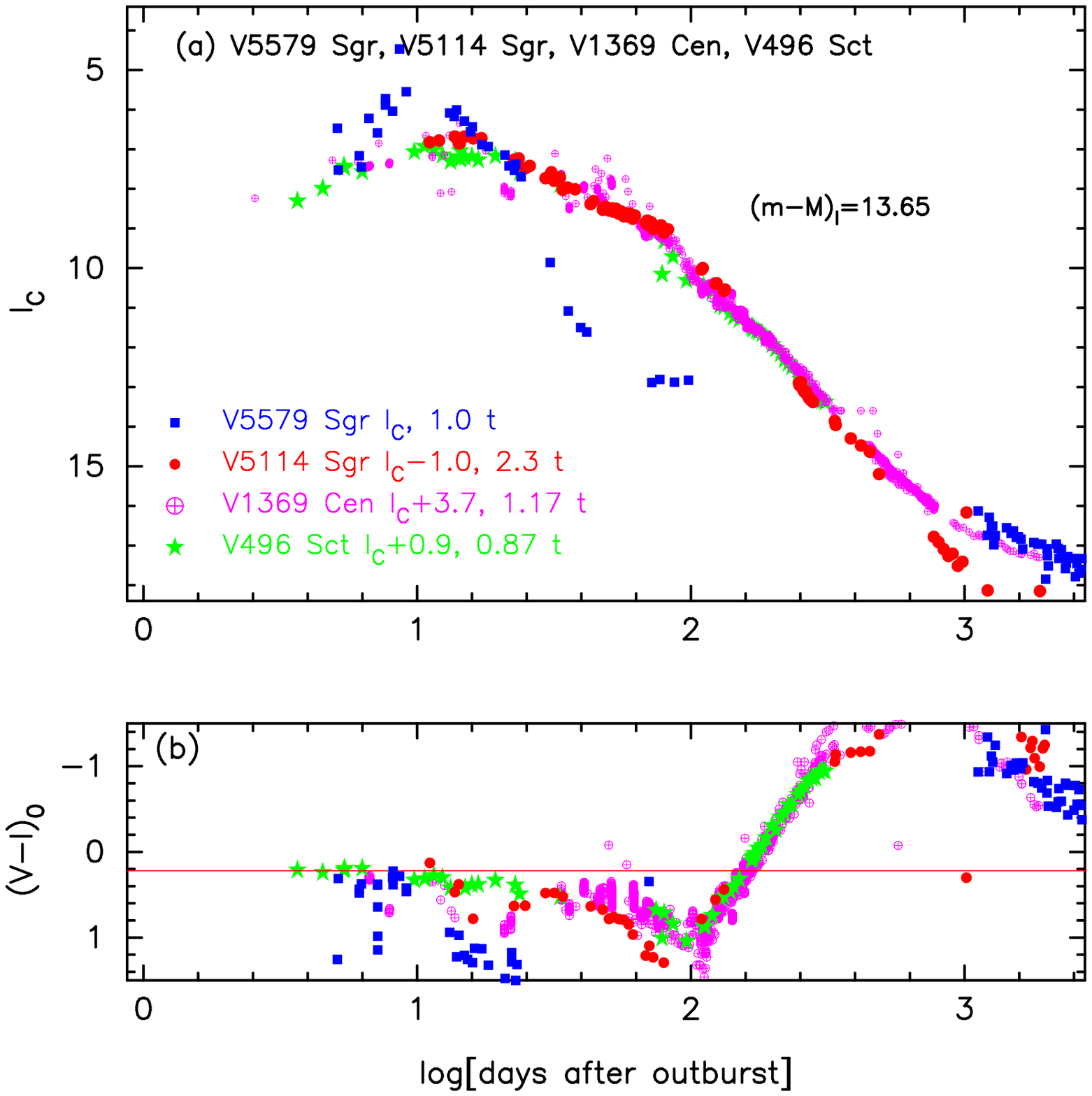}
\caption{
The (a) $I_{\rm C}$ light curve and (b) $(V-I_{\rm C})_0$ color curve
of V5579~Sgr as well as those of V5114~Sgr, V1369~Cen, and V496~Sct.
\label{v5579_sgr_v5114_sgr_v1369_cen_v496_sct_i_vi_color_logscale}}
\end{figure}


\begin{figure}
\plotone{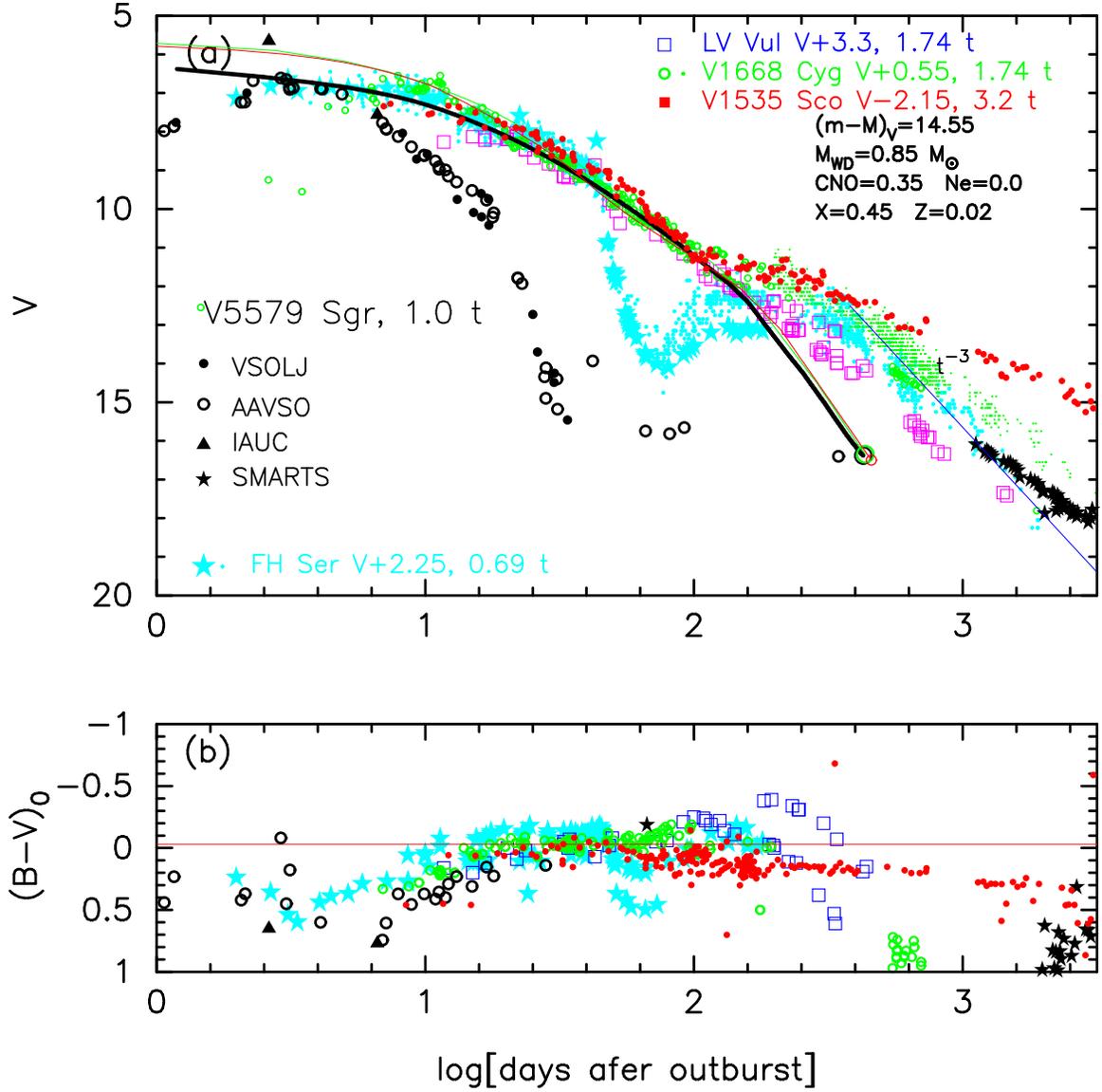}
\caption{
The (a) $V$ light curve and (b) $(B-V)_0$ color curve
of V5579~Sgr as well as those of FH~Ser, LV~Vul, V1668~Cyg, and V1535~Sco.
In panel (a), we show a $0.85~M_\sun$ WD $V$ model 
(CO3, solid black line) for V5579~Sgr as well as   
a $0.98~M_\sun$ WD model (CO3, solid green line) for V1668~Cyg
and $1.20~M_\sun$ WD model (Ne2, solid red line) for V1535~Sco. 
\label{v5579_sgr_lv_vul_v1668_cyg_v1535_sco_v_bv_ub_color_logscale_no2}}
\end{figure}


\begin{figure*}
\gridline{\fig{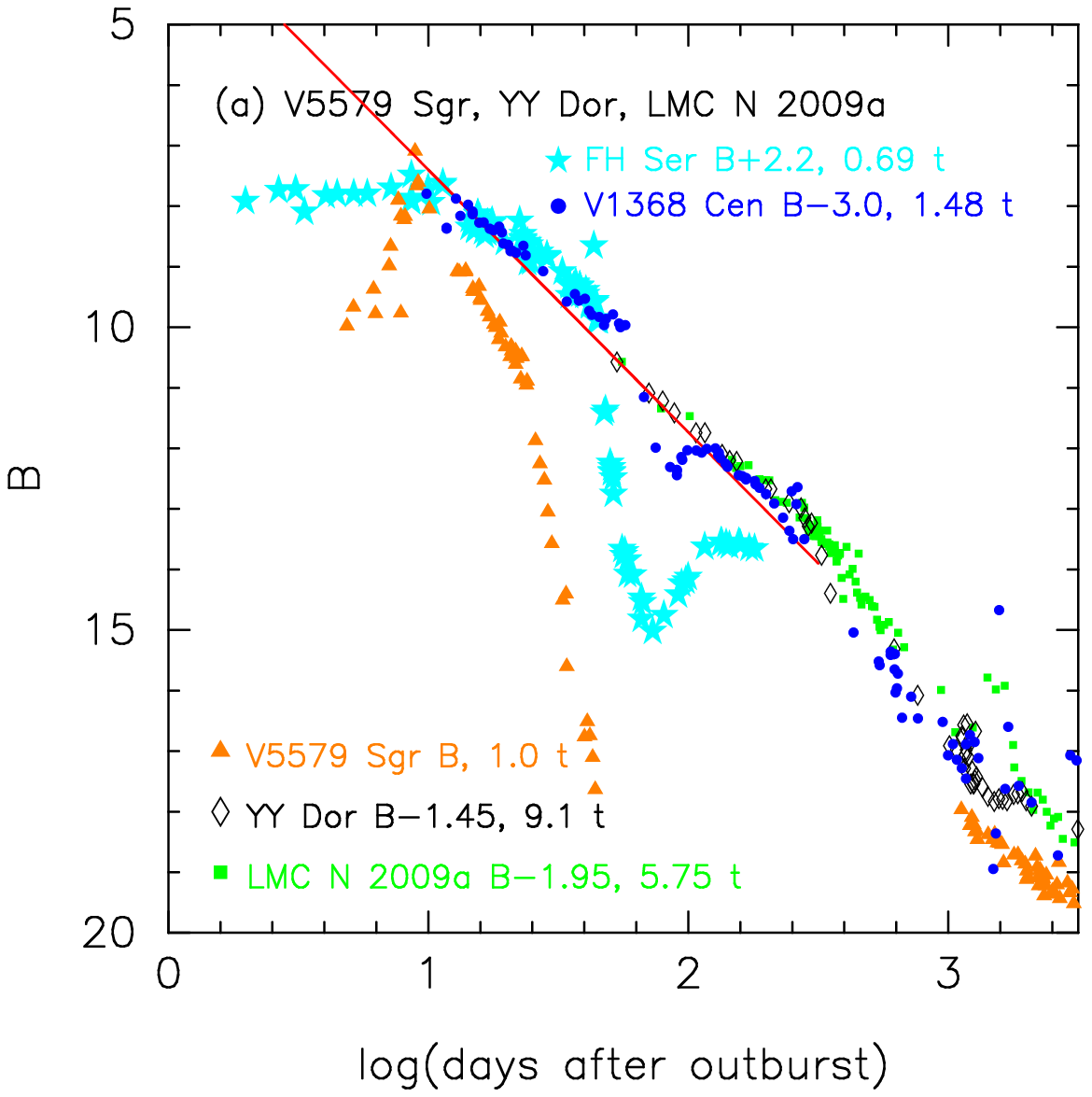}{0.4\textwidth}{(a)}
          \fig{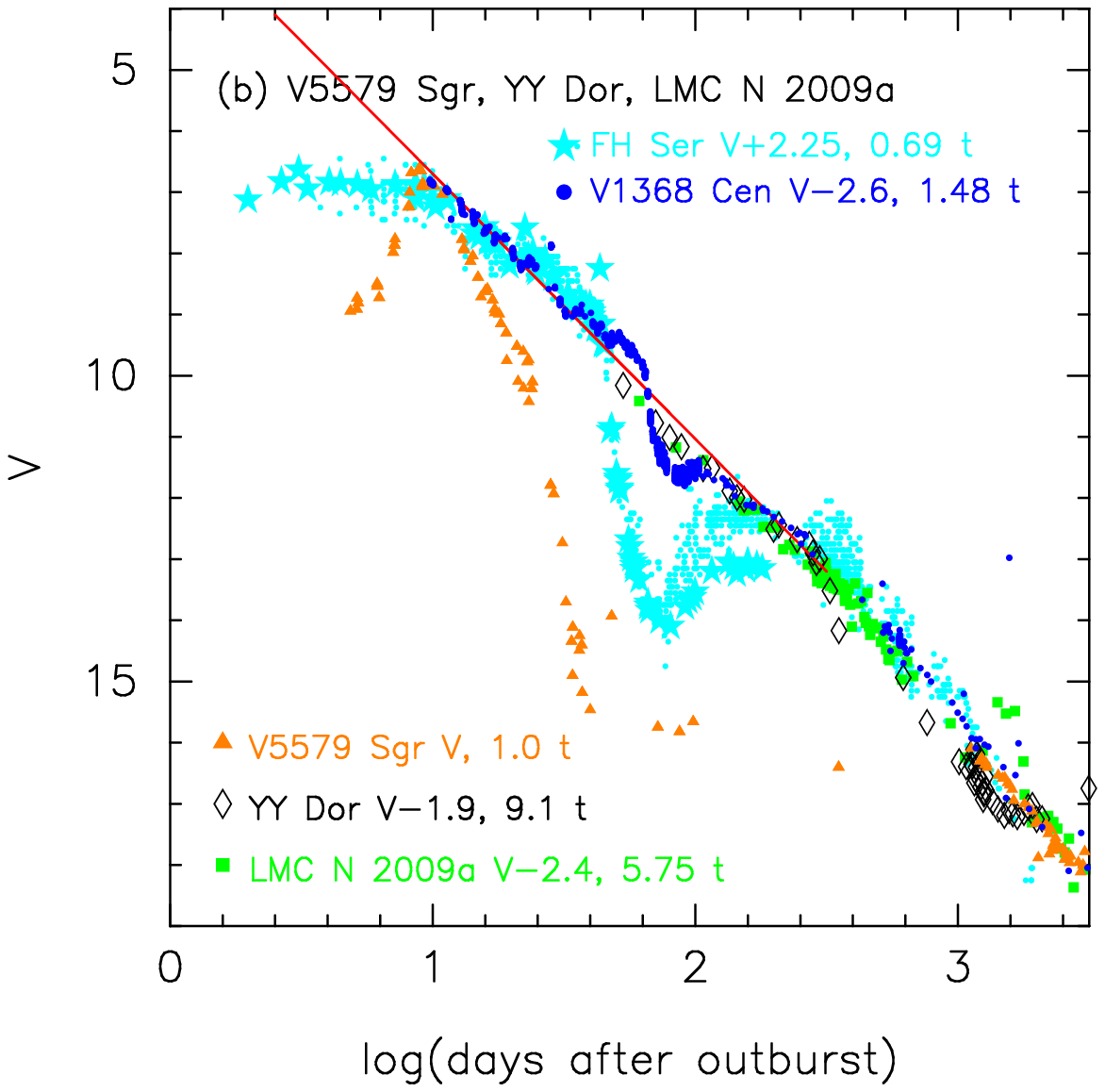}{0.4\textwidth}{(b)}
          }
\gridline{\fig{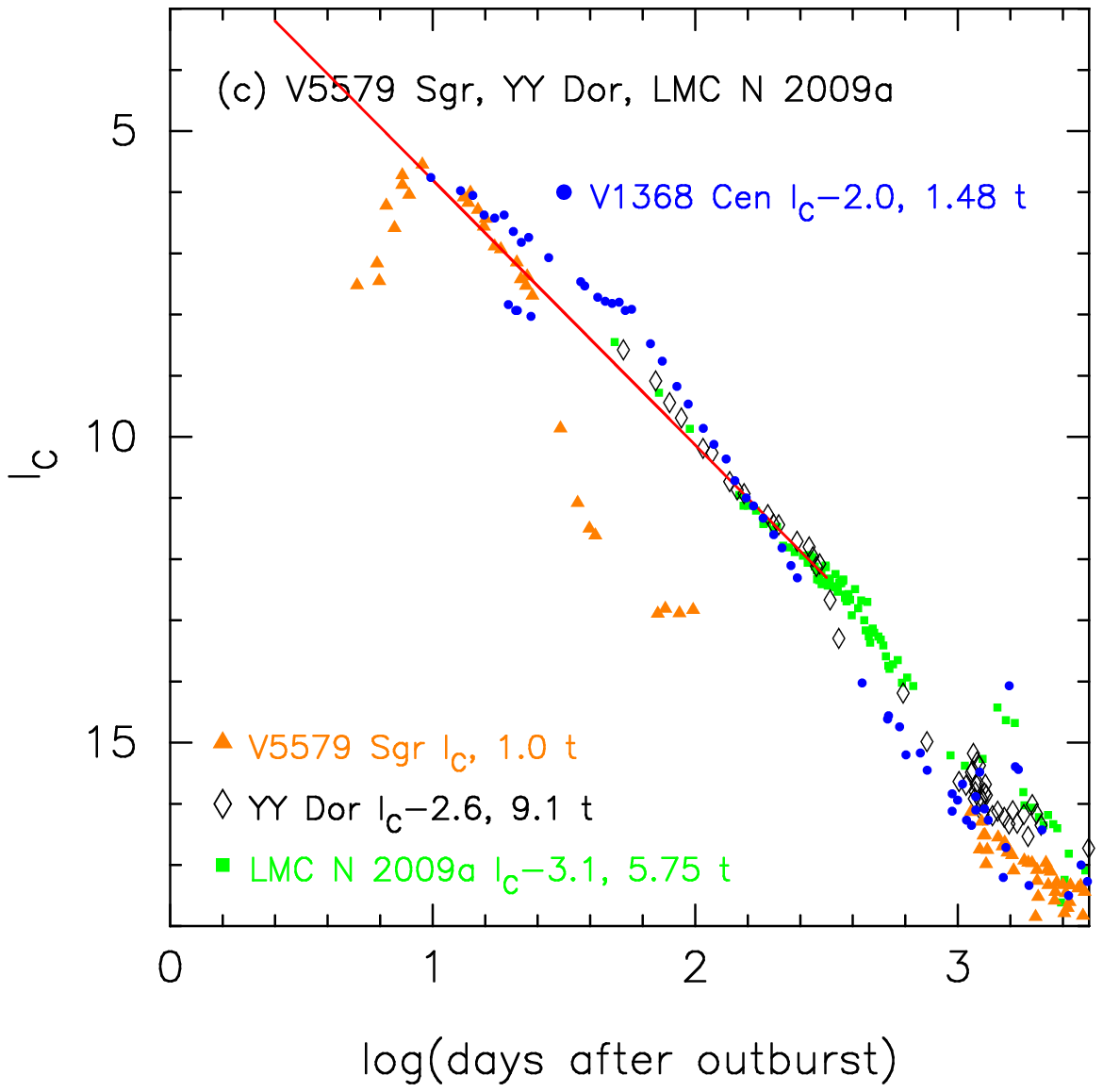}{0.4\textwidth}{(c)}
          \fig{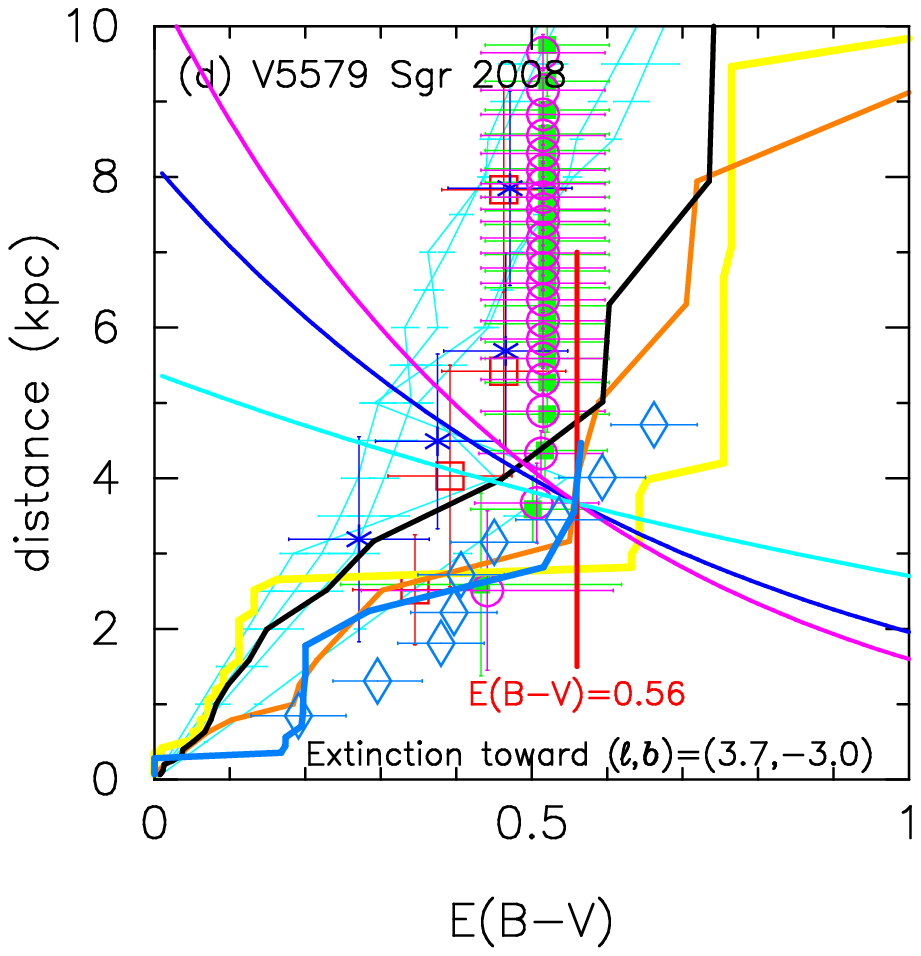}{0.4\textwidth}{(d)}
          }
\caption{
The (a) $B$, (b) $V$, and (c) $I_{\rm C}$ light curves of V5579~Sgr
as well as FH~Ser, V1368~Cen, YY~Dor, and LMC~N~2009a.
The $BVI_{\rm C}$ data of V5579~Sgr are taken from AAVSO, VSOLJ, SMARTS,
and IAU Circular.
(d) Various distance-reddening relations toward V5579~Sgr.
The thin solid lines of magenta, blue, and cyan denote the distance-reddening
relations given by $(m-M)_B= 15.12$, $(m-M)_V= 14.55$, 
and $(m-M)_I= 13.67$, respectively.
\label{distance_reddening_v5579_sgr_bvi_xxxxxx}}
\end{figure*}

\subsection{V5579~Sgr 2008}
\label{v5579_sgr_bvi}
We have reanalyzed the $BVI_{\rm C}$ multi-band light/color curves
of V5579~Sgr based on the time-stretching method.  
The important revised point is the color excess of $E(B-V)$,
which is changed from the previous $E(B-V)= 0.82$ to
the present $E(B-V)= 0.56$ in order to overlap the $(V-I_{\rm C})_0$
color curves of V5579~Sgr with other novae as shown in Figure
\ref{v5579_sgr_v5114_sgr_v1369_cen_v496_sct_i_vi_color_logscale}(b).
Figure \ref{v5579_sgr_v5114_sgr_v1369_cen_v496_sct_i_vi_color_logscale}
shows the (a) $I_{\rm C}$ light and (b) $(V-I_{\rm C})_0$ color curves
of V5579~Sgr as well as V5114~Sgr, V1369~Cen, and V496~Sct.
The $BVI_{\rm C}$ data of V5579~Sgr are taken from AAVSO, VSOLJ, SMARTS,
and IAU Circular No.8937, 8930, and 8930.
Adopting the color excess of $E(B-V)= 0.56$ as mentioned below,
we redetermine the timescaling factor $\log f_{\rm s}= +0.24$ 
against that of LV~Vul.
This is because the $(V-I)_0$ color evolution of V5579~Sgr overlaps with
the other novae as much as possible.
Then, we apply Equation (8) of \citet{hac19ka} for the $I$ band to Figure
\ref{v5579_sgr_v5114_sgr_v1369_cen_v496_sct_i_vi_color_logscale}(a)
and obtain
\begin{eqnarray}
(m&-&M)_{I, \rm V5579~Sgr} \cr
&=& ((m - M)_I + \Delta I_{\rm C})
_{\rm V5114~Sgr} - 2.5 \log 2.3 \cr
&=& 15.55 - 1.0\pm0.2 - 0.9 = 13.65\pm0.2 \cr
&=& ((m - M)_I + \Delta I_{\rm C})
_{\rm V1369~Cen} - 2.5 \log 1.17 \cr
&=& 10.11 + 3.7\pm0.2 - 0.175 = 13.64\pm0.2 \cr
&=& ((m - M)_I + \Delta I_{\rm C})
_{\rm V496~Sct} - 2.5 \log 0.87 \cr
&=& 12.9 + 0.9\pm0.2 - 0.15 = 13.65\pm0.2,
\label{distance_modulus_i_vi_v5579_sgr}
\end{eqnarray}
where we adopt
$(m-M)_{I, \rm V5114~Sgr}=15.55$ in Appendix \ref{v5114_sgr_ubvi},
$(m-M)_{I, \rm V1369~Cen}=10.11$ from \citet{hac19ka}, and
$(m-M)_{I, \rm V496~Sct}=12.9$ in Appendix \ref{v496_sct_bvi}.
Thus, we obtain $(m-M)_{I, \rm V5579~Sgr}= 13.65\pm0.2$.

Figure 
\ref{v5579_sgr_lv_vul_v1668_cyg_v1535_sco_v_bv_ub_color_logscale_no2}
shows the (a) $V$ and (b) $(B-V)_0$ evolutions of V5579~Sgr
as well as those of FH~Ser, LV~Vul, V1668~Cyg, and V1535~Sco.
The data of V1535~Sco are the same as those in Section \ref{v1535_sco_vi}
and Appendix \ref{v1535_sco_bvi}.
Applying Equation (4) of \citet{hac19ka} for the $V$ band to them,
we have the relation
\begin{eqnarray}
(m&-&M)_{V, \rm V5579~Sgr} \cr
&=& ((m - M)_V + \Delta V)_{\rm FH~Ser} - 2.5 \log 0.69 \cr
&=& 11.9 + 2.25\pm0.2 + 0.4 = 14.55\pm0.2 \cr
&=& ((m - M)_V + \Delta V)_{\rm LV~Vul} - 2.5 \log 1.74 \cr
&=& 11.85 + 3.3\pm0.2 - 0.6 = 14.55\pm0.2 \cr
&=& ((m - M)_V + \Delta V)_{\rm V1668~Cyg} - 2.5 \log 1.74 \cr
&=& 14.6 + 0.55\pm0.2 - 0.6 = 14.55\pm0.2 \cr
&=& ((m - M)_V + \Delta V)_{\rm V1535~Sco} - 2.5 \log 3.2 \cr
&=& 17.95 - 2.15\pm0.2 - 1.25 = 14.55\pm0.2,
\label{distance_modulus_v_bv_v5579_sgr}
\end{eqnarray}
where we adopt $(m-M)_{V, \rm LV~Vul}=11.85$ and
$(m-M)_{V, \rm V1668~Cyg}=14.6$ both from \citet{hac19ka}, and
$(m-M)_{V, \rm V1535~Sco}=17.95$ in Appendix \ref{v1535_sco_bvi}.
Here, we have redetermined $(m-M)_{V, \rm FH~Ser}=11.9$ and 
$\log f_{\rm s}= +0.40$ against LV~Vul from the fittings in Figure
\ref{v5579_sgr_lv_vul_v1668_cyg_v1535_sco_v_bv_ub_color_logscale_no2}. 
Thus, we obtain $(m-M)_{V, \rm V5579~Sgr}= 14.55\pm0.1$ 
and $\log f_{\rm s}= \log 1.74 = +0.24$ against LV~Vul.

We also plot the $B$, $V$, and $I_{\rm C}$ light curves of V5579~Sgr
together with FH~Ser, V1368~Cen, and the LMC novae,
YY~Dor and LMC~N~2009a, in Figure
\ref{distance_reddening_v5579_sgr_bvi_xxxxxx}(a)(b)(c).
We apply Equation (7) of \citet{hac19ka} for the $B$ band to 
Figure \ref{distance_reddening_v5579_sgr_bvi_xxxxxx}(a) and obtain
\begin{eqnarray}
(m&-&M)_{B, \rm V5579~Sgr} \cr
&=& ((m - M)_B + \Delta B)_{\rm FH~Ser} - 2.5 \log 0.69 \cr
&=& 12.5 + 2.2\pm0.2 + 0.4 = 15.1\pm0.2 \cr
&=& ((m - M)_B + \Delta B)_{\rm V1368~Cen} - 2.5 \log 1.48 \cr
&=& 18.58 - 3.0\pm0.2 - 0.425 = 15.15\pm0.2 \cr
&=& ((m - M)_B + \Delta B)_{\rm YY~Dor} - 2.5 \log 9.1 \cr
&=& 18.98 - 1.45\pm0.2 - 2.4 = 15.13\pm0.2 \cr
&=& ((m - M)_B + \Delta B)_{\rm LMC~N~2009a} - 2.5 \log 5.75 \cr
&=& 18.98 - 1.95\pm0.2 - 1.9 = 15.13\pm0.2.
\label{distance_modulus_b_v5579_sgr_yy_dor_lmcn2009a}
\end{eqnarray}
Thus, we have $(m-M)_{B, \rm V5579~Sgr}= 15.12\pm0.1$.

For the $V$ light curves in Figure
\ref{distance_reddening_v5579_sgr_bvi_xxxxxx}(b),
we similarly obtain
\begin{eqnarray}
(m&-&M)_{V, \rm V5579~Sgr} \cr
&=& ((m - M)_V + \Delta V)_{\rm FH~Ser} - 2.5 \log 0.69 \cr
&=& 11.9 + 2.25\pm0.2 + 0.4 = 14.55\pm0.2 \cr
&=& ((m - M)_V + \Delta V)_{\rm V1368~Cen} - 2.5 \log 1.48 \cr
&=& 17.6 - 2.6\pm0.2 - 0.425 = 14.58\pm0.2 \cr
&=& ((m - M)_V + \Delta V)_{\rm YY~Dor} - 2.5 \log 9.1 \cr
&=& 18.86 - 1.9\pm0.2 - 2.4 = 14.56\pm0.2 \cr
&=& ((m - M)_V + \Delta V)_{\rm LMC~N~2009a} - 2.5 \log 5.75 \cr
&=& 18.86 - 2.4\pm0.2 - 1.9 = 14.56\pm0.2.
\label{distance_modulus_v_v5579_sgr_yy_dor_lmcn2009a}
\end{eqnarray}
We have $(m-M)_{V, \rm V5579~Sgr}= 14.56\pm0.1$, which is
consistent with Equation (\ref{distance_modulus_v_bv_v5579_sgr}).

We apply Equation (8) of \citet{hac19ka} for
the $I_{\rm C}$ band to Figure
\ref{distance_reddening_v5579_sgr_bvi_xxxxxx}(c) and obtain
\begin{eqnarray}
(m&-&M)_{I, \rm V5579~Sgr} \cr
&=& ((m - M)_I + \Delta I_C)_{\rm V1368~Cen} - 2.5 \log 1.48 \cr
&=& 16.1 - 2.0\pm0.2 - 0.425 = 13.67\pm 0.2 \cr
&=& ((m - M)_I + \Delta I_C)_{\rm YY~Dor} - 2.5 \log 9.1 \cr
&=& 18.67 - 2.6\pm0.2 - 2.4 = 13.67\pm 0.2 \cr
&=& ((m - M)_I + \Delta I_C)_{\rm LMC~N~2009a} - 2.5 \log 5.75 \cr
&=& 18.67 - 3.1\pm0.2 - 1.9 = 13.67\pm 0.2.
\label{distance_modulus_i_v5579_sgr_yy_dor_lmcn2009a}
\end{eqnarray}
Thus, we have $(m-M)_{I, \rm V5579~Sgr}= 13.67\pm0.1$.

Figure \ref{distance_reddening_v5579_sgr_bvi_xxxxxx}(d) depicts the
three distance moduli in $B$, $V$, and $I_{\rm C}$ bands.
These three lines cross at $d=3.6$~kpc and $E(B-V)=0.56$.
The crossing point is consistent with the distance-reddening relations
given by \citet[][orange line]{gre18}, \citet[][cyan-blue line]{chen19},
and \citet[][unfilled cyan-blue diamonds]{ozd18}.


\begin{figure}
\plotone{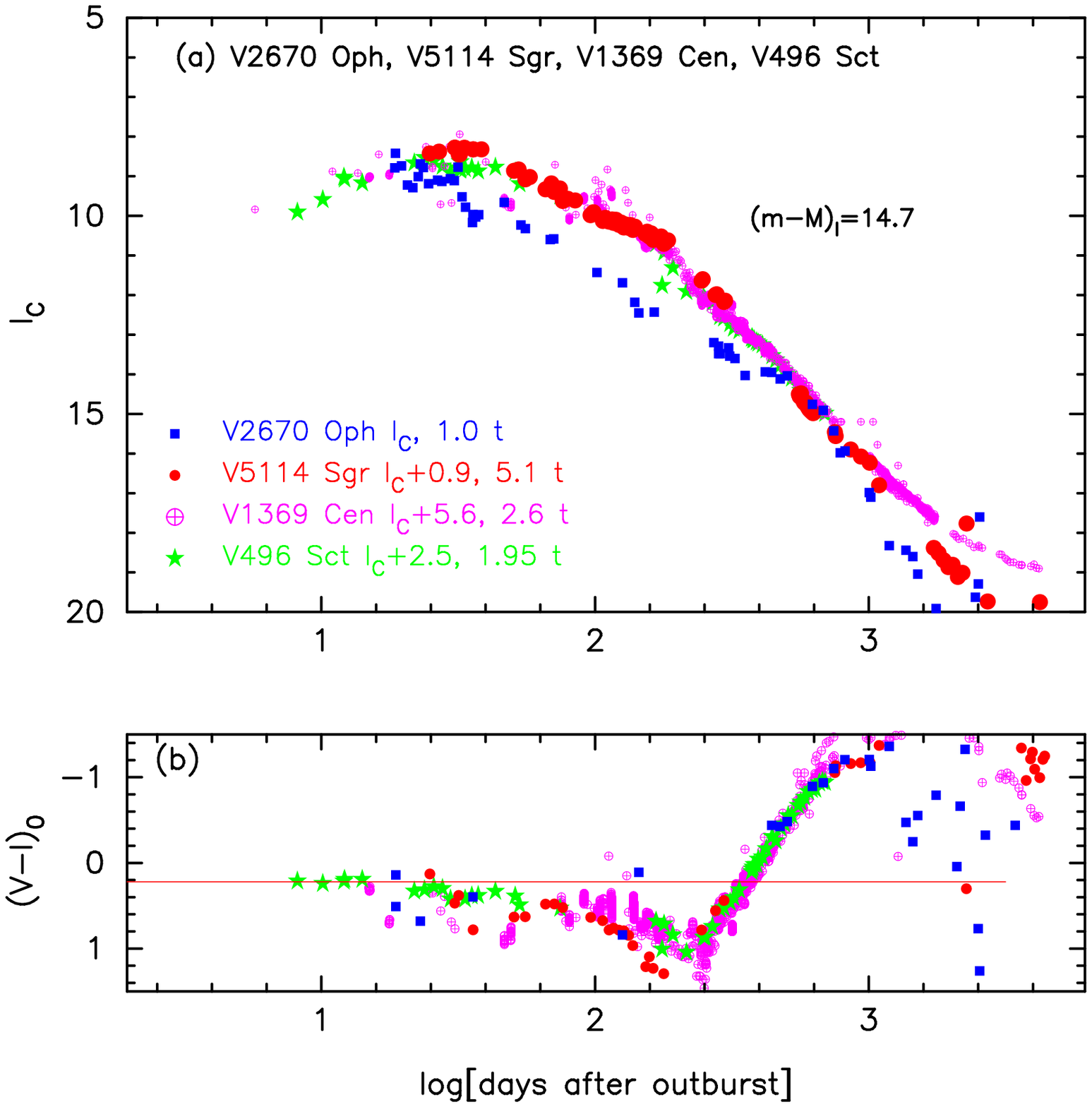}
\caption{
The (a) $I_{\rm C}$ light curve and (b) $(V-I_{\rm C})_0$ color curve
of V2670~Oph as well as those of V5114~Sgr, V1369~Cen, and V496~Sct.
\label{v2670_oph_v5114_sgr_v1369_cen_v496_sct_i_vi_color_logscale}}
\end{figure}


\begin{figure}
\plotone{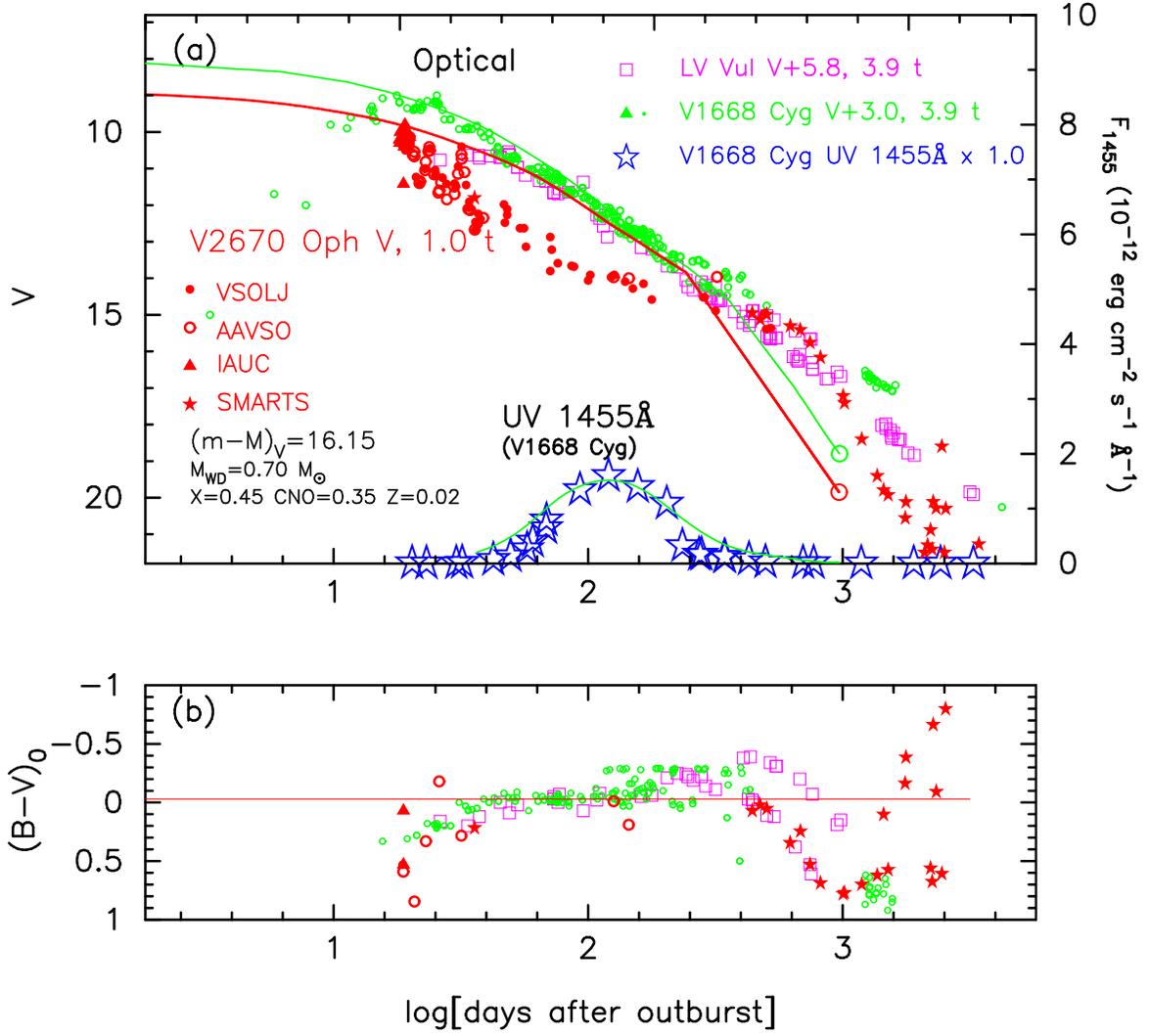}
\caption{
The (a) $V$ light and (b) $(B-V)_0$ color curves of V2670~Oph (red symbols)
as well as those of LV~Vul and V1668~Cyg.
The data of V2670~Oph are taken from IAU Circular, AAVSO, VSOLJ, and SMARTS.
In panel (a), we add a $0.70~M_\sun$ WD model (CO3, solid red lines)
for V2670~Oph as well as a $0.98~M_\sun$ WD model (CO3, solid green lines)
for V1668~Cyg.
\label{v2670_oph_v1668_cyg_lv_vul_v_bv_x45z02c10o15_logscale_no2}}
\end{figure}


\begin{figure*}
\gridline{\fig{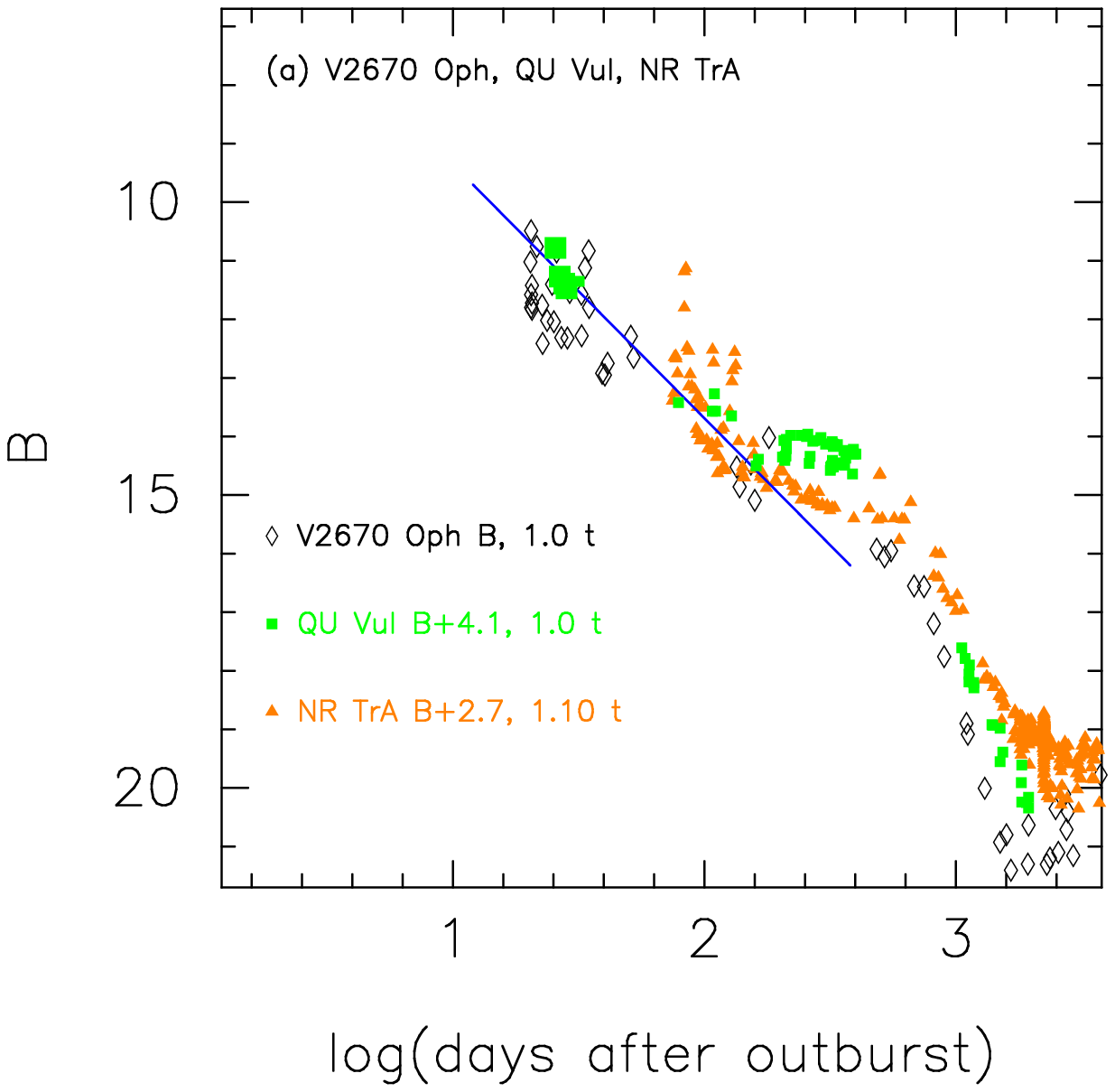}{0.4\textwidth}{(a)}
          \fig{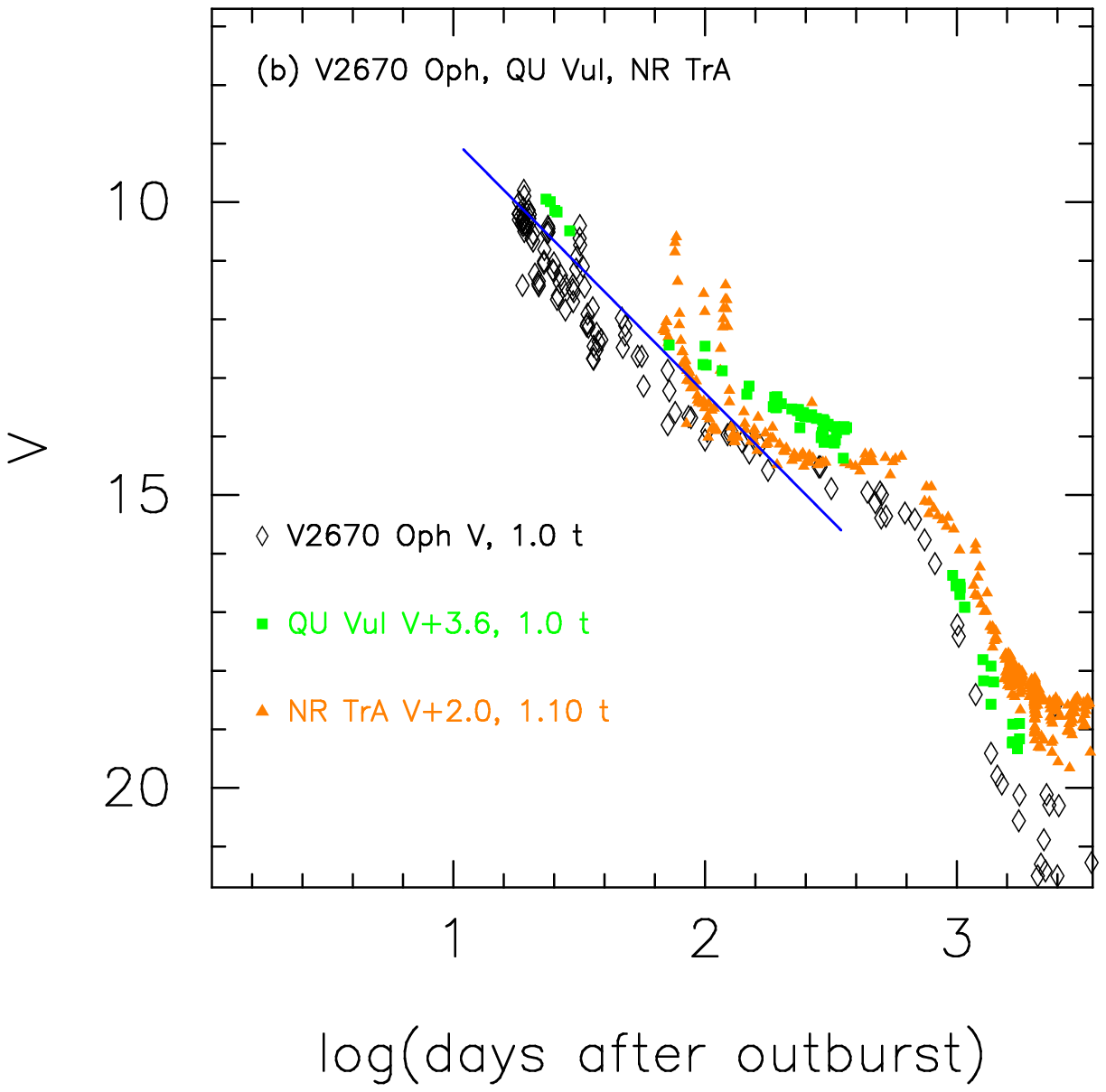}{0.4\textwidth}{(b)}
          }
\gridline{\fig{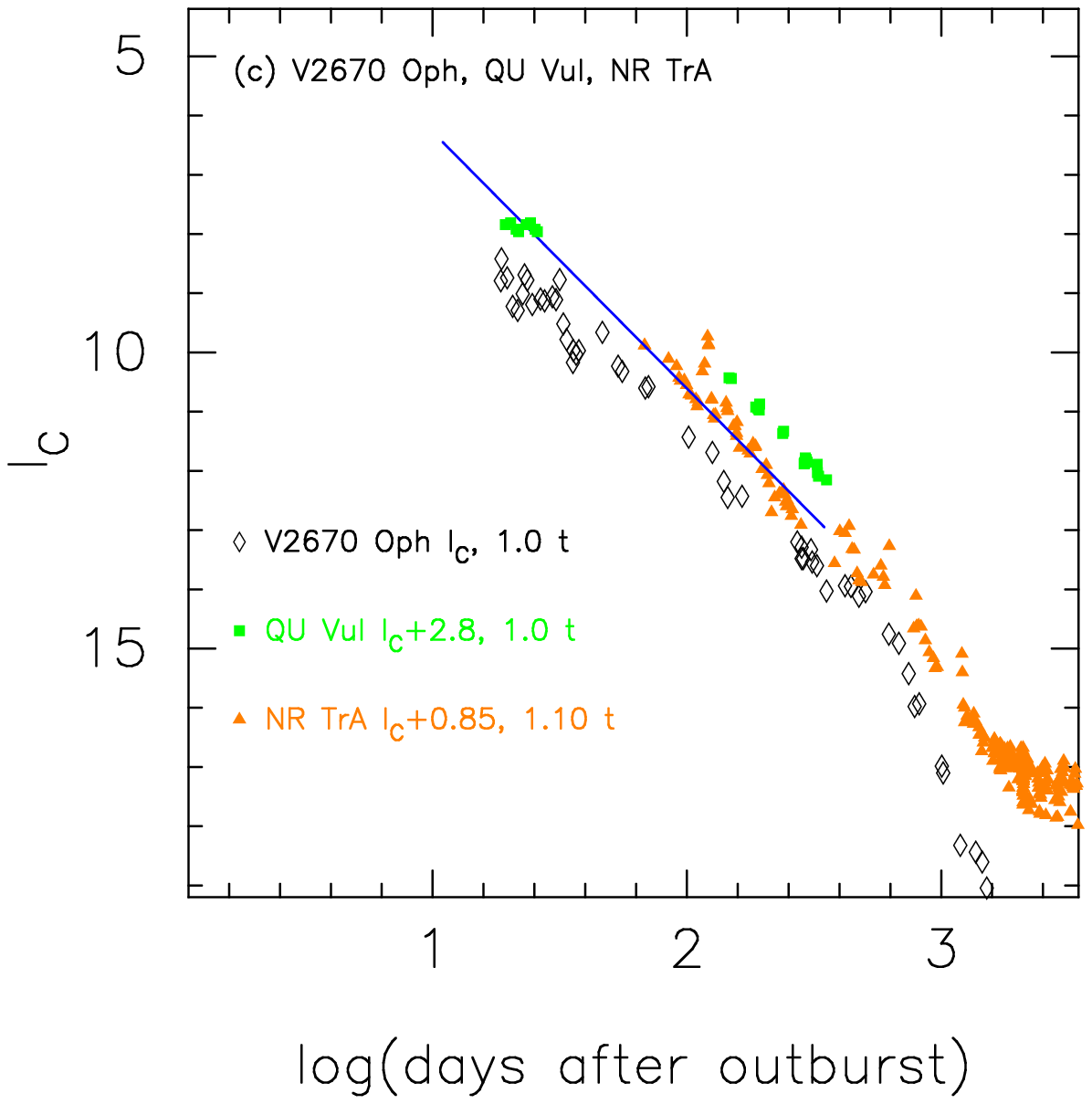}{0.4\textwidth}{(c)}
          \fig{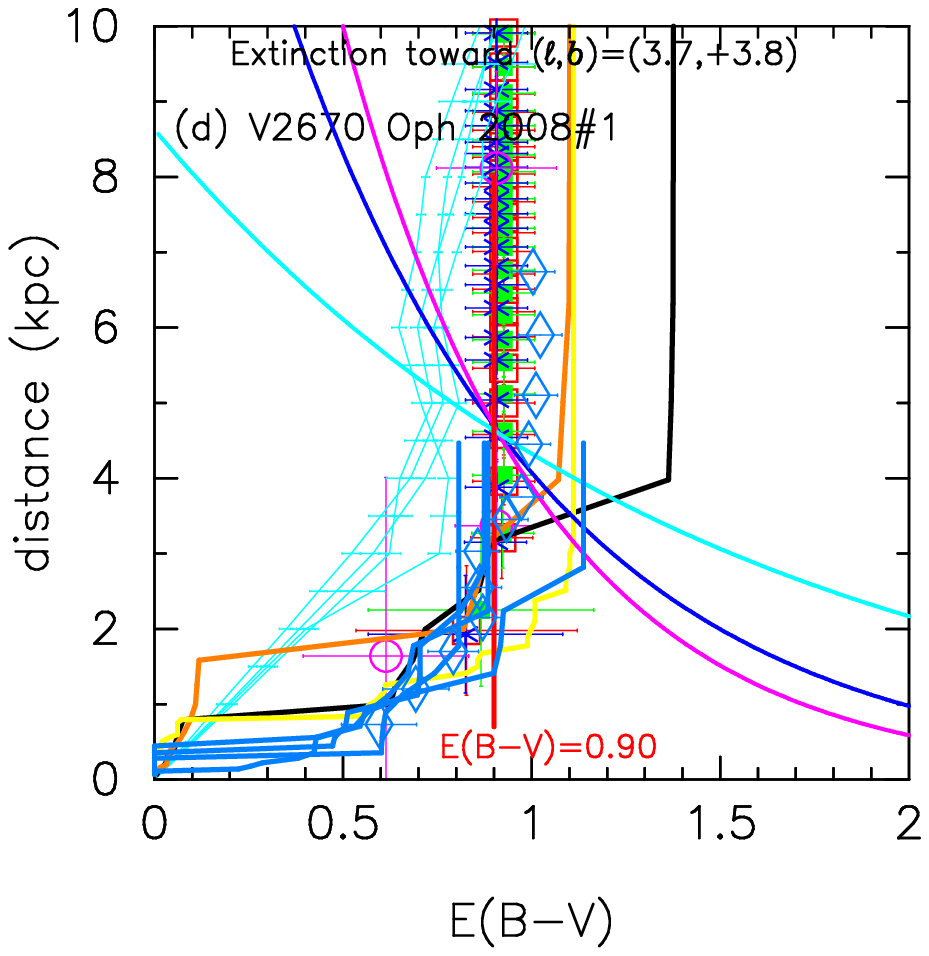}{0.4\textwidth}{(d)}
          }
\caption{
The (a) $B$, (b) $V$, and (c) $I_{\rm C}$ light curves of V2670~Oph
as well as QU~Vul and NR~TrA.
The $BVI_{\rm C}$ data of V2670~Oph are taken from AAVSO, VSOLJ, and SMARTS.
(d) Various distance-reddening relations toward V2670~Oph.
The thin solid lines of magenta, blue, and cyan denote the distance-reddening
relations given by $(m-M)_B= 17.05$, $(m-M)_V= 16.15$, 
and $(m-M)_I= 14.68$, respectively.
\label{distance_reddening_v2670_oph_bvi_xxxxxx}}
\end{figure*}

\subsection{V2670~Oph 2008}
\label{v2670_oph_bvi}
We have reanalyzed the $BVI_{\rm C}$ multi-band light/color curves
of V2670~Oph based on the time-stretching method.  
The important revised points are the reddening of $E(B-V)=0.90$,
distance modulus in $V$ band of $(m-M)_V=16.15$, and timescaling
factor of $\log f_{\rm s}=+0.59$, which are changed from the previous
$E(B-V)=1.05$, $(m-M)_V=17.6$, and $\log f_{\rm s}=+0.33$.
Figure \ref{v2670_oph_v5114_sgr_v1369_cen_v496_sct_i_vi_color_logscale}
shows the (a) $I_{\rm C}$ light and (b) $(V-I_{\rm C})_0$ color curves
of V2670~Oph as well as V5114~Sgr, V1369~Cen, and V496~Sct.
The $BVI_{\rm C}$ data of V2670~Oph are taken from AAVSO, VSOLJ, and SMARTS.
Adopting the color excess of $E(B-V)= 0.90$ as mentioned below,
we determine the timescaling factor $\log f_{\rm s}= +0.59$ 
against that of LV~Vul from Figure
\ref{v2670_oph_v5114_sgr_v1369_cen_v496_sct_i_vi_color_logscale}(b).
This value is larger than that determined by \citet{hac19kb}.
We apply Equation (8) of \citet{hac19ka} for the $I$ band to Figure
\ref{v2670_oph_v5114_sgr_v1369_cen_v496_sct_i_vi_color_logscale}(a)
and obtain
\begin{eqnarray}
(m&-&M)_{I, \rm V2670~Oph} \cr
&=& ((m - M)_I + \Delta I_{\rm C})
_{\rm V5114~Sgr} - 2.5 \log 5.1 \cr
&=& 15.55 + 0.9\pm0.2 - 1.78 = 14.67\pm0.2 \cr
&=& ((m - M)_I + \Delta I_{\rm C})
_{\rm V1369~Cen} - 2.5 \log 2.6 \cr
&=& 10.11 + 5.6\pm0.2 - 1.05 = 14.66\pm0.2 \cr
&=& ((m - M)_I + \Delta I_{\rm C})
_{\rm V496~Sct} - 2.5 \log 1.95 \cr
&=& 12.9 + 2.5\pm0.2 - 0.725 = 14.68\pm0.2,
\label{distance_modulus_i_vi_v2670_oph}
\end{eqnarray}
where we adopt
$(m-M)_{I, \rm V5114~Sgr}=15.55$ from Appendix \ref{v5114_sgr_ubvi},
$(m-M)_{I, \rm V1369~Cen}=10.11$ from \citet{hac19ka}, and
$(m-M)_{I, \rm V496~Sct}=12.9$ in Appendix \ref{v496_sct_bvi}.
Thus, we obtain $(m-M)_{I, \rm V2670~Oph}= 14.67\pm0.2$.


Figure \ref{v2670_oph_v1668_cyg_lv_vul_v_bv_x45z02c10o15_logscale_no2}
shows the (a) $V$ light and (b) $(B-V)_0$ color curves for
V2670~Oph, LV~Vul, and V1668~Cyg.
Applying Equation (4) of \citet{hac19ka} to them, we have the relation
\begin{eqnarray}
(m-M)_{V, \rm V2670~Oph}
&=& ((m - M)_V + \Delta V)_{\rm LV~Vul} - 2.5 \log 3.9 \cr
&=& 11.85 + 5.8\pm0.3 - 1.48 = 16.17\pm0.3 \cr
&=& ((m - M)_V + \Delta V)_{\rm V1668~Cyg} - 2.5 \log 3.9 \cr
&=& 14.6 + 3.0\pm0.3 - 1.48 = 16.12\pm0.3,
\label{distance_modulus_v_bv_v2670_oph_lv_vul}
\end{eqnarray}
where we adopt $(m-M)_{V, \rm LV~Vul}=11.85$ and
$(m-M)_{V, \rm V1668~Cyg}=14.6$, both from \citet{hac19ka}.
Thus, we have $(m-M)_{V, \rm V2670~Oph}=16.15$.

Figure \ref{distance_reddening_v2670_oph_bvi_xxxxxx}(a)(b)(c) shows
the $B$, $V$, and $I_{\rm C}$ light curves of V2670~Oph together with
those of QU~Vul and NR~TrA.  We apply Equation (7) for the
$B$ band to Figure \ref{distance_reddening_v2670_oph_bvi_xxxxxx}(a)
and obtain
\begin{eqnarray}
(m&-&M)_{B, \rm V2670~Oph} \cr
&=& ((m - M)_B + \Delta B)_{\rm QU~Vul} - 2.5 \log 1.0 \cr
&=& 12.95 + 4.1\pm0.3 - 0.0 = 17.05\pm0.3 \cr
&=& ((m - M)_B + \Delta B)_{\rm NR~TrA} - 2.5 \log 1.10 \cr
&=& 14.45 + 2.7\pm0.3 - 0.1 = 17.05\pm0.3,
\label{distance_modulus_b_v2670_oph_qu_vul_nr_tra}
\end{eqnarray}
where we adopt $(m - M)_{B, \rm QU~Vul}= 12.55 + 0.4= 12.95$ from Appendix
\ref{qu_vul_ubvi} and $(m - M)_{B, \rm NR~TrA}= 14.45$ in Appendix
\ref{nr_tra_bvi}.  We obtain $(m-M)_{B, \rm V2670~Oph}= 17.05\pm0.2$.

For the $V$ light curves in Figure
\ref{distance_reddening_v2670_oph_bvi_xxxxxx}(b),
we obtain
\begin{eqnarray}
(m&-&M)_{V, \rm V2670~Oph} \cr
&=& ((m - M)_V + \Delta V)_{\rm QU~Vul} - 2.5 \log 1.0 \cr
&=& 12.55 + 3.6\pm0.3 - 0.0 = 16.15\pm0.3 \cr
&=& ((m - M)_V + \Delta V)_{\rm NR~TrA} - 2.5 \log 1.10 \cr
&=& 14.25 + 2.0\pm0.3 - 0.1 = 16.15\pm0.3,
\label{distance_modulus_v_v2670_oph_qu_vul_nr_tra}
\end{eqnarray}
where we adopt $(m-M)_{V, \rm QU~Vul}= 12.55$ in Appendix \ref{qu_vul_ubvi}
and $(m-M)_{V, \rm NR~TrA}=14.25$ in Appendix \ref{nr_tra_bvi}.
We obtain $(m-M)_{V, \rm V2670~Oph}= 16.15\pm0.2$, which is
consistent with Equation (\ref{distance_modulus_v_bv_v2670_oph_lv_vul}).

From the $I_{\rm C}$ data in Figure
\ref{distance_reddening_v2670_oph_bvi_xxxxxx}(c),  we obtain
\begin{eqnarray}
(m&-&M)_{I, \rm V2670~Oph} \cr
&=& ((m - M)_I + \Delta I_C)_{\rm QU~Vul} - 2.5 \log 1.0 \cr
&=& 11.91 + 2.8\pm0.3 - 0.0 = 14.71\pm0.3 \cr
&=& ((m - M)_I + \Delta I_C)_{\rm NR~TrA} - 2.5 \log 1.10 \cr
&=& 13.95 + 0.85\pm0.3 - 0.1 = 14.68\pm0.3,
\label{distance_modulus_i_v2670_oph_qu_vul_nr_tra}
\end{eqnarray}
where we adopt $(m-M)_{I, \rm QU~Vul}=11.91$ in Appendix \ref{qu_vul_ubvi}
and $(m-M)_{I, \rm NR~TrA}= 13.95$ in Appendix \ref{nr_tra_bvi}.
We obtain $(m-M)_{I, \rm V2670~Oph}= 14.7\pm0.2$.

We plot $(m-M)_B=17.05$, $(m-M)_V=16.15$, and $(m-M)_I=14.7$,
which broadly cross at $d=4.7$~kpc and $E(B-V)=0.90$,
in Figure \ref{distance_reddening_v2670_oph_bvi_xxxxxx}(d).
The crossing point is consistent with the distance-reddening relations
given by \citet[][cyan-blue lines]{chen19}
and \citet[][blue asterisks]{mar06}.
Thus, we have $d=4.7\pm0.6$~kpc and $E(B-V)=0.90\pm0.1$.


\begin{figure}
\plotone{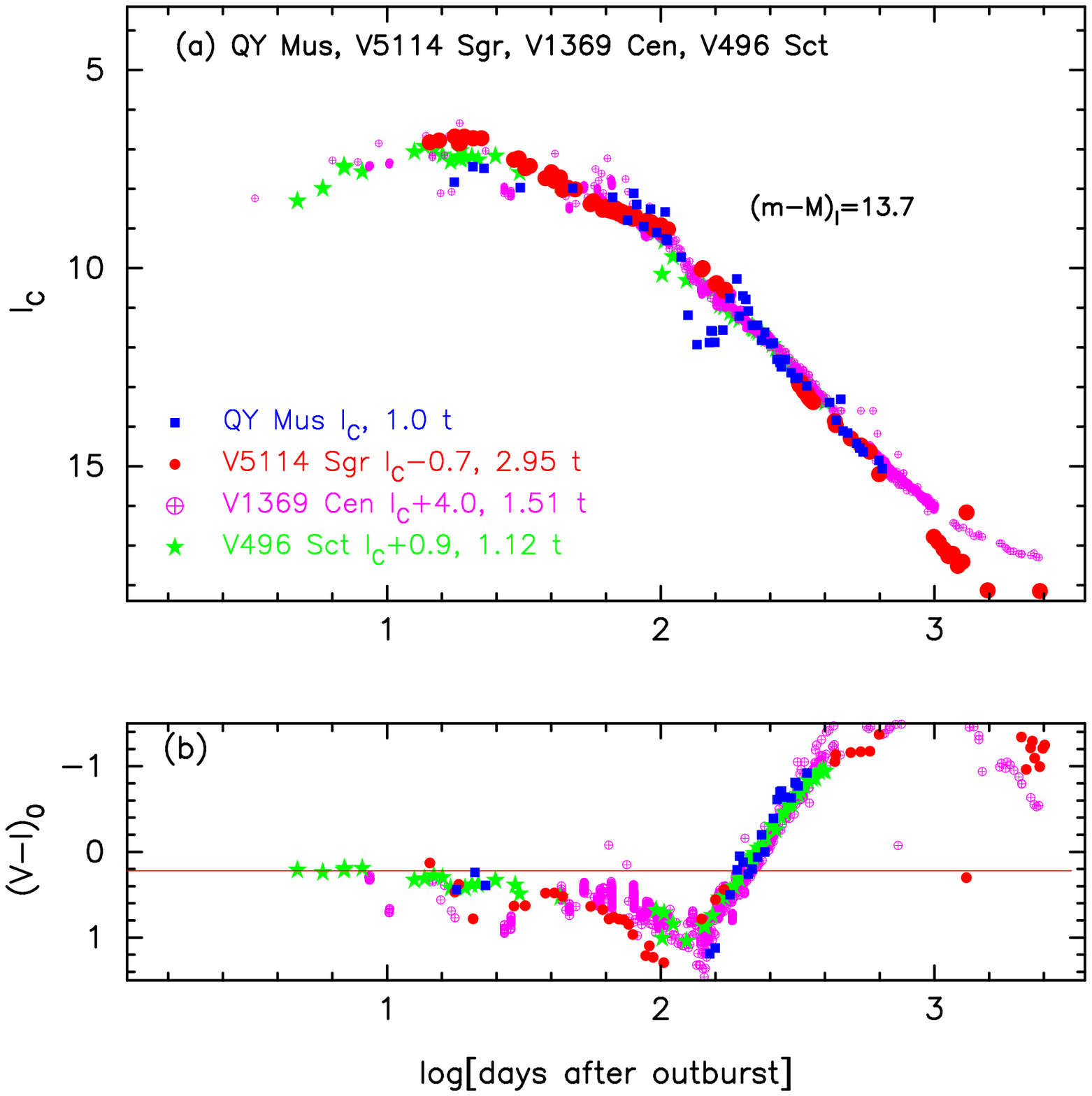}
\caption{
The (a) $I_{\rm C}$ light curve and (b) $(V-I_{\rm C})_0$ color curve
of QY~Mus as well as those of V5114~Sgr, V1369~Cen, and V496~Sct.
\label{qy_mus_v5114_sgr_v1369_cen_v496_sct_i_vi_color_logscale}}
\end{figure}

\subsection{QY~Mus 2008}
\label{qy_mus_bvi}
We have reanalyzed the $BVI_{\rm C}$ multi-band light/color curves
of QY~Mus based on the time-stretching method.  
Figure \ref{qy_mus_v5114_sgr_v1369_cen_v496_sct_i_vi_color_logscale}
shows the (a) $I_{\rm C}$ light and (b) $(V-I_{\rm C})_0$ color curves
of QY~Mus as well as V5114~Sgr, V1369~Cen, and V496~Sct.
The $BVI_{\rm C}$ data of QY~Mus are taken from VSOLJ.
We adopt the color excess of $E(B-V)= 0.58$ after \citet{hac19kb}.
We apply Equation (8) of \citet{hac19ka} for the $I$ band to Figure
\ref{qy_mus_v5114_sgr_v1369_cen_v496_sct_i_vi_color_logscale}(a)
and obtain
\begin{eqnarray}
(m&-&M)_{I, \rm QY~Mus} \cr
&=& ((m - M)_I + \Delta I_{\rm C})
_{\rm V5114~Sgr} - 2.5 \log 2.95 \cr
&=& 15.55 - 0.7\pm0.2 - 1.175 = 13.68\pm0.2 \cr
&=& ((m - M)_I + \Delta I_{\rm C})
_{\rm V1369~Cen} - 2.5 \log 1.51 \cr
&=& 10.11 + 4.0\pm0.2 - 0.45 = 13.66\pm0.2 \cr
&=& ((m - M)_I + \Delta I_{\rm C})
_{\rm V496~Sct} - 2.5 \log 1.12 \cr
&=& 12.9 +0.9\pm0.2 - 0.125 = 13.67\pm0.2,
\label{distance_modulus_i_vi_qy_mus}
\end{eqnarray}
where we adopt
$(m-M)_{I, \rm V5114~Sgr}=15.55$ from Appendix \ref{v5114_sgr_ubvi},
$(m-M)_{I, \rm V1369~Cen}=10.11$ from \citet{hac19ka}, and
$(m-M)_{I, \rm V496~Sct}=12.9$ in Appendix \ref{v496_sct_bvi}.
Thus, we obtain $(m-M)_{I, \rm QY~Mus}= 13.67\pm0.2$.
This result is consistent with the parameter set of
$E(B-V)= 0.58$, $(m-M)_V= 14.65$, $d= 3.7$~kpc, and 
$\log f_{\rm s}= +0.35$ obtained by \citet{hac19kb}.


\begin{figure}
\plotone{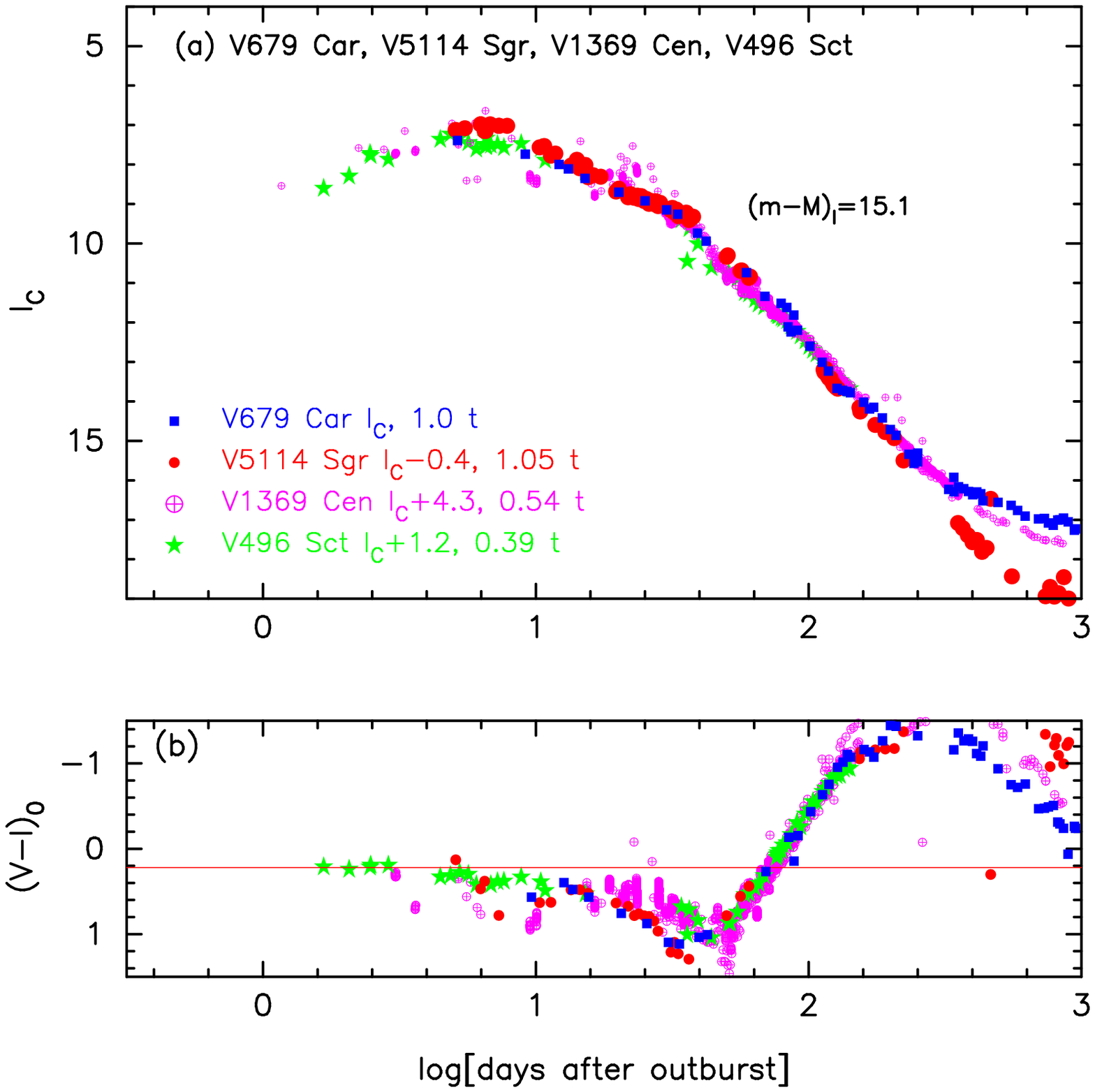}
\caption{
The (a) $I_{\rm C}$ light curve and (b) $(V-I_{\rm C})_0$ color curve
of V679~Car as well as those of V5114~Sgr, V1369~Cen, and V496~Sct.
\label{v679_car_v5114_sgr_v1369_cen_v496_sct_i_vi_color_logscale}}
\end{figure}


\begin{figure}
\plotone{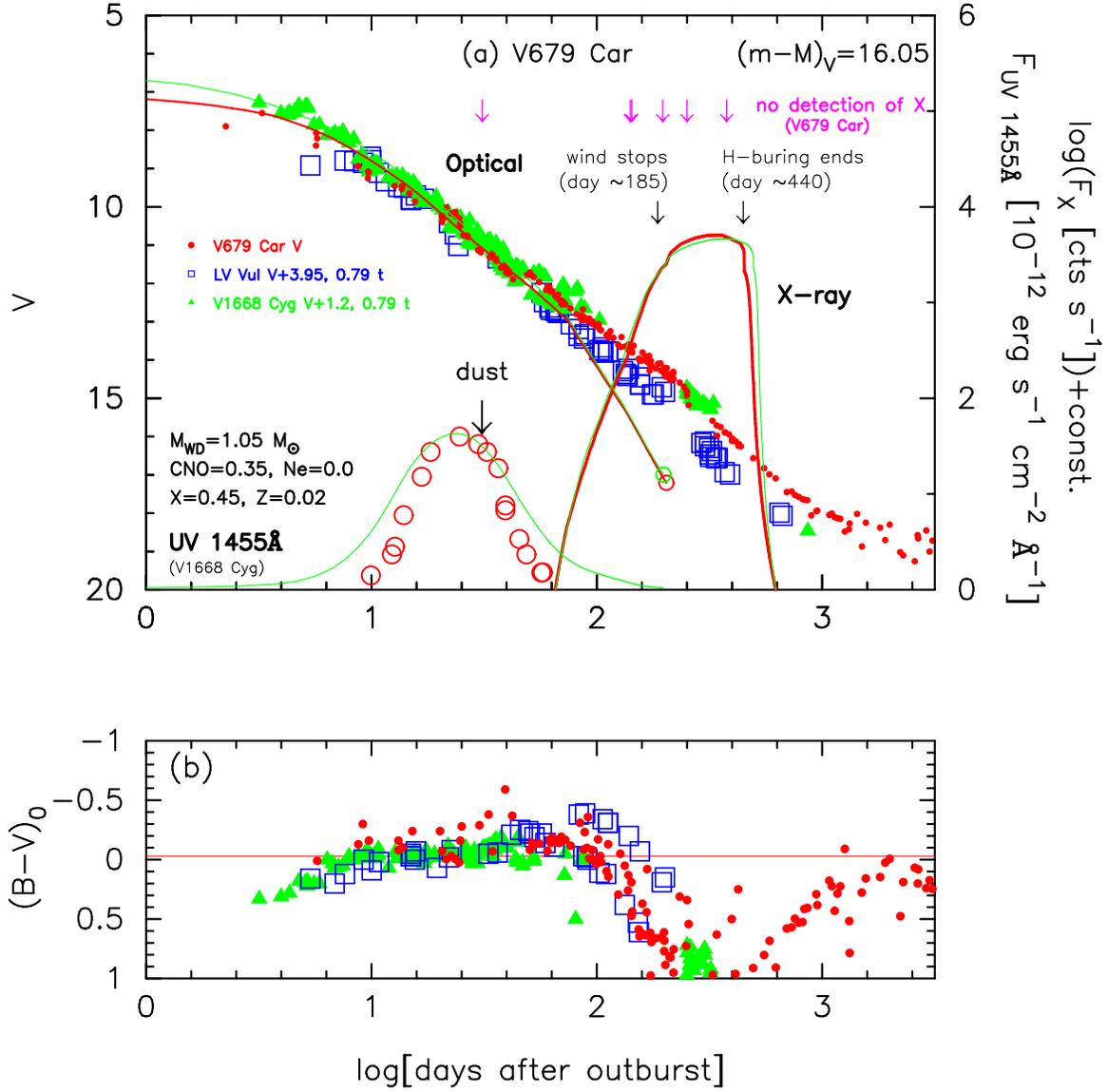}
\caption{
The (a) $V$ light and (b) $(B-V)_0$ color curves of V679~Car as well as
those of LV~Vul and V1668~Cyg.
No supersoft X-rays of V679~Car were detected with {\it Swift}, 
the epochs of which were indicated by downward magenta arrows \citep{schw11}.
We add a $1.05~M_\sun$ WD model (CO3, solid red lines) for V679~Car as
well as a $0.98~M_\sun$ WD model (CO3, solid green lines)
for V1668~Cyg.
\label{v679_car_lv_vul_v1668_cyg_v_bv_logscale_no2}}
\end{figure}


\begin{figure*}
\plottwo{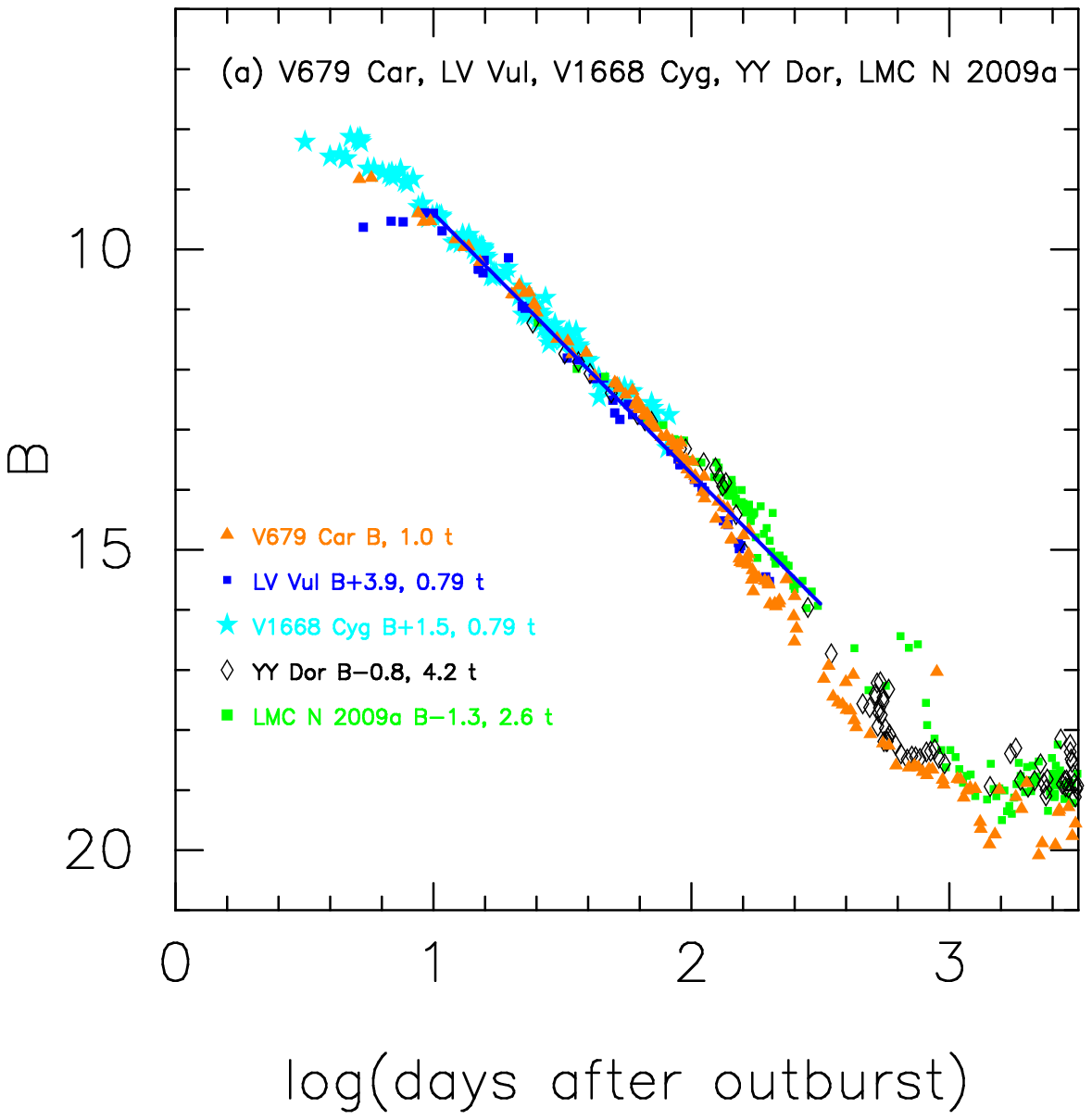}{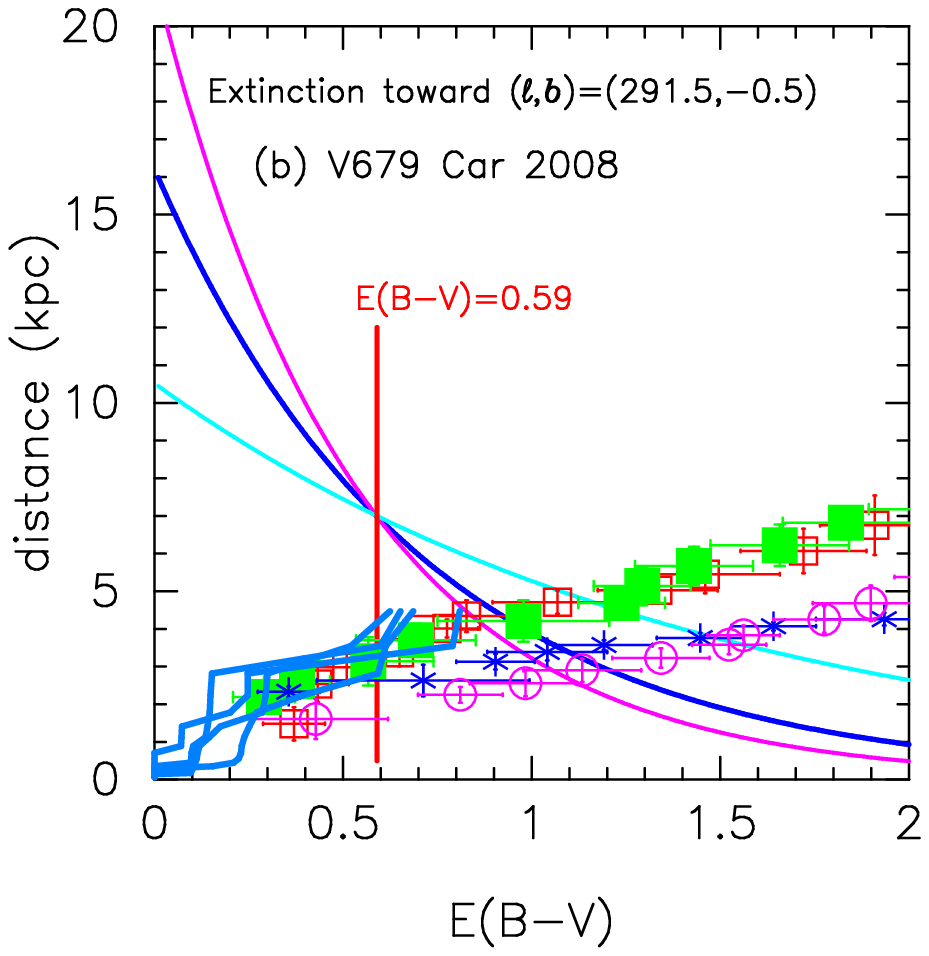}
\caption{
(a) The $B$ light curve of V679~Car as well as LV~Vul, V1668~Cyg,
YY~Dor, and LMC~N~2009a.
(b) Various distance-reddening relations toward V679~Car.
The thin solid lines of magenta, blue, and cyan denote the distance-reddening
relations given by $(m-M)_B= 16.64$, $(m-M)_V= 16.05$, 
and $(m-M)_I= 15.11$, respectively.
\label{distance_reddening_v679_car_bvi_xxxxxx}}
\end{figure*}

\subsection{V679~Car 2008}
\label{v679_car_bvi}
We have reanalyzed the $BVI_{\rm C}$ multi-band light/color curves
of V679~Car based on the time-stretching method.  
Figure \ref{v679_car_v5114_sgr_v1369_cen_v496_sct_i_vi_color_logscale}
shows the (a) $I_{\rm C}$ light and (b) $(V-I_{\rm C})_0$ color curves
of V679~Car as well as V5114~Sgr, V1369~Cen, and V496~Sct.
The $BVI_{\rm C}$ data of V679~Car are taken from VSOLJ and SMARTS.
We adopt the color excess of $E(B-V)= 0.59$ in order to overlap
the $(V-I)_0$ color curve of V679~Car with the other novae, as shown in
Figure \ref{v679_car_v5114_sgr_v1369_cen_v496_sct_i_vi_color_logscale}(b).
We apply Equation (8) of \citet{hac19ka} for the $I$ band to Figure
\ref{v679_car_v5114_sgr_v1369_cen_v496_sct_i_vi_color_logscale}(a)
and obtain
\begin{eqnarray}
(m&-&M)_{I, \rm V679~Car} \cr
&=& ((m - M)_I + \Delta I_{\rm C})
_{\rm V5114~Sgr} - 2.5 \log 1.05 \cr
&=& 15.55 - 0.4\pm0.2 - 0.05 = 15.1\pm0.2 \cr
&=& ((m - M)_I + \Delta I_{\rm C})
_{\rm V1369~Cen} - 2.5 \log 0.54 \cr
&=& 10.11 + 4.3\pm0.2 + 0.675 = 15.08\pm0.2 \cr
&=& ((m - M)_I + \Delta I_{\rm C})
_{\rm V496~Sct} - 2.5 \log 0.39 \cr
&=& 12.9 + 1.2\pm0.2 + 1.0 = 15.1\pm0.2,
\label{distance_modulus_i_vi_v679_car}
\end{eqnarray}
where we adopt
$(m-M)_{I, \rm V5114~Sgr}=15.55$ from Appendix \ref{v5114_sgr_ubvi},
$(m-M)_{I, \rm V1369~Cen}=10.11$ from \citet{hac19ka}, and
$(m-M)_{I, \rm V496~Sct}=12.9$ in Appendix \ref{v496_sct_bvi}.
Thus, we obtain $(m-M)_{I, \rm V679~Car}= 15.1\pm0.2$.

Figure \ref{v679_car_lv_vul_v1668_cyg_v_bv_logscale_no2} shows the
(a) $V$ light and (b) $(B-V)_0$ color curves of V679~Car together with
those of LV~Vul and V1668~Cyg.
Applying Equation (4) of \citet{hac19ka} to Figure
\ref{v679_car_lv_vul_v1668_cyg_v_bv_logscale_no2}(a),
we have the relation of
\begin{eqnarray}
(m &-& M)_{V, \rm V679~Car} \cr
&=& (m-M)_{V, \rm LV~Vul} + \Delta V - 2.5 \log 0.79 \cr
&=& 11.85 + 3.95\pm 0.2 + 0.25 = 16.05\pm 0.2 \cr
&=& (m-M)_{V, \rm V1668~Cyg} + \Delta V - 2.5 \log 0.79 \cr
&=& 14.6 + 1.2\pm 0.2 + 0.25 = 16.05\pm 0.2.
\label{distance_modulus_v679_car_lv_vul_v1668_cyg_v}
\end{eqnarray}
where we adopt $(m-M)_{V, \rm LV~Vul}=11.85$ and
$(m-M)_{V, \rm V1668~Cyg}=14.6$ both from \citet{hac19ka}.
Thus, we obtain $\log f_{\rm s}= \log 0.79 = -0.10$ against LV~Vul
and $(m-M)_V=16.05\pm0.1$ for V679~Car.

Figure \ref{distance_reddening_v679_car_bvi_xxxxxx}(a) shows the
$B$ light curve of V679~Car as well as LV~Vul, V1668~Cyg, and
the LMC novae YY~Dor and LMC~N~2009a. 
We apply Equation (7) of \citet{hac19ka} for the $B$ band to Figure
\ref{distance_reddening_v679_car_bvi_xxxxxx}(a) and obtain
\begin{eqnarray}
(m&-&M)_{B, \rm V679~Car} \cr
&=& ((m - M)_B + \Delta B)_{\rm LV~Vul} - 2.5 \log 0.79 \cr
&=& 12.45 + 3.9\pm0.2 + 0.25 = 16.6\pm0.2 \cr
&=& ((m - M)_B + \Delta B)_{\rm V1668~Cyg} - 2.5 \log 0.79 \cr
&=& 14.9 + 1.5\pm0.2 + 0.25 = 16.65\pm0.2 \cr
&=& ((m - M)_B + \Delta B)_{\rm YY~Dor} - 2.5 \log 4.2 \cr
&=& 18.98 - 0.8\pm0.2 - 1.55 = 16.63\pm0.2 \cr
&=& ((m - M)_B + \Delta B)_{\rm LMC~N~2009a} - 2.5 \log 2.6 \cr
&=& 18.98 -1.3\pm0.2 - 1.05 = 16.63\pm0.2, 
\label{distance_modulus_b_v679_car_yy_dor_lmcn2009a}
\end{eqnarray}
Thus, we obtain $(m-M)_{B, \rm V679~Car}= 16.63\pm0.1$.

We plot the three distance moduli 
in Figure \ref{distance_reddening_v679_car_bvi_xxxxxx}(b).
These three lines cross at $E(B-V)= 0.59$ and $d= 7.0$~kpc.
Thus, we confirm the color excess of $E(B-V)= 0.59$ mentioned above.


\begin{figure}
\plotone{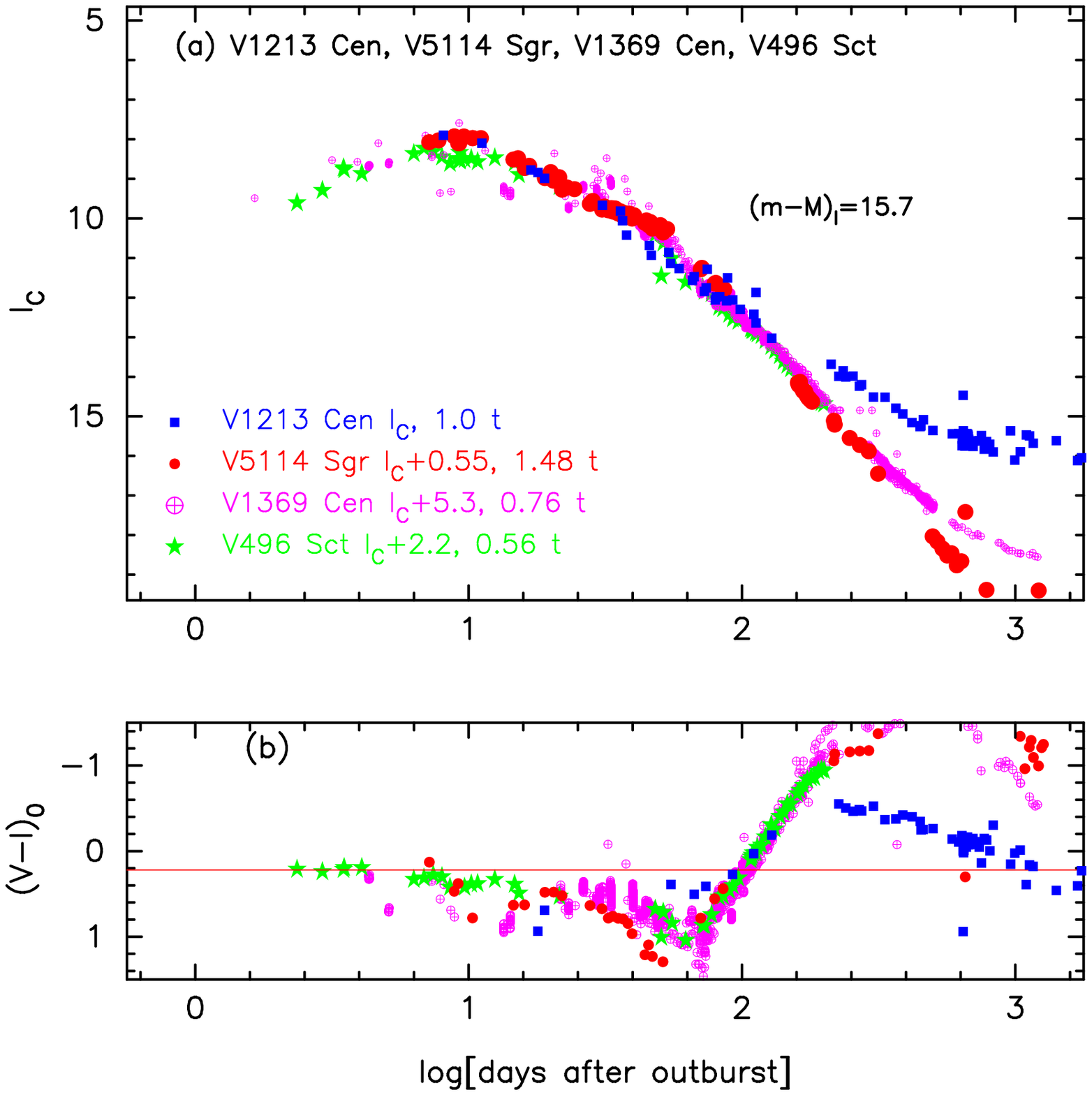}
\caption{
The (a) $I_{\rm C}$ light curve and (b) $(V-I_{\rm C})_0$ color curve
of V1213~Cen as well as those of V5114~Sgr, V1369~Cen, and V496~Sct.
\label{v1213_cen_v5114_sgr_v1369_cen_v496_sct_i_vi_color_logscale}}
\end{figure}


\begin{figure}
\epsscale{1.0}
\plotone{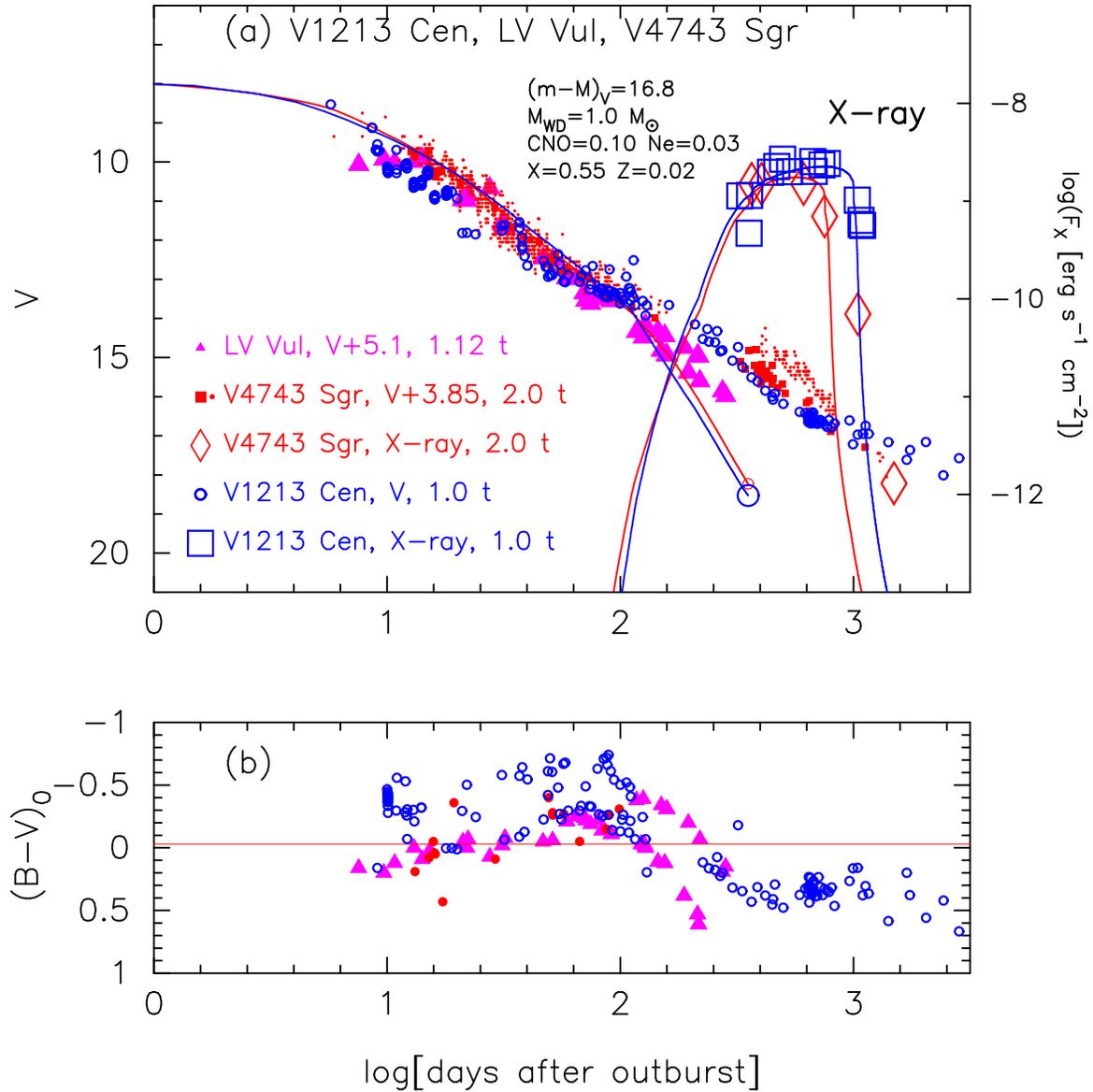}
\caption{
The (a) $V$ light and (b) $(B-V)_0$ color curves of V1213~Cen as well as
those of LV~Vul and V4743~Sgr.
The data of V1213~Cen are taken from AAVSO, VSOLJ, and SMARTS.
The data of V4743~Sgr
are the same as those in Figures 18 of \citet{hac10k}.
In panel (a), we add model light curves (solid blue lines) of a
$1.0~M_\sun$ WD \citep[Ne2;][]{hac10k} both for the $V$
and X-ray (0.2--2.0~keV), assuming that $(m-M)_V=16.8$ for V1213~Cen.
The solid red lines denote the model light curves of
a $1.15~M_\sun$ WD (Ne2),
assuming $(m-M)_V=13.7$ for V4743~Sgr \citep{hac10k}.
\label{v1213_cen_lv_vul_v4743_sgr_v_bv_ub_color_curve_logscale}}
\end{figure}


\begin{figure}
\epsscale{1.0}
\plottwo{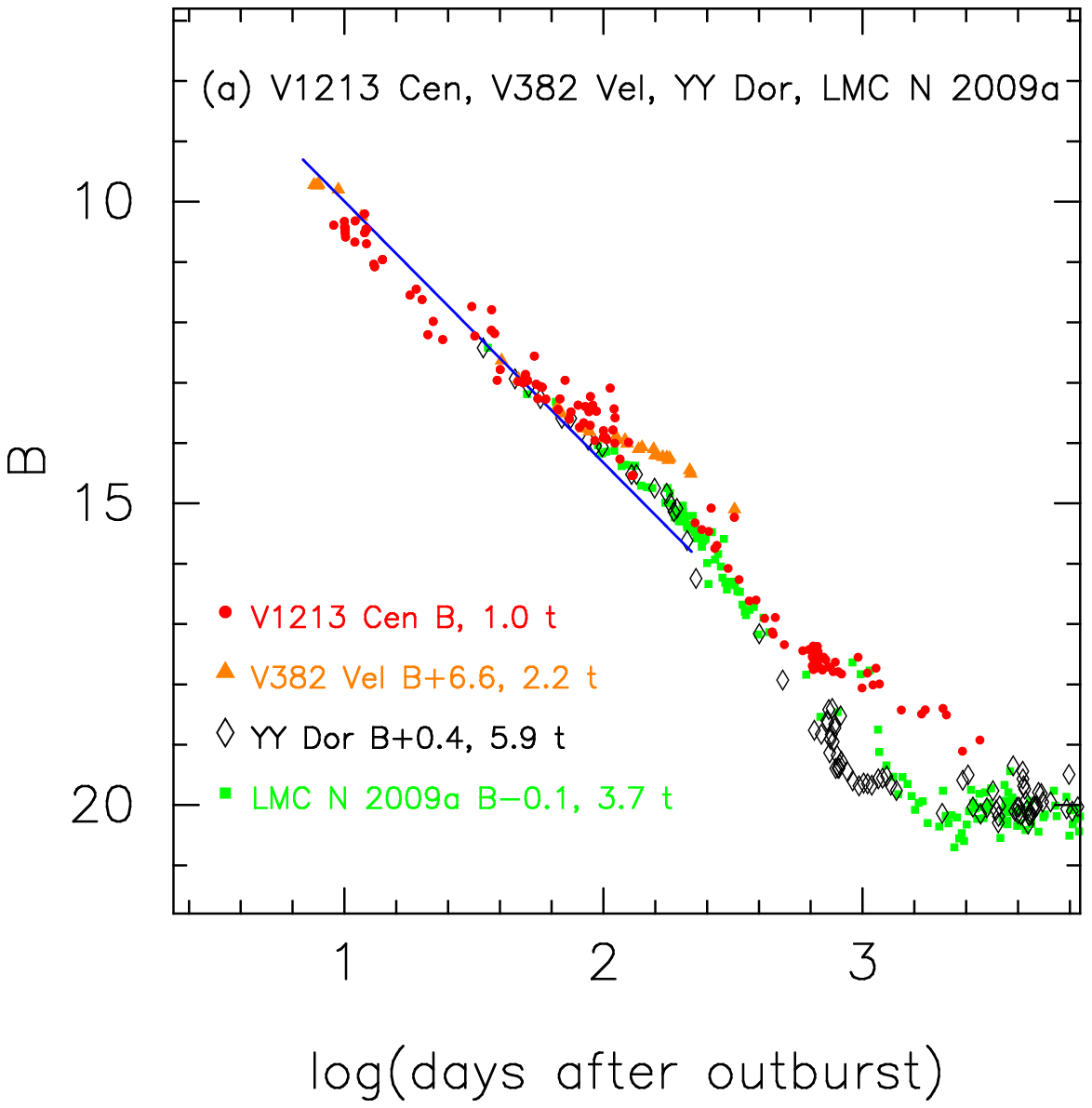}{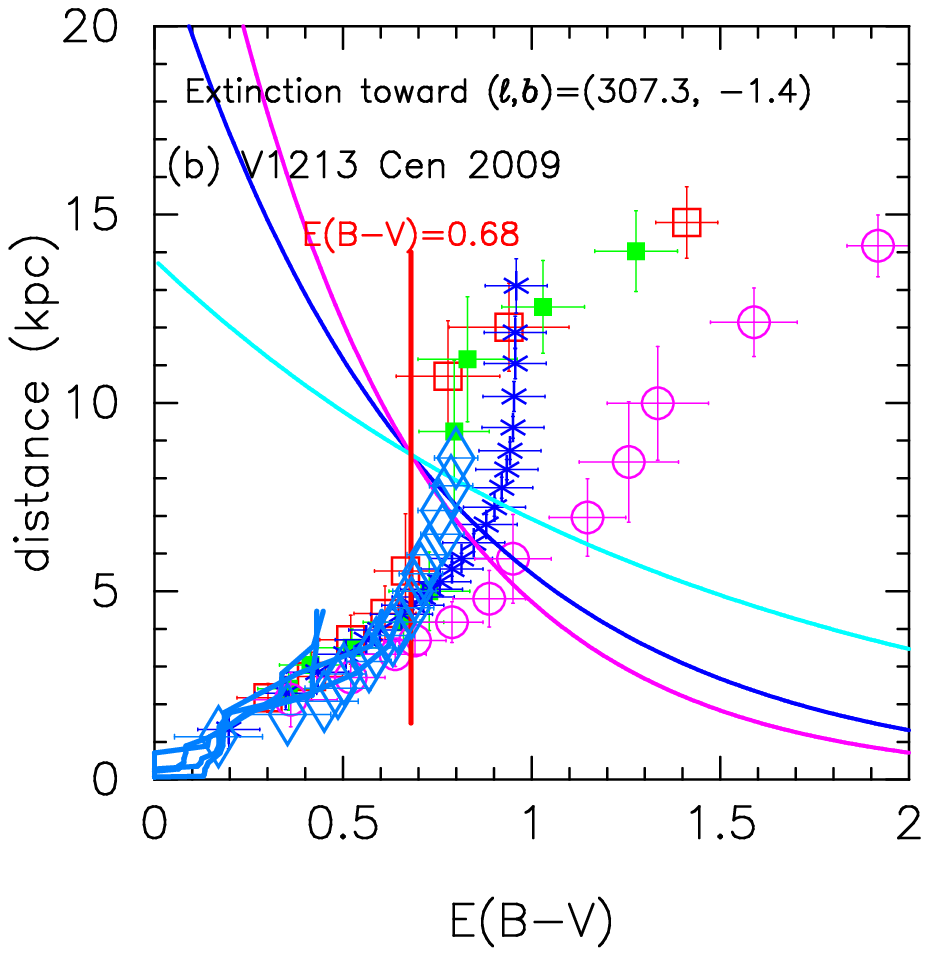}
\caption{
(a) The $B$ light curve of V1213~Cen as well as V382~Vel, YY~Dor,
and LMC~N~2009a.  The $BVI_{\rm C}$ data of V1213~Cen are taken from
AAVSO, VSOLJ, and SMARTS.
(b) Various distance-reddening relations toward V1213~Cen.
The thin solid lines of magenta, blue, and cyan denote the distance-reddening
relations given by $(m-M)_B= 17.47$, $(m-M)_V= 16.8$,
and $(m-M)_I= 15.71$, respectively.
\label{distance_reddening_v1213_cen_bvi_xxxxxx}}
\end{figure}

\subsection{V1213~Cen 2009}
\label{v1213_cen_bvi}
We have reanalyzed the $BVI_{\rm C}$ multi-band light/color curves
of V1213~Cen based on the time-stretching method.  
Figure \ref{v1213_cen_v5114_sgr_v1369_cen_v496_sct_i_vi_color_logscale}
shows the (a) $I_{\rm C}$ light and (b) $(V-I_{\rm C})_0$ color curves
of V1213~Cen as well as V5114~Sgr, V1369~Cen, and V496~Sct.
The $BVI_{\rm C}$ data of V1213~Cen are taken from AAVSO, VSOLJ, and SMARTS.
We adopt the color excess of $E(B-V)= 0.68$ as mentioned below.
We apply Equation (8) of \citet{hac19ka} for the $I$ band to Figure
\ref{v1213_cen_v5114_sgr_v1369_cen_v496_sct_i_vi_color_logscale}(a)
and obtain
\begin{eqnarray}
(m&-&M)_{I, \rm V1213~Cen} \cr
&=& ((m - M)_I + \Delta I_{\rm C})
_{\rm V5114~Sgr} - 2.5 \log 1.48 \cr
&=& 15.55 + 0.55\pm0.2 - 0.425 = 15.68\pm0.2 \cr
&=& ((m - M)_I + \Delta I_{\rm C})
_{\rm V1369~Cen} - 2.5 \log 0.76 \cr
&=& 10.11 + 5.3\pm0.2 + 0.3 = 15.71\pm0.2 \cr
&=& ((m - M)_I + \Delta I_{\rm C})
_{\rm V496~Sct} - 2.5 \log 0.56 \cr
&=& 12.9 + 2.2\pm0.2 + 0.625 = 15.72\pm0.2,
\label{distance_modulus_i_vi_v1213_cen}
\end{eqnarray}
where we adopt
$(m-M)_{I, \rm V5114~Sgr}=15.55$ from Appendix \ref{v5114_sgr_ubvi},
$(m-M)_{I, \rm V1369~Cen}=10.11$ from \citet{hac19ka}, and
$(m-M)_{I, \rm V496~Sct}=12.9$ in Appendix \ref{v496_sct_bvi}.
Thus, we obtain $(m-M)_{I, \rm V1213~Cen}= 15.71\pm0.2$.

Figure \ref{v1213_cen_lv_vul_v4743_sgr_v_bv_ub_color_curve_logscale}
shows the (a) $V$ light and (b) $(B-V)_0$ color curves of V1213~Cen
as well as those of LV~Vul and V4743~Sgr. 
Applying Equation (4) of \citet{hac19ka} to 
Figure \ref{v1213_cen_lv_vul_v4743_sgr_v_bv_ub_color_curve_logscale}(a),
we have
\begin{eqnarray}
(m&-&M)_{V, \rm V1213~Cen} \cr
&=& (m - M + \Delta V)_{V, \rm LV~Vul} - 2.5 \log 1.12 \cr
&=& 11.85 + 5.1\pm0.3 - 0.13 = 16.82\pm0.3 \cr
&=& (m - M + \Delta V)_{V, \rm V4743~Sgr} - 2.5 \log 2.0 \cr
&=& 13.7 + 3.85\pm0.3 - 0.75 = 16.8\pm0.3,
\label{distance_modulus_v1213_cen}
\end{eqnarray}
where we adopt $(m-M)_{V, \rm LV~Vul}=11.85$ from \citet{hac19ka}
and $(m-M)_{V, \rm V4743~Sgr}=13.7$ from \cite{hac10k}.
Thus, we obtain $(m-M)_V=16.81\pm0.2$ and $f_{\rm s}=1.12$ against LV~Vul.

Figure \ref{distance_reddening_v1213_cen_bvi_xxxxxx}(a) shows the $B$
light curves of V1213~Cen as well as those of V382~Vel, YY~Dor, and
LMC~N~2009a.  We apply Equation (7) of \citet{hac19ka} 
for the $B$ band to them and obtain
\begin{eqnarray}
(m&-&M)_{B, \rm V1213~Cen} \cr
&=& ((m - M)_B + \Delta B)_{\rm V382~Vel} - 2.5 \log 2.2 \cr
&=& 11.72 + 6.6\pm0.2 - 0.85 = 17.47\pm0.2 \cr
&=& ((m - M)_B + \Delta B)_{\rm YY~Dor} - 2.5 \log 5.9 \cr
&=& 18.98 + 0.4\pm0.2 - 1.92 = 17.46\pm0.2 \cr
&=& ((m - M)_B + \Delta B)_{\rm LMC~N~2009a} - 2.5 \log 3.7 \cr
&=& 18.98 - 0.1\pm0.2 - 1.42 = 17.46\pm0.2,
\label{distance_modulus_b_v1213_cen_v382_vel_yy_dor_lmcn2009a}
\end{eqnarray}
where we adopt $(m-M)_{B, \rm V382~Vel}= 11.6 + 0.12= 11.72$
and $\log f_{\rm s}= -0.29$ against LV~Vul for V382~Vel in Appendix
\ref{v382_vel_ubvi}.
Thus, we have $(m-M)_{B, \rm V1213~Cen}= 17.47\pm0.1$.

We plot $(m-M)_B= 17.47$, $(m-M)_V= 16.8$, and $(m-M)_I= 15.7$,
which broadly cross at $d=8.6$~kpc and $E(B-V)=0.68$,  in Figure
\ref{distance_reddening_v1213_cen_bvi_xxxxxx}(b).
Thus, we have $E(B-V)=0.68\pm0.05$ and $d=8.6\pm1$~kpc.


\begin{figure}
\plotone{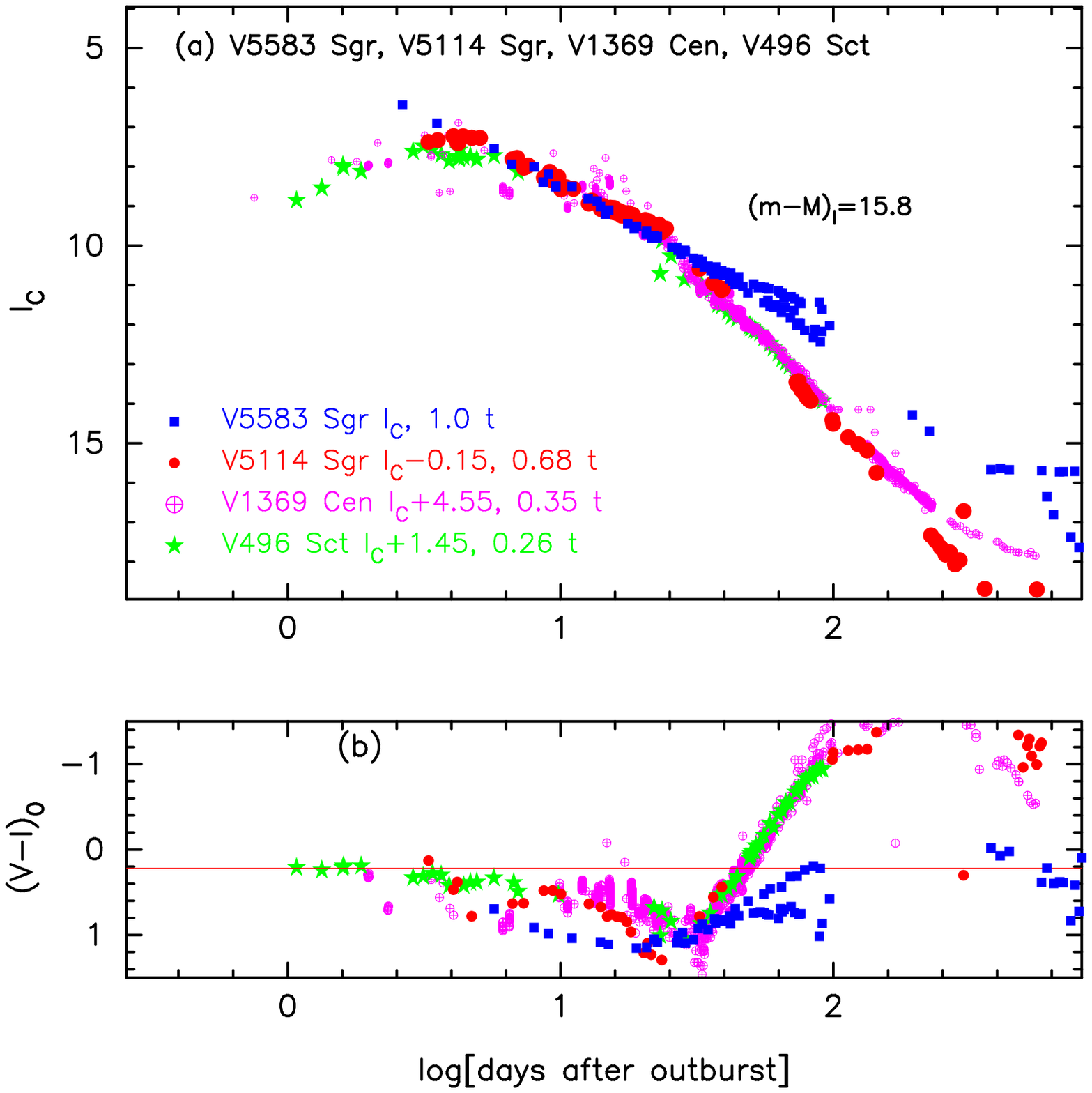}
\caption{
The (a) $I_{\rm C}$ light curve and (b) $(V-I_{\rm C})_0$ color curve
of V5583~Sgr as well as those of V5114~Sgr, V1369~Cen, and V496~Sct.
\label{v5583_sgr_v5114_sgr_v1369_cen_v496_sct_i_vi_color_logscale}}
\end{figure}

\subsection{V5583~Sgr 2009\#3}
\label{v5583_sgr_bvi}
We have reanalyzed the $BVI_{\rm C}$ multi-band light/color curves
of V5583~Sgr based on the time-stretching method.  
Figure \ref{v5583_sgr_v5114_sgr_v1369_cen_v496_sct_i_vi_color_logscale}
shows the (a) $I_{\rm C}$ light and (b) $(V-I_{\rm C})_0$ color curves
of V5583~Sgr as well as V5114~Sgr, V1369~Cen, and V496~Sct.
The $BVI_{\rm C}$ data of V5583~Sgr are taken from AAVSO, VSOLJ, and SMARTS.
We adopt the color excess of $E(B-V)= 0.30$ after \citet{hac19kb}.
We apply Equation (8) of \citet{hac19ka} for the $I$ band to Figure
\ref{v5583_sgr_v5114_sgr_v1369_cen_v496_sct_i_vi_color_logscale}(a)
and obtain
\begin{eqnarray}
(m&-&M)_{I, \rm V5583~Sgr} \cr
&=& ((m - M)_I + \Delta I_{\rm C})
_{\rm V5114~Sgr} - 2.5 \log 0.68 \cr
&=& 15.55 - 0.15\pm0.2 + 0.425 =  15.82\pm0.2 \cr
&=& ((m - M)_I + \Delta I_{\rm C})
_{\rm V1369~Cen} - 2.5 \log 0.35 \cr
&=& 10.11 + 4.55\pm0.2 + 1.15 = 15.81\pm0.2 \cr
&=& ((m - M)_I + \Delta I_{\rm C})
_{\rm V496~Sct} - 2.5 \log 0.26 \cr
&=& 12.9 + 1.45\pm0.2 + 1.475 = 15.82\pm0.2,
\label{distance_modulus_i_vi_v5583_sgr}
\end{eqnarray}
where we adopt
$(m-M)_{I, \rm V5114~Sgr}=15.55$ from Appendix \ref{v5114_sgr_ubvi},
$(m-M)_{I, \rm V1369~Cen}=10.11$ from \citet{hac19ka}, and
$(m-M)_{I, \rm V496~Sct}=12.9$ in Appendix \ref{v496_sct_bvi}.
Thus, we obtain $(m-M)_{I, \rm V5583~Sgr}= 15.82\pm0.2$.
This result is consistent with the parameter set of
$E(B-V)= 0.30$, $(m-M)_V= 16.3$, $d= 12$~kpc, and 
$\log f_{\rm s}= -0.29$ obtained by \citet{hac19kb}.


\begin{figure}
\plotone{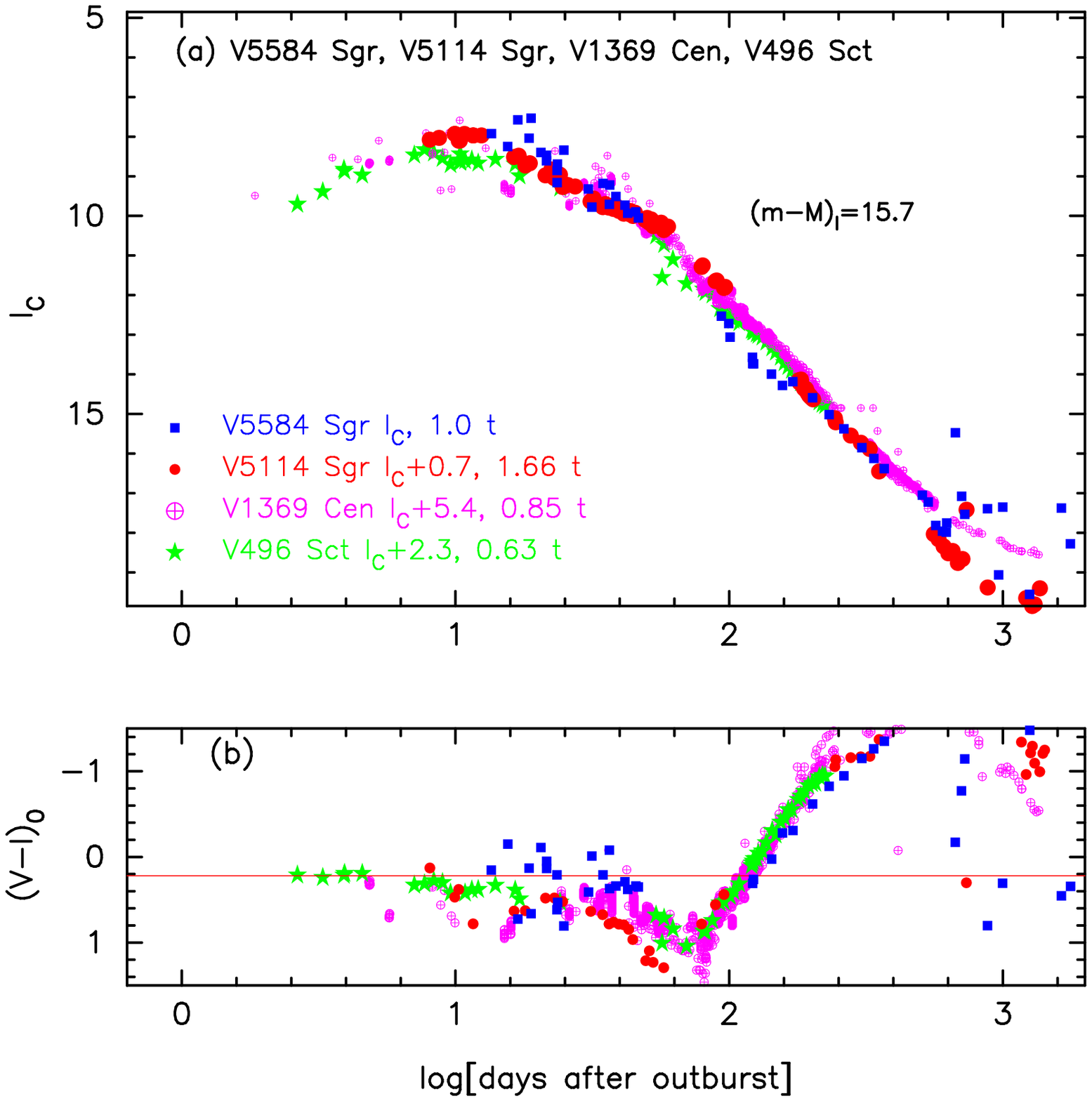}
\caption{
The (a) $I_{\rm C}$ light curve and (b) $(V-I_{\rm C})_0$ color curve
of V5584~Sgr as well as those of V5114~Sgr, V1369~Cen, and V496~Sct.
\label{v5584_sgr_v5114_sgr_v1369_cen_v496_sct_i_vi_color_logscale}}
\end{figure}


\begin{figure}
\plotone{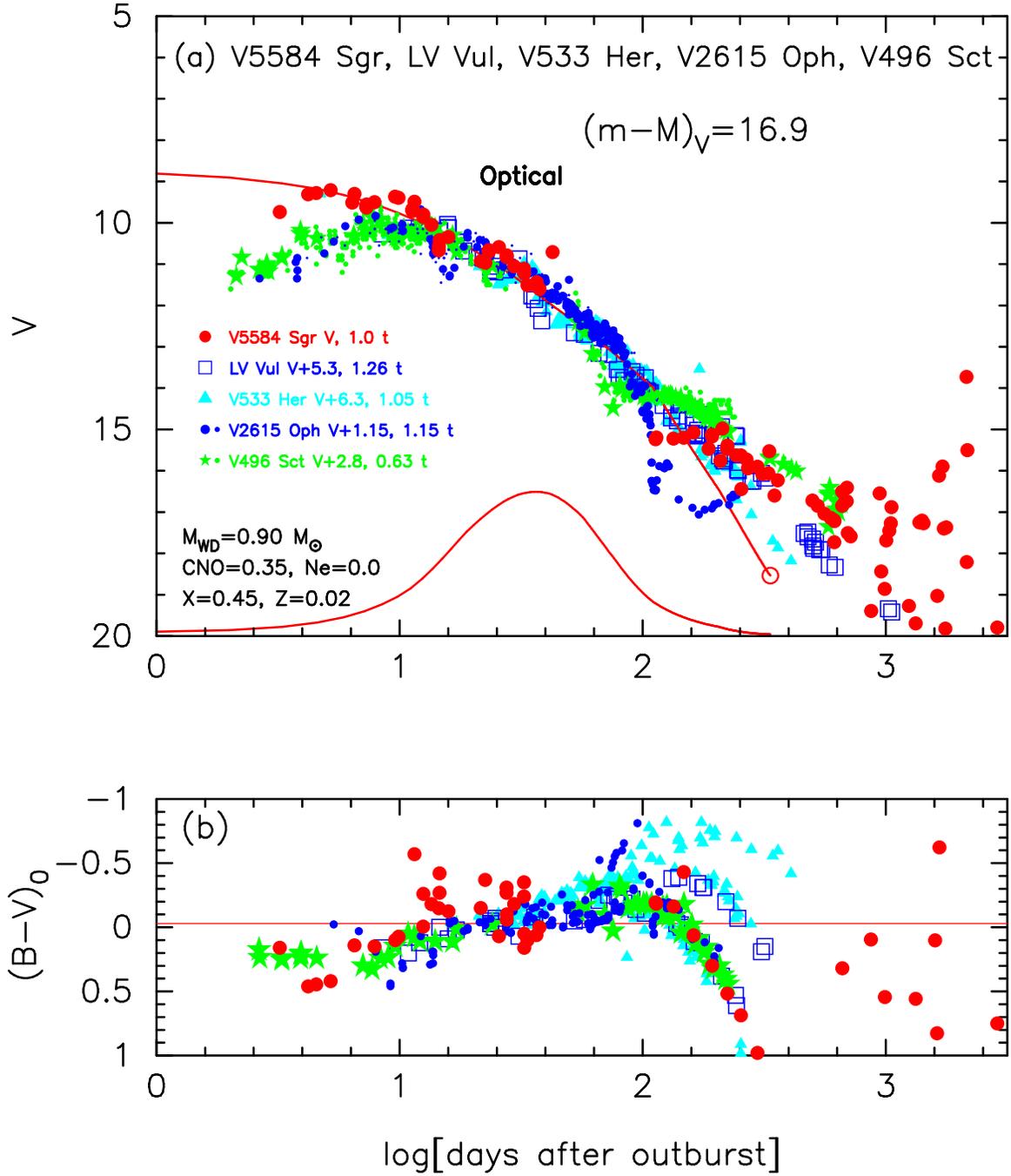}
\caption{
The (a) $V$ light and (b) $(B-V)_0$ color curves of V5584~Sgr 
as well as those of LV~Vul, V533~Her, V2615~Oph, and V496~Sct.
In panel (a), we add a $0.90~M_\sun$ WD model (CO3, solid red lines)
for V5584~Sgr.
\label{v5584_sgr_v2615_oph_v496_sct_v_bv_logscale_no2}}
\end{figure}


\begin{figure*}
\plottwo{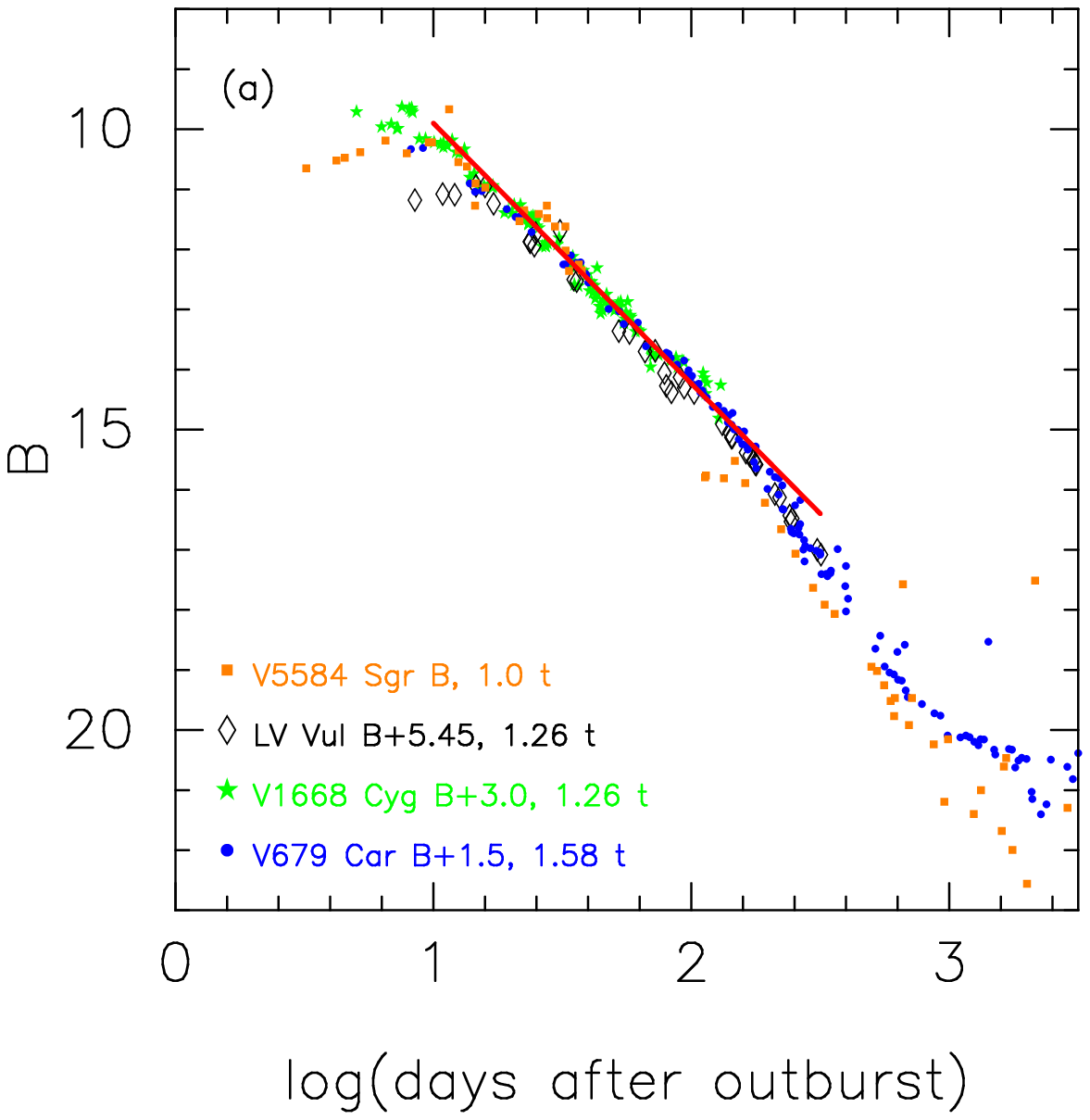}{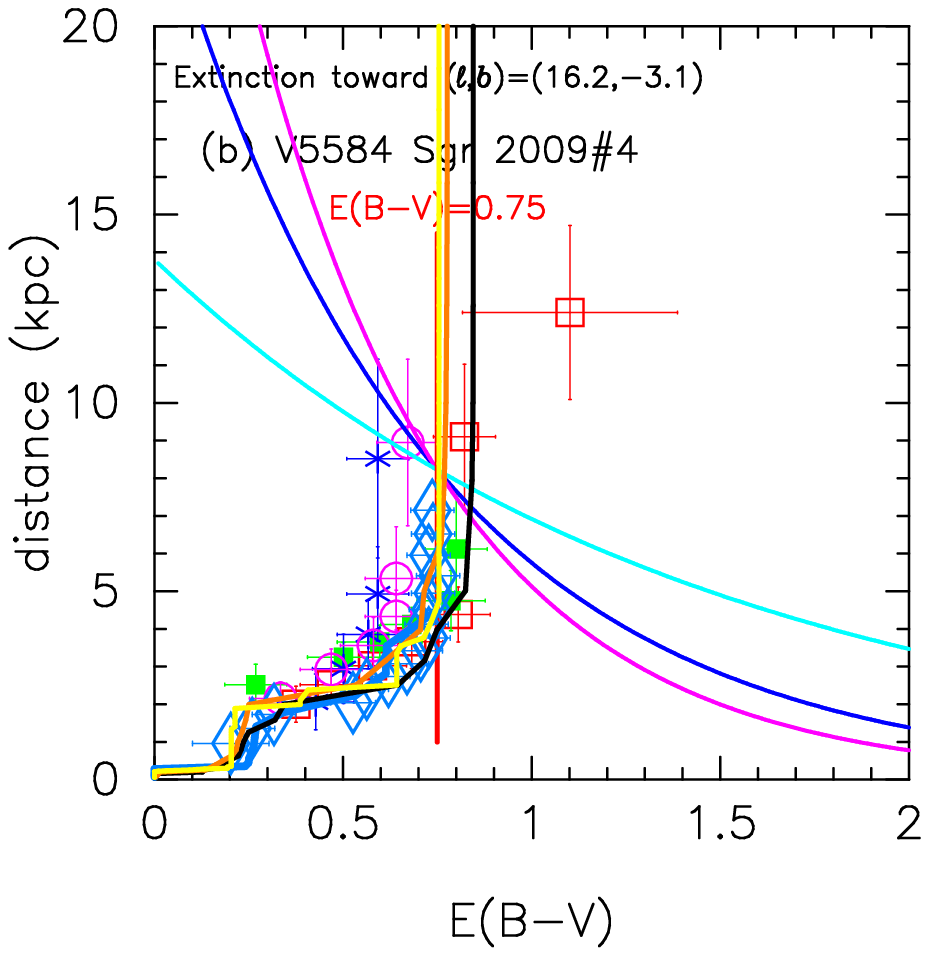}
\caption{
(a) The $B$ light curves of V5584~Sgr
as well as those of LV~Vul, V1668~Cyg, and V679~Car.
(b) Various distance-reddening relations toward V5584~Sgr.
The thin solid lines of magenta, blue, and cyan denote the distance-reddening
relations given by  $(m-M)_B= 17.65$, $(m-M)_V= 16.9$, and $(m-M)_I= 15.7$,
respectively.  
\label{distance_reddening_v5584_sgr_bvi_xxxxxx}}
\end{figure*}

\subsection{V5584~Sgr 2009\#4}
\label{v5584_sgr_bvi}
We have reanalyzed the $BVI_{\rm C}$ multi-band light/color curves
of V5584~Sgr based on the time-stretching method.  
Figure \ref{v5584_sgr_v5114_sgr_v1369_cen_v496_sct_i_vi_color_logscale}
shows the (a) $I_{\rm C}$ light and (b) $(V-I_{\rm C})_0$ color curves
of V5584~Sgr as well as V5114~Sgr, V1369~Cen, and V496~Sct.
The $BVI_{\rm C}$ data of V5584~Sgr are taken from \citet{mun09},
AAVSO, VSOLJ, and SMARTS.
We adopt the color excess of $E(B-V)= 0.75$ as mentioned below.
We apply Equation (8) of \citet{hac19ka} for the $I$ band to Figure
\ref{v5584_sgr_v5114_sgr_v1369_cen_v496_sct_i_vi_color_logscale}(a)
and obtain
\begin{eqnarray}
(m&-&M)_{I, \rm V5584~Sgr} \cr
&=& ((m - M)_I + \Delta I_{\rm C})
_{\rm V5114~Sgr} - 2.5 \log 1.66  \cr
&=& 15.55 + 0.7\pm0.2 - 0.55 =  15.7\pm0.2 \cr
&=& ((m - M)_I + \Delta I_{\rm C})
_{\rm V1369~Cen} - 2.5 \log 0.85 \cr
&=& 10.11 + 5.4\pm0.2 + 0.175 = 15.68\pm0.2 \cr
&=& ((m - M)_I + \Delta I_{\rm C})
_{\rm V496~Sct} - 2.5 \log 0.63 \cr
&=& 12.9 + 2.3\pm0.2 + 0.5 = 15.7\pm0.2,
\label{distance_modulus_i_vi_v5584_sgr}
\end{eqnarray}
where we adopt
$(m-M)_{I, \rm V5114~Sgr}=15.55$ from Appendix \ref{v5114_sgr_ubvi},
$(m-M)_{I, \rm V1369~Cen}=10.11$ from \citet{hac19ka}, and
$(m-M)_{I, \rm V496~Sct}=12.9$ in Appendix \ref{v496_sct_bvi}.
Thus, we obtain $(m-M)_{I, \rm V5584~Sgr}= 15.7\pm0.2$.

Figure \ref{v5584_sgr_v2615_oph_v496_sct_v_bv_logscale_no2} shows
(a) the $V$ and (b) $(B-V)_0$ evolutions of V5584~Sgr.
Figure \ref{v5584_sgr_v2615_oph_v496_sct_v_bv_logscale_no2} also shows the
light/color curves of LV~Vul, V533~Her, V2615~Oph, and V496~Sct.
Applying Equation (4) of \citet{hac19ka} to them,
we have the relation
\begin{eqnarray}
(m&-&M)_{V, \rm V5584~Sgr} \cr
&=& ((m - M)_V + \Delta V)_{\rm LV~Vul} - 2.5 \log 1.26 \cr
&=& 11.85 + 5.3\pm0.2 - 0.25 = 16.9\pm0.2 \cr
&=& ((m - M)_V + \Delta V)_{\rm V533~Her} - 2.5 \log 1.05 \cr
&=& 10.65 + 6.3\pm0.2 - 0.05 = 16.9\pm0.2 \cr
&=& (m-M)_{V, \rm V2615~Oph} + \Delta V - 2.5 \log 1.15 \cr
&=& 15.9 + 1.15\pm 0.2 - 0.15 = 16.9\pm0.2 \cr
&=& ((m - M)_V + \Delta V)_{\rm V496~Sct} - 2.5 \log 0.63 \cr
&=& 13.6 + 2.8\pm0.2 + 0.5 = 16.9\pm0.2,
\label{distance_modulus_v_bv_v5584_sgr}
\end{eqnarray}
where we adopt $(m-M)_{V, \rm LV~Vul}=11.85$ and
$(m-M)_{V, \rm V533~Her}=10.65$ both from \citet{hac19ka}, and
$(m-M)_{V, \rm V496~Sct}=13.6$ in Appendix \ref{v496_sct_bvi}, and
$(m-M)_{V, \rm V2615~Oph}=15.9$ in Section \ref{v2615_oph_vi} and
Appendix \ref{v2615_oph_bvi}.
Thus, we obtain $(m-M)_{V, \rm V5584~Sgr}=16.9\pm0.1$ 
and $\log f_{\rm s}= \log 1.26 = +0.10$ against LV~Vul.

Figure \ref{distance_reddening_v5584_sgr_bvi_xxxxxx}(a) shows
the $B$ light curve of V5584~Sgr together with
those of LV~Vul, V1668~Cyg, and V679~Car.
We apply Equation (7) of \citet{hac19ka} for the $B$ band to Figure
\ref{distance_reddening_v5584_sgr_bvi_xxxxxx}(a)
and obtain
\begin{eqnarray}
(m&-&M)_{B, \rm V5584~Sgr} \cr
&=& ((m - M)_B + \Delta B)_{\rm LV~Vul} - 2.5 \log 1.26 \cr
&=& 12.45 + 5.45\pm0.2 - 0.25 = 17.65\pm0.2 \cr
&=& ((m - M)_B + \Delta B)_{\rm V1668~Cyg} - 2.5 \log 1.26 \cr
&=& 14.9 + 3.0\pm0.2 - 0.25 = 17.65\pm0.2 \cr
&=& ((m - M)_B + \Delta B)_{\rm V679~Car} - 2.5 \log 1.58 \cr
&=& 16.64 + 1.5\pm0.2 - 0.50 = 17.64\pm0.2,
\label{distance_modulus_b_v5584_sgr_lv_vul_v1668_cyg}
\end{eqnarray}
where we adopt $(m-M)_{B, \rm LV~Vul}= 11.85 + 0.60= 12.45$,
$(m-M)_{B, \rm V1668~Cyg}= 14.6 + 0.30= 14.9$, both from \citet{hac19ka},
and $(m-M)_{B, \rm V679~Car}= 16.05 + 0.59= 16.64$ from Appendix
\ref{v679_car_bvi}.  We have $(m-M)_{B, \rm V5584~Sgr}= 17.65\pm0.1$.

We plot $(m-M)_B= 17.65$, $(m-M)_V= 16.9$, and $(m-M)_I= 15.7$,
which cross at $d=8.2$~kpc and $E(B-V)=0.75$, in Figure
\ref{distance_reddening_v5584_sgr_bvi_xxxxxx}(b).
The crossing point is consistent with the distance-reddening relations
given by \citet[][orange and yellow lines]{gre18, gre19},
\citet[][cyan-blue lines]{chen19}
and \citet[][unfilled cyan-blue diamonds]{ozd18}.
Thus, we obtain $E(B-V)=0.75\pm0.05$ and $d=8.2\pm1$~kpc.


\begin{figure}
\plotone{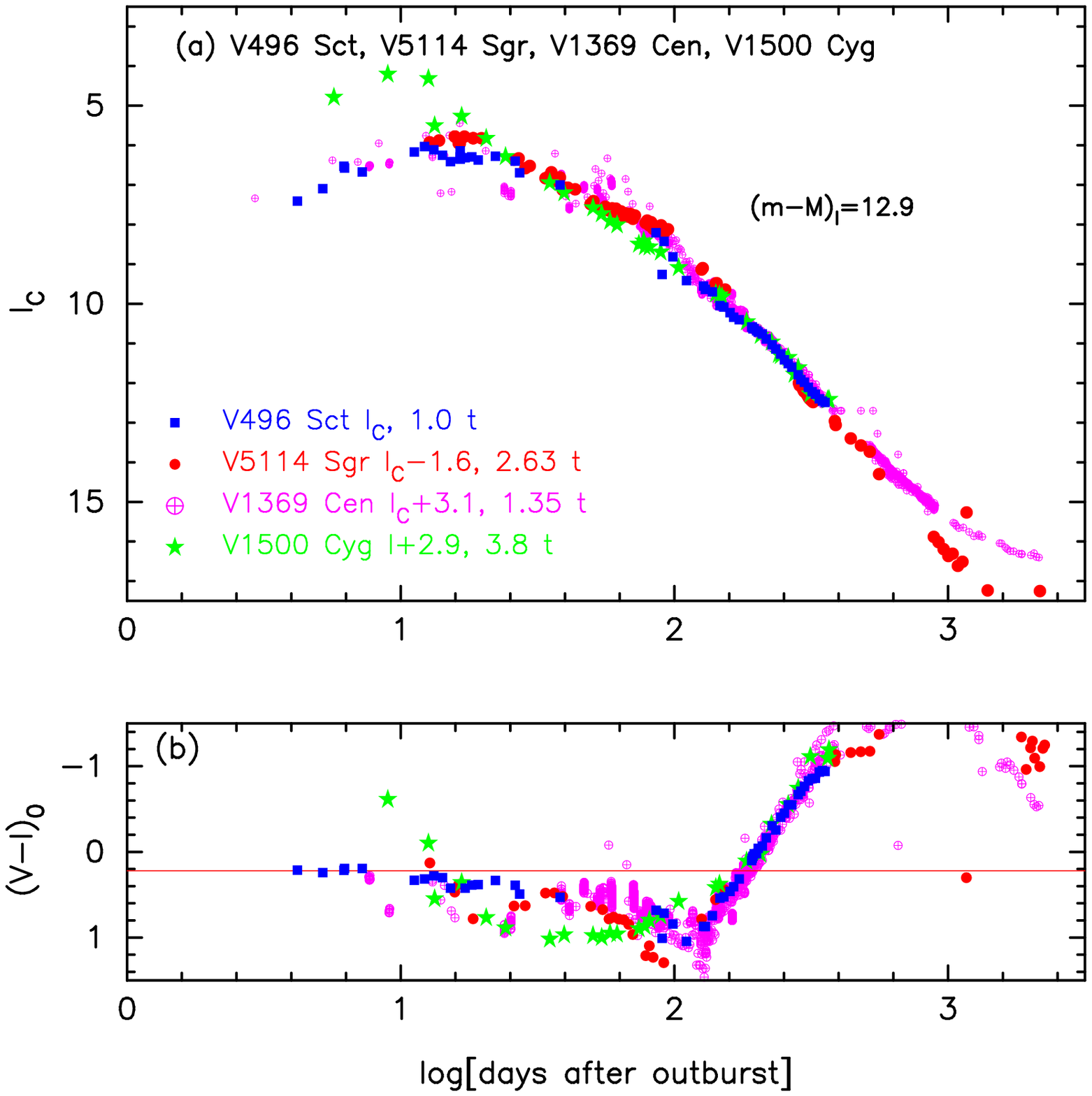}
\caption{
The (a) $I_{\rm C}$ light curve and (b) $(V-I_{\rm C})_0$ color curve
of V496~Sct as well as those of V5114~Sgr, V1369~Cen, and V1500~Cyg.
\label{v496_sct_v1500_cyg_v5114_sgr_v1369_cen_i_vi_logscale}}
\end{figure}


\begin{figure}
\plotone{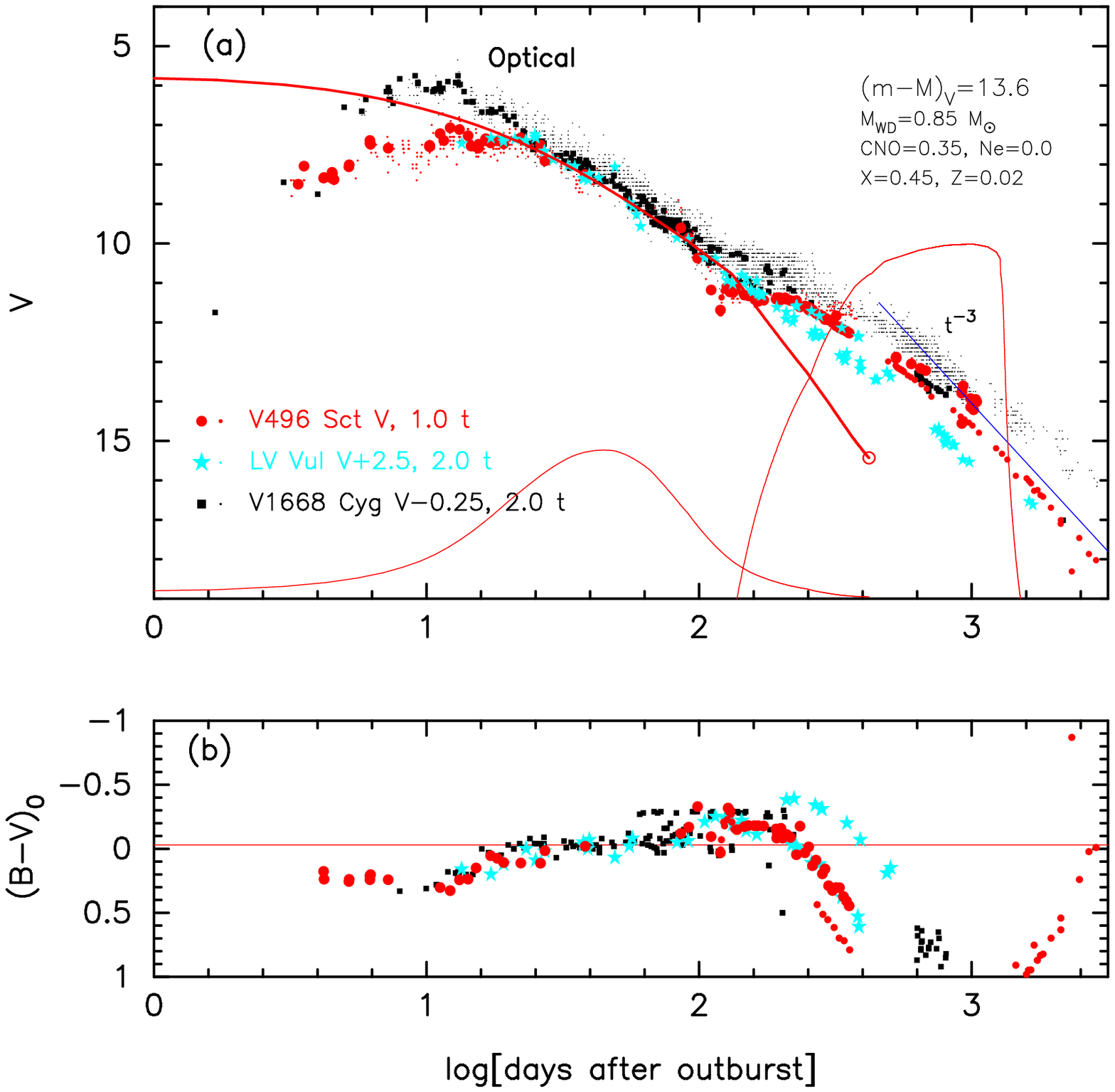}
\caption{
The (a) $V$ light curve and (b) $(B-V)_0$ color curve of V496~Sct 
(filled red circles for $V$ and small red dots for visual).
as well as LV~Vul and V1668~Cyg.  The data of V496~Sct are the same as
those in Figure 73 of \citet[][Paper II]{hac16kb}.
We plot a $0.85~M_\sun$ WD model (solid red lines)
with the chemical composition of CO nova 3 \citep{hac16k},
taking $(m-M)_V=13.6$ for V496~Sct.
We add the UV~1455\AA\  flux (left thin solid red line)
and supersoft X-ray flux (right thin solid red line) of
the $0.85~M_\sun$ WD model.
See the text for more details.
\label{v496_sct_lv_vul_v1668_cyg_v_bv_logscale_no2}}
\end{figure}


\begin{figure*}
\plottwo{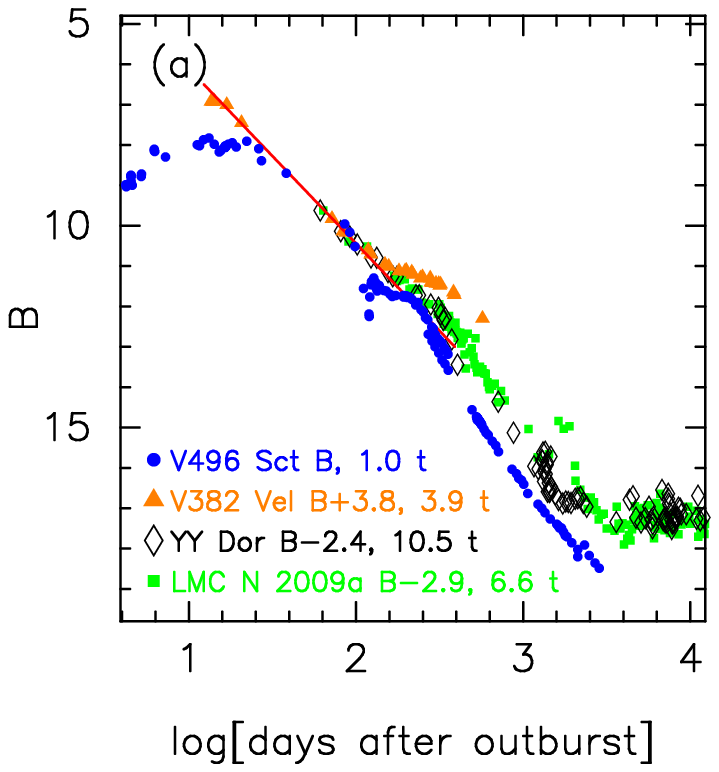}{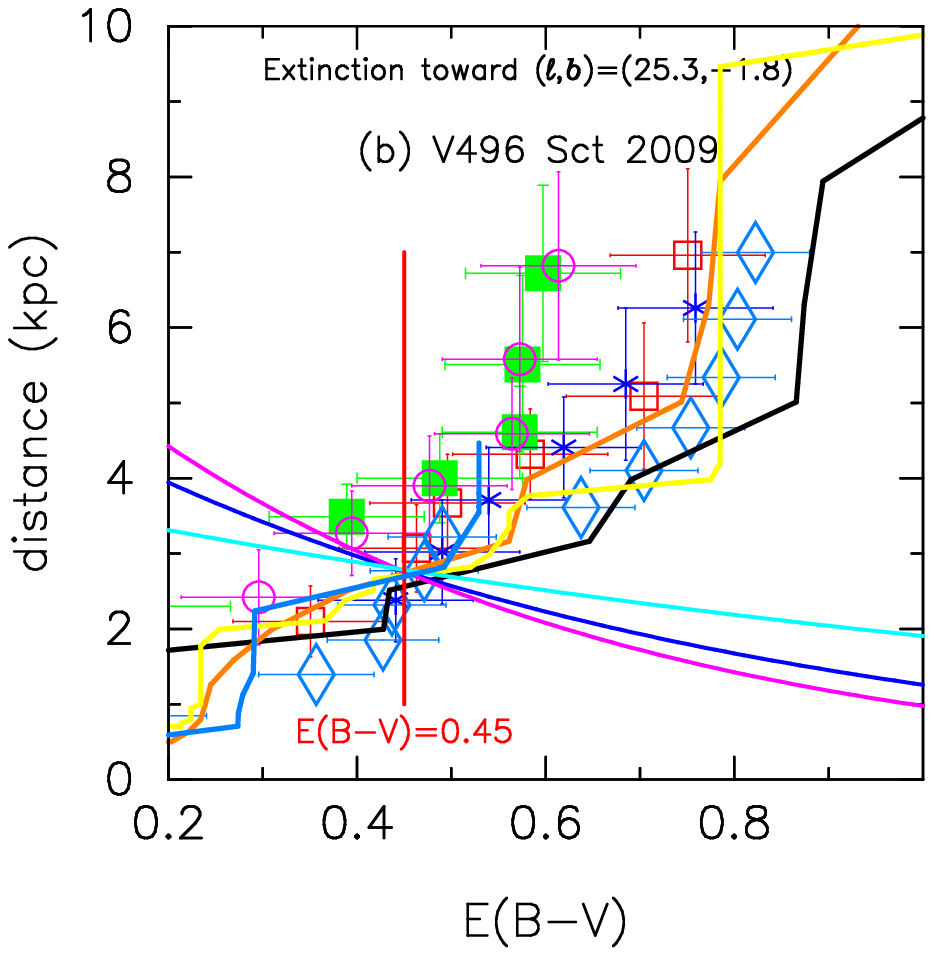}
\caption{
(a) The $B$ light curve of V496~Sct
as well as those of V382~Vel, YY~Dor, and LMC~N~2009a.
The $B$ data of V496~Sct are taken from \citet{raj12},
AAVSO, VSOLJ, and SMARTS.
(b) Various distance-reddening relations toward V496~Sct.
The thin solid lines of magenta, blue, and cyan denote the distance-reddening
relations given by  $(m-M)_B= 14.05$, $(m-M)_V= 13.6$, and $(m-M)_I= 12.9$,
respectively.
\label{distance_reddening_v496_sct_bvi_xxxxxx}}
\end{figure*}

\subsection{V496~Sct 2009}
\label{v496_sct_bvi}
We have reanalyzed the $BVI_{\rm C}$ multi-band 
light/color curves of V496~Sct based on the time-stretching method.  
We plot the time-stretched 
(a) $I_{\rm C}$ light and (b) $(V-I_{\rm C})_0$ color curves 
of V496~Sct as well as V5114~Sgr, V1369~Cen, and V1500~Cyg in
Figure \ref{v496_sct_v1500_cyg_v5114_sgr_v1369_cen_i_vi_logscale}.
The $BVI_{\rm C}$ data of V496~Sct are taken from \citet{raj12},
AAVSO, VSOLJ, and SMARTS.
Using the color excess $E(B-V)= 0.45$ and the timescaling factor 
$\log f_{\rm s}= +0.30$ taken from \citet{hac19ka}, we are able to overlap
the $(V-I)_0$ color curve of V496~Sct with the other novae, as shown in
Figure \ref{v496_sct_v1500_cyg_v5114_sgr_v1369_cen_i_vi_logscale}(b).
We apply Equation (8) of \citet{hac19ka} for the $I$ band to Figure
\ref{v496_sct_v1500_cyg_v5114_sgr_v1369_cen_i_vi_logscale}(a)
and obtain
\begin{eqnarray}
(m&-&M)_{I, \rm V496~Sct} \cr
&=& ((m - M)_I + \Delta I_{\rm C})
_{\rm V5114~Sgr} - 2.5 \log 2.63 \cr
&=& 15.55 - 1.6\pm0.2 - 1.05 = 12.9\pm0.2 \cr
&=& ((m - M)_I + \Delta I_{\rm C})
_{\rm V1369~Cen} - 2.5 \log 1.35 \cr
&=& 10.11 + 3.1\pm0.2 - 0.325 = 12.89\pm0.2 \cr
&=& ((m - M)_I + \Delta I_{\rm C})
_{\rm V1500~Cyg} - 2.5 \log 3.8 \cr
&=& 11.45 + 2.9\pm0.2 - 1.45 = 12.9\pm0.2,
\label{distance_modulus_i_vi_v496_sct}
\end{eqnarray}
where we adopt
$(m-M)_{I, \rm V5114~Sgr}=15.55$ from Appendix \ref{v5114_sgr_ubvi},
$(m-M)_{I, \rm V1369~Cen}=10.11$ from \citet{hac19ka}, and
$(m-M)_{I, \rm V1500~Cyg}=11.45$ from Appendix \ref{v1500_cyg_ubvi}.
Thus, we obtain $(m-M)_{I, \rm V496~Sct}= 12.9\pm0.2$ and the 
timescaling factor of $\log f_{\rm s}= +0.30$.
These parameters are all consistent with the previous values of
$(m-M)_I= 13.0\pm 0.2$ and $\log f_{\rm s}= +0.30$ obtained
by \citet{hac19ka}.  

Figure \ref{v496_sct_lv_vul_v1668_cyg_v_bv_logscale_no2} shows the $V$
light and $(B-V)_0$ color curves of V496~Sct as well as those of
LV~Vul and V1668~Cyg. 
Based on the time-stretching method, we have the relation of
\begin{eqnarray}
(m&-&M)_{V, \rm V496~Sct} \cr
&=& (m - M + \Delta V)_{V, \rm LV~Vul} - 2.5 \log 2.0 \cr
&=& 11.85 + 2.5\pm0.3 - 0.75 = 13.6\pm0.3 \cr
&=& (m - M + \Delta V)_{V, \rm V1668~Cyg} - 2.5 \log 2.0 \cr
&=& 14.6 - 0.25\pm0.3 - 0.75 = 13.6 \pm0.3 .
\label{distance_modulus_v496_sct_lv_vul_v1668_cyg}
\end{eqnarray}
Thus, we obtain $f_{\rm s}=2.0$ against the template nova LV~Vul
and $(m-M)_V=13.6\pm0.2$.

Figure \ref{distance_reddening_v496_sct_bvi_xxxxxx}(a) shows
the $B$ light curve of V496~Sct as well as those of V382~Vel, YY~Dor,
and LMC~N~2009a.
We apply Equation (7) of \citet{hac19ka} for the $B$ band
to \ref{distance_reddening_v496_sct_bvi_xxxxxx}(a) and obtain
\begin{eqnarray}
(m&-&M)_{B, \rm V496~Sct} \cr
&=& ((m - M)_B + \Delta B)_{\rm YY~Dor} - 2.5 \log 10.5 \cr
&=& 18.98 - 2.4\pm0.2 - 2.55 = 14.03\pm0.2 \cr
&=& ((m - M)_B + \Delta B)_{\rm LMC~N~2009a} - 2.5 \log 6.6 \cr
&=& 18.98 - 2.9\pm0.2 - 2.05 = 14.03\pm0.2 \cr
&=& ((m - M)_B + \Delta B)_{\rm V382~Vel} - 2.5 \log 3.9 \cr
&=& 11.72 + 3.8\pm0.2 - 1.47 = 14.05\pm0.2.
\label{distance_modulus_b_v496_sct_v382_vel_yy_dor_lmcn2009a}
\end{eqnarray}
Thus, we obtain $(m-M)_{B, \rm V496~Sct}= 14.05\pm0.1$,

We plot $(m-M)_B= 14.05$, $(m-M)_V= 13.6$, and $(m-M)_I= 12.9$,
which broadly cross at $d=2.76$~kpc and $E(B-V)=0.45$, in Figure
\ref{distance_reddening_v496_sct_bvi_xxxxxx}(b).
The crossing point is consistent with the distance-reddening relations
given by \citet[][orange and yellow lines]{gre18, gre19},
\citet[][cyan-blue lines]{chen19}.
Thus, we have $E(B-V)=0.45\pm0.05$ and $d=2.76\pm0.2$~kpc.


\begin{figure}
\plotone{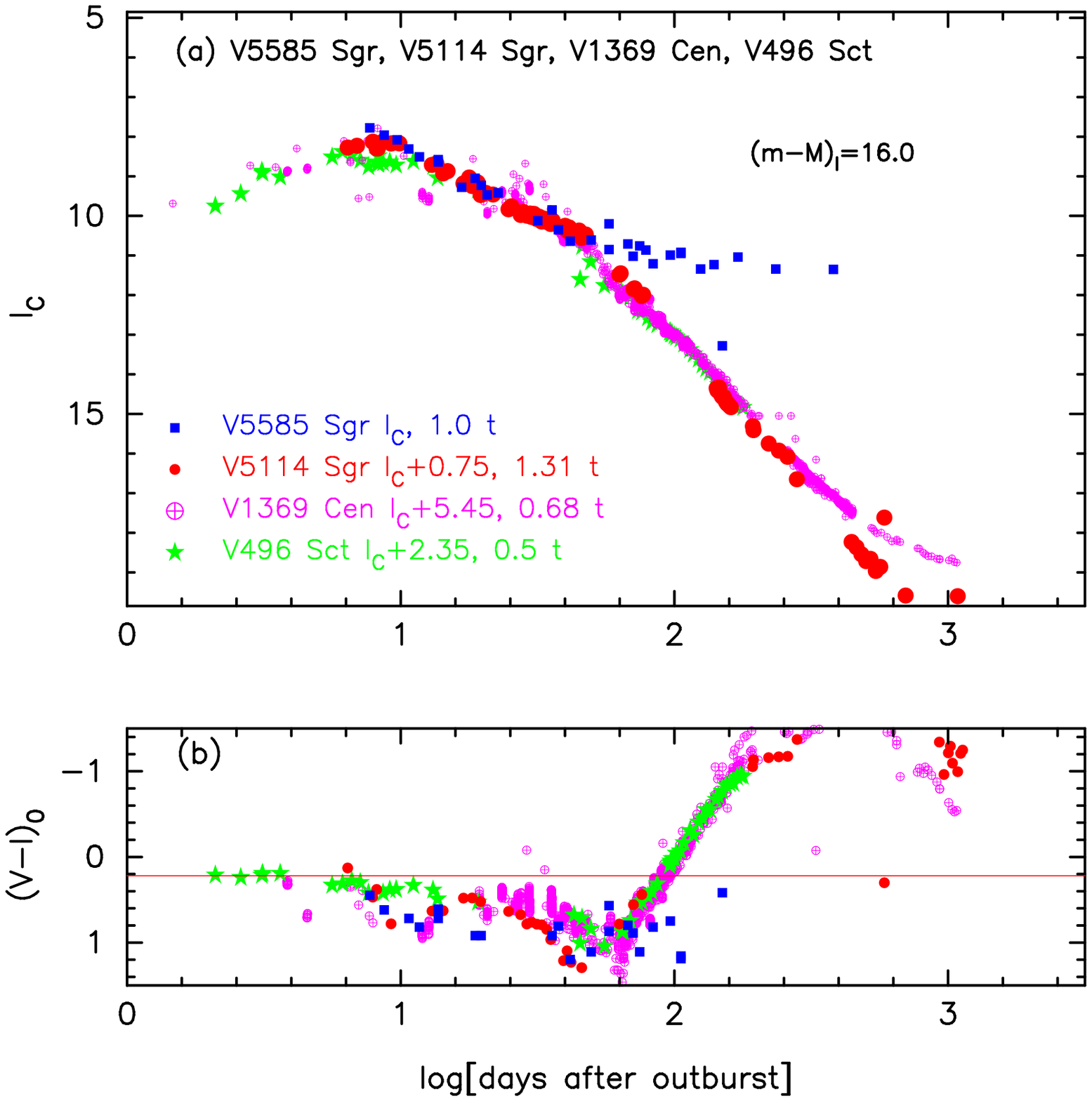}
\caption{
The (a) $I_{\rm C}$ light curve and (b) $(V-I_{\rm C})_0$ color curve
of V5585~Sgr as well as those of V5114~Sgr, V1369~Cen, and V496~Sct.
\label{v5585_sgr_v5114_sgr_v1369_cen_v496_sct_i_vi_color_logscale}}
\end{figure}


\begin{figure}
\plotone{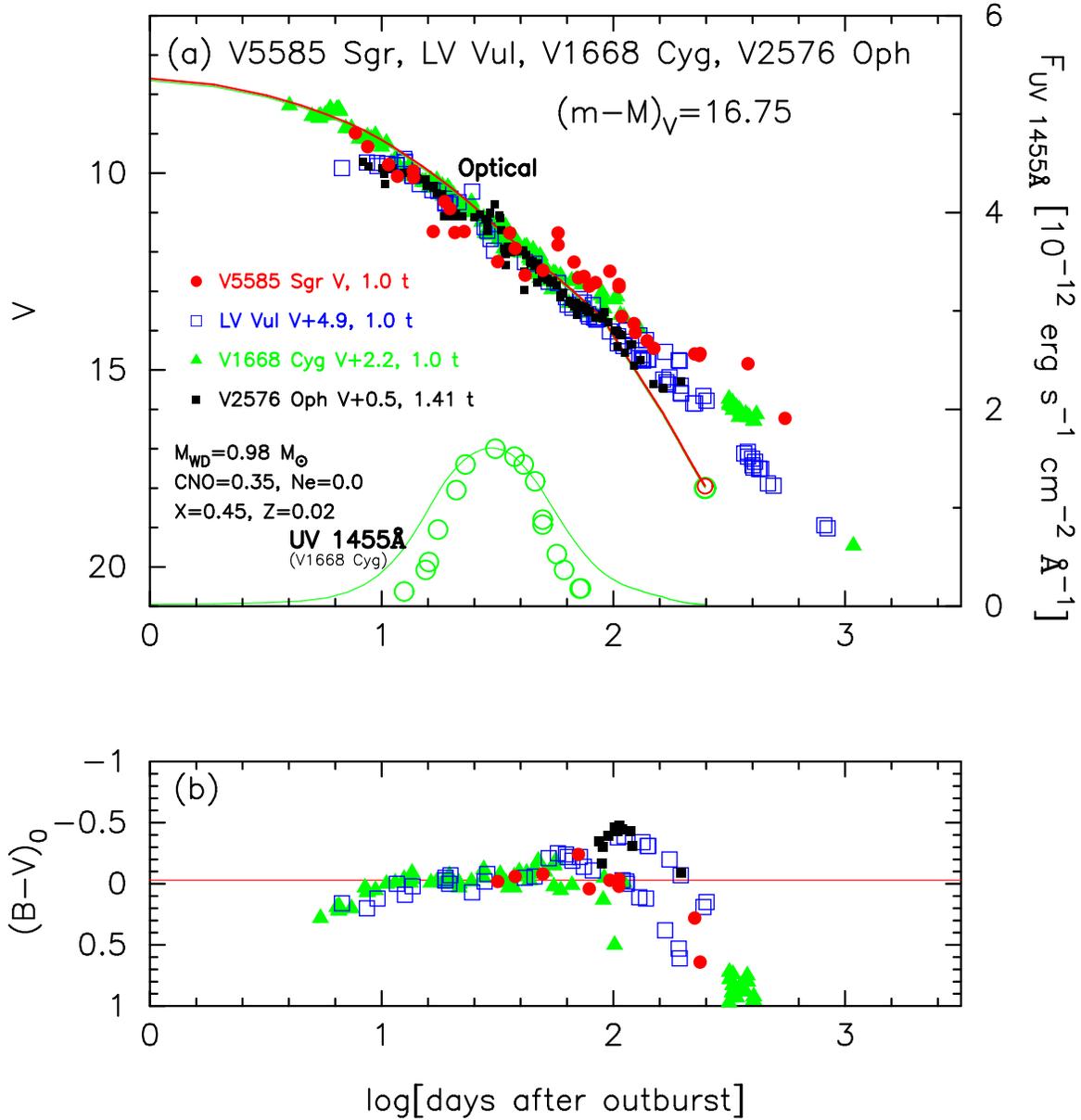}
\caption{
The (a) $V$ light curve and (b) $(B-V)_0$ color curve of V5585~Sgr
as well as LV~Vul, V1668~Cyg, and V2576~Oph.
In panel (a), we add a $0.98~M_\sun$ WD model (CO3, solid red line)
for V5585~Sgr as well as the same $0.98~M_\sun$ WD model 
(CO3, solid green lines) for V1668~Cyg.
\label{v5585_sgr_v2576_oph_v1668_cyg_lv_vul_v_bv_logscale_no2}}
\end{figure}


\begin{figure*}
\plottwo{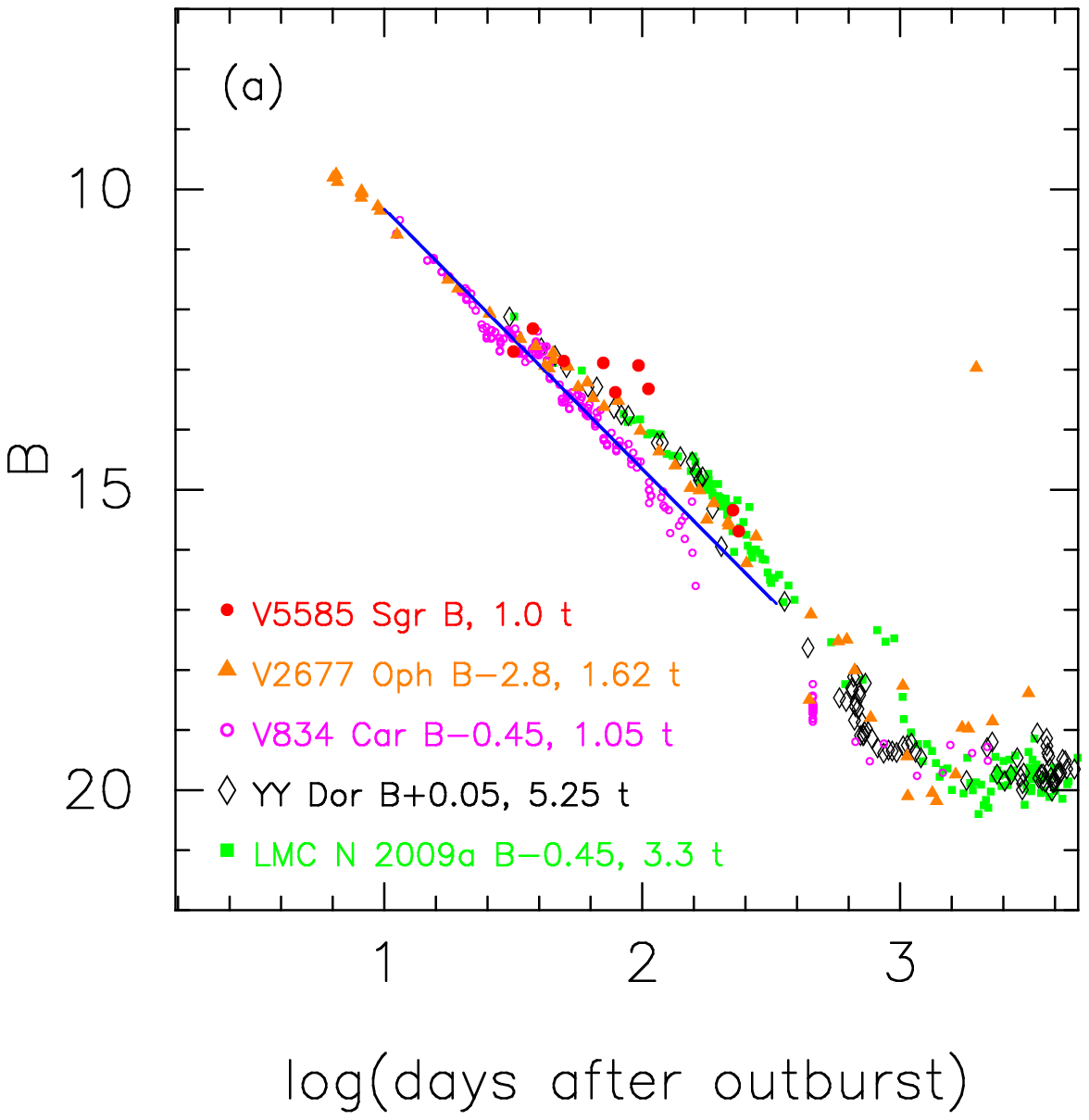}{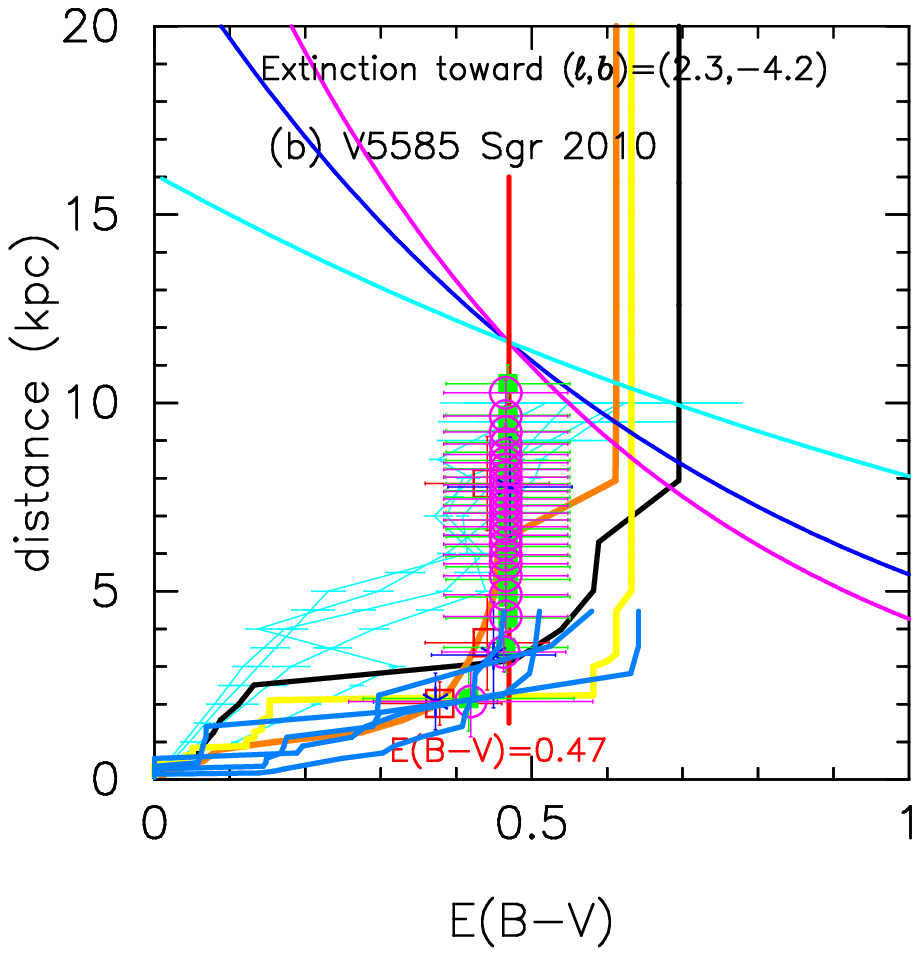}
\caption{
(a) The $B$ light curve of V5585~Sgr
as well as those of V2677~Oph, V834~Car, YY~Dor, and LMC~N~2009a.
The $B$ data of V5585~Sgr are taken from VSOLJ.
(b) Various distance-reddening relations toward V5585~Sgr.
The thin solid lines of magenta, blue, and cyan denote the distance-reddening
relations given by  $(m-M)_B= 17.25$, $(m-M)_V= 16.78$, and $(m-M)_I= 16.01$,
respectively.  
\label{distance_reddening_v5585_sgr_bvi_xxxxxx}}
\end{figure*}

\subsection{V5585~Sgr 2010}
\label{v5585_sgr_bvi}
We have reanalyzed the $BVI_{\rm C}$ multi-band light/color curves
of V5585~Sgr based on the time-stretching method.  
The main differences are the timescaling factor of $\log f_{\rm s}= 
+0.0$ (the previous value of $+0.10$).
Figure \ref{v5585_sgr_v5114_sgr_v1369_cen_v496_sct_i_vi_color_logscale}
shows the (a) $I_{\rm C}$ light and (b) $(V-I_{\rm C})_0$ color curves
of V5585~Sgr as well as V5114~Sgr, V1369~Cen, and V496~Sct.
The $BVI_{\rm C}$ data of V5585~Sgr are taken from VSOLJ.
We adopt the color excess of $E(B-V)= 0.47$ as mentioned below.
We apply Equation (8) of \citet{hac19ka} for the $I$ band to Figure
\ref{v5585_sgr_v5114_sgr_v1369_cen_v496_sct_i_vi_color_logscale}(a)
and obtain
\begin{eqnarray}
(m&-&M)_{I, \rm V5585~Sgr} \cr
&=& ((m - M)_I + \Delta I_{\rm C})
_{\rm V5114~Sgr} - 2.5 \log 1.31 \cr
&=& 15.55 + 0.75\pm0.2 - 0.3 =  16.0\pm0.2 \cr
&=& ((m - M)_I + \Delta I_{\rm C})
_{\rm V1369~Cen} - 2.5 \log 0.68 \cr
&=& 10.11 + 5.45\pm0.2 + 0.425 = 15.99\pm0.2 \cr
&=& ((m - M)_I + \Delta I_{\rm C})
_{\rm V496~Sct} - 2.5 \log 0.50 \cr
&=& 12.9 + 2.35\pm0.2 + 0.75 = 16.0\pm0.2,
\label{distance_modulus_i_vi_v5585_sgr}
\end{eqnarray}
where we adopt
$(m-M)_{I, \rm V5114~Sgr}=15.55$ from Appendix \ref{v5114_sgr_ubvi},
$(m-M)_{I, \rm V1369~Cen}=10.11$ from \citet{hac19ka}, and
$(m-M)_{I, \rm V496~Sct}=12.9$ in Appendix \ref{v496_sct_bvi}.
Thus, we obtain $(m-M)_{I, \rm V5585~Sgr}= 16.0\pm0.2$.

Figure \ref{v5585_sgr_v2576_oph_v1668_cyg_lv_vul_v_bv_logscale_no2} shows
the light/color curves of V5585~Sgr, LV~Vul, V1668~Cyg, and V2576~Oph.
Applying Equation (4) of \citet{hac19ka} to them,
we have the relation
\begin{eqnarray}
(m&-&M)_{V, \rm V5585~Sgr} \cr
&=& ((m - M)_V + \Delta V)_{\rm LV~Vul} - 2.5 \log 1.0 \cr
&=& 11.85 + 4.9\pm0.3 - 0.0 = 16.75\pm0.3 \cr
&=& ((m - M)_V + \Delta V)_{\rm V1668~Cyg} - 2.5 \log 1.0 \cr
&=& 14.6 + 2.2\pm0.3 - 0.0 = 16.8\pm0.3 \cr
&=& ((m - M)_V + \Delta V)_{\rm V2576~Oph} - 2.5 \log 1.41 \cr
&=& 16.65 + 0.5\pm0.2 - 0.375 = 16.78\pm0.2,
\label{distance_modulus_v_bv_v5585_sgr}
\end{eqnarray}
where we adopt $(m-M)_{V, \rm LV~Vul}=11.85$ and
$(m-M)_{V, \rm V1668~Cyg}=14.6$, both from \citet{hac19ka},
and $(m-M)_{V, \rm V2576~Oph}=16.65$ from \citet{hac19kb}.
Thus, we obtain $(m-M)_{V, \rm V5585~Sgr}=16.78\pm0.2$ and 
$\log f_{\rm s}= \log 1.0 = +0.0$ against LV~Vul.

Figure \ref{distance_reddening_v5585_sgr_bvi_xxxxxx}(a)
shows the $B$ light curve of V5585~Sgr
together with those of V2677~Oph, V834~Car, YY~Dor, and LMC~N~2009a.
We apply Equation (7) of \citet{hac19ka} for the $B$ band to Figure
\ref{distance_reddening_v5585_sgr_bvi_xxxxxx}(a) and obtain
\begin{eqnarray}
(m&-&M)_{B, \rm V5585~Sgr} \cr
&=& ((m - M)_B + \Delta B)_{\rm V2677~Oph} - 2.5 \log 1.62 \cr
&=& 20.6 - 2.8\pm0.2 - 0.525 = 17.27\pm0.2 \cr
&=& ((m - M)_B + \Delta B)_{\rm V834~Car} - 2.5 \log 1.05 \cr
&=& 17.75 - 0.45\pm0.2 - 0.05 = 17.25\pm0.2 \cr
&=& ((m - M)_B + \Delta B)_{\rm YY~Dor} - 2.5 \log 5.25 \cr
&=& 18.98 + 0.05\pm0.2 - 1.8 = 17.23\pm0.2 \cr
&=& ((m - M)_B + \Delta B)_{\rm LMC~N~2009a} - 2.5 \log 3.3 \cr
&=& 18.98 - 0.45\pm0.2 - 1.3 = 17.23\pm0.2,
\label{distance_modulus_b_v5585_sgr_v2677_oph_v834_car_yy_dor_lmcn2009a}
\end{eqnarray}
where we adopt $(m-M)_{B, \rm V2677~Oph}= 19.2 + 1.4= 20.6$ in Appendix
\ref{v2677_oph_bvi}, and $(m-M)_{B, \rm V834~Car}= 17.25 + 0.50= 17.75$
in Appendix \ref{v834_car_bvi}.
We have $(m-M)_{B, \rm V5585~Sgr}= 17.25\pm0.1$.

We plot $(m-M)_B= 17.25$, $(m-M)_V= 16.78$, and $(m-M)_I= 16.01$,
which broadly cross at $d=11.6$~kpc and $E(B-V)=0.47$, in Figure
\ref{distance_reddening_v5585_sgr_bvi_xxxxxx}(b).
The crossing point is consistent with the distance-reddening relations
given by \citet{mar06} 
and \citet[][cyan-blue lines]{chen19}.
Thus, we have $E(B-V)=0.47\pm0.05$ and $d=11.6\pm2$~kpc.


\begin{figure}
\plotone{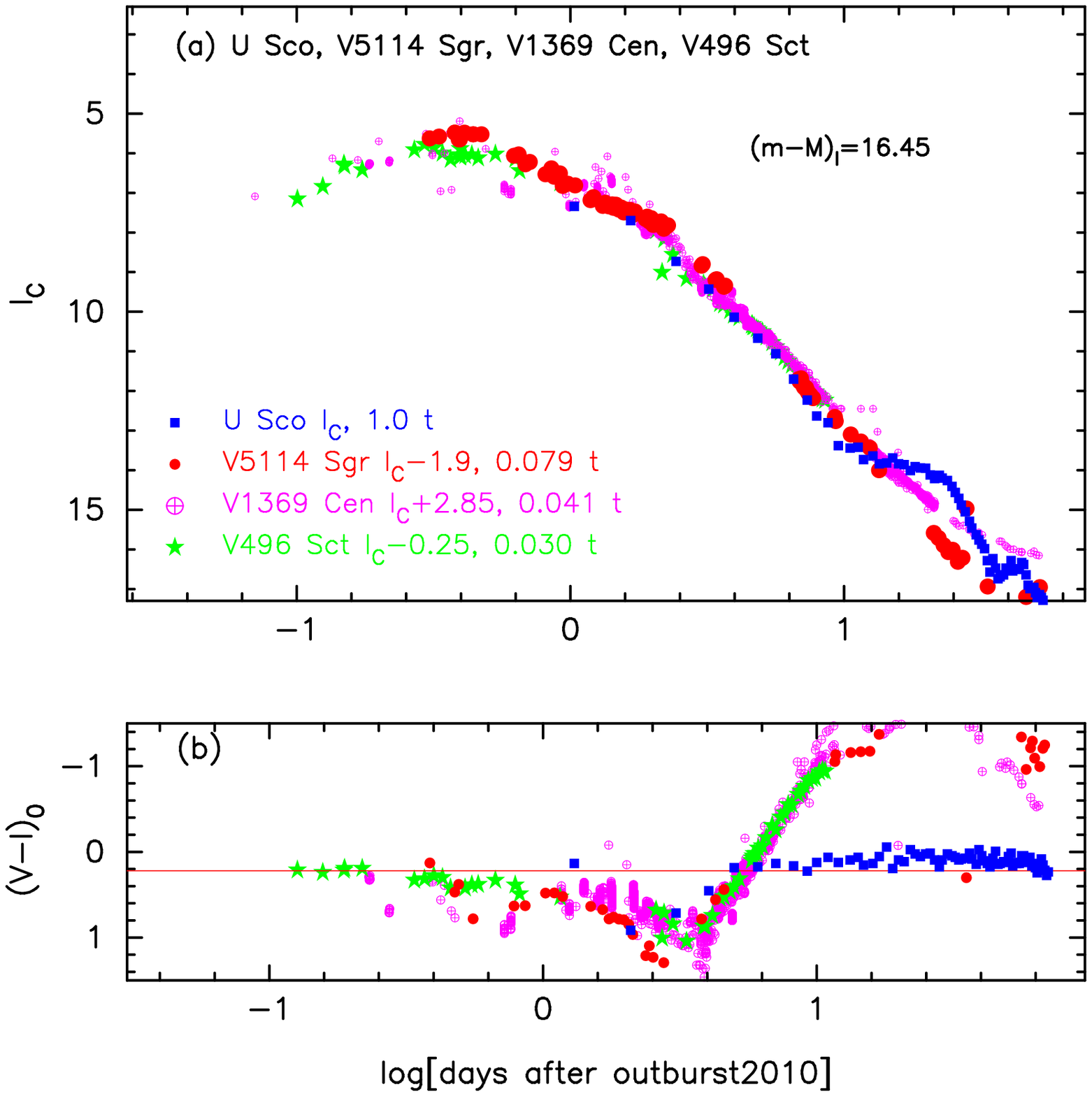}
\caption{
The (a) $I_{\rm C}$ light curve and (b) $(V-I_{\rm C})_0$ color curve
of U~Sco as well as those of V5114~Sgr, V1369~Cen, and V496~Sct.
\label{u_sco_v5114_sgr_v1369_cen_v496_sct_i_vi_color_logscale}}
\end{figure}


\begin{figure}
\plotone{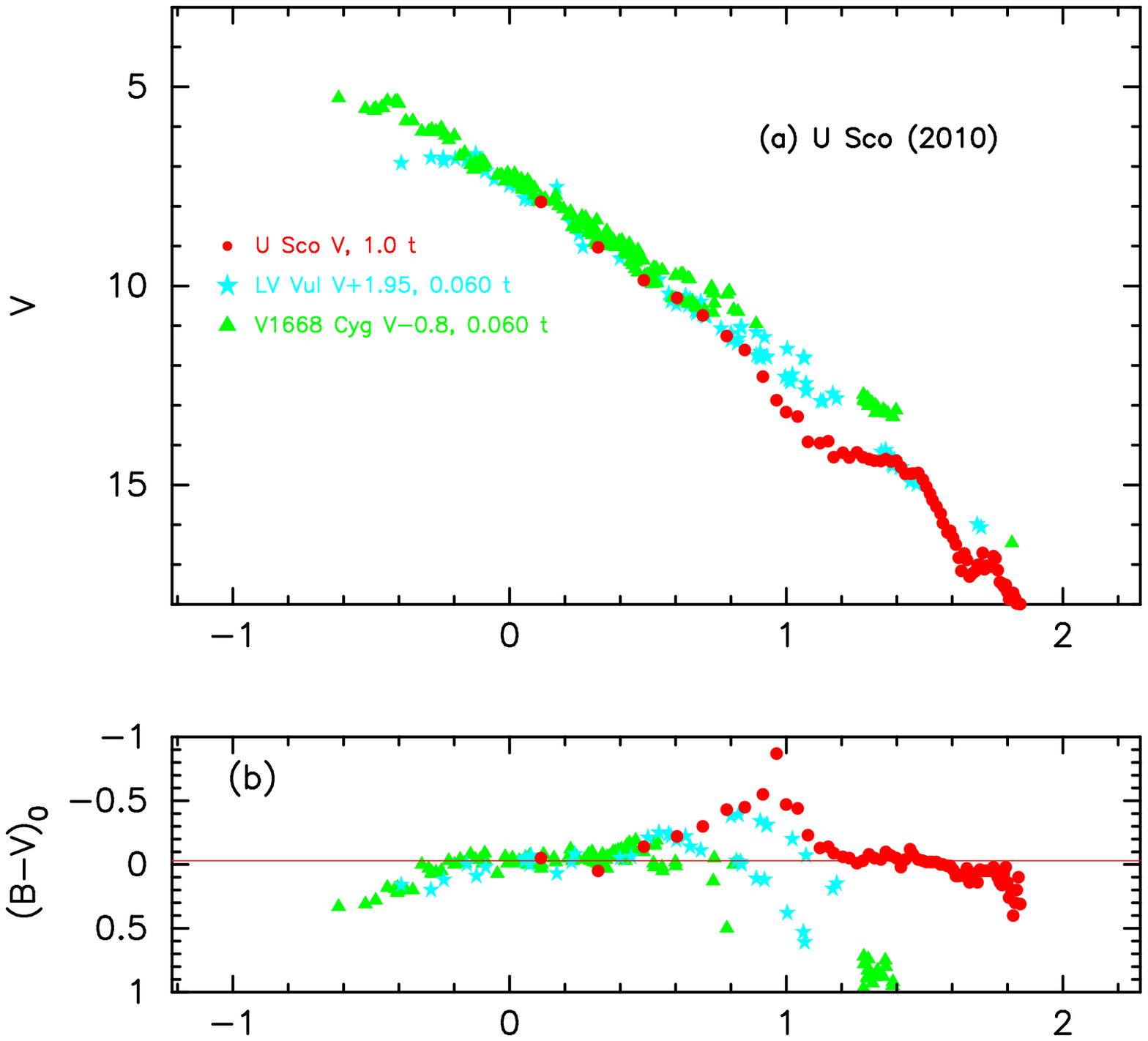}
\caption{
The (a) $V$ light curve and (b) $(B-V)_0$ color curves of U~Sco 2010
outburst as well as LV~Vul and V1668~Cyg.
\label{u_sco_lv_vul_v1668_cyg_v_bv_logscale_no2}}
\end{figure}


\begin{figure}
\plotone{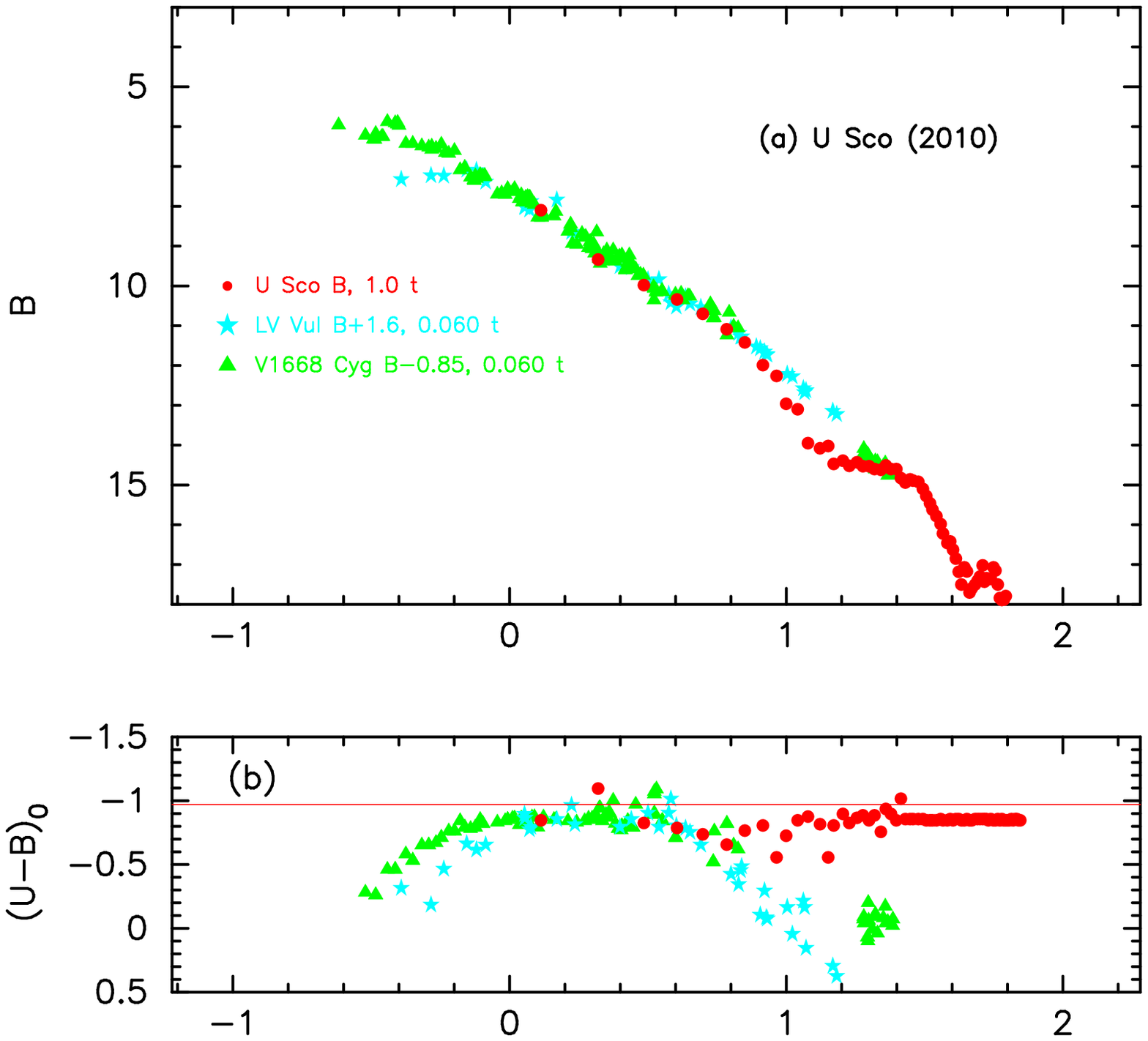}
\caption{
The (a) $B$ light curve and (b) $(U-B)_0$ color curves of U~Sco 2010
outburst as well as LV~Vul and V1668~Cyg.
\label{u_sco_lv_vul_v1668_cyg_b_ub_logscale_no2}}
\end{figure}


\begin{figure*}
\plottwo{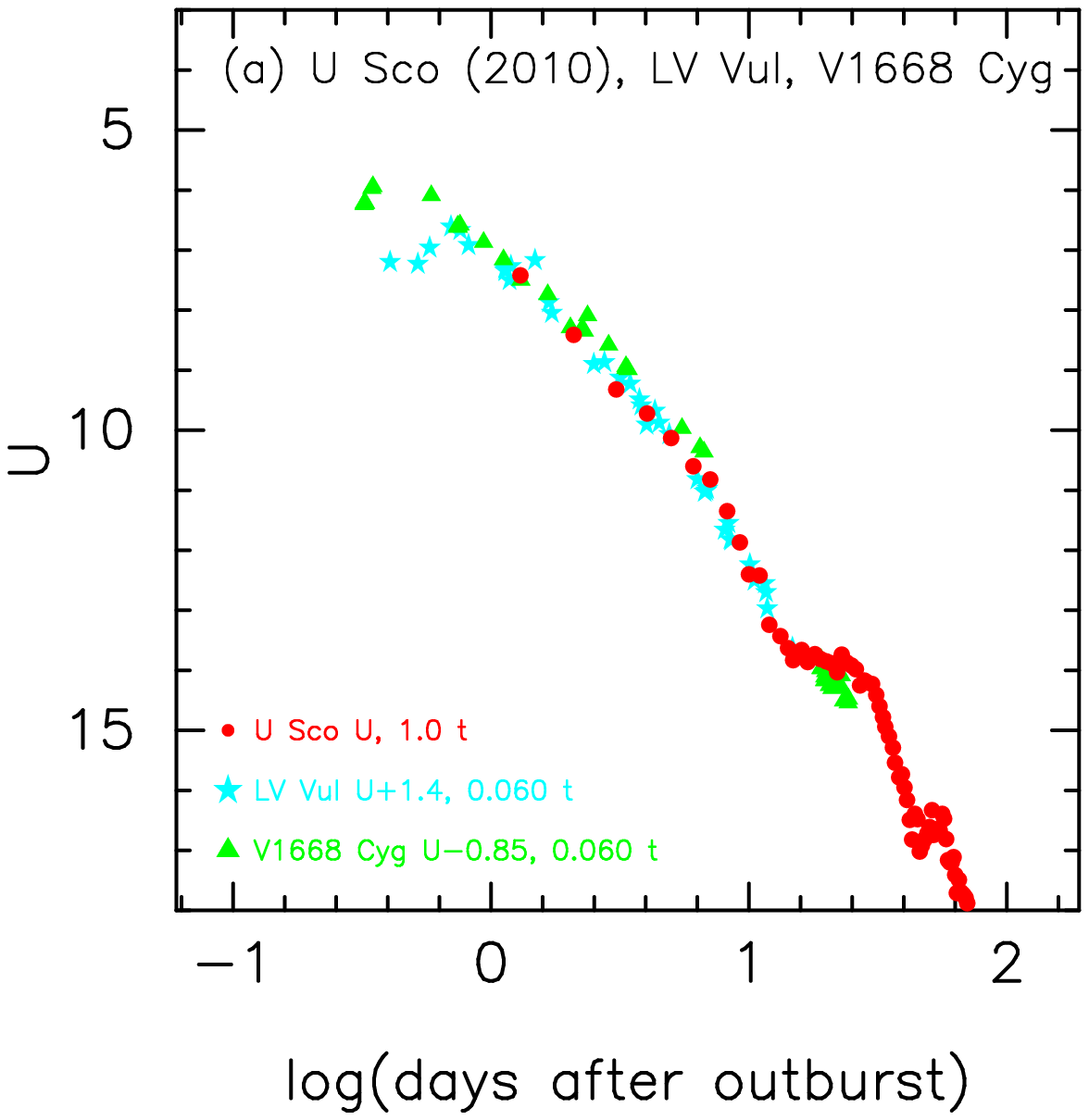}{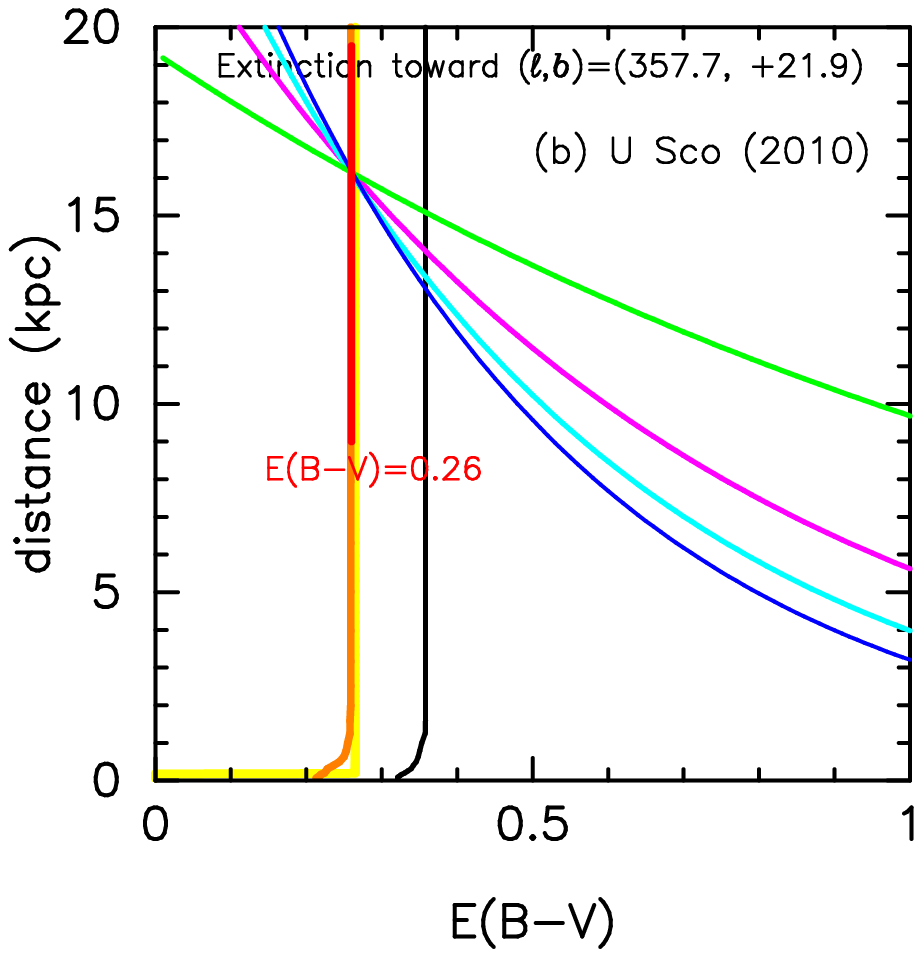}
\caption{
(a) The $U$ light curve of the U~Sco 2010 outburst 
as well as LV~Vul and V1668~Cyg.
(b) Various distance-reddening relations toward U~Sco.
The solid blue, cyan, magenta, and green lines denote the distance-reddening
relations given by  $(m-M)_U= 17.3$, $(m-M)_B= 17.1$, 
$(m-M)_V= 16.85$, and $(m-M)_I= 16.45$,
respectively.  
\label{distance_reddening_u_sco_ubvik_xxxxxx}}
\end{figure*}

\subsection{U~Sco 2010}
\label{u_sco_ubvik}
We have reanalyzed the $UBVI_{\rm C}$ multi-band light/color curves
of U~Sco based on the time-stretching method.  The $I_{\rm C}$ and 
$(V-I_{\rm C})$ data are not included in our previous work. 
Figure \ref{u_sco_v5114_sgr_v1369_cen_v496_sct_i_vi_color_logscale}
shows the (a) $I_{\rm C}$ light and (b) $(V-I_{\rm C})_0$ color curves
of U~Sco as well as V5114~Sgr, V1369~Cen, and V496~Sct.
The $UBVI_{\rm C}$ data of U~Sco are taken from \citet{pagnotta15}.
We adopt the color excess of $E(B-V)= 0.26$ as mentioned below.
We apply Equation (8) of \citet{hac19ka} for the $I$ band to Figure
\ref{u_sco_v5114_sgr_v1369_cen_v496_sct_i_vi_color_logscale}(a)
and obtain
\begin{eqnarray}
(m&-&M)_{I, \rm U~Sco} \cr
&=& ((m - M)_I + \Delta I_{\rm C})
_{\rm V5114~Sgr} - 2.5 \log 0.079 \cr
&=& 15.55 - 1.9\pm0.2 + 2.75 =  16.4\pm0.2 \cr
&=& ((m - M)_I + \Delta I_{\rm C})
_{\rm V1369~Cen} - 2.5 \log 0.041 \cr
&=& 10.11 + 2.85\pm0.2 + 3.475 = 16.44\pm0.2 \cr
&=& ((m - M)_I + \Delta I_{\rm C})
_{\rm V496~Sct} - 2.5 \log 0.030 \cr
&=& 12.9 - 0.25\pm0.2 + 3.8 = 16.45\pm0.2,
\label{distance_modulus_i_vi_u_sco}
\end{eqnarray}
where we adopt
$(m-M)_{I, \rm V5114~Sgr}=15.55$ from Appendix \ref{v5114_sgr_ubvi},
$(m-M)_{I, \rm V1369~Cen}=10.11$ from \citet{hac19ka}, and
$(m-M)_{I, \rm V496~Sct}=12.9$ in Appendix \ref{v496_sct_bvi}.
Thus, we obtain $(m-M)_{I, \rm U~Sco}= 16.43\pm0.2$.

We plot the (a) $V$ light and (b) $(B-V)_0$ color curves of three novae, 
U~Sco, LV~Vul, and V1668~Cyg, in Figure
\ref{u_sco_lv_vul_v1668_cyg_v_bv_logscale_no2},
to obtain the distance modulus of U~Sco.
Applying Equation (4) of \citet{hac19ka} to them,
we have the relation
\begin{eqnarray}
(m-M)_{V, \rm U~Sco}
&=& ((m - M)_V + \Delta V)_{\rm LV~Vul} - 2.5 \log 0.060 \cr
&=& 11.85 + 1.95\pm0.2 + 3.05 = 16.85\pm0.2 \cr
&=& ((m - M)_V + \Delta V)_{\rm V1668~Cyg} - 2.5 \log 0.060 \cr
&=& 14.6 - 0.8\pm0.2 + 3.05 = 16.85\pm0.2,
\label{distance_modulus_v_bv_u_sco}
\end{eqnarray}
where we adopt $(m-M)_{V, \rm LV~Vul}=11.85$ and
$(m-M)_{V, \rm V1668~Cyg}=14.6$ both from \citet{hac19ka}.
Thus, we adopt $(m-M)_{V, \rm U~Sco}=16.85\pm0.2$.


In Figure \ref{u_sco_lv_vul_v1668_cyg_b_ub_logscale_no2},
we plot the (a) $B$ light and (b) $(U-B)_0$ color curves of
three novae, U~Sco, LV~Vul, and V1668~Cyg,
to obtain the distance modulus of U~Sco.
Applying Equation (7) of \citet{hac19ka} to them,
we have the relation
\begin{eqnarray}
(m-M)_{B, \rm U~Sco}
&=& ((m - M)_B + \Delta B)_{\rm LV~Vul} - 2.5 \log 0.060 \cr
&=& 12.45 + 1.6\pm0.2 + 3.05 = 17.1\pm0.2 \cr
&=& ((m - M)_B + \Delta B)_{\rm V1668~Cyg} - 2.5 \log 0.060 \cr
&=& 14.9 - 0.85\pm0.2 + 3.05 = 17.1\pm0.2,
\label{distance_modulus_b_ub_u_sco}
\end{eqnarray}
where we adopt $(m-M)_{B, \rm LV~Vul}=12.45$ and
$(m-M)_{B, \rm V1668~Cyg}=14.9$ both from \citet{hac19ka}.
Thus, we adopt $(m-M)_{B, \rm U~Sco}=17.1\pm0.2$.

We plot the $U$ light curves of three novae, U~Sco, LV~Vul, and V1668~Cyg
in Figure \ref{distance_reddening_u_sco_ubvik_xxxxxx}(a),
to obtain the distance modulus of U~Sco.
Applying Equation (6) of \citet{hac19ka} to them,
we have the relation
\begin{eqnarray}
(m-M)_{U, \rm U~Sco}
&=& (m - M + \Delta U)_{U, \rm LV~Vul} - 2.5 \log 0.060 \cr
&=& 12.85 + 1.4\pm0.2 + 3.05 = 17.3\pm0.2 \cr
&=& (m - M + \Delta U)_{U, \rm V1668~Cyg} - 2.5 \log 0.060 \cr
&=& 15.10 - 0.85\pm0.2 + 3.05 = 17.3\pm0.2,
\label{distance_modulus_u_u_sco_lv_vul_v1668_cyg}
\end{eqnarray}
where we adopt $(m-M)_{U, \rm LV~Vul}=12.85$ and
$(m-M)_{U, \rm V1668~Cyg}=15.10$ both from \citet{hac19ka}.
Thus, we adopt $(m-M)_{U, \rm U~Sco}=17.3\pm0.2$.


In Figure \ref{distance_reddening_u_sco_ubvik_xxxxxx}(b), we plot 
$(m-M)_U= 17.3$, $(m-M)_B= 17.1$, $(m-M)_V= 16.85$, and $(m-M)_I= 16.45$,
which broadly cross at $d=16.2$~kpc and $E(B-V)=0.26$.
The crossing point is consistent with the distance-reddening relations
given by
\citet[][orange and yellow lines]{gre18, gre19}.
Thus, we have $E(B-V)=0.26\pm0.05$ and $d=16.2\pm2$~kpc for U~Sco.


\begin{figure}
\plotone{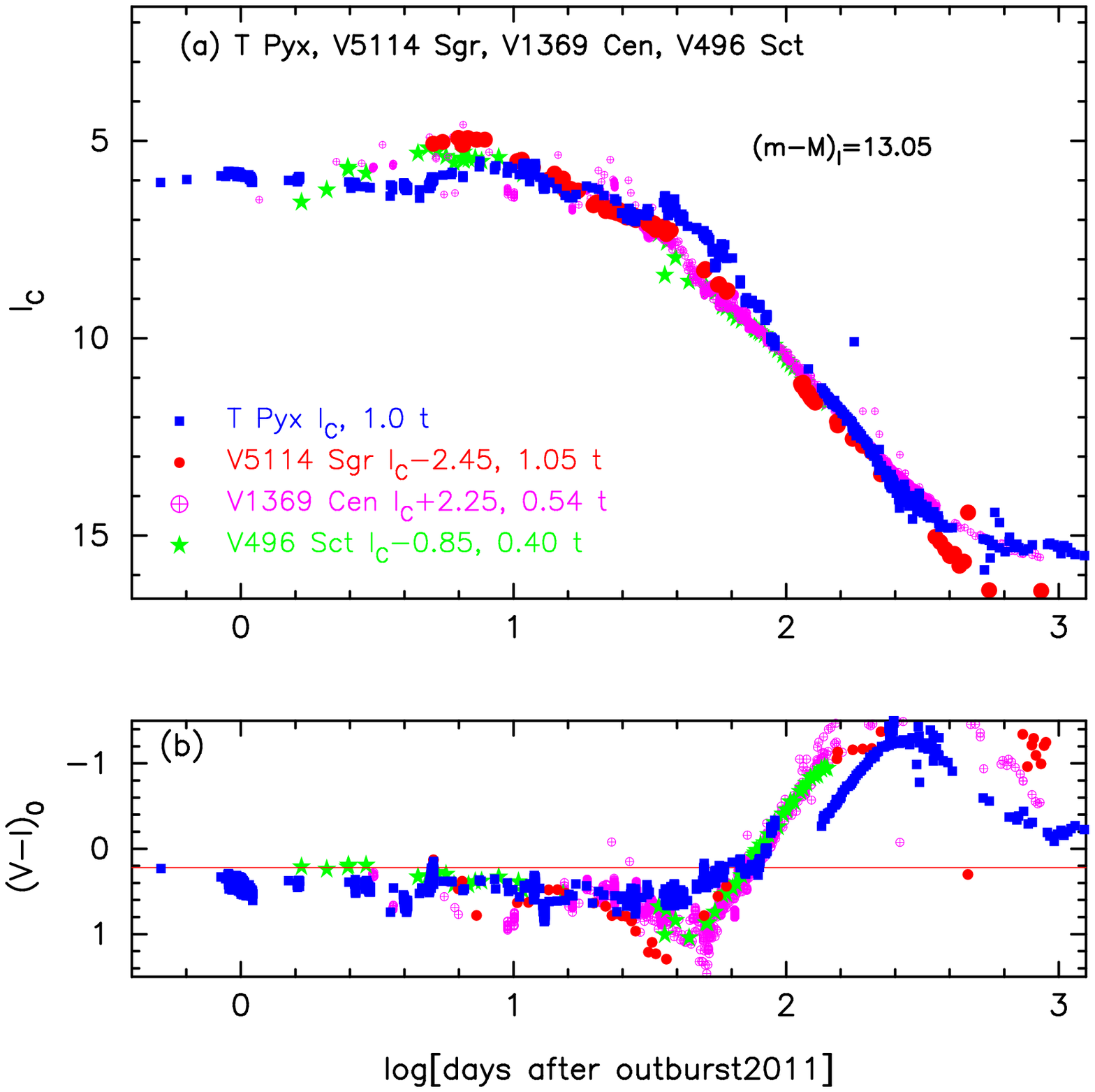}
\caption{
The (a) $I_{\rm C}$ light curve and (b) $(V-I_{\rm C})_0$ color curve
of T~Pyx (2011) as well as those of V5114~Sgr, V1369~Cen, and V496~Sct.
\label{t_pyx_v5114_sgr_v1369_cen_v496_sct_i_vi_color_logscale}}
\end{figure}


\begin{figure}
\plotone{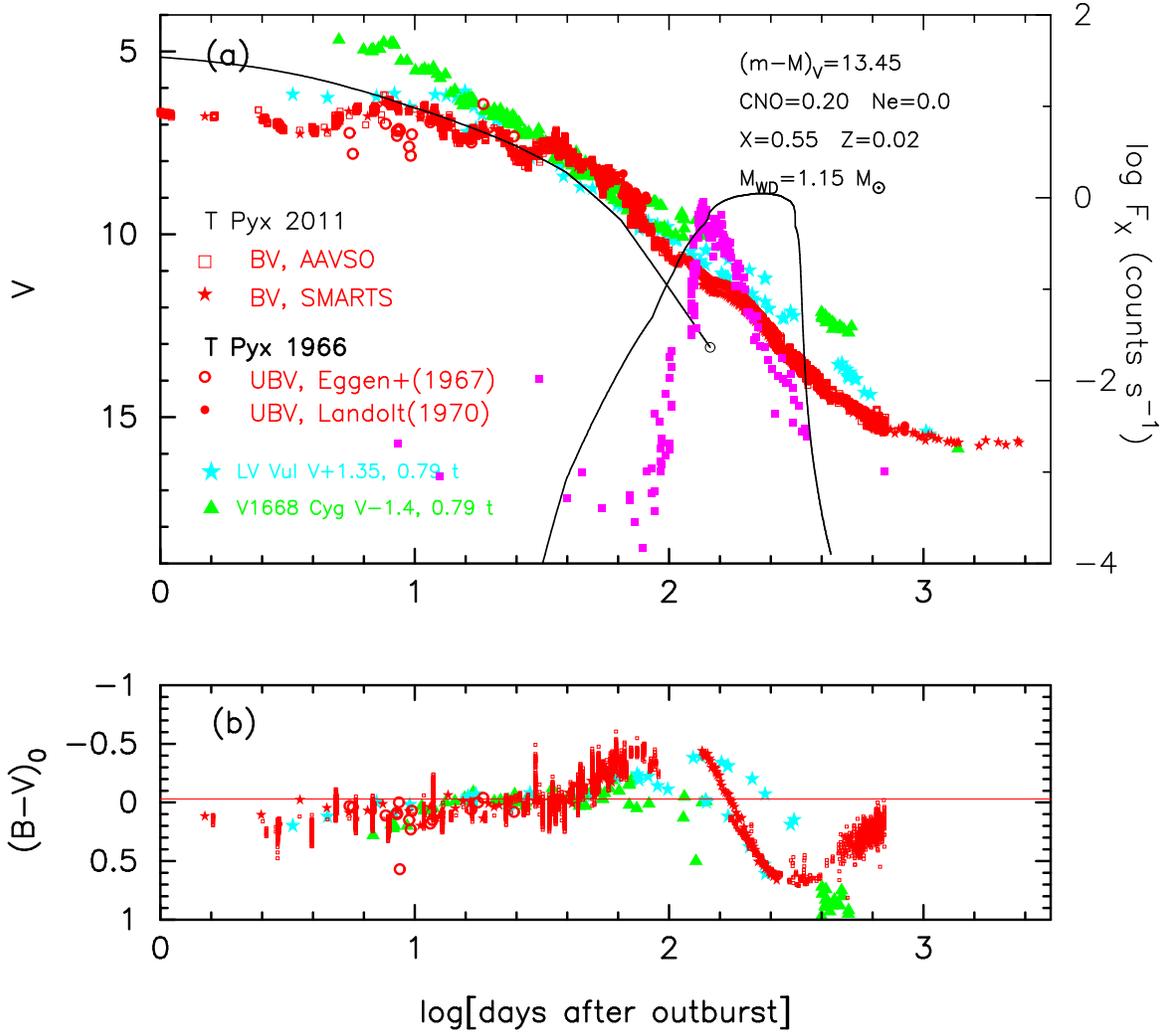}
\caption{
The (a) $V$ light and (b) $(B-V)_0$ color curves of the T~Pyx 2011 outburst
as well as LV~Vul and V1668~Cyg.  We add the data of the T~Pyx 1966 outburst.
In panel (a), we add a $1.15~M_\sun$ WD model (CO4, thin solid black lines)
for T~Pyx.  The filled magenta squares depict the soft X-ray count rate
taken from the {\it Swift} website \citep{eva09}.
\label{t_pyx_lv_vul_v1668_cyg_v_bv_ub_color_logscale_no6}}
\end{figure}


\begin{figure}
\plotone{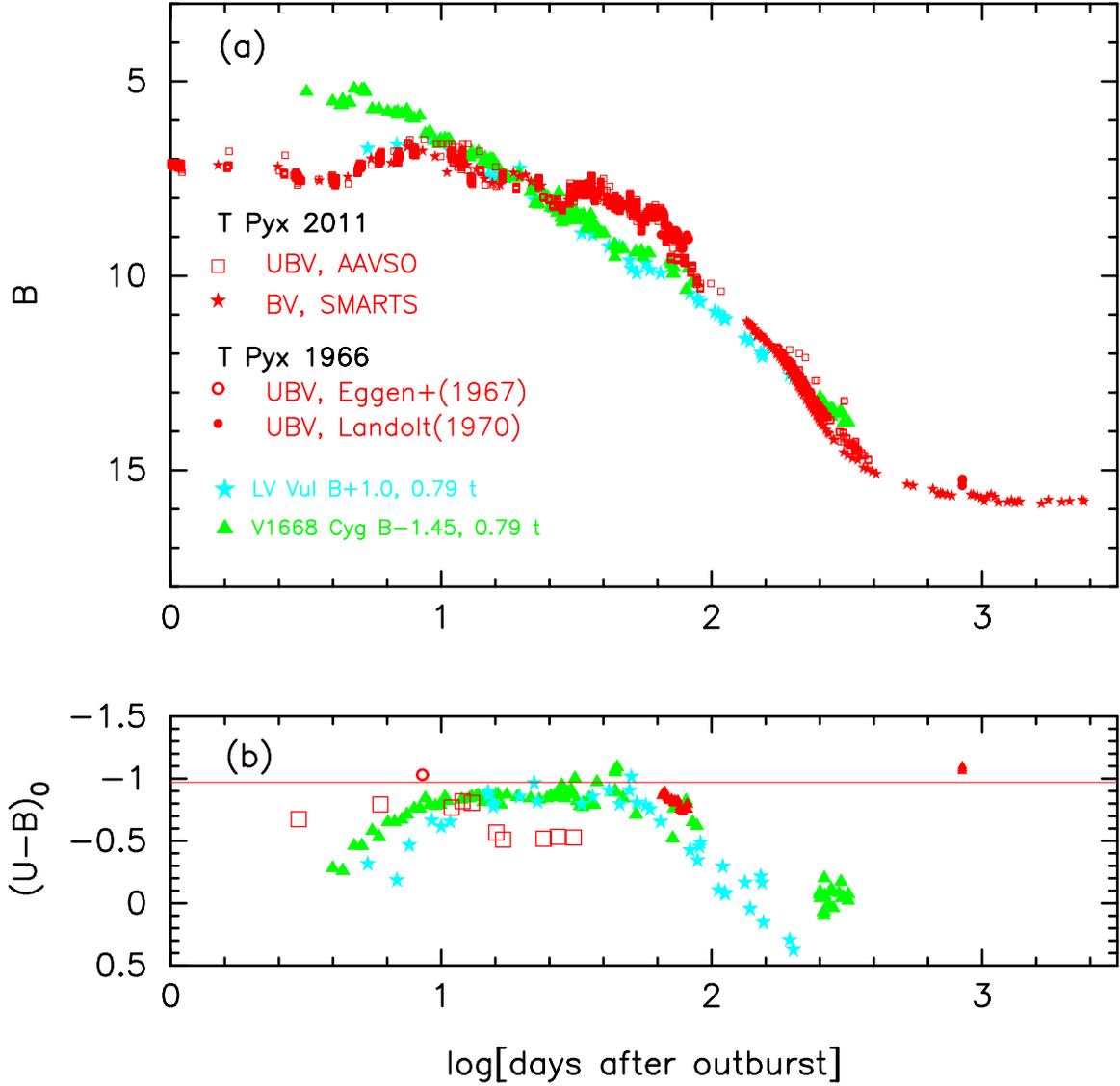}
\caption{
The (a) $B$ light curve and (b) $(U-B)_0$ color curve of the T~Pyx 2011
outburst as well as LV~Vul and V1668~Cyg.
We also add the data of the T~Pyx 1966 outburst.
\label{t_pyx_lv_vul_v1668_cyg_b_ub_color_logscale_no6}}
\end{figure}


\begin{figure*}
\plottwo{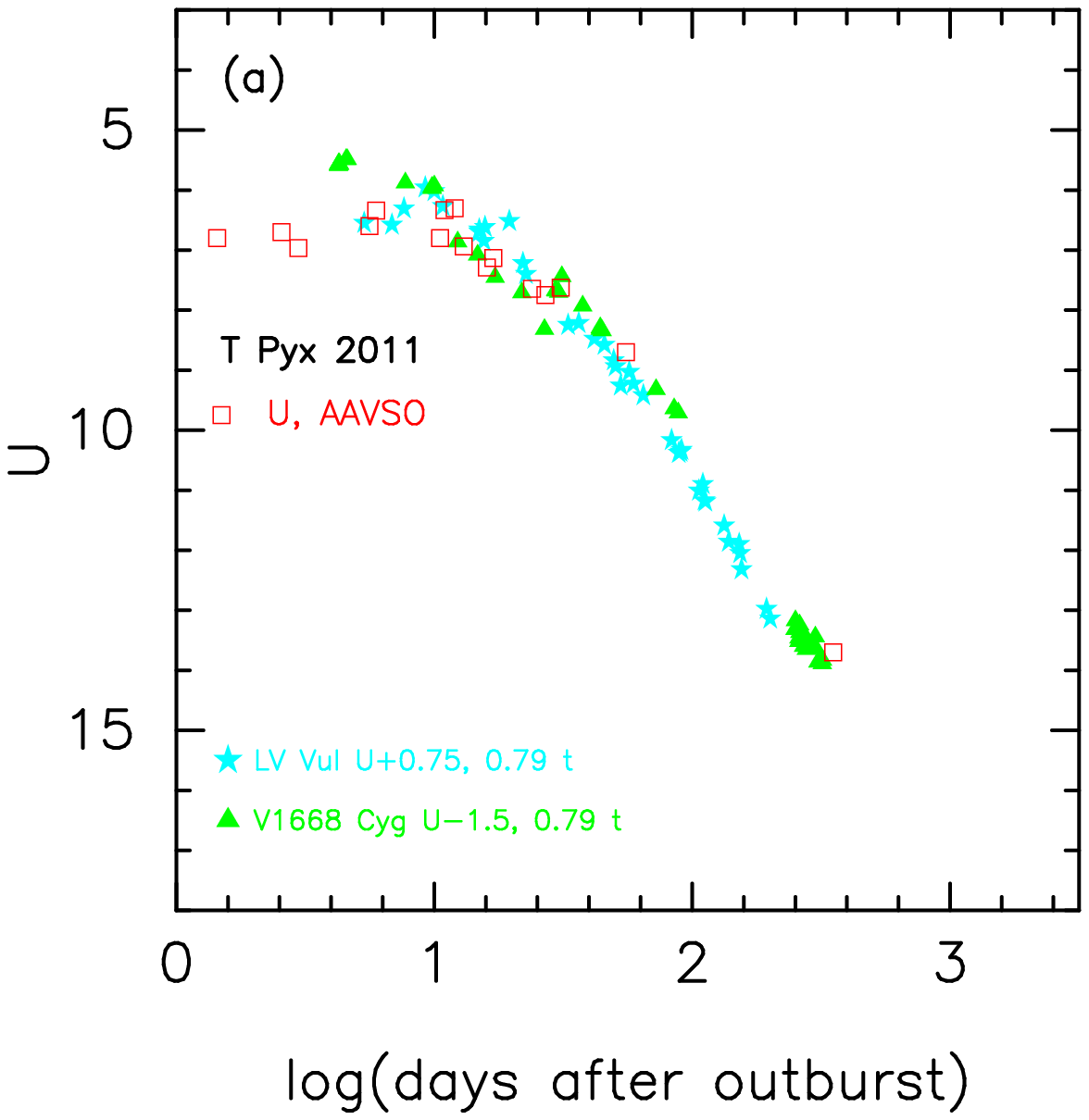}{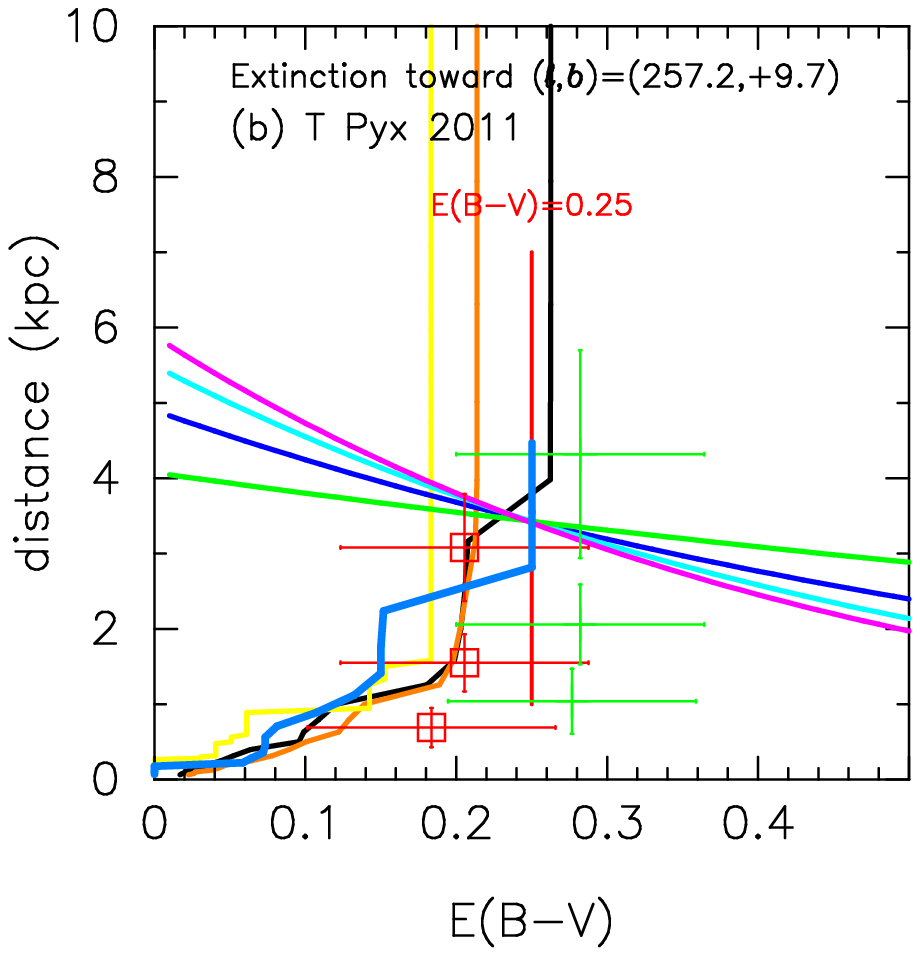}
\caption{
(a) The $U$ light curve of the T~Pyx 2011 outburst 
as well as LV~Vul and V1668~Cyg.
(b) Various distance-reddening relations toward T~Pyx.
The four thin lines of magenta, cyan, blue, and green
denote the distance-reddening relations given by $(m-M)_U=13.85$,
$(m-M)_B= 13.70$, $(m-M)_V= 13.45$, and $(m-M)_I= 13.05$, respectively.  
\label{distance_reddening_t_pyx_ubvi_xxxxxx}}
\end{figure*}

\subsection{T~Pyx 2011}
\label{t_pyx_ubvi}
We have reanalyzed the $UBVI_{\rm C}$ multi-band 
light/color curves of T~Pyx based on the time-stretching method.  
Figure \ref{t_pyx_v5114_sgr_v1369_cen_v496_sct_i_vi_color_logscale}
shows the (a) $I_{\rm C}$ light and (b) $(V-I_{\rm C})_0$ color curves
of the T~Pyx 2011 outburst as well as V5114~Sgr, V1369~Cen, and V496~Sct.
The $UBVI_{\rm C}$ data of T~Pyx are taken from AAVSO, VSOLJ, and SMARTS.
We adopt the color excess of $E(B-V)= 0.25$ in order to overlap
the $(V-I)_0$ color curve of T~Pyx with the other novae, as shown in
Figure \ref{t_pyx_v5114_sgr_v1369_cen_v496_sct_i_vi_color_logscale}(b).
We apply Equation (8) of \citet{hac19ka} for the $I$ band to Figure
\ref{t_pyx_v5114_sgr_v1369_cen_v496_sct_i_vi_color_logscale}(a)
and obtain
\begin{eqnarray}
(m&-&M)_{I, \rm T~Pyx} \cr
&=& ((m - M)_I + \Delta I_{\rm C})
_{\rm V5114~Sgr} - 2.5 \log 1.05 \cr
&=& 15.55 - 2.45\pm0.2 - 0.05 = 13.05\pm0.2 \cr
&=& ((m - M)_I + \Delta I_{\rm C})
_{\rm V1369~Cen} - 2.5 \log 0.54 \cr
&=& 10.11 + 2.25\pm0.2 + 0.675 = 13.04\pm0.2 \cr
&=& ((m - M)_I + \Delta I_{\rm C})
_{\rm V496~Sct} - 2.5 \log 0.40 \cr
&=& 12.9 - 0.85\pm0.2 + 1.0 = 13.05\pm0.2,
\label{distance_modulus_i_vi_t_pyx}
\end{eqnarray}
where we adopt
$(m-M)_{I, \rm V5114~Sgr}=15.55$ from Appendix \ref{v5114_sgr_ubvi},
$(m-M)_{I, \rm V1369~Cen}=10.11$ from \citet{hac19ka}, and
$(m-M)_{I, \rm V496~Sct}=12.9$ in Appendix \ref{v496_sct_bvi}.
Thus, we obtain $(m-M)_{I, \rm T~Pyx}= 13.05\pm0.2$.

We plot the (a) $V$ light and (b) $(B-V)_0$ color curves of three novae,
T~Pyx, LV~Vul, and V1668~Cyg, in Figure
\ref{t_pyx_lv_vul_v1668_cyg_v_bv_ub_color_logscale_no6},
to obtain the distance modulus of T~Pyx.  Applying Equation (4) of
\citet{hac19ka} to them, we have the relation
\begin{eqnarray}
(m-M)_{V, \rm T~Pyx}
&=& ((m - M)_V + \Delta V)_{\rm LV~Vul} - 2.5 \log 0.79 \cr
&=& 11.85 + 1.35\pm0.2 + 0.25 = 13.45\pm0.2 \cr
&=& ((m - M)_V + \Delta V)_{\rm V1668~Cyg} - 2.5 \log 0.79 \cr
&=& 14.6 - 1.4\pm0.2 + 0.25 = 13.45\pm0.2,
\label{distance_modulus_v_bv_t_pyx}
\end{eqnarray}
where we adopt $(m-M)_{V, \rm LV~Vul}=11.85$ and
$(m-M)_{V, \rm V1668~Cyg}=14.6$ both from \citet{hac19ka}.
Thus, we adopt $(m-M)_{V, \rm T~Pyx}=13.45\pm0.2$.


We plot the (a) $B$ light and (b) $(U-B)_0$ color curves of three novae,
T~Pyx, LV~Vul, and V1668~Cyg, in Figure
\ref{t_pyx_lv_vul_v1668_cyg_b_ub_color_logscale_no6}, 
to obtain the distance modulus in $B$ band of T~Pyx.  
Applying Equation (7) of \citet{hac19ka} to them, we have the relation
\begin{eqnarray}
(m-M)_{B, \rm T~Pyx}
&=& ((m - M)_B + \Delta B)_{\rm LV~Vul} - 2.5 \log 0.79 \cr
&=& 12.45 + 1.0\pm0.2 + 0.25 = 13.70\pm0.2 \cr
&=& ((m - M)_B + \Delta B)_{\rm V1668~Cyg} - 2.5 \log 0.79 \cr
&=& 14.9 - 1.45\pm0.2 + 0.25 = 13.70\pm0.2,
\label{distance_modulus_b_ub_t_pyx}
\end{eqnarray}
where we adopt $(m-M)_{B, \rm LV~Vul}=12.45$ and
$(m-M)_{B, \rm V1668~Cyg}=14.9$ both from \citet{hac19ka}.
Thus, we adopt $(m-M)_{B, \rm T~Pyx}=13.70\pm0.2$.

We further plot the $U$ light curves of three novae, T~Pyx, LV~Vul,
and V1668~Cyg in Figure \ref{distance_reddening_t_pyx_ubvi_xxxxxx}(a).
Applying Equation (6) of \citet{hac19ka} to them, we have the relation
\begin{eqnarray}
(m-M)_{U, \rm T~Pyx}
&=& (m - M + \Delta U)_{U, \rm LV~Vul} - 2.5 \log 0.79 \cr
&=& 12.85 + 0.75\pm0.2 + 0.25 = 13.85\pm0.2 \cr
&=& (m - M + \Delta U)_{U, \rm V1668~Cyg} - 2.5 \log 0.79 \cr
&=& 15.10 - 1.5\pm0.2 + 0.25 = 13.85\pm0.2,
\label{distance_modulus_u_t_pyx_lv_vul_v1668_cyg}
\end{eqnarray}
where we adopt $(m-M)_{U, \rm LV~Vul}=12.85$ and
$(m-M)_{U, \rm V1668~Cyg}=15.10$ both from \citet{hac19ka}.
Thus, we adopt $(m-M)_{U, \rm T~Pyx}=13.85\pm0.2$.

Figure \ref{distance_reddening_t_pyx_ubvi_xxxxxx}(b) shows
the distance-reddening relations toward T~Pyx.  
The four thin solid lines of $(m-M)_U= 13.85$, $(m-M)_B= 13.70$, 
$(m-M)_V= 13.45$, and $(m-M)_I= 13.05$ 
broadly cross at $d=3.4$~kpc and $E(B-V)=0.25$.
The crossing point is consistent with the distance-reddening relations
given by
\citet{mar06} 
and \citet[][cyan-blue lines]{chen19}.
Thus, we have $E(B-V)=0.25\pm0.05$ and $d=3.4\pm0.5$~kpc for T~Pyx.


\begin{figure}
\plotone{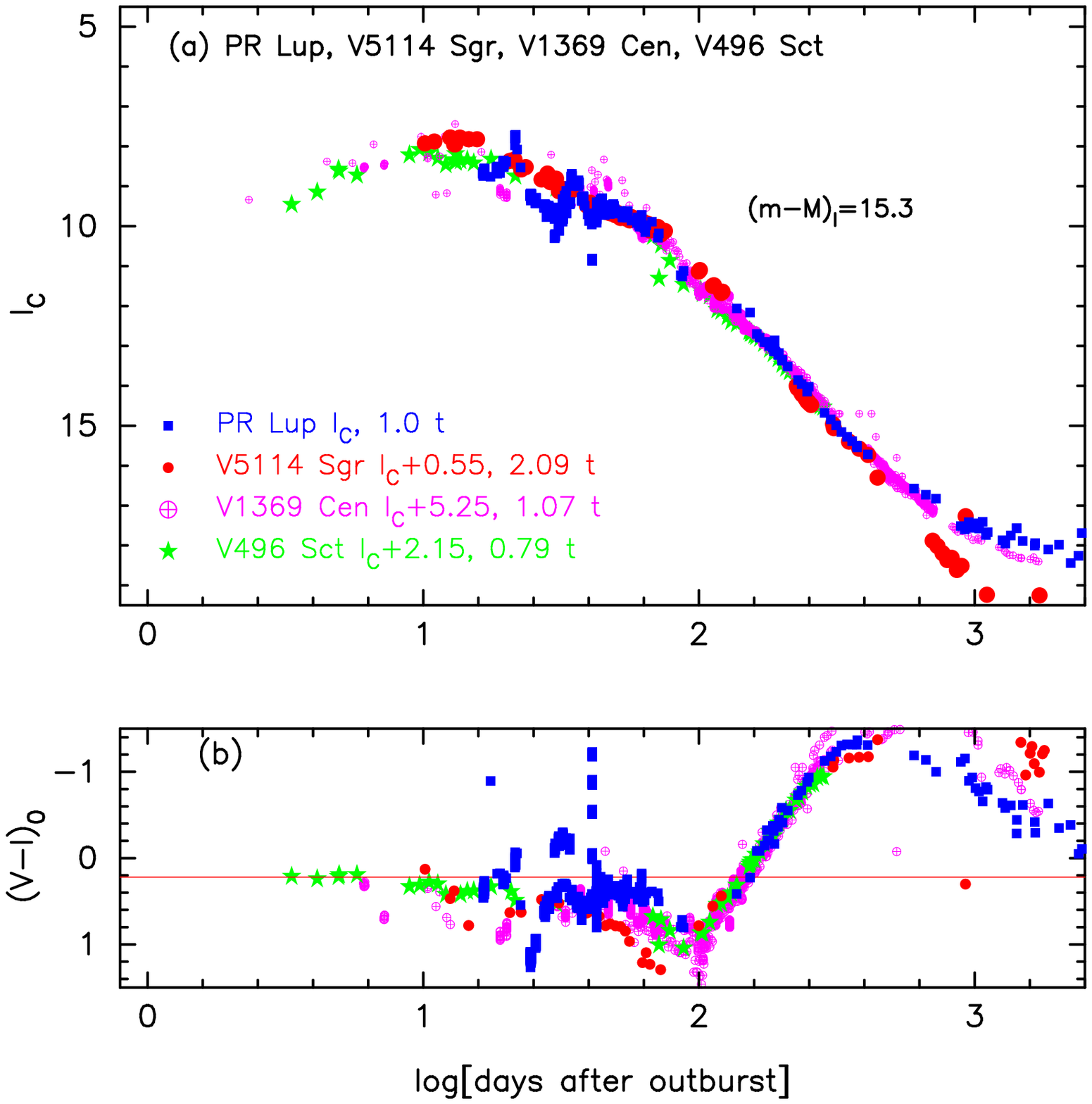}
\caption{
The (a) $I_{\rm C}$ light curve and (b) $(V-I_{\rm C})_0$ color curve
of PR~Lup as well as those of V5114~Sgr, V1369~Cen, and V496~Sct.
\label{pr_lup_v5114_sgr_v1369_cen_v496_sct_i_vi_color_logscale}}
\end{figure}


\begin{figure}
\plotone{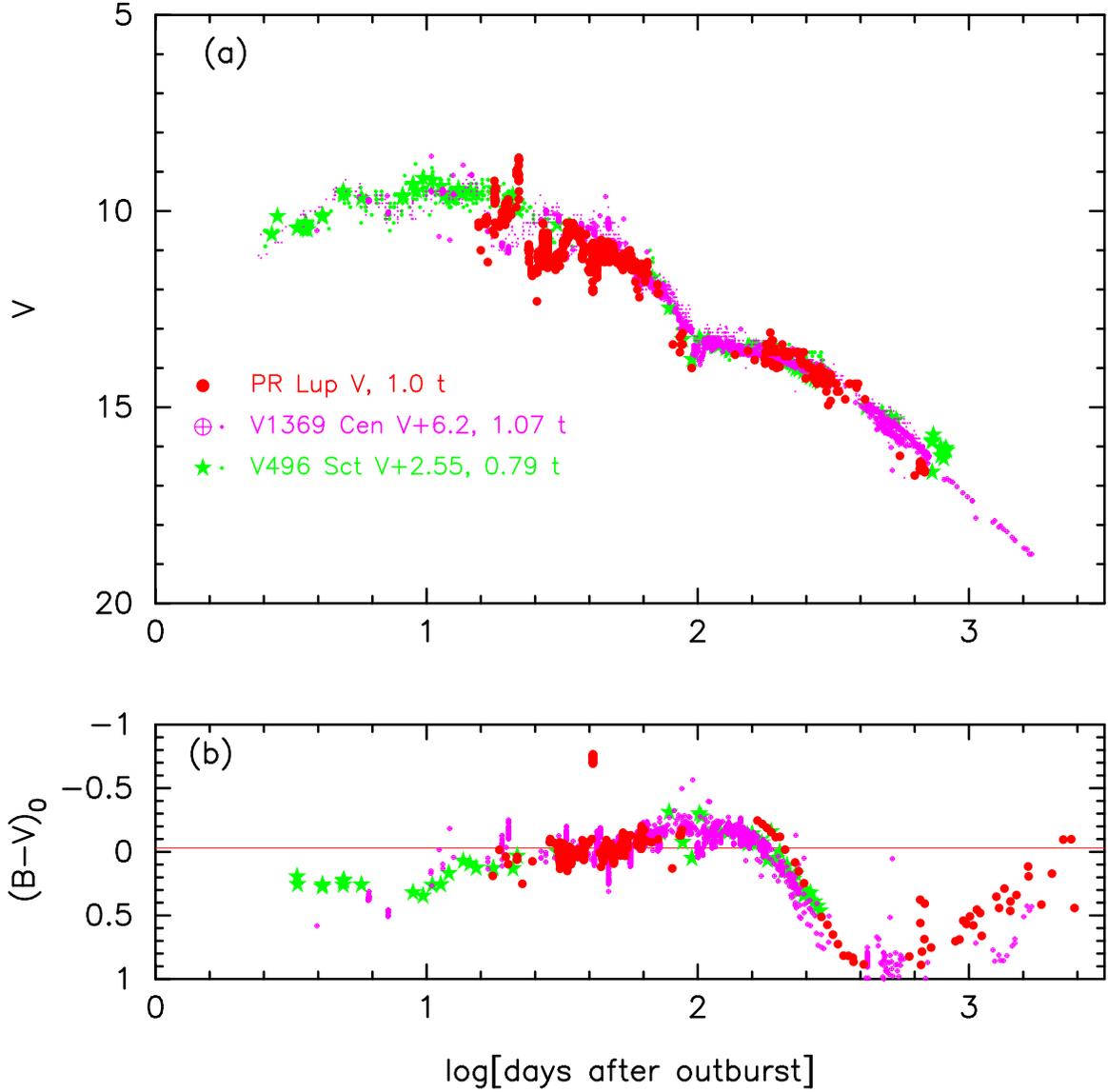}
\caption{
The (a) $V$ light curve and (b) $(B-V)_0$ color curve of
PR~Lup as well as those of V1369~Cen and V496~Sct.
The data of PR~Lup are taken from AAVSO, VSOLJ, and SMARTS.
\label{pr_lup_v1369_cen_v496_sct_v_bv_color_logscale_no2}}
\end{figure}


\begin{figure*}
\plottwo{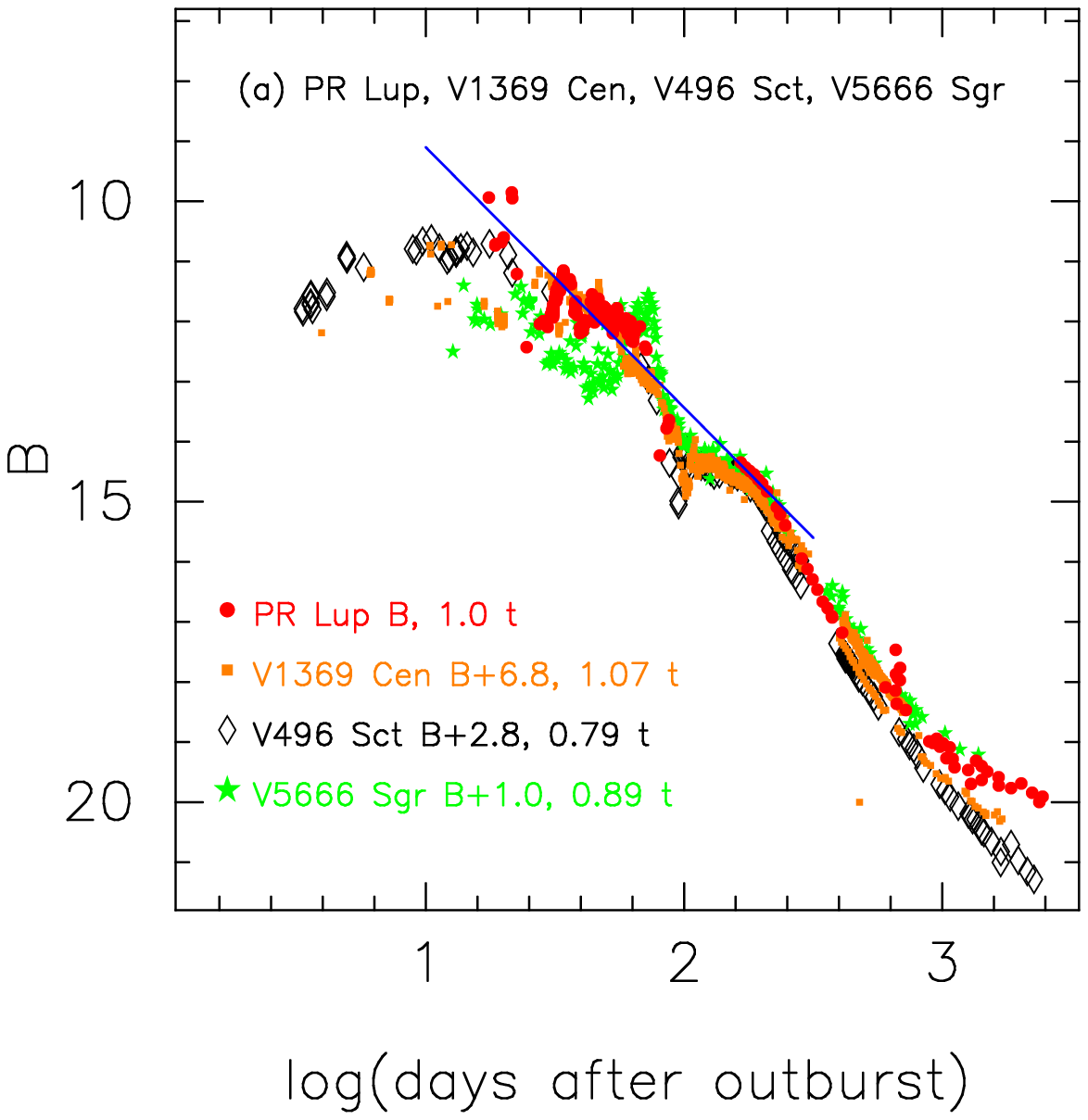}{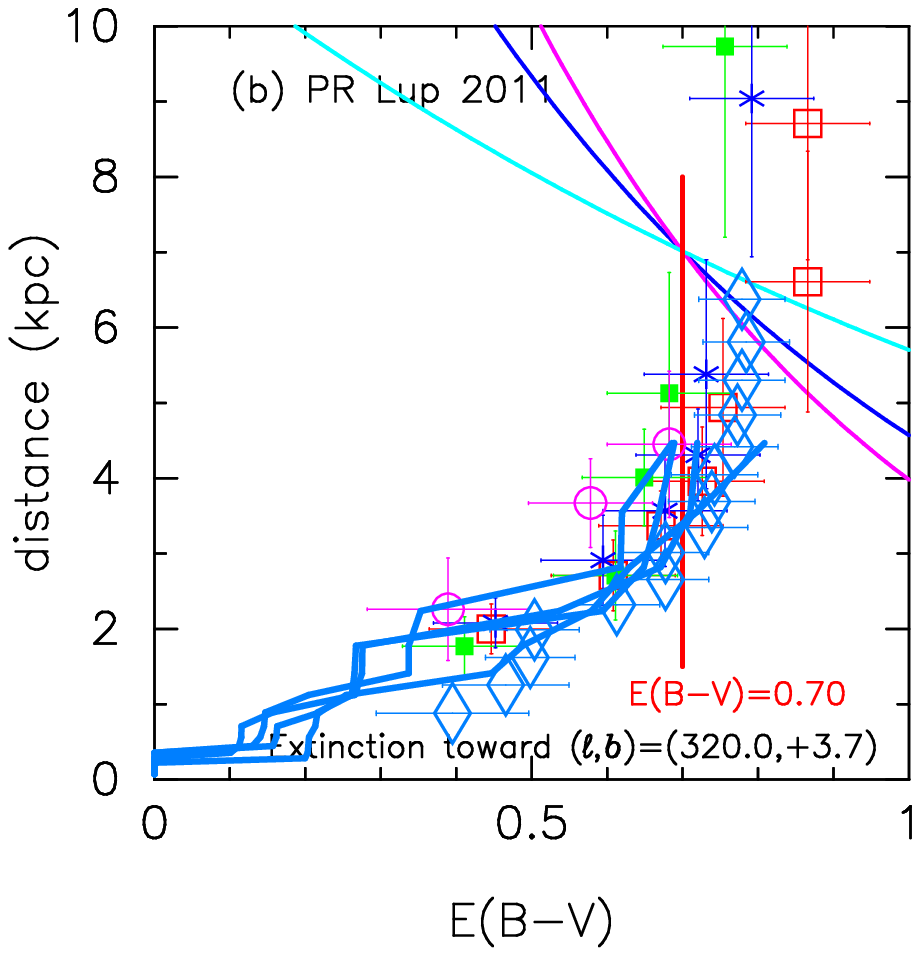}
\caption{
(a) The $B$ light curve of PR~Lup
as well as those of V1369~Cen, V496~Sct, and V5666~Sgr.
The $BVI_{\rm C}$ data of PR~Lup
are taken from AAVSO, VSOLJ, and SMARTS.
(b) Various distance-reddening relations toward PR~Lup.
The thin solid lines of magenta, blue, and cyan denote the distance-reddening
relations given by  $(m-M)_B= 17.1$, $(m-M)_V= 16.4$, and $(m-M)_I= 15.3$,
respectively.  
\label{distance_reddening_pr_lup_bvi_xxxxxx}}
\end{figure*}

\subsection{PR~Lup 2011}
\label{pr_lup_bvi}
We have reanalyzed the $BVI_{\rm C}$ multi-band 
light/color curves of PR~Lup based on the time-stretching method.  
Figure \ref{pr_lup_v5114_sgr_v1369_cen_v496_sct_i_vi_color_logscale}
shows the (a) $I_{\rm C}$ light and (b) $(V-I_{\rm C})_0$ color curves
of PR~Lup as well as V5114~Sgr, V1369~Cen, and V496~Sct.
The $BVI_{\rm C}$ data of PR~Lup are taken from AAVSO, VSOLJ, and SMARTS.
Here we assume that PR~Lup outbursted on JD~2,455,766.0 (Day 0).
We adopt the color excess of $E(B-V)= 0.70$ as mentioned below.
We apply Equation (8) of \citet{hac19ka} for the $I$ band to Figure
\ref{pr_lup_v5114_sgr_v1369_cen_v496_sct_i_vi_color_logscale}(a)
and obtain
\begin{eqnarray}
(m&-&M)_{I, \rm PR~Lup} \cr
&=& ((m - M)_I + \Delta I_{\rm C})
_{\rm V5114~Sgr} - 2.5 \log 2.09 \cr
&=& 15.55 + 0.55\pm0.2 - 0.8 = 15.3\pm0.2 \cr
&=& ((m - M)_I + \Delta I_{\rm C})
_{\rm V1369~Cen} - 2.5 \log 1.07 \cr
&=& 10.11 + 5.25\pm0.2 - 0.075 = 15.28\pm0.2 \cr
&=& ((m - M)_I + \Delta I_{\rm C})
_{\rm V496~Sct} - 2.5 \log 0.79 \cr
&=& 12.9 + 2.15\pm0.2 + 0.25 = 15.3\pm0.2,
\label{distance_modulus_i_vi_pr_lup}
\end{eqnarray}
where we adopt
$(m-M)_{I, \rm V5114~Sgr}=15.55$ from Appendix \ref{v5114_sgr_ubvi},
$(m-M)_{I, \rm V1369~Cen}=10.11$ from \citet{hac19ka}, and
$(m-M)_{I, \rm V496~Sct}=12.9$ in Appendix \ref{v496_sct_bvi}.
Thus, we obtain $(m-M)_{I, \rm PR~Lup}= 15.3\pm0.2$.

Figure \ref{pr_lup_v1369_cen_v496_sct_v_bv_color_logscale_no2} shows
the (a) $V$ light and (b) $(B-V)_0$ color curves of PR~Lup, V1369~Cen,
and V496~Sct.  These $V$ light and $(B-V)_0$ color curves overlap
each other.  Applying Equation (4) of \citet{hac19ka} for the $V$ band 
to them, we have the relation
\begin{eqnarray}
(m&-&M)_{V, \rm PR~Lup} \cr
&=& (m-M + \Delta V)_{V, \rm V1369~Cen} - 2.5 \log 1.07 \cr
&=& 10.25 + 6.2\pm0.2 - 0.075 = 16.38\pm0.2 \cr
&=& (m-M + \Delta V)_{V, \rm V496~Sct} - 2.5 \log 0.79 \cr
&=& 13.6 + 2.55\pm0.2 + 0.25 = 16.4\pm0.2,
\label{distance_modulus_pr_lup}
\end{eqnarray}
where we adopt
$(m-M)_{V, \rm V1369~Cen}=10.25$ from \citet{hac19ka}, and
$(m-M)_{V, \rm V496~Sct}=13.6$ in Appendix \ref{v496_sct_bvi}.
Thus, we obtain $(m-M)_{V, \rm PR~Lup}=16.4\pm0.1$.

Figure \ref{distance_reddening_pr_lup_bvi_xxxxxx}(a)
shows the $B$ light curves of PR~Lup
together with those of V1369~Cen, V496~Sct, and V5666~Sgr.
Applying Equation (7) of \citet{hac19ka} for the $B$ band to Figure
\ref{distance_reddening_pr_lup_bvi_xxxxxx}(a), we have the relation
\begin{eqnarray}
(m&-&M)_{B, \rm PR~Lup} \cr
&=& \left( (m-M)_B + \Delta B\right)_{\rm V1369~Cen} - 2.5 \log 1.07 \cr
&=& 10.36 + 6.8\pm0.2 - 0.075 = 17.09\pm0.2 \cr
&=& \left( (m-M)_B + \Delta B\right)_{\rm V496~Sct} - 2.5 \log 0.79 \cr
&=& 14.05 + 2.8\pm0.2 + 0.25 = 17.1\pm0.2 \cr
&=& \left( (m-M)_B + \Delta B\right)_{\rm V5666~Sgr} - 2.5 \log 0.89 \cr
&=& 16.0 + 1.0\pm0.2 + 0.125 = 17.12\pm0.2,
\label{distance_modulus_pr_lup_v1369_cen_v496_sct_v5666_sgr_b}
\end{eqnarray}
where we adopt $(m-M)_{B, \rm V1369~Cen}= 10.36$ in Appendix 
\ref{v1369_cen_bvi}, $(m-M)_{B, \rm V496~Sct}= 14.05$ in Appendix
\ref{v496_sct_bvi}, and $(m-M)_{B, \rm V5666~Sgr}= 16.0$ in Appendix
\ref{v5666_sgr_bvi}.
We have $(m-M)_{B, \rm PR~Lup}=17.1\pm0.1$.

We plot $(m-M)_B= 17.1$, $(m-M)_V= 16.4$, and $(m-M)_I= 15.32$, 
which broadly cross at $d=7.0$~kpc and $E(B-V)=0.70$, in Figure
\ref{distance_reddening_pr_lup_bvi_xxxxxx}(b).
The crossing point is consistent with the distance-reddening relations
given by \citet[][filled green squares]{mar06} 
and \citet[][cyan-blue lines]{chen19}.
Thus, we have $E(B-V)=0.70\pm0.05$ and $d=7.0\pm0.7$~kpc for PR~Lup.


\begin{figure}
\plotone{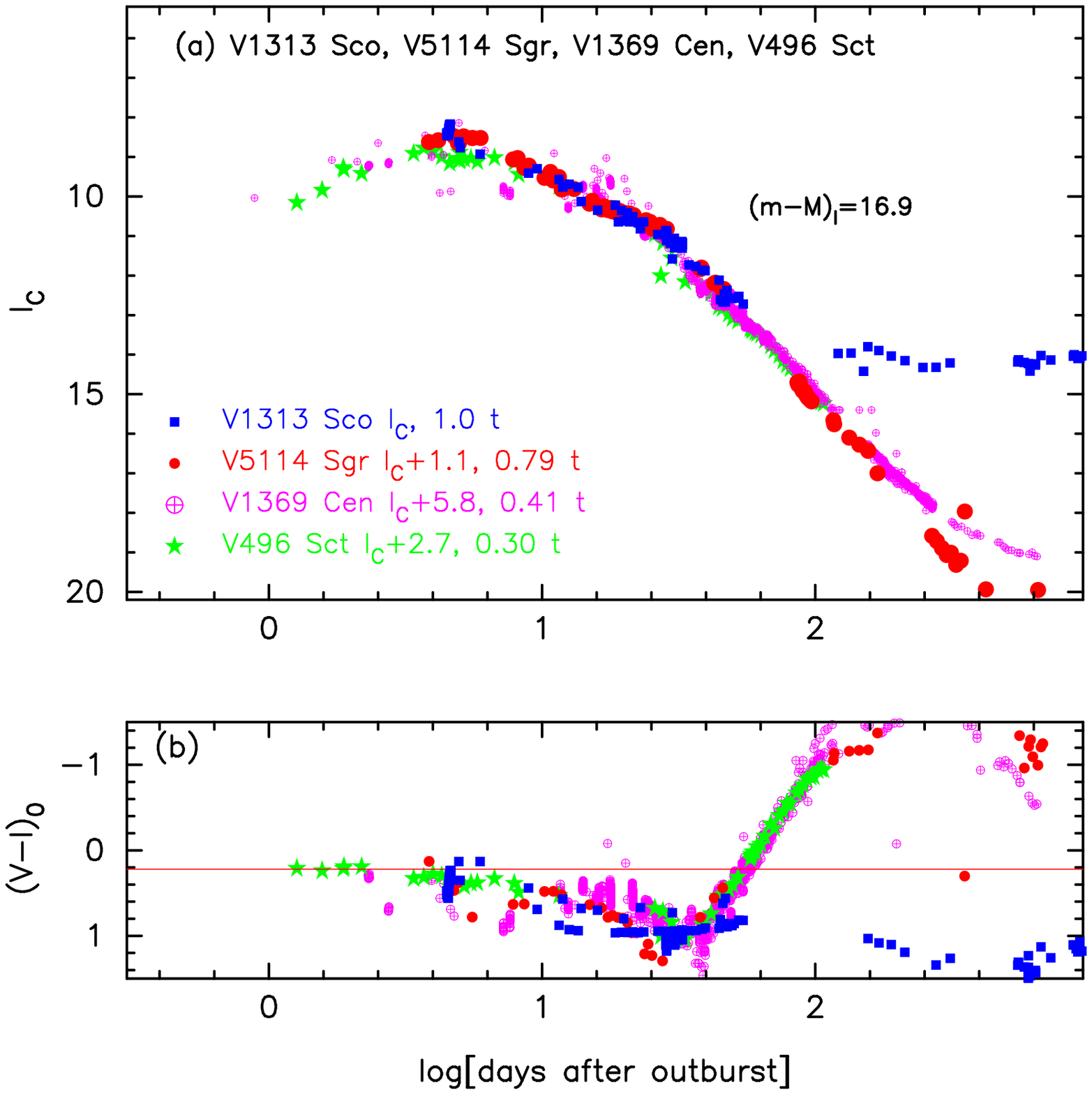}
\caption{
The (a) $I_{\rm C}$ light curve and (b) $(V-I_{\rm C})_0$ color curve
of V1313~Sco as well as those of V5114~Sgr, V1369~Cen, and V496~Sct.
\label{v1313_sco_v5114_sgr_v1369_cen_v496_sct_i_vi_color_logscale}}
\end{figure}


\begin{figure}
\plotone{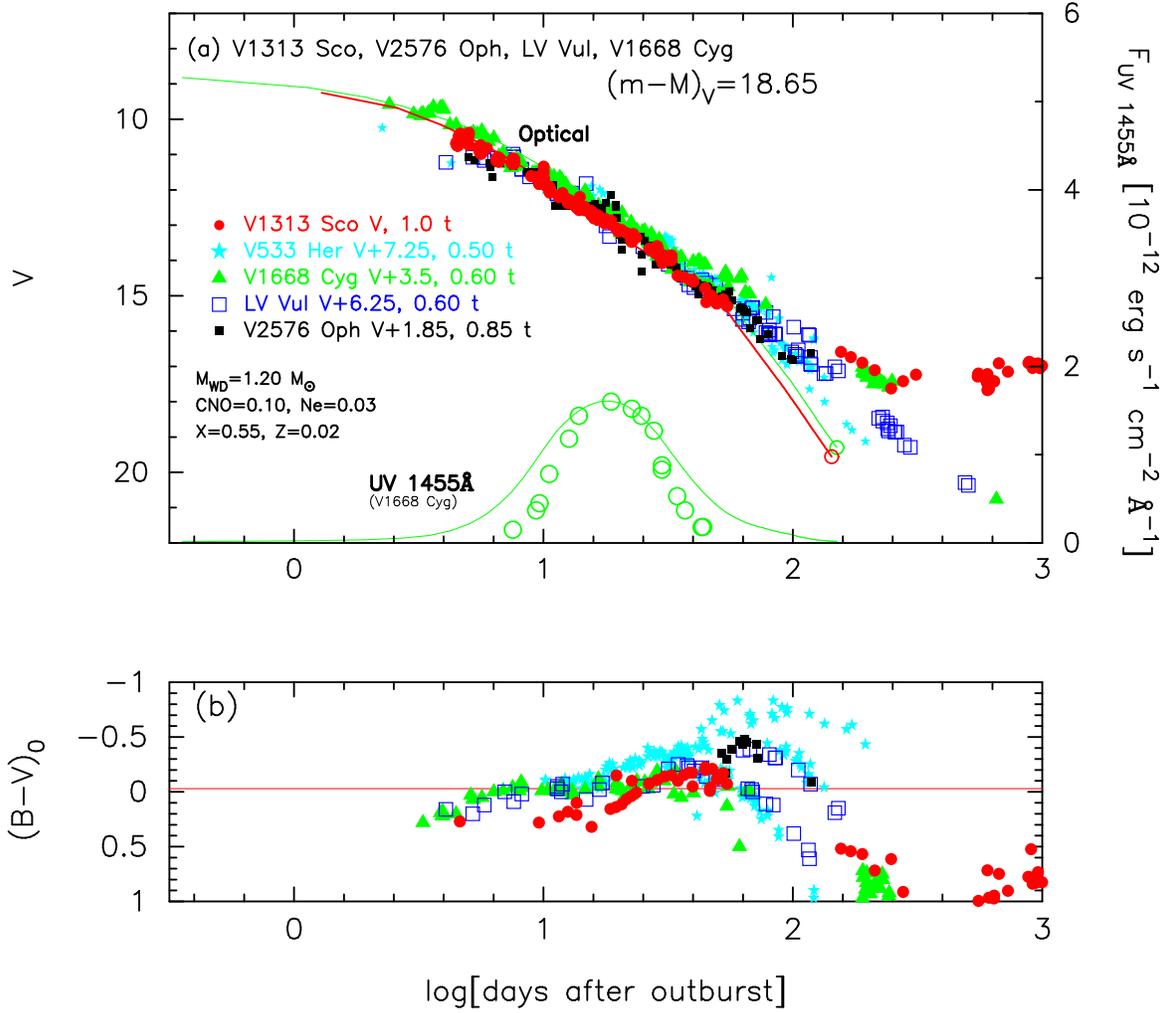}
\caption{
The (a) $V$ light curve and (b) $(B-V)_0$ color curve of
V1313~Sco as well as those of V533~Her, V1668~Cyg, LV~Vul, and V2576~Oph.
The data of V1313~Sco are taken from AAVSO, VSOLJ, and SMARTS.
\label{v1313_sco_v1312_sco_v2576_oph_v1668_cyg_lv_vul_v_bv_logscale}}
\end{figure}


\begin{figure*}
\plottwo{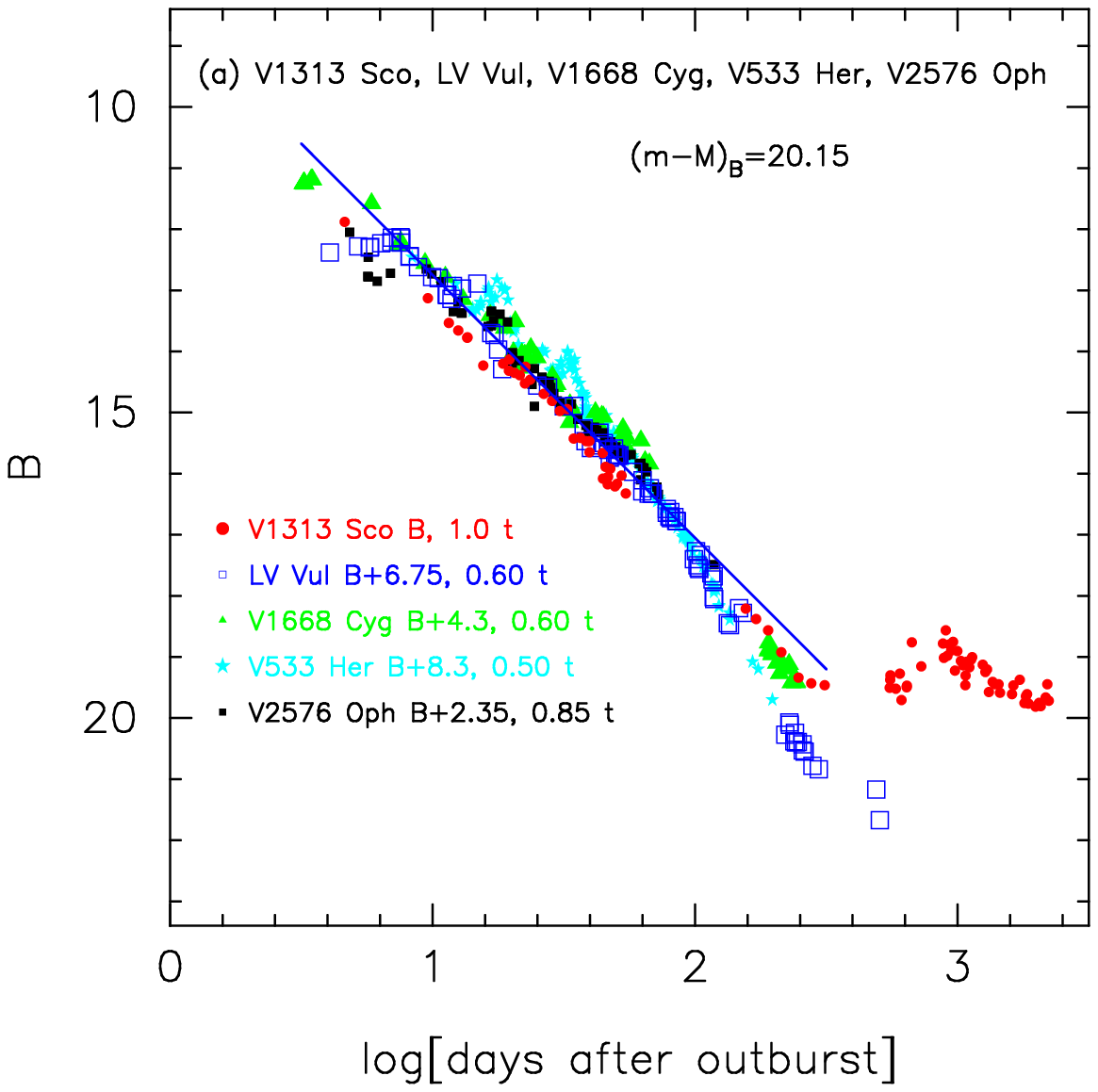}{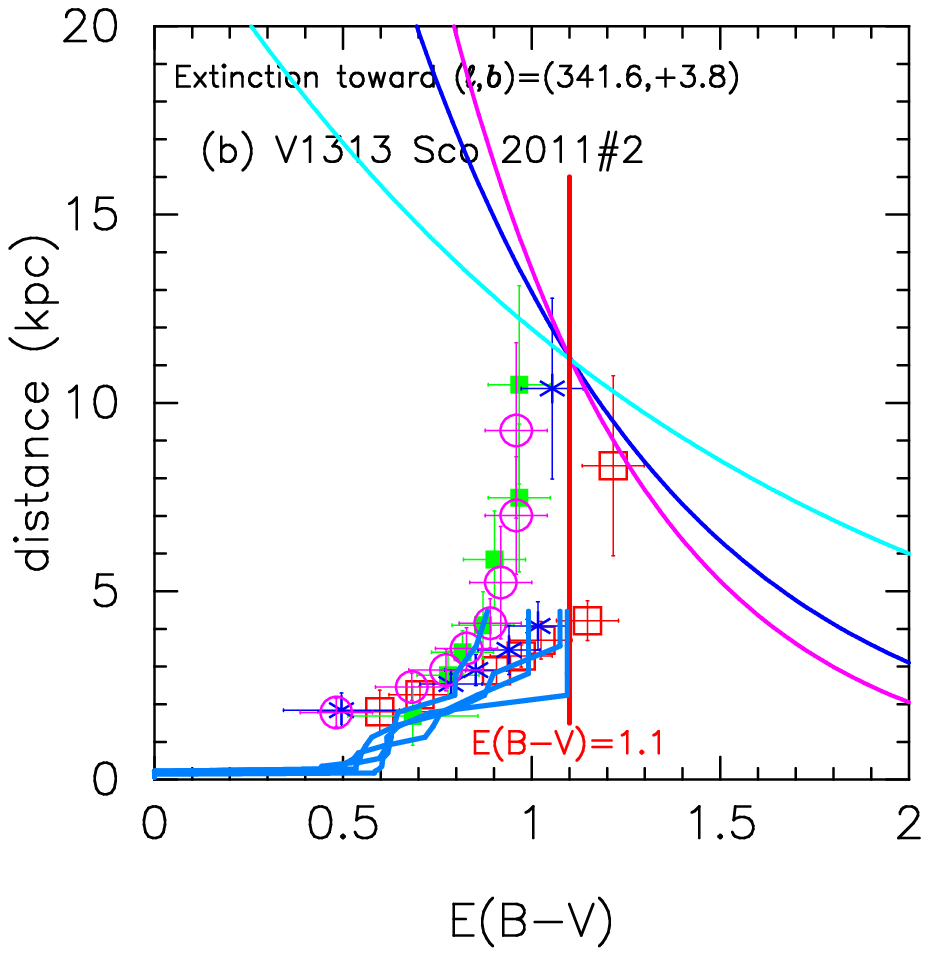}
\caption{
(a) The $B$ light curve of V1313~Sco
as well as those of LV~Vul, V1668~Cyg, V533~Her, and V2576~Oph.
The $BVI_{\rm C}$ data of PR~Lup
are taken from AAVSO, VSOLJ, and SMARTS.
(b) Various distance-reddening relations toward V1313~Sco.
The thin solid lines of magenta, blue, and cyan denote the distance-reddening
relations given by  $(m-M)_B=19.76$, $(m-M)_V=18.66$, 
and $(m-M)_I=16.89$, respectively.  
\label{distance_reddening_v1313_sco_bvi_xxxxxx}}
\end{figure*}

\subsection{V1313~Sco 2011\#2}
\label{v1313_sco_bvi}
We have reanalyzed the $BVI_{\rm C}$ multi-band 
light/color curves of V1313~Sco based on the time-stretching method.  
Figure \ref{v1313_sco_v5114_sgr_v1369_cen_v496_sct_i_vi_color_logscale}
shows the (a) $I_{\rm C}$ light and (b) $(V-I_{\rm C})_0$ color curves
of V1313~Sco as well as V5114~Sgr, V1369~Cen, and V496~Sct.
The $BVI_{\rm C}$ data of V1313~Sco are taken from AAVSO, VSOLJ, and SMARTS.
We adopt the color excess of $E(B-V)= 1.1$ as mentioned below.
We apply Equation (8) of \citet{hac19ka} for the $I$ band to Figure
\ref{v1313_sco_v5114_sgr_v1369_cen_v496_sct_i_vi_color_logscale}(a)
and obtain
\begin{eqnarray}
(m&-&M)_{I, \rm V1313~Sco} \cr
&=& ((m - M)_I + \Delta I_{\rm C})
_{\rm V5114~Sgr} - 2.5 \log 0.79 \cr
&=& 15.55 + 1.1\pm0.2 + 0.25 = 16.9\pm0.2 \cr
&=& ((m - M)_I + \Delta I_{\rm C})
_{\rm V1369~Cen} - 2.5 \log 0.41 \cr
&=& 10.11 + 5.8\pm0.2 + 0.975 = 16.88\pm0.2 \cr
&=& ((m - M)_I + \Delta I_{\rm C})
_{\rm V496~Sct} - 2.5 \log 0.30 \cr
&=& 12.9 + 2.7\pm0.2 + 1.3 = 16.9\pm0.2,
\label{distance_modulus_i_vi_v1313_sco}
\end{eqnarray}
where we adopt
$(m-M)_{I, \rm V5114~Sgr}=15.55$ from Appendix \ref{v5114_sgr_ubvi},
$(m-M)_{I, \rm V1369~Cen}=10.11$ from \citet{hac19ka}, and
$(m-M)_{I, \rm V496~Sct}=12.9$ in Appendix \ref{v496_sct_bvi}.
Thus, we obtain $(m-M)_{I, \rm V1313~Sco}= 16.9\pm0.2$.

Figure \ref{v1313_sco_v1312_sco_v2576_oph_v1668_cyg_lv_vul_v_bv_logscale}
shows the light/color curves of V1313~Sco, LV~Vul, V1668~Cyg, V533~Her, and
V2576~Oph.  Here, $(B-V)_0$ are dereddened with $E(B-V)=1.1$.
We have the relation
\begin{eqnarray}
(m&-&M)_{V, \rm V1313~Sco} \cr
&=& (m - M + \Delta V)_{V, \rm LV~Vul} - 2.5 \log 0.60 \cr
&=& 11.85 + 6.25\pm0.2 + 0.55 = 18.65\pm0.2 \cr
&=& (m - M + \Delta V)_{V, \rm V1668~Cyg} - 2.5 \log 0.60 \cr
&=& 14.6 + 3.5\pm0.2 + 0.55 = 18.65\pm0.2 \cr
&=& (m - M + \Delta V)_{V, \rm V533~Her} - 2.5 \log 0.50 \cr
&=& 10.65 + 7.25\pm0.2 + 0.75 = 18.65\pm0.2 \cr
&=& (m - M + \Delta V)_{V, \rm V2576~Oph} - 2.5 \log 0.85 \cr
&=& 16.65 + 1.85\pm0.2 + 0.18 = 18.68\pm0.2,
\label{distance_modulus_v1313_sco}
\end{eqnarray}
where we adopt $(m-M)_{V, \rm LV~Vul}=11.85$, $(m-M)_{V, \rm V1668~Cyg}=14.6$, 
and $(m-M)_{V, \rm V533~Her}=10.65$ from \citet{hac19ka},
and $(m-M)_{V, \rm V2576~Oph}=16.65$  from \citet{hac19kb}.
Thus, we obtain $(m-M)_V=18.66\pm0.1$ and 
$\log f_{\rm s}= \log 0.60 = -0.22$ against LV~Vul.

Figure \ref{distance_reddening_v1313_sco_bvi_xxxxxx}(a)
shows the $B$ light curves of V1313~Sco
together with those of LV~Vul, V1668~Cyg, V533~Her, and V2576~Oph.
We obtain
\begin{eqnarray}
(m&-&M)_{B, \rm V1313~Sco} \cr
&=& ((m - M)_B + \Delta B)_{\rm LV~Vul} - 2.5 \log 0.60 \cr
&=& 12.45 + 6.75\pm0.2 + 0.55 = 19.75\pm0.2 \cr
&=& ((m - M)_B + \Delta B)_{\rm V1668~Cyg} - 2.5 \log 0.60 \cr
&=& 14.9 + 4.3\pm0.2 + 0.55 = 19.75\pm0.2 \cr
&=& ((m - M)_B + \Delta B)_{\rm V533~Her} - 2.5 \log 0.50 \cr
&=& 10.69 + 8.3\pm0.2 + 0.75 = 19.74\pm0.2 \cr
&=& ((m - M)_B + \Delta B)_{\rm V2576~Oph} - 2.5 \log 0.85 \cr
&=& 17.25 + 2.35\pm0.2 + 0.18 = 19.78\pm0.2.
\label{distance_modulus_b_v1313_sco_lv_vul_v1668_cyg}
\end{eqnarray}
We have $(m-M)_{B, \rm V1313~Sco}= 19.76\pm0.2$.

We plot the distance-reddening relations of
$(m-M)_B=19.76$, $(m-M)_V=18.66$, and $(m-M)_I=16.89$ in Figure
\ref{distance_reddening_v1313_sco_bvi_xxxxxx}(b).
The three thin solid lines of magenta, blue, and cyan 
consistently cross at $d=11.2$~kpc and $E(B-V)=1.1$.  
This point is also consistent with the distance-reddening relation
given by \citet{mar06}.


\begin{figure}
\plotone{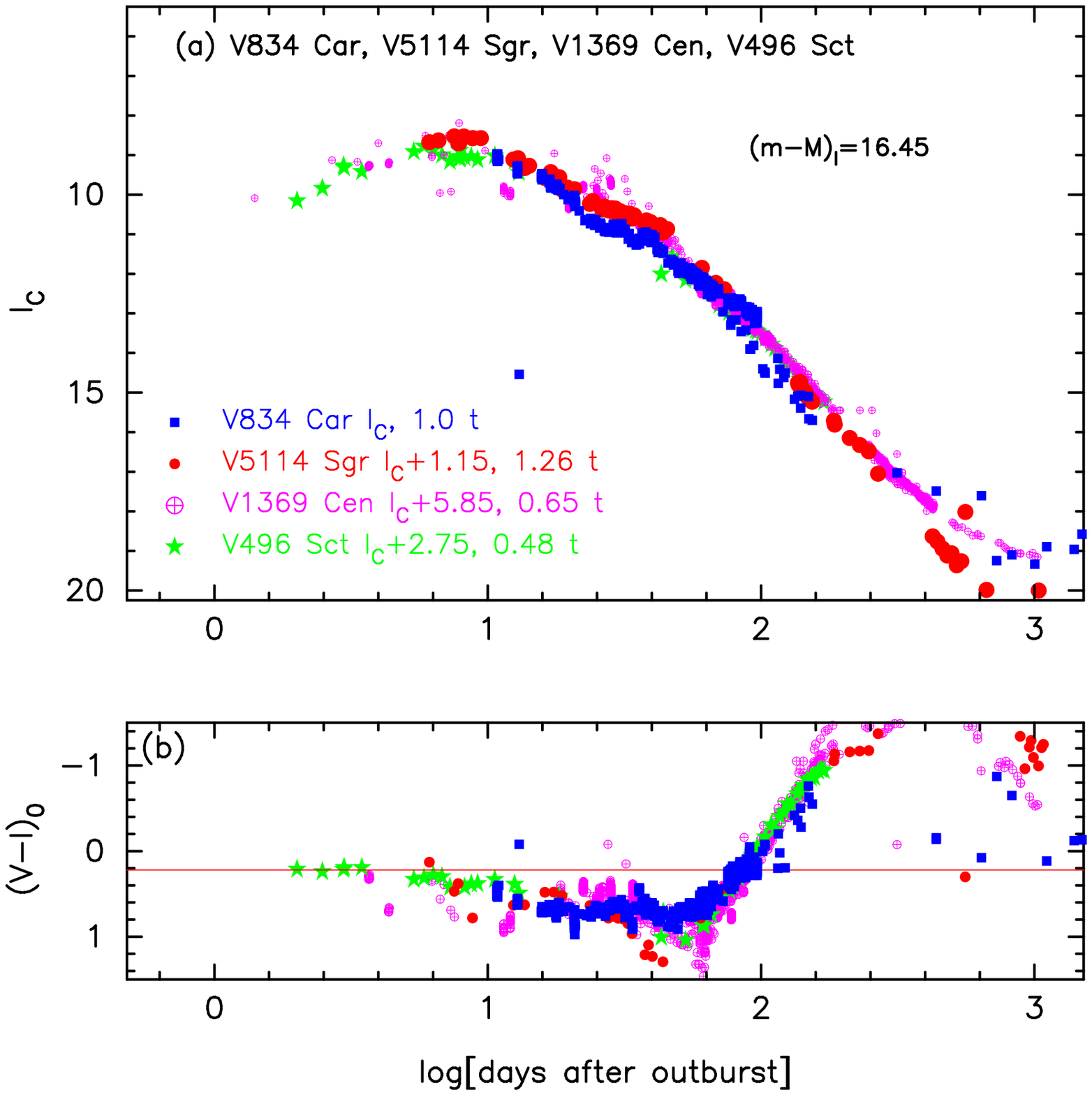}
\caption{
The (a) $I_{\rm C}$ light curve and (b) $(V-I_{\rm C})_0$ color curve
of V834~Car as well as those of V5114~Sgr, V1369~Cen, and V496~Sct.
\label{v834_car_v5114_sgr_v1369_cen_v496_sct_i_vi_color_logscale}}
\end{figure}


\begin{figure}
\plotone{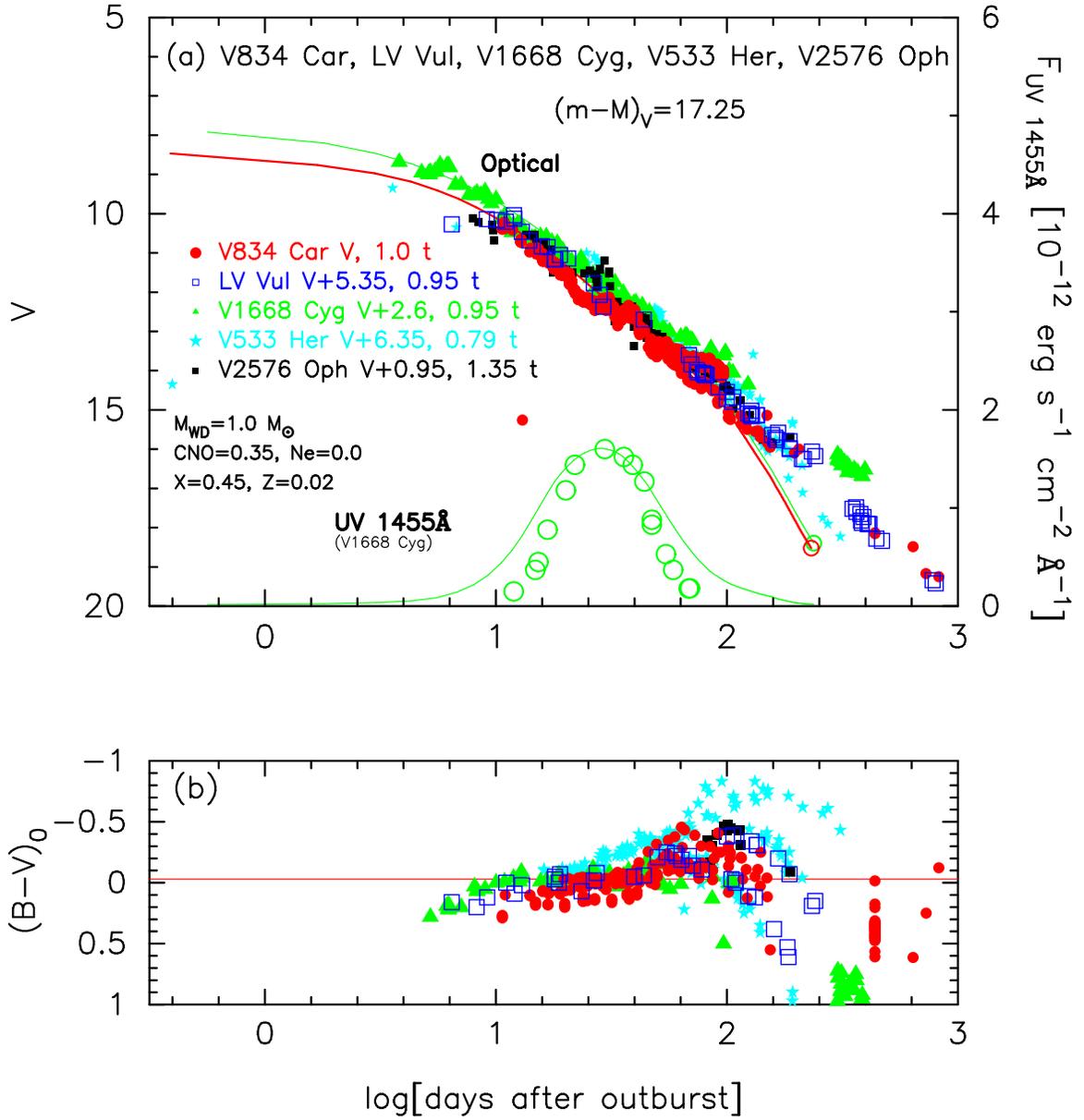}
\caption{
The (a) $V$ light curve and (b) $(B-V)_0$ color curve
of V834~Car as well as those of LV~Vul, V1668~Cyg, V533~Her, and V2576~Oph.
In panel (a), we add a $1.0~M_\sun$ WD model (CO3, solid red line) for V834~Car
as well as a $0.98~M_\sun$ WD model (CO3, solid green lines) for V1668~Cyg.
\label{v834_car_v2576_oph_v1668_cyg_lv_vul_v_bv_logscale_no2}}
\end{figure}


\begin{figure}
\epsscale{0.55}
\plotone{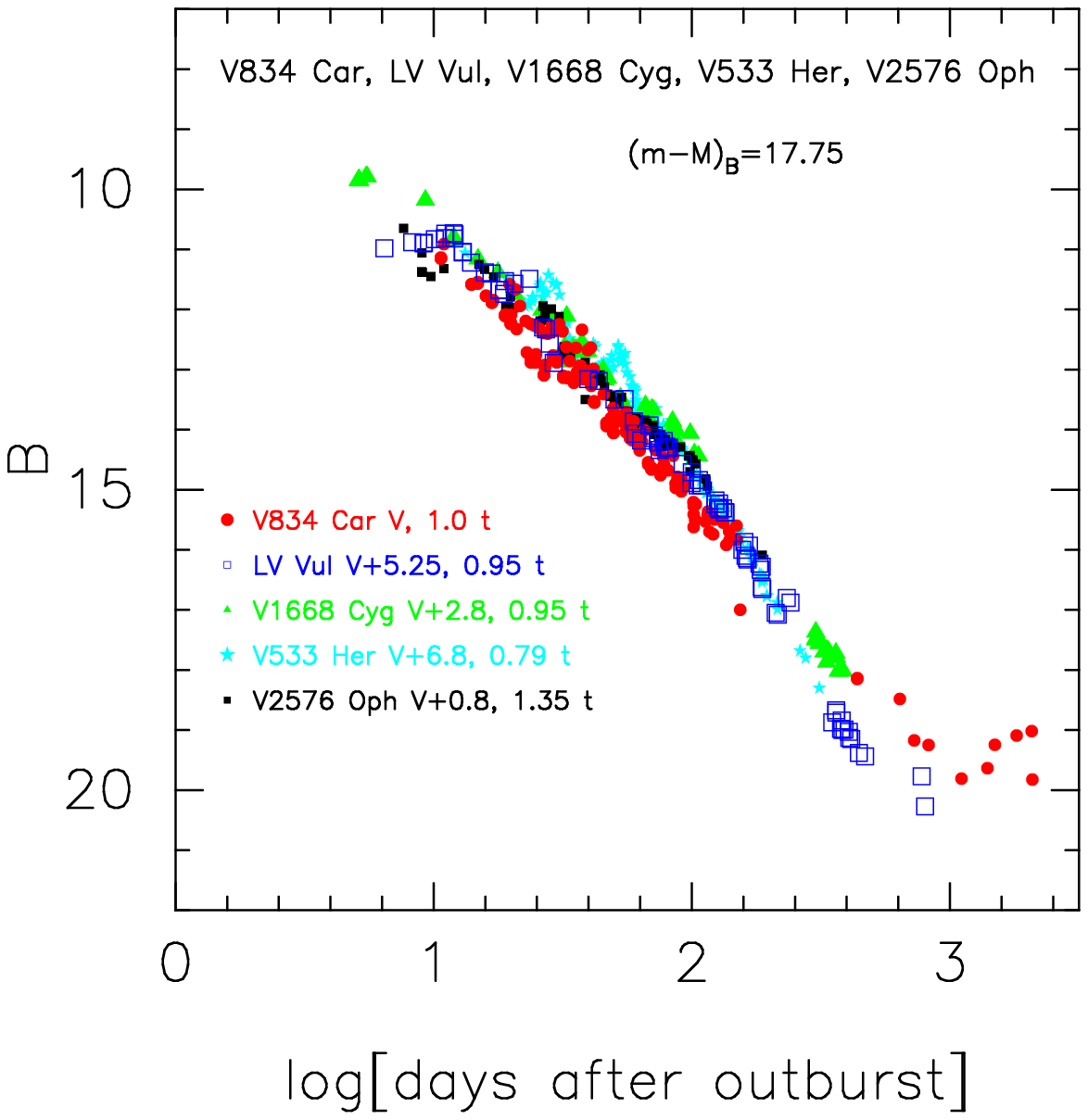}
\caption{
The $B$ light curve of V834~Car together with
LV~Vul, V1668~Cyg, V533~Her, and V2576~Oph. 
The $BVI_{\rm C}$ data of V834~Car are taken from AAVSO, VSOLJ, and SMARTS.
\label{v834_car_v2576_oph_v1668_cyg_lv_vul_b_only_logscale}}
\end{figure}

\subsection{V834~Car 2012}
\label{v834_car_bvi}
We have reanalyzed the $BVI_{\rm C}$ multi-band 
light/color curves of V834~Car based on the time-stretching method.  
Figure \ref{v834_car_v5114_sgr_v1369_cen_v496_sct_i_vi_color_logscale}
shows the (a) $I_{\rm C}$ light and (b) $(V-I_{\rm C})_0$ color curves
of V834~Car as well as V5114~Sgr, V1369~Cen, and V496~Sct.
The $BVI_{\rm C}$ data of V834~Car are taken from AAVSO, VSOLJ, and SMARTS.
We adopt the color excess of $E(B-V)= 0.50$ after \citet{hac19kb}
and determine the timescaling factor of $\log f_{\rm s}= -0.02$
in order to overlap the $(V-I)_0$ color curve of V834~Car
with the other novae, as shown in
Figure \ref{v834_car_v5114_sgr_v1369_cen_v496_sct_i_vi_color_logscale}(b).
We apply Equation (8) of \citet{hac19ka} for the $I$ band to Figure
\ref{v834_car_v5114_sgr_v1369_cen_v496_sct_i_vi_color_logscale}(a)
and obtain
\begin{eqnarray}
(m&-&M)_{I, \rm V834~Car} \cr
&=& ((m - M)_I + \Delta I_{\rm C})
_{\rm V5114~Sgr} - 2.5 \log 1.26 \cr
&=& 15.55 + 1.15\pm0.2 - 0.25 = 16.45\pm0.2 \cr
&=& ((m - M)_I + \Delta I_{\rm C})
_{\rm V1369~Cen} - 2.5 \log 0.65 \cr
&=& 10.11 + 5.85\pm0.2 + 0.475 = 16.44\pm0.2 \cr
&=& ((m - M)_I + \Delta I_{\rm C})
_{\rm V496~Sct} - 2.5 \log 0.48 \cr
&=& 12.9 + 2.75\pm0.2 + 0.8 = 16.45\pm0.2,
\label{distance_modulus_i_vi_v834_car}
\end{eqnarray}
where we adopt
$(m-M)_{I, \rm V5114~Sgr}=15.55$ from Appendix \ref{v5114_sgr_ubvi},
$(m-M)_{I, \rm V1369~Cen}=10.11$ from \citet{hac19ka}, and
$(m-M)_{I, \rm V496~Sct}=12.9$ in Appendix \ref{v496_sct_bvi}.
Thus, we obtain $(m-M)_{I, \rm V834~Car}= 16.45\pm0.2$.

Figure \ref{v834_car_v2576_oph_v1668_cyg_lv_vul_v_bv_logscale_no2} shows
the (a) $V$ light and (b) $(B-V)_0$ color curves of V834~Car, LV~Vul, 
V1668~Cyg, V533~Her, and V2576~Oph.  
Applying Equation (4) of \citet{hac19ka} to them, we have the relation
\begin{eqnarray}
(m&-&M)_{V, \rm V834~Car} \cr
&=& ((m - M)_V + \Delta V)_{\rm LV~Vul} - 2.5 \log 0.95 \cr
&=& 11.85 + 5.35\pm0.2 + 0.05 = 17.25\pm0.2 \cr
&=& ((m - M)_V + \Delta V)_{\rm V1668~Cyg} - 2.5 \log 0.95 \cr
&=& 14.6 + 2.6\pm0.2 + 0.05 = 17.25\pm0.2 \cr
&=& ((m - M)_V + \Delta V)_{\rm V533~Her} - 2.5 \log 0.79 \cr
&=& 10.65 + 6.35\pm0.2 + 0.25 = 17.25\pm0.2 \cr
&=& ((m - M)_V + \Delta V)_{\rm V2576~Oph} - 2.5 \log 1.35 \cr
&=& 16.65 + 0.95\pm0.2 - 0.325 = 17.27\pm0.2,
\label{distance_modulus_v834_car}
\end{eqnarray}
where we adopt $(m-M)_{V, \rm LV~Vul}=11.85$,
$(m-M)_{V, \rm V1668~Cyg}=14.6$, and
$(m-M)_{V, \rm V533~Her}=10.65$ from \citet{hac19ka},
and $(m-M)_{V, \rm V2576~Oph}=16.65$ from \citet{hac19kb}.
Thus, we obtain $(m-M)_{V, \rm V834~Car}=17.25\pm0.1$ 
and $\log f_{\rm s}= \log 0.95 = -0.02$ against LV~Vul.

Figure \ref{v834_car_v2576_oph_v1668_cyg_lv_vul_b_only_logscale} shows
the $B$ light curve of V834~Car
together with those of LV~Vul, V1668~Cyg, V533~Her, and V2576~Oph.
We apply Equation (7) of \citet{hac19ka} for the $B$ band to Figure
\ref{v834_car_v2576_oph_v1668_cyg_lv_vul_b_only_logscale}
and obtain
\begin{eqnarray}
(m&-&M)_{B, \rm V834~Car} \cr
&=& ((m - M)_B + \Delta B)_{\rm LV~Vul} - 2.5 \log 0.95 \cr
&=& 12.45 + 5.25\pm0.2 + 0.05 = 17.75\pm0.2 \cr
&=& ((m - M)_B + \Delta B)_{\rm V1668~Cyg} - 2.5 \log 0.95 \cr
&=& 14.9 + 2.8\pm0.2 + 0.05 = 17.75\pm0.2 \cr
&=& ((m - M)_B + \Delta B)_{\rm V533~Her} - 2.5 \log 0.79 \cr
&=& 10.69 + 6.8\pm0.2 + 0.25 = 17.74\pm0.2 \cr
&=& ((m - M)_B + \Delta B)_{\rm V2576~Oph} - 2.5 \log 1.35 \cr
&=& 17.25 + 0.8\pm0.2 - 0.325 = 17.73\pm0.2.
\label{distance_modulus_b_v834_car_yy_dor_lmcn2009a}
\end{eqnarray}
We have $(m-M)_{B, \rm V834~Car}= 17.75\pm0.1$.

We obtained $(m-M)_B=17.75$, $(m-M)_V=17.25$, and $(m-M)_I=16.45$,
which cross at $d=14$~kpc and $E(B-V)=0.50$.  These three distance
moduli are the same as those obtained by \citet{hac19kb}.
The three relations were already plotted in Figure
13(d) of \citet{hac19kb}.  Only the difference from the previous
results is the timescaling factor of $\log f_{\rm s}= -0.02$ 
(the previous value is $\log f_{\rm s}= -0.19$).


\begin{figure}
\plotone{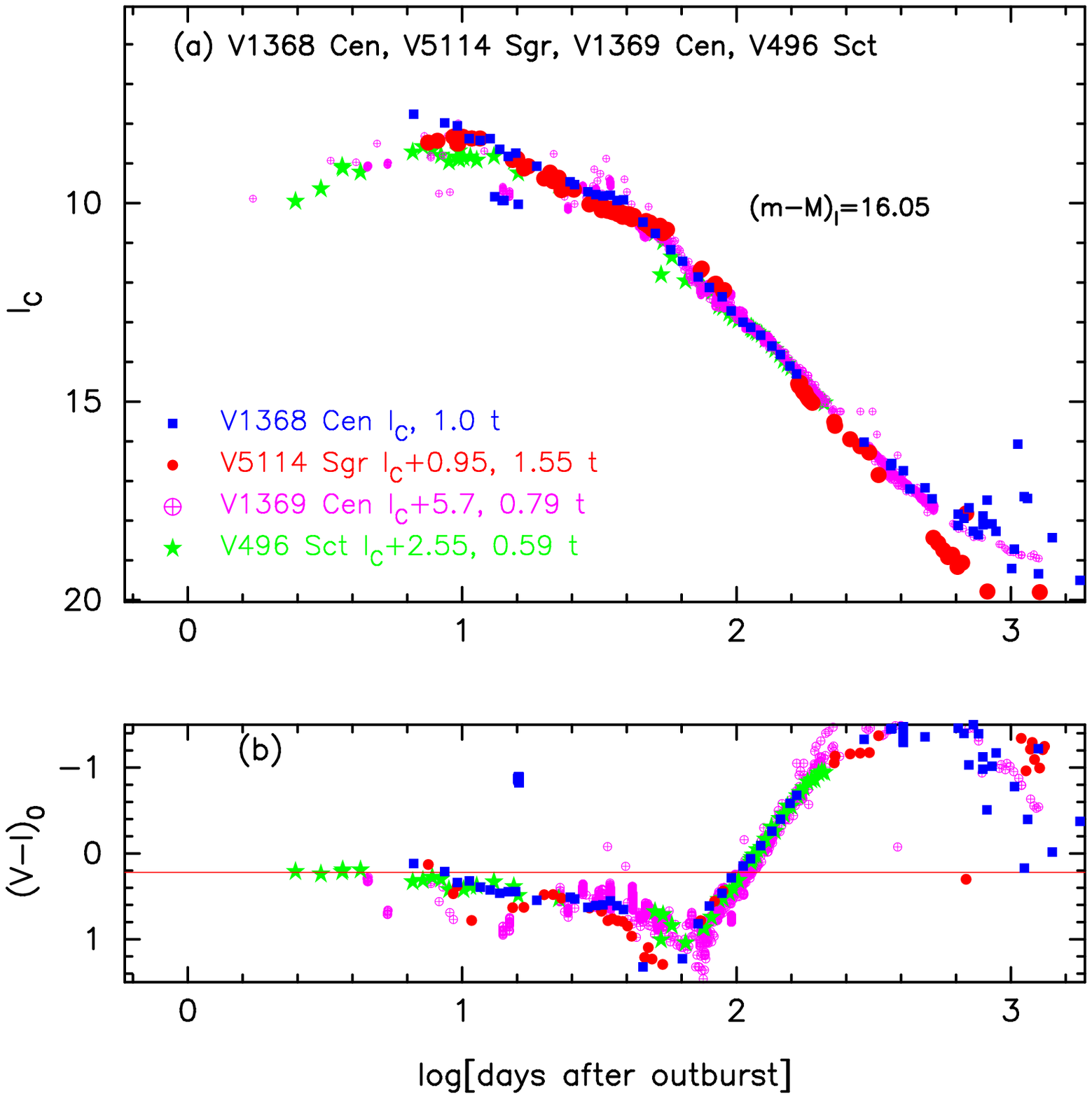}
\caption{
The (a) $I_{\rm C}$ light curve and (b) $(V-I_{\rm C})_0$ color curve
of V1368~Cen as well as those of V5114~Sgr, V1369~Cen, and V496~Sct.
\label{v1368_cen_v5114_sgr_v1369_cen_v496_sct_i_vi_color_logscale}}
\end{figure}


\begin{figure}
\epsscale{1.0}
\plotone{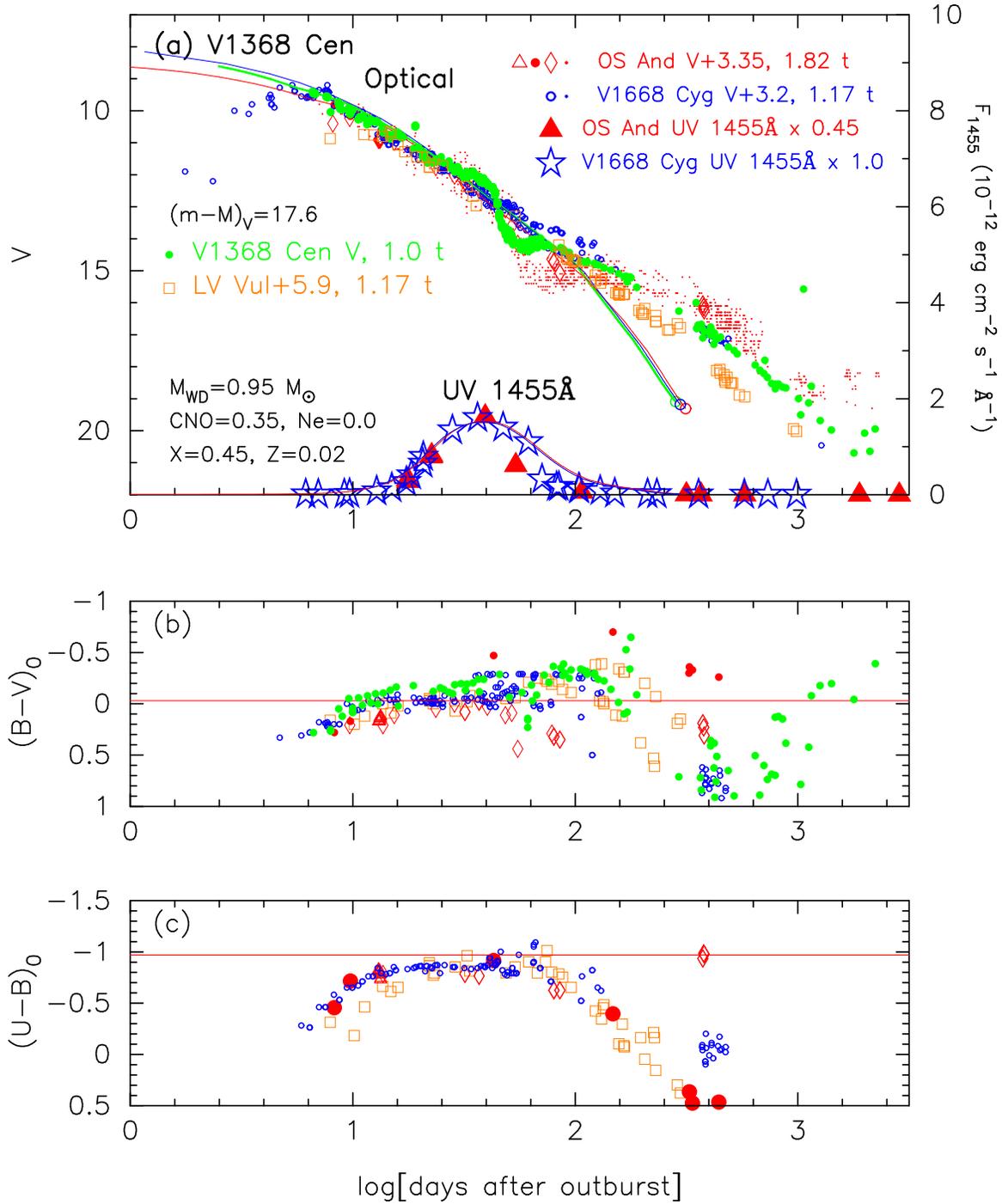}
\caption{
The (a) $V$ light curve, (b) $(B-V)_0$, and (c) $(U-B)_0$ color curves
of V1368~Cen as well as those of LV~Vul, V1668~Cyg, and OS~And.
The data of V1368~Cen are taken from AAVSO and SMARTS.
In panel (a), we plot a $0.95~M_\sun$ WD model (CO3, solid green lines)
for V1368~Cen as well as a $1.05~M_\sun$ WD model (CO3, solid red lines)
for OS~And and $0.98~M_\sun$ WD model (CO3, solid blue lines) for V1668~Cyg.
\label{v1368_cen_lv_vul_v1668_cyg_os_and_v_bv_ub_logscale}}
\end{figure}


\begin{figure*}
\plottwo{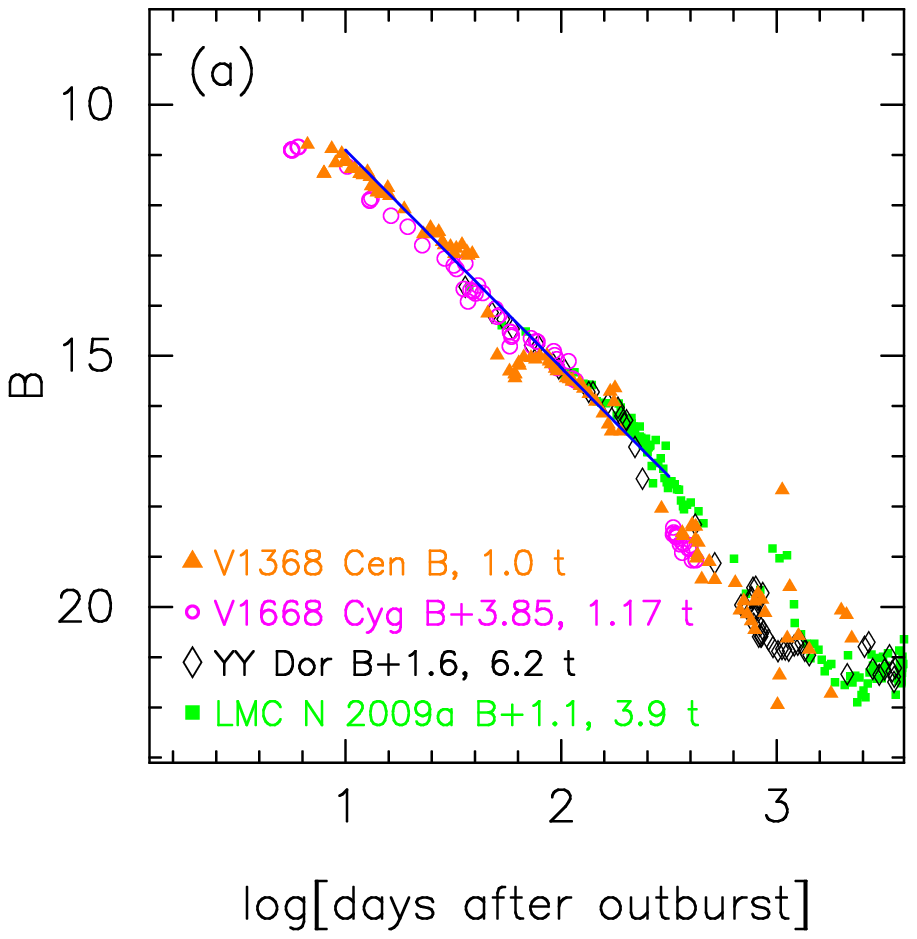}{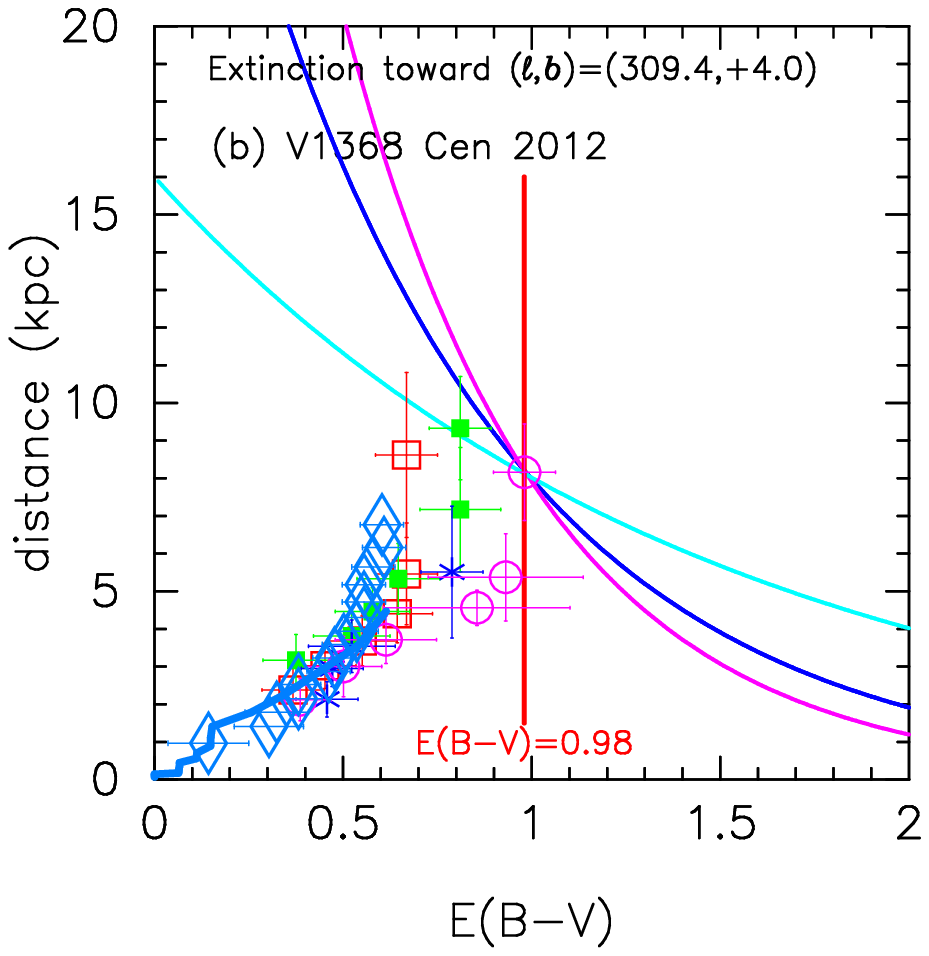}
\caption{
(a) The $B$ light curve of V1368~Cen as well as 
V1668~Cyg, YY~Dor, and LMC~N~2009a.
(b) Various distance-reddening relations toward V1368~Cen.
The thin solid lines of magenta, blue, and cyan denote the distance-reddening
relations given by $(m-M)_B= 18.6$, $(m-M)_V= 17.6$, and $(m-M)_I= 16.04$,
respectively.
\label{distance_reddening_v1368_cen_bvi_xxxxxx}}
\end{figure*}

\subsection{V1368~Cen 2012}
\label{v1368_cen_bvi}
We have reanalyzed the $BVI_{\rm C}$ multi-band 
light/color curves of V1368~Cen based on the time-stretching method.  
Figure \ref{v1368_cen_v5114_sgr_v1369_cen_v496_sct_i_vi_color_logscale}
shows the (a) $I_{\rm C}$ light and (b) $(V-I_{\rm C})_0$ color curves
of V1368~Cen as well as V5114~Sgr, V1369~Cen, and V496~Sct.
The $BVI_{\rm C}$ data of V1368~Cen are taken from AAVSO, VSOLJ, and SMARTS.
We adopt the color excess of $E(B-V)= 0.98$ as mentioned below.
We apply Equation (8) of \citet{hac19ka} for the $I$ band to Figure
\ref{v1368_cen_v5114_sgr_v1369_cen_v496_sct_i_vi_color_logscale}(a)
and obtain
\begin{eqnarray}
(m&-&M)_{I, \rm V1368~Cen} \cr
&=& ((m - M)_I + \Delta I_{\rm C})
_{\rm V5114~Sgr} - 2.5 \log 1.55 \cr
&=& 15.55 + 0.95\pm0.2 - 0.475 = 16.02\pm0.2 \cr
&=& ((m - M)_I + \Delta I_{\rm C})
_{\rm V1369~Cen} - 2.5 \log 0.79 \cr
&=& 10.11 + 5.7\pm0.2 + 0.25 = 16.06\pm0.2 \cr
&=& ((m - M)_I + \Delta I_{\rm C})
_{\rm V496~Sct} - 2.5 \log 0.59 \cr
&=& 12.9 + 2.55\pm0.2 + 0.575 = 16.02\pm0.2,
\label{distance_modulus_i_vi_v1368_cen}
\end{eqnarray}
where we adopt
$(m-M)_{I, \rm V5114~Sgr}=15.55$ from Appendix \ref{v5114_sgr_ubvi},
$(m-M)_{I, \rm V1369~Cen}=10.11$ from \citet{hac19ka}, and
$(m-M)_{I, \rm V496~Sct}=12.9$ in Appendix \ref{v496_sct_bvi}.
Thus, we obtain $(m-M)_{I, \rm V1368~Cen}= 16.03\pm0.2$.

Figure \ref{v1368_cen_lv_vul_v1668_cyg_os_and_v_bv_ub_logscale} shows
the light/color curves of V1368~Cen, LV~Vul, V1668~Cyg, and OS~And.
They overlap each other.
Applying Equation (4) of \citet{hac19ka} to them,
we have the relation
\begin{eqnarray}
(m&-&M)_{V, \rm V1368~Cen} \cr
&=& (m - M + \Delta V)_{V, \rm LV~Vul} - 2.5 \log 1.17 \cr
&=& 11.85 + 5.9\pm0.2 - 0.175 = 17.58\pm0.2 \cr
&=& (m - M + \Delta V)_{V, \rm V1668~Cyg} - 2.5 \log 1.17 \cr
&=& 14.6 + 3.2\pm0.2 - 0.175 = 17.62\pm0.2 \cr
&=& (m - M + \Delta V)_{V, \rm OS~And} - 2.5 \log 1.82 \cr
&=& 14.8 +3.35\pm0.2 - 0.55 = 17.6\pm0.2,
\label{distance_modulus_v_bv_v1368_cen}
\end{eqnarray}
where we adopt $(m-M)_{V, \rm LV~Vul}=11.85$ and
$(m-M)_{V, \rm V1668~Cyg}=14.6$ from \citet{hac19ka}, and
$(m-M)_{V, \rm OS~And}=14.8$ from \citet{hac16kb}.
Thus, we obtain $(m-M)_V=17.6\pm0.1$ and $f_{\rm s}=1.17$ against LV~Vul.


Figure \ref{distance_reddening_v1368_cen_bvi_xxxxxx}(a)
shows the $B$ light curves of V1368~Cen
together with those of V1668~Cyg, YY~Dor, and LMC~N~2009a.
We apply Equation (7) of \citet{hac19ka} for the $B$ band 
to Figure \ref{distance_reddening_v1368_cen_bvi_xxxxxx}(a)
and obtain
\begin{eqnarray}
(m&-&M)_{B, \rm V1368~Cen} \cr
&=& ((m - M)_B + \Delta B)_{\rm V1668~Cyg} - 2.5 \log 1.17 \cr
&=& 14.9 + 3.85\pm0.2 - 0.175 = 18.58\pm0.2 \cr
&=& ((m - M)_B + \Delta B)_{\rm YY~Dor} - 2.5 \log 6.2 \cr
&=& 18.98 + 1.6\pm0.2 - 1.975 2.05 = 18.6\pm0.2 \cr
&=& ((m - M)_B + \Delta B)_{\rm LMC~N~2009a} - 2.5 \log 3.9 \cr
&=& 18.98 + 1.1\pm0.2 - 1.475 = 18.6\pm0.2.
\label{distance_modulus_b_v1368_cen_v834_car_yy_dor_lmcn2009a}
\end{eqnarray}
We have $(m-M)_{B, \rm V1368~Cen}= 18.6\pm0.1$.

We plot $(m-M)_B= 18.6$, $(m-M)_V= 17.6$, and $(m-M)_I= 16.04$,
which broadly cross at $d=8.2$~kpc and $E(B-V)=0.98$, in Figure
\ref{distance_reddening_v1368_cen_bvi_xxxxxx}(b).
The crossing point is consistent with the distance-reddening relation
given by \citet[][unfilled magenta circles]{mar06}.
Thus, we obtain $E(B-V)=0.98\pm0.05$ and $d=8.2\pm1$~kpc.


\begin{figure}
\plotone{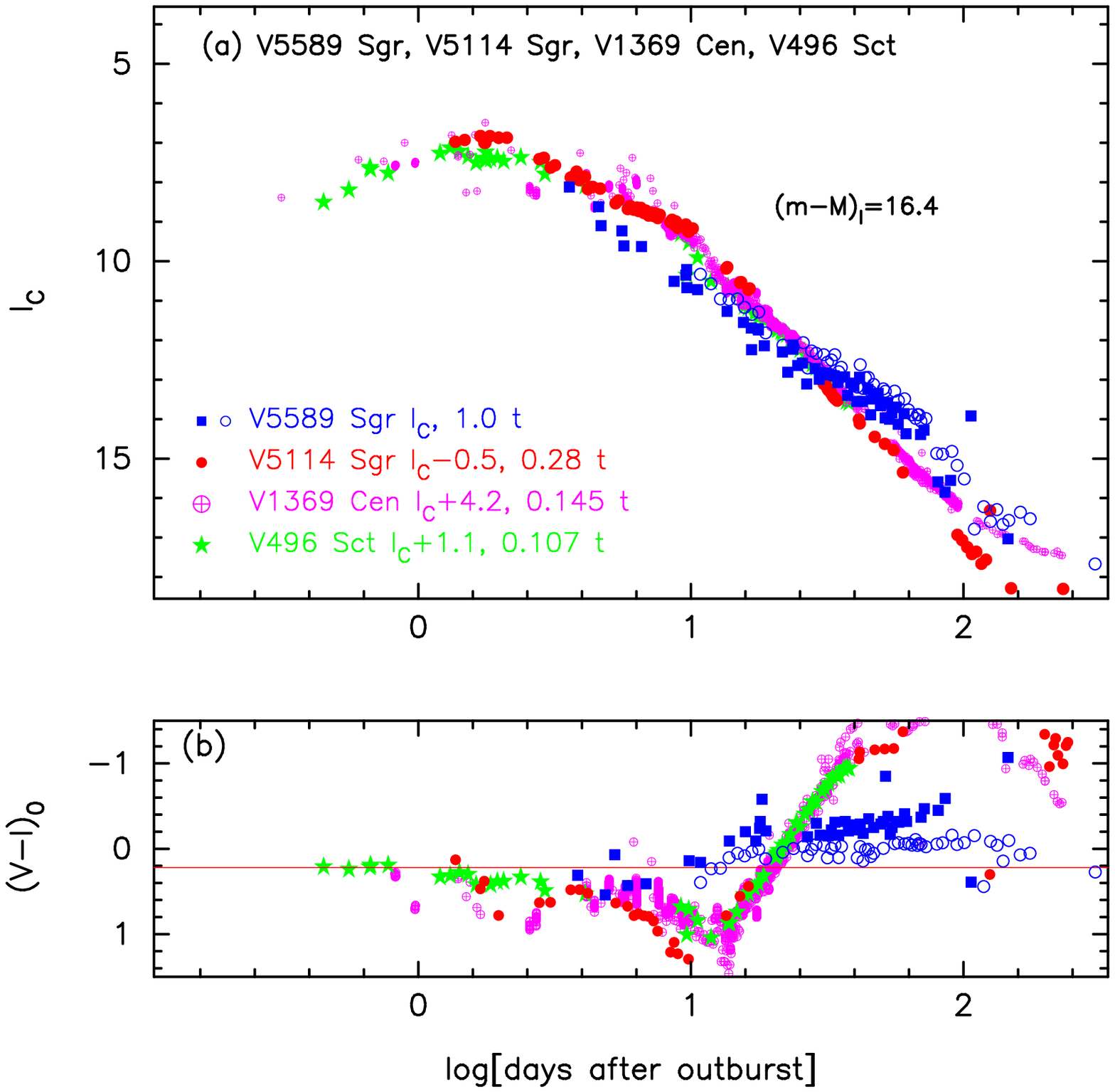}
\caption{
The (a) $I_{\rm C}$ light curve and (b) $(V-I_{\rm C})_0$ color curve
of V5589~Sgr as well as those of V5114~Sgr, V1369~Cen, and V496~Sct.
The $BVI_{\rm C}$ data of V5589~Sgr are taken from VSOLJ (filled blue
squares) and SMARTS (unfilled blue circles).
\label{v5589_sgr_v5114_sgr_v1369_cen_v496_sct_i_vi_color_logscale}}
\end{figure}


\begin{figure}
\plotone{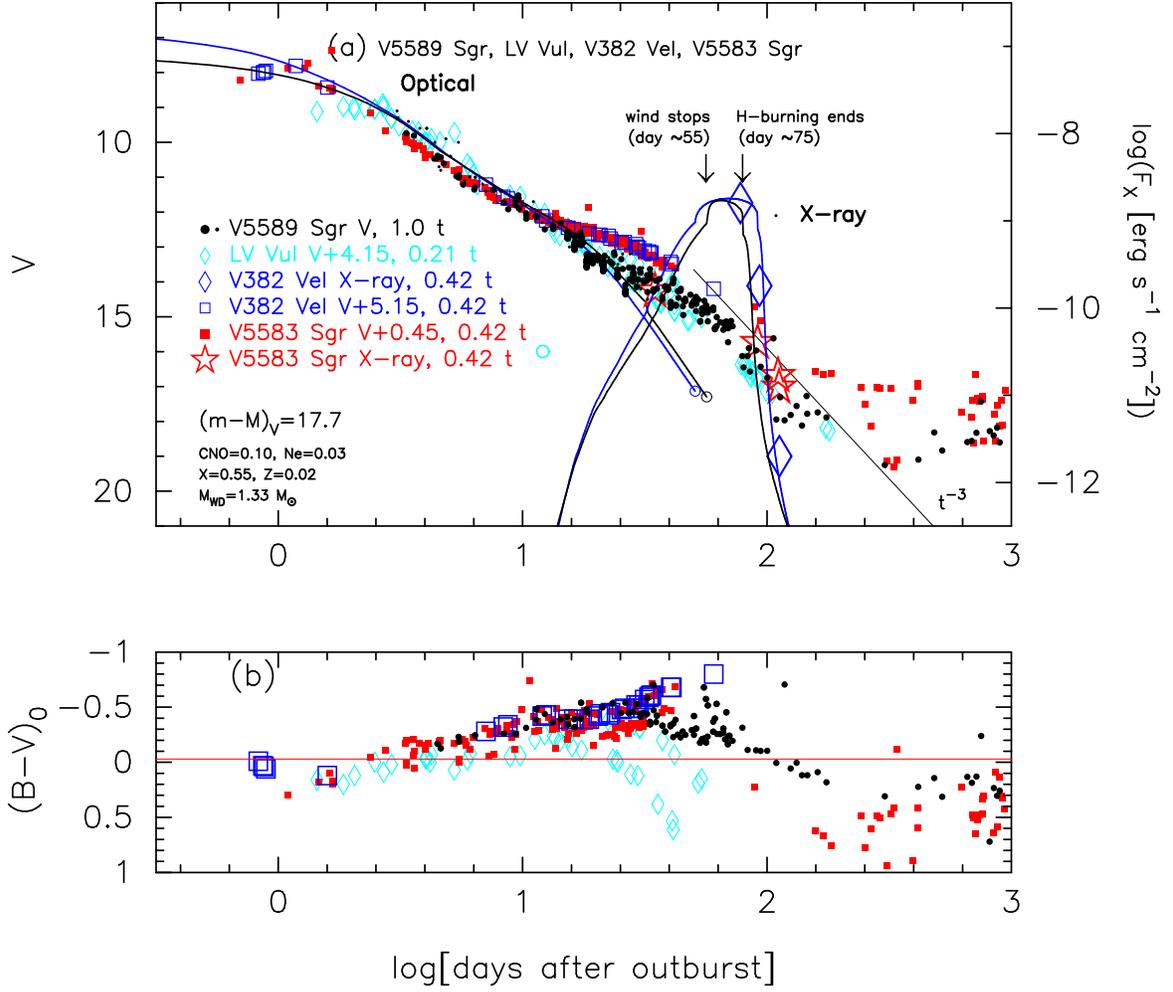}
\caption{
The (a) $V$ light curve and (b) $(B-V)_0$ color curve
of V5589~Sgr as well as those of LV~Vul, V382~Vel, and V5583~Sgr.
The data of V5589~Sgr are taken from AAVSO, VSOLJ, and SMARTS.
In panel (a), we show a $1.33~M_\sun$ WD model (Ne2, solid black lines)
for V5589~Sgr as well as a $1.23~M_\sun$ WD model (Ne2, solid blue lines)
for V382~Vel.
\label{v5589_sgr_v5583_sgr_v382_vel_lv_vul_v_bv_logscale_no2}}
\end{figure}


\begin{figure*}
\plottwo{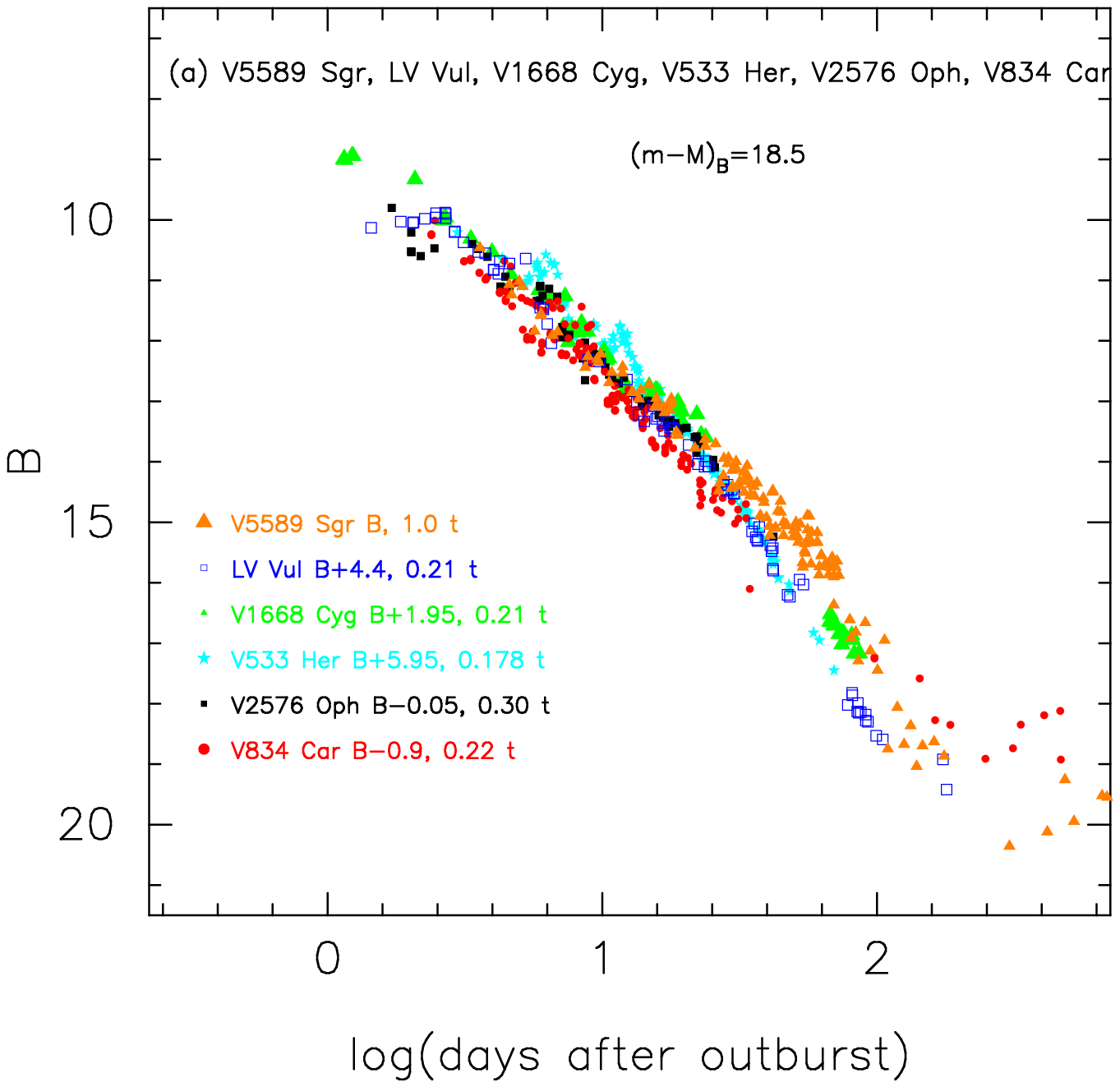}{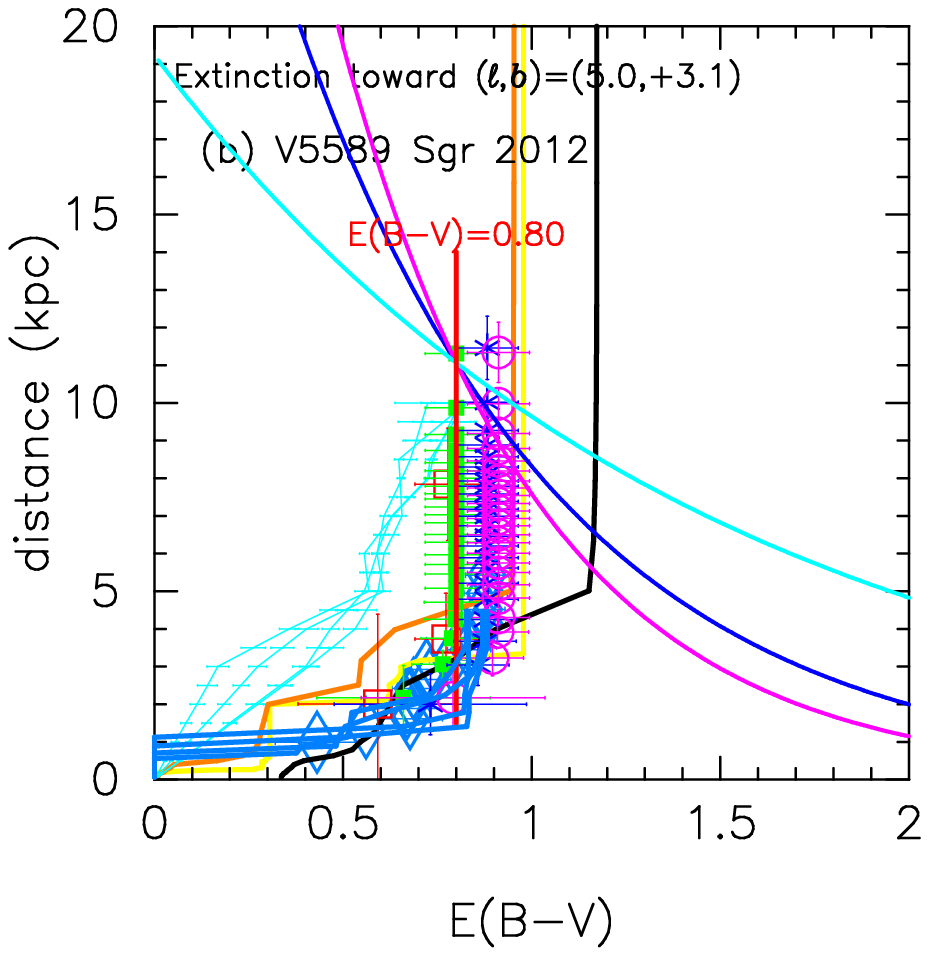}
\caption{
(a) The $B$ light curve of V5589~Sgr as well as 
LV~Vul, V1668~Cyg, V533~Her, V2576~Oph, and V834~Car.
(b) Various distance-reddening relations toward V5589~Sgr.
The thin solid lines of magenta, blue, and cyan denote the distance-reddening
relations given by $(m-M)_B=18.5$, $(m-M)_V=17.7$, and $(m-M)_I=16.42$,
respectively.  
\label{distance_reddening_v5589_sgr_bvi_xxxxxx}}
\end{figure*}

\subsection{V5589~Sgr 2012}
\label{v5589_sgr_bvi}
We have reanalyzed the $BVI_{\rm C}$ multi-band light/color curves
of V5589~Sgr based on the time-stretching method.  
Figure \ref{v5589_sgr_v5114_sgr_v1369_cen_v496_sct_i_vi_color_logscale}
shows the (a) $I_{\rm C}$ light and (b) $(V-I_{\rm C})_0$ color curves
of V5589~Sgr as well as V5114~Sgr, V1369~Cen, and V496~Sct.
The $BVI_{\rm C}$ data of V5589~Sgr are taken from AAVSO, VSOLJ, SMARTS.
We have determined the timescaling factor of $\log f_{\rm s}= -0.67$
to overlap the color evolution of V5589~Sgr with the other novae
as much as possible both in the $(V-I_{\rm C})_0$ and $(B-V)_0$ colors.
Then, we apply Equation (8) of \citet{hac19ka} for the $I$ band to Figure
\ref{v5589_sgr_v5114_sgr_v1369_cen_v496_sct_i_vi_color_logscale}(a)
and obtain
\begin{eqnarray}
(m&-&M)_{I, \rm V5589~Sgr} \cr
&=& ((m - M)_I + \Delta I_{\rm C})
_{\rm V5114~Sgr} - 2.5 \log 0.28 \cr
&=& 15.55 - 0.5\pm0.2 + 1.375 = 16.43\pm0.2 \cr
&=& ((m - M)_I + \Delta I_{\rm C})
_{\rm V1369~Cen} - 2.5 \log 0.145 \cr
&=& 10.11 + 4.2\pm0.2 + 2.1 = 16.41\pm0.2 \cr
&=& ((m - M)_I + \Delta I_{\rm C})
_{\rm V496~Sct} - 2.5 \log 0.107 \cr
&=& 12.9 + 1.1\pm0.2 + 2.425 = 16.42\pm0.2,
\label{distance_modulus_i_vi_v5589_sgr}
\end{eqnarray}
where we adopt
$(m-M)_{I, \rm V5114~Sgr}=15.55$ from Appendix \ref{v5114_sgr_ubvi},
$(m-M)_{I, \rm V1369~Cen}=10.11$ from \citet{hac19ka}, and
$(m-M)_{I, \rm V496~Sct}=12.9$ in Appendix \ref{v496_sct_bvi}.
Thus, we obtain $(m-M)_{I, \rm V5589~Sgr}= 16.42\pm0.2$.

Figure \ref{v5589_sgr_v5583_sgr_v382_vel_lv_vul_v_bv_logscale_no2}
shows the (a) visual and $V$, and (b) $(B-V)_0$ evolutions of V5589~Sgr
as well as LV~Vul, V382~Vel, and V5583~Sgr.
Applying Equation (4) of \citet{hac19ka} for the $V$ band to 
Figure \ref{v5589_sgr_v5583_sgr_v382_vel_lv_vul_v_bv_logscale_no2}(a),
we have the relation
\begin{eqnarray}
(m&-&M)_{V, \rm V5589~Sgr} \cr
&=& ((m - M)_V + \Delta V)_{\rm LV~Vul} - 2.5 \log 0.21 \cr
&=& 11.85 + 4.15\pm0.2 + 1.675 = 17.68\pm0.2 \cr
&=& ((m - M)_V + \Delta V)_{\rm V382~Vel} - 2.5 \log 0.42 \cr
&=& 11.6 + 5.15\pm0.2 + 0.95 = 17.7\pm0.2 \cr
&=& ((m - M)_V + \Delta V)_{\rm V5583~Sgr} - 2.5 \log 0.42 \cr
&=& 16.3 + 0.45\pm0.2 + 0.95 = 17.7\pm0.2,
\label{distance_modulus_v5589_sgr}
\end{eqnarray}
where we adopt $(m-M)_{V, \rm LV~Vul}=11.85$ from \citet{hac19ka}, 
$(m-M)_{V, \rm V382~Vel}=11.6$ in Appendix \ref{v382_vel_ubvi},
and $(m-M)_{V, \rm V5583~Sgr}=16.3$ in Appendix \ref{v5583_sgr_bvi}.
Thus, we adopt $(m-M)_{V, \rm V5589~Sgr}=17.7\pm0.1$.

Figure \ref{distance_reddening_v5589_sgr_bvi_xxxxxx}(a) shows
the $B$ light curve of V5589~Sgr
together with those of LV~Vul, V1668~Cyg, V533~Her, V2576~Oph, and V834~Car.
We apply Equation (7) of \citet{hac19ka} for the $B$ band to Figure
\ref{distance_reddening_v5589_sgr_bvi_xxxxxx}(a) and obtain
\begin{eqnarray}
(m&-&M)_{B, \rm V5589~Sgr} \cr
&=& ((m - M)_B + \Delta B)_{\rm LV~Vul} - 2.5 \log 0.21 \cr
&=& 12.45 + 4.4\pm0.2 + 1.675 = 18.52\pm0.2 \cr
&=& ((m - M)_B + \Delta B)_{\rm V1668~Cyg} - 2.5 \log 0.21 \cr
&=& 14.9 + 1.95\pm0.2 + 1.675 = 18.52\pm0.2 \cr
&=& ((m - M)_B + \Delta B)_{\rm V533~Her} - 2.5 \log 0.178 \cr
&=& 10.69 + 5.95\pm0.2 + 1.875 = 18.51\pm0.2 \cr
&=& ((m - M)_B + \Delta B)_{\rm V2576~Oph} - 2.5 \log 0.30 \cr
&=& 17.25 - 0.05\pm0.2 + 1.3 = 18.5\pm0.2 \cr
&=& ((m - M)_B + \Delta B)_{\rm V834~Car} - 2.5 \log 0.22 \cr
&=& 17.75 - 0.9\pm0.2 + 1.625 = 18.48\pm0.2.
\label{distance_modulus_b_v5589_sgr_lv_vul_v1668_cyg}
\end{eqnarray}
We have $(m-M)_{B, \rm V5589~Sgr}= 18.5\pm0.2$.

We plot the three distance moduli
in Figure \ref{distance_reddening_v5589_sgr_bvi_xxxxxx}(b),
which cross at $d=11.0$~kpc and $E(B-V)=0.80$.  
The crossing point is consistent with the distance-reddening relation
given by \citet[][filled green squares]{mar06}.
Thus, we obtained $d=11.0\pm2$~kpc and $E(B-V)=0.80\pm0.05$ for V5589~Sgr.


\begin{figure}
\plotone{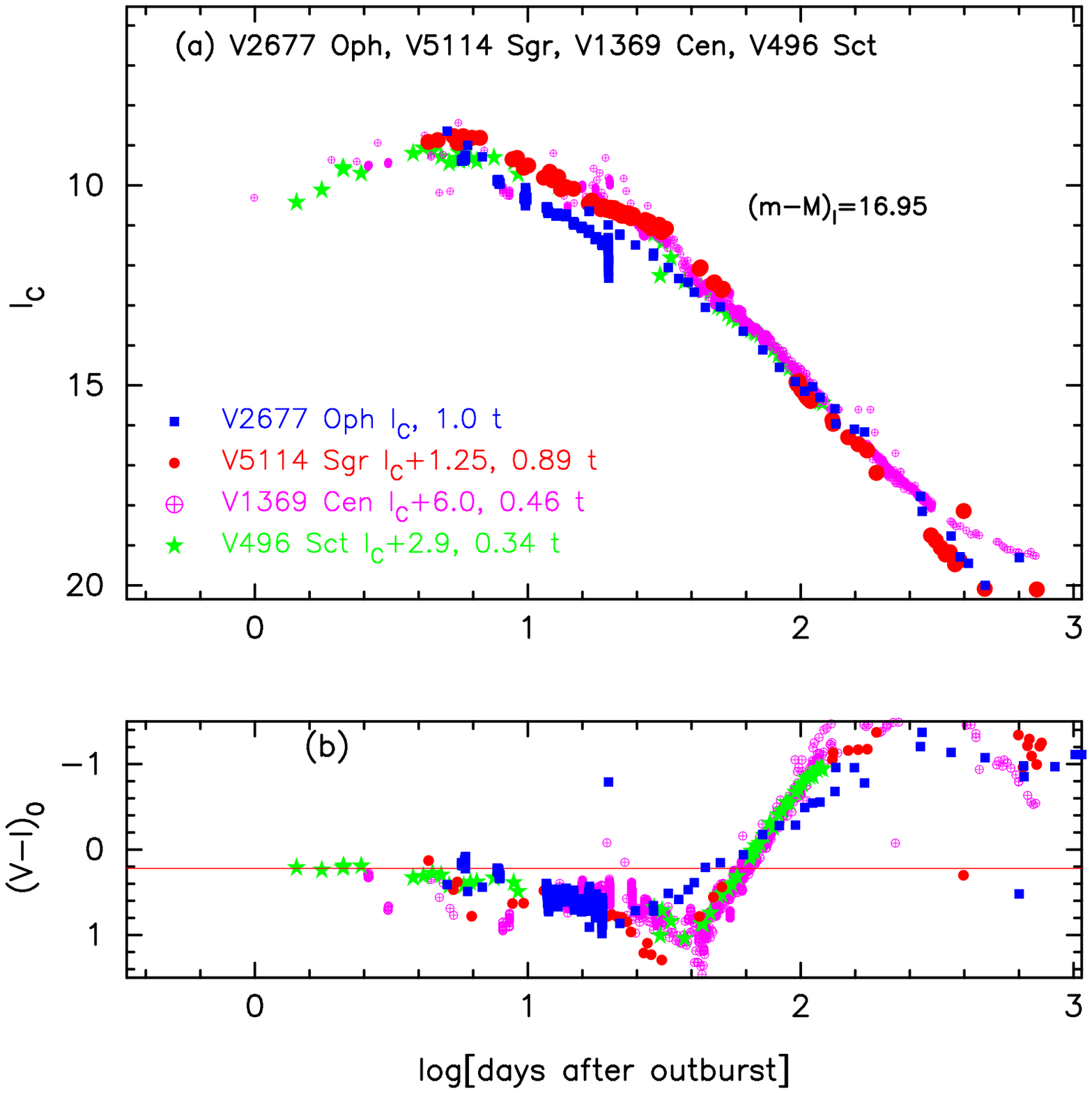}
\caption{
The (a) $I_{\rm C}$ light curve and (b) $(V-I_{\rm C})_0$ color curve
of V2677~Oph as well as those of V5114~Sgr, V1369~Cen, and V496~Sct.
\label{v2677_oph_v5114_sgr_v1369_cen_v496_sct_i_vi_color_logscale}}
\end{figure}


\begin{figure}
\plotone{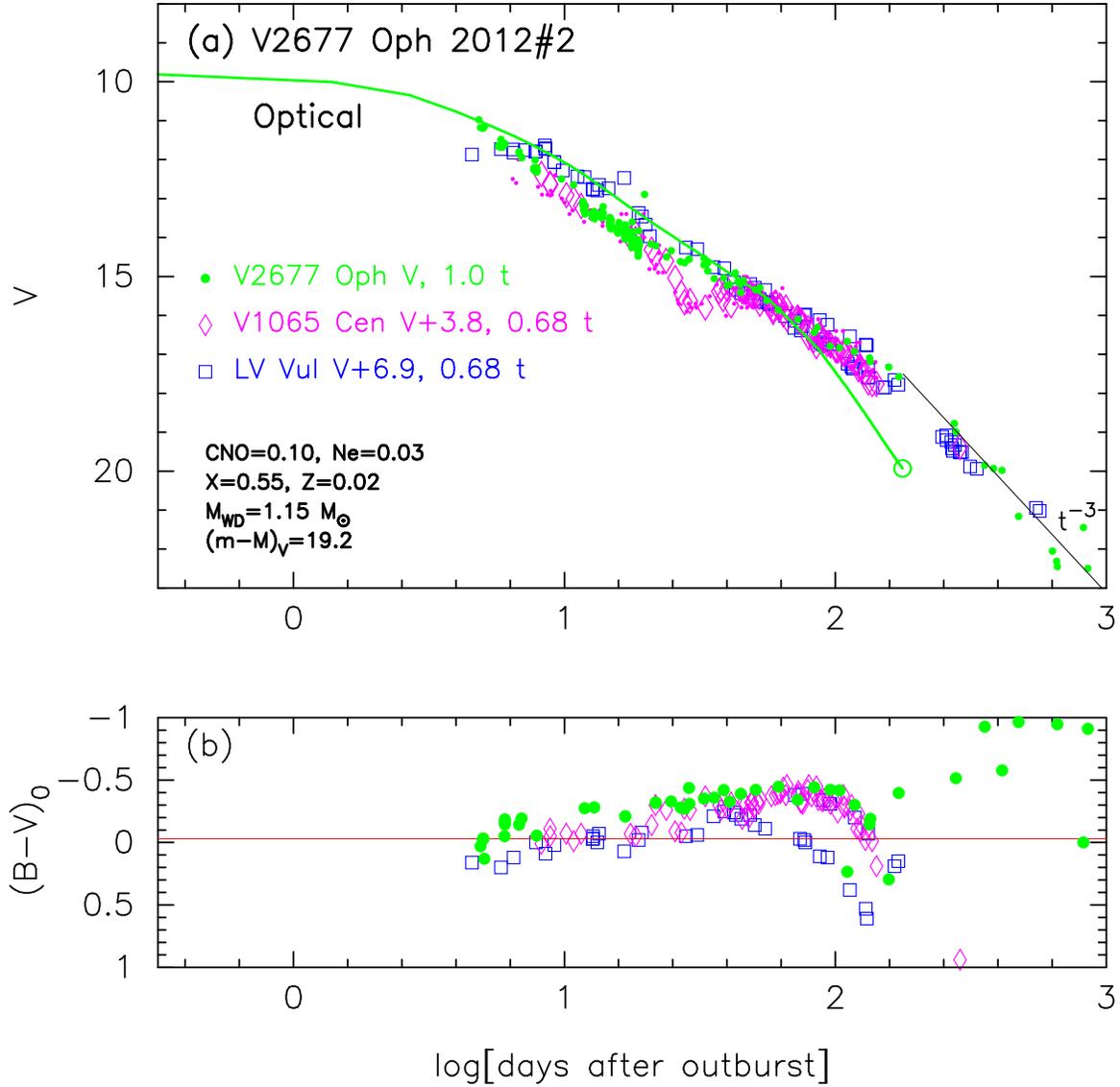}
\caption{
The (a) $V$ light curve and (b) $(B-V)_0$ color curve
of V2677~Oph as well as those of LV~Vul and V1065~Cen.
The data of V2677~Oph are taken from AAVSO and SMARTS.
In panel (a), we show a $1.15~M_\sun$ WD model (Ne2, solid green line)
for V2677~Oph.
\label{v2677_oph_v1065_cen_lv_vul_v_bv_logscale}}
\end{figure}


\begin{figure*}
\plottwo{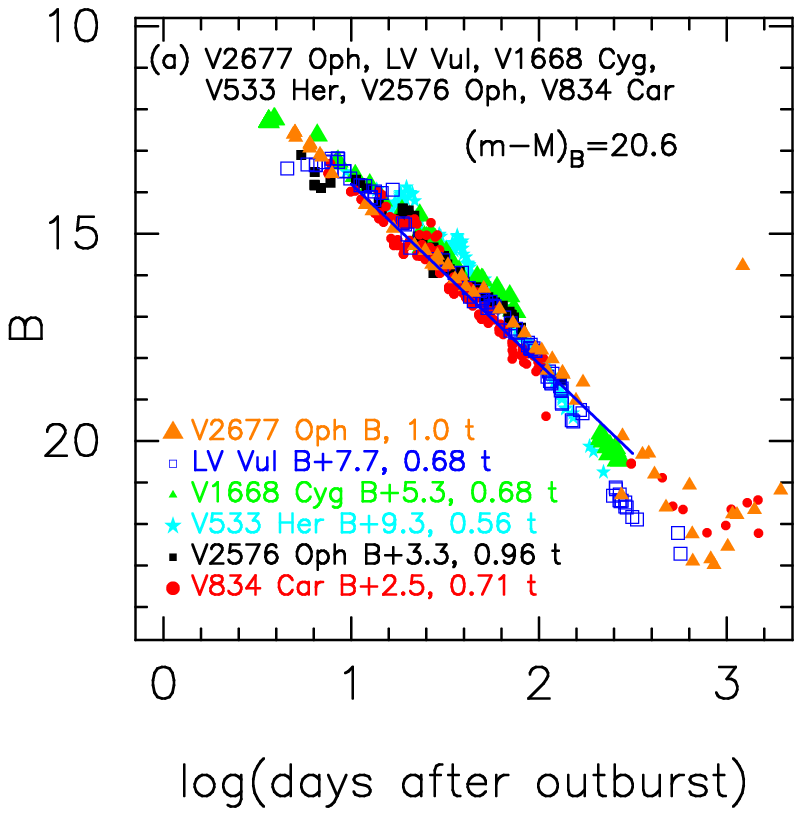}{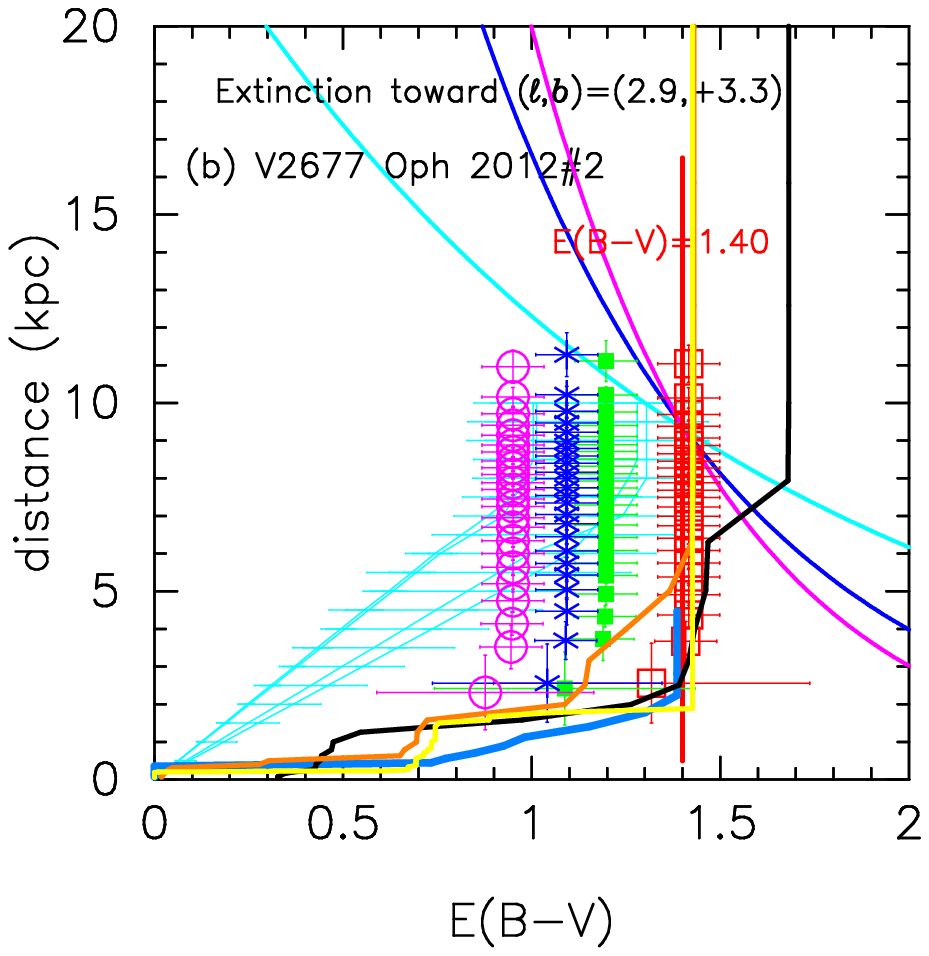}
\caption{
(a) The $B$ light curve of V2677~Oph as well as 
LV~Vul, V1668~Cyg, V533~Her, V2576~Oph, and V834~Car.
(b) Various distance-reddening relations toward V2677~Oph.
The thin solid lines of magenta, blue, and cyan denote the distance-reddening
relations given by $(m-M)_B=20.6$, $(m-M)_V=19.2$, and $(m-M)_I=16.95$,
respectively.  
\label{distance_reddening_v2677_oph_bvi_xxxxxx}}
\end{figure*}

\subsection{V2677~Oph 2012\#2}
\label{v2677_oph_bvi}
We have reanalyzed the $BVI_{\rm C}$ multi-band 
light/color curves of V2677~Oph based on the time-stretching method.  
Figure \ref{v2677_oph_v5114_sgr_v1369_cen_v496_sct_i_vi_color_logscale}
shows the (a) $I_{\rm C}$ light and (b) $(V-I_{\rm C})_0$ color curves
of V2677~Oph as well as V5114~Sgr, V1369~Cen, and V496~Sct.
The $BVI_{\rm C}$ data of V2677~Oph are taken from AAVSO and SMARTS.
We adopt the color excess of $E(B-V)= 1.40$ as mentioned below.
We apply Equation (8) of \citet{hac19ka} for the $I$ band to Figure
\ref{v2677_oph_v5114_sgr_v1369_cen_v496_sct_i_vi_color_logscale}(a)
and obtain
\begin{eqnarray}
(m&-&M)_{I, \rm V2677~Oph} \cr
&=& ((m - M)_I + \Delta I_{\rm C})
_{\rm V5114~Sgr} - 2.5 \log 0.89 \cr
&=& 15.55 + 1.25\pm0.2 + 0.125 = 16.93\pm0.2 \cr
&=& ((m - M)_I + \Delta I_{\rm C})
_{\rm V1369~Cen} - 2.5 \log 0.46 \cr
&=& 10.11 + 6.0\pm0.2 + 0.85 = 16.96\pm0.2 \cr
&=& ((m - M)_I + \Delta I_{\rm C})
_{\rm V496~Sct} - 2.5 \log 0.34 \cr
&=& 12.9 + 2.9\pm0.2 + 1.175 = 16.97\pm0.2,
\label{distance_modulus_i_vi_v2677_oph}
\end{eqnarray}
where we adopt
$(m-M)_{I, \rm V5114~Sgr}=15.55$ from Appendix \ref{v5114_sgr_ubvi},
$(m-M)_{I, \rm V1369~Cen}=10.11$ from \citet{hac19ka}, and
$(m-M)_{I, \rm V496~Sct}=12.9$ in Appendix \ref{v496_sct_bvi}.
Thus, we obtain $(m-M)_{I, \rm V2677~Oph}= 16.95\pm0.2$.

Figure \ref{v2677_oph_v1065_cen_lv_vul_v_bv_logscale} shows
the light/color curves of V2677~Oph, LV~Vul, and V1065~Cen.
We have the relation
\begin{eqnarray}
(m&-&M)_{V, \rm V2677~Oph} \cr
&=& (m - M + \Delta V)_{V, \rm LV~Vul} - 2.5 \log 0.68 \cr
&=& 11.85 + 6.9\pm0.2 + 0.425 = 19.18\pm0.2 \cr
&=& (m - M + \Delta V)_{V, \rm V1065~Cen} - 2.5 \log 0.68 \cr
&=& 15.0 + 3.8\pm0.2 + 0.425 = 19.23\pm0.2,
\label{distance_modulus_v2677_oph}
\end{eqnarray}
where we adopt
$(m-M)_{V, \rm LV~Vul}=11.85$ from \citet{hac19ka} and
$(m-M)_{V, \rm V1065~Cen}=15.0$ from \citet{hac18k}.
Thus, we obtain $(m-M)_V=19.2\pm0.1$ and $f_{\rm s}=0.68$ against LV~Vul.

Figure \ref{distance_reddening_v2677_oph_bvi_xxxxxx}(a) shows
the $B$ light curve of V2677~Oph
together with those of LV~Vul, V1668~Cyg, V533~Her, V2576~Oph, and V834~Car.
We obtain
\begin{eqnarray}
(m&-&M)_{B, \rm V2677~Oph} \cr
&=& ((m - M)_B + \Delta B)_{\rm LV~Vul} - 2.5 \log 0.68 \cr
&=& 12.45 + 7.7\pm0.2 + 0.425 = 20.58\pm0.2 \cr
&=& ((m - M)_B + \Delta B)_{\rm V1668~Cyg} - 2.5 \log 0.68 \cr
&=& 14.9 + 5.3\pm0.2 + 0.425 = 20.63\pm0.2 \cr
&=& ((m - M)_B + \Delta B)_{\rm V533~Her} - 2.5 \log 0.56 \cr
&=& 10.69 + 9.3\pm0.2 + 0.625 = 20.62\pm0.2 \cr
&=& ((m - M)_B + \Delta B)_{\rm V2576~Oph} - 2.5 \log 0.95 \cr
&=& 17.25 + 3.3\pm0.2 + 0.05 = 20.6\pm0.2 \cr
&=& ((m - M)_B + \Delta B)_{\rm V834~Car} - 2.5 \log 0.71 \cr
&=& 17.75 + 2.5\pm0.2 + 0.375 = 20.63\pm0.2.
\label{distance_modulus_b_v2677_oph_lv_vul_v1668_cyg}
\end{eqnarray}
We have $(m-M)_{B, \rm V2677~Oph}= 20.6\pm0.2$.

We plot the three distance moduli
in Figure \ref{distance_reddening_v2677_oph_bvi_xxxxxx}(b),
which cross at $d=9.4$~kpc and $E(B-V)=1.40$.  
Thus, we obtained $d=9.4\pm1$~kpc and $E(B-V)=1.40\pm0.05$ for V2677~Oph.  
This crossing point is consistent with the distance-reddening relations
of \citet[][unfilled red squares]{mar06}, 
\citet[][thick solid orange and yellow lines]{gre18, gre19},
and \citet[][thick solid cyan-blue line]{chen19}.


\begin{figure}
\plotone{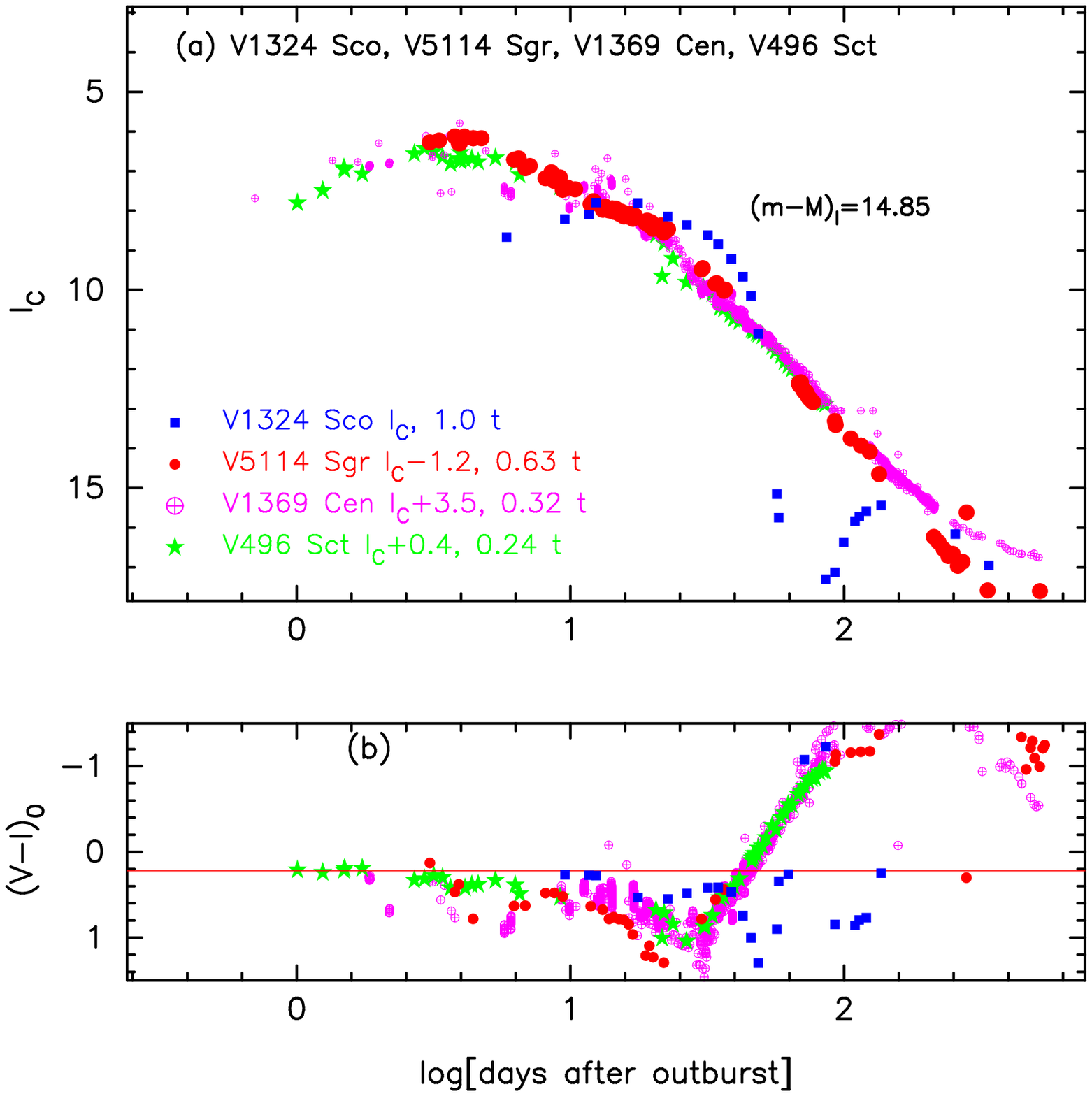}
\caption{
The (a) $I_{\rm C}$ light curve and (b) $(V-I_{\rm C})_0$ color curve
of V1324~Sco as well as those of V5114~Sgr, V1369~Cen, and V496~Sct.
\label{v1324_sco_v5114_sgr_v1369_cen_v496_sct_i_vi_color_logscale}}
\end{figure}

\subsection{V1324~Sco 2012\#2}
\label{v1324_sco_bvi}
We have reanalyzed the $BVI_{\rm C}$ multi-band 
light/color curves of V1324~Sco based on the time-stretching method.  
Figure \ref{v1324_sco_v5114_sgr_v1369_cen_v496_sct_i_vi_color_logscale}
shows the (a) $I_{\rm C}$ light and (b) $(V-I_{\rm C})_0$ color curves
of V1324~Sco as well as V5114~Sgr, V1369~Cen, and V496~Sct.  
The $BVI_{\rm C}$ data of V1324~Sco are taken from AAVSO, SMARTS, and 
\citet{mun15wh}.
We adopt the color excess of $E(B-V)= 1.32$ after \citet{hac19kb}.
We apply Equation (8) of \citet{hac19ka} for the $I$ band to Figure
\ref{v1324_sco_v5114_sgr_v1369_cen_v496_sct_i_vi_color_logscale}(a)
and obtain
\begin{eqnarray}
(m&-&M)_{I, \rm V1324~Sco} \cr
&=& ((m - M)_I + \Delta I_{\rm C})
_{\rm V5114~Sgr} - 2.5 \log 0.63 \cr
&=& 15.55 - 1.2\pm0.2 + 0.5 = 14.85\pm0.2 \cr
&=& ((m - M)_I + \Delta I_{\rm C})
_{\rm V1369~Cen} - 2.5 \log 0.32 \cr
&=& 10.11 + 3.5\pm0.2 + 1.225 = 14.84\pm0.2 \cr
&=& ((m - M)_I + \Delta I_{\rm C})
_{\rm V496~Sct} - 2.5 \log 0.24 \cr
&=& 12.9 + 0.4\pm0.2 + 1.55 = 14.85\pm0.2,
\label{distance_modulus_i_vi_v1324_sco}
\end{eqnarray}
where we adopt
$(m-M)_{I, \rm V5114~Sgr}=15.55$ from Appendix \ref{v5114_sgr_ubvi},
$(m-M)_{I, \rm V1369~Cen}=10.11$ from \citet{hac19ka}, and
$(m-M)_{I, \rm V496~Sct}=12.9$ in Appendix \ref{v496_sct_bvi}.
Thus, we obtain $(m-M)_{I, \rm V1324~Sco}= 14.85\pm0.2$.
These parameters are all consistent with the previous results of
$(m-M)_I= 14.78\pm0.2$ obtained by \citet{hac19kb}  
except the timescaling factor of $\log f_{\rm s}= +0.32$
(the previous value of $\log f_{\rm s}= +0.28$ in Table 1 of
\citet{hac19kb} is a typographical error
as explained in Section \ref{v1324_sco_vi}).


\begin{figure}
\plotone{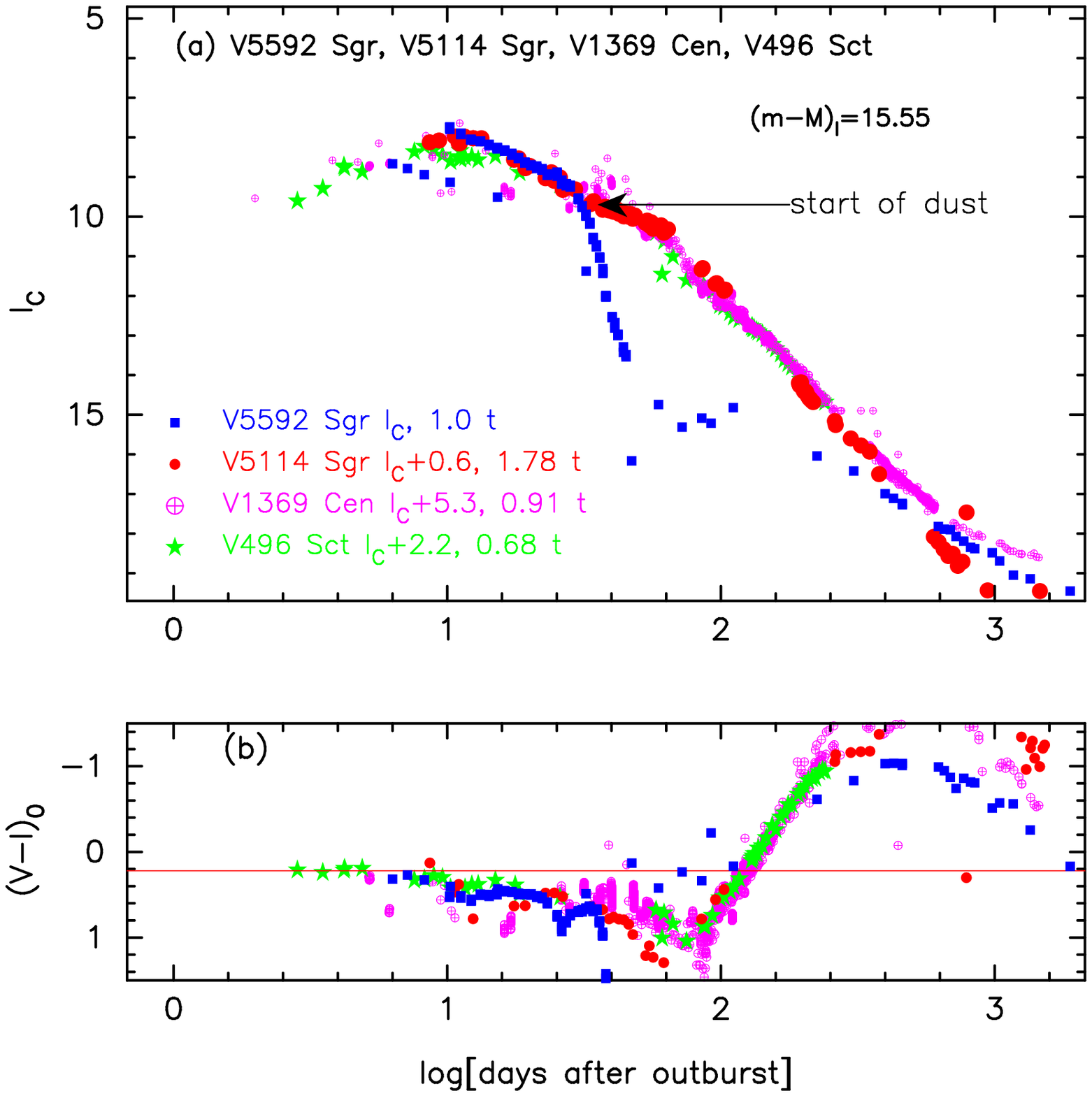}
\caption{
The (a) $I_{\rm C}$ light curve and (b) $(V-I_{\rm C})_0$ color curve
of V5592~Sgr as well as those of V5114~Sgr, V1369~Cen, and V496~Sct.
\label{v5592_sgr_v5114_sgr_v1369_cen_v496_sct_i_vi_color_logscale}}
\end{figure}

\subsection{V5592~Sgr 2012\#4}
\label{v5592_sgr_bvi}
We have reanalyzed the $BVI_{\rm C}$ multi-band 
light/color curves of V5592~Sgr based on the time-stretching method.  
Figure \ref{v5592_sgr_v5114_sgr_v1369_cen_v496_sct_i_vi_color_logscale}
shows the (a) $I_{\rm C}$ light and (b) $(V-I_{\rm C})_0$ color curves
of V5592~Sgr as well as V5114~Sgr, V1369~Cen, and V496~Sct.
The $BVI_{\rm C}$ data of V5592~Sgr are taken from AAVSO and SMARTS.
We adopt the color excess of $E(B-V)= 0.33$ after \citet{hac19kb}.
We apply Equation (8) of \citet{hac19ka} for the $I$ band to Figure
\ref{v5592_sgr_v5114_sgr_v1369_cen_v496_sct_i_vi_color_logscale}(a)
and obtain
\begin{eqnarray}
(m&-&M)_{I, \rm V5592~Sgr} \cr
&=& ((m - M)_I + \Delta I_{\rm C})
_{\rm V5114~Sgr} - 2.5 \log 1.78 \cr
&=& 15.55 + 0.6\pm0.2 - 0.625 = 15.53\pm0.2 \cr
&=& ((m - M)_I + \Delta I_{\rm C})
_{\rm V1369~Cen} - 2.5 \log 0.91 \cr
&=& 10.11 + 5.3\pm0.2 + 0.1 = 15.51\pm0.2 \cr
&=& ((m - M)_I + \Delta I_{\rm C})
_{\rm V496~Sct} - 2.5 \log 0.68 \cr
&=& 12.9 + 2.2\pm0.2 + 0.425 = 15.52\pm0.2,
\label{distance_modulus_i_vi_v5592_sgr}
\end{eqnarray}
where we adopt
$(m-M)_{I, \rm V5114~Sgr}=15.55$ from Appendix \ref{v5114_sgr_ubvi},
$(m-M)_{I, \rm V1369~Cen}=10.11$ from \citet{hac19ka}, and
$(m-M)_{I, \rm V496~Sct}=12.9$ in Appendix \ref{v496_sct_bvi}.
Thus, we obtain $(m-M)_{I, \rm V5592~Sgr}= 15.52\pm0.2$.
These parameters are all consistent with the previous results of
$(m-M)_I= 15.59\pm0.2$ obtained by \citet{hac19kb}.

\begin{figure}
\plotone{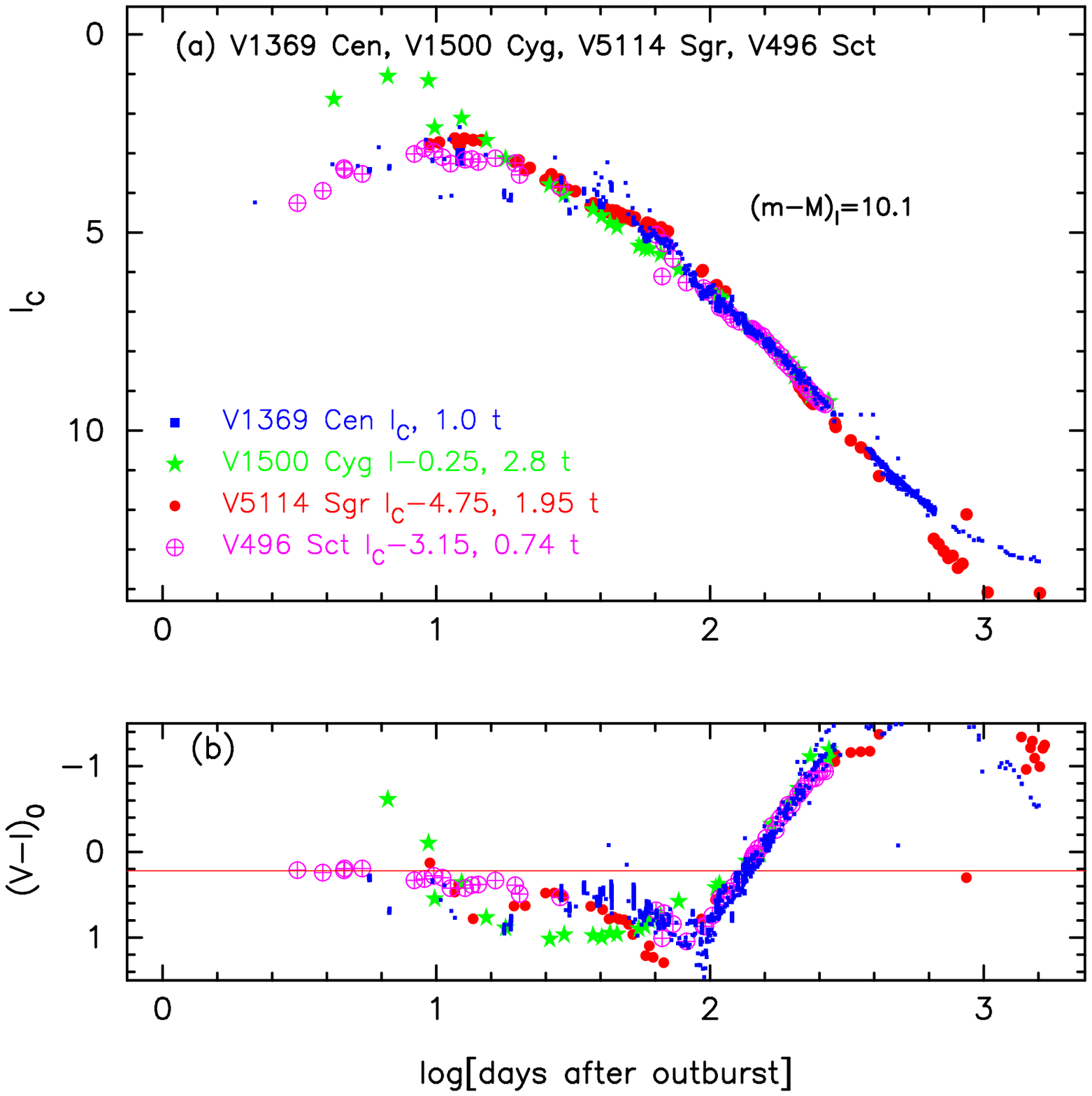}
\caption{
The (a) $I_{\rm C}$ light curve and (b) $(V-I_{\rm C})_0$ color curve
of V1369~Cen as well as those of V1500~Cyg, V5114~Sgr, and V496~Sct.
\label{v1369_cen_v1500_cyg_v5114_sgr_v496_sct_i_vi_logscale}}
\end{figure}

\subsection{V1369~Cen 2013}
\label{v1369_cen_bvi}
We have reanalyzed the $BVI_{\rm C}$ multi-band 
light/color curves of V1369~Cen based on the time-stretching method.  
Adopting the color excess $E(B-V)= 0.11$, distance modulus in 
$I_{\rm C}$ band $(m-M)_I= 10.1$, and timescaling factor
$\log f_{\rm s}= +0.17$ from \citet{hac19ka}, we plot the
time-stretched (a) $I_{\rm C}$ light and (b) $(V-I_{\rm C})_0$ color
curves of V1369~Cen as well as V1500~Cyg, V5114~Sgr, and V496~Sct in
Figure \ref{v1369_cen_v1500_cyg_v5114_sgr_v496_sct_i_vi_logscale}.
The $BVI_{\rm C}$ data of V1369~Cen are taken from AAVSO, VSOLJ, and SMARTS.
Using the color excess $E(B-V)= 0.11$ and timescaling factor 
$\log f_{\rm s}= +0.17$, we are able to overlap
the $(V-I)_0$ color curve of V1369~Cen with the other novae, as shown in
Figure \ref{v1369_cen_v1500_cyg_v5114_sgr_v496_sct_i_vi_logscale}(b).
We apply Equation (8) of \citet{hac19ka} for the $I$ band to Figure
\ref{v1369_cen_v1500_cyg_v5114_sgr_v496_sct_i_vi_logscale}(a)
and obtain
\begin{eqnarray}
(m&-&M)_{I, \rm V1369~Cen} \cr
&=& ((m - M)_I + \Delta I_{\rm C})
_{\rm V1500~Cyg} - 2.5 \log 2.8 \cr
&=& 11.45 - 0.25\pm0.2 - 1.125 = 10.08\pm0.2 \cr
&=& ((m - M)_I + \Delta I_{\rm C})
_{\rm V5114~Sgr} - 2.5 \log 1.95 \cr
&=& 15.55 - 4.75\pm0.2 - 0.725 = 10.08\pm0.2 \cr
&=& ((m - M)_I + \Delta I_{\rm C})
_{\rm V496~Sct} - 2.5 \log 0.74 \cr
&=& 12.9 - 3.15\pm0.2 + 0.325 = 10.08\pm0.2,
\label{distance_modulus_i_vi_v1369_cen}
\end{eqnarray}
where we adopt
$(m-M)_{I, \rm V1500~Cyg}=11.45$ in Appendix \ref{v1500_cyg_ubvi}, 
$(m-M)_{I, \rm V5114~Sgr}=15.55$ from Appendix \ref{v5114_sgr_ubvi}, and
$(m-M)_{I, \rm V496~Sct}= 12.9$ in Appendix \ref{v496_sct_bvi}. 
Thus, we obtain $(m-M)_{I, \rm V1369~Cen}= 10.08\pm0.2$ and 
$\log f_{\rm s}= +0.17$ against LV~Vul.
These parameters are all consistent with the previous vales of
$(m-M)_I= 10.1$ and $\log f_{\rm s}= +0.17$ obtained by \citet{hac19ka}.


\begin{figure}
\plotone{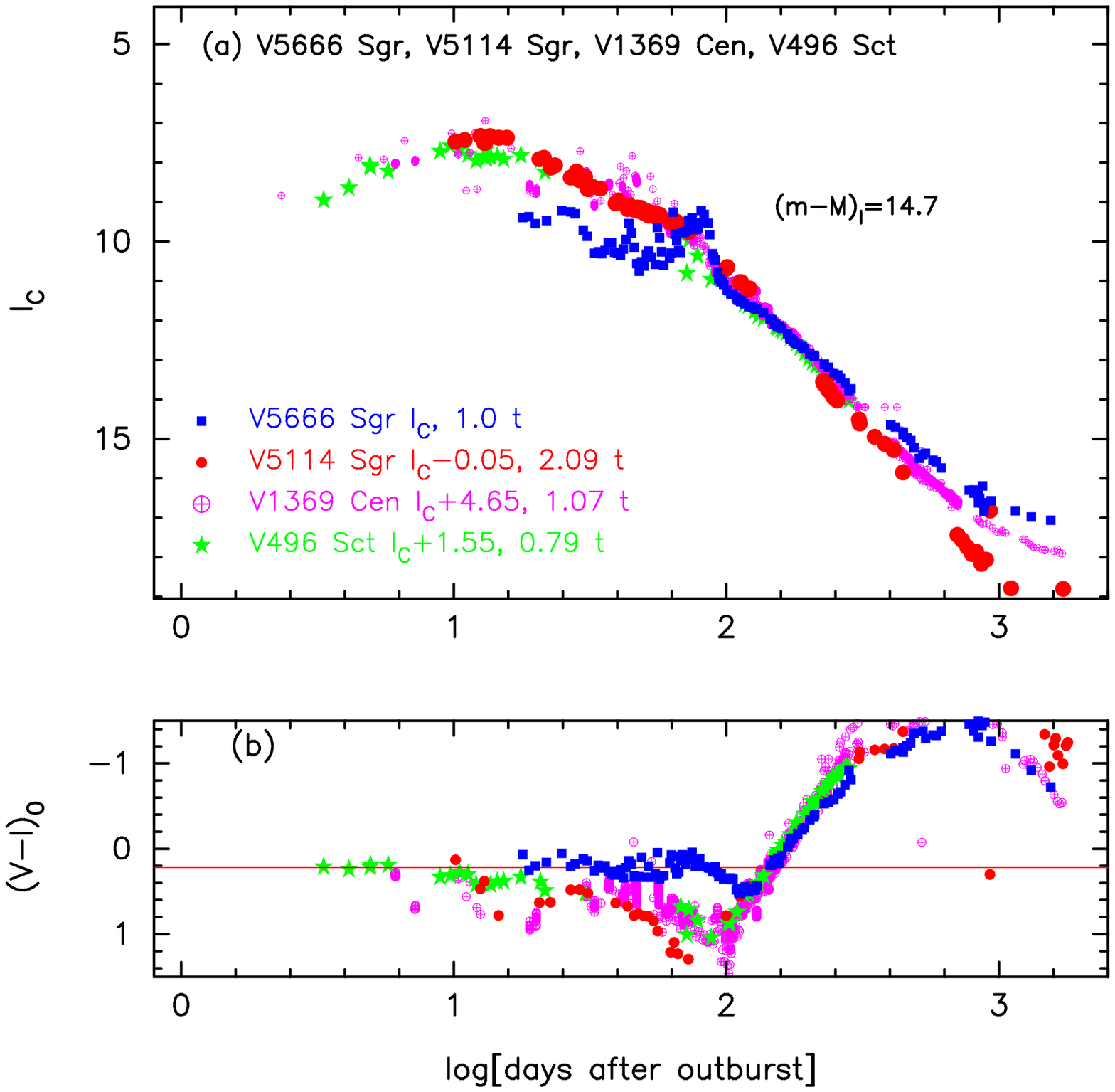}
\caption{
The (a) $I_{\rm C}$ light curve and (b) $(V-I_{\rm C})_0$ color curve
of V5666~Sgr as well as those of V5114~Sgr, V1369~Cen, and V496~Sct.
\label{v5666_sgr_v5114_sgr_v1369_cen_v496_sct_i_vi_color_logscale}}
\end{figure}


\begin{figure}
\plotone{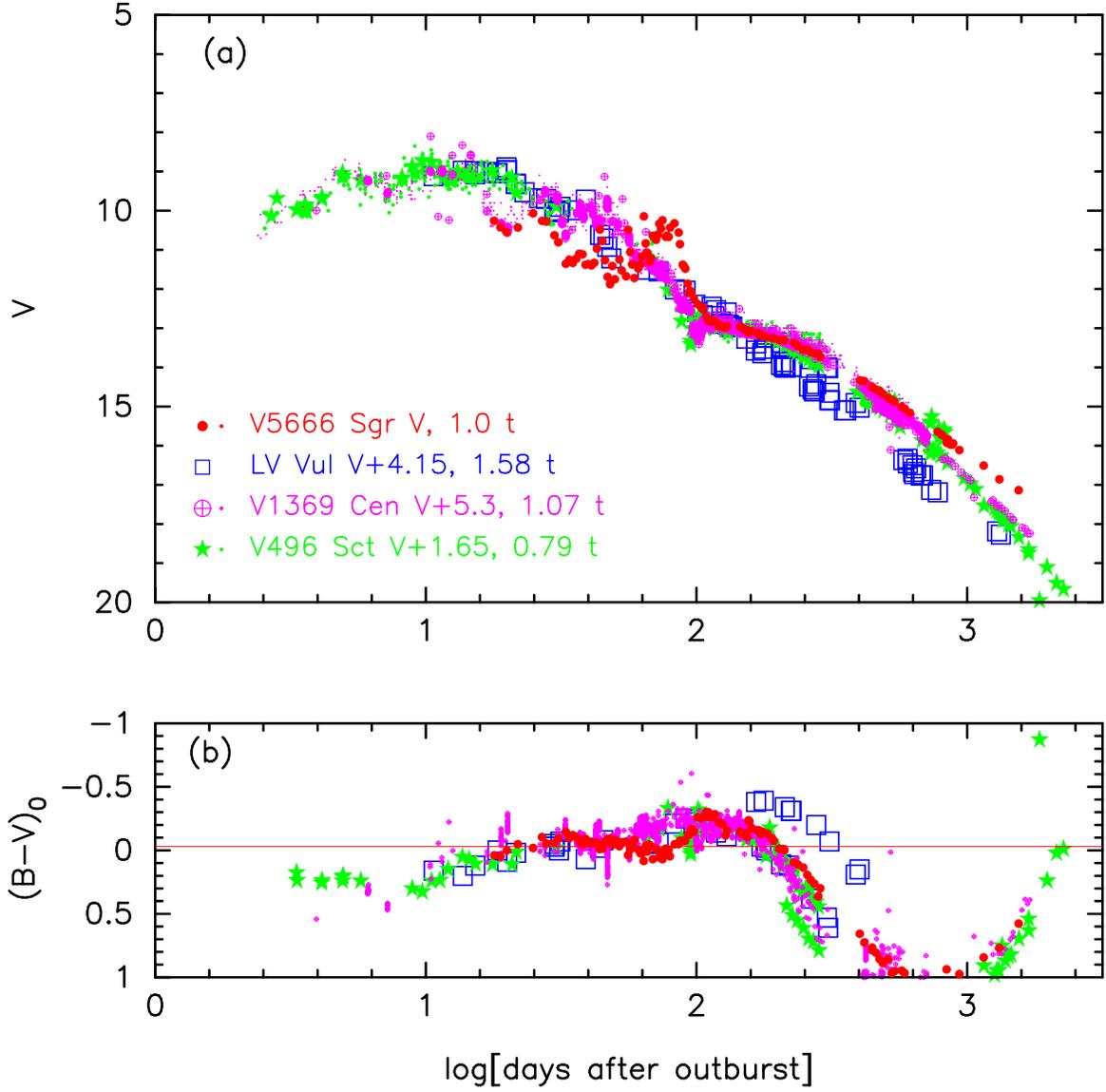}
\caption{
The (a) $V$ light and (b) $(B-V)_0$ color curves of V5666~Sgr
as well as those of LV~Vul, V1369~Cen, and V496~Sct.
\label{v5666_sgr_v1369_cen_v496_sct_lv_vul_v_bv_color_logscale_no2}}
\end{figure}


\begin{figure}
\plottwo{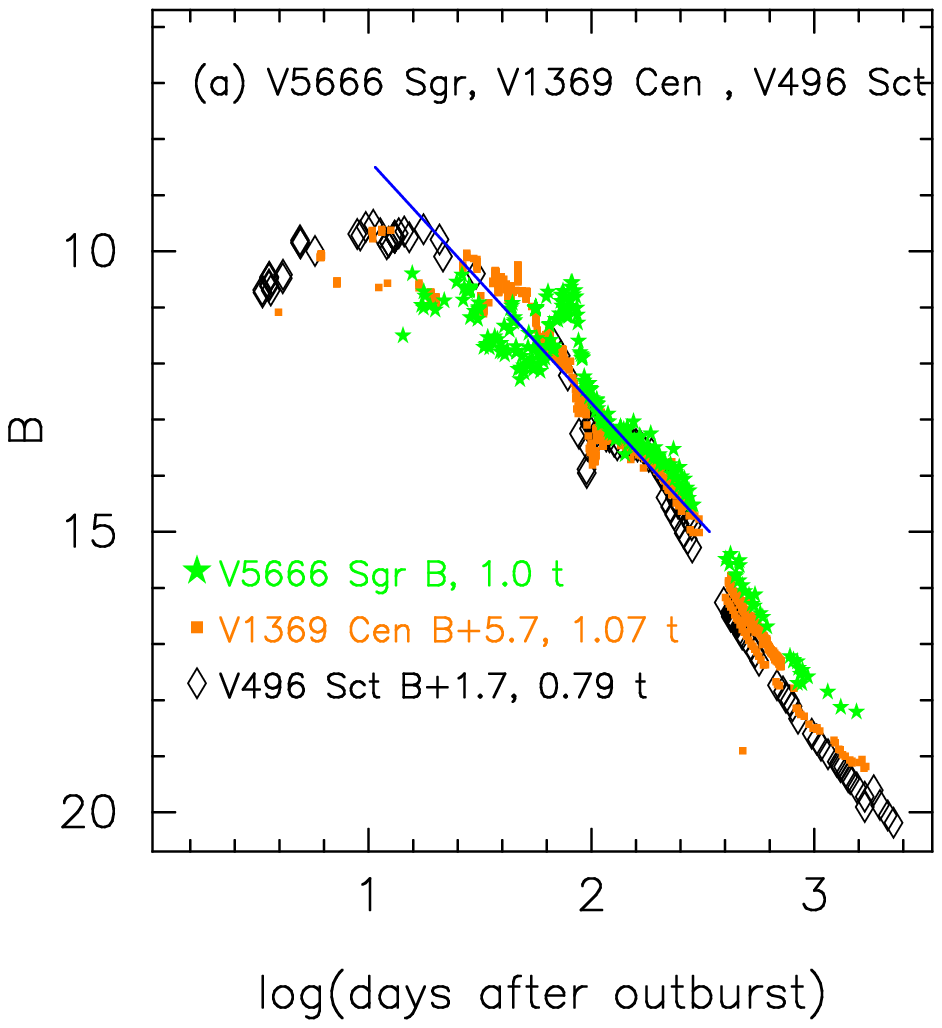}{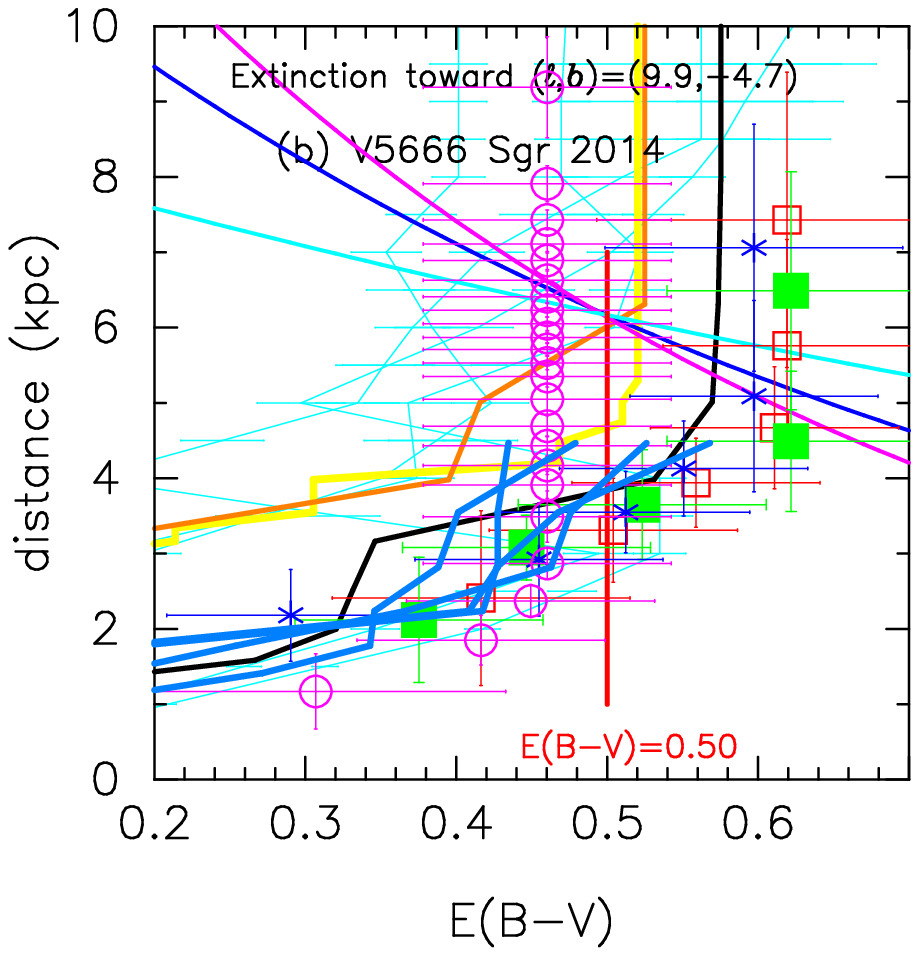}
\caption{
(a) The time-stretched $B$ light curves of V5666~Sgr, V1369~Cen,
and V496~Sct.
(b) Various distance-reddening relations toward V5666~Sgr.
The thin solid lines of magenta, blue, and cyan denote the distance-reddening
relations given by $(m-M)_B= 16.0$, $(m-M)_V= 15.5$, and $(m-M)_I= 14.7$, 
respectively.  
\label{v5666_sgr_v1369_cen_v496_sct_b_only_logscale}}
\end{figure}

\subsection{V5666~Sgr 2014}
\label{v5666_sgr_bvi}
We have reanalyzed the $BVI_{\rm C}$ multi-band 
light/color curves of V5666~Sgr based on the time-stretching method.  
Figure \ref{v5666_sgr_v5114_sgr_v1369_cen_v496_sct_i_vi_color_logscale}
shows the (a) $I_{\rm C}$ light and (b) $(V-I_{\rm C})_0$ color curves
of V5666~Sgr as well as V5114~Sgr, V1369~Cen, and V496~Sct.
The $BVI_{\rm C}$ data of V5666~Sgr are taken from AAVSO, VSOLJ, and SMARTS.
We adopt the color excess of $E(B-V)= 0.50$ after \citet{hac19ka}
in order to overlap the $(V-I)_0$ color curve of V5666~Sgr
with the other novae, as shown in Figure 
\ref{v5666_sgr_v5114_sgr_v1369_cen_v496_sct_i_vi_color_logscale}(b).
We apply Equation (8) of \citet{hac19ka} for the $I$ band to Figure
\ref{v5666_sgr_v5114_sgr_v1369_cen_v496_sct_i_vi_color_logscale}(a)
and obtain
\begin{eqnarray}
(m&-&M)_{I, \rm V5666~Sgr} \cr
&=& ((m - M)_I + \Delta I_{\rm C})
_{\rm V5114~Sgr} - 2.5 \log 2.09 \cr
&=& 15.55 - 0.05\pm0.2 - 0.8 = 14.7\pm0.2 \cr
&=& ((m - M)_I + \Delta I_{\rm C})
_{\rm V1369~Cen} - 2.5 \log 1.07 \cr
&=& 10.11 + 4.65\pm0.2 - 0.075 = 14.68\pm0.2 \cr
&=& ((m - M)_I + \Delta I_{\rm C})
_{\rm V496~Sct} - 2.5 \log 0.79 \cr
&=& 12.9 + 1.55\pm0.2 + 0.25 = 14.7\pm0.2,
\label{distance_modulus_i_vi_v5666_sgr}
\end{eqnarray}
where we adopt
$(m-M)_{I, \rm V5114~Sgr}=15.55$ from Appendix \ref{v5114_sgr_ubvi},
$(m-M)_{I, \rm V1369~Cen}=10.11$ from \citet{hac19ka}, and
$(m-M)_{I, \rm V496~Sct}=12.9$ in Appendix \ref{v496_sct_bvi}. 
Thus, we obtain $(m-M)_{I, \rm V5666~Sgr}= 14.7\pm0.2$.

Figure \ref{v5666_sgr_v1369_cen_v496_sct_lv_vul_v_bv_color_logscale_no2}
shows (a) the $V$ light curves of V5666~Sgr, LV~Vul, V1369~Cen, and V496~Sct,
and (b) their $(B-V)_0$ color curves.
From Equation (4) of \citet{hac19ka}, we have the relation of
\begin{eqnarray}
(m&-&M)_{V, \rm V5666~Sgr} \cr
&=& (m-M + \Delta V)_{V, \rm LV~Vul} - 2.5 \log 1.58 \cr
&=& 11.85 + 4.15\pm0.2 - 0.5 = 15.5\pm0.2 \cr
&=& (m-M + \Delta V)_{V, \rm V1369~Cen} - 2.5 \log 1.07 \cr
&=& 10.25 + 5.3\pm0.2 - 0.075  = 15.48\pm0.2 \cr
&=& (m-M + \Delta V)_{V, \rm V496~Sct} - 2.5 \log 0.79 \cr
&=& 13.6 + 1.65\pm0.2 + 0.25 = 15.5\pm0.2,
\label{distance_modulus_v5666_sgr_v}
\end{eqnarray}
Thus, we obtained $(m-M)_{V, \rm V5666~Sgr}=15.5\pm0.1$.

Figure \ref{v5666_sgr_v1369_cen_v496_sct_b_only_logscale}(a) shows
the time-stretched $B$ light curves of V5666~Sgr, V1369~Cen, and V496~Sct. 
Applying Equation (7) of \citet{hac19ka} for the $B$ band to Figure
\ref{v5666_sgr_v1369_cen_v496_sct_b_only_logscale}(a),
we have the relation of
\begin{eqnarray}
(m&-&M)_{B, \rm V5666~Sgr} \cr
&=& \left( (m-M)_B + \Delta B\right)_{\rm V1369~Cen} - 2.5 \log 1.07 \cr
&=& 10.36 + 5.7\pm0.2 - 0.075 = 15.98\pm0.2 \cr
&=& \left( (m-M)_B + \Delta B\right)_{\rm V496~Sct} - 2.5 \log 0.79 \cr
&=& 14.05 + 1.7\pm0.2 + 0.25 = 16.0\pm0.2,
\label{distance_modulus_v5666_sgr_v496_sct_v1369_cen_b}
\end{eqnarray}
where we adopt $(m-M)_{B, \rm V1369~Cen}=10.25 + 1.0\times 0.11=10.36$
from \citet{hac19ka} and $(m-M)_{B, \rm V496~Sct}=13.6 + 1.0\times 0.45=
14.05$ from Appendix \ref{v496_sct_bvi}.
Thus, we obtain $(m-M)_B=15.99\pm0.1$ for V5666~Sgr.

Figure \ref{v5666_sgr_v1369_cen_v496_sct_b_only_logscale}(b) shows
the three distance moduli.
These three lines cross at $d=6.2$~kpc and $E(B-V)=0.50$.
This crossing point is consistent with the distance-reddening relations
given by \citet{mar06} and \citet[][thick solid orange line]{gre18}.


\begin{figure}
\plotone{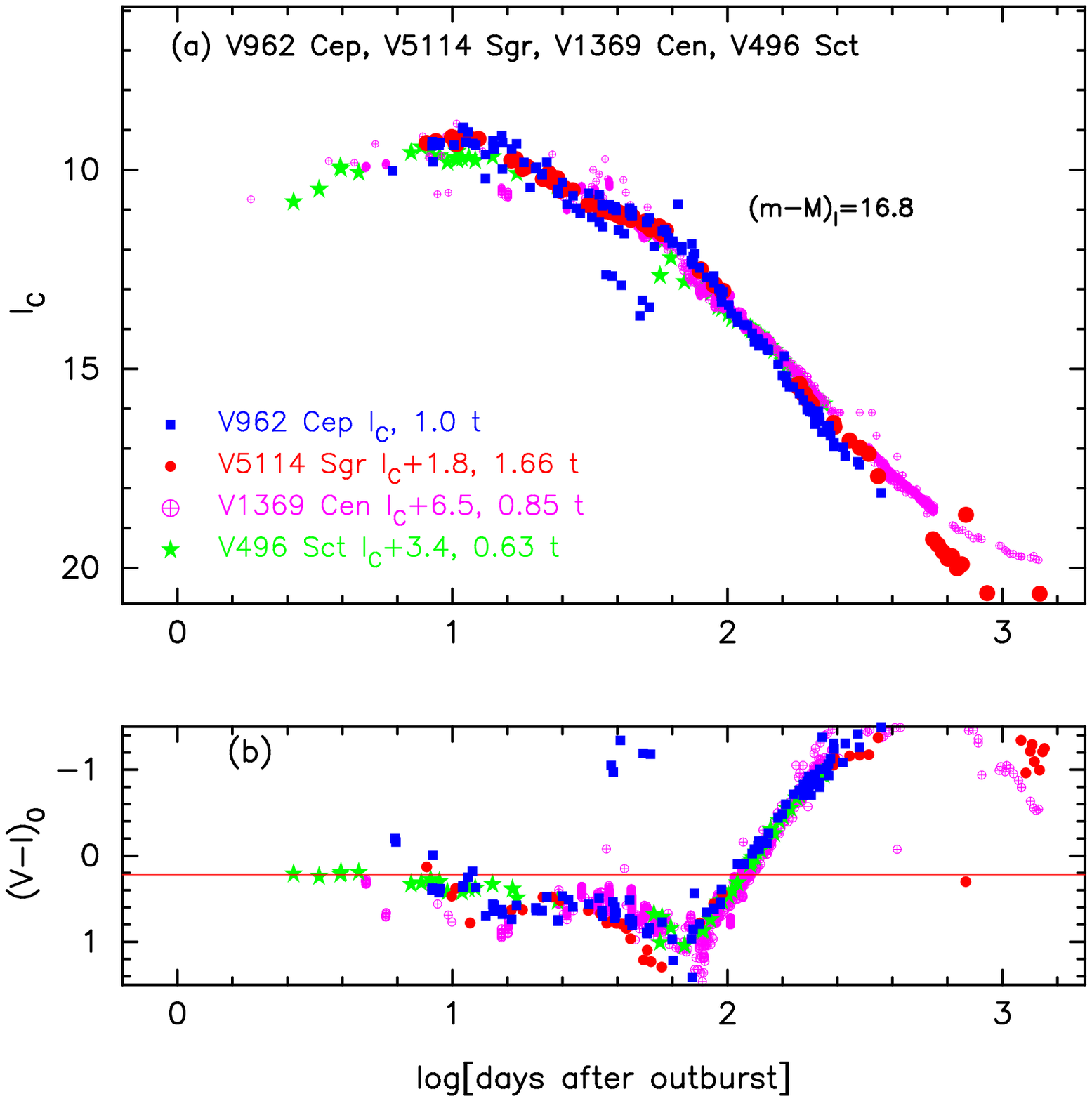}
\caption{
The (a) $I_{\rm C}$ light curve and (b) $(V-I_{\rm C})_0$ color curve
of V962~Cep as well as those of V5114~Sgr, V1369~Cen, and V496~Sct.
\label{v962_cep_v5114_sgr_v1369_cen_v496_sct_i_vi_color_logscale}}
\end{figure}


\begin{figure}
\plotone{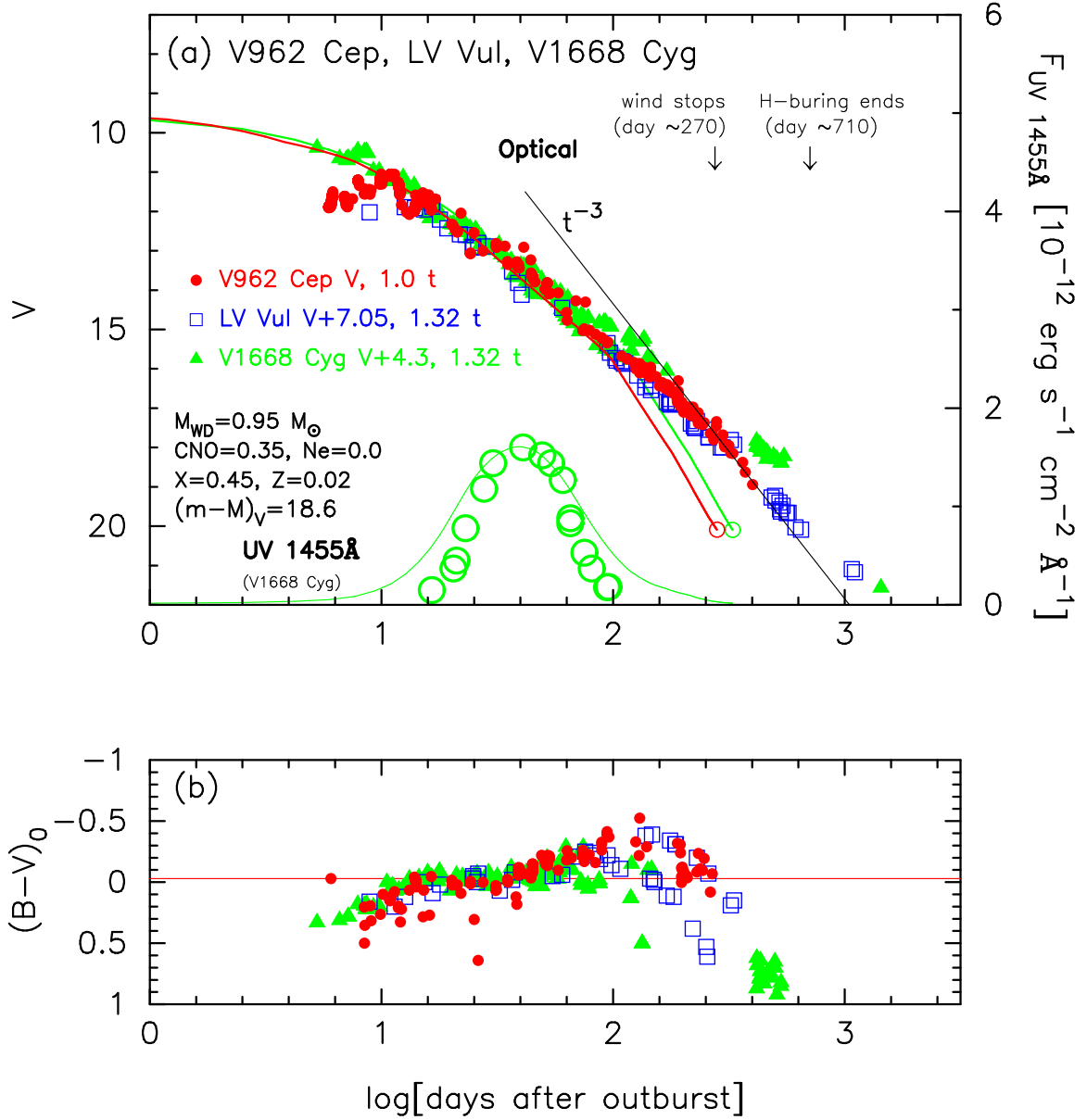}
\caption{
The (a) $V$ light and (b) $(B-V)_0$ color curves of V962~Cep
as well as those of LV~Vul and V1668~Cyg.
In panel (a), we add a $0.95~M_\sun$ WD model (CO3, solid red line)
for V962~Cep as well as a $0.98~M_\sun$ WD model (CO3, solid green lines)
for V1668~Cyg.
\label{v962_cep_v1668_cyg_lv_vul_v_bv_logscale}}
\end{figure}


\begin{figure}
\plottwo{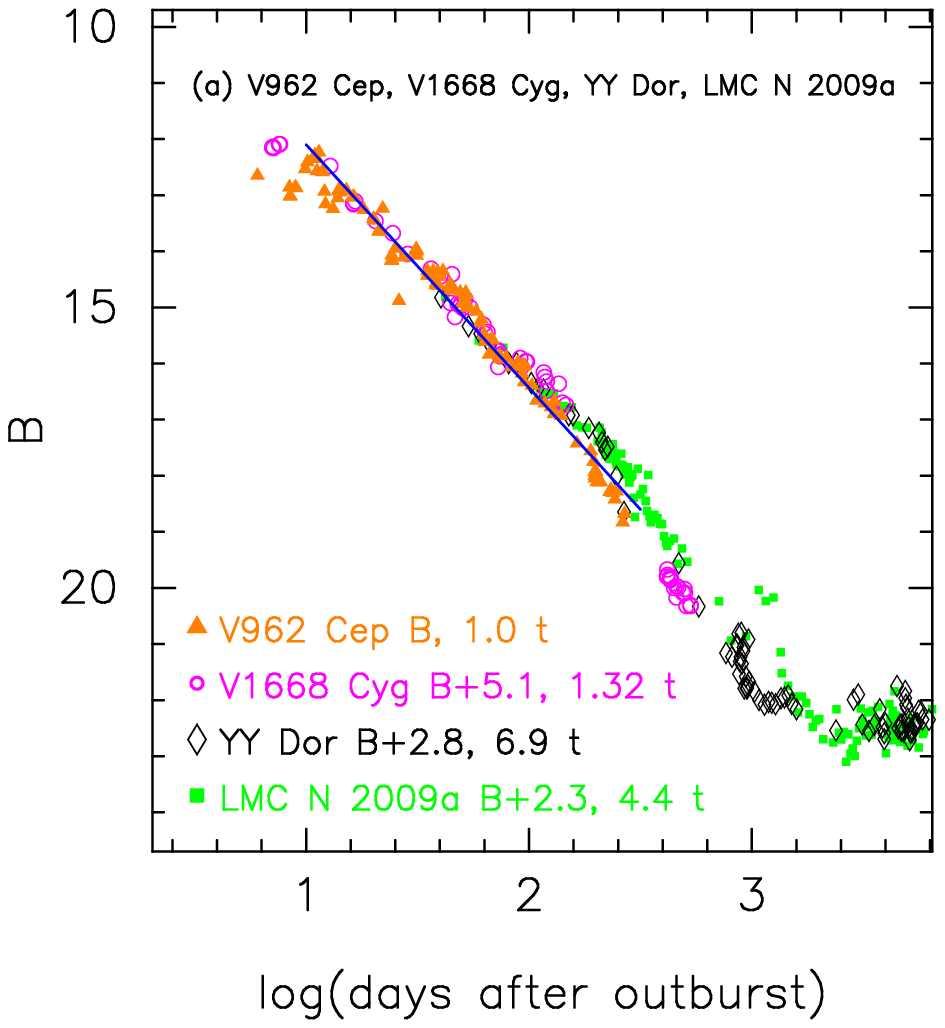}{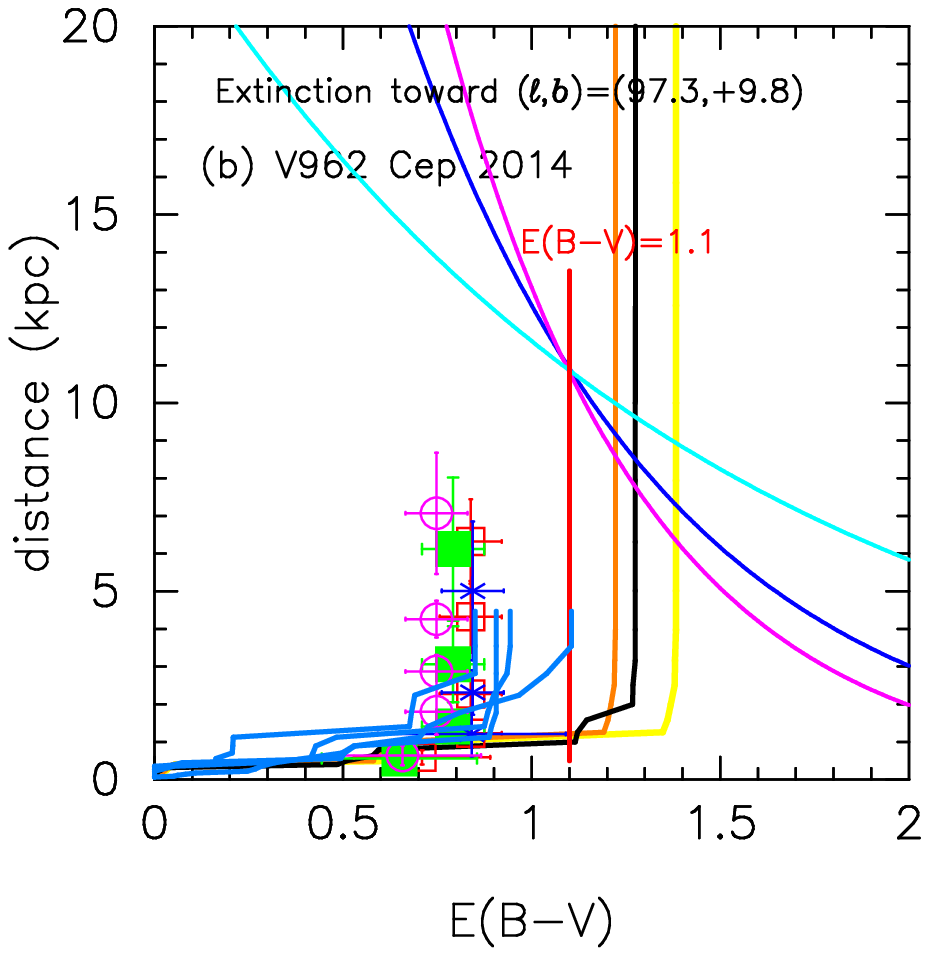}
\caption{
(a) The time-stretched $B$ light curves of V962~Cep as well as 
those of V1668~Cyg, YY~Dor, and LMC~N~2009a.
(b) Various distance-reddening relations toward V962~Cep.
The thin solid lines of magenta, blue, and cyan denote the distance-reddening
relations given by $(m-M)_B= 19.68$, $(m-M)_V= 18.6$, and $(m-M)_I= 16.81$, 
respectively.  
\label{v962_cep_v1668_cyg_yy_dor_lmcn_2009a_b_only}}
\end{figure}

\subsection{V962~Cep 2014}
\label{v962_cep_bvi}
We have reanalyzed the $BVI_{\rm C}$ multi-band 
light/color curves of V962~Cep based on the time-stretching method.  
Figure \ref{v962_cep_v5114_sgr_v1369_cen_v496_sct_i_vi_color_logscale}
shows the (a) $I_{\rm C}$ light and (b) $(V-I_{\rm C})_0$ color curves
of V962~Cep as well as V5114~Sgr, V1369~Cen, and V496~Sct.
The $BVI_{\rm C}$ data of V962~Cep are taken from AAVSO and VSOLJ.
We adopt the color excess of $E(B-V)= 1.10$ after \citet{hac19kb}.
We apply Equation (8) of \citet{hac19ka} for the $I$ band to Figure
\ref{v962_cep_v5114_sgr_v1369_cen_v496_sct_i_vi_color_logscale}(a)
and obtain
\begin{eqnarray}
(m&-&M)_{I, \rm V962~Cep} \cr
&=& ((m - M)_I + \Delta I_{\rm C})
_{\rm V5114~Sgr} - 2.5 \log 1.66 \cr
&=& 15.55 + 1.8\pm0.2 - 0.55 = 16.8\pm0.2 \cr
&=& ((m - M)_I + \Delta I_{\rm C})
_{\rm V1369~Cen} - 2.5 \log 0.85 \cr
&=& 10.11 + 6.5\pm0.2 + 0.175 = 16.79\pm0.2 \cr
&=& ((m - M)_I + \Delta I_{\rm C})
_{\rm V496~Sct} - 2.5 \log 0.63 \cr
&=& 12.9 + 3.4\pm0.2 + 0.5 = 16.8\pm0.2,
\label{distance_modulus_i_vi_v962_cep}
\end{eqnarray}
where we adopt
$(m-M)_{I, \rm V5114~Sgr}=15.55$ from Appendix \ref{v5114_sgr_ubvi},
$(m-M)_{I, \rm V1369~Cen}=10.11$ from \citet{hac19ka}, and
$(m-M)_{I, \rm V496~Sct}=12.9$ in Appendix \ref{v496_sct_bvi}.
Thus, we obtain $(m-M)_{I, \rm V962~Cep}= 16.8\pm0.2$.

Figure \ref{v962_cep_v1668_cyg_lv_vul_v_bv_logscale} shows
the (a) $V$ light and (b) $(B-V)_0$ color curves of V962~Cep 
as well as those of LV~Vul and V1668~Cyg.
From Equation (4) of \citet{hac19ka}, we have the relation of
\begin{eqnarray}
(m&-&M)_{V, \rm V962~Cep} \cr
&=& (m - M + \Delta V)_{V, \rm LV~Vul} - 2.5 \log 1.32 \cr
&=& 11.85 + 7.05\pm0.2 - 0.30 = 18.6\pm0.2 \cr
&=& (m - M + \Delta V)_{V, \rm V1668~Cyg} - 2.5 \log 1.32 \cr
&=& 14.6 + 4.3\pm0.2 - 0.30 = 18.6\pm0.2,
\label{distance_modulus_v_bv_v962_cep}
\end{eqnarray}
where we adopt $(m-M)_{V, \rm LV~Vul}=11.85$ and
$(m-M)_{V, \rm V1668~Cyg}=14.6$ both from \citet{hac19ka}.
Thus, we obtain $(m-M)_{V, \rm V962~Cep}=18.6\pm0.1$ 
and $\log f_{\rm s}= \log 1.32 = +0.12$ against LV~Vul.

Figure \ref{v962_cep_v1668_cyg_yy_dor_lmcn_2009a_b_only}(a)
shows the $B$ light curve of V962~Cep
together with those of V1668~Cyg, YY~Dor, and LMC~N~2009a.
We apply Equation (7) of \citet{hac19ka} for the $B$ band to Figure
\ref{v962_cep_v1668_cyg_yy_dor_lmcn_2009a_b_only}(a)
and obtain
\begin{eqnarray}
(m&-&M)_{B, \rm V962~Cep} \cr
&=& ((m - M)_B + \Delta B)_{\rm V1668~Cyg} - 2.5 \log 1.32 \cr
&=& 14.9 + 5.1\pm0.2 - 0.3 = 19.7\pm0.2 \cr
&=& ((m - M)_B + \Delta B)_{\rm YY~Dor} - 2.5 \log 6.9 \cr
&=& 18.98 + 2.8\pm0.2 - 2.1 = 19.68\pm0.2 \cr
&=& ((m - M)_B + \Delta B)_{\rm LMC~N~2009a} - 2.5 \log 4.4 \cr
&=& 18.98 + 2.3\pm0.2 - 1.6 = 19.68\pm0.2.
\label{distance_modulus_b_v962_cep}
\end{eqnarray}
We have $(m-M)_{B, \rm V962~Cep}= 19.68\pm0.1$.

We plot $(m-M)_B= 19.68$, $(m-M)_V= 18.6$, and $(m-M)_I= 16.81$,
which cross at $d=10.9$~kpc and $E(B-V)=1.10$, in Figure
\ref{v962_cep_v1668_cyg_yy_dor_lmcn_2009a_b_only}(b).
This crossing point is consistent with the distance-reddening relations
given by \citet[][thick solid cyan-blue lines]{chen19}.
Thus, we obtain $E(B-V)=1.10\pm0.10$ and $d=10.9\pm2$~kpc.


\begin{figure}
\plotone{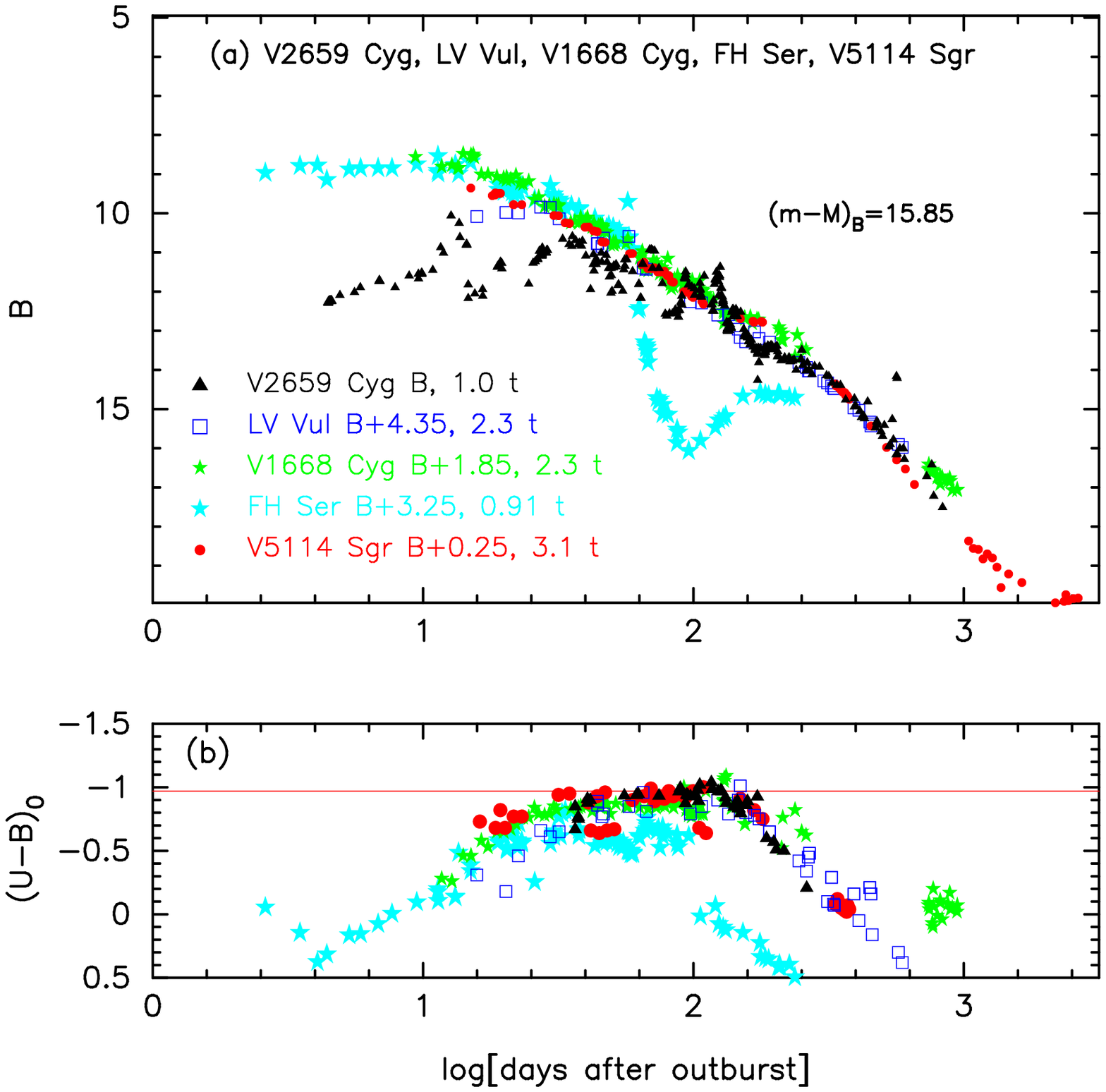}
\caption{
The (a) $B$ light and (b) $(U-B)_0$ color curves of V2659~Cyg
as well as those of LV~Vul, V1668~Cyg, FH~Ser, and V5114~Sgr.
The $UBV$ data of V2659~Cyg are taken from \citet{bur15}.
\label{v2659_cyg_v5114_sgr_v1668_cyg_lv_vul_b_ub_color_logscale_no2}}
\end{figure}


\begin{figure}
\plotone{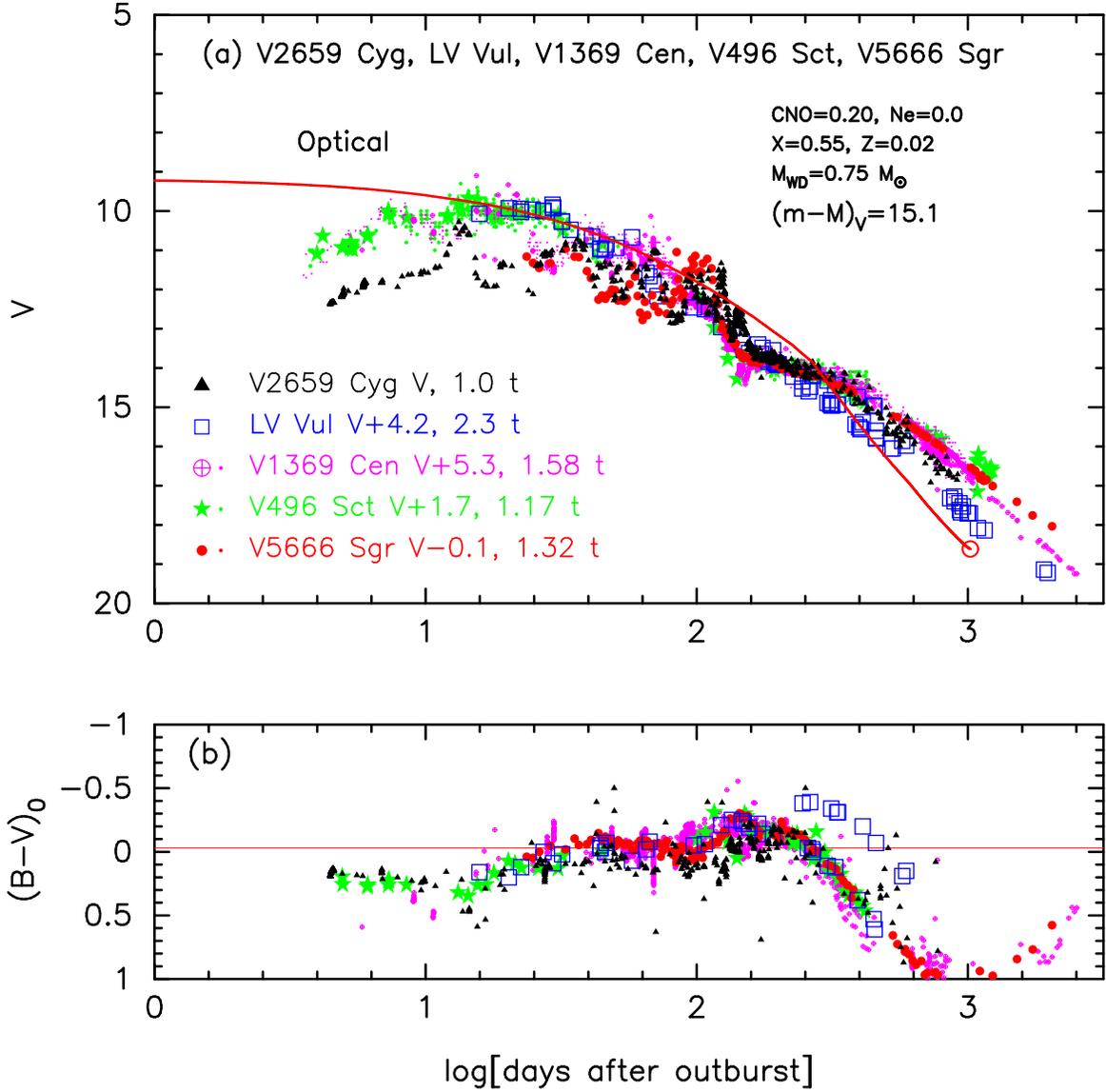}
\caption{
The (a) $V$ light and (b) $(B-V)_0$ color curves of V2659~Cyg
as well as those of LV~Vul, V1369~Cen, V496~Sct, and V5666~Sgr.
We add a $0.75~M_\sun$ WD model (CO4, solid red line) for V2659~Cyg.
\label{v2659_cyg_v5666_sgr_v1369_cen_v496_sct_lv_vul_v_bv_color_logscale_no2}}
\end{figure}


\begin{figure}
\plotone{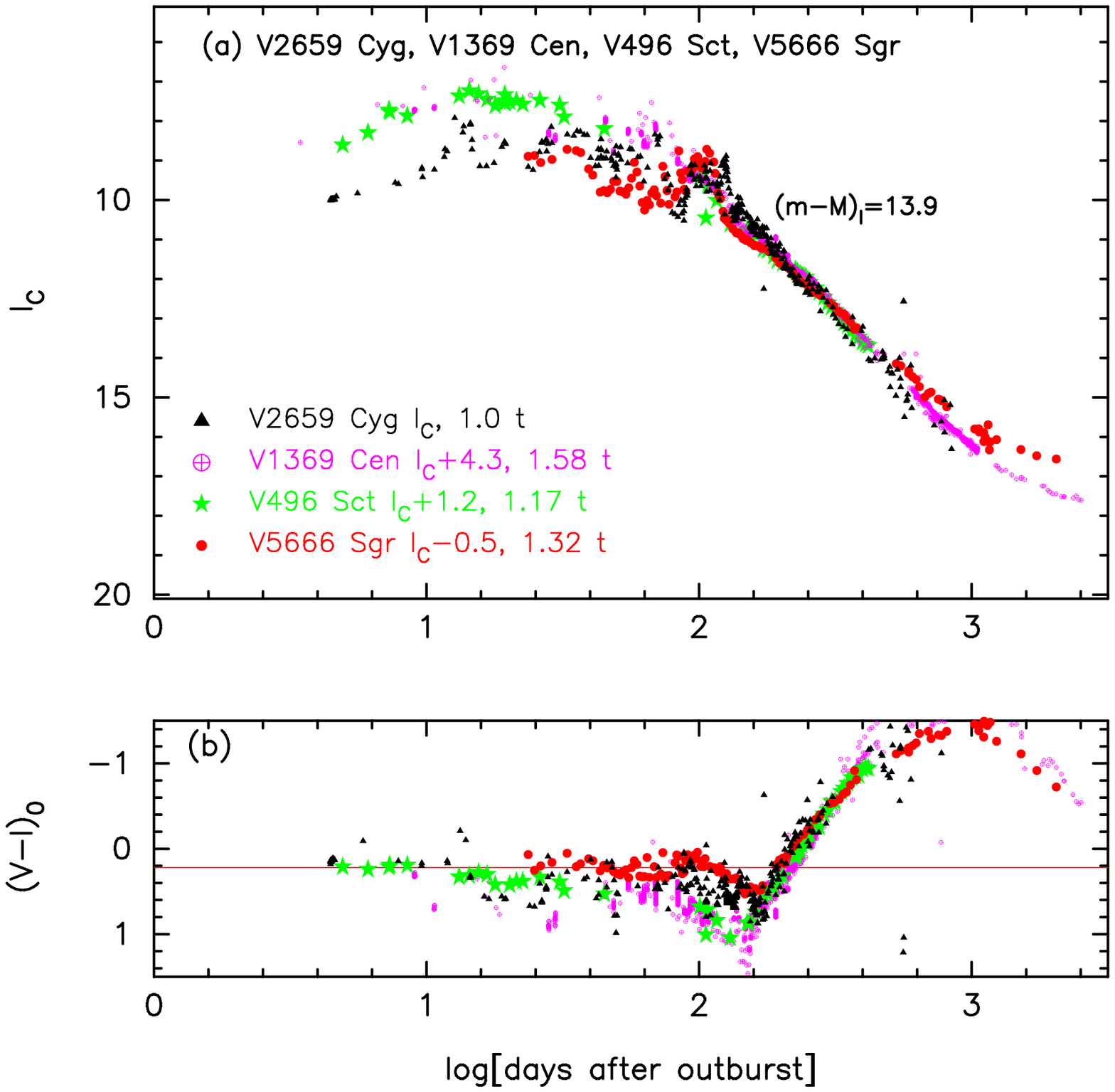}
\caption{
The (a) $I_{\rm C}$ light and (b) $(V-I_{\rm C})_0$ color curves
of V2659~Cyg as well as those of V1369~Cen, V496~Sct, and V5666~Sgr.
\label{v2659_cyg_v5666_sgr_v1369_cen_v496_sct_lv_vul_i_vi_color_logscale}}
\end{figure}

\subsection{V2659~Cyg 2014}
\label{v2659_cyg_ubvi}
\citet{hac19kb} analyzed the $BVI_{\rm C}$ light curves of V2659~Cyg
and obtained the color excess, distance moduli 
in $BVI_{\rm C}$ bands, distance, and timescaling factor.
The $UBV$ light curves are now available in 
\citet{bur15}, so we reanalyze the above various parameters
based on the $UBV$ data.   Figure 
\ref{v2659_cyg_v5114_sgr_v1668_cyg_lv_vul_b_ub_color_logscale_no2}
shows the (a) $B$ light and (b) $(U-B)_0$ color curves of V2659~Cyg
as well as those of LV~Vul, V1668~Cyg, FH~Ser, and V5114~Sgr.
Applying Equation (7) of \citet{hac19ka} for the $B$ band to Figure
\ref{v2659_cyg_v5114_sgr_v1668_cyg_lv_vul_b_ub_color_logscale_no2}(a), 
we obtain
\begin{eqnarray}
(m&-&M)_{B, \rm V2659~Cyg} \cr
&=& ((m - M)_B + \Delta B)_{\rm LV~Vul} - 2.5 \log 2.3 \cr
&=& 12.45 + 4.35\pm0.2 - 0.925 = 15.87\pm0.2 \cr
&=& ((m - M)_B + \Delta B)_{\rm V1668~Cyg} - 2.5 \log 2.3 \cr
&=& 14.9 + 1.85\pm0.2 - 0.925 = 15.82\pm0.2 \cr 
&=& ((m - M)_B + \Delta B)_{\rm FH~Ser} - 2.5 \log 0.91 \cr
&=& 12.5 + 3.25\pm0.2 + 0.1 = 15.85\pm0.2 \cr 
&=& ((m - M)_B + \Delta B)_{\rm V5114~Sgr} - 2.5 \log 3.1 \cr
&=& 16.85 + 0.25\pm0.2 - 1.225 = 15.87\pm0.2,
\label{distance_modulus_b_v2659_cyg}
\end{eqnarray}
where we adopt 
$(m-M)_{B, \rm LV~Vul}= 12.45$ and $(m-M)_{B, \rm V1668~Cyg}= 14.9$
both from \citet{hac19ka}, and
$(m-M)_{B, \rm V5114~Sgr}= 16.85$ in Appendix \ref{v5114_sgr_ubvi}. 
Here, we have redetermined
$(m-M)_{B, \rm FH~Ser}= 12.5$ from the fitting in Figure
\ref{v2659_cyg_v5114_sgr_v1668_cyg_lv_vul_b_ub_color_logscale_no2}.
Thus, we obtain $(m-M)_{B, \rm V2659~Sgr}= 15.85\pm0.2$.

Figure 
\ref{v2659_cyg_v5666_sgr_v1369_cen_v496_sct_lv_vul_v_bv_color_logscale_no2}
shows the (a) $V$ light and (b) $(B-V)_0$ color curves of V2659~Cyg
as well as LV~Vul, V1369~Cen, V496~Sct, and V5666~Sgr.
From Equation (4) of \citet{hac19ka}, we obtain
\begin{eqnarray}
(m&-&M)_{V, \rm V2659~Cyg} \cr
&=& ((m - M)_V + \Delta V)_{\rm LV~Vul} - 2.5 \log 2.3 \cr
&=& 11.85 + 4.2\pm0.2 - 0.925 = 15.13\pm0.2 \cr
&=& ((m - M)_V + \Delta V)_{\rm V1369~Cen} - 2.5 \log 1.58 \cr
&=& 10.25 + 5.3\pm0.2 - 0.5 = 15.05\pm0.2 \cr
&=& ((m - M)_V + \Delta V)_{\rm V496~Sct} - 2.5 \log 1.17 \cr
&=& 13.6 + 1.7\pm0.2 - 0.175 = 15.13\pm0.2 \cr
&=& ((m - M)_V + \Delta V)_{\rm V5666~Sgr} - 2.5 \log 1.32 \cr
&=& 15.5 - 0.1\pm0.2 - 0.3 = 15.1\pm0.2.
\label{distance_modulus_v_v2659_cyg}
\end{eqnarray}
Thus, we obtain $(m-M)_{V, \rm V2659~Cyg}= 15.1\pm0.2$.

Figure
\ref{v2659_cyg_v5666_sgr_v1369_cen_v496_sct_lv_vul_i_vi_color_logscale}
shows the (a) $I_{\rm C}$ light and (b) $(V-I)_0$ color curves of
V2659~Cyg as well as V1369~Cen, V496~Sct, and V5666~Sgr.
Applying Equation (8) of \citet{hac19ka} for the $I$ band to Figure
\ref{v2659_cyg_v5666_sgr_v1369_cen_v496_sct_lv_vul_i_vi_color_logscale}(a), 
we obtain
\begin{eqnarray}
(m&-&M)_{I, \rm V2659~Cyg} \cr
&=& ((m - M)_I + \Delta V)_{\rm V1369~Cen} - 2.5 \log 1.58 \cr
&=& 10.11 + 4.3\pm0.2 - 0.5 = 13.91\pm0.2 \cr
&=& ((m - M)_I + \Delta V)_{\rm V496~Sct} - 2.5 \log 1.17 \cr
&=& 12.9 + 1.2\pm0.2 - 0.175 = 13.92\pm0.2 \cr
&=& ((m - M)_I + \Delta V)_{\rm V5666~Sgr} - 2.5 \log 1.32 \cr
&=& 14.7 - 0.5\pm0.2 - 0.3 = 13.9\pm0.2. 
\label{distance_modulus_i_v2659_cyg}
\end{eqnarray}
Thus, we obtain $(m-M)_{I, \rm V2659~Cyg}= 13.91\pm0.2$.

\begin{figure*}
\plottwo{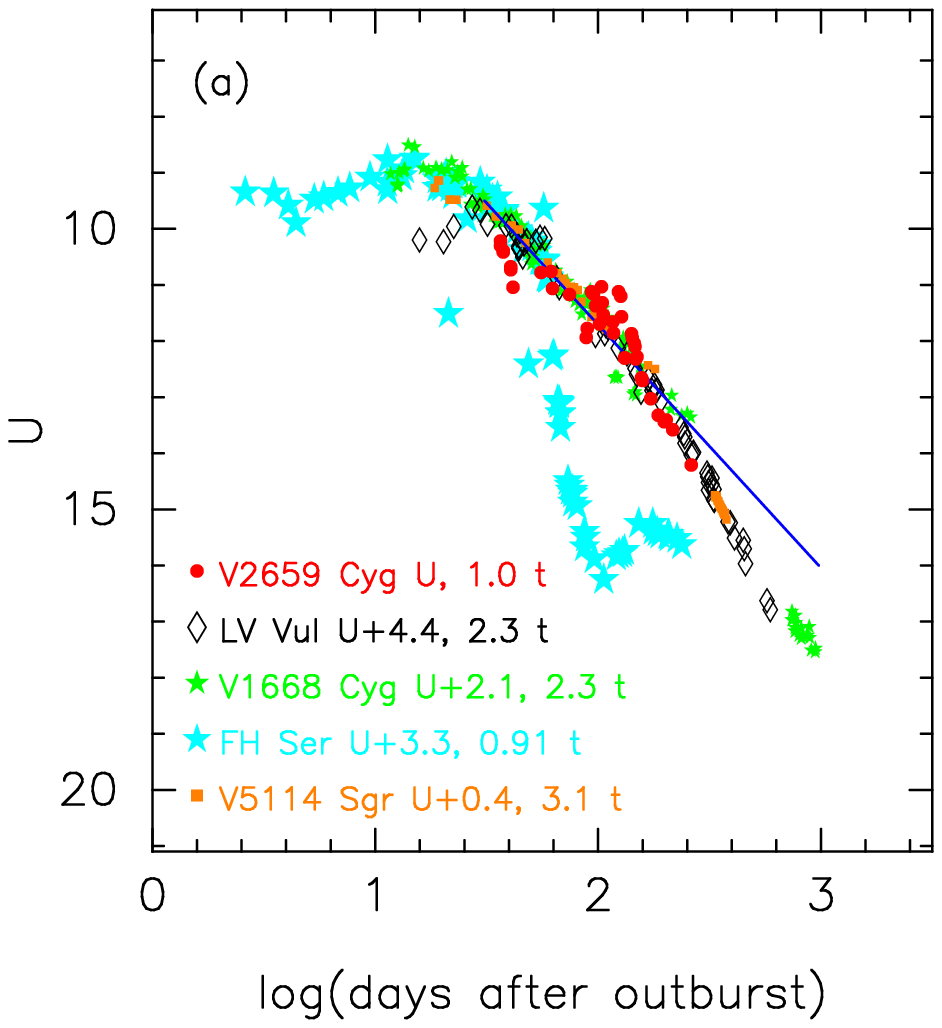}{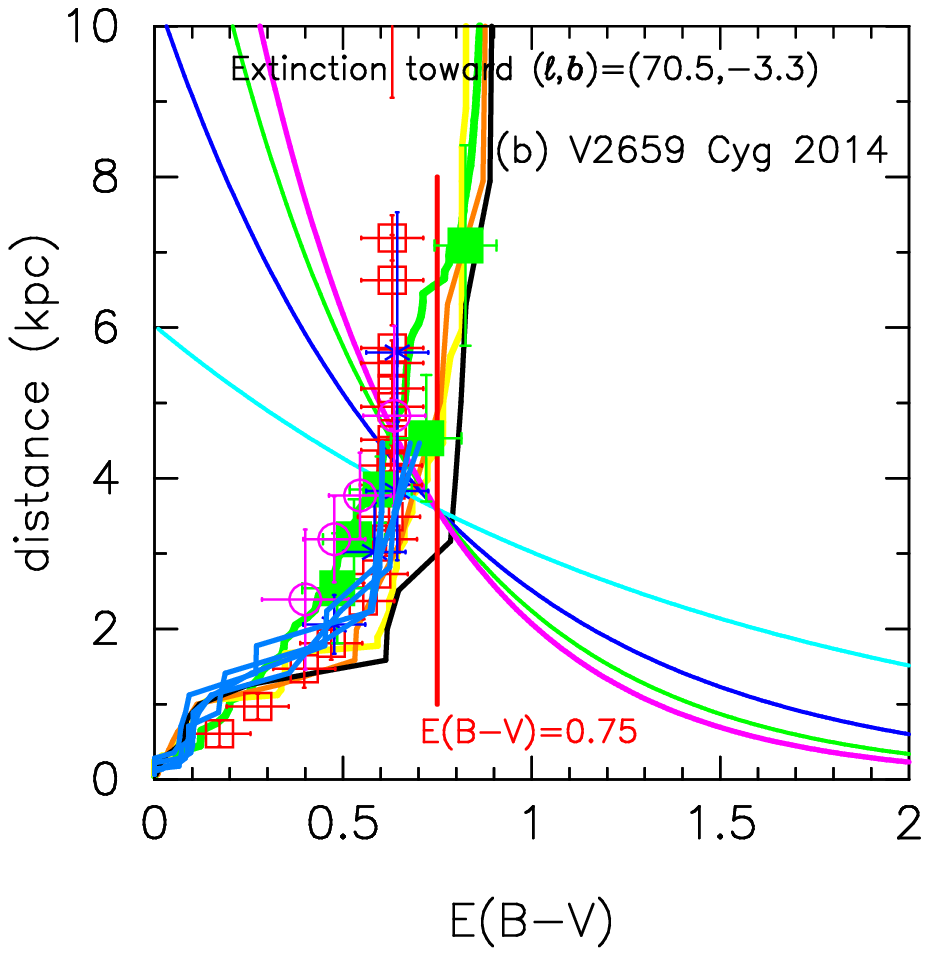}
\caption{
(a) The $U$ light curves of V2659~Cyg as well as 
those of LV~Vul, V1668~Cyg, FH~Ser, and V5114~Sgr.  
The $UBV$ data of V2659~Cyg are taken from \citet{bur15}.  
(b) Various distance-reddening relations toward V2659~Cyg.
The thin solid lines of magenta, green, blue, and cyan denote 
the distance-reddening relations given by $(m-M)_U= 16.31$, 
$(m-M)_B= 15.85$, $(m-M)_V= 15.1$, and $(m-M)_I= 13.9$, respectively.  
\label{distance_reddening_xxxxxx_v2659_cyg}}
\end{figure*}

We further plot the $U$ light curves of V2659~Cyg together
with those of LV~Vul, V1668~Cyg, FH~Ser, and V5114~Sgr in Figure 
\ref{distance_reddening_xxxxxx_v2659_cyg}(a).
We apply Equation (6) of \citet{hac19ka} for the $U$ band to
Figure \ref{distance_reddening_xxxxxx_v2659_cyg}(a)
and obtain
\begin{eqnarray}
(m&-&M)_{U, \rm V2659~Cyg} \cr
&=& ((m - M)_U + \Delta U)_{\rm LV~Vul} - 2.5 \log 2.3 \cr
&=& 12.85 + 4.4\pm0.2 - 0.925 = 16.33\pm0.2 \cr
&=& ((m - M)_U + \Delta U)_{\rm V1668~Cyg} - 2.5 \log 2.3 \cr
&=& 15.1 + 2.1\pm0.2 - 0.925 = 16.28\pm0.2 \cr 
&=& ((m - M)_U + \Delta U)_{\rm FH~Ser} - 2.5 \log 0.91 \cr
&=& 12.9 + 3.3\pm0.2 + 0.1 = 16.3\pm0.2 \cr 
&=& ((m - M)_U + \Delta U)_{\rm V5114~Sgr} - 2.5 \log 3.1 \cr
&=& 17.15 + 0.4\pm0.2 - 1.225 = 16.33\pm0.2, 
\label{distance_modulus_u_v2659_cyg}
\end{eqnarray}
where we adopt 
$(m-M)_{U, \rm LV~Vul}= 12.85$ and
$(m-M)_{U, \rm V1668~Cyg}= 15.10$ both from \citet{hac19ka},
and $(m-M)_{U, \rm V5114~Sgr}= 17.15$ in Appendix \ref{v5114_sgr_ubvi}.
Here, we have redetermined $(m-M)_{U, \rm FH~Ser}= 12.9$ from the
fitting in Figure \ref{distance_reddening_xxxxxx_v2659_cyg}(a).
Thus, we obtain $(m-M)_{U, \rm V2659~Cyg}= 16.31\pm0.2$.

Figure \ref{distance_reddening_xxxxxx_v2659_cyg}(b) shows various 
distance-reddening relations toward V2659~Cyg.
We plot the four distance moduli in $U$, $B$, $V$, and $I_{\rm C}$ bands.
These four lines cross at $d=3.6$~kpc and $E(B-V)=0.75$.
This crossing point is between the distance-reddening relations
given by \citet[][thick solid black line]{gre15} and
\citet[][thick solid orange and yellow lines]{gre18, gre19}.


\begin{figure}
\plotone{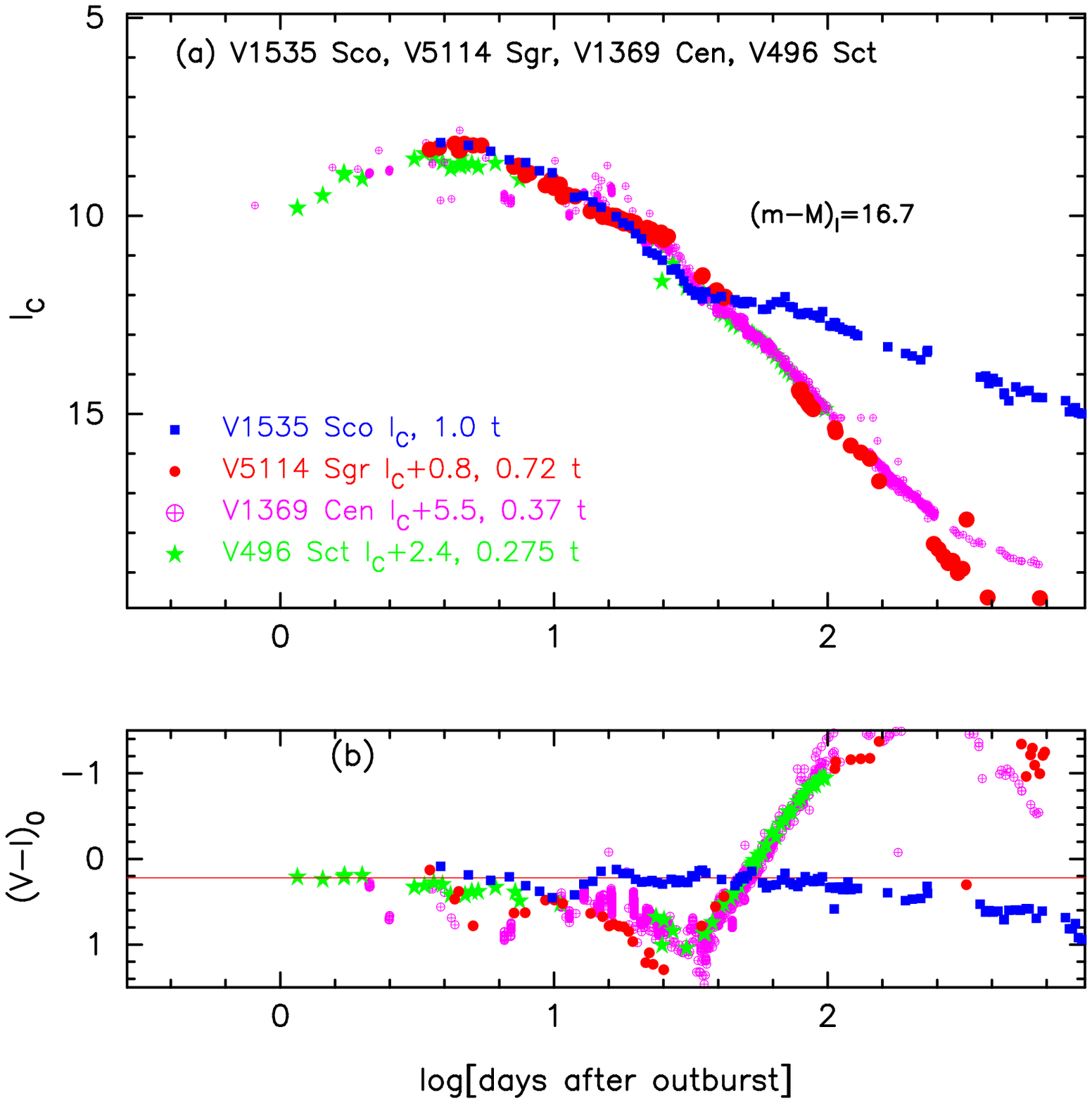}
\caption{
The (a) $I_{\rm C}$ light curve and (b) $(V-I_{\rm C})_0$ color curve
of V1535~Sco as well as those of V5114~Sgr, V1369~Cen, and V496~Sct.
\label{v1535_sco_v5114_sgr_v1369_cen_v496_sct_i_vi_color_logscale}}
\end{figure}


\begin{figure}
\plotone{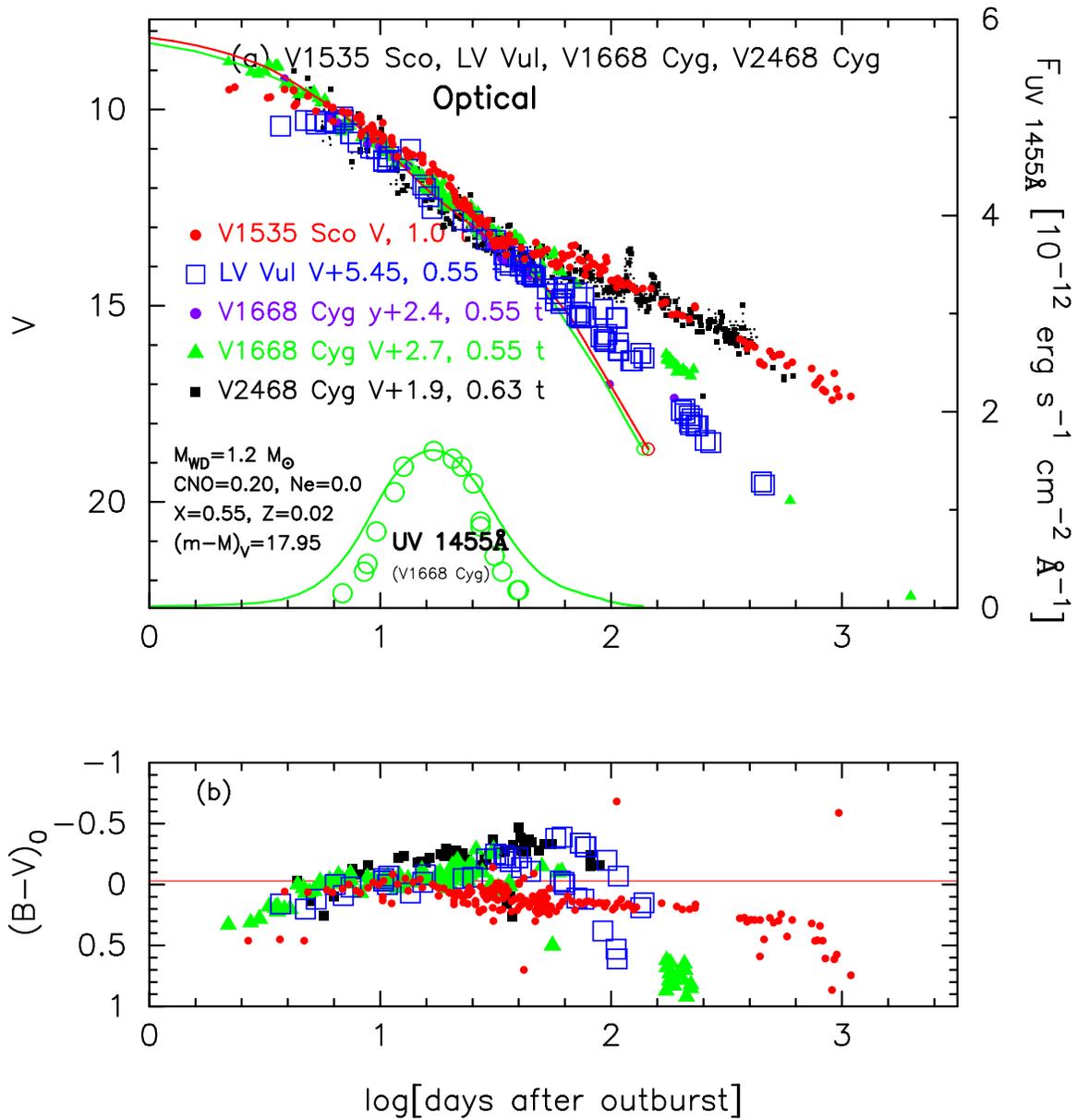}
\caption{
The (a) $V$ light curve and (b) $(B-V)_0$ color curve of V1535~Sco
as well as those of LV~Vul, V1668~Cyg, and V2468~Cyg.  In panel (a),
we add a $1.2~M_\sun$ WD model (CO4, solid red line) for V1535~Sco
as well as a $0.98~M_\sun$ WD model (CO3, solid green lines) for V1668~Cyg.
\label{v1535_sco_lv_vul_v1668_cyg_v2468_cyg_v_bv_logscale_no2}}
\end{figure}


\begin{figure*}
\plottwo{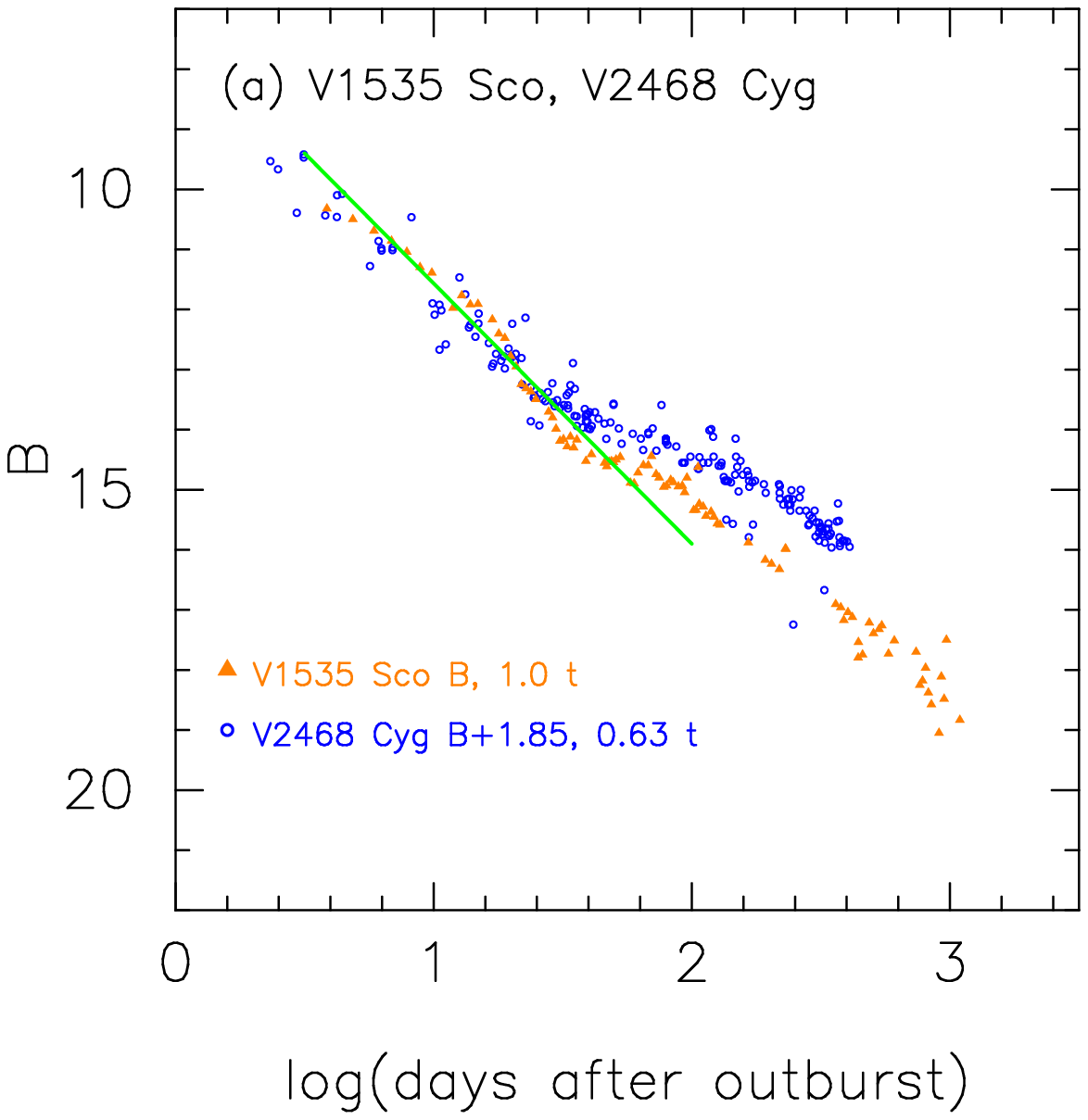}{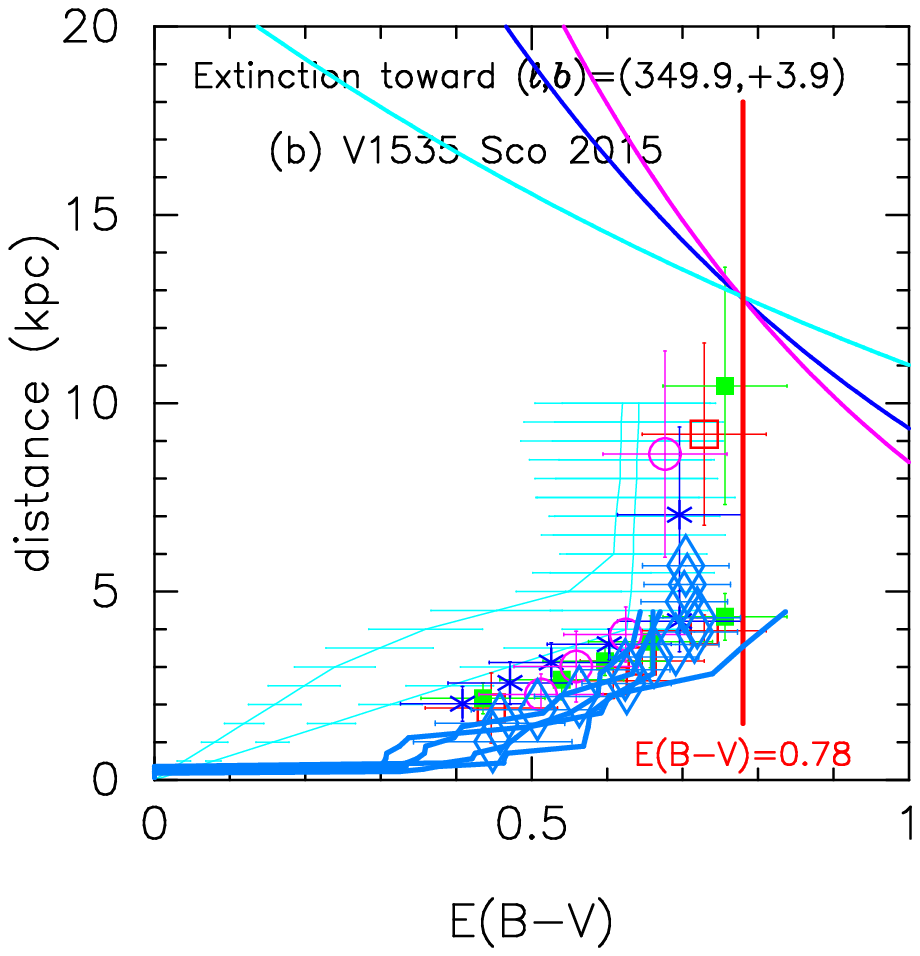}
\caption{
(a) The $B$ light curves of V1535~Sco as well as those of V2468~Cyg.
The $B$ data of V1535~Sco are taken from SMARTS.
(b) Various distance-reddening relations toward V1535~Sco.
The thin solid lines of magenta, blue, and cyan denote the distance-reddening
relations given by  $(m-M)_B= 18.73$, $(m-M)_V= 17.95$, and $(m-M)_I= 16.71$,
respectively.  
\label{distance_reddening_v1535_sco_bvi_xxxxxx}}
\end{figure*}

\subsection{V1535~Sco 2015}
\label{v1535_sco_bvi}
We have reanalyzed the $BVI_{\rm C}$ multi-band 
light/color curves of V1535~Sco based on the time-stretching method.  
Figure \ref{v1535_sco_v5114_sgr_v1369_cen_v496_sct_i_vi_color_logscale}
shows the (a) $I_{\rm C}$ light and (b) $(V-I_{\rm C})_0$ color curves
of V1535~Sco as well as V5114~Sgr, V1369~Cen, and V496~Sct.
The $BVI_{\rm C}$ data of V1535~Sco are taken from SMARTS.
We adopt the color excess of $E(B-V)= 0.78$ after \citet{hac19kb}.
We apply Equation (8) of \citet{hac19ka} for the $I$ band to Figure
\ref{v1535_sco_v5114_sgr_v1369_cen_v496_sct_i_vi_color_logscale}(a)
and obtain
\begin{eqnarray}
(m&-&M)_{I, \rm V1535~Sco} \cr
&=& ((m - M)_I + \Delta I_{\rm C})
_{\rm V5114~Sgr} - 2.5 \log 0.72 \cr
&=& 15.55 + 0.8\pm0.2 + 0.35 = 16.7\pm0.2 \cr
&=& ((m - M)_I + \Delta I_{\rm C})
_{\rm V1369~Cen} - 2.5 \log 0.37 \cr
&=& 10.11 + 5.5\pm0.2 + 1.075 = 16.69\pm0.2 \cr
&=& ((m - M)_I + \Delta I_{\rm C})
_{\rm V496~Sct} - 2.5 \log 0.275 \cr
&=& 12.9 + 2.4\pm0.2 + 1.4 = 16.7\pm0.2,
\label{distance_modulus_i_vi_v1535_sco}
\end{eqnarray}
where we adopt
$(m-M)_{I, \rm V5114~Sgr}=15.55$ from Appendix \ref{v5114_sgr_ubvi},
$(m-M)_{I, \rm V1369~Cen}=10.11$ from \citet{hac19ka}, and
$(m-M)_{I, \rm V496~Sct}=12.9$ in Appendix \ref{v496_sct_bvi}.
Thus, we obtain $(m-M)_{I, \rm V1535~Sco}= 16.7\pm0.2$.

Figure \ref{v1535_sco_lv_vul_v1668_cyg_v2468_cyg_v_bv_logscale_no2}
shows the (a) $V$ light and (b) $(B-V)_0$ color curves of V1535~Sco.
Applying Equation (4) of \citet{hac19ka} to them,
we have the relation
\begin{eqnarray}
(m&-&M)_{V, \rm V1535~Sco} \cr
&=& ((m - M)_V + \Delta V)_{\rm LV~Vul} - 2.5 \log 0.55 \cr
&=& 11.85 + 5.45\pm0.2 + 0.65 = 17.95\pm0.2 \cr
&=& ((m - M)_V + \Delta V)_{\rm V1668~Cyg} - 2.5 \log 0.55 \cr
&=& 14.6 + 2.7\pm0.2 + 0.65 = 17.95\pm0.2 \cr
&=& ((m - M)_V + \Delta V)_{\rm V2468~Cyg} - 2.5 \log 0.63 \cr
&=& 15.55 + 1.9\pm0.2 + 0.5 = 17.95\pm0.2,
\label{distance_modulus_v_bv_v1535_sco}
\end{eqnarray}
where we adopt $(m-M)_{V, \rm LV~Vul}=11.85$ and
$(m-M)_{V, \rm V1668~Cyg}=14.6$, both from \citet{hac19ka},
and $(m-M)_{V, \rm V2468~Cyg}=15.55$ in Appendix \ref{v2468_cyg_bvi}. 
Thus, we obtain $(m-M)_{V, \rm V1535~Sco}=17.95\pm0.1$ 
and $\log f_{\rm s}= \log 0.55 = -0.26$ against LV~Vul.

Figure \ref{distance_reddening_v1535_sco_bvi_xxxxxx}(a)
shows the $B$ light curves of V1535~Sco and V2468~Cyg.
Applying Equation (7) of \citet{hac19ka} to them,
we have the relation
\begin{eqnarray}
(m&-&M)_{B, \rm V1535~Sco} \cr
&=& ((m - M)_B + \Delta B)_{\rm V2468~Cyg} - 2.5 \log 0.63 \cr
&=& 16.38 + 1.85\pm0.2 + 0.5 = 18.73\pm0.2,
\label{distance_modulus_v1535_sco}
\end{eqnarray}
where we adopt 
$(m-M)_{B, \rm V2468~Cyg}=16.38$ in Appendix \ref{v2468_cyg_bvi}. 
Thus, we obtain $(m-M)_{B, \rm V1535~Sco}=18.73\pm0.1$.

We plot $(m-M)_B= 18.73$, $(m-M)_V= 17.95$, and $(m-M)_I= 16.71$, 
which broadly cross at $d=12.8$~kpc and $E(B-V)=0.78$, in Figure
\ref{distance_reddening_v1535_sco_bvi_xxxxxx}(b).
Thus, we have $E(B-V)=0.78\pm0.05$ and $d=12.8\pm1$~kpc.
This crossing point is consistent with the distance-reddening relations
given by \citet[][filled green squares]{mar06}.


\begin{figure}
\plotone{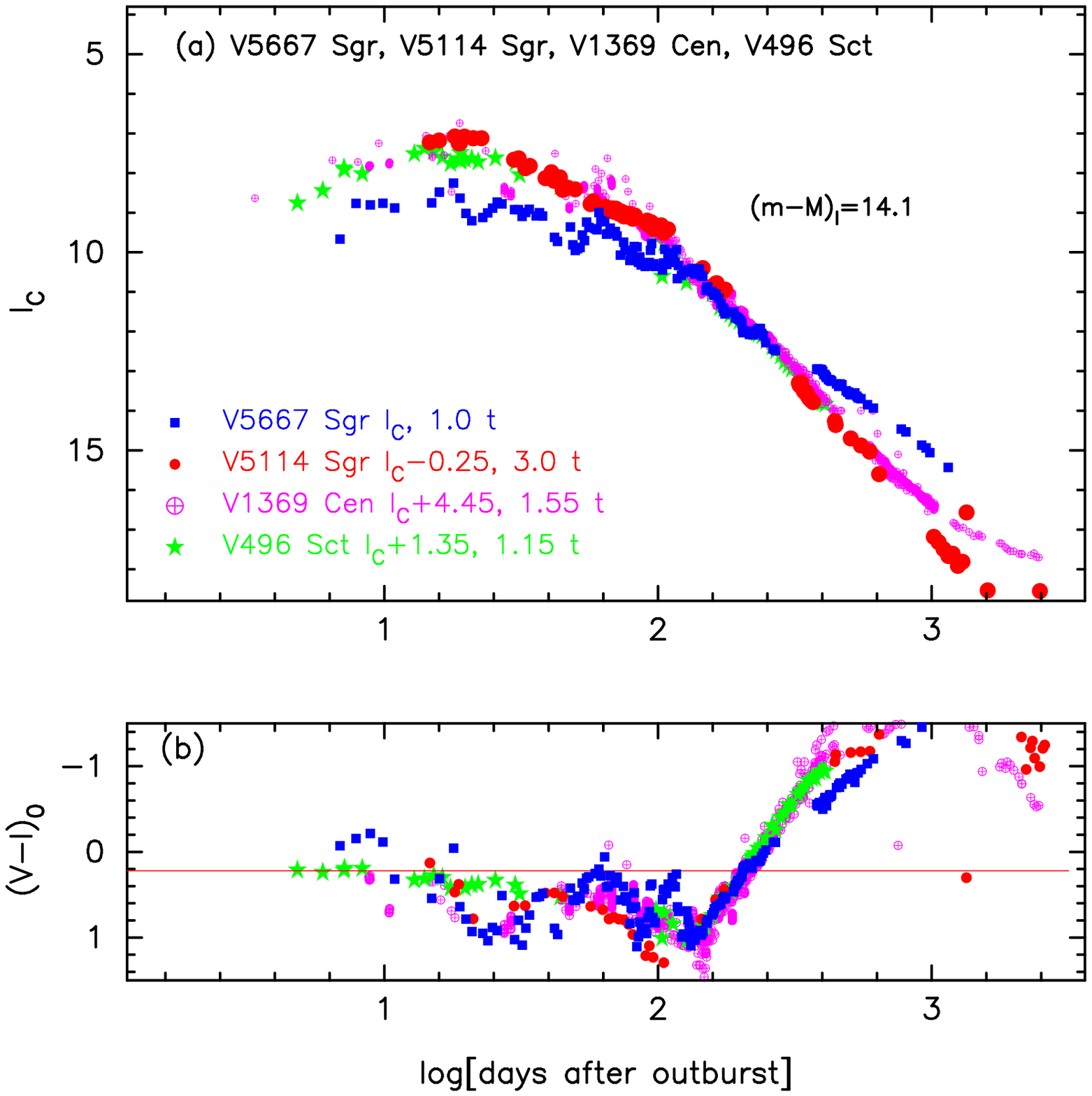}
\caption{
The (a) $I_{\rm C}$ light curve and (b) $(V-I_{\rm C})_0$ color curve
of V5667~Sgr as well as those of V5114~Sgr, V1369~Cen, and V496~Sct.
\label{v5667_sgr_v5114_sgr_v1369_cen_v496_sct_i_vi_color_logscale}}
\end{figure}


\begin{figure}
\plotone{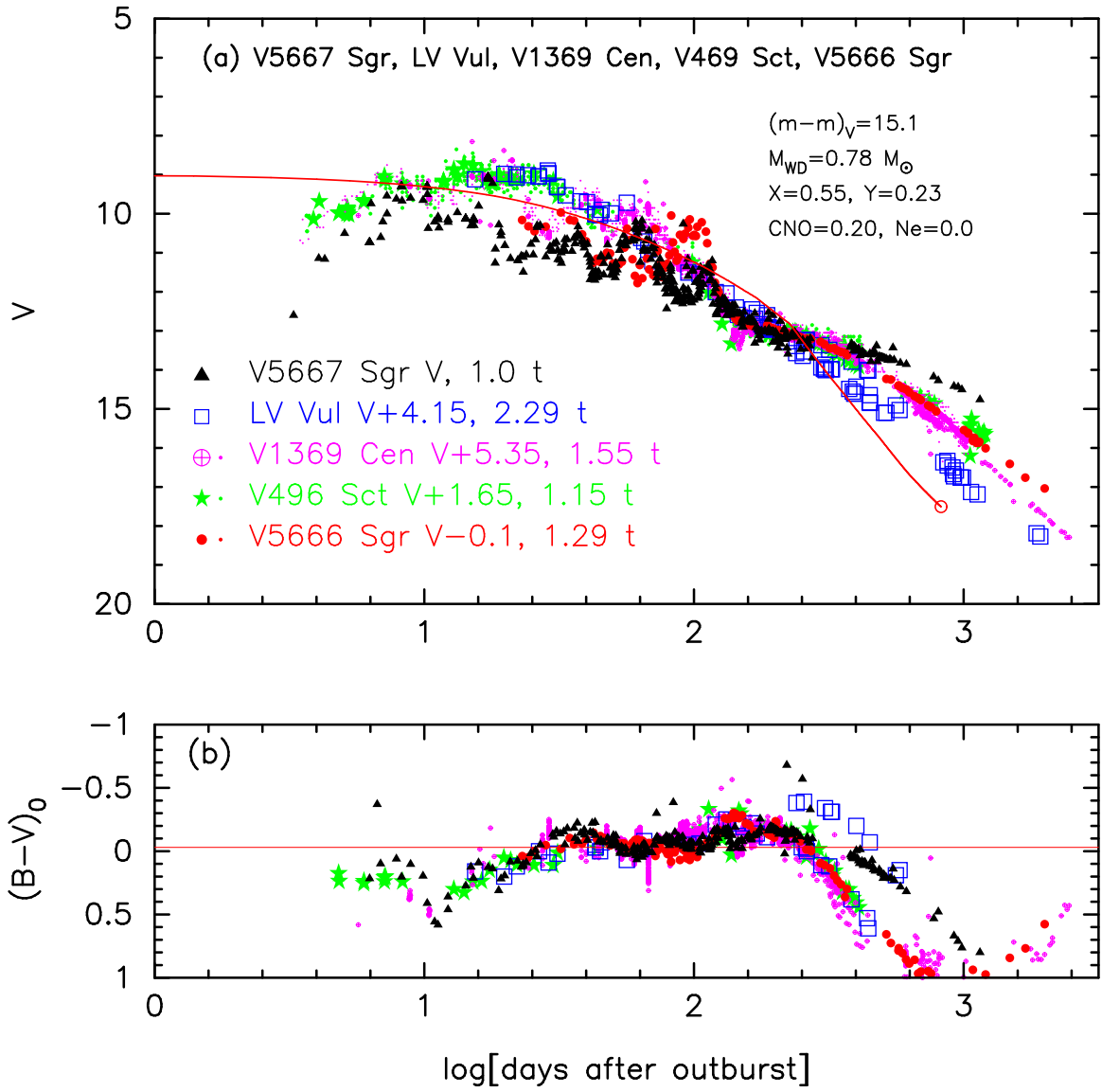}
\caption{
The (a) $V$ light curve and (b) $(B-V)_0$ color curve
of V5667~Sgr (filled black triangles) as well as those of LV~Vul,
V1369~Cen, V496~Sct, and V5666~Sgr.  In panel (a),
we add a $0.78~M_\sun$ WD model (CO4, solid red line) for V5667~Sgr.
\label{v5667_sgr_lv_vul_v5666_sgr_v1369_cen_v496_sct_v_bv_logscale_no2}}
\end{figure}


\begin{figure*}
\plottwo{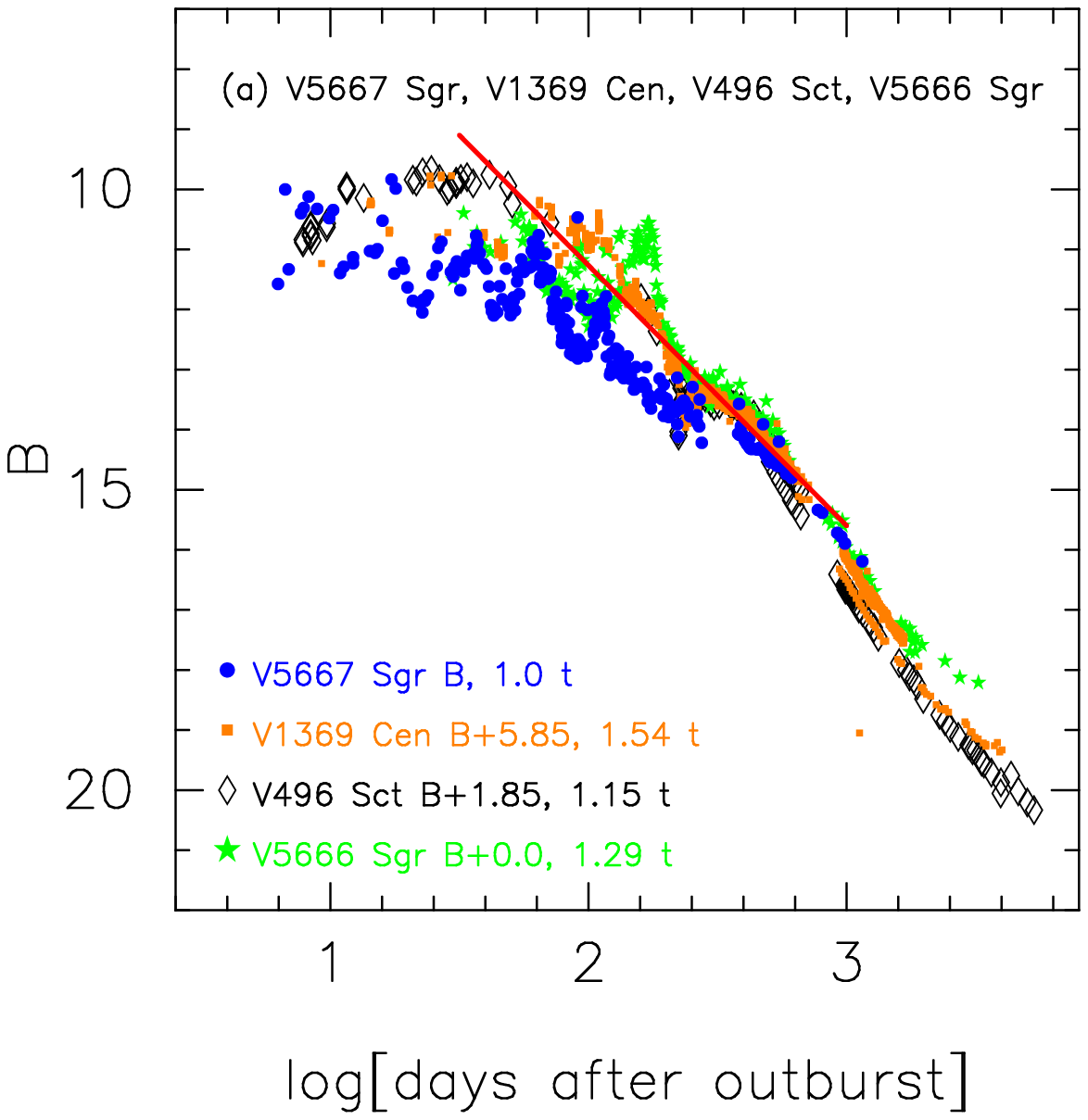}{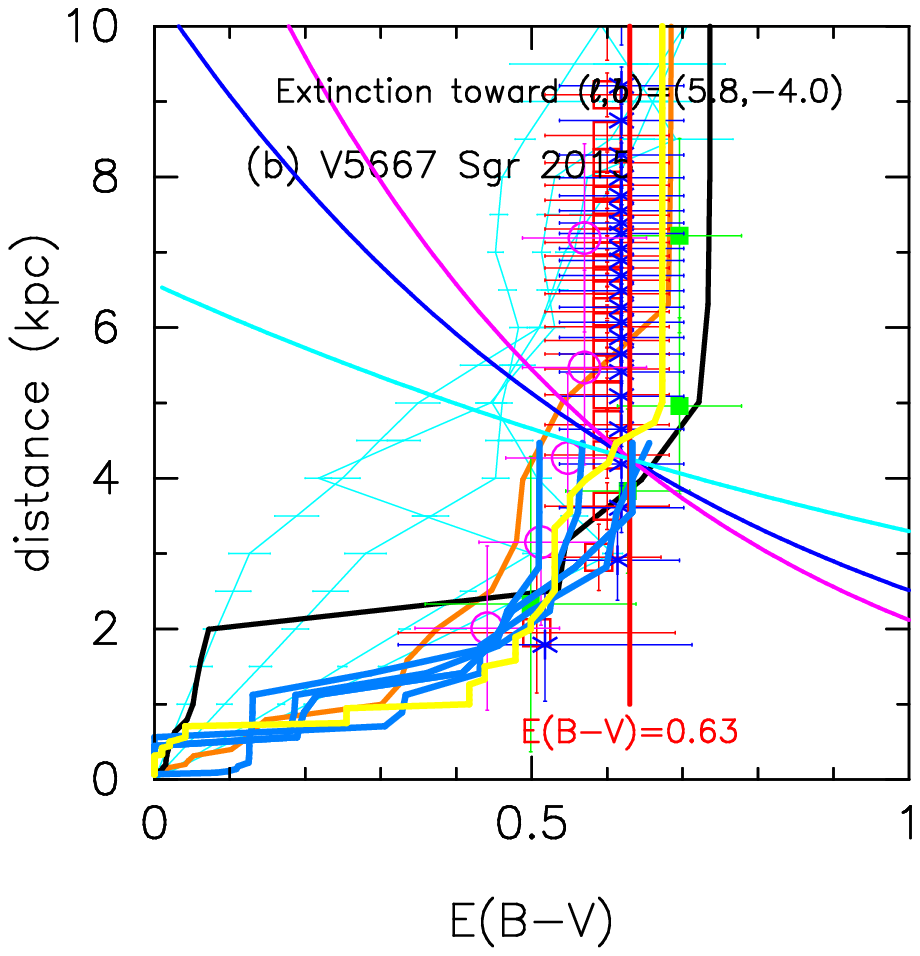}
\caption{
(a) The $B$ light curve of V5667~Sgr as well as those of 
V1369~Cen, V496~Sct, and V5666~Sgr.
(b) Various distance-reddening relations toward V5667~Sgr.
The thin solid lines of magenta, blue, and cyan denote the distance-reddening
relations given by  $(m-M)_B= 15.73$, $(m-M)_V= 15.1$, and $(m-M)_I= 14.09$,
respectively.  
\label{distance_reddening_v5667_sgr_bvi_xxxxxx}}
\end{figure*}

\subsection{V5667~Sgr 2015\#1}
\label{v5667_sgr_bvi}
We have reanalyzed the $BVI_{\rm C}$ multi-band 
light/color curves of V5667~Sgr based on the time-stretching method.  
Figure \ref{v5667_sgr_v5114_sgr_v1369_cen_v496_sct_i_vi_color_logscale}
shows the (a) $I_{\rm C}$ light and (b) $(V-I_{\rm C})_0$ color curves
of V5667~Sgr as well as V5114~Sgr, V1369~Cen, and V496~Sct.
The $BVI_{\rm C}$ data of V5667~Sgr are taken from SMARTS.
We adopt the color excess of $E(B-V)= 0.63$ after \citet{hac19kb}.
We apply Equation (8) of \citet{hac19ka} for the $I$ band to Figure
\ref{v5667_sgr_v5114_sgr_v1369_cen_v496_sct_i_vi_color_logscale}(a)
and obtain
\begin{eqnarray}
(m&-&M)_{I, \rm V5667~Sgr} \cr
&=& ((m - M)_I + \Delta I_{\rm C})
_{\rm V5114~Sgr} - 2.5 \log 3.0 \cr
&=& 15.55 - 0.25\pm0.2 - 1.2 = 14.1\pm0.2 \cr
&=& ((m - M)_I + \Delta I_{\rm C})
_{\rm V1369~Cen} - 2.5 \log 1.55 \cr
&=& 10.11 + 4.45\pm0.2 - 0.475 = 14.09\pm0.2 \cr
&=& ((m - M)_I + \Delta I_{\rm C})
_{\rm V496~Sct} - 2.5 \log 1.15 \cr
&=& 12.9 + 1.35\pm0.2 - 0.15 = 14.1\pm0.2,
\label{distance_modulus_i_vi_v5667_sgr}
\end{eqnarray}
where we adopt
$(m-M)_{I, \rm V5114~Sgr}=15.55$ from Appendix \ref{v5114_sgr_ubvi},
$(m-M)_{I, \rm V1369~Cen}=10.11$ from \citet{hac19ka}, and
$(m-M)_{I, \rm V496~Sct}=12.9$ in Appendix \ref{v496_sct_bvi}.
Thus, we obtain $(m-M)_{I, \rm V5667~Sgr}= 14.1\pm0.2$.

Figure
\ref{v5667_sgr_lv_vul_v5666_sgr_v1369_cen_v496_sct_v_bv_logscale_no2}
shows the (a) $V$ light and (b) $(B-V)_0$ color curves of V5667~Sgr
as well as those of LV~Vul, V1369~Cen, V496~Sct, and V5666~Sgr.
Applying Equation (4) of \citet{hac19ka} for the $V$ band to them,
we have the relation
\begin{eqnarray}
(m&-&M)_{V, \rm V5667~Sgr} \cr
&=& (m-M + \Delta V)_{V, \rm LV~Vul} - 2.5 \log 2.29 \cr
&=& 11.85 + 4.15\pm0.3 - 0.9 = 15.1\pm0.3 \cr
&=& (m-M + \Delta V)_{V, \rm V1369~Cen} - 2.5 \log 1.55 \cr
&=& 10.25 + 5.35\pm0.3 - 0.475  = 15.125\pm0.3 \cr
&=& (m-M + \Delta V)_{V, \rm V496~Sct} - 2.5 \log 1.15 \cr
&=& 13.6 + 1.65\pm0.3 - 0.15 = 15.1\pm0.3 \cr
&=& (m-M + \Delta V)_{V, \rm V5666~Sgr} - 2.5 \log 1.29 \cr
&=& 15.5 - 0.1\pm0.3 - 0.275 = 15.125\pm0.3,
\label{distance_modulus_v_bv_v5667_sgr}
\end{eqnarray}
where we adopt $(m-M)_{V, \rm LV~Vul}=11.85$ and
$(m-M)_{V, \rm V1369~Cen}=10.25$ both from \citet{hac19ka},
$(m-M)_{V, \rm V5666~Sgr}=15.5$ in Appendix \ref{v5666_sgr_bvi},
and $(m-M)_{V, \rm V496~Sct}=13.6$ in Appendix \ref{v496_sct_bvi}.
Thus, we obtained $\log f_{\rm s} = \log 2.29 = +0.36$ against LV~Vul and
$(m-M)_{V, \rm V5667~Sgr}=15.1\pm0.2$.

Figure \ref{distance_reddening_v5667_sgr_bvi_xxxxxx}(a)
shows the $B$ light curves of V5667~Sgr
together with those of V1369~Cen, V496~Sct, and V5666~Sgr.
Applying Equation (7) of \citet{hac19ka} for the $B$ band to Figure
\ref{distance_reddening_v5667_sgr_bvi_xxxxxx}(a),
we have the relation
\begin{eqnarray}
(m&-&M)_{B, \rm V5667~Sgr} \cr
&=& \left( (m-M)_B + \Delta B\right)_{\rm V1369~Cen} - 2.5 \log 1.54 \cr
&=& 10.36 + 5.85\pm0.3 - 0.475 = 15.73\pm0.3 \cr
&=& \left( (m-M)_B + \Delta B\right)_{\rm V496~Sct} - 2.5 \log 1.15 \cr
&=& 14.05 + 1.85\pm0.3 - 0.15 = 15.75\pm0.3 \cr
&=& \left( (m-M)_B + \Delta B\right)_{\rm V5666~Sgr} - 2.5 \log 1.29 \cr
&=& 16.0 + 0.0\pm0.3 - 0.275 = 15.73\pm0.3,
\label{distance_modulus_v5667_sgr_v1369_cen_v496_sct_v5666_sgr_b}
\end{eqnarray}
where we adopt $(m-M)_{B, \rm V1369~Cen}= 10.36$ from \citet{hac19ka},
$(m-M)_{B, \rm V496~Sct}= 14.05$ in Appendix \ref{v496_sct_bvi}, and
$(m-M)_{B, \rm V5666~Sgr}= 16.0$ in Appendix \ref{v5666_sgr_bvi}.
We have $(m-M)_{B, \rm V5667~Sgr}=15.73\pm0.2$.

We plot $(m-M)_B=15.73$, $(m-M)_V=15.1$, and $(m-M)_I=14.09$,
which cross at $d=4.3$~kpc and $E(B-V)=0.63$, in Figure
\ref{distance_reddening_v5667_sgr_bvi_xxxxxx}(b).
This crossing point is consistent with the distance-reddening relations
given by \citet[][filled green squares]{mar06},
\citet[][thick solid black and yellow lines]{gre15, gre19}, and
\citet[][thick solid cyan-blue lines]{chen19}.
Thus, we obtain $d=4.3\pm0.5$~kpc and $E(B-V)=0.63\pm0.05$.


\begin{figure}
\plotone{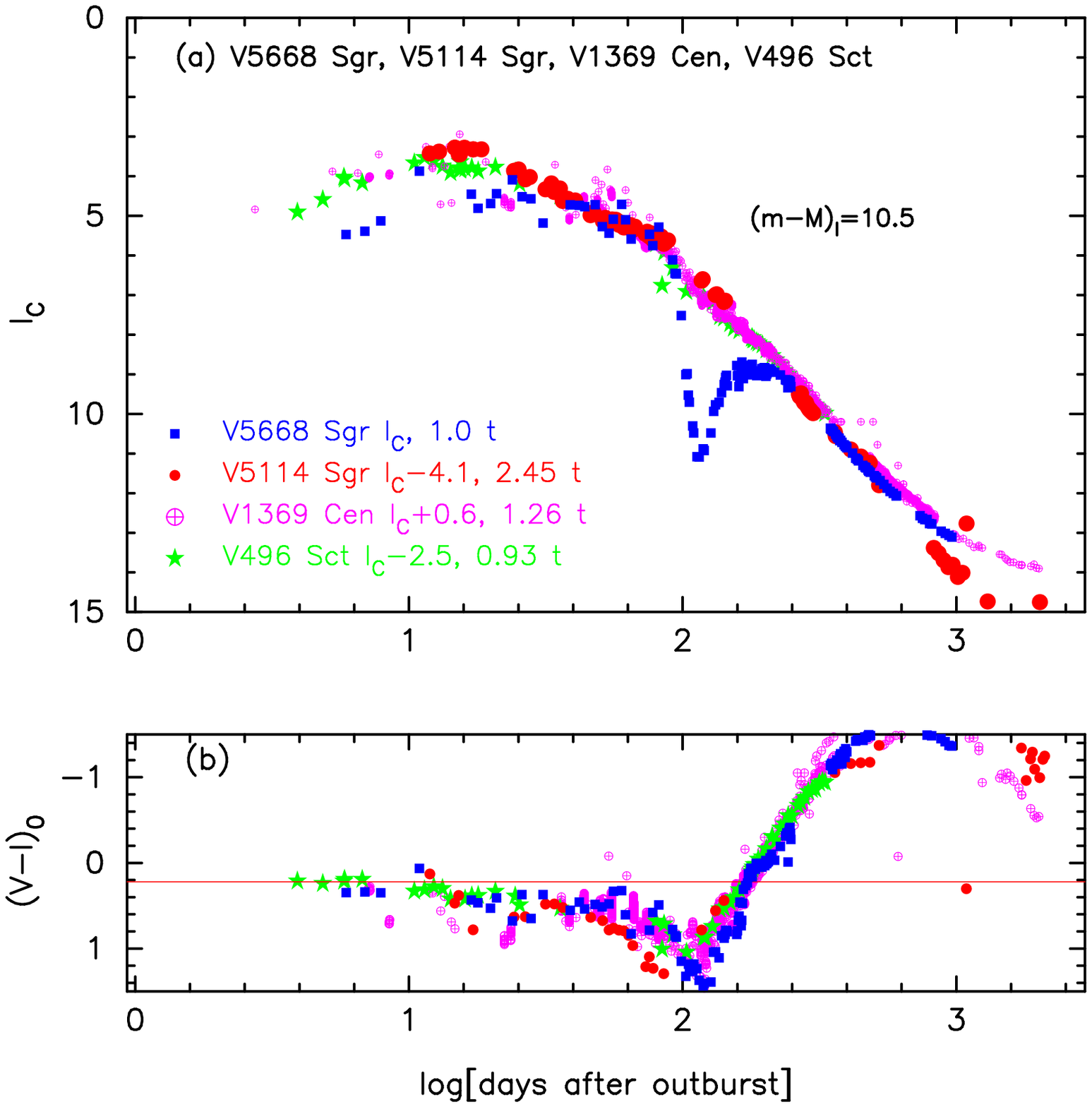}
\caption{
The (a) $I_{\rm C}$ light curve and (b) $(V-I_{\rm C})_0$ color curve
of V5668~Sgr as well as those of V5114~Sgr, V1369~Cen, and V496~Sct.
\label{v5668_sgr_v5114_sgr_v1369_cen_v496_sct_i_vi_color_logscale}}
\end{figure}


\begin{figure}
\plotone{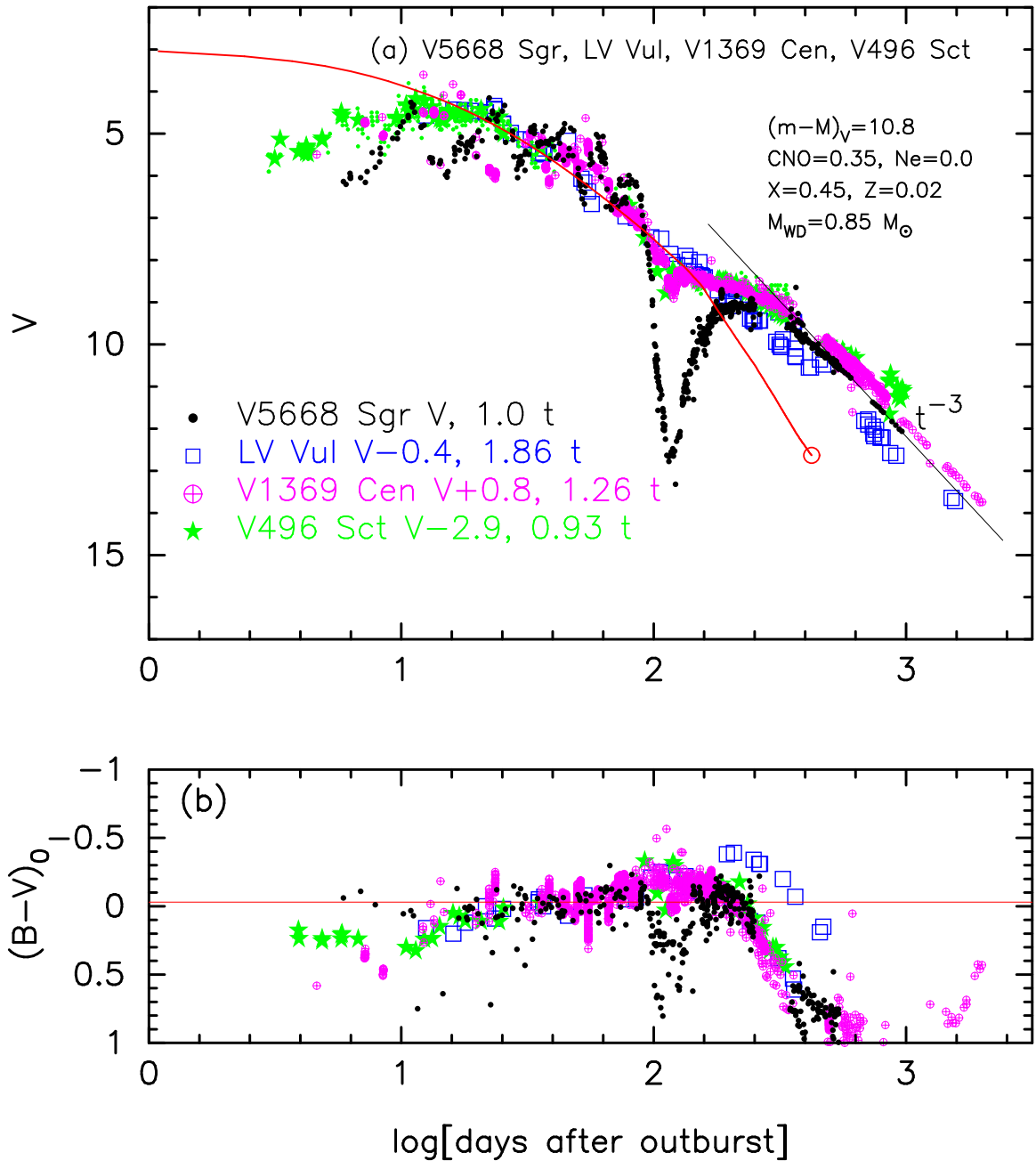}
\caption{
The (a) $V$ light curve and (b) $(B-V)_0$ color curve
of V5668~Sgr as well as those of LV~Vul, V1369~Cen, and V496~Sct.
The data of V5668~Sgr are taken from SMARTS.
In panel (a), we add a $0.85~M_\sun$ WD model (CO3, solid red line)
for V5668~Sgr.
\label{v5668_sgr_lv_vul_v496_sct_v1369_cen_v_bv_ub_color_logscale_no2}}
\end{figure}


\begin{figure*}
\plottwo{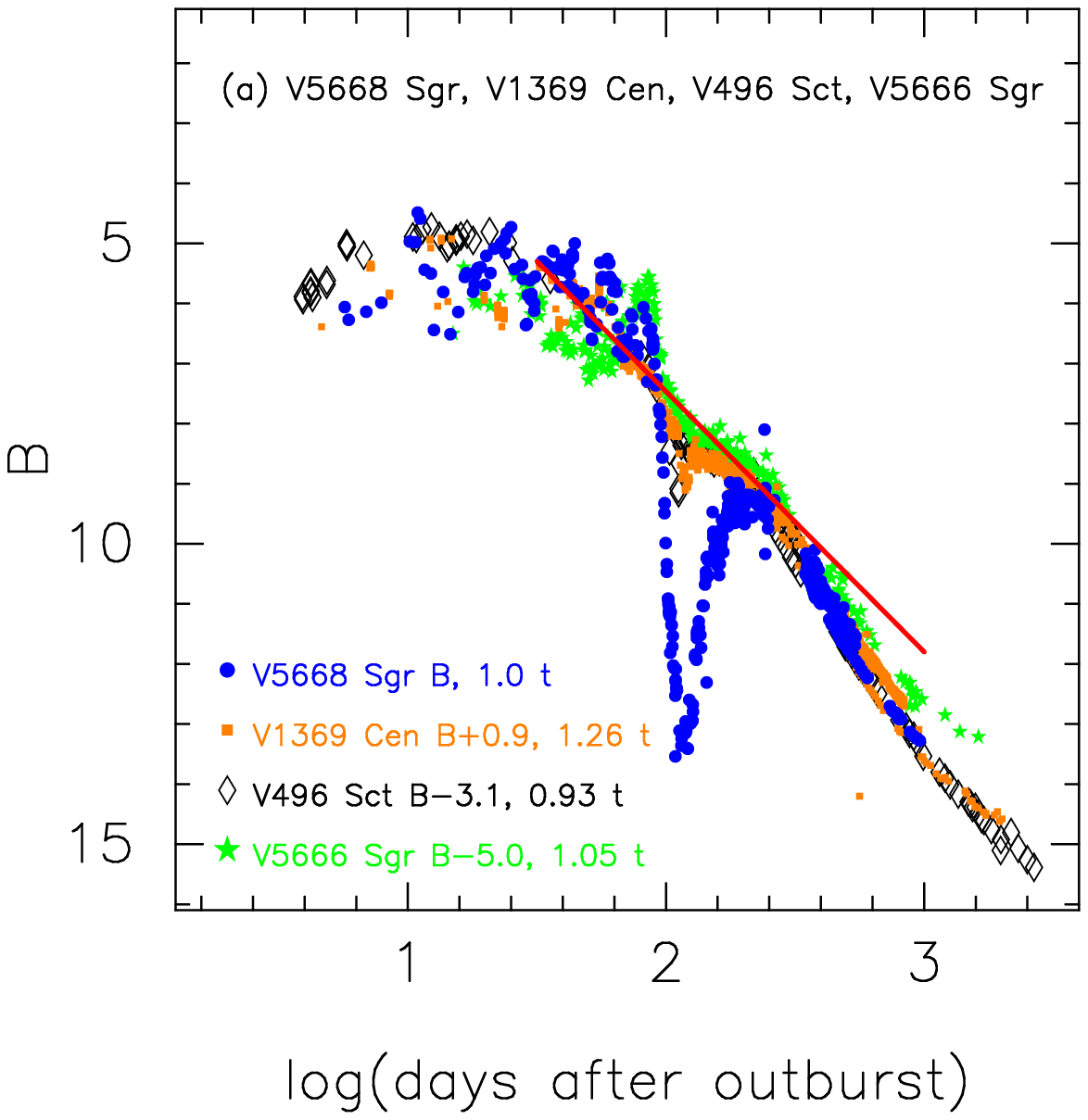}{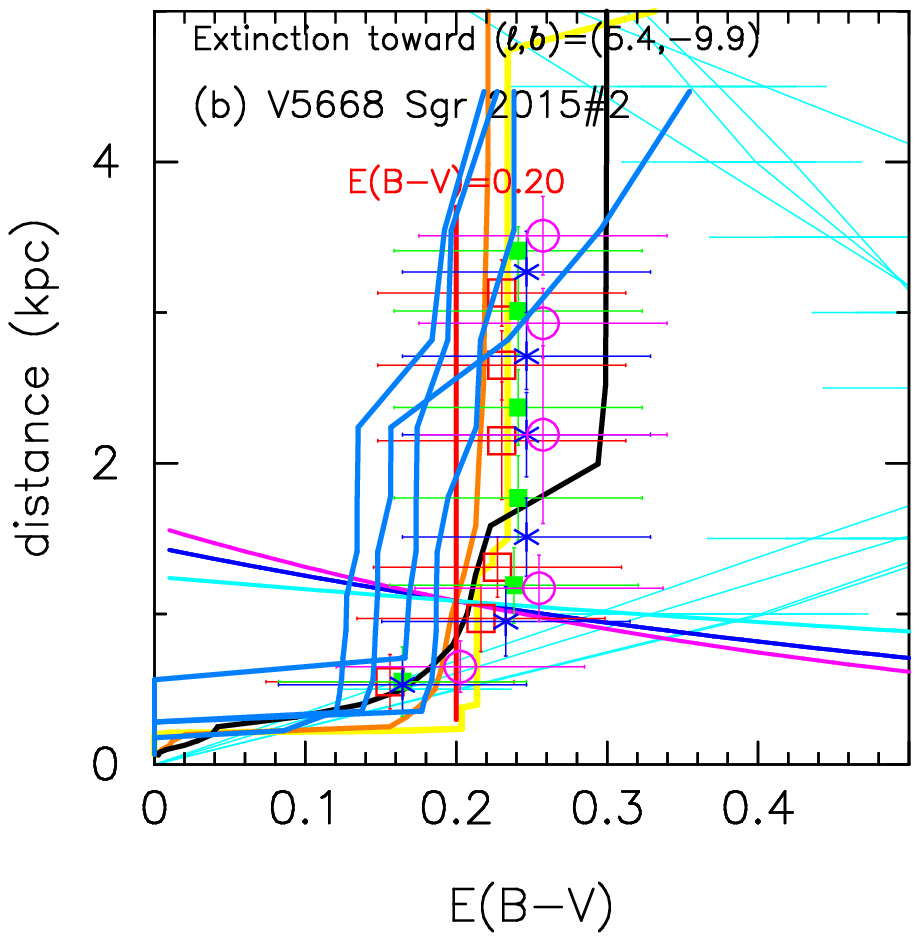}
\caption{
(a) The $B$ light curves of V5668~Sgr
as well as those of V1369~Cen, V496~Sct, and V5666~Sgr.
The $BV$ data of V5668~Sgr are taken from AAVSO, VSOLJ, and SMARTS.
(b) Various distance-reddening relations toward V5668~Sgr.
The thin solid lines of magenta, blue, and cyan denote the distance-reddening
relations given by  $(m-M)_B= 11.0$, $(m-M)_V= 10.8$, and $(m-M)_I= 10.48$,
respectively.  
\label{distance_reddening_v5668_sgr_bvi_xxxxxx}}
\end{figure*}

\subsection{V5668~Sgr 2015\#2}
\label{v5668_sgr_bvi}
We have reanalyzed the $BVI_{\rm C}$ multi-band 
light/color curves of V5668~Sgr based on the time-stretching method.  
Figure \ref{v5668_sgr_v5114_sgr_v1369_cen_v496_sct_i_vi_color_logscale}
shows the (a) $I_{\rm C}$ light and (b) $(V-I_{\rm C})_0$ color curves of
V5668~Sgr as well as V5114~Sgr, V1369~Cen, and V496~Sct.
The $BVI_{\rm C}$ data of V5668~Sgr are taken from SMARTS.
We adopt the color excess of $E(B-V)= 0.20$ after \citet{hac19kb}.
We apply Equation (8) of \citet{hac19ka} for the $I$ band to Figure
\ref{v5668_sgr_v5114_sgr_v1369_cen_v496_sct_i_vi_color_logscale}(a)
and obtain
\begin{eqnarray}
(m&-&M)_{I, \rm V5668~Sgr} \cr
&=& ((m - M)_I + \Delta I_{\rm C})
_{\rm V5114~Sgr} - 2.5 \log 2.45 \cr
&=& 15.55 - 4.1\pm0.2 - 0.975 = 10.48\pm0.2 \cr
&=& ((m - M)_I + \Delta I_{\rm C})
_{\rm V1369~Cen} - 2.5 \log 1.26 \cr
&=& 10.11 + 0.6\pm0.2 - 0.25 = 10.46\pm0.2 \cr
&=& ((m - M)_I + \Delta I_{\rm C})
_{\rm V496~Sct} - 2.5 \log 0.93 \cr
&=& 12.9 - 2.5\pm0.2 + 0.075 = 10.48\pm0.2,
\label{distance_modulus_i_vi_v5668_sgr}
\end{eqnarray}
where we adopt
$(m-M)_{I, \rm V5114~Sgr}=15.55$ from Appendix \ref{v5114_sgr_ubvi},
$(m-M)_{I, \rm V1369~Cen}=10.11$ from \citet{hac19ka}, and
$(m-M)_{I, \rm V496~Sct}=12.9$ in Appendix \ref{v496_sct_bvi}.
Thus, we obtain $(m-M)_{I, \rm V5668~Sgr}= 10.48\pm0.2$.

Figure \ref{v5668_sgr_lv_vul_v496_sct_v1369_cen_v_bv_ub_color_logscale_no2}
shows the (a) $V$ light and (b) $(B-V)_0$ color curves of V5668~Sgr
as well as those of LV~Vul, V1369~Cen, and V496~Sct.
Applying Equation (4) of \citet{hac19ka} for the $V$ band to them,
we have the relation
\begin{eqnarray}
(m&-&M)_{V, \rm V5668~Sgr} \cr
&=& ((m - M)_V + \Delta V)_{\rm LV~Vul} - 2.5 \log 1.86 \cr
&=& 11.85 - 0.4\pm0.2 - 0.68 = 10.77\pm0.2 \cr
&=& ((m - M)_V + \Delta V)_{\rm V1369~Cen} - 2.5 \log 1.26 \cr
&=& 10.25 + 0.8\pm0.2 - 0.25 = 10.8\pm0.2 \cr
&=& ((m - M)_V + \Delta V)_{\rm V496~Sct} - 2.5 \log 0.93 \cr
&=& 13.6 - 2.9\pm0.2 + 0.08 = 10.78\pm0.2, 
\label{distance_modulus_v5568_sgr_lv_vul}
\end{eqnarray}
where we adopt $(m-M)_{V, \rm LV~Vul}=11.85$ and 
$(m-M)_{V, \rm V1369~Cen}=10.25$ both from \citet{hac19ka},
and $(m-M)_{V, \rm V496~Sct}=13.6$ in Appendix \ref{v496_sct_bvi}.
Thus, we obtain $(m-M)_{V, \rm V5668~Sgr}=10.8\pm0.1$ 
and $\log f_{\rm s} = \log 1.86 = +0.27$ against LV~Vul.

Figure \ref{distance_reddening_v5668_sgr_bvi_xxxxxx}(a)
shows the $B$ light curves of V5668~Sgr
together with those of V1369~Cen, V496~Sct, and V5666~Sgr.
Applying Equation (7) of \citet{hac19ka} for the $B$ band to Figure
\ref{distance_reddening_v5668_sgr_bvi_xxxxxx}(a),
we have the relation
\begin{eqnarray}
(m&-&M)_{B, \rm V5668~Sgr} \cr
&=& \left( (m-M)_B + \Delta B\right)_{\rm V1369~Cen} - 2.5 \log 1.26 \cr
&=& 10.36 + 0.9\pm0.3 - 0.25 = 11.01\pm0.3 \cr
&=& \left( (m-M)_B + \Delta B\right)_{\rm V496~Sct} - 2.5 \log 0.93 \cr
&=& 14.05 - 3.1\pm0.3 + 0.08 = 11.03\pm0.3 \cr
&=& \left( (m-M)_B + \Delta B\right)_{\rm V5666~Sgr} - 2.5 \log 1.05 \cr
&=& 16.0 - 5.0\pm0.3 - 0.05 = 10.95\pm0.3,
\label{distance_modulus_v5668_sgr_v1369_cen_v496_sct_v5666_sgr_b}
\end{eqnarray}
where we adopt $(m-M)_{B, \rm V1369~Cen}= 10.36$ from \citet{hac19ka}, 
$(m-M)_{B, \rm V5666~Sgr}= 16.0$ in Appendix \ref{v5666_sgr_bvi},
and $(m-M)_{B, \rm V496~Sct}= 14.05$ in Appendix \ref{v496_sct_bvi}.
We have $(m-M)_{B, \rm V5668~Sgr}=11.0\pm0.2$.

We plot $(m-M)_B=11.0$, $(m-M)_V=10.8$, and $(m-M)_I=10.48$,
which cross at $d=1.1$~kpc and $E(B-V)=0.20$, in Figure
\ref{distance_reddening_v5668_sgr_bvi_xxxxxx}(b).
This crossing point is broadly consistent with the distance-reddening
relations given by \citet{mar06},
\citet[][thick solid black, orange, and yellow lines]{gre15, gre18, gre19},
and \citet[][thick solid cyan-blue lines]{chen19}.
Thus, we obtain $d=1.1\pm0.1$~kpc and $E(B-V)=0.20\pm0.03$.


\begin{figure}
\plotone{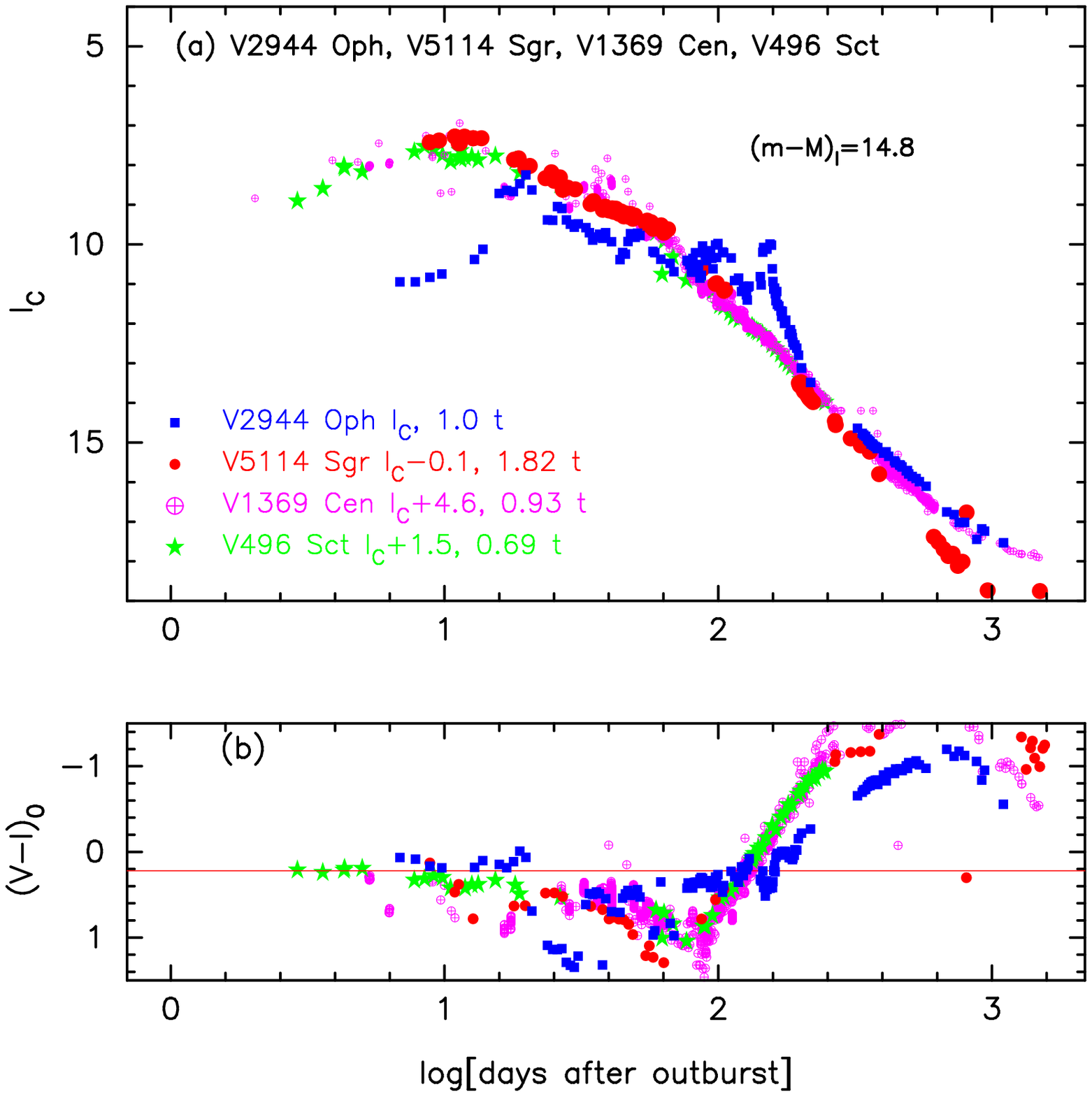}
\caption{
The (a) $I_{\rm C}$ light curve and (b) $(V-I_{\rm C})_0$ color curve
of V2944~Oph as well as those of V5114~Sgr, V1369~Cen, and V496~Sct.
\label{v2944_oph_v5114_sgr_v1369_cen_v496_sct_i_vi_color_logscale}}
\end{figure}


\begin{figure}
\plotone{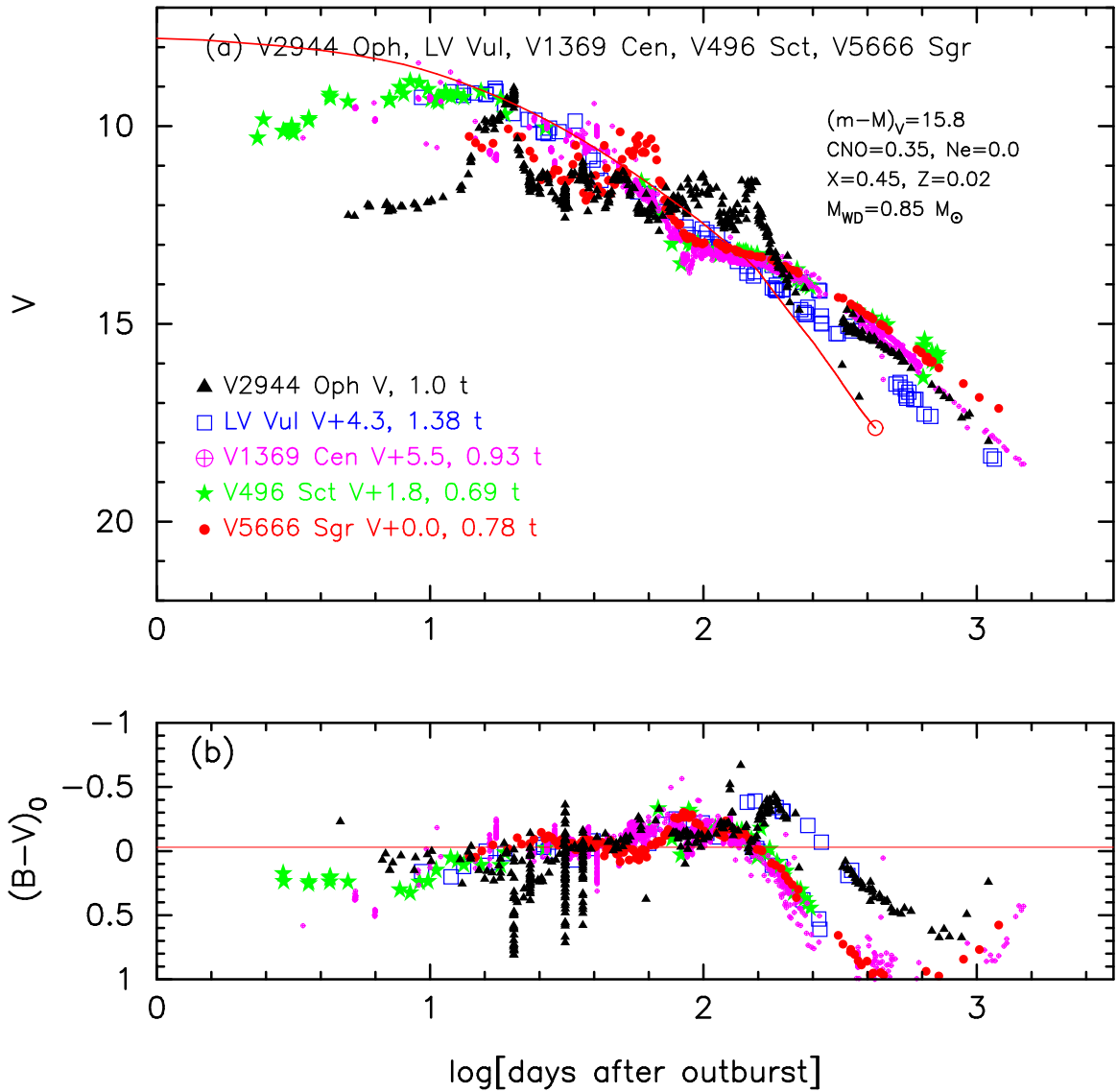}
\caption{
The (a) $V$ light curve and (b) $(B-V)_0$ color curve
of V2944~Oph as well as those of LV~Vul, V1369~Cen, V496~Sct, and V5666~Sgr.  
The data of V2944~Oph are taken from AAVSO, VSOLJ, and SMARTS.
In panel (a), we add a $0.85~M_\sun$ WD model (CO3, solid red line)
for V2944~Oph.
\label{v2944_oph_v5666_sgr_v1369_cen_v496_sct_v_bv_logscale_no2}}
\end{figure}


\begin{figure*}
\plottwo{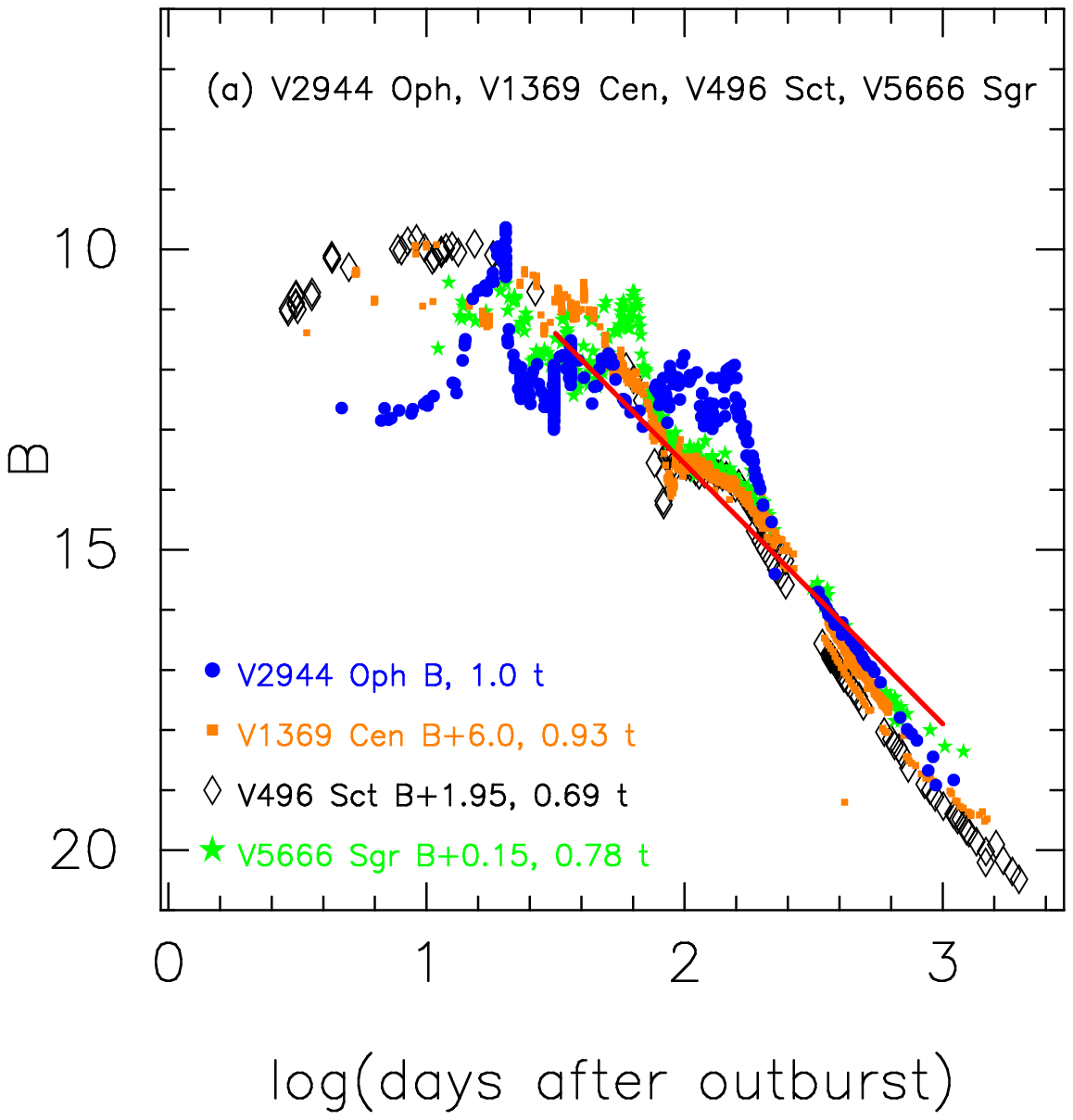}{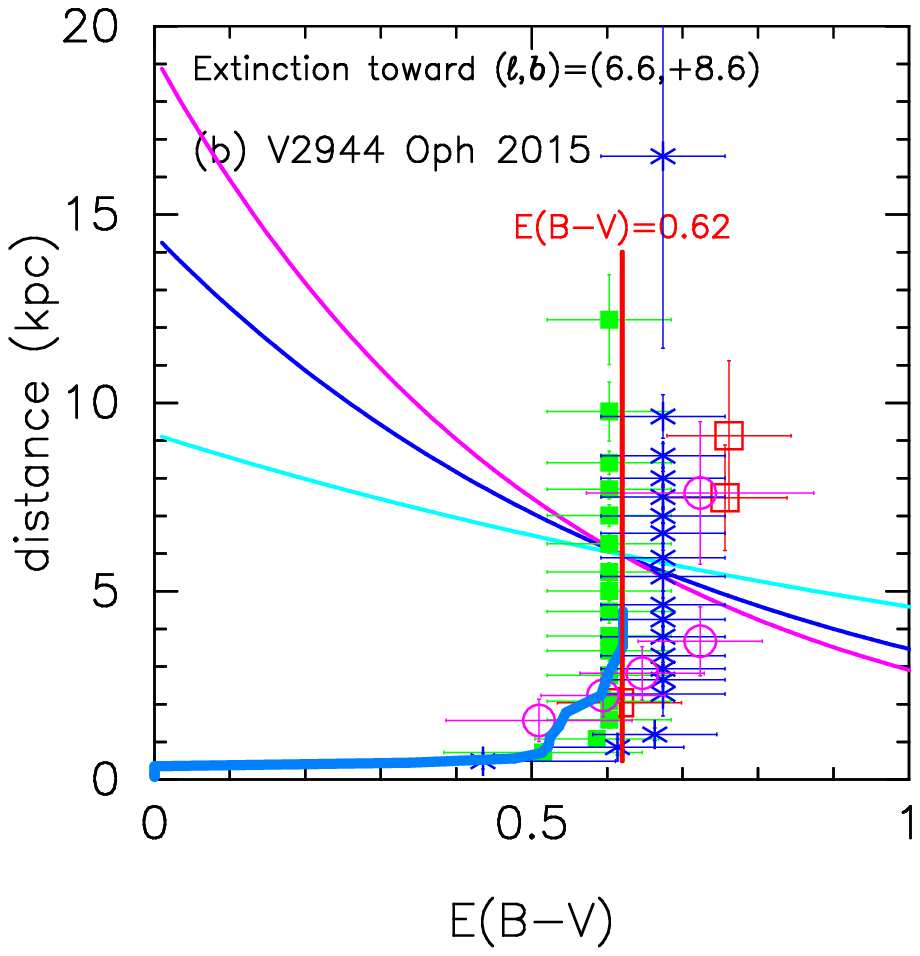}
\caption{
(a) The $B$ light curve of V2944~Oph
as well as those of V1369~Cen, V496~Sct, and V5666~Sgr.
The $BV$ data of V2944~Oph are taken from AAVSO, VSOLJ, and SMARTS.
(b) Various distance-reddening relations toward V2944~Oph.
The thin solid lines of magenta, blue, and cyan denote the distance-reddening
relations given by $(m-M)_B=16.42$, $(m-M)_V=15.8$, and $(m-M)_I=14.81$,
respectively.  
\label{distance_reddening_v2944_oph_bvi_xxxxxx}}
\end{figure*}

\subsection{V2944~Oph 2015}
\label{v2944_oph_bvi}
We have reanalyzed the $BVI_{\rm C}$ multi-band 
light/color curves of V2944~Oph based on the time-stretching method.  
Figure \ref{v2944_oph_v5114_sgr_v1369_cen_v496_sct_i_vi_color_logscale}
shows the (a) $I_{\rm C}$ light and (b) $(V-I_{\rm C})_0$ color curves
of V2944~Oph as well as V5114~Sgr, V1369~Cen, and V496~Sct.
The $BVI_{\rm C}$ data of V2944~Oph are taken from SMARTS.
We adopt the color excess of $E(B-V)= 0.62$ after \citet{hac19kb}.
We apply Equation (8) of \citet{hac19ka} for the $I$ band to Figure
\ref{v2944_oph_v5114_sgr_v1369_cen_v496_sct_i_vi_color_logscale}(a)
and obtain
\begin{eqnarray}
(m&-&M)_{I, \rm V2944~Oph} \cr
&=& ((m - M)_I + \Delta I_{\rm C})
_{\rm V5114~Sgr} - 2.5 \log 1.82 \cr
&=& 15.55 - 0.1\pm0.2 - 0.65 = 14.8\pm0.2 \cr
&=& ((m - M)_I + \Delta I_{\rm C})
_{\rm V1369~Cen} - 2.5 \log 0.93 \cr
&=& 10.11 + 4.6\pm0.2 + 0.075 = 14.79\pm0.2 \cr
&=& ((m - M)_I + \Delta I_{\rm C})
_{\rm V496~Sct} - 2.5 \log 0.69 \cr
&=& 12.9 + 1.5\pm0.2 + 0.4 = 14.8\pm0.2,
\label{distance_modulus_i_vi_v2944_oph}
\end{eqnarray}
where we adopt
$(m-M)_{I, \rm V5114~Sgr}=15.55$ from Appendix \ref{v5114_sgr_ubvi},
$(m-M)_{I, \rm V1369~Cen}=10.11$ from \citet{hac19ka}, and
$(m-M)_{I, \rm V496~Sct}=12.9$ in Appendix \ref{v496_sct_bvi}.
Thus, we obtain $(m-M)_{I, \rm V2944~Oph}= 14.8\pm0.2$.

Figure \ref{v2944_oph_v5666_sgr_v1369_cen_v496_sct_v_bv_logscale_no2}
shows the (a) $V$ light and (b) $(B-V)_0$ color curves
of V2944~Oph, LV~Vul, V1369~Cen, V496~Sct, and V5666~Sgr.
Applying Equation (4) of \citet{hac19ka} for the $V$ band to them,
we have the relation
\begin{eqnarray}
(m&-&M)_{V, \rm V2944~Oph} \cr
&=& (m-M + \Delta V)_{V, \rm LV~Vul} - 2.5 \log 1.38 \cr
&=& 11.85 + 4.3\pm0.3 - 0.35 = 15.8\pm0.3 \cr
&=& (m-M + \Delta V)_{V, \rm V1369~Cen} - 2.5 \log 0.93 \cr
&=& 10.25 + 5.5\pm0.3 + 0.075  = 15.82\pm0.3 \cr
&=& (m-M + \Delta V)_{V, \rm V496~Sct} - 2.5 \log 0.69 \cr
&=& 13.6 + 1.8\pm0.3 + 0.4 = 15.8\pm0.3 \cr
&=& (m-M + \Delta V)_{V, \rm V5666~Sgr} - 2.5 \log 0.78 \cr
&=& 15.5 + 0.0\pm0.3 + 0.275 = 15.78\pm0.3,
\label{distance_modulus_v_bv_v2944_oph}
\end{eqnarray}
where we adopt $(m-M)_{V, \rm LV~Vul}=11.85$,
$(m-M)_{V, \rm V1369~Cen}=10.25$ from \citet{hac19ka},
and $(m-M)_{V, \rm V5666~Sgr}=15.5$ in Appendix \ref{v5666_sgr_bvi},
and $(m-M)_{V, \rm V496~Sct}=13.6$ in Appendix \ref{v496_sct_bvi}.
Thus, we obtained $\log f_{\rm s}= \log 1.38 = +0.14$ against LV~Vul and
$(m-M)_{V, \rm V2944~Oph}=15.8\pm0.2$.

Figure \ref{distance_reddening_v2944_oph_bvi_xxxxxx}(a)
shows the $B$ light curves of V2944~Oph
together with those of V1369~Cen, V496~Sct, and V5666~Sgr.
Applying Equation (7) of \citet{hac19ka} for the $B$ band to Figure
\ref{distance_reddening_v2944_oph_bvi_xxxxxx}(a), we have the relation
\begin{eqnarray}
(m&-&M)_{B, \rm V2944~Oph} \cr
&=& \left( (m-M)_B + \Delta B\right)_{\rm V1369~Cen} - 2.5 \log 0.93 \cr
&=& 10.36 + 6.0\pm0.3 + 0.075 = 16.43\pm0.3 \cr
&=& \left( (m-M)_B + \Delta B\right)_{\rm V496~Sct} - 2.5 \log 0.69 \cr
&=& 14.05 + 1.95\pm0.3 + 0.4 = 16.4\pm0.3 \cr
&=& \left( (m-M)_B + \Delta B\right)_{\rm V5666~Sgr} - 2.5 \log 0.78 \cr
&=& 16.0 + 0.15\pm0.3 + 0.275 = 17.42\pm0.3,
\label{distance_modulus_v2944_oph_v1369_cen_v496_sct_v5666_sgr_b}
\end{eqnarray}
where we adopt $(m-M)_{B, \rm V1369~Cen}= 10.36$ from \citet{hac19ka},
and $(m-M)_{B, \rm V5666~Sgr}= 16.0$ in Appendix \ref{v5666_sgr_bvi}, and
$(m-M)_{B, \rm V496~Sct}= 14.05$ in Appendix \ref{v496_sct_bvi}.
We have $(m-M)_{B, \rm V2944~Oph}=16.42\pm0.2$.

We plot $(m-M)_B=16.42$, $(m-M)_V=15.8$, and $(m-M)_I=14.81$,
which cross at $d=6.0$~kpc and $E(B-V)=0.62$, in Figure
\ref{distance_reddening_v2944_oph_bvi_xxxxxx}(b).
This crossing point is consistent with the distance-reddening
relations given by \citet[][filled green squares]{mar06} 
and \citet[][thick solid cyan-blue line]{chen19}.
Thus, we obtain $d=6.0\pm1$~kpc and $E(B-V)=0.62\pm0.05$.

\clearpage







\end{document}